\documentclass[10pt, conference]{IEEEtran}

\usepackage{xspace}
\usepackage{enumitem}
\usepackage[most]{tcolorbox}
\usepackage{tabularx}
\usepackage{booktabs}
\usepackage{wrapfig}
\usepackage{graphicx}
\usepackage{subcaption}
\usepackage{array}
\usepackage{longtable}
\usepackage{placeins}
\usepackage{cite}
\usepackage{multibib}
\newcites{app}{Appendix References}
\usepackage{hyperref}
\usepackage{xurl}
\usepackage{listings}
\lstdefinestyle{prompt}{basicstyle=\footnotesize\ttfamily,breaklines=true,breakindent=0pt,%
  columns=fullflexible,frame=single,framesep=4pt,xleftmargin=3pt,xrightmargin=3pt,%
  aboveskip=4pt,belowskip=4pt,literate={’}{'}1 {‘}{`}1 {–}{-}1 {—}{---}1}

\AtBeginDocument{%
  }

\newcommand{\GH}{{\sc GitHub}\xspace}
\newcommand{\eg}[0]{e.g.,\xspace}

\newcommand{\etal}[0]{et al.\xspace}
\newcommand{\cf}[0]{cf.\xspace}

\newcommand{\mysec}[1]{\smallskip\noindent\textbf{\textit{#1.}}}

\definecolor{violetRay}{HTML}{8800FF}
\definecolor{hyperMagenta}{HTML}{C400EA}
\definecolor{neonRose}{HTML}{FF00D4}
\definecolor{neonPink}{HTML}{FF007B}
\definecolor{red}{HTML}{FF0000}
\definecolor{crimsonCarrot}{HTML}{FF4400}
\definecolor{darkOrange}{HTML}{FF8800}
\definecolor{brightGold}{HTML}{FFE100}
\definecolor{bogdan}{HTML}{6699CC}

\usepackage{soul}
\definecolor{quotehl}{HTML}{FF8800}
\sethlcolor{quotehl!10!white}
\newcommand{\qt}[2]{{\color{darkOrange!80!white}\textit{``#1''}~({#2})}}

\newcounter{prop}
\newcommand{\prop}[2][]{\refstepcounter{prop}\def\proptmp{#1}\ifx\proptmp\empty\else\label{#1}\fi\begin{propbox}\textbf{P\theprop.}~#2\end{propbox}}

\newtcolorbox{keyInsightsBox}{
  enhanced,
  sharp corners,
  colback=darkOrange!20!white,
  colframe=darkOrange!20!white,
  boxrule=0.5pt,
  left=3pt,
  right=3pt,
  top=3pt,
  bottom=3pt,
  before skip=3pt,
  after skip=3pt,
  fonttitle=\bfseries\small,
  coltitle=black,
  attach title to upper={},
  separator sign={\ },
}

\newtcolorbox{propbox}{
  enhanced,
  breakable,
  sharp corners,
  colback=darkOrange!20!white,
  colframe=darkOrange!20!white,
  boxrule=0.5pt,
  boxsep= 1 pt,
  left=1pt, right=1pt, top=1pt, bottom=1pt,
  before skip = 1 pt, after skip=1pt
}

\begin{document}
\pagestyle{plain}

\title{3100 Opinions on Code Review in an AI World: \\ Building Causal Theory from Practitioner Discourse}

\author{
  \IEEEauthorblockN{Shyam Agarwal, Courtney Miller, Christian K\"astner, and Bogdan Vasilescu}
  \IEEEauthorblockA{Carnegie Mellon University\\
  shyamaga@andrew.cmu.edu, courtneymiller@cmu.edu, vasilescu@cmu.edu}
}

\maketitle

\begin{abstract}
Coding agents now author entire pull requests, and practitioners sharply disagree about what this does to code review: whether it becomes the bottleneck, whether human review is still necessary, and whether it quietly erodes the understanding that it once built. Repository-mining studies measure surface trends but seldom explain the mechanisms beneath them, and the trends themselves prove unstable. A motivating observational analysis of public \GH{} activity finds that agent-authored pull requests are reviewed less often, merged several times faster, and discussed less than human-authored ones, yet the direction of these trends flips under different but equally defensible analysis choices, so the traces establish \emph{what} is changing without explaining \emph{why}. To recover the mechanisms, we synthesize practitioner discourse at scale into an \emph{explanatory theory}: we collect 38{,}709 grey-literature documents (engineering blogs and Reddit threads), filter to those substantively about code review, and code a stratified random sample of 3{,}100 with an LLM-assisted pipeline, from which we build a causal model of 26 constructs and 67 relationships (64 directed, 3 contested). Its organizing claim is that review is the control point through which a coding agent's effect on software is decided, and that AI does not fix the sign of that effect: the team sets it, through the expertise its humans bring and how it structures the review process. The theory makes the competing positions explicit and turns ``AI is changing code review'' into falsifiable propositions with named constructs and moderators. As a secondary contribution, we offer the underlying LLM-assisted, grey-literature theory-building method as a scalable template for software-engineering research, with a public implementation.
\end{abstract}

\section{Introduction}
\label{sec:introduction}

\textbf{Code review is under pressure from AI,} but there is little agreement on how exactly it is changing or what the consequences will be.
As coding agents can now author entire pull requests, practitioners have staked out sharply different positions on how this does and should affect review~\cite{gl:latent-killreview, gl:swarmia-humans, gl:graphite-humanreview}.
A common refrain is that agents now produce code faster than humans can review it, so that review becomes the delivery bottleneck~\cite{gl:levelup-fork, gl:devclass-scale}.
Some worry about the human cost, that the work shifts people away from an activity many enjoy, coding, toward one many do not, reviewing, with disengagement and burnout risks~\cite{gl:sciam-hours, gl:evilmartians-burnout, gl:atomic-fatigue}.
In open source, there is concern about large amounts of AI-generated low-quality contributions flooding projects, overwhelming maintainers and externalizing the cost of review onto them~\cite{gl:newstack-curl, gl:register-curl, gl:signadot-scale}.
Others argue the opposite, that human review may no longer be necessary at all and can be delegated to AI reviewers or folded into the agent's own loop as it checks its own work~\cite{gl:roby-agents, gl:adwaitx-bugbot, gl:aicorner-multiagent}.
Still others warn of subtler and possibly delayed harms, a creeping cognitive debt or cognitive surrender as developers stop building the understanding that review once demanded~\cite{gl:osmani-comprehension, gl:alexcloudstar-debt, gl:oleg-thinclient}.

These effects are not well understood, and they are difficult to study.
Repository-mining work has been measuring how AI is changing development activity~\cite{he2025speed,agarwal2026ai, liu2026debt}, including code review, but the results so far are fragmentary and at times conflicting, reporting opposite signs for basic quantities such as how large agent-authored changes are and whether they are merged more or less readily than human ones~\cite{li2025rise,popescu2026investigating,Watanabeetal2025,pham2026code,branco2026lgtm,yu2026habituation}.
In our own analysis of code review of AI-authored pull requests on GitHub (Section~\ref{sec:motivating_study}), we find trends, but they are not always stable under different but equally defensible operationalizations, and they do not explain what is happening underneath the surface or adjudicate between the competing causal stories that could produce the same numbers.

Understanding what is actually happening requires causal mechanisms, not just surface trends: How raising one factor changes another, but mediated by a third and moderated by a fourth, and how some of these relationships may feed back on themselves to form virtuous or vicious cycles.
Studying mechanisms at that level first requires a theory of them~\cite{pearl-book-of-why}, an account of the forces that shape code review in the presence of AI, the variables those forces act on, and how the variables relate.
\textit{Building such a theory is the goal of this paper.}

We start from what practitioners have already written publicly about code review in an AI world, whether at length in blog posts and long-form articles or more informally in forums such as Reddit (usually referred to as \emph{grey literature}~\cite{garousi2019grey}).
We analyze these practitioner accounts to construct a theory of how coding agents reshape code review, centered on factors such as review load, how thoroughly review is performed (its efficiency, depth, and effectiveness), code quality, comprehension debt, and reviewer skill, and on the causal pathways and feedback loops that connect them.
The resulting theory (Figure~\ref{fig:rq2_model}) can inform hypotheses that future studies can test, and, more importantly, a framework for designing those studies: A shared vocabulary for defining variables, study designs (for example, causal inference using mediators and colliders), and research questions.
We envision that this framework will let researchers untangle the competing underlying effects that are not visible from a surface-level analysis of trends.

We also intend a secondary contribution in \emph{how} we build the theory, using \emph{grey-literature analysis at an unprecedented scale} with LLMs to help handle and code large amounts of data.
Rather than conducting a small number of interviews, we collected 38{,}709 public documents discussing code review (7{,}630 web articles and 31{,}079 Reddit threads).
After filtering to those substantively about code review, we coded a stratified random sample of 3{,}100 documents (1{,}263 web articles and 1{,}837 Reddit threads, about 12{,}000 pages of text; about 500 words/page) using LLMs, inspired by recent work on LLM-driven thematic analysis~\cite{qiao2025thematiclm}.
We then constructed the theory from the coded data in a largely manual but LLM-assisted process, using LLMs to organize codes in different ways and to search among codes and their corresponding quotes.
We believe this could be a model for future repository-mining and theory-building research, making theory construction more accessible and scaling it from a few interviews or a small literature review to the wide range of practitioner perspectives now available at scale.

\smallskip
In summary, this paper makes the following contributions:
\begin{itemize}[leftmargin=*]
  \item (primary) an explanatory theory of how coding agents reshape code review, with 26 constructs and 67 relationships, 
  that makes competing mechanisms explicit and turns them into falsifiable hypotheses;
  \item (secondary) a scalable, LLM-assisted methodology for grey-literature theory building that we offer as a template for future work, with a publicly shared, scalable implementation; 
  \item (tertiary) a motivating observational study of how agent-authored pull requests are reviewed in open source today, whose trends corroborate prior findings yet shift direction under defensible analysis choices (Section~\ref{sec:motivating_study}).
\end{itemize}

\section{Trends in Code Review after the AI adoption}
\label{sec:relatedwork}

\subsection{Code Review in the Pre-AI Era}

Code review is a key practice, with a long history~\cite{fagan76}; the lightweight, tool-mediated ``modern code review'' version dominates today across open source and industry~\cite{bacchelli13,gousios16,rigby16}.
Review serves multiple purposes beyond defect detection: knowledge transfer, onboarding, team awareness, and governance~\cite{bacchelli13,Sadowskietal2017,Davilaetal2021,Pascarellaetal2018}, and mining studies link review participation to downstream quality~\cite{mcintosh14}.
A recurring theme in this literature is that review is attention-limited and cognitively demanding~\cite{Unterkalmsteineretal2024,Goncalvesetal2025}, and that the structural properties of a change (its size, dispersion, and presentation) shape reviewer load and defect detection~\cite{Rametal2018,Fregnanetal2022}; metrics such as change entropy and churn concentration characterize how edits distribute across files~\cite{hassan09,nagappan05}, a tradition our own motivating study (Section~\ref{sec:motivating_study}) extends to agent-associated changes).
Pull-request threads have also long mixed human discussion with machine activity from CI services, dependency updaters, and suggestion bots~\cite{Shihabetal2022,Wesseletal2023,Palvannanetal2023,Rahmanetal2024}, and this bot activity can inflate apparent review effort if machine comments are not separated from human scrutiny~\cite{Wesseletal2020,Wesseletal_12021,Golzadehetal2020,Olewickietal2024}, a measurement hazard that only intensifies once the \emph{author} is also a machine.

\subsection{A Motivating Study: Review Trends in Agent-Authored PRs}
\label{sec:motivating_study}
Our theory work grew out of our struggle to interpret observational data. We ran a longitudinal study of how agent-authored pull requests are changing code review. The trends were clear enough and largely matched prior work, but we could not say what they meant: the same measurements support opposite conclusions about human oversight depending on defensible analysis choices, and they conflict with other recent studies. This pushed us to build a theory of the underlying mechanisms that future, better studies can draw from.

We analyzed the public \emph{Agents in the Wild} corpus~\cite{logicstar-agentsinthewild}, re-scraping each repository's full PR history for a longitudinal view (Appendix~\ref{app:rq1} gives sampling, author labeling, operationalization, and other details).
Consistent with recent snapshot studies (deferred to Section~\ref{sec:related-after-AI-code}), agent-authored PRs are merged several times faster than human PRs~\cite{li2025rise,popescu2026investigating}, attract less review discussion in absolute terms and per line changed~\cite{li2025rise,Watanabeetal2025,DBLP:journals/corr/abs-2605-02273}, and are reviewed less independently, with 40.1\% examined only by the developer who invoked the agent versus 21.5\% of human PRs~\cite{DBLP:journals/corr/abs-2605-02273,peralta2026agentic}.

\begin{figure}[t]
  \centering
  \includegraphics[width=0.7\linewidth]{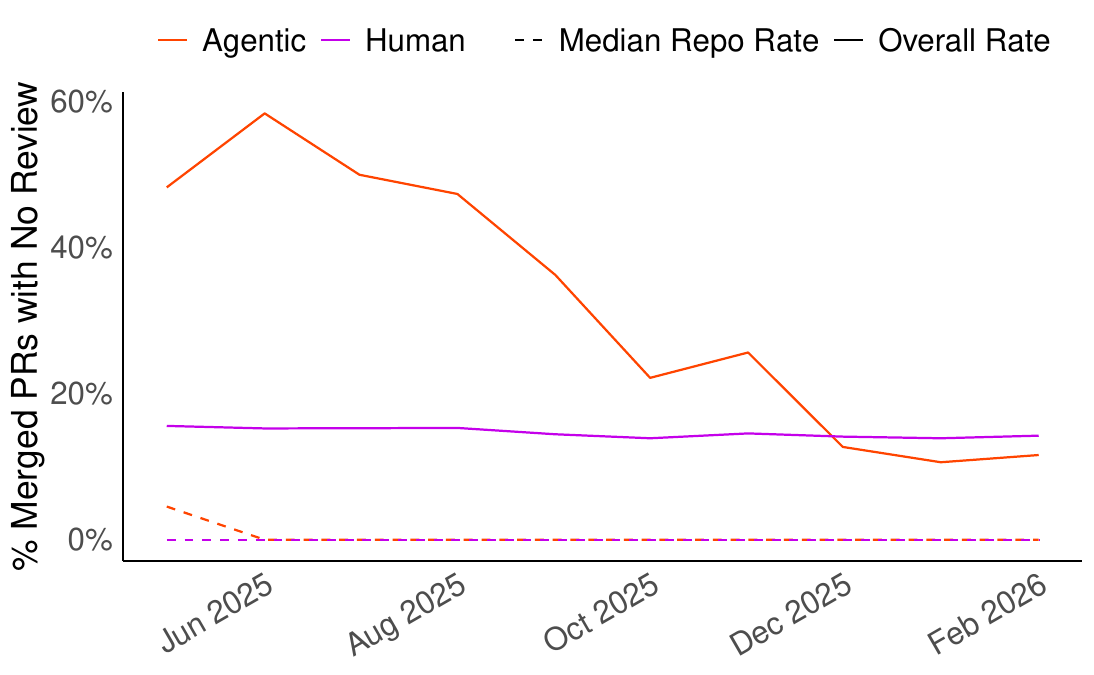}
  \caption{Proportion of merged PRs with no human review over time, in our sample (solid line: overall rate; dashed line: median of per-project rate). The early agentic no-review rate in mid-2025 falls toward the stable human rate by early 2026.}
  \label{fig:motivNoReview}
\end{figure}

Re-scraping full histories also surfaced over-time trends not previously reported.
Most strikingly, the share of merged agent-authored PRs receiving no human review falls from over 50\% in mid-2025 toward the steady human baseline of about 14\% by early 2026 (Figure~\ref{fig:motivNoReview}): an initial willingness to merge agent output unchecked gives way to reviewing it much as human PRs are reviewed.
This runs opposite to the within-reviewer habituation reported by Yu \etal{}~\cite{yu2026habituation}, where oversight weakens over time as approval rises and commenting falls.
On the basic question of whether human oversight is consolidating or eroding, studies disagree.

The instability is also within our own data: a finding's direction can reverse under defensible analysis choices.
Whether the developer who invoked an agent counts as an independent reviewer or as the author reviewing their own work decides whether independent review of agent PRs sits above or below the human rate.
Such studies establish what is changing, not why, and cannot settle which reading is correct.
This motivates our theory: a model that locates the direction of an effect in specific practices and context factors rather than the technology, and that guides what to study beyond surface observations about drivers and outcomes.

\subsection{Related Work on Theory Building in Software Engineering}

Causal inference should begin with a theory. As Pearl puts it, ``data are profoundly dumb''~\cite{pearl-book-of-why}: data can show that two quantities move together, but not why, and no volume of data answers a causal question without a model of the mechanism that produced it. That model must come from outside the data: prior work, domain knowledge, or expert elicitation. Constructing such a model for AI-era code review is our task.

What counts as a theory, and how one is built and judged, is itself a subject of methodological guidance. Gregor distinguishes five types; ours is a \emph{theory for explaining} (her Type II), accounting for how and why rather than predicting what will happen~\cite{gregor06nature}, though we cast its claims as falsifiable propositions to seed future studies. Sj{\o}berg \etal{} and Stol and Fitzgerald set out how theories are constructed and what makes them useful~\cite{sjoberg08theories, stol15theory}; grounded-theory methodology is standard for  bottom-up theory building~\cite{charmaz06, corbin14, stol16grounded}. We follow this guidance, presenting our result as constructs and their relationships and judging it by credibility, resonance, and usefulness rather than statistical generalization.

While theory building is well established in our field, the resulting theories are typically distilled from a small, deeply analyzed evidence base: a few dozen interviews or observations, or a handful of teams (for example, a theory of architectural technical debt from 18 interviews~\cite{architectualtechdebt}, or of debugging from seven observed developers~\cite{coblenz}), and the few that draw on the literature screen at most a few hundred papers~\cite{2106.01885v1}. None engages practitioner discourse at anything close to our scale.
We thus make a secondary methodological contribution: showing how to scale theory work to thousands of documents, using LLMs for some mechanical steps.

\section{Building a Theory of Post-AI Code Review}
\label{sec:rq2}

\subsection{Methods}
\label{sec:rq2_methods}

We build an explanatory theory of how coding agents reshape code review from \emph{grey literature}: public, practitioner-authored writing about review in the age of AI. We follow the coding stages of grounded theory in the Straussian tradition~\cite{corbin14,stol16grounded}, with a deliberate division of labor: an LLM pipeline does the mechanical work that scale makes infeasible by hand (constructing a corpus, filtering for relevance, and \emph{open-coding} it into a shared codebook), while the authors perform the theory construction, the \emph{axial} and \emph{selective} coding that decides the constructs and their relationships. We code only practitioner documents, not academic (Sec.~\ref{sec:related-after-AI-code}); coding both into one codebook would conflate what practitioners believe with what researchers have measured.%

\mysec{Corpus construction}
We collect from two registers, tagged with the same AI and review lexicon. The first is \textbf{Reddit developer discussion} from the public \emph{arctic\_shift} archive~\cite{arcticshift} (so the frame is fixed and reproducible). Under a stratified purposive design we scan the full 2020--2026 history of 43 subreddits in seven strata: code review (r/codereview, kept in full); version control (r/git, r/github, r/gitlab); general software engineering, the largest (r/ExperiencedDevs, r/programming, r/cscareerquestions); a critical/satirical community; nine language ecosystems (\eg r/Python, r/rust); AI coding tools (r/ChatGPTCoding, r/ClaudeAI, r/cursor); and broader AI forums (r/OpenAI, r/LocalLLaMA). A thread is kept if its title or body contains \emph{review}, \emph{reviews}, \emph{reviewing}, \emph{reviewer}, \emph{reviewers}, or \emph{reviewed} at a word boundary, with no popularity filter (which would suppress the minority views thematic analysis needs) and only minimal discussion required; each becomes one Markdown document with its comment tree and provenance (\textbf{31{,}079 threads}). The second is \textbf{edited long-form writing} (engineering blogs, vendor and practitioner write-ups, tech-press analysis), retrieved through the Exa neural search engine~\cite{exa} by semantic similarity; to avoid tilting the corpus we issue 68 natural-language queries, each pairing a directional query with its opposite pole (\eg ``How AI coding assistants are changing the code review process'' vs.\ ``Why code review has not fundamentally changed despite AI''), over three passes: the open web, the entire Feedspot ranking of software-engineering blogs, and a category-balanced set of institutional domains (engineering-practice media, agent vendors, AI review-tool vendors, developer-productivity research groups). Reversible filters then apply (URL deduplication, removal of very short and long pages, and a block list for document \emph{types} that are never practitioner discourse: academic databases, PDF and book dumps, registries, content farms), yielding \textbf{7{,}630 articles} across 1{,}437 domains. Together this gives one corpus of \textbf{38{,}709 documents} spanning 2020--2026; Appendix~\ref{app:rq2corpus} gives the full sampling frame, vocabularies, three-pass design, and corpus-composition figures.

\mysec{Relevance filtering}
The gates are recall-oriented and admit false positives (a performance or product review, ``review'' meaning \emph{revise}), so we tighten precision in two steps. First, a Reddit keyword gate: outside r/codereview, a thread is kept only if it contains \emph{code} with a review word; web articles, already semantically retrieved, are kept in full. Second, an LLM judge (Google's \emph{Gemini 2.5 Flash}, temperature 0) applies a single neutral, versioned, reasoning-first rubric to the \textbf{23{,}631} candidates (16{,}001 Reddit threads, 7{,}630 web articles), retaining \textbf{13{,}469} (57\%); a stronger model re-judging a random sample agrees at Cohen's $\kappa = 0.75$ (substantial; Appendix~\ref{app:rq2judge}). Because agent-related discussion is overwhelmingly recent (the AI-mentioning share of threads rises from under 1\% before 2023 to roughly two-thirds in 2026), we restrict coding to \textbf{2025--2026}, when agent-authored pull requests became a live concern, leaving about \textbf{9{,}100} documents.

\mysec{Coding to a shared codebook}
From this pool we draw a \textbf{source-stratified random sample of 3{,}100 documents} and code them with \emph{Thematic-LM}~\cite{qiao2025thematiclm}, a previously-evaluated multi-agent LLM realization of inductive thematic analysis~\cite{braun06} whose \emph{adaptive} codebook is built and revised as data arrive rather than fixed in advance. This is the \emph{open coding} of grounded theory~\cite{corbin14,stol16grounded}, the part we delegate to the LLMs because it is the most labor-intensive and because each code stays anchored to a verbatim quote a human can check. The coding unit differs by register: each Reddit thread is kept whole to preserve conversational context, while each web article is split into topical segments by an LLM segmenter. Three \emph{coder} agents independently attach one to three short codes, each with a verbatim quote, to every segment; rather than treat coder identity as generic perspective diversity, we instantiate three lenses for the polarized debate: a \emph{neutral} inductive lens, a \emph{critical} lens attentive to how agents and automated review may erode review, and an \emph{appreciative} lens attentive to how they may strengthen it, so optimistic and pessimistic readings enter the codebook by design rather than by the model's default. An \emph{aggregator} then merges near-duplicate codes, and a \emph{reviewer} maintains a single versioned codebook, using sentence-embedding similarity to decide whether each incoming code extends or merges into an existing one. Coding runs in batches of 100 documents (31 in total), each finalizing a codebook revision, and is resumable and tolerant of per-segment failures; all calls use \emph{Gemini 2.5 Flash}. The resulting codebook has \textbf{4{,}838 codes} grounded in \textbf{109{,}951 quotes} from \textbf{2{,}669} sampled documents (1{,}454 Reddit, 1{,}215 web). We release a scalable, open-source reimplementation of this pipeline in our replication package; the research context, the three coder identities, and every prompt are in Appendix~\ref{app:rq2coding}.

\mysec{From codebook to theory}
This is where we draw the line between machine and researcher. Thematic-LM would develop themes in a second automated pass over the codebook, but that presumes a codebook small enough to reason about at once, and with nearly five thousand codes the assumption does not hold. More fundamentally, the step from codes to a causal theory (the \emph{axial} coding that relates categories to one another and the \emph{selective} coding that organizes them around a core~\cite{corbin14,stol16grounded}) is the interpretive heart of the study, and we did not delegate it. The authors performed it by hand over many iterations, using the model only as an instrument for organizing and retrieving evidence at scale, in the spirit of recent mixed-initiative qualitative analysis~\cite{Baltesetal2025,Carlsenetal2022,Paoli2023,corbin14}.

The construction proceeded in stages. We first asked an LLM to propose the themes latent in the codebook, an initial scaffold rather than a result; the authors then validated and redefined them by hand, splitting overlapping themes, sharpening definitions, and adding constructs we had observed while reading the codebook but the model had missed. A second LLM pass assigned every code to a theme, separating codes that bear on different concerns, and within each theme a further pass proposed sub-themes that the authors again validated, aggregating the codes that make the same claim. These sub-themes gave us an initial vocabulary of constructs and candidate relationships among them. From that draft the authors built the theory through repeated revision (the central work of this study and the part we performed manually), deciding across many passes which constructs to keep, split, or merge and which relationships to assert, reverse, or drop, until the model stabilized. Throughout, we used the codebook as a \emph{search engine}, querying it through the LLM for the practitioner text bearing on a given construct or relationship (for example, every quote about comprehension debt) under varied retrieval strategies, and either confirming a relationship or revising it; at times we hypothesized a relationship from prior understanding (including the empirical findings reviewed in Section~\ref{sec:relatedwork}, ~\ref{sec:related-after-AI-code}), and queried the codebook for the text that would support or contradict it. This interplay of emerging categories and constant comparison against the data, and its willingness to check a posited relationship rather than only let it emerge, is characteristic of the Straussian variant~\cite{corbin14,suddaby06}. A closing author audit merges near-duplicate relationships and attaches, to every relationship, its supporting practitioner text and, for the contested relationships, the opposing text as well. The result is the explanatory theory of \textbf{26 constructs and 67 relationships} presented next (Figure~\ref{fig:rq2_model}); Appendix~\ref{app:rq2theory} gives these prompts and the evidence-tier definitions.

\mysec{Reproducibility and traceability}
Both collection frames (the subreddit strata and inclusion rule; the query list, source passes, and filter specification) and the judge rubric are versioned in configuration, making the corpus reproducible, and an append-only manifest traces every quote in the theory back to its originating document. Each grounded quotation is cited inline by a stable handle (\eg G2935), and all 3{,}100 coded documents appear in the \emph{Supplementary Document Index} with each one's handle, dataset identifier, title, and source URL.

\subsubsection{Limitations}
Three threats are specific to this design. First, our sources are written in the age of generative AI, and some may be partly or wholly LLM-generated; we study \emph{discourse} about review, not verified firsthand practice, and read the corpus as practitioner argument rather than ground truth. Second, LLM coding and filtering can miscode, miss, or hallucinate; we mitigate with three independent coders, a verbatim quote behind every code, a separate relevance judge validated against a stronger model, and an author audit of every relationship, though residual error remains. Third, the move from codebook to theory is author judgment at each structural step (by design, since axial and selective coding are the interpretive core of grounded-theory work~\cite{corbin14}), so a different team could have drawn the constructs and relationships differently; we constrain this by grounding every one in coded text and attaching opposing evidence to the contested edges, but the result is a \emph{proposed} explanatory theory to be tested, not validated. Finally, grey literature over-represents vocal early adopters and carries vendor advocacy and post-incident hindsight, which we counter through the breadth of thousands of independent sources rather than eliminate.

\section{Results: The Theory}
\begin{figure*}[t]
  \centering
  \caption{An explanatory theory of how coding agents reshape code review: \textbf{26 constructs} and \textbf{67 relationships} (64 directed, 3 contested) synthesized from practitioner discourse. The graph reads top to bottom: the \emph{drivers} AI introduces (top), through the reviewer's dispositions, the team's responses, and the five-construct \emph{review act} that the rest of this section unpacks, down to the \emph{outcomes} practitioners care about (bottom). A node's \emph{shape} encodes its role (rectangle: a \emph{driver}; ellipse: a \emph{review-dynamics} construct, i.e., a review-act, reviewer, or team-response construct; hexagon: an \emph{outcome}), and an edge's \emph{line style} encodes the sign of the relationship (solid: increases; dashed: decreases; bold: \emph{contested}, where the sign is set by a moderator).}
  \label{fig:rq2_model}
  \includegraphics[width=\textwidth]{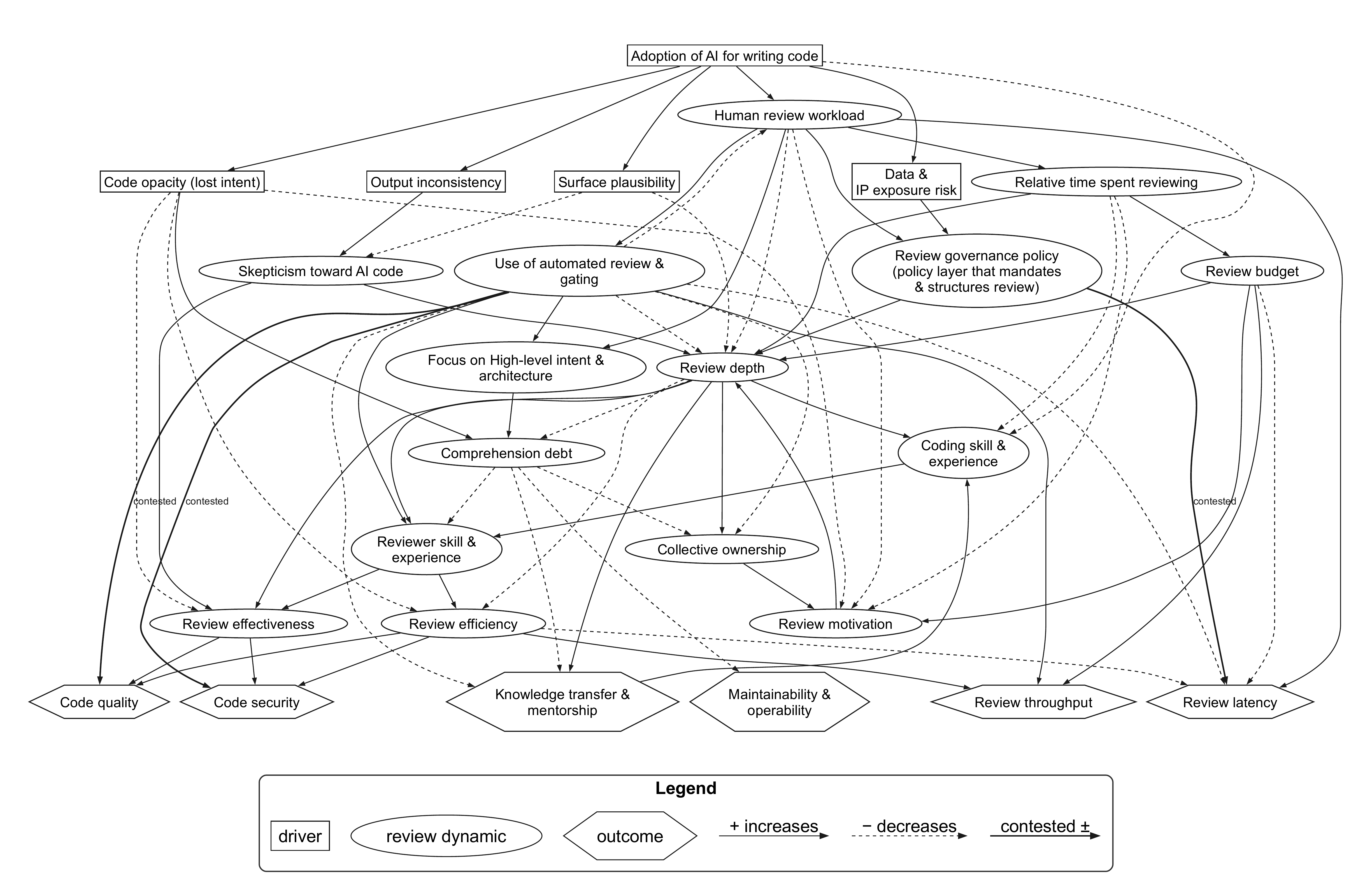}
\end{figure*}
We report the theory not as a list of isolated observations but as a \emph{conceptual theory}: a vocabulary of constructs and the relationships among them that together explain how code review changes when coding agents author pull requests, and what practitioners believe is at stake~\cite{stol15theory,sjoberg08theories,baltes18expertise}. In Gregor's taxonomy the result is a \emph{theory for explaining} (Type II), which says \emph{how} and \emph{why} rather than \emph{what will happen}~\cite{gregor06nature}: we make the forces and mechanisms explicit and state them as falsifiable claims, but do not test them or use them for prediction here. Following conventions for reporting qualitative theories in software engineering~\cite{stol16grounded,baltes18expertise}, we present the theory as \emph{constructs} (the boxes of Figure~\ref{fig:rq2_model}) and \emph{relationships} (its directed edges, each a force that fosters or hinders its target), and ground every relationship in coded practitioner text. 
The theory (Fig.~\ref{fig:rq2_model}) comprises \textbf{26 constructs} linked by \textbf{67 relationships} (64 directed, 3 contested), distilled from roughly 2{,}700 practitioner sources and 4{,}800 codes (\cf Section~\ref{sec:rq2_methods}). 

In our setting, relationships are directed, causal claims in the form of \emph{`an increase in construct A \emph{increases} (solid line) or \emph{decreases} (dashed line) construct B.'} In the following description, we highlight a subset of the relationships as numbered, falsifiable \emph{propositions} (P1\,\ldots) and provide corresponding practitioner quotes -- these propositions are the key claims that future work should test and quantify with empirical data, ideally using causal estimands controlling for other constructs in the theory where possible.

\mysec{Constructs}
Constructs in the theory (Fig.~\ref{fig:rq2_model}) can be divided roughly into three groups:
\begin{itemize}
  \item\textbf{Drivers:} On the one side (top), \emph{drivers} are properties of agent-authored code that practitioners treat as exogenous (coming from the outside, not easily influenced), such as increasing \emph{volume and velocity} of code production due to AI, the \emph{surface plausibility} of AI generated code that reads as clean and idiomatic, the \emph{opacity} of AI generated code due to lost design rationale when no human author formed the intent, and the \emph{data and IP risk} of AI use.
  \item \textbf{Outcomes:} On the other side (bottom), \emph{outcomes} are the consequences practitioners care about, including \emph{review throughput}, \emph{review latency}, \emph{code quality}, \emph{code security},  and \emph{maintainability}.
  \item \textbf{Review dynamics:} In between, the theory describes various constructs related to \emph{review dynamics}, which practitioners can usually influence directly or indirectly. For example, practitioners can choose (within limits) of how deeply to perform a review (\emph{review depth}), how much time to spend on a review (\emph{review efficiency}), or how to perform reviews (\emph{review policies}, \emph{use of automated review}). They can also influence their own \emph{review motivation} and \emph{reviewer skill}.
\end{itemize}

These constructs relate to concerns that practitioners talk about, though not always in their words. We often decomposed constructs to emphasize how different forces influence the same concept through different mechanisms -- for example, \emph{review throughput} may be increased because a team has more \emph{review experience}, because they lower \emph{review depth} (`rubber-stamping'), or because they bypass manual review with \emph{automated code review}. 
Each of these factors may again be influenced by several others and multiple constructs have opposite influences on another or form feedback loops. Decomposing this conceptually is a key contribution of the theory.

\mysec{Increased code volumes are the key driver}
Practitioner accounts usually start with the same premise of how coding agents
decouple the rate at which code is produced from the rate at which it can be reviewed: \qt{Reviewer throughput grows linearly with headcount. AI generation throughput grows multiplicatively per developer}{G2802}. If human peer review is pursued for all this code, it raises the \emph{human review load}; without an increase in the \emph{review budget}, this puts pressure on \emph{review efficiency} to review more code in the same time.

\begin{wrapfigure}{r}{0pt}
  \includegraphics[trim=34 34 34 34, clip, scale=0.17]{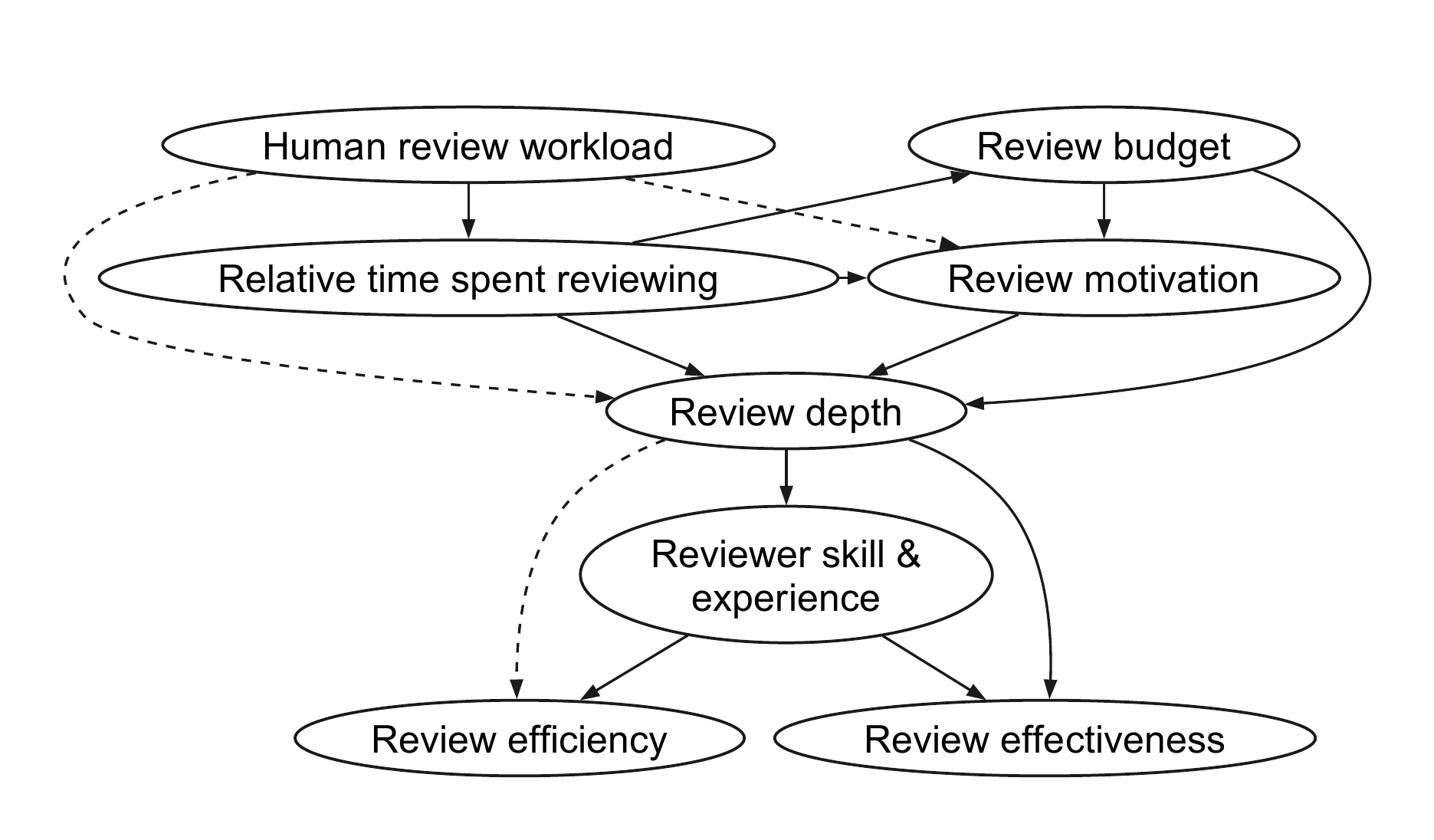}
\end{wrapfigure}
The effects of this pressure are commonly discussed among practitioners and usually go through one of two mechanisms:
The most common reaction is to perform reviews with less \emph{review depth} (e.g., less attention, looking for fewer problems) so
it can be performed quicker, for example, skimming instead of reading: \qt{if that engineer is reviewing 200 AI-generated PRs a sprint instead of writing code themselves, they're skimming, not reading}{G2048}.
Some practitioners also report that the increased load lowers \emph{review motivation}, the willingness to do the job well, as sustained load shades into fatigue: \qt{Reviewers might keep up for a sprint, but they'll burn out [or] worse, start rubber-stamping reviews}{G2935}. While it speeds up reviewing, it negatively impacts \emph{review effectiveness} and various outcomes we care about, like \emph{code quality}.

\prop[prop:load-depth]{Higher review load decreases review depth, which impacts efficiency and effectiveness.}
\prop[prop:load-motivation]{Higher review load decreases review motivation, which impacts review depth.}
Another commonly discussed mechanism is that practitioners increase the \emph{relative time spent reviewing} rather than coding to increase the \emph{review budget} -- \qt{Many software engineers already spend more time reviewing AI-generated code than writing it manually}{G2045}. The worry is that code production consumes the additional review budget even if the team switches most of their time to reviewing (which may further demotivate developers, \qt{The thing nobody warned me about is the review fatigue. After a few hours reading through AI-generated code, I'm more drained than I'd be after twice as long actually writing code}{G439}): \qt{increase code generation speed by 10 to 25 times without proportionally increasing review capacity}{G2394}, while \qt{The number of reviewers hasn't changed, and neither has the human attention available}{G2991}. Several practitioners predict that human review becomes either shallow or abandoned: \qt{if your team generates more PRs than your reviewers can thoughtfully evaluate, you either add an automated first-pass layer or you accept that human review becomes performative}{G3088}.

\prop[prop:load-reltime]{Higher review load increases relative time spent reviewing, which decreases review motivation.}

\mysec{Polished AI code makes reviewing harder}
AI code is often polished, clean, and idiomatic -- which we call \emph{surface plausibility}. Practitioners report that surface plausibility makes it harder to review as it can disarm the reviewer's defenses: \qt{polish can lower the reviewer's guard}{G2222} and \qt{seniors are rubber-stamping code that looks idiomatic but hides subtle bugs}{G2208}. It lowers effectiveness by camouflage, since the defect is dressed as correct code, \qt{Plausible violations are harder to catch than obvious ones}{G2600}. \begin{wrapfigure}{r}{0pt}
  \includegraphics[trim=34 34 34 34, clip, scale=0.21]{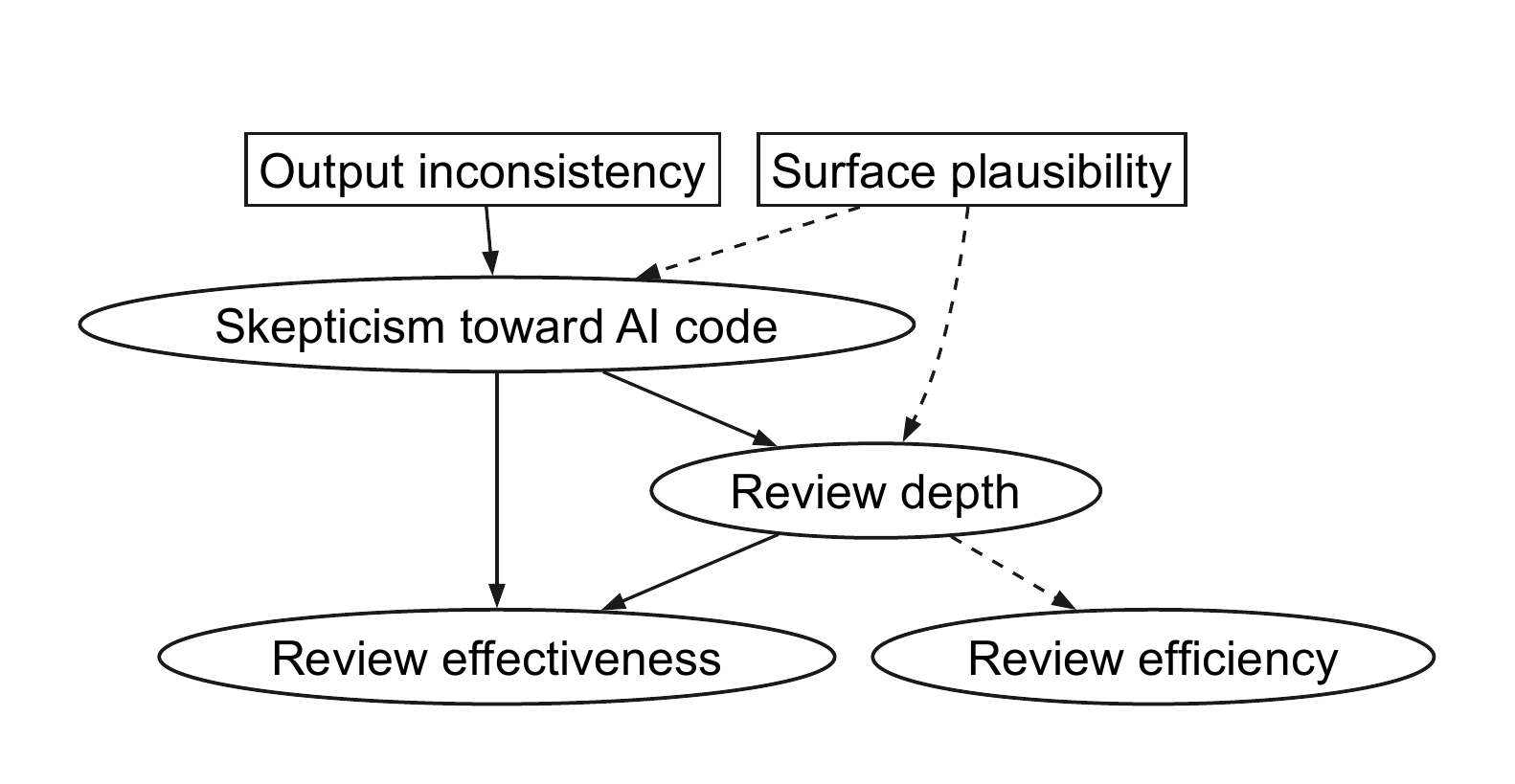}
\end{wrapfigure} Polished code also clears faster, but that speed is a consequence of the dropped attention rather than a separate effect, \qt{When AI submits polished code, reviewers pattern-match against `looks professional' and approve faster}{G2613}: the reviewer approves faster because they are scrutinizing less. We posit that \emph{surface plausibility} lowers both \emph{review depth} and thus indirectly \emph{review effectiveness}.

\prop[prop:plausibility-depth]{Surface plausibility decreases review depth, which decreases review effectiveness.}

The mirror is \emph{skepticism}, the reviewer's disposition to treat AI code as particularly suspicious. Where plausibility disarms, a skeptical reviewer attends harder, \qt{I've started treating them like code from a junior dev who's super confident, but only moderately competent. I read every line like it's a puzzle}{G3043}, and catches more, because an adversarial frame surfaces what a trusting read passes over, \qt{the problem is that auth code `looks correct' to anyone who isn't specifically thinking adversarially}{G103}. The increased depth also makes the skeptic slower, but that slowness is the labor of the added attention, not a separate effect, \qt{Developers end up spending more time validating that the AI's code actually works}{G2558}.

\prop[prop:skepticism-depth]{Skepticism toward AI code increases review depth, which increases review effectiveness and decreases review efficiency.}

Skepticism may be shaped by external forces. On the one hand \emph{surface plausibility} lowers it, because polish reads as a safety signal, \qt{when they see a polished code block, they assume it's safe unless something obvious jumps out}{G2693}. However, observing \emph{output inconsistency}, where AI output differs from run to run, invites an adversarial stance, treating AI like an \qt{unpredictably unreliable junior engineer whose output you must review with 100\% scrutiny, every time, forever}{G2250}. 

\prop[prop:skepticism-compensate]{Skepticism can compensate for surface plausibility and is influenced by awareness of output inconsistency.}

\mysec{Reviewing code without understanding intent}
\begin{wrapfigure}{r}{0pt}
  \includegraphics[trim=34 34 34 34, clip, scale=0.21]{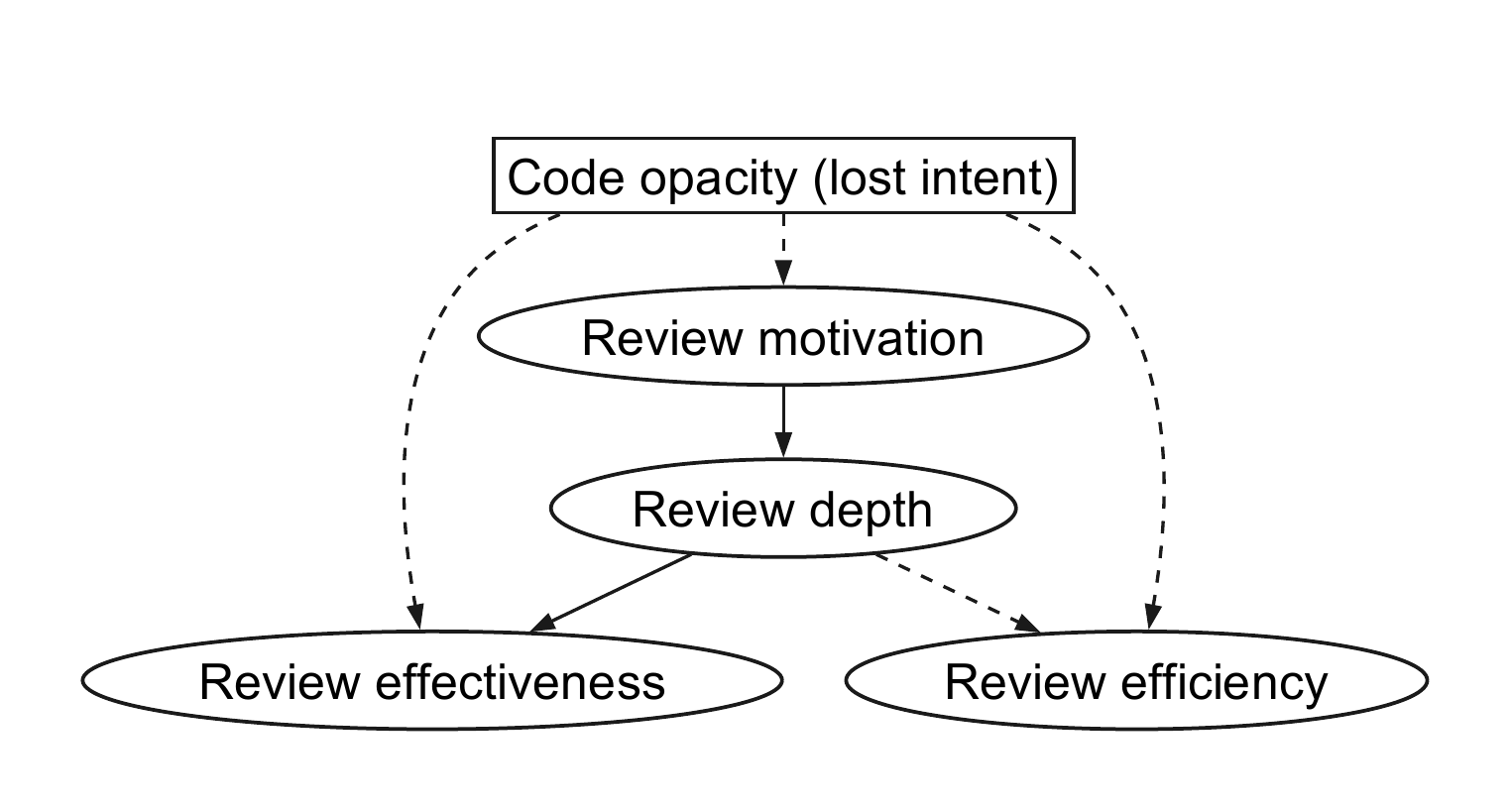}
\end{wrapfigure}
Practitioners often discuss how AI-written code arrives without a design rationale, since no human author formed the intent and there is no one to ask why the code is the way it is. This \emph{code opacity} degrades review along three channels. First, it directly lowers \emph{review efficiency}, because an engaged reviewer must reconstruct the intent rather than check against it, \qt{One reason AI-generated code is harder to review is that you're reconstructing intent}{G2066}. Second, it lowers \emph{review effectiveness} even at full effort, because the yardstick for judging correctness, the intent itself, is gone, \qt{You can't meaningfully review code you didn't write and don't understand}{G2413}. And finally, it lowers motivation, because reviewing code whose rationale can never be recovered is draining, \qt{reviewing 5k lines of AI-generated code after the fact is miserable}{G978}.

\prop[prop:opacity]{Code opacity decreases review efficiency, review effectiveness, and review motivation.}

\mysec{The impact of AI for code review is controversial}
\includegraphics[trim=34 34 34 34, clip, scale=0.21]{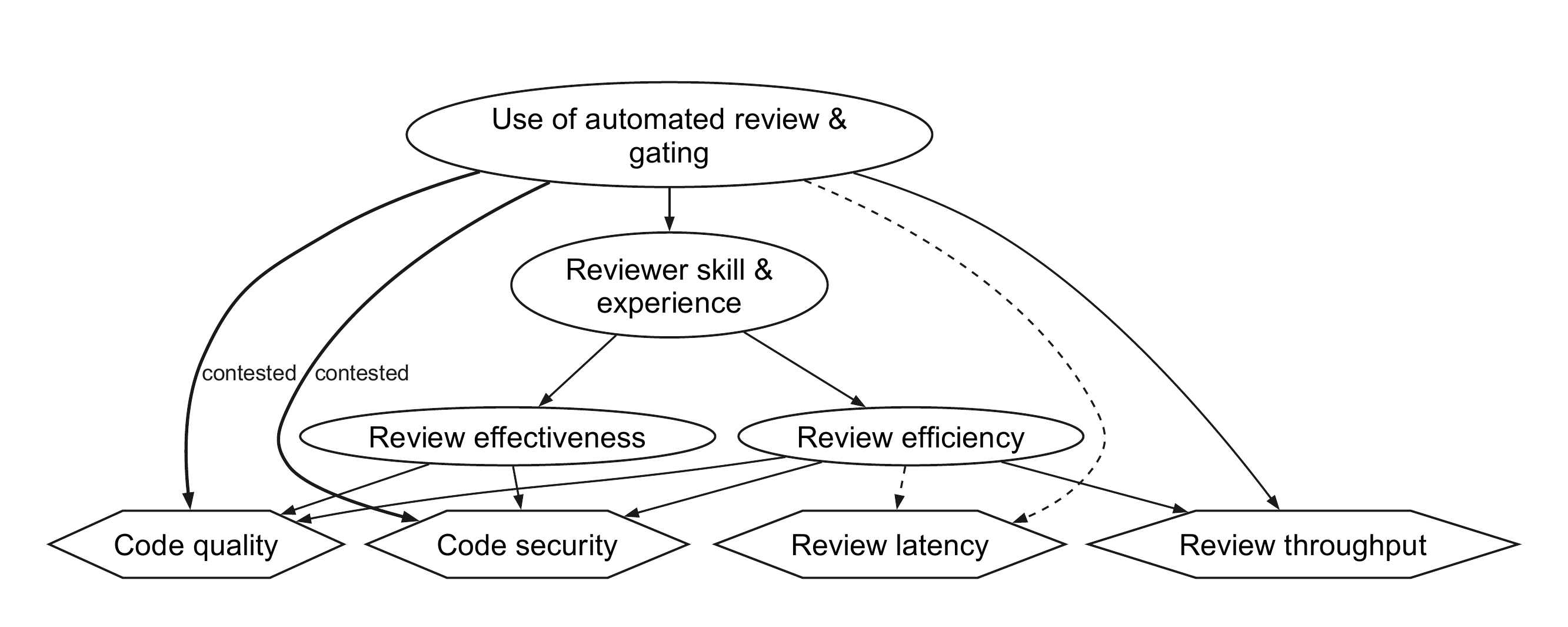}
It is not possible to separate discussions of reviewing AI-generated code entirely from the rise in AI use for code review as they both evolve rapidly at the same time. As a consequence, \emph{automated code review} has many relations in our theory. AI reviewing can take different forms, from assisting human reviewers, to reviewing for a subset of concerns, to full automation, we need to untangle.
First, in our theory most outcomes that practitioners care about, including \emph{review throughput}, \emph{review latency}, \emph{code quality}, and \emph{code security} are influenced through \emph{review efficiency} and \emph{review effectiveness} in the obvious ways (we omit supporting quotes).
Fully automating review bypasses the qualities of human review to increase the overall (combined human and automated) \emph{review throughput} and decrease \emph{review latency} (e.g., \qt{uReview's feedback appears minutes after the commit is posted for review}{G2304}).
The effect of \emph{automated code review} on \emph{code quality} and \emph{code security} is contested.
Many argue that given context, the automation catches real defects, \qt{CodeRabbit catches approximately 82\% of planted bugs with a 15\% false positive rate}{G2836}, and is strongest where humans struggle, \qt{CodeRabbit's cross-file contextual analysis is particularly strong for detecting injection vulnerabilities}{G2027}. However, others argue that automated reviews miss what matters and add noise, \qt{Automated review catches style issues. It misses the `this will cause a race condition under load' kind of problems}{G1957}, and can manufacture false confidence or raise false alarms, \qt{LLMs can confidently flag a `critical SQL injection' in a parameterized query that's perfectly safe}{G1891}.

\prop[prop:auto-throughput]{Fully automating code review increases the overall review throughput and decreases review latency.}
\prop[prop:auto-quality]{Fully automating code review influences code quality and code security outcomes (direction contested).}

Rather than full automation, AI can also be used to assist developers in reviewing code. We describe this in our theory as empowerment of \emph{reviewer skills} (similar to past improvements with better tooling for code review~\cite{bacchelli13}), with indirect effects on \emph{review efficiency} and \emph{review effectiveness} -- that is, the same reviewer is more effective at the same time spent or more efficient at the same defect detection rate. Under volume, and with automation handling the line-level pass, the reviewer climbs the abstraction ladder, from reading lines to auditing intent, \qt{When agents produce thousands of lines per hour, line-by-line review does not scale}{G2869}, until \qt{You are no longer reading every line. You are now an auditor}{G2517}. The automation helps by clearing routine work before the developer can focus on intent, \qt{focus on higher-value tasks such as architectural design}{G2087}.
A too high-level focus has risks though, leaving the line-level understanding unbuilt, \qt{We are not reviewing the recipe. We are tasting the dish}{G2392}, so engineers \qt{are increasingly reviewing outputs instead of understanding behavior end to end}{G2161}, and \emph{comprehension debt} accrues (discussed next) even when attention is high.

\prop[prop:assist-skill]{AI-assisted code review increases reviewer skills, with positive influence on review efficiency and effectiveness.}
\prop[prop:focus-debt]{Greater focus on high-level intent and architecture, driven by volume and automated review, increases comprehension debt even when review depth is high.}

\mysec{The human costs: comprehension, ownership, and skill}
\noindent\includegraphics[trim=34 34 34 34, clip, width=\linewidth]{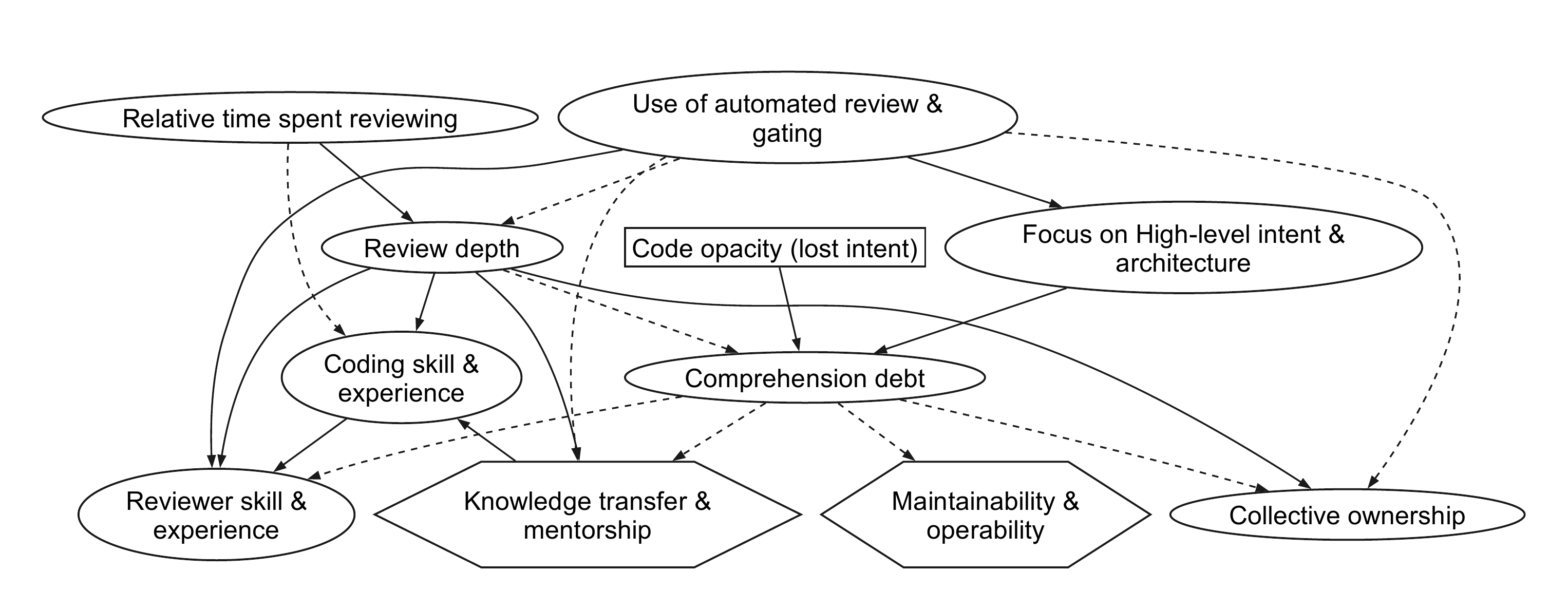}
Many concepts in our theory are influenced directly or indirectly by reviewer skills or motivation, but those are not fixed and static concepts and we need to pay attention to potential feedback loops.

\noindent First, \emph{review skill} evolves over time through practice. Higher \emph{review depth} provides more opportunity to learn \qt{Reviewing AI-generated pull requests develops a critical eye for subtle bugs}{G2356}. Similarly, many argue that \emph{review skills} are downstream from \emph{coding skills} (e.g., \qt{How can you review what you do not yet know how to build?}{G2384}), which are honed by spending more time programming  (i.e., less \emph{relative time spent reviewing}; \qt{skill atrophy from reduced practice writing code manually}{G2501}) and getting constructive feedback (i.e., more \emph{review depth} from others). That is, degrading code review due to pressures from high volume AI-generated code, can undermine the ability to review that code in the future, in a negative feedback loop.

\prop[prop:depth-skill]{Lower review depth limits growth in reviewer skills.}
\prop[prop:coding-skill]{Lower coding skills undermined by less time coding and  receiving low quality reviews reduce review skills.}

Second, \emph{review depth} improves \emph{collective ownership} and \emph{knowledge transfer}. The act of attentive review is itself how a team comes to own a change, \qt{When no human has deeply reviewed the logic, nobody owns it}{G2538}, and how knowledge moves from senior to junior \qt{The thread between a senior and a junior engineer inside a PR shapes how that junior engineer approaches the next twenty PRs they write}{G3024}. A codebase can thus end up formally merged by many but genuinely owned by no one, with little learning imparted on junior team members. Similarly, automating review with AI undercuts that same human capital in two ways. By lowering \emph{review depth}, the attention that builds \emph{collective ownership}, \emph{knowledge transfer}, and \emph{reviewer skill}, it weakens all three at their source; and by shortcutting the human review arrangement that carried them, it removes them directly, \qt{AI agent reviewers rarely provide knowledge transfer suggestions}{G2932}.

\prop[prop:depth-ownership]{Review depth raises collective ownership and knowledge transfer; automated review undermines them.}

Third, beyond direct impacts, the act of low-depth reviews creates \emph{comprehension debt}, undermining the developers' future \emph{review skills} for the code base. Attentive review is how a reviewer forms a mental model of a change, \qt{That knowledge doesn't come from writing prompts. It comes from reviewing carefully enough to understand}{G2066}; \qt{juniors who skip the `read the diff carefully' step never build the mental model}{G2810}. Opacity widens the gap, since part of agent code's rationale was never anyone's to understand, \qt{Nobody has a working mental model of why the code is structured the way it is}{G2847}. The resulting comprehension debt erodes \emph{maintainability}, \qt{your ability to debug it, extend it, or refactor it tomorrow is compromised}{G2026}, further hollows \emph{collective ownership}, \qt{If it is not understood, it is not owned}{G2161}, and further breaks \emph{knowledge transfer} that could teach others and build collective ownership, \qt{If you're merging PRs you don't fully understand, the bus factor of the project climbs to dangerous levels}{G2570}.

\prop[prop:debt]{Low review depth and opacity increase comprehension debt, undermining review skill, maintainability, collective ownership, and knowledge transfer.}

Here we see the potential for several negative feedback loops, where shallow reviews (e.g., due to volume pressure from AI-generated code) increasingly undermine the ability to perform skilled in-depth reviews.
This is particularly important, since AI coding does not just undermine the direct form of building a mental model during coding, but its pressures can also undermine the secondary path of building a mental model during code review.

\mysec{Governance: Policy levers likely have many direct and indirect effects}
As a reaction to changes in code volume and code review, many organizations, both in open source and industry, explore how to formalize behavior expectations with various \emph{review governance policies}, e.g., \qt{The nine rules below define a modern code review culture designed for high-speed, high-volume change}{G2552}, or mandating human gates on sensitive changes like \qt{If the code touches security, authentication, or core data structures, it must be written by a human and reviewed by a lead engineer}{G2339}. In addition, also data and intellectual property concerns trigger policy discussions, for example, since agent output \qt{frequently includes real API keys or test credentials directly in generated code}{G1967}.

\begin{wrapfigure}{r}{0pt}
  \includegraphics[trim=34 34 34 34, clip, scale=0.21]{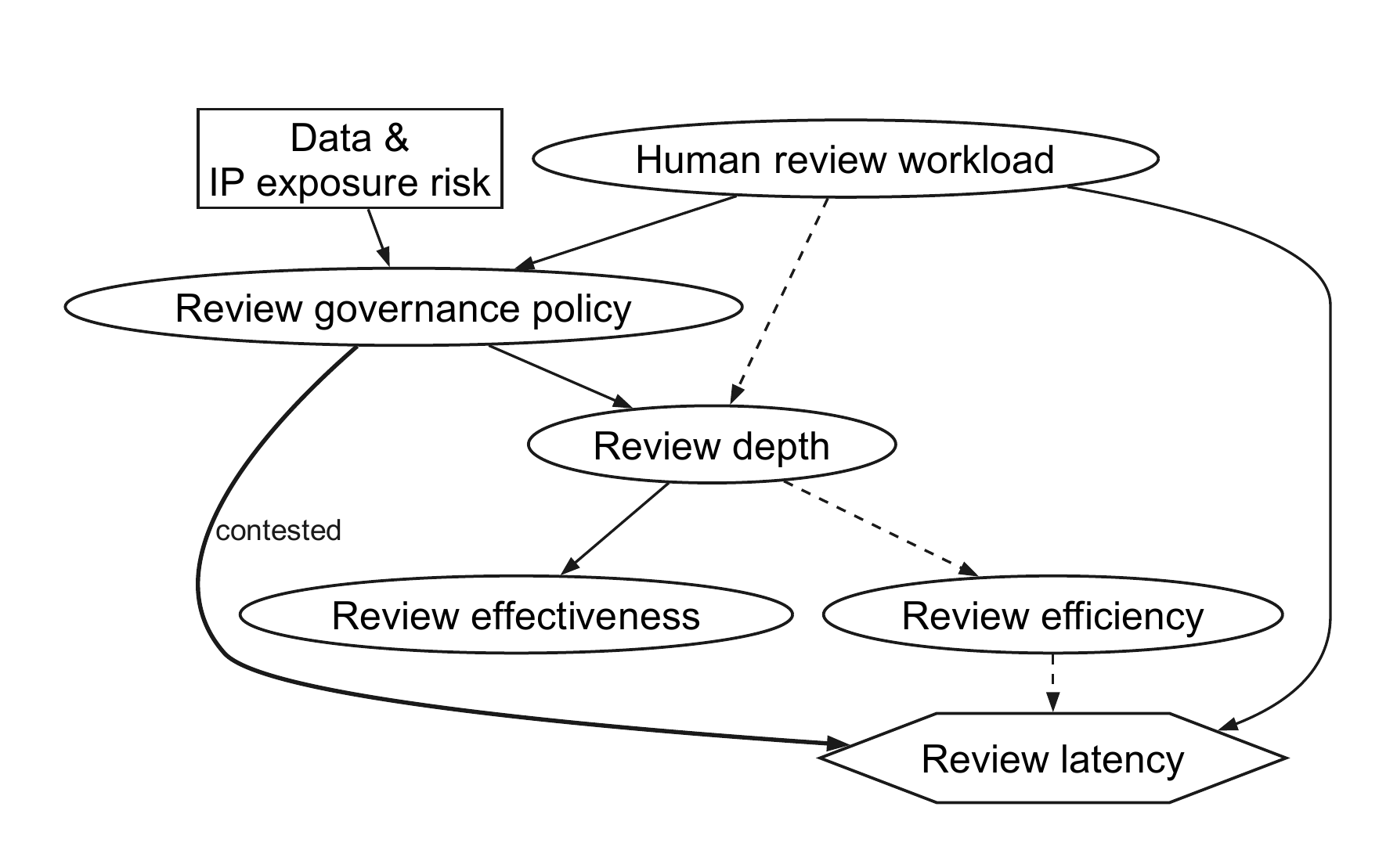}
\end{wrapfigure}

A \emph{review governance policy} sets what must be reviewed, by whom, and against what checklist. Its most direct effect is to raise the rigor of review: a policy that demands a careful pass deepens \emph{review depth}, which catches more but takes longer, so it raises \emph{review effectiveness} and lowers \emph{review efficiency}. A reviewer \qt{spends 45 minutes reviewing it -- checking for bugs, edge cases, and code that doesn't match team standards}{G2527}. Beyond mandating steps, a policy can deepen review by assigning accountability: when a team keeps responsibility with people rather than ceding it to the tool, reviewers attend more carefully because they own the output, \qt{If your team bans `the AI did it' as an excuse, review quality improves because authors know they own the output}{G1872}.

\prop[prop:gov-depth]{Review governance policy raises review depth, which raises review effectiveness and lowers review efficiency.}

A policy also sets which changes a human reviews at all, and this is where its effect on \emph{review latency}, the time a change waits to clear, splits, set by the policy's calibration. The aim is to be selective rather than exhaustive, \qt{The goal isn't to put human eyes on all the code. It's to direct attention where it matters}{G3058}, reserving human sign-off for the changes that need it, \qt{A few hours of focused audit against the right checklist will find more than weeks of general code review}{G1893}. A risk-tiered policy that gates only material changes and sends routine ones through automated validation clears the bulk of work quickly, \qt{risk-based change categorisation: only material changes require human sign-off; routine changes flow through automated validation}{G2335}, while a blanket policy that gates every change adds delay, \qt{Does stricter governance slow down AI code generation? Initially, yes. Enforcement and review add process}{G2175}.

\prop[prop:gov-latency]{The effect of review governance policy on review latency is contested: its sign is set by policy calibration, lower latency under a risk-tiered policy that gates only material changes, higher latency under a blanket policy that gates every change.}

\mysec{The shape of the theory}
The value of the theory lies in the interconnected whole, rather than in any single edge. By decomposing what makes review effective, what makes it efficient, and what lets a change bypass review entirely, we can separate the forces at work and identify the mechanism behind each one. This decomposition reveals a small central core. Most paths run through \emph{review depth} and \emph{reviewer skill}, the attention a reviewer actually applies to a change and the ability they have in doing that effectively. They are the busiest construct in the model, pushed and pulled by many forces and feeding many others, and concepts like \emph{surface plausibility}, \emph{collective code ownership}, and \emph{coding skill} all reach throughput and quality through them and their effects on \emph{review efficiency} and \emph{review effectiveness}. 
The core also feeds back on itself, through the loops described earlier, such as rubberstamping eroding the foundation for effective future review. A few edges stay genuinely contested, especially regarding what effect AI use has during code review itself. 

\mysec{Status of the theory}
Consistent with theory-building in software engineering~\cite{stol15theory,baltes18expertise}, this is a \emph{proposed} explanatory theory, not a validated one. Its relationships are grounded in what practitioners report and argue, filtered through the lens of the researchers. Its value is conceptual: It supplies a shared vocabulary, makes competing mechanisms explicit and contestable, and turns ``AI is changing code review'' into a structured set of falsifiable claims and the moderators that would decide them. 

\section{Related Work on Code Review of AI Code}
\label{sec:related-after-AI-code}
We deferred related work on code review after AI adoption until after letting practitioners speak, since, as we now show, studies already differ on drivers and outcomes and rarely explore the why.

On the \emph{drivers} of change (top nodes in our theory): coding agents generate code rapidly, with OpenAI Codex alone opening over 400{,}000 pull requests on \GH{} within two months of release~\cite{chowdhury2026industry}, raising concern about whether review can keep up~\cite{li2025rise}. AI-authored code is more verbose~\cite{mao2026large} and its commit messages more descriptive than rationale-capturing~\cite{pham2026code}, but studies conflict on whether AI-authored PRs are larger~\cite{Watanabeetal2025,popescu2026investigating} or smaller and more focused~\cite{popescu2026investigating,pham2026code}.

On review \emph{outcomes} (bottom nodes in our theory), studies again diverge: agent PRs are accepted at \emph{lower} rates than human PRs in some~\cite{li2025rise,Watanabeetal2025,chowdhury2026industry}, at \emph{higher} rates for Claude and Codex in others~\cite{popescu2026investigating}, and auto-merged more often but bimodally in yet others~\cite{branco2026lgtm}. Change size is an established moderator (larger PRs attract more friction and merge less~\cite{destefanis2026dose,DBLP:journals/corr/abs-2601-15195}), as are task type (e.g., documentation, CI, bug fixes)~\cite{DBLP:journals/corr/abs-2601-15195,DBLP:journals/corr/abs-2602-02345} and author follow-up~\cite{DBLP:journals/corr/abs-2602-19441}.

On the review process, most work asks whether PRs are reviewed at all (most are not~\cite{li2025rise,DBLP:journals/corr/abs-2605-02273}) and characterizes comment patterns (fewer comments~\cite{DBLP:journals/corr/abs-2601-13754}, more bot reviews~\cite{li2025rise,chowdhury2026industry}, and more steering than evaluating~\cite{DBLP:journals/corr/abs-2605-02273}). Several taxonomize the output of code-review agents such as CodeRabbit~\cite{DBLP:journals/corr/abs-2601-19287,fatima2026footprints,chowdhury2026industry}, finding reviews of agent PRs concentrate on style over deeper concerns~\cite{DBLP:journals/corr/abs-2601-19287}; norms and policies are still evolving, with a recent overview by Holterhoff~\cite{HolterhoffUnknown-hh}. Explorations of deeper mechanisms are rare: studies of rejected agentic PRs find little data to work with~\cite{peralta2026agentic,nakashima2026agentic}, and those of agentic PRs merged without comments find no explanation in code quality~\cite{hasan2026quiet,cynthia2026beyond}.

Other concepts in our theory have been studied, though mostly in isolation rather than as a connected process: overtrusting AI suggestions~\cite{sabouri2025trust,perry2023users}, judging code by readability, plausibility, or passing tests rather than actual correctness~\cite{sabouri2025trust,DBLP:journals/corr/abs-2601-21276,dhanorkar2026human}, and a trend toward decreasing vigilance or increasing rubber-stamping~\cite{yu2026habituation}. Our theory puts such factors into context.

Taken together, this literature is remarkable for its volume, fragmentation, and disagreement. Much of it draws on the same dataset (AIDev)~\cite{li2025rise} yet reports opposite signs for change size, acceptance, and review effort depending on which agents, repositories, and thresholds fall in scope. Where studies do reach for explanation, they do so locally and tentatively: a fast merge \emph{might} signal superficial review or confidence~\cite{li2025rise,yu2026habituation}; light feedback \emph{might} reflect trust or inattention. That gap is what makes the surface trends so easy to read in opposite directions, and supplying the missing theory was the goal of this paper.

\section{Discussion}
\label{sec:discussion}

\mysec{Theory turns a polarized debate into mechanisms that future studies can test}
We are not the first to argue that theory must precede causal inference~\cite{pearl-book-of-why, sjoberg08theories, stol15theory}, and our own attempts to study this change without one make the case: in the motivating study (Sec.~\ref{sec:motivating_study}) the same data supported opposite conclusions under defensible analysis choices, and the literature was no clearer, reporting opposite signs for quantities as basic as change size, acceptance, and review effort (cf.\ Sec.~\ref{sec:related-after-AI-code}). We can establish what is changing but only speculate why, and our speculation shifted with whichever result we were explaining.

The public narrative is polarized: agents supposedly make review either obsolete or indispensable, faster or more dangerous. Our theory replaces that debate with mechanisms: the effect of agents on review is not monocausal, as external forces pressure the review system without fixing any single outcome. We confirm none of these relationships but lay a foundation for future studies, supplying concepts to operationalize or control for so that a surface metric (a fast merge, few review comments) can be read through the mechanisms beneath it and untangled from the confounders, colliders, mediators, and feedback loops that also shape it. These mechanisms likely explain practitioners' divergent experiences: a reviewer who finds agents have made review faster and one who finds they have made it a rubber stamp need not disagree about the facts, only about which pathway dominated. 

\mysec{Developers have the power to shape the review process in an AI world, but it's not monocausal}
The theory shows how AI's effects on peer review can be steered. Most paths to the outcomes practitioners care about run through two central levers, \emph{reviewer skill} and \emph{review depth}: the ability a reviewer brings to a change and the attention they apply to it. Both start from a team's existing position (how well it understands its codebase, how motivated it is to review deeply) but neither is fixed; each is shaped through practice and deliberate decisions. The most consequential is the \emph{use of automated review}, governed by \emph{review policy}: the same lever can focus human attention where it matters or bypass human review entirely. Developers are thus not passive recipients of AI interrupting their work but have agency to shape it at predictable leverage points.

That agency is constrained, however, by an interconnected system in which decisions interact and can create feedback loops: buying throughput by lowering review depth may work in the moment but erodes the reviewer skill, ownership, and knowledge transfer that effective future review depends on, and automation carries the same double edge, accruing comprehension debt even when attention stays high. Whether feedback turns virtuous (skepticism, deeper review, sharper reviewers) or vicious (rubber-stamping, skill atrophy, still shallower review) likely depends on many interacting decisions. The question is thus not whether \emph{AI will ruin or fix code review?} but \emph{which decisions, under which conditions, push the system toward good rather than bad outcomes?}

To design interventions beyond individual decisions like per-project policies, future work could draw on a system-theory lens~\cite{meadows2008thinking}. System dynamics diagrams resemble our causal models, relating concepts (``stocks'') through ``flows,'' but add a time component (delayed flows) and center on feedback loops and equilibria rather than causal inference. System theory also ranks leverage points: adjusting local parameters and regulating negative feedback loops are least effective, while driving positive feedback loops and changing rules and goals are most effective overall~\cite{meadows2008thinking}. Our theory offers a strong starting point for such a system-thinking approach.

\mysec{Experienced benefits and limitations of scaling grey literature analysis with AI}
This is, in part, a method paper in disguise: scaling grey-literature analysis with AI was itself an experiment. Our primary contribution is the theory of the previous section, but we regard the way we built it, scaling grey-literature analysis far beyond past theory-building studies, as a secondary one. Ideally, every future repository-mining study could cheaply survey the grey literature and build a theory to inform it; we are not there yet, but our approach offers a template for others heading that way, and we share all the infrastructure we created.

Scale was an important goal. First, it increases reach and breadth of voice: thousands of independent sources spanning years and registers, rather than academics or practitioners with reach, or interviewees sampled from a personal network, and that reach comes without the recruitment and interview-time cost to researchers and practitioners.

The costs lie in two places: the tokens for analysis and codebook (about US\$0.35 per document across all steps) and the effort of turning the resulting codes and quotes into a theory. Automation worked reasonably well for coding (though merging similar codes could have worked better), but automating the back half, building the theory itself, proved ineffective. To gauge how far automation could go, we extracted causal statements directly from the codebook in one bottom-up pass (prompt in Appendix~\ref{app:rq2theory}); it yielded \textbf{15{,}029 statements}, but shallow, redundant, and overlapping ones, so we built the theory by hand from the sub-theme claims instead. This is unsurprising: practitioners use inconsistent terms, and the same term for different things, which we had to untangle; at one point a single \emph{review rigor} construct conflated what is now \emph{review efficiency}, \emph{review effectiveness}, \emph{review depth}, and \emph{reviewer skill}, and separating them took sustained interpretive work. The distinction is genuinely useful but not inherent in the documents; we had to impose it by reading in context, keeping interpretation with the human analyst. 

Two further notes. On saturation, we suspect 3{,}100 coded documents exceeded what was needed and far fewer would have sufficed; we scaled because we could. Because the theory-building step was manual over the large codebook rather than automated, we cannot establish after the fact where saturation set in. Second, we coded only the grey literature and brought the academic studies of Section~\ref{sec:relatedwork} and Section ~\ref{sec:related-after-AI-code} in afterward, as comparison, reflecting their different role: grey literature is our object of study, whereas the empirical papers are prior scientific claims that already codify concepts and operationalizations, though our infrastructure could code them just as well.

\section{Conclusion}
\label{sec:conclusion}

We synthesized practitioner discourse at scale into a conceptual explanatory theory of how coding agents reshape code review. Its central claim is that review has become the control point through which an agent's effect on software is decided, and that AI does not set the sign of that effect. The team does, through the expertise its reviewers bring, which AI amplifies, and the review process it adapts in response, which AI rewards; we capture this in three moderators: reviewer expertise and disposition, automated-reviewer capability, and process adaptation. The theory makes the agent-era review debate's competing claims explicit and contestable, surfaces a set of under-attended human costs (comprehension debt, eroded ownership and knowledge transfer, and reviewer deskilling) that compound through reinforcing loops, and turns ``AI is changing code review'' into a structured set of falsifiable propositions together with the moderators that would let future studies decide them.

\textbf{Data Availability}
\label{sec:dataAvailability}
A fully anonymized replication package~\url{https://anonymous.4open.science/r/code-review-theory-replication-package-1230} contains the motivation study, collection and relevance-judging tool, the full document-codebook-theory pipeline, the 3{,}100 coded documents with grounding quotes, the theory model, all figures, Appendix 1--5, and \emph{Supplementary Document Index} (also part of Appendix 6).

\bibliographystyle{IEEEtran}
\bibliography{sabib}

\clearpage
\appendices
\renewcommand{\appendixname}{Appendix}
\gdef\thesection{\arabic{section}}
\gdef\thesectiondis{\arabic{section}}

\section{Motivating Study: Methods and Full Results}
\label{app:rq1}

This appendix gives the full data-collection, methods, and results for the motivating observational study summarized in Section~\ref{sec:motivating_study}.

\mysec{Data source and repository sampling}
The study draws on the public \emph{Agents in the Wild} corpus,\footnote{\url{https://insights.logicstar.ai/}} which links \GH{} pull requests opened between May~15 and November~12, 2025 to accounts associated with coding agents (\eg GitHub Copilot, Devin, Cursor, Claude Code, OpenHands) alongside human and bot activity. We retain non-fork repositories with at least ten stars that contain at least one agent-authored and at least one human-authored PR in that window, so that every comparison is made \emph{within} a repository exhibiting both modes of authoring; this leaves \textbf{16{,}317 eligible repositories} from which we sample. Because pull requests nest inside repositories whose baseline activity and agent adoption differ by orders of magnitude, we sample at the \emph{repository} level under a stratified design: repositories are cross-classified by volume bands (quartiles of total PR count, with a fixed-width fallback under extreme skew) and agent-share bins (0--5\%, 5--25\%, 25--50\%, 50--75\%, 75--100\% agent-associated PRs); a fixed target of 3{,}000 repositories is allocated across strata in proportion to stratum size by the largest-remainder rule, never exceeding a stratum's population, and repositories are drawn without replacement under a fixed seed. The design favors breadth of coverage across activity and agent-exposure regimes over an exact match to every PR-level marginal. Table~\ref{tab:repoPopSampleDescriptives} contrasts univariate repository metadata (stars, forks, subscribers, age) for the eligible list versus the drawn sample; the two track closely on central tendency, while star extremes reflect the heavy tails typical of open-source popularity.

\begin{table*}[t]
\small
\centering
\caption{Univariate repository metadata for the eligible repository list (\textbf{Elig.}) versus the stratified sample (\textbf{Samp.}), restricted to repositories with at least ten stars (consistent with study inclusion). Subscribers are \GH{} watcher counts.}
\label{tab:repoPopSampleDescriptives}
\setlength{\tabcolsep}{4pt}
\begin{tabular}{@{}l rr rr rr rr rr@{}}
\toprule
 & \multicolumn{2}{c}{Min} & \multicolumn{2}{c}{Mean} & \multicolumn{2}{c}{Median} & \multicolumn{2}{c}{Mode} & \multicolumn{2}{c}{Max} \\
\cmidrule(lr){2-3} \cmidrule(lr){4-5} \cmidrule(lr){6-7} \cmidrule(lr){8-9} \cmidrule(lr){10-11}
 & Elig. & Samp. & Elig. & Samp. & Elig. & Samp. & Elig. & Samp. & Elig. & Samp. \\
\midrule
Stars & 10 & 10 & 2{,}706.1 & 2{,}657.4 & 115 & 116.5 & 10 & 10 & 431{,}791 & 182{,}688 \\
Forks & 0 & 0 & 510.4 & 494.0 & 29 & 29 & 1 & 1 & 243{,}387 & 46{,}118 \\
Subscribers & 0 & 0 & 47.9 & 47.3 & 6 & 6 & 1 & 1 & 9{,}789 & 3{,}563 \\
Age (days) & 62 & 136 & 1{,}730 & 1{,}776 & 1{,}331 & 1{,}349 & 385 & 136 & 6{,}643 & 6{,}643 \\
\bottomrule
\end{tabular}
\vspace{-0.5\baselineskip}
\end{table*}

\mysec{Ingest}
The \emph{Agents in the Wild} window fixes only which repositories are eligible; for the analysis itself we re-scrape each sampled repository's full pull-request history via the \GH{} API spanning January 2020 through February 2026, so the contrasts are not confined to that window. Although the scrape reaches back to 2020, agent-associated PRs are essentially absent before early 2025, so the agent-versus-human contrasts are driven by activity from roughly May 2025 onward. Of the 3{,}000 sampled repositories, 2{,}860 could be retrieved and traced; the remaining 140 could not be located at scrape time (deleted, made private, or renamed since sampling) and are excluded. Each PR is materialized as one row carrying repository linkage, identifiers, timestamps, merge state, changed files, and structured text for commits, comments, and reviews; churn metrics, PR-type labels, and review-derived features are computed in later offline passes. The stratified scrape retains over two and a half million PRs with churn fields and reviewer-side metadata where available.

\mysec{Rule-based author labeling}
Each PR receives an author bucket from deterministic rules rather than a learned classifier. Tool-specific labels are assigned when head-branch naming matches known product conventions, when the author login matches a known integration account, or when first-commit metadata matches known agent or bot patterns; vendor buckets include \texttt{codex}, \texttt{jules}, \texttt{copilot}, \texttt{devin}, \texttt{cursor}, \texttt{claude}, \texttt{openhands}, \texttt{codegen}, \texttt{cosine}, and \texttt{tembo}. Actors \GH{} marks as bots without a tool match are labeled \texttt{bot}; all remaining PRs default to \texttt{human}. The scheme is fast and auditable but imperfect: it can mislabel borderline automation, attribute multi-author work to one login, and under-count agents that avoid these signatures, so ``agent'' should be read as agent-\emph{associated under our rules} rather than verified end-to-end autonomy.

\mysec{Measures}
The independent variable is author type (agent-associated vs.\ human). The outcomes describe \emph{whether and how} a pull request is reviewed rather than its internal structure. \emph{Review coverage} records whether a merged PR received any human review at all. \emph{Latency} is time to merge. \emph{Discussion} is the volume of review commentary, reported both raw and normalized by change size as comments per thousand lines modified, since a fixed comment count means something different on a one-line and a thousand-line patch~\citeapp{bacchelli13}. \emph{Interaction pattern} categorizes who takes part in a PR's review thread under a deterministic rule over participants: no human review, author-only (the only human in the thread is the PR author, alone or alongside a bot), a single reviewer other than the author, or multiple people; this separates review \emph{coverage} from review \emph{independence}. Finally, each PR carries a purpose label (feature, bug fix, refactor, test, CI, and related categories in the spirit of Levin and Yehudai~\citeapp{levin17}) so that coverage can be read by change type.

\mysec{Estimation}
Because pull requests nest within repositories whose baseline activity and agent exposure differ by orders of magnitude, we summarize each outcome at the \emph{repository} level first and then aggregate across repositories, rather than pooling every PR into a single corpus-wide rate that the largest projects would dominate. Coverage and interaction-pattern outcomes are reported as means of per-project proportions; latency is reported as the median of per-project medians alongside the pooled median. For temporal trends we recompute the same summaries at repository-month granularity: a repository counts as an \emph{adopter} from the first month it records an agent-classified PR, and the headline agent-share series averages repository-level monthly agent fractions rather than a corpus-wide fraction. Our claims are descriptive; the one inferential test we report is a paired Wilcoxon signed-rank test on per-project author-only proportions, which holds repository identity constant by pairing on project. Where an outcome admits more than one defensible operationalization (for example, conditioning on all merged versus only reviewed PRs, or summarizing by mean versus distribution), we report the alternatives side by side rather than privileging one~\citeapp{gelman16, silberzahn18}.

\mysec{Limitations}
Our operationalization of ``agent-authored'' PRs relies on account classification, which may not capture all forms of AI assistance (\eg human PRs with substantial Copilot-generated code). Our review measures (coverage, comment volume, and thread participants) capture observable traces of review but may not reflect its depth, the difficulty of the change under review, or review that took place off-platform. Because the analysis is observational and cross-sectional, unobserved confounding factors may influence both agent adoption and the review outcomes we measure, so the study supports descriptive contrasts rather than causal claims. Our sample is limited to public \GH{} repositories, may not generalize to private or enterprise contexts, and reflects a specific time period given the rapidly evolving capabilities of coding agents. Finally, the sample is built entirely from repositories that already contain agent-authored PRs, and is stratified to span activity and agent-exposure regimes rather than to match every PR-level marginal; it therefore lacks a baseline control group of non-adopting projects, so aggregate contrasts should be read as descriptive of agent-exposed repositories rather than of \GH{} as a whole.

\mysec{Full results}
The findings from our empirical analysis reveal several trends with opposing implications, each of which we outline below. We deliberately scraped each sampled repository's full history back to 2020 to span three eras, pre-AI, non-agentic AI, and agentic AI, but across a range of operationalizations we do not discuss here we found no trends of interest before the agentic era, so we restrict the results to it, roughly May 2025 onward.

\begin{figure}[t]
  \centering
  \includegraphics[width=\linewidth]{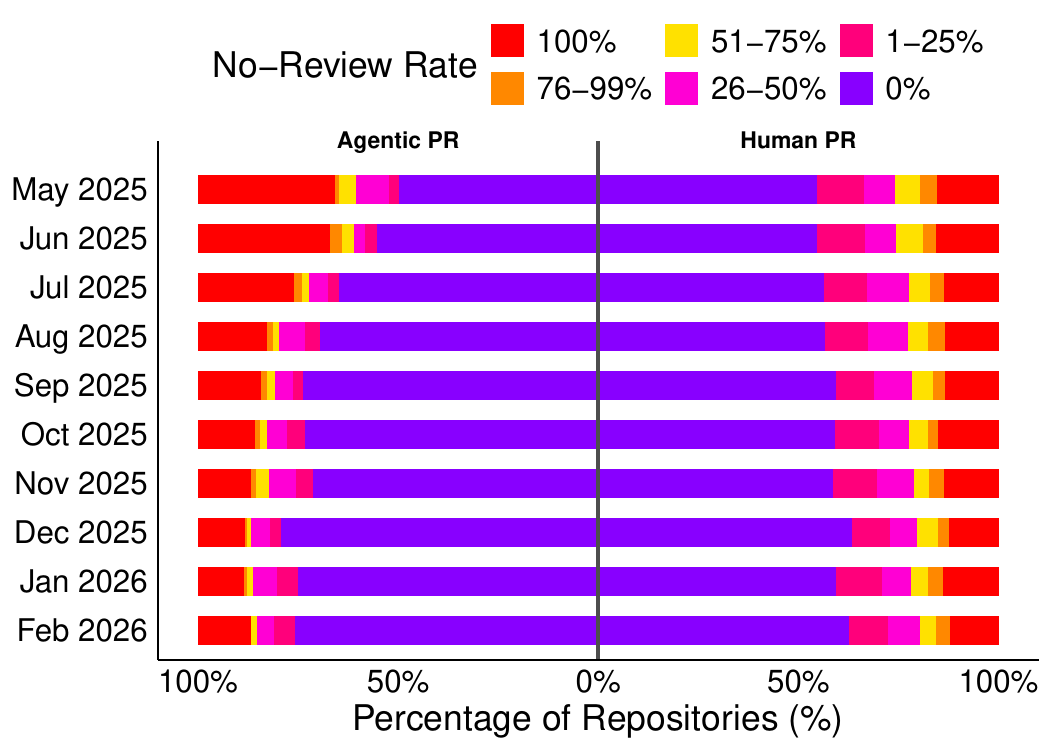}
  \caption{Distribution of per-project no-review rate over time.}
  \label{fig:PerRepoNoReviewRateDistributionOverTime}
\end{figure}

\noindent\textbf{No-Review Rate.}
The share of repositories that merge all agentic PRs without review decreased steadily over the observation period (\cf Figure~\ref{fig:PerRepoNoReviewRateDistributionOverTime}). In May 2025, 34\% of repositories with agentic PRs merged all of them with no review; but by February 2026 that share had fallen to 13\%, with the majority of the mass shifting into the 0\% no-review bin indicating a shift in practice towards reviewing agentic PRs.

Despite the clear decreasing trend in no-review rates for agentic PRs across the projects in our sample, it is also worth noting that merging PRs with no review appears to be a minority behavior to begin with generally speaking: the median per-project no-review rate is near 0\% for both author groups in every month (\cf Figure~\ref{fig:motivNoReview}, dashed lines), and the overall agentic no-review rate falls from over 50\% early in the window to roughly 12\% by February 2026, approaching the stable human rate of about 14\% (\cf Figure~\ref{fig:motivNoReview}, solid lines). The human no-review rate holds relatively stable throughout the observation period.

Among projects that merge PRs without review, no-review rates appear to differ by PR type: agentic PR no-review rates vary widely by PR type (test 69\%, refactor 41\%, bug fix 25\%) while human rates are nearly flat (feature 8\% -- CI 14\%), suggesting that projects triage which agentic changes to review by perceived risk rather than skipping review indiscriminately.

\emph{Takeaway.} Review coverage of agentic PRs is lower but improving over time, while review coverage of human PRs remains stable.

\begin{figure}[t]
  \centering
  \includegraphics[width=\linewidth]{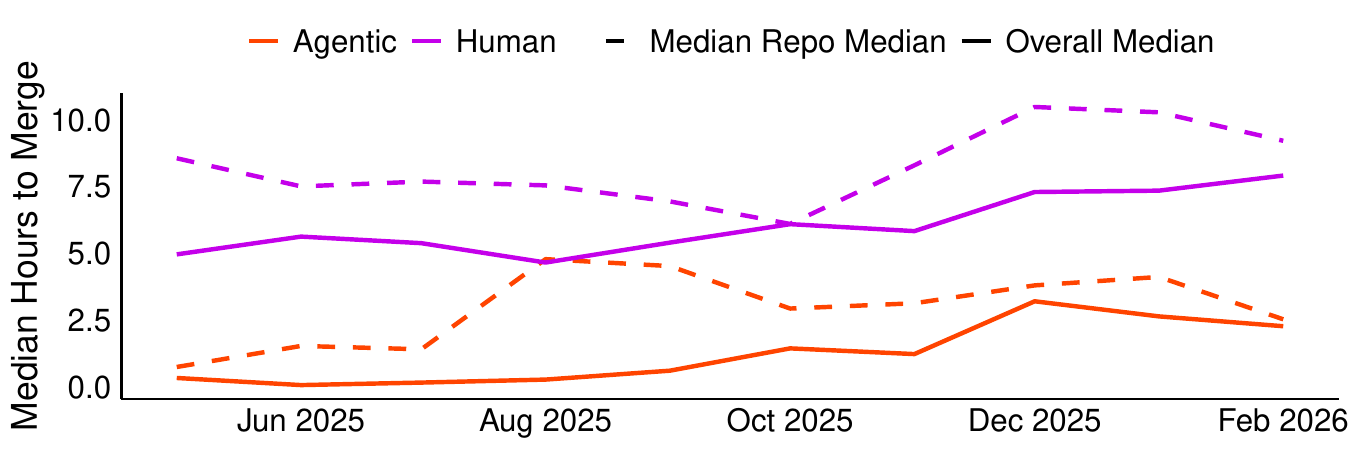}
  \caption{Median time to merge for human versus agentic PRs over time. Solid line = overall median, dashed line = median of per-project median.}
  \label{fig:timeToMergeByGroupOverTime}
\end{figure}

\noindent\textbf{Review Speed and Thoroughness.}
In every month of the observation period, the median time to merge is lower for agentic PRs than for human PRs, both pooled across all merged PRs and at the per-project level (\cf Figure~\ref{fig:timeToMergeByGroupOverTime}). Merge times lengthen for both groups across the window: the agentic median climbs from minutes early in the window (roughly 20 minutes in May 2025) to 2--3 hours by early 2026, while the human median rises from about 5 to 8 hours. The relative gap therefore narrows substantially, from more than tenfold to roughly threefold, even as the absolute gap of around 5 hours persists.

Agentic PRs also accumulate less discussion on their way to merging. Half of agentic PRs (50.2\%) receive no review comments at all, compared to 37.9\% of human PRs, and normalizing comment counts by change size, human PRs are more often heavily discussed (34.5\% versus 27.8\% exceed 20 comments per thousand lines modified). Taken together, agentic PRs spend less time in review and attract less commentary while there, both in absolute terms and per line changed.

\emph{Takeaway.} Agentic PRs move through review substantially faster, with less discussion per line changed, though the speed of agentic PR review is slowing down.

\begin{figure}[t]
  \centering
  \includegraphics[width=\linewidth]{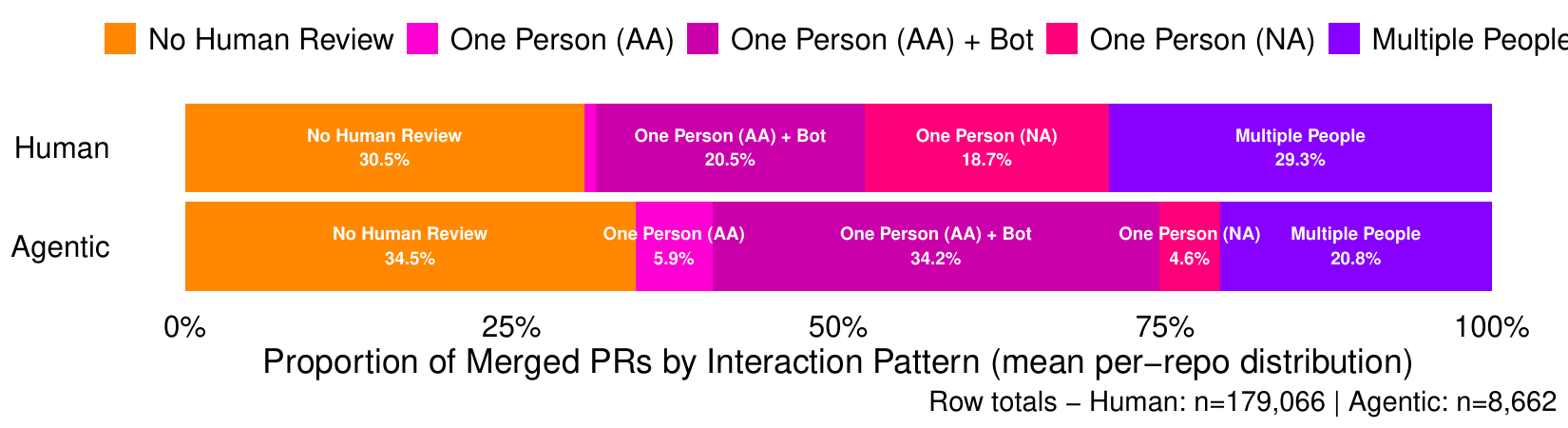}
  \caption{Mean per-project interaction pattern distribution by PR author group.}
  \label{fig:interactionPatternDistributionBarPlot}
\end{figure}

\begin{figure}[t]
  \centering
  \includegraphics[width=\linewidth]{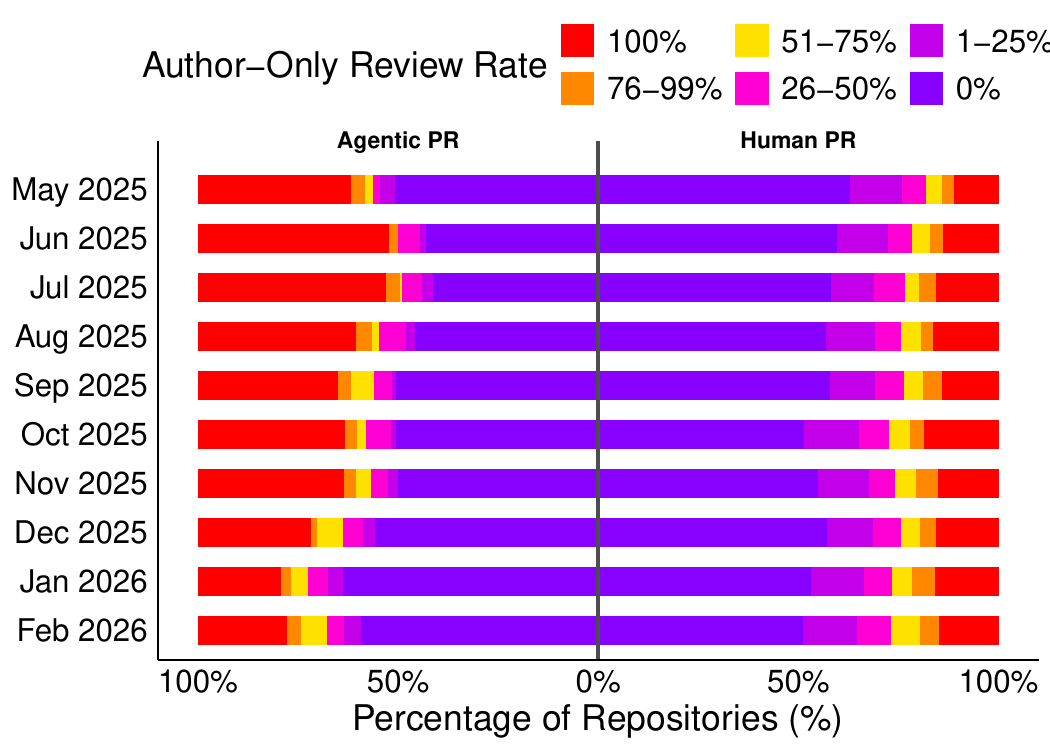}
  \caption{Distribution of per-project-month author-only review rate among PRs that received any human review, over time.}
  \label{fig:PerRepoauthorOnlyReviewRateDistributionOverTime}
\end{figure}

\noindent\textbf{Independent Review and Interaction Patterns.}
Averaging per-project proportions of each interaction pattern, review coverage is similar across author groups: on average, 34.5\% of a project's agentic PRs and 30.5\% of its human PRs are merged with no human review (\cf Figure~\ref{fig:interactionPatternDistributionBarPlot}). This per-project, whole-window average sits above the pooled end-of-window no-review rate reported earlier: it weights every project equally rather than by PR volume, and it averages over the entire window, including the early months when most agentic PRs went unreviewed. However, the composition of the review that does occur appears to differ.

Author-only review, in which the only human in the thread is the PR author, alone or alongside a bot, accounts for 40.1\% of agentic PRs versus 21.5\% of human PRs, while review by a single person other than the author accounts for 4.6\% versus 18.7\%, and review involving multiple people for 20.8\% versus 29.3\%. These results indicate review of agentic contributions appears less collaborative relatively speaking. While human PRs routinely bring a second person into the discussion, agentic PRs are most often examined only by the person who invoked the agent. A paired Wilcoxon signed-rank test on per-project author-only proportions confirms the difference ($k = 1{,}553$, $V = 418{,}564$, $p < 0.001$, $r = 0.42$); pairing on project holds repository identity constant, so the gap is not attributable to between-project differences.

However, within the subset of merged PRs with any human review, the share of repositories in which every reviewed agentic PR involves only its author also declines modestly over the window, though far less sharply than the no-review rate (\cf Figure~\ref{fig:PerRepoauthorOnlyReviewRateDistributionOverTime}): independent review of agentic code is growing, but most of the apparent movement appears to reflect more agentic PRs being reviewed at all rather than more people joining the reviews that happen.

Notably, Figures~\ref{fig:interactionPatternDistributionBarPlot} and~\ref{fig:PerRepoauthorOnlyReviewRateDistributionOverTime} support different headline conclusions from the same data. When averaging per-project proportions (Figure~\ref{fig:interactionPatternDistributionBarPlot}), only about a quarter of agentic PRs involve any human beyond the author. Yet when conditioning on reviewed PRs and binning per-project monthly rates (Figure~\ref{fig:PerRepoauthorOnlyReviewRateDistributionOverTime}), roughly half of project-months have no author-only reviews at all, suggesting projects that review agentic code routinely involve a second person. Both operationalizations are defensible; they differ in conditioning (all merged PRs versus reviewed PRs), unit of analysis (project versus project-month), and summary statistic (mean versus distribution). The ambiguity extends to the construct itself: if the agent is treated as the author, the human who invoked it is a second, independent set of eyes, and independent review has become more common in agentic PRs than it was for human-authored PRs according to Figure~\ref{fig:interactionPatternDistributionBarPlot}. However, if the agent is treated as a tool, that same human is the author reviewing their own work, and independent review has sharply declined.

\emph{Takeaway.} The review agentic PRs receive is less independent and involves less between-human collaboration.

\noindent\emph{Summary.} Many of the identified trends oppose one another. Review coverage of agentic PRs improves steadily while independent review of them grows only modestly; agentic PRs continue to merge several times faster than human PRs even as merge times lengthen for both groups and the relative gap narrows; review discussion on agentic PRs is thinner both in absolute terms and per line of code. Which story the data tells depends on defensible analysis choices~\citeapp{gelman16, silberzahn18}; some operationalizations indicate projects are folding agentic contributions into existing review practices and workflows, while others indicate that most agentic code is reviewed by the agent's operator, or by no humans at all. The traces establish what is changing, but not why, and the data alone cannot adjudicate between these scenarios; doing so requires a theory of what review of agent-authored code is for, what counts as review, and which of these changes carry practical consequences for the defect-detection and knowledge-sharing functions practitioners value review for~\citeapp{bacchelli13}.

\begin{figure}[t]
  \centering
  \includegraphics[width=\linewidth]{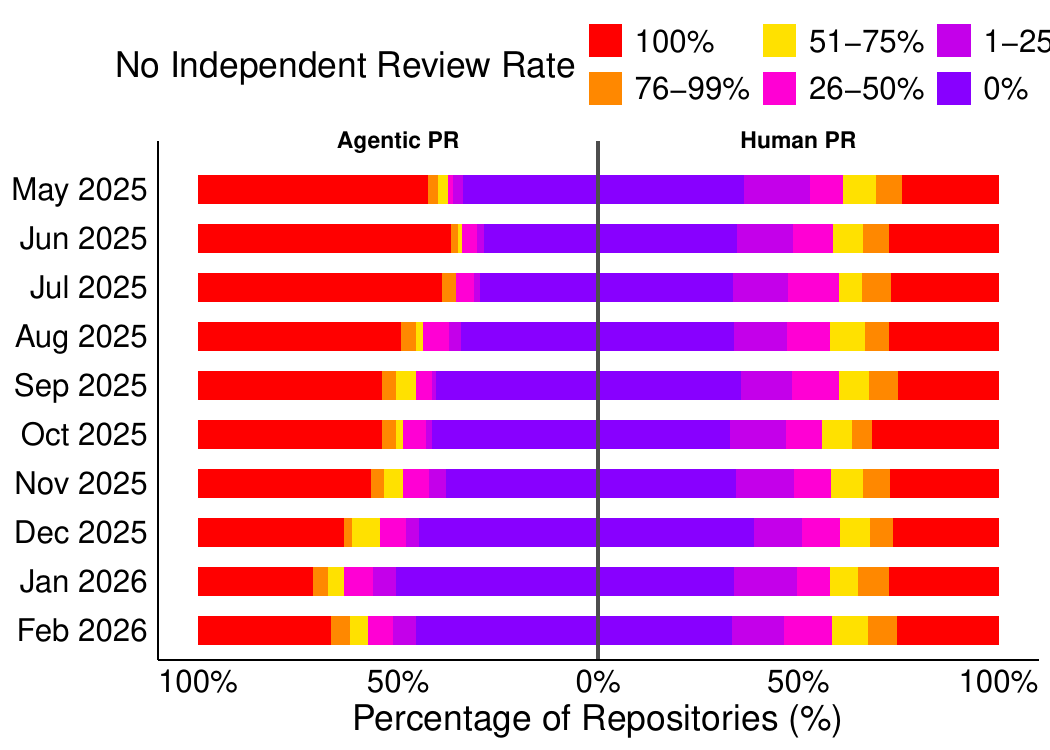}
  \caption{Distribution of the per-project no-independent-review rate over time (the share of a project's reviewed agentic PRs examined only by their author).}
  \label{fig:app_noindep}
\end{figure}

\begin{figure}[t]
  \centering
  \includegraphics[width=\linewidth]{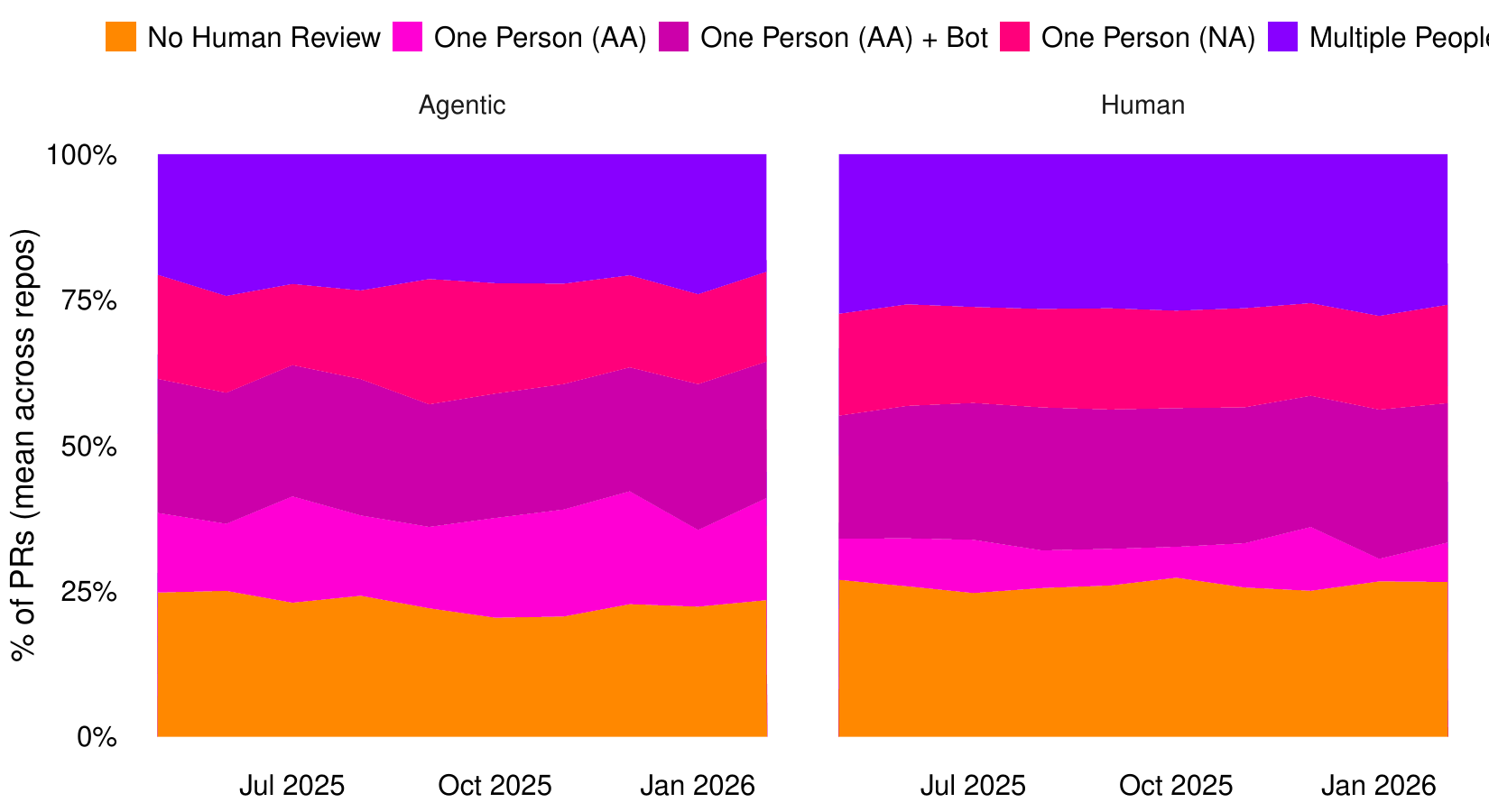}
  \caption{Mean per-project interaction-pattern distribution by PR author group over time.}
  \label{fig:app_interpattern_area}
\end{figure}

\begin{figure}[t]
  \centering
  \includegraphics[width=\linewidth]{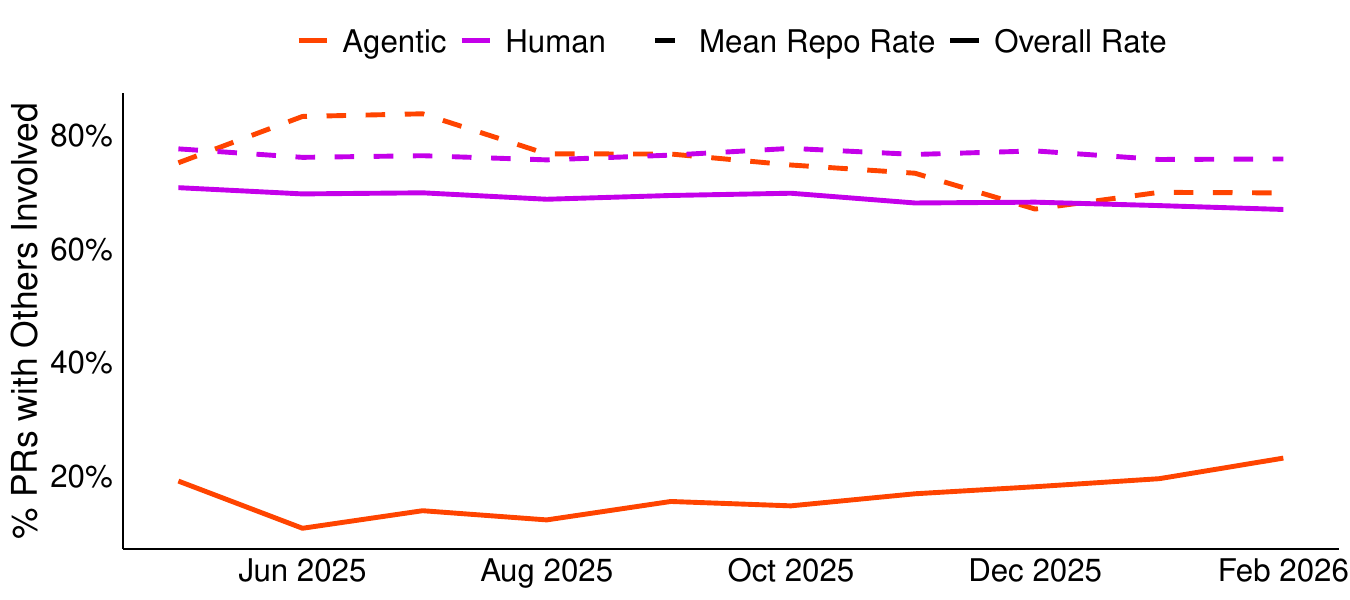}
  \caption{Proportion of merged PRs with a participant beyond the original author over time. Solid line: overall rate; dashed line: mean of per-project rate.}
  \label{fig:app_others}
\end{figure}

\mysec{Additional temporal figures}
Figures~\ref{fig:app_noindep}--\ref{fig:app_others} give three repository-month views that complement the cross-sectional contrasts in Section~\ref{sec:motivating_study}: the per-project no-independent-review rate over time, the mean per-project interaction-pattern mix over time, and the rate of PRs with a participant beyond the author over time. They are descriptive and not referenced by a specific claim in the main text.

\section{Theory Building: Corpus Construction}
\label{app:rq2corpus}

\mysec{Reddit sampling frame}
Table~\ref{tab:app_strata} lists the seven strata, their subreddits, and the threads collected from each under the review-word inclusion gate. A thread is kept if its title or body contains any of \emph{review}, \emph{reviews}, \emph{reviewing}, \emph{reviewer}, \emph{reviewers}, or \emph{reviewed} at a word boundary (r/codereview is kept in full). Whether a thread mentions AI is recorded as a tag, never used to include or exclude, using a vocabulary of tool and concept terms (Copilot, Cursor, Claude, Codex, Devin, Windsurf, Codeium, Aider, CodeRabbit, ChatGPT, Gemini, LLM, AI-generated, coding agent, agentic, vibe coding, and related forms).

\begin{table}[t]
\small\centering
\caption{Reddit sampling frame: seven strata, 43 subreddits, threads collected under the review-word gate (2020--2026).}
\label{tab:app_strata}
\setlength{\tabcolsep}{4pt}
\begin{tabularx}{\columnwidth}{@{}lXr@{}}
\toprule
Stratum & Representative subreddits & Threads \\
\midrule
code-review native & r/codereview & 541 \\
VCS workflow & r/git, r/github, r/gitlab & 904 \\
SE practice & r/ExperiencedDevs, r/programming, r/cscareerquestions, r/devops, r/webdev, $\ldots$ (13) & 14{,}717 \\
SE critical & r/programmingcirclejerk & 25 \\
language ecosystems & r/Python, r/rust, r/golang, r/javascript, $\ldots$ (9) & 3{,}772 \\
AI coding tools & r/ChatGPTCoding, r/ClaudeAI, r/cursor, r/GithubCopilot, r/vibecoding, r/AI\_Agents, $\ldots$ (8) & 7{,}660 \\
AI general & r/OpenAI, r/LocalLLaMA, r/ArtificialInteligence, r/singularity, $\ldots$ (8) & 3{,}460 \\
\midrule
\textbf{Total} & \textbf{43 subreddits} & \textbf{31{,}079} \\
\bottomrule
\end{tabularx}
\end{table}

\mysec{Web retrieval design}
The web register is retrieved through the Exa neural search engine~\citeapp{exa} with 68 natural-language queries, each directional query balanced by its opposite pole. Every query is run through three passes, each with its own 100-result budget: \emph{broad} (the open web, no domain restriction; 2{,}509 kept), \emph{feedspot} (the entire Feedspot ``Best Software Engineering Blogs'' ranking of 52 domains, taken wholesale; 2{,}658 kept), and \emph{curated} (39 institutional domains chosen by category: engineering-practice media, the agent vendors under review, AI review-tool vendors, and developer-productivity research groups; 2{,}463 kept). Raw results are written first; a reversible filter then deduplicates by URL, drops pages under 150 or over 15{,}000 words, and blocks document types that are never practitioner discourse (academic databases, PDF and book dumps, registries, content farms). This yields 7{,}630 articles across 1{,}437 domains.

\begin{figure}[t]
  \centering
  \includegraphics[width=\linewidth]{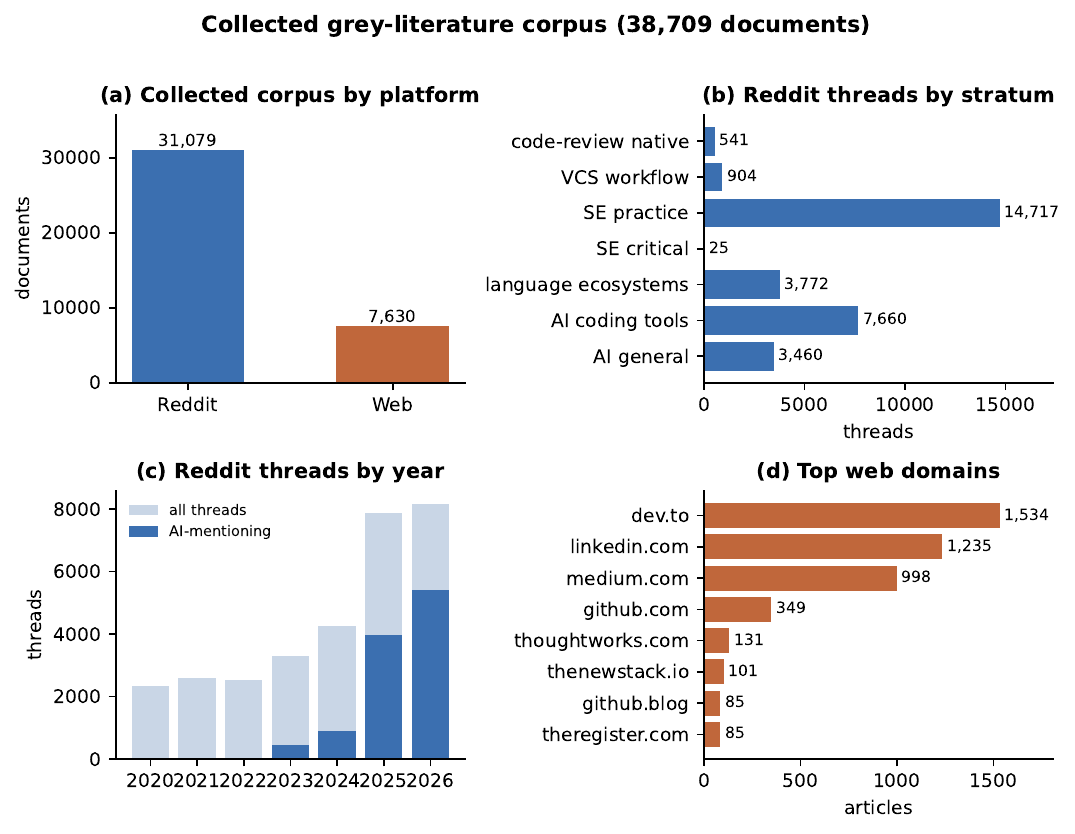}
  \caption{Collected grey-literature corpus (38{,}709 documents): composition by platform, Reddit stratum, year, and top web domain.}
  \label{fig:app_corpus}
\end{figure}

\begin{figure}[t]
  \centering
  \includegraphics[width=\linewidth]{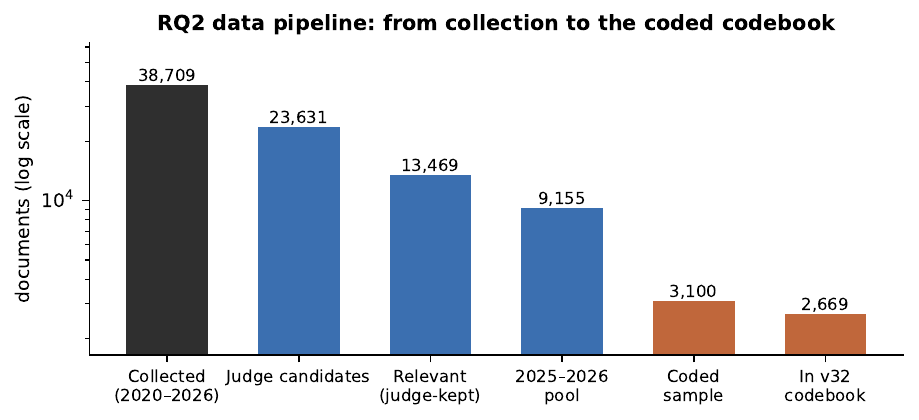}
  \caption{Theory-building data pipeline (log scale): collection, the relevance-judge candidate and kept sets, the 2025--2026 pool, the coded sample, and the documents represented in the final codebook.}
  \label{fig:app_funnel}
\end{figure}

\mysec{Corpus composition}
Figure~\ref{fig:app_corpus} shows the collected corpus by platform, Reddit stratum, year (AI-mentioning vs.\ all), and top web domain; Figure~\ref{fig:app_funnel} shows the full pipeline from collection to the coded codebook.

\section{Theory Building: Relevance Judge}
\label{app:rq2judge}

The precision filter (Section~\ref{sec:rq2_methods}) runs the prompt below over the 23{,}631 candidate documents (the Reddit code+review subset plus all web articles) with \emph{Gemini 2.5 Flash} at temperature 0, retaining the verdict, rationale, and full response per document. It keeps 13{,}469 documents (57.0\%: Reddit 9{,}241 of 16{,}001; web 4{,}228 of 7{,}630). A re-judgment of 885 documents by a stronger model agrees at Cohen's $\kappa = 0.75$ (substantial; 87.9\% raw agreement).

\begin{lstlisting}[style=prompt]
You are an expert empirical software-engineering researcher and a precise data
classifier. For the document below, decide whether it is relevant to the topic of
CODE REVIEW.

Code review is the software-engineering practice in which a developer's proposed or
written code is deliberately examined (by other engineers, by the author, or by
automated and AI tools) to assess its correctness, quality, and maintainability and
to give feedback, typically before the change is merged into a shared codebase. It
is fundamentally a collaborative quality-assurance and knowledge-sharing activity.
As a topic it encompasses the act of reviewing code, the reviewers and authors
involved, the process, norms, and tooling that surround it, and how the practice is
changing and with what consequences.

Mark the document RELEVANT (true) only when it substantively engages with code
review in this sense: code review, or some facet of it, is the document's focus or a
substantive part of its discussion, rather than a passing or incidental mention.
Apply this judgement regardless of stance, length, or format, and whether or not
automated or AI tools are involved.

Mark it NOT RELEVANT (false) when the document is essentially about something else,
even if the words "code" or "review" appear in another sense, for instance a
performance or product review, "review" used to mean revising or studying, generic
coding or debugging help with no review angle, or text that merely mentions code
review in passing.

Judge only from the text provided (it may be truncated). Respond with a single JSON
object containing, in this order: "rationale" -- one sentence explaining the
decision; "relevant" -- true or false; and "confidence" -- one of "high", "medium",
or "low" in that decision.
\end{lstlisting}

\section{Theory Building: Coding Pipeline}
\label{app:rq2coding}

\begin{figure}[t]
  \centering
  \includegraphics[width=\linewidth]{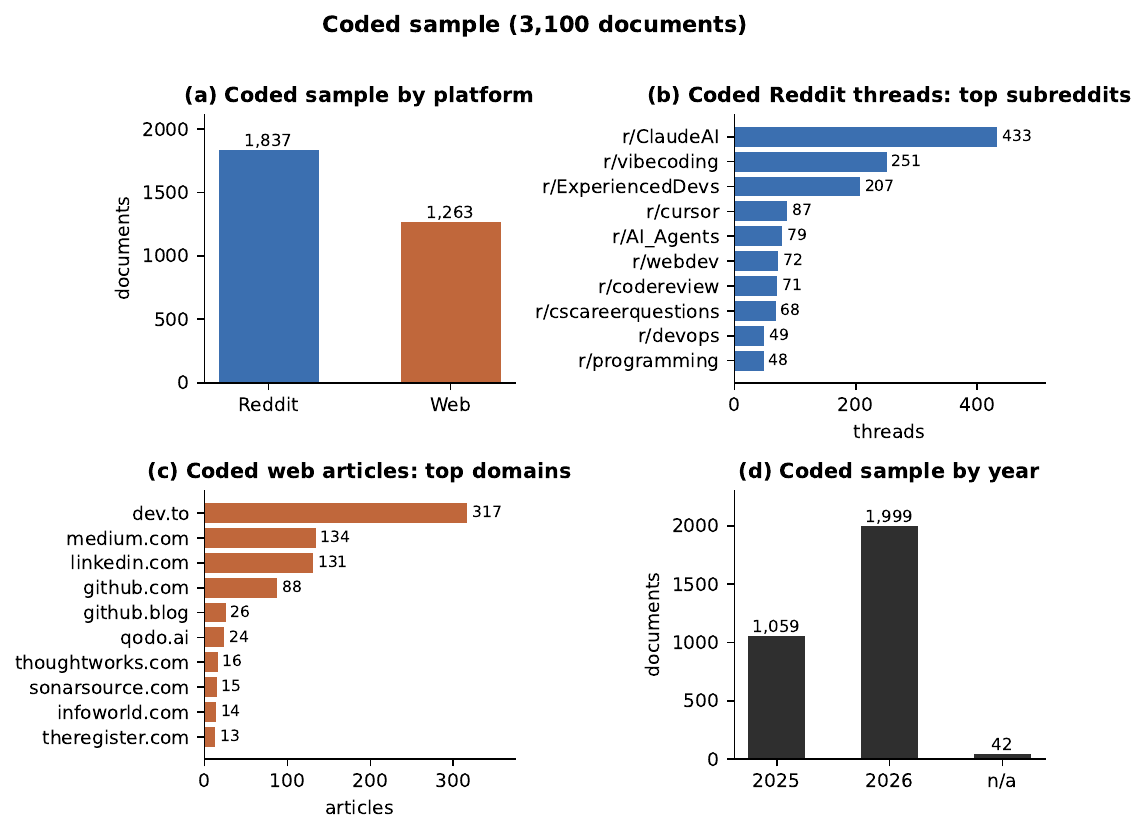}
  \caption{Coded sample (3{,}100 documents): composition by platform, top subreddits, top web domains, and year.}
  \label{fig:app_coded}
\end{figure}

We code a source-stratified random sample of 3{,}100 documents (1{,}837 Reddit threads, 1{,}263 web articles; 2025: 1{,}059, 2026: 1{,}999; about 12{,}000 pages) with the Thematic-LM multi-agent design~\citeapp{qiao2025thematiclm}. Each Reddit thread is one coding unit; each web article is split into topical segments by an LLM-driven segmenter. Figure~\ref{fig:app_coded} shows the coded sample's composition. All coding-stage LLM calls use \emph{Gemini 3.1 Flash Lite}; the segmenter uses \emph{Gemini 2.5 Flash Lite}. Coding runs in batches of 100 documents (31 batches), finalizing a codebook revision after each; the final codebook holds 4{,}838 distinct codes grounded in 109{,}951 quotes from 2{,}669 of the sampled documents.

\mysec{Research context given to every agent}
\begin{lstlisting}[style=prompt]
This is a thematic analysis of how the code review process is changing, especially
with the arrival of AI coding agents, and what the consequences of that change are.

Code review is the software-engineering practice in which a developer's proposed or
written code is deliberately examined (by other engineers, by the author, or by
automated and AI tools) to assess its correctness, quality, and maintainability and
to give feedback, typically before the change is merged into a shared codebase. It
is fundamentally a collaborative quality-assurance and knowledge-sharing activity.
As a topic it encompasses the act of reviewing code, the reviewers and authors
involved, the process, norms, and tooling that surround it, and how the practice is
changing and with what consequences.

Code from any period is in scope, whether or not it mentions AI, because the aim is
to characterize this change over time and its consequences rather than to study AI
for its own sake. Material that does not bear on the practice of code review is out
of scope.
\end{lstlisting}

\mysec{The three coder identities}
The three coders read every segment through deliberately contrasting analytic lenses:
\begin{lstlisting}[style=prompt]
[Neutral] An empirical software engineering researcher coding inductively in the
grounded-theory tradition, building codes bottom-up from the participants' own
words, without preconceived categories or any prior view on whether AI helps or
harms code review, attentive both to the process and practice of code review
(workflow, effort, norms, standards, roles, tooling, the speed and volume of
changes) and to its human and social dimensions (trust, accountability, skill,
learning, collaboration, and the relationships among the people and tools
involved), and to how all of these are shifting.

[Critical] An empirical software engineering researcher reading the data through a
critical lens, attentive to ways AI coding agents and automated review may erode
code review, including its quality, the depth of understanding it provides, trust,
and accountability.

[Appreciative] An empirical software engineering researcher reading the data through
an appreciative lens, attentive to ways AI coding agents and AI review tools
strengthen and augment code review, including speed, coverage, leverage, and new
capabilities.
\end{lstlisting}

\mysec{Segmenter prompt (web articles)}
\begin{lstlisting}[style=prompt]
You segment documents into topical sections for qualitative analysis. You receive a
document with each line prefixed by its line number. Identify the line numbers where
major topic shifts occur. Each segment must be at least paragraph-length (several
sentences); prefer fewer, larger segments over many small ones. Shorter documents
have 1-5 segments, longer documents can have 10 to 50 segments, rarely much more
than that. A boundary marks a genuine shift in what is being discussed, not a minor
turn within the same topic. The first segment must start at line 1. For each
segment, give its start_line and a 3-8 word title describing its topic.
\end{lstlisting}

\mysec{Coder prompt}
\begin{lstlisting}[style=prompt]
You are a coder in thematic analysis. When given a text segment, write 1-3 codes for
the segment. The code should capture concepts or ideas with the most analytical
interest, relevant to the research focus.

For each code, provide a short description (one sentence) of what the concept means
as a general analytic category, and extract one or more quotes from the segment
corresponding to the code. Each quote needs to be an extract from a sentence --
copied verbatim from the segment, not paraphrased.

When an existing code in the codebook fits, reuse its exact label.

If the segment does not address the research focus, return an empty list of codes.
Do not invent codes to cover off-topic material.
\end{lstlisting}

\mysec{Aggregator prompt}
\begin{lstlisting}[style=prompt]
You are an expert qualitative researcher responsible for organizing codes from
multiple coders. Your task is to identify codes with similar meanings that should be
merged, while retaining codes that represent distinct concepts.

Guidelines for Code Aggregation:
1. Semantic similarity: Merge codes that capture the same underlying concept
2. Preserve nuance: Keep codes separate if they capture different aspects
3. Create clear labels: When merging, create a clear label for the merged code
4. Document decisions: Explain why codes were merged or kept separate

Respond with JSON listing merge groups (each with merged_code, original_codes,
rationale) and retain_codes (kept separate, unchanged).
\end{lstlisting}

\mysec{Reviewer prompt}
\begin{lstlisting}[style=prompt]
You are an expert qualitative researcher responsible for reviewing codes and
maintaining codebook quality. Your task is to evaluate aggregated codes and decide
whether to add them to the codebook, merge them with existing codes, or reject them.

Guidelines: (1) Quality check -- codes are clear, meaningful, well-defined;
(2) Consistency with existing entries; (3) Redundancy -- does it duplicate existing
concepts; (4) Relevance to the research questions.

Decision options: add_new, merge_existing (specify target), or reject. Respond with
JSON: decision, merge_target (if merging), rationale.
\end{lstlisting}

\section{Theory Building: From Codebook to Theory}
\label{app:rq2theory}

Theme development is manual but LLM-assisted (Section~\ref{sec:rq2_methods}). For each analytic lens, an LLM proposes candidate themes as causal \emph{claims} over the full codebook; the authors validate and redefine them; a further pass derives \emph{sub-themes} within each theme and assigns every code to one; and the resulting sub-theme claims are then grounded by querying the codebook as a search engine. The theme-proposal and sub-theme prompts that drive this are below. We separately tested a more direct route, extracting causal statements straight from the codebook in a single bottom-up pass over the codes, whose prompt and outcome are given at the end of this appendix.

\mysec{Theme-proposal prompt}
\begin{lstlisting}[style=prompt]
You adopt the following analyst perspective (verbatim, the same perspective used
during coding): {lens}

From that perspective, you are doing reflexive thematic analysis of a large corpus
(Reddit discussion and web articles, 2025-2026) to answer this research question:
{research question}

Below is the full codebook from an earlier coding pass: every code, with the number
of supporting quotes in brackets (a rough salience signal). Read it as a whole.

Propose 6 to 10 candidate THEMES that answer the research question from your lens.

What a theme is, and is NOT:
- A theme is a PATTERN OF SHARED MEANING organized around a central concept -- an
  analytic CLAIM that says something interesting about how review is changing and
  with what consequence. Write each title as a claim/argument, e.g. "The bottleneck
  shifts from writing code to reviewing it", NOT a topic label like "AI tooling".
- A theme CUTS ACROSS many codes; it is not a single code renamed, and not a
  category of similar codes.
- Capture the CHANGE as the data describes it (a from -> to shift) and its
  CONSEQUENCE. The corpus is all 2025-2026, so the change is what participants
  RECOUNT/EXPERIENCE, not a time trend.

For each theme give: a claim-style title; a one-to-two sentence claim; the central
organizing concept; the from->to shift; the consequence; and a brief note of which
parts of the codebook it draws on. Use only what is supported by the codebook.
\end{lstlisting}

\mysec{Sub-theme prompt}
Within each theme's codes, an LLM derives the sub-theme \emph{claims}; a short follow-up pass then assigns each code to its single best-fitting sub-theme.
\begin{lstlisting}[style=prompt]
You are analysing the codes within ONE category of a thematic analysis of
practitioner discourse (Reddit + web, 2025-2026). Research question (scope):
{research question}

Category: "{category name}" -- {category description}

Below are the codes assigned to this category (with supporting-quote counts).
Identify the recurring SUB-THEMES -- the distinct claims or patterns these codes
collectively express within this category.

Rules:
- A sub-theme is a specific CLAIM or pattern the codes make (e.g. "AI introduces new
  security vulnerabilities", or "AI-generated code adheres better to standards").
- COMPETING or OPPOSING claims must be kept as SEPARATE sub-themes -- do not merge
  "AI improves X" with "AI harms X". Surface both sides exactly as the codes express
  them; do not impose a direction or balance.
- Aim for a handful (about 4 to 8) of sub-themes that together cover the codes; each
  substantive and distinct.
- Derive them strictly from what the codes say.

For each sub-theme: a stable id (s1, s2, ...); a clear claim-style statement; a
one-sentence description; and inclusion criteria for assigning codes to it.

CODES:
{codes with supporting-quote counts}
\end{lstlisting}

\mysec{Direct causal extraction (tested alternative)}
As an alternative to the theme and sub-theme route above, we tested extracting causal statements directly from the codebook in a single bottom-up pass over the codes. It produced 15{,}029 statements, but with many shallow, redundant, and overlapping claims, so we built the final model from the sub-theme claims above rather than from this direct extraction.
\begin{lstlisting}[style=prompt]
We study ONE phenomenon: software teams increasingly use AI coding agents to WRITE
code, and we want the impact on the CODE REVIEW PROCESS. Keep two settings SEPARATE:
- AUTOMATED code review: AI / tools acting as the reviewer (AI reviewing code).
- MANUAL code review: humans reviewing code, especially AI-written code.

Below are practitioner-derived CODES (each a recurring idea distilled from online
discourse, 2025-2026) with supporting verbatim QUOTES. For EACH code, extract EVERY
distinct CAUSAL STATEMENT it genuinely supports ABOUT THE CODE REVIEW PROCESS or its
consequences.

A causal statement is: SUBJECT (the cause/driver) --RELATION--> OBJECT (the affected
aspect of review, or a downstream consequence).

For each statement provide:
- code_id (the code it came from)
- subject, relation (a short causal verb phrase: increases, reduces, erodes, enables,
  triggers, shifts, requires, degrades, improves, replaces, obscures, accelerates),
  object (concise noun phrases)
- review_type: "automated" | "manual" | "both" | "general"
- topic: one of [process, reviewer-load, trust-verification, rigor, defect-quality,
  security, speed-throughput, knowledge-transfer, team-awareness, reviewer-skill,
  accountability, governance, tooling, cost, reviewer-role]
- polarity: "benefit" | "harm" | "neutral"
- quote: ONE verbatim supporting quote, copied exactly from the code's quotes.

Rules:
- ONLY statements about the code review process or its consequences (NOT generic
  commentary about AI or coding in general).
- Ground every statement in the quotes; never invent. If a code supports no causal
  statement about review, omit it.
- A single code may yield MULTIPLE statements, INCLUDING OPPOSING ones, keep them all.
- Be specific and faithful to what the code/quotes actually say.

CODES:
{codes with verbatim quotes}
\end{lstlisting}

\mysec{Evidence tiers}
Every relationship in the theory is tagged by how it is grounded: \emph{quote-grounded} (stated outright in practitioner text), \emph{read-into} (a ceteris-paribus reading in which the variables move together but the link is not stated as such), and \emph{by-definition} (the internal wiring of the five-construct review act, which holds by how the terms are defined). The three contested relationships additionally carry opposing practitioner text, since their sign is set by a moderator rather than fixed.

\clearpage
\onecolumn
\renewcommand{\arraystretch}{1.15}
\section{Supplementary Document Index}
\label{app:docindex}
This index lists the 3100 coded grey-literature documents that ground the theory (RQ2). Each carries a stable handle \texttt{G\#}; in-text grounded quotations cite these handles. The \emph{doc id} column is the identifier in the released dataset. Sorted by \emph{doc id}. Non-ASCII characters in titles are transliterated.

\smallskip
\noindent Web sources were collected through the Exa neural search index, which captures each page's text content at retrieval time. For a small number of sources, most notably LinkedIn posts, this captured text may not be directly visible to a visitor following the live URL, since the page can render its body dynamically or gate it behind a sign-in. In such cases the URL identifies the source document, while the exact quoted text comes from the content indexed at collection time rather than the page's live rendered view.
\medskip
\footnotesize
\begin{longtable}{@{}r r p{0.44\textwidth} p{0.30\textwidth}@{}}
\textbf{G\#} & \textbf{doc id} & \textbf{Title} & \textbf{URL} \\ \hline \endhead
G1 & 4 & Has anyone used AI style auto code review tools for GitHub PRs? & \url{https://www.reddit.com/r/webdev/comments/1mod7ry/has_anyone_used_ai_style_auto_code_review_tools/} \\
G2 & 7 & GitHub - zulrah93/wave\_t: A C++ header only for examining and writing .WAV PCM audio files & \url{https://www.reddit.com/r/cpp/comments/1lwlj1k/github_zulrah93wave_t_a_c_header_only_for/} \\
G3 & 9 & Why do Indian developers like calling and setting up meetings & \url{https://www.reddit.com/r/cscareerquestions/comments/1hy0xdc/why_do_indian_developers_like_calling_and_setting/} \\
G4 & 10 & Anyone actually running multiple AI agents in parallel for coding? & \url{https://www.reddit.com/r/AI_Agents/comments/1o0rvdq/anyone_actually_running_multiple_ai_agents_in/} \\
G5 & 13 & I really worry that ChatGPT/AI is producing very bad and very lazy junior engineers & \url{https://www.reddit.com/r/ExperiencedDevs/comments/1lb5ktb/i_really_worry_that_chatgptai_is_producing_very/} \\
G6 & 18 & Appropriate Setup for Claude in Enterprise & \url{https://www.reddit.com/r/ClaudeAI/comments/1sg44zj/appropriate_setup_for_claude_in_enterprise/} \\
G7 & 21 & [ Removed by moderator ] & \url{https://www.reddit.com/r/devops/comments/1suhoss/do_you_trust_ai_agents_running_code_on_your/} \\
G8 & 22 & Council of Dragons - A prompt engineering framework for Claude Code that keeps context from drifting off a cliff & \url{https://www.reddit.com/r/ClaudeAI/comments/1qb0rlz/council_of_dragons_a_prompt_engineering_framework/} \\
G9 & 23 & Doing code review on the 10,000 lines Claude Code wrote & \url{https://www.reddit.com/r/ClaudeAI/comments/1pc6thq/doing_code_review_on_the_10000_lines_claude_code/} \\
G10 & 24 & Should I leave my job?? & \url{https://www.reddit.com/r/ExperiencedDevs/comments/1rdqt6z/should_i_leave_my_job/} \\
G11 & 41 & I built an AI-powered GitHub App to automate PR review and issue triageI've been experimenting with automating some common GitHub repository workflows. & \url{https://www.reddit.com/r/github/comments/1rmkt34/i_built_an_aipowered_github_app_to_automate_pr/} \\
G12 & 43 & Strict(strong)-typing in Rust. & \url{https://www.reddit.com/r/rust/comments/1skpua5/strictstrongtyping_in_rust/} \\
G13 & 52 & Never Seen This Before on Claude Sonnet 4 & \url{https://www.reddit.com/r/ClaudeAI/comments/1nejzdx/never_seen_this_before_on_claude_sonnet_4/} \\
G14 & 57 & Is it just me or is reviewing PRs getting exponentially harder? & \url{https://www.reddit.com/r/programming/comments/1rddoyn/is_it_just_me_or_is_reviewing_prs_getting/} \\
G15 & 59 & The orcs got an upgrade We taught them to plan, gave them notifications, and now they mass-ship reviewed PRs while I drink coffee. I keep checking if the PRs ar & \url{https://www.reddit.com/r/ClaudeAI/comments/1s1y08s/the_orcs_got_an_upgrade_we_taught_them_to_plan/} \\
G16 & 60 & Reviewing ai slop & \url{https://www.reddit.com/r/ExperiencedDevs/comments/1lq5nnv/reviewing_ai_slop/} \\
G17 & 66 & Claude's new Feature "Code Review AI" with confidence-based filtering: Is this the end of manual "nitpicking" in PRs? & \url{https://www.reddit.com/r/ClaudeAI/comments/1rqwgn3/claudes_new_feature_code_review_ai_with/} \\
G18 & 68 & Ai coding tools prediction for upcoming 4 years & \url{https://www.reddit.com/r/vibecoding/comments/1qcqlen/ai_coding_tools_prediction_for_upcoming_4_years/} \\
G19 & 70 & Experiment: commit-time LLM code reviews using Git hooks (diff-scoped analysis) & \url{https://www.reddit.com/r/ArtificialInteligence/comments/1t0xwyx/experiment_committime_llm_code_reviews_using_git/} \\
G20 & 79 & Diff Synapse - A new way to understand code changes & \url{https://www.reddit.com/r/ChatGPTCoding/comments/1lua73k/diff_synapse_a_new_way_to_understand_code_changes/} \\
G21 & 81 & Bringing Code Review to Claude Code & \url{https://www.reddit.com/r/artificial/comments/1rqatlf/bringing_code_review_to_claude_code/} \\
G22 & 83 & I did security review of vibe coded apps and they are a nightmare when it comes to security & \url{https://www.reddit.com/r/vibecoding/comments/1seowva/i_did_security_review_of_vibe_coded_apps_and_they/} \\
G23 & 85 & [AskJS] Would you actually use this? I'm building a code review assistant that understands your app like this. & \url{https://www.reddit.com/r/javascript/comments/1jxples/askjs_would_you_actually_use_this_im_building_a/} \\
G24 & 91 & Roast my first C++ project: An N-Body Gravity Simulator. Looking for ruthless code review and architecture feedback! & \url{https://www.reddit.com/r/cpp/comments/1s8mxv0/roast_my_first_c_project_an_nbody_gravity/} \\
G25 & 96 & Take your team to the next level 8 & \url{https://www.reddit.com/r/windsurf/comments/1kgcraf/take_your_team_to_the_next_level_8/} \\
G26 & 105 & Looking for a Code Review - Tanstack Start with E2E Encryption & \url{https://www.reddit.com/r/codereview/comments/1ra6jor/looking_for_a_code_review_tanstack_start_with_e2e/} \\
G27 & 107 & AGENTS.md trick that stopped Codex from doing dumb work at premium rates & \url{https://www.reddit.com/r/OpenAI/comments/1t3jsot/agentsmd_trick_that_stopped_codex_from_doing_dumb/} \\
G28 & 111 & : Claude + Codex & \url{https://www.reddit.com/r/ClaudeAI/comments/1rti39t/_claude_codex_/} \\
G29 & 114 & I built an open source AI workflow orchestrator with GitOps-friendly YAML DSL & \url{https://www.reddit.com/r/devops/comments/1mslngo/i_built_an_open_source_ai_workflow_orchestrator/} \\
G30 & 119 & Why are almost all AI code audit skills just smarter linters? & \url{https://www.reddit.com/r/AI_Agents/comments/1sjmi4m/why_are_almost_all_ai_code_audit_skills_just/} \\
G31 & 128 & 7 Claude Code Power Tips Nobody's Talking About & \url{https://www.reddit.com/r/ClaudeAI/comments/1qstcb9/7_claude_code_power_tips_nobodys_talking_about/} \\
G32 & 129 & I kept fixing the same AI skill problems, so I built a governance framework in Markdown. & \url{https://www.reddit.com/r/ClaudeAI/comments/1s6q6rw/i_kept_fixing_the_same_ai_skill_problems_so_i/} \\
G33 & 130 & Is Pro Plan worth it ? & \url{https://www.reddit.com/r/ClaudeAI/comments/1pjqifm/is_pro_plan_worth_it/} \\
G34 & 132 & 3 reasons why vibe coding can't survive production & \url{https://www.reddit.com/r/ArtificialInteligence/comments/1ojf0xe/3_reasons_why_vibe_coding_cant_survive_production/} \\
G35 & 133 & Loving GPT 5.1 (high reasoning) !! & \url{https://www.reddit.com/r/windsurf/comments/1padz87/loving_gpt_51_high_reasoning/} \\
G36 & 137 & How do you evaluate engineers when everyone's using AI coding tools now & \url{https://www.reddit.com/r/ExperiencedDevs/comments/1pn39gb/how_do_you_evaluate_engineers_when_everyones/} \\
G37 & 140 & Have you tried any AI code/PR review tools? & \url{https://www.reddit.com/r/AI_Agents/comments/1p33dgd/have_you_tried_any_ai_codepr_review_tools/} \\
G38 & 143 & Help me improve my coding workflow & \url{https://www.reddit.com/r/vibecoding/comments/1tioh6z/help_me_improve_my_coding_workflow/} \\
G39 & 151 & Amazon service was taken down by AI coding bot & \url{https://www.reddit.com/r/webdev/comments/1r9pqkz/amazon_service_was_taken_down_by_ai_coding_bot/} \\
G40 & 157 & Devs using Cursor how do you handle testing? & \url{https://www.reddit.com/r/cursor/comments/1rzsy40/devs_using_cursor_how_do_you_handle_testing/} \\
G41 & 160 & Security and Maintainability Guide for Vibe-Coders & \url{https://www.reddit.com/r/vibecoding/comments/1tlvpjr/security_and_maintainability_guide_for_vibecoders/} \\
G42 & 164 & How we got amazing results from Claude Code (it's all about the prompting strategy) & \url{https://www.reddit.com/r/ClaudeAI/comments/1os5khh/how_we_got_amazing_results_from_claude_code_its/} \\
G43 & 168 & welcome to the new PXGH & \url{https://www.reddit.com/r/ClaudeAI/comments/1suxbg5/welcome_to_the_new_pxgh/} \\
G44 & 169 & Agentic Engineering is just Vibe coding & \url{https://www.reddit.com/r/ExperiencedDevs/comments/1syrcgw/agentic_engineering_is_just_vibe_coding/} \\
G45 & 170 & Role play as an absolute savant of an engineer & \url{https://www.reddit.com/r/cursor/comments/1iq86u3/role_play_as_an_absolute_savant_of_an_engineer/} \\
G46 & 171 & ISO: Tool for viewing individual "performance" on teams (free to .edu or open source?) & \url{https://www.reddit.com/r/github/comments/1p3e8l6/iso_tool_for_viewing_individual_performance_on/} \\
G47 & 172 & Tips for babysitting a vibecoded app? & \url{https://www.reddit.com/r/ExperiencedDevs/comments/1sskdxm/tips_for_babysitting_a_vibecoded_app/} \\
G48 & 175 & I hired a senior dev to review my code and this is what he said & \url{https://www.reddit.com/r/vibecoding/comments/1s3ut9k/i_hired_a_senior_dev_to_review_my_code_and_this/} \\
G49 & 182 & Building a marketplace with vibe code & \url{https://www.reddit.com/r/vibecoding/comments/1s3feme/building_a_marketplace_with_vibe_code/} \\
G50 & 184 & Manager causing problems & \url{https://www.reddit.com/r/ExperiencedDevs/comments/1nd9bob/manager_causing_problems/} \\
G51 & 186 & Opinion on AI code review tools? Any good ones? & \url{https://www.reddit.com/r/vibecoding/comments/1p5keus/opinion_on_ai_code_review_tools_any_good_ones/} \\
G52 & 188 & Peers MCP server - let AI coding sessions talk to each other & \url{https://www.reddit.com/r/ClaudeAI/comments/1s8k7dt/peers_mcp_server_let_ai_coding_sessions_talk_to/} \\
G53 & 191 & Tell your AI to block XSS attacks or hackers will thank you later & \url{https://www.reddit.com/r/vibecoding/comments/1kpwym2/tell_your_ai_to_block_xss_attacks_or_hackers_will/} \\
G54 & 197 & Lightweight RAG Claude can query? & \url{https://www.reddit.com/r/LocalLLaMA/comments/1ncfp6q/lightweight_rag_claude_can_query/} \\
G55 & 205 & Comparing two most popular AI Code Review tools: Bito vs CodeRabbit & \url{https://www.reddit.com/r/codereview/comments/1otitlr/comparing_two_most_popular_ai_code_review_tools/} \\
G56 & 206 & Vibe coding without understanding the language felt wrong and that's actually a good sign & \url{https://www.reddit.com/r/vibecoding/comments/1to6rcz/vibe_coding_without_understanding_the_language/} \\
G57 & 221 & Tips/advice for using Claude Code with PR reviews? & \url{https://www.reddit.com/r/ClaudeAI/comments/1ltbrxu/tipsadvice_for_using_claude_code_with_pr_reviews/} \\
G58 & 225 & Ever walked away from your code out of frustration... and wished someone debugged it while you were gone? & \url{https://www.reddit.com/r/vibecoding/comments/1o2ap16/ever_walked_away_from_your_code_out_of/} \\
G59 & 227 & Would you be interested in a Slack bot that helps teams using GitLab (and optionally GCP) manage merge requests, releases, code review requests more easily? & \url{https://www.reddit.com/r/gitlab/comments/1mcll1a/would_you_be_interested_in_a_slack_bot_that_helps/} \\
G60 & 229 & I built a open-source high-performance IP geolocation service using Go & \url{https://www.reddit.com/r/programming/comments/1p772ie/i_built_a_opensource_highperformance_ip/} \\
G61 & 234 & Rust backend authentication module -- code review & \url{https://www.reddit.com/r/rust/comments/1qwkw7s/rust_backend_authentication_module_code_review/} \\
G62 & 240 & Code review reality - How to learn best practices outside work? & \url{https://www.reddit.com/r/webdev/comments/1pd4292/code_review_reality_how_to_learn_best_practices/} \\
G63 & 243 & made one rule for PRs: no diagram means no review. reviews got way faster. & \url{https://www.reddit.com/r/devops/comments/1qr131v/made_one_rule_for_prs_no_diagram_means_no_review/} \\
G64 & 244 & Tea app & \url{https://www.reddit.com/r/vibecoding/comments/1md6jsu/tea_app/} \\
G65 & 245 & To All Software Engineers In The AI Age, Do You Still Like Your Job ? & \url{https://www.reddit.com/r/SoftwareEngineering/comments/1tnx2c2/to_all_software_engineers_in_the_ai_age_do_you/} \\
G66 & 246 & We built an AI to review your pull requests & \url{https://www.reddit.com/r/github/comments/1kybvmb/we_built_an_ai_to_review_your_pull_requests/} \\
G67 & 247 & I built a registry of 156 production-ready skills for Claude Code - think "plugins" that teach it domain expertise & \url{https://www.reddit.com/r/ClaudeAI/comments/1rv1wsc/i_built_a_registry_of_156_productionready_skills/} \\
G68 & 249 & Why traditional tooling beats AGENTS.md - and how to combine the two & \url{https://www.reddit.com/r/ExperiencedDevs/comments/1tils8z/why_traditional_tooling_beats_agentsmd_and_how_to/} \\
G69 & 251 & What's your secret weapon for efficient code reviews? & \url{https://www.reddit.com/r/codereview/comments/1j9y7wj/whats_your_secret_weapon_for_efficient_code/} \\
G70 & 255 & Vibe coded my own chess website this year while unemployed. & \url{https://www.reddit.com/r/vibecoding/comments/1of0qx7/vibe_coded_my_own_chess_website_this_year_while/} \\
G71 & 256 & I deleted 18k token worth of Claude code because it was not easy for me to debug even though it worked. & \url{https://www.reddit.com/r/ClaudeAI/comments/1ovwwrz/i_deleted_18k_token_worth_of_claude_code_because/} \\
G72 & 262 & How do you handle rare but possible exhaustion cases while doing code review ? & \url{https://www.reddit.com/r/codereview/comments/1mjqfso/how_do_you_handle_rare_but_possible_exhaustion/} \\
G73 & 263 & How are teams vibe coding? & \url{https://www.reddit.com/r/vibecoding/comments/1oyu69z/how_are_teams_vibe_coding/} \\
G74 & 264 & Tailwind isn't broken, but AI broke the social contract around it & \url{https://www.reddit.com/r/webdev/comments/1tntgf0/tailwind_isnt_broken_but_ai_broke_the_social/} \\
G75 & 266 & I am on Max Plan. How to use Claude Code with Gemini CLI. I have tried few times using Gemini to provide valuable input to Claude.. I like both of them working  & \url{https://www.reddit.com/r/ClaudeAI/comments/1lr2ovr/i_am_on_max_plan_how_to_use_claude_code_with/} \\
G76 & 269 & How are you integrating AI agents into your QA workflow? Looking for real-world experiences & \url{https://www.reddit.com/r/QualityAssurance/comments/1swx050/how_are_you_integrating_ai_agents_into_your_qa/} \\
G77 & 270 & Have you noticed AI being a bad influence on junior devs? & \url{https://www.reddit.com/r/ExperiencedDevs/comments/1j9ryjx/have_you_noticed_ai_being_a_bad_influence_on/} \\
G78 & 273 & Code review/mentor tool & \url{https://www.reddit.com/r/codereview/comments/1p5iw0s/code_reviewmentor_tool/} \\
G79 & 277 & I've been trying to work towards "Level 5" AI-assisted development for a year. I'm not quite there, but this is the flow that's working for me. & \url{https://www.reddit.com/r/ExperiencedDevs/comments/1sg1glq/ive_been_trying_to_work_towards_level_5/} \\
G80 & 278 & Has AI actually changed your day-to-day work yet? & \url{https://www.reddit.com/r/cscareerquestions/comments/1qh068f/has_ai_actually_changed_your_daytoday_work_yet/} \\
G81 & 279 & I built PatchParty.ai -- a security platform for AI-generated apps (how I built it + lessons learned) & \url{https://www.reddit.com/r/vibecoding/comments/1slqpxg/i_built_patchpartyai_a_security_platform_for/} \\
G82 & 282 & This Nov 2025 AI-Powered Python Development Bible I created has made a BIG difference for me. It gets Claude up to date with cutting edge workflows AND cutting  & \url{https://www.reddit.com/r/ClaudeAI/comments/1oq1ixp/this_nov_2025_aipowered_python_development_bible/} \\
G83 & 286 & On learning to become disciplined with Git & \url{https://www.reddit.com/r/vibecoding/comments/1sckv6m/on_learning_to_become_disciplined_with_git/} \\
G84 & 287 & Am I screwed because of this confusing process? & \url{https://www.reddit.com/r/cscareerquestions/comments/1okw8r7/am_i_screwed_because_of_this_confusing_process/} \\
G85 & 295 & GitHub Copilot agentic memory system in Copilot CLI, Copilot coding agent, and Copilot code review now in Public Preview & \url{https://www.reddit.com/r/GithubCopilot/comments/1qdybnu/github_copilot_agentic_memory_system_in_copilot/} \\
G86 & 302 & Am I Vibe Coding? & \url{https://www.reddit.com/r/cursor/comments/1ky0l4u/am_i_vibe_coding/} \\
G87 & 310 & Would you trust AI to review your AI code? & \url{https://www.reddit.com/r/codereview/comments/1npmf69/would_you_trust_ai_to_review_your_ai_code/} \\
G88 & 312 & How to Do Thoughtful Code Reviews Like a Human & \url{https://www.reddit.com/r/programming/comments/1j3egno/how_to_do_thoughtful_code_reviews_like_a_human/} \\
G89 & 314 & Been building a tool that remembers WHY you wrote that code 4 days ago & \url{https://www.reddit.com/r/SoftwareEngineering/comments/1obcwze/been_building_a_tool_that_remembers_why_you_wrote/} \\
G90 & 317 & How do you catch auth bypass risks in generated code that looks completely correct & \url{https://www.reddit.com/r/ChatGPTCoding/comments/1rw787p/how_do_you_catch_auth_bypass_risks_in_generated/} \\
G91 & 321 & I am somehow no worried about the vibe coders anymore & \url{https://www.reddit.com/r/webdev/comments/1p6r833/i_am_somehow_no_worried_about_the_vibe_coders/} \\
G92 & 324 & Built an open-source harness so I can delegate a backlog to an AI agent and actually trust what comes back & \url{https://www.reddit.com/r/AI_Agents/comments/1thrdvi/built_an_opensource_harness_so_i_can_delegate_a/} \\
G93 & 326 & cursor rule prompt I like & \url{https://www.reddit.com/r/cursor/comments/1kjz3ja/cursor_rule_prompt_i_like/} \\
G94 & 328 & Looking for a Technical Consultant for AI Mental Health Counseling App & \url{https://www.reddit.com/r/ChatGPTCoding/comments/1iredn1/looking_for_a_technical_consultant_for_ai_mental/} \\
G95 & 329 & Okay I admit, Cursor Ultra plan is pretty good & \url{https://www.reddit.com/r/cursor/comments/1mi7mxi/okay_i_admit_cursor_ultra_plan_is_pretty_good/} \\
G96 & 330 & My journey with Claude Code and research & \url{https://www.reddit.com/r/ArtificialInteligence/comments/1s37cw4/my_journey_with_claude_code_and_research/} \\
G97 & 334 & Why does Go's for-range loop return indexes, not values & \url{https://www.reddit.com/r/golang/comments/1kifznk/why_does_gos_forrange_loop_return_indexes_not/} \\
G98 & 335 & Do PR reviews slow everything down for anyone else? & \url{https://www.reddit.com/r/ExperiencedDevs/comments/1pqeowv/do_pr_reviews_slow_everything_down_for_anyone_else/} \\
G99 & 337 & How do you feel about using AI at work? & \url{https://www.reddit.com/r/ExperiencedDevs/comments/1os5col/how_do_you_feel_about_using_ai_at_work/} \\
G100 & 346 & Agentic development is pretty good after all & \url{https://www.reddit.com/r/ExperiencedDevs/comments/1q21lcr/agentic_development_is_pretty_good_after_all/} \\
G101 & 348 & Sub Agent / Multi-Agent Claude Code Commands for Refactoring, Testing, and Optimisation (Watch Your Tokens Disappear and Use Sparingly) & \url{https://www.reddit.com/r/ClaudeAI/comments/1lf5gwp/sub_agent_multiagent_claude_code_commands_for/} \\
G102 & 350 & Questions about GitHub's reputability? & \url{https://www.reddit.com/r/github/comments/1ijxrtm/questions_about_githubs_reputability/} \\
G103 & 354 & A Lovable app leaked 18,000 users' data last week. I scanned 20 more. Here's every security issue I found & \url{https://www.reddit.com/r/vibecoding/comments/1rl2mk0/a_lovable_app_leaked_18000_users_data_last_week_i/} \\
G104 & 357 & Code review: async mmaped file & \url{https://www.reddit.com/r/rust/comments/1i061x1/code_review_async_mmaped_file/} \\
G105 & 364 & Code reviews: what's the most valuable feedback you've received? & \url{https://www.reddit.com/r/webdev/comments/1lkkv64/code_reviews_whats_the_most_valuable_feedback/} \\
G106 & 365 & We're building a new Git collab platform (in Go) on top of the AT Protocol! & \url{https://www.reddit.com/r/golang/comments/1jsptwo/were_building_a_new_git_collab_platform_in_go_on/} \\
G107 & 367 & Building an app with Claude Code as a non-developer for 6+ months -- looking for advice from experienced engineers & \url{https://www.reddit.com/r/vibecoding/comments/1tk4tmp/building_an_app_with_claude_code_as_a/} \\
G108 & 371 & I shipped an entire feature without touching my keyboard. This is insane. & \url{https://www.reddit.com/r/cursor/comments/1qjrc3s/i_shipped_an_entire_feature_without_touching_my/} \\
G109 & 374 & Remote Software Engineer Intern | Built scalable systems and fixed security bugs & \url{https://www.reddit.com/r/devops/comments/1ogng9o/remote_software_engineer_intern_built_scalable/} \\
G110 & 383 & be honest, would you yourself buy a vibe coded app? & \url{https://www.reddit.com/r/vibecoding/comments/1n43ud2/be_honest_would_you_yourself_buy_a_vibe_coded_app/} \\
G111 & 385 & Trouble with understanding how to contribute to an existing project & \url{https://www.reddit.com/r/github/comments/1kw7nc2/trouble_with_understanding_how_to_contribute_to/} \\
G112 & 394 & Journey documented - launching my first iOS app into Beta. & \url{https://www.reddit.com/r/vibecoding/comments/1savo9b/journey_documented_launching_my_first_ios_app/} \\
G113 & 397 & Quite disappointed by vibe coding. & \url{https://www.reddit.com/r/webdev/comments/1tr43kj/quite_disappointed_by_vibe_coding/} \\
G114 & 405 & Offering free month of AI code review in exchange for feedback & \url{https://www.reddit.com/r/codereview/comments/1pvvlss/offering_free_month_of_ai_code_review_in_exchange/} \\
G115 & 406 & I built an open-source Action based on problems I wish I'd solved earlier & \url{https://www.reddit.com/r/github/comments/1r21t7a/i_built_an_opensource_action_based_on_problems_i/} \\
G116 & 410 & HELP I am an imposter ? & \url{https://www.reddit.com/r/ExperiencedDevs/comments/1tjqgx6/help_i_am_an_imposter/} \\
G117 & 415 & Wharton researchers just proved why "just review the AI output" doesn't work. Our brains literally give up. & \url{https://www.reddit.com/r/ArtificialInteligence/comments/1s154ig/wharton_researchers_just_proved_why_just_review/} \\
G118 & 418 & I'm shipping a bullet hell game on Steam. Entirely vibe-coded with Claude Code. Here's what nobody tells you about the gap between prototype and production. & \url{https://www.reddit.com/r/ClaudeAI/comments/1rego5o/im_shipping_a_bullet_hell_game_on_steam_entirely/} \\
G119 & 419 & I built a tool where Claude and GPT review each other's code; built using itself & \url{https://www.reddit.com/r/ClaudeAI/comments/1r24ttq/i_built_a_tool_where_claude_and_gpt_review_each/} \\
G120 & 423 & This Trivy Compromise is Insane. & \url{https://www.reddit.com/r/devops/comments/1s1ygvn/this_trivy_compromise_is_insane/} \\
G121 & 424 & Why everyone hate Cursor atm ? & \url{https://www.reddit.com/r/cursor/comments/1oatijs/why_everyone_hate_cursor_atm/} \\
G122 & 426 & What if Copilot was a pipeline instead of a pair programmer & \url{https://www.reddit.com/r/GithubCopilot/comments/1rnoggx/what_if_copilot_was_a_pipeline_instead_of_a_pair/} \\
G123 & 435 & I built an open-source coding agent that makes context visible and editable -- you curate exactly what the LLM sees & \url{https://www.reddit.com/r/AI_Agents/comments/1tt077k/i_built_an_opensource_coding_agent_that_makes/} \\
G124 & 438 & Software development became boring with Claude Code & \url{https://www.reddit.com/r/ClaudeAI/comments/1pxvd0g/software_development_became_boring_with_claude/} \\
G125 & 441 & "My Agent got dumber" -- No, try the Columbo Method. Here's my workflow. & \url{https://www.reddit.com/r/cursor/comments/1qs2wgt/my_agent_got_dumber_no_try_the_columbo_method/} \\
G126 & 446 & My teammates are generating enormous test suites now & \url{https://www.reddit.com/r/ExperiencedDevs/comments/1po8uud/my_teammates_are_generating_enormous_test_suites/} \\
G127 & 447 & Getting the most from your manager 1on1's & \url{https://www.reddit.com/r/cscareerquestions/comments/1ig0u8v/getting_the_most_from_your_manager_1on1s/} \\
G128 & 449 & "Looks Good to Me" Constructive Code Reviews * Adrienne Braganza Tacke \& Paul Slaughter & \url{https://www.reddit.com/r/programming/comments/1iff4sq/looks_good_to_me_constructive_code_reviews/} \\
G129 & 450 & Anyone here using Qodo for AI-powered code reviews? & \url{https://www.reddit.com/r/codereview/comments/1mi0z28/anyone_here_using_qodo_for_aipowered_code_reviews/} \\
G130 & 452 & I launched a real product using Github Copilot. & \url{https://www.reddit.com/r/GithubCopilot/comments/1nr5mzg/i_launched_a_real_product_using_github_copilot/} \\
G131 & 455 & Code protection secretly removed from an embedded software by a local LLM & \url{https://www.reddit.com/r/LocalLLaMA/comments/1sieqrf/code_protection_secretly_removed_from_an_embedded/} \\
G132 & 456 & I Built an iOS Fitness App with Claude. Here's the 3 AM Bug That Almost Broke Me. & \url{https://www.reddit.com/r/ClaudeAI/comments/1pejqny/i_built_an_ios_fitness_app_with_claude_heres_the/} \\
G133 & 457 & I vibe-coded a complete Chrome extension development pipeline & \url{https://www.reddit.com/r/vibecoding/comments/1tk9big/i_vibecoded_a_complete_chrome_extension/} \\
G134 & 458 & Built an AI PR review workflow that generates actual fix PRs instead of just comments & \url{https://www.reddit.com/r/AI_Agents/comments/1te6gvo/built_an_ai_pr_review_workflow_that_generates/} \\
G135 & 463 & First time reviewing an applicant's code exercise & \url{https://www.reddit.com/r/ExperiencedDevs/comments/1mnuyvn/first_time_reviewing_an_applicants_code_exercise/} \\
G136 & 470 & What are your experiences reviewing code your colleagues use ai to write & \url{https://www.reddit.com/r/AskProgramming/comments/1o7w8nw/what_are_your_experiences_reviewing_code_your/} \\
G137 & 471 & Accidental vibe coder, now what? & \url{https://www.reddit.com/r/vibecoding/comments/1tufghu/accidental_vibe_coder_now_what/} \\
G138 & 472 & AI writes HTML fast but does it write HTML well, genuinely not sure anymore & \url{https://www.reddit.com/r/programming/comments/1rb7ahr/ai_writes_html_fast_but_does_it_write_html_well/} \\
G139 & 473 & How and why are companies using AI agents to review codes? & \url{https://www.reddit.com/r/ExperiencedDevs/comments/1razy16/how_and_why_are_companies_using_ai_agents_to/} \\
G140 & 477 & Vibe coders are the script kiddies of programming & \url{https://www.reddit.com/r/vibecoding/comments/1n6nfyh/vibe_coders_are_the_script_kiddies_of_programming/} \\
G141 & 478 & Are you guys even reviewing your own code anymore? & \url{https://www.reddit.com/r/cscareerquestions/comments/1savuwm/are_you_guys_even_reviewing_your_own_code_anymore/} \\
G142 & 484 & How we use Claude Code and Cursor at work & \url{https://www.reddit.com/r/ClaudeAI/comments/1ppxvgv/how_we_use_claude_code_and_cursor_at_work/} \\
G143 & 486 & I built an "agentic Jira" for startups -- it auto-creates PRDs, tasks, and GitHub issues from your repo. Would you pay \$20/mo? & \url{https://www.reddit.com/r/AI_Agents/comments/1o4v09o/i_built_an_agentic_jira_for_startups_it/} \\
G144 & 487 & Building Diffly: a keyboard-first code review desktop app (looking for feedback) & \url{https://www.reddit.com/r/codereview/comments/1npjhvo/building_diffly_a_keyboardfirst_code_review/} \\
G145 & 488 & How Engineers using AI Coders & \url{https://www.reddit.com/r/vibecoding/comments/1nrodg1/how_engineers_using_ai_coders/} \\
G146 & 491 & Claude is just awesome & \url{https://www.reddit.com/r/ClaudeAI/comments/1rj4fu4/claude_is_just_awesome/} \\
G147 & 492 & How far into a learning project do you go & \url{https://www.reddit.com/r/Python/comments/1ps39bq/how_far_into_a_learning_project_do_you_go/} \\
G148 & 494 & Spawned agents & \url{https://www.reddit.com/r/ClaudeAI/comments/1titt2o/spawned_agents/} \\
G149 & 500 & How do I get my team to actually conduct Pull Request Reviews? & \url{https://www.reddit.com/r/ExperiencedDevs/comments/1i10rl2/how_do_i_get_my_team_to_actually_conduct_pull/} \\
G150 & 504 & Need opinions from devs about AI coding. I have stakeholders all in on this mode of working on multiple levels... & \url{https://www.reddit.com/r/ExperiencedDevs/comments/1qc9ih0/need_opinions_from_devs_about_ai_coding_i_have/} \\
G151 & 505 & Code Review Now Available in Source Control!?!?! & \url{https://www.reddit.com/r/windsurf/comments/1qs95dw/code_review_now_available_in_source_control/} \\
G152 & 507 & I ran the same security audit 3 ways on the same codebase. The difference was surprising. & \url{https://www.reddit.com/r/ClaudeAI/comments/1s3beks/i_ran_the_same_security_audit_3_ways_on_the_same/} \\
G153 & 510 & Big company, big project, unclear and changing rules & \url{https://www.reddit.com/r/ExperiencedDevs/comments/1i1lpto/big_company_big_project_unclear_and_changing_rules/} \\
G154 & 522 & I executed a single code review prompt using Claude Code Sonnet 4.5.... & \url{https://www.reddit.com/r/ClaudeAI/comments/1pn6ttp/i_executed_a_single_code_review_prompt_using/} \\
G155 & 525 & LeetCode is a Corpse--Stop Worshipping It, You Masochists & \url{https://www.reddit.com/r/ExperiencedDevs/comments/1jb9btk/leetcode_is_a_corpsestop_worshipping_it_you/} \\
G156 & 529 & Sanity check: using git to make LLM-assisted work accumulate over time & \url{https://www.reddit.com/r/ChatGPTCoding/comments/1sqwmx9/sanity_check_using_git_to_make_llmassisted_work/} \\
G157 & 530 & Anthropic Dev: "Claude Code wrote 80\% of its own code" & \url{https://www.reddit.com/r/accelerate/comments/1kihdfy/anthropic_dev_claude_code_wrote_80_of_its_own_code/} \\
G158 & 534 & Are too many commits in a code review bad? & \url{https://www.reddit.com/r/ExperiencedDevs/comments/1rfjuum/are_too_many_commits_in_a_code_review_bad/} \\
G159 & 539 & Opencode-power-pack - Claude Code skills ported to OpenCode & \url{https://www.reddit.com/r/LocalLLaMA/comments/1swf37n/opencodepowerpack_claude_code_skills_ported_to/} \\
G160 & 544 & Move Fast and Break things is not working for Cursor & \url{https://www.reddit.com/r/cursor/comments/1p3zrag/move_fast_and_break_things_is_not_working_for/} \\
G161 & 546 & How to deal with a guy who allways tries enforce his style and almost never backs down during code reviews? & \url{https://www.reddit.com/r/ExperiencedDevs/comments/1kiogra/how_to_deal_with_a_guy_who_allways_tries_enforce/} \\
G162 & 550 & Senior backend dev struggling with "just ship tickets" culture after working in a strong engineering team & \url{https://www.reddit.com/r/ExperiencedDevs/comments/1rlezlv/senior_backend_dev_struggling_with_just_ship/} \\
G163 & 552 & logicutils -- composable Unix-style utilities for logic-enhanced builds (Rust 2024, BSD-2-Clause) & \url{https://www.reddit.com/r/rust/comments/1swvn0i/logicutils_composable_unixstyle_utilities_for/} \\
G164 & 553 & Should I take a year off to attend Apple Academy and pivot to iOS dev? & \url{https://www.reddit.com/r/cscareerquestions/comments/1mx2lcq/should_i_take_a_year_off_to_attend_apple_academy/} \\
G165 & 555 & Most AI Coding Tools Are Overcomplicated & \url{https://www.reddit.com/r/AskProgramming/comments/1tet9bg/most_ai_coding_tools_are_overcomplicated/} \\
G166 & 557 & What happens when you put Claude, GPT, Gemini, Copilot and Codex in the same chatroom and give them a bug list? [video] & \url{https://www.reddit.com/r/ClaudeAI/comments/1r5zryx/what_happens_when_you_put_claude_gpt_gemini/} \\
G167 & 558 & Lines of code and velocity actually dead as devprod metrics? & \url{https://www.reddit.com/r/devops/comments/1jz0zc8/lines_of_code_and_velocity_actually_dead_as/} \\
G168 & 559 & About to give up on Claude Code & \url{https://www.reddit.com/r/ClaudeAI/comments/1o5xmt4/about_to_give_up_on_claude_code/} \\
G169 & 560 & This tool made our engineering team faster without writing a single extra line of code & \url{https://www.reddit.com/r/artificial/comments/1krnjfp/this_tool_made_our_engineering_team_faster/} \\
G170 & 563 & Cursor users! Bring Your AI Agent: I Opened a Public MCP Server and I Want to See What Happens & \url{https://www.reddit.com/r/cursor/comments/1ro9wdl/cursor_users_bring_your_ai_agent_i_opened_a/} \\
G171 & 567 & Low cortisol portfolio -- vibecoded the whole thing and my stress levels never left the floor & \url{https://www.reddit.com/r/vibecoding/comments/1rxlpfb/low_cortisol_portfolio_vibecoded_the_whole_thing/} \\
G172 & 568 & Claude Code - Easiest way to achieve simple "orchestration" where one agent is coder and another is reviewer? & \url{https://www.reddit.com/r/ClaudeAI/comments/1qcubn0/claude_code_easiest_way_to_achieve_simple/} \\
G173 & 572 & Increased context switching code & \url{https://www.reddit.com/r/ClaudeAI/comments/1l84n9k/increased_context_switching_code/} \\
G174 & 573 & Quick tip: remote SSH control of my personal PC to run Claude Code and guide it while on the go & \url{https://www.reddit.com/r/ClaudeAI/comments/1l6dt99/quick_tip_remote_ssh_control_of_my_personal_pc_to/} \\
G175 & 574 & When two LLMs love each other very much... & \url{https://www.reddit.com/r/ClaudeAI/comments/1stmdlo/when_two_llms_love_each_other_very_much/} \\
G176 & 575 & Did you ever stay too long? & \url{https://www.reddit.com/r/ExperiencedDevs/comments/1qg1veg/did_you_ever_stay_too_long/} \\
G177 & 577 & Am I Vibe Coding? & \url{https://www.reddit.com/r/vibecoding/comments/1tvqxmo/am_i_vibe_coding/} \\
G178 & 578 & How I shifted my team into Spec-Driven Development (and why it works) & \url{https://www.reddit.com/r/cursor/comments/1op8a5s/how_i_shifted_my_team_into_specdriven_development/} \\
G179 & 586 & Are you using AI to review code? & \url{https://www.reddit.com/r/ExperiencedDevs/comments/1plrf2k/are_you_using_ai_to_review_code/} \\
G180 & 587 & I finally tried vibe coding and it was meh & \url{https://www.reddit.com/r/ExperiencedDevs/comments/1n4akjd/i_finally_tried_vibe_coding_and_it_was_meh/} \\
G181 & 589 & First weekend "vibe coding" with Claude Code + ChatGPT -- shipped demotest.io live & \url{https://www.reddit.com/r/ClaudeAI/comments/1mpr0r5/first_weekend_vibe_coding_with_claude_code/} \\
G182 & 590 & The real bottleneck in our team is not coding speed, it is pull requests sitting for three days waiting on reviews. & \url{https://www.reddit.com/r/AskProgramming/comments/1t65tuo/the_real_bottleneck_in_our_team_is_not_coding/} \\
G183 & 592 & Boss just announced plan to move to AI First Dev & \url{https://www.reddit.com/r/ExperiencedDevs/comments/1rae1po/boss_just_announced_plan_to_move_to_ai_first_dev/} \\
G184 & 594 & how obvious is this retry logic bug to you? & \url{https://www.reddit.com/r/Python/comments/1p1yzzd/how_obvious_is_this_retry_logic_bug_to_you/} \\
G185 & 596 & How to tell your coding agent to ship features in small reviewable chunks & \url{https://www.reddit.com/r/ClaudeAI/comments/1ro12vz/how_to_tell_your_coding_agent_to_ship_features_in/} \\
G186 & 598 & How much do you use AI for coding? & \url{https://www.reddit.com/r/cscareerquestions/comments/1os9o9h/how_much_do_you_use_ai_for_coding/} \\
G187 & 601 & [ Removed by moderator ] & \url{https://www.reddit.com/r/ClaudeAI/comments/1rp2e6i/i_built_a_free_claude_skill_that_scans_other/} \\
G188 & 603 & I got tired of vibe-coded PRs getting destroyed in code review. Built 24 slash commands that enforce quality gates on every commit. Works with Claude Code, Curs & \url{https://www.reddit.com/r/vibecoding/comments/1srfprn/i_got_tired_of_vibecoded_prs_getting_destroyed_in/} \\
G189 & 606 & Claude is an incredible builder. It will also leave your app completely exposed and not say a word about it. & \url{https://www.reddit.com/r/ClaudeAI/comments/1tvv0vw/claude_is_an_incredible_builder_it_will_also/} \\
G190 & 610 & Whats the craziest code review you had with a junior? Were you surprised positively or negatively? & \url{https://www.reddit.com/r/ExperiencedDevs/comments/1q7s3ny/whats_the_craziest_code_review_you_had_with_a/} \\
G191 & 611 & React, AI, and the Future of Front-End Development & \url{https://www.reddit.com/r/cursor/comments/1ncizyj/react_ai_and_the_future_of_frontend_development/} \\
G192 & 616 & Anthropic just launched Claude Security in public beta AI that scans your codebase, validates its own findings, and proposes fixes. Here's what actually matters & \url{https://www.reddit.com/r/ClaudeAI/comments/1t12l3t/anthropic_just_launched_claude_security_in_public/} \\
G193 & 618 & Claude Code messing up my git diffs & \url{https://www.reddit.com/r/ClaudeAI/comments/1lj50e9/claude_code_messing_up_my_git_diffs/} \\
G194 & 620 & Honest question about a non-coder/developer making apps with Claude & \url{https://www.reddit.com/r/ClaudeAI/comments/1tgkc4o/honest_question_about_a_noncoderdeveloper_making/} \\
G195 & 623 & Stop telling Claude Code "build me an app." Spec your features like an engineer and it becomes a completely different tool & \url{https://www.reddit.com/r/ClaudeAI/comments/1rsi5lo/stop_telling_claude_code_build_me_an_app_spec/} \\
G196 & 624 & I vibe-coded an iOS app with Claude. From zero knowledge to 10k+ App Store downloads in 2.5 months & \url{https://www.reddit.com/r/vibecoding/comments/1qjqjkz/i_vibecoded_an_ios_app_with_claude_from_zero/} \\
G197 & 628 & Am I overspending on AI dev tools? \$150/month for Claude, Warp, CodeRabbit, etc & \url{https://www.reddit.com/r/vibecoding/comments/1o5kjzq/am_i_overspending_on_ai_dev_tools_150month_for/} \\
G198 & 630 & ChatML - Open-source desktop app for orchestrating parallel Claude Code agents & \url{https://www.reddit.com/r/artificial/comments/1rmorsh/chatml_opensource_desktop_app_for_orchestrating/} \\
G199 & 632 & 2030 Claude Sweatshops & \url{https://www.reddit.com/r/ClaudeAI/comments/1qd8q0y/2030_claude_sweatshops/} \\
G200 & 642 & 45-test benchmark around my homelab use cases and testing 19 local LLMs (incl. Gemma 4 and Qwen 3.5) on a Strix Halo & \url{https://www.reddit.com/r/LocalLLaMA/comments/1sbpuri/45test_benchmark_around_my_homelab_use_cases_and/} \\
G201 & 645 & i made an mcp server that connects chatgpt, claude, gemini and perplexity -- all 4 running together inside your ide at the same time & \url{https://www.reddit.com/r/vibecoding/comments/1sdjwkq/i_made_an_mcp_server_that_connects_chatgpt_claude/} \\
G202 & 647 & [Project] My first complete GUI app - File organizer with duplicate detection & \url{https://www.reddit.com/r/Python/comments/1qtjk1e/project_my_first_complete_gui_app_file_organizer/} \\
G203 & 654 & Tried GPT-5.2/Pro vs Opus 4.5 vs Gemini 3 on 3 coding tasks, here's the output & \url{https://www.reddit.com/r/ClaudeAI/comments/1po8fiq/tried_gpt52pro_vs_opus_45_vs_gemini_3_on_3_coding/} \\
G204 & 655 & 8 years as a dev, just built my first vibe coding project - here's what actually surprised me & \url{https://www.reddit.com/r/vibecoding/comments/1rmcs61/8_years_as_a_dev_just_built_my_first_vibe_coding/} \\
G205 & 656 & How are you leveling up your AI coding workflow? & \url{https://www.reddit.com/r/vibecoding/comments/1tcwpeo/how_are_you_leveling_up_your_ai_coding_workflow/} \\
G206 & 661 & PSA: Found identical malicious code in two separate projects -- only common factor is GitHub Copilot & \url{https://www.reddit.com/r/github/comments/1rndc5q/psa_found_identical_malicious_code_in_two/} \\
G207 & 663 & The Claude SKILLS files that fixed my vibe coded mess by adding in linters & \url{https://www.reddit.com/r/vibecoding/comments/1t0mgti/the_claude_skills_files_that_fixed_my_vibe_coded/} \\
G208 & 666 & How do you review your code? & \url{https://www.reddit.com/r/rust/comments/1o4145i/how_do_you_review_your_code/} \\
G209 & 668 & AI Orchestrator & \url{https://www.reddit.com/r/ChatGPTCoding/comments/1mnay5h/ai_orchestrator/} \\
G210 & 673 & Show: an agent that writes code from a Jira story and opens a draft PR & \url{https://www.reddit.com/r/vibecoding/comments/1tmwvth/show_an_agent_that_writes_code_from_a_jira_story/} \\
G211 & 675 & How can you make sure a GitHub project is safe to use if you have little or no coding experience? & \url{https://www.reddit.com/r/github/comments/1t66lk2/how_can_you_make_sure_a_github_project_is_safe_to/} \\
G212 & 676 & Claude Code now has Automated Security Reviews & \url{https://www.reddit.com/r/ClaudeAI/comments/1mjc40q/claude_code_now_has_automated_security_reviews/} \\
G213 & 677 & AI Code Reviewer CodeAnt founds Critical Authentication Bypass in pac4j-jwt - Using Only a Public Key (CVSS 10)CVE-2026-29000 & \url{https://www.reddit.com/r/programming/comments/1rlb1zp/ai_code_reviewer_codeant_founds_critical/} \\
G214 & 680 & check run agents - customizable AI agents for code review & \url{https://www.reddit.com/r/codereview/comments/1t159tb/check_run_agents_customizable_ai_agents_for_code/} \\
G215 & 681 & I Didn't Let an LLM Write My Project -- Here's What I Actually Did & \url{https://www.reddit.com/r/rust/comments/1rym1to/i_didnt_let_an_llm_write_my_project_heres_what_i/} \\
G216 & 682 & Is it just me or you too? Vibe coding is cool... until one point. & \url{https://www.reddit.com/r/vibecoding/comments/1sxbv9e/is_it_just_me_or_you_too_vibe_coding_is_cool/} \\
G217 & 683 & Context and Memory Management Tips & \url{https://www.reddit.com/r/ClaudeAI/comments/1oq0esn/context_and_memory_management_tips/} \\
G218 & 685 & PR comments that doesn't really have any impact & \url{https://www.reddit.com/r/ExperiencedDevs/comments/1k1eadn/pr_comments_that_doesnt_really_have_any_impact/} \\
G219 & 686 & SuperBeads Universal Framework (Wiggum flavour) - Open Source & \url{https://www.reddit.com/r/ClaudeAI/comments/1qbsiw0/superbeads_universal_framework_wiggum_flavour/} \\
G220 & 687 & Daily AI Archive 8/27/2025 & \url{https://www.reddit.com/r/accelerate/comments/1n20m47/daily_ai_archive_8272025/} \\
G221 & 693 & A bash loop named "Ralph Wiggum" accidentally revealed why subagents are the future of AI-assisted dev & \url{https://www.reddit.com/r/ClaudeAI/comments/1r6hk6z/a_bash_loop_named_ralph_wiggum_accidentally/} \\
G222 & 694 & Can Claude Code (terminal) launch a GPT-5.4 reviewer subagent (via Cursor or cursor cli) & \url{https://www.reddit.com/r/ClaudeAI/comments/1sriqgg/can_claude_code_terminal_launch_a_gpt54_reviewer/} \\
G223 & 700 & I created Lanes, a VS Code extension for AI Project Management & \url{https://www.reddit.com/r/ClaudeAI/comments/1qdchwf/i_created_lanes_a_vs_code_extension_for_ai/} \\
G224 & 703 & New to Go: Built a Google Chat App with 150k Users - Looking for Feedback on My Code & \url{https://www.reddit.com/r/golang/comments/1ia9k9g/new_to_go_built_a_google_chat_app_with_150k_users/} \\
G225 & 705 & My answer to the great IDE debate, both VS and VSCode & \url{https://www.reddit.com/r/dotnet/comments/1rk763s/my_answer_to_the_great_ide_debate_both_vs_and/} \\
G226 & 706 & I've been using the new 3.7 model for about an hour and a half and I hate it & \url{https://www.reddit.com/r/ClaudeAI/comments/1ixcxtu/ive_been_using_the_new_37_model_for_about_an_hour/} \\
G227 & 711 & One week of letting Codex do everything & \url{https://www.reddit.com/r/vibecoding/comments/1nmbz13/one_week_of_letting_codex_do_everything/} \\
G228 & 713 & I built a custom 2D ECS Game Engine in Rust \& wGPU that compiles to WASM. (Live Demo + Source) & \url{https://www.reddit.com/r/rust/comments/1r6qq02/i_built_a_custom_2d_ecs_game_engine_in_rust_wgpu/} \\
G229 & 720 & Coincidence, or not? & \url{https://www.reddit.com/r/ClaudeAI/comments/1kx604j/coincidence_or_not/} \\
G230 & 722 & Now that even Linus Torvalds is vibe coding his Linux kernel changes, what value do we provide? & \url{https://www.reddit.com/r/cscareerquestions/comments/1qar3fx/now_that_even_linus_torvalds_is_vibe_coding_his/} \\
G231 & 723 & Here's how we've been getting fewer "how did this make it to QA" bugs lately. Your recommendations? & \url{https://www.reddit.com/r/QualityAssurance/comments/1m7hkqr/heres_how_weve_been_getting_fewer_how_did_this/} \\
G232 & 725 & Which is the best way to vibe code ? & \url{https://www.reddit.com/r/vibecoding/comments/1rj3fl1/which_is_the_best_way_to_vibe_code/} \\
G233 & 728 & Make markdown files executable with Claude Code (shebang support) & \url{https://www.reddit.com/r/ClaudeAI/comments/1q44kkd/make_markdown_files_executable_with_claude_code/} \\
G234 & 732 & A founder vibe-coded his entire SaaS with AI. Hackers found API keys in the frontend and stole \$87,500. & \url{https://www.reddit.com/r/webdev/comments/1rt4gxr/a_founder_vibecoded_his_entire_saas_with_ai/} \\
G235 & 734 & 10 Claude Code tips from Boris, the creator of Claude Code, summarized & \url{https://www.reddit.com/r/ClaudeAI/comments/1qspcip/10_claude_code_tips_from_boris_the_creator_of/} \\
G236 & 736 & After a year of using Cursor, Claude Code, Antigravity, and Copilot daily -- I think AI tools are making a lot of devs slower, not faster. Here's why. & \url{https://www.reddit.com/r/dotnet/comments/1rbhblj/after_a_year_of_using_cursor_claude_code/} \\
G237 & 737 & [Contributor Request] We hit 10k+ nodes on a local-first P2P mesh - Seeking help to scale the "Sovereign Workhorse" & \url{https://www.reddit.com/r/AI_Agents/comments/1syayqg/contributor_request_we_hit_10k_nodes_on_a/} \\
G238 & 738 & Is it true in some places dev get offended during coding review or meeting with end users and fight physically each other that's why it's called 1 on 1? & \url{https://www.reddit.com/r/cscareerquestions/comments/1j19f5k/is_it_true_in_some_places_dev_get_offended_during/} \\
G239 & 741 & I built an HTTP/1.1 server from scratch in go & \url{https://www.reddit.com/r/golang/comments/1nqvq6c/i_built_an_http11_server_from_scratch_in_go/} \\
G240 & 744 & Am I the only who feels this way abou AI generated code? & \url{https://www.reddit.com/r/ExperiencedDevs/comments/1tifd5z/am_i_the_only_who_feels_this_way_abou_ai/} \\
G241 & 745 & Personalized persistent working memory for Claude Code -- every correction you make becomes ingrained and permanent & \url{https://www.reddit.com/r/ClaudeAI/comments/1sg8y46/personalized_persistent_working_memory_for_claude/} \\
G242 & 748 & [Code Review Request] Memory leak working with streams & \url{https://www.reddit.com/r/csharp/comments/1o6dxzw/code_review_request_memory_leak_working_with/} \\
G243 & 755 & Are QA Engineers Just Becoming Automation Developers with a Different Job Title? & \url{https://www.reddit.com/r/QualityAssurance/comments/1oe01td/are_qa_engineers_just_becoming_automation/} \\
G244 & 759 & Fired from New Role -- Help Me Reflect & \url{https://www.reddit.com/r/ExperiencedDevs/comments/1od73g8/fired_from_new_role_help_me_reflect/} \\
G245 & 763 & Codelane: Agentic Development Environment with Code Review & \url{https://www.reddit.com/r/rust/comments/1s37ovy/codelane_agentic_development_environment_with/} \\
G246 & 765 & 5.2 Appreciation & \url{https://www.reddit.com/r/OpenAI/comments/1pmzymg/52_appreciation/} \\
G247 & 767 & Death of a thousand nits: the gentle art of code review & \url{https://www.reddit.com/r/programming/comments/1iri0l4/death_of_a_thousand_nits_the_gentle_art_of_code/} \\
G248 & 770 & Prepared a collection of 100+ production-ready Claude Code subagents & \url{https://www.reddit.com/r/AI_Agents/comments/1mi5ih7/prepared_a_collection_of_100_productionready/} \\
G249 & 777 & If an AI agent opened a PR for you, what would you want to see first? & \url{https://www.reddit.com/r/ClaudeAI/comments/1tg4za3/if_an_ai_agent_opened_a_pr_for_you_what_would_you/} \\
G250 & 778 & How I eliminated context-switch fatigue when working with multiple AI agents in parallel & \url{https://www.reddit.com/r/vibecoding/comments/1ro01gv/how_i_eliminated_contextswitch_fatigue_when/} \\
G251 & 781 & Claude Code Review is \$15-25 per PR. That sounds crazy. Is it worth it for enterprises? & \url{https://www.reddit.com/r/LocalLLaMA/comments/1rqvong/claude_code_review_is_1525_per_pr_that_sounds/} \\
G252 & 782 & 100 production-ready AI agent configs that actually run (not demos, not concepts) & \url{https://www.reddit.com/r/artificial/comments/1roq2su/100_productionready_ai_agent_configs_that/} \\
G253 & 785 & I pay 5x more for Claude than Codex. I only use it 2x more. & \url{https://www.reddit.com/r/vibecoding/comments/1sl4tz7/i_pay_5x_more_for_claude_than_codex_i_only_use_it/} \\
G254 & 786 & Recursive Self-Improvement Loop & \url{https://www.reddit.com/r/ClaudeAI/comments/1stlflt/recursive_selfimprovement_loop/} \\
G255 & 788 & Things I taught my interns about AI coding that I wish someone told me earlier & \url{https://www.reddit.com/r/cscareerquestions/comments/1sx2z1e/things_i_taught_my_interns_about_ai_coding_that_i/} \\
G256 & 789 & The Missing Bridge Between Vibe Prototypes and Production Systems & \url{https://www.reddit.com/r/vibecoding/comments/1snev9u/the_missing_bridge_between_vibe_prototypes_and/} \\
G257 & 790 & Pulled our full dependency tree after six months of heavy Copilot use and there are packages in there I genuinely cannot account for & \url{https://www.reddit.com/r/webdev/comments/1rwz98t/pulled_our_full_dependency_tree_after_six_months/} \\
G258 & 791 & [Code Review Request] Capstone Project - Streamlit App for Box Office Prediction & \url{https://www.reddit.com/r/codereview/comments/1jojaaq/code_review_request_capstone_project_streamlit/} \\
G259 & 793 & Opus said there were no problems. I ran 5 Sonnets in parallel. They found 21. & \url{https://www.reddit.com/r/vibecoding/comments/1sn42m4/opus_said_there_were_no_problems_i_ran_5_sonnets/} \\
G260 & 794 & Would this pass your Code Review? & \url{https://www.reddit.com/r/dotnet/comments/1r61hyt/would_this_pass_your_code_review/} \\
G261 & 795 & CEOs fired developer on my team based on unsubstantiated "slacking off" rumors, leaving me to pick up their slack & \url{https://www.reddit.com/r/ExperiencedDevs/comments/1nszrp2/ceos_fired_developer_on_my_team_based_on/} \\
G262 & 796 & How to ACTUALLY make your (vibe coded) apps secure (from an actual hacker) & \url{https://www.reddit.com/r/ChatGPTCoding/comments/1nyuh8b/how_to_actually_make_your_vibe_coded_apps_secure/} \\
G263 & 807 & For senior engineers using LLMs: are we gaining leverage or losing the craft? how much do you rely on LLMs for implementation vs design and review? how are LLMs & \url{https://www.reddit.com/r/cursor/comments/1qy4ztm/for_senior_engineers_using_llms_are_we_gaining/} \\
G264 & 808 & This week Ai has killed one more thing, my passion to mentor interns & \url{https://www.reddit.com/r/ExperiencedDevs/comments/1rzp29a/this_week_ai_has_killed_one_more_thing_my_passion/} \\
G265 & 811 & AI-generated code is silently wrecking dev teams & \url{https://www.reddit.com/r/vibecoding/comments/1kulppm/aigenerated_code_is_silently_wrecking_dev_teams/} \\
G266 & 814 & Change My Mind: LLMs bad for Production quality code & \url{https://www.reddit.com/r/windsurf/comments/1mzidcf/change_my_mind_llms_bad_for_production_quality/} \\
G267 & 824 & AI code reviewers worth trying? & \url{https://www.reddit.com/r/AI_Agents/comments/1qcmp5m/ai_code_reviewers_worth_trying/} \\
G268 & 830 & Annoying Code Review -- Unit Tests & \url{https://www.reddit.com/r/csharp/comments/1m5n43r/annoying_code_review_unit_tests/} \\
G269 & 837 & Would you use an AI-powered platform for practicing real-world software engineering (not just algorithms)? & \url{https://www.reddit.com/r/cscareerquestions/comments/1s2d65k/would_you_use_an_aipowered_platform_for/} \\
G270 & 840 & Using GitHub Flow with Claude to add a feature to a React app (issue -> branch -> PR) & \url{https://www.reddit.com/r/github/comments/1qdej73/using_github_flow_with_claude_to_add_a_feature_to/} \\
G271 & 841 & Feeling like a fraud, I use AI to do most of my job at my new Frontend position & \url{https://www.reddit.com/r/ExperiencedDevs/comments/1oglvl8/feeling_like_a_fraud_i_use_ai_to_do_most_of_my/} \\
G272 & 845 & Code review on HTTP1.1 server implementation & \url{https://www.reddit.com/r/codereview/comments/1jk63we/code_review_on_http11_server_implementation/} \\
G273 & 846 & How To Review Code & \url{https://www.reddit.com/r/coding/comments/1rkiiwh/how_to_review_code/} \\
G274 & 850 & I vibe-coded almost half-a-million lines of code and lived to tell the tale & \url{https://www.reddit.com/r/vibecoding/comments/1n61dc3/i_vibecoded_almost_halfamillion_lines_of_code_and/} \\
G275 & 851 & "Vibe Coding" Is Everywhere -- Is Traditional Programming on Its Way Out? & \url{https://www.reddit.com/r/ArtificialInteligence/comments/1nce63n/vibe_coding_is_everywhere_is_traditional/} \\
G276 & 853 & Teach and stay humble in code reviews & \url{https://www.reddit.com/r/ExperiencedDevs/comments/1mtl60u/teach_and_stay_humble_in_code_reviews/} \\
G277 & 858 & Do you really need a smarter model than Gemini 3 Flash? & \url{https://www.reddit.com/r/GithubCopilot/comments/1q8z9pp/do_you_really_need_a_smarter_model_than_gemini_3/} \\
G278 & 864 & Requesting a React Native code review & \url{https://www.reddit.com/r/codereview/comments/1mywocl/requesting_a_react_native_code_review/} \\
G279 & 865 & Seeking help learning the modern industry standards of c++ & \url{https://www.reddit.com/r/codereview/comments/1ndlfrf/seeking_help_learning_the_modern_industry/} \\
G280 & 868 & Can someone build this dream vibe coding orchestrator now that it's 2026? & \url{https://www.reddit.com/r/vibecoding/comments/1q2105p/can_someone_build_this_dream_vibe_coding/} \\
G281 & 869 & What Models do you use for coding, code review, planning/spec design, feature work, implementation, etc? & \url{https://www.reddit.com/r/cursor/comments/1q2nzkw/what_models_do_you_use_for_coding_code_review/} \\
G282 & 871 & Improved markdown quality, code intelligence for 248 formats, and more in Kreuzberg v4.7.0 & \url{https://www.reddit.com/r/dotnet/comments/1scvunp/improved_markdown_quality_code_intelligence_for/} \\
G283 & 873 & LazySpecKit: SpecKit without babysitting & \url{https://www.reddit.com/r/GithubCopilot/comments/1rclrnj/lazyspeckit_speckit_without_babysitting/} \\
G284 & 879 & Check how crazy is my CLAUDE.md & \url{https://www.reddit.com/r/ClaudeAI/comments/1lhto28/check_how_crazy_is_my_claudemd/} \\
G285 & 886 & Is it hard to read your teammates code? Could source code maintained in natural language improve this? & \url{https://www.reddit.com/r/AskProgramming/comments/1lpeeeh/is_it_hard_to_read_your_teammates_code_could/} \\
G286 & 887 & What if your code reviewer knew the whole repo, not just the latest diff? & \url{https://www.reddit.com/r/programming/comments/1krpmep/what_if_your_code_reviewer_knew_the_whole_repo/} \\
G287 & 888 & We wrote a "human-first" guide to code reviews (PDF, free with email) & \url{https://www.reddit.com/r/ExperiencedDevs/comments/1lfb26m/we_wrote_a_humanfirst_guide_to_code_reviews_pdf/} \\
G288 & 891 & Pointless PR Comments - going insane. & \url{https://www.reddit.com/r/ExperiencedDevs/comments/1mkwph8/pointless_pr_comments_going_insane/} \\
G289 & 892 & AI Sucks at Code Reviews & \url{https://www.reddit.com/r/coding/comments/1k8a2x9/ai_sucks_at_code_reviews/} \\
G290 & 894 & I just built a skill to prompt Codex from Claude Code. It's super convenient. & \url{https://www.reddit.com/r/ClaudeAI/comments/1ocoahc/i_just_built_a_skill_to_prompt_codex_from_claude/} \\
G291 & 896 & What AI projects deliver real ROI? & \url{https://www.reddit.com/r/AI_Agents/comments/1pmmdmq/what_ai_projects_deliver_real_roi/} \\
G292 & 897 & AI code and code reviews & \url{https://www.reddit.com/r/SoftwareEngineering/comments/1lhpktz/ai_code_and_code_reviews/} \\
G293 & 898 & Database vibing is here. We built a "Cursor for Databases" with a full undo button. & \url{https://www.reddit.com/r/vibecoding/comments/1r1eqsf/database_vibing_is_here_we_built_a_cursor_for/} \\
G294 & 905 & I got tired of Claude generating "we are passionate about innovation" on every landing page. Here's what fixed it. & \url{https://www.reddit.com/r/ClaudeAI/comments/1syv1qp/i_got_tired_of_claude_generating_we_are/} \\
G295 & 914 & I created a GitHub repo with battle-tested Claude prompt templates (free, open-source) & \url{https://www.reddit.com/r/ClaudeAI/comments/1p1dd7l/i_created_a_github_repo_with_battletested_claude/} \\
G296 & 917 & Code Review: For high frequency trading application & \url{https://www.reddit.com/r/codereview/comments/1p52m67/code_review_for_high_frequency_trading_application/} \\
G297 & 918 & SKILL.md is quietly becoming the standard for teaching AI agents new capabilities - here's what's happening & \url{https://www.reddit.com/r/AI_Agents/comments/1stcu8e/skillmd_is_quietly_becoming_the_standard_for/} \\
G298 & 919 & Anyone else finding AI code review tools useless once you hit 10+ microservices? & \url{https://www.reddit.com/r/devops/comments/1q9tup1/anyone_else_finding_ai_code_review_tools_useless/} \\
G299 & 920 & I ran 3 Claude Code agents in parallel against a production codebase. Here's what broke and what I built to fix it. & \url{https://www.reddit.com/r/ClaudeAI/comments/1r4tuu9/i_ran_3_claude_code_agents_in_parallel_against_a/} \\
G300 & 922 & GPT-4.5 and debugging & \url{https://www.reddit.com/r/ChatGPTPro/comments/1jqidzk/gpt45_and_debugging/} \\
G301 & 925 & Opus 4.6 might be the cheapest model to use & \url{https://www.reddit.com/r/vibecoding/comments/1rcqnz6/opus_46_might_be_the_cheapest_model_to_use/} \\
G302 & 926 & When I ask Claude Code to give me a code review & \url{https://www.reddit.com/r/ClaudeAI/comments/1lqb3fw/when_i_ask_claude_code_to_give_me_a_code_review/} \\
G303 & 927 & [Showcase] Nikui: A Forensic Technical Debt Analyzer (Hotspots = Stench Churn) & \url{https://www.reddit.com/r/Python/comments/1rp7gtk/showcase_nikui_a_forensic_technical_debt_analyzer/} \\
G304 & 933 & What practices help you ensure code quality during rapid development cycles? & \url{https://www.reddit.com/r/ExperiencedDevs/comments/1rdjp8r/what_practices_help_you_ensure_code_quality/} \\
G305 & 939 & What is the best AI-assisted development setup for building client applications? & \url{https://www.reddit.com/r/cscareerquestions/comments/1tobics/what_is_the_best_aiassisted_development_setup_for/} \\
G306 & 940 & Could a company become very strict as a means to justify firing rather than laying people off? & \url{https://www.reddit.com/r/cscareerquestions/comments/1thq85m/could_a_company_become_very_strict_as_a_means_to/} \\
G307 & 945 & I built an OS-level sandbox so Claude Code can run with full shell access without touching my real filesystem. & \url{https://www.reddit.com/r/ClaudeAI/comments/1s2v9g4/i_built_an_oslevel_sandbox_so_claude_code_can_run/} \\
G308 & 948 & What is a software job actually like these days in the age of AI? Is it just repeatedly asking AI to "fix this code?" & \url{https://www.reddit.com/r/cscareerquestions/comments/1tq7s9h/what_is_a_software_job_actually_like_these_days/} \\
G309 & 950 & I had my agent use autoresearch over 8 iterations to improve my CLAUDE.md, measuring each version against tasks from real PRs. The best one still regressed on a & \url{https://www.reddit.com/r/ClaudeAI/comments/1tpghlh/i_had_my_agent_use_autoresearch_over_8_iterations/} \\
G310 & 952 & Copy-pasting between CC Opus 4.5 and GPT 5.1 Codex has 10'ed my vibecoding & \url{https://www.reddit.com/r/ClaudeAI/comments/1pcmime/copypasting_between_cc_opus_45_and_gpt_51_codex/} \\
G311 & 954 & ASANMOD: A System That Prevents Cursor AI Errors - A 2-Month Enterprise Project Story & \url{https://www.reddit.com/r/cursor/comments/1pvb509/asanmod_a_system_that_prevents_cursor_ai_errors_a/} \\
G312 & 956 & Qwen3.5 27B refusals on vuln research tasks & \url{https://www.reddit.com/r/LocalLLaMA/comments/1rqdby6/qwen35_27b_refusals_on_vuln_research_tasks/} \\
G313 & 961 & 10 brutal lessons from a year of vibe coding & \url{https://www.reddit.com/r/vibecoding/comments/1oj0id9/10_brutal_lessons_from_a_year_of_vibe_coding/} \\
G314 & 962 & Hey Rustaceans! Got a question? Ask here (42/2025)! & \url{https://www.reddit.com/r/rust/comments/1o5drn4/hey_rustaceans_got_a_question_ask_here_422025/} \\
G315 & 963 & What is the best way to replicate the Cursor code review flow in Claude Code? & \url{https://www.reddit.com/r/ClaudeAI/comments/1pukfbl/what_is_the_best_way_to_replicate_the_cursor_code/} \\
G316 & 964 & For local coders & \url{https://www.reddit.com/r/vibecoding/comments/1sw1ais/for_local_coders/} \\
G317 & 965 & That nginx 18 year vuln has me rethinking how we review infra code & \url{https://www.reddit.com/r/devops/comments/1tf08o8/that_nginx_18_year_vuln_has_me_rethinking_how_we/} \\
G318 & 968 & AI Engineer is the best engineer & \url{https://www.reddit.com/r/ExperiencedDevs/comments/1qf2nsr/ai_engineer_is_the_best_engineer/} \\
G319 & 969 & A code review tool designed to understand your project, rather than perform a static analysis & \url{https://www.reddit.com/r/codereview/comments/1twykg9/a_code_review_tool_designed_to_understand_your/} \\
G320 & 970 & Code Review: First Ever Go App? & \url{https://www.reddit.com/r/golang/comments/1ildtxf/code_review_first_ever_go_app/} \\
G321 & 972 & How to turn a 5-minute Al prompt into 48 hours of work for your team & \url{https://www.reddit.com/r/ArtificialInteligence/comments/1rw0jvp/how_to_turn_a_5minute_al_prompt_into_48_hours_of/} \\
G322 & 973 & Feeling stuck & \url{https://www.reddit.com/r/SoftwareEngineering/comments/1n4nzbr/feeling_stuck/} \\
G323 & 975 & I built an app to efficiently use my Pro plan tokens & \url{https://www.reddit.com/r/ClaudeAI/comments/1r8nwxs/i_built_an_app_to_efficiently_use_my_pro_plan/} \\
G324 & 978 & Governed AgentChain for Claude -- Modular Sub-Agent Loop with Test Integrity, Emotion Modulation, and Safe Output Delivery [M87 Studio] & \url{https://www.reddit.com/r/ClaudeAI/comments/1lskmaw/governed_agentchain_for_claude_modular_subagent/} \\
G325 & 982 & Claude Code Vs Gemini CLI - Initial Agentic Impressions & \url{https://www.reddit.com/r/ClaudeAI/comments/1lkew5x/claude_code_vs_gemini_cli_initial_agentic/} \\
G326 & 984 & I built an open source tool to help teams incorporate code reviews into interviews: CodeRev.app & \url{https://www.reddit.com/r/webdev/comments/1iksm5x/i_built_an_open_source_tool_to_help_teams/} \\
G327 & 986 & How Do You Guys Are Using Plan Mode? & \url{https://www.reddit.com/r/ClaudeAI/comments/1scn8ln/how_do_you_guys_are_using_plan_mode/} \\
G328 & 987 & icedream/testctxlint: Golang linter to ensure use of test context & \url{https://www.reddit.com/r/golang/comments/1m5d4vx/icedreamtestctxlint_golang_linter_to_ensure_use/} \\
G329 & 990 & Roo Code 3.29.0 Release Updates | Cloud Agent | Intelligent file reading | Browser-use for image models + fixes & \url{https://www.reddit.com/r/ChatGPTCoding/comments/1offm4r/roo_code_3290_release_updates_cloud_agent/} \\
G330 & 991 & Production ready pull request automation sounds great in theory but does it work in practice? & \url{https://www.reddit.com/r/webdev/comments/1rawe06/production_ready_pull_request_automation_sounds/} \\
G331 & 993 & We are building an agent-to-agent research hub. & \url{https://www.reddit.com/r/AI_Agents/comments/1s8jvz7/we_are_building_an_agenttoagent_research_hub/} \\
G332 & 996 & Spent 7\% of my Max20 weekly tokens A/B testing oh-my-claudecode hooks so you don't have to & \url{https://www.reddit.com/r/ClaudeAI/comments/1rsh7oi/spent_7_of_my_max20_weekly_tokens_ab_testing/} \\
G333 & 997 & My 1-on-1s with a senior engineer went from mentorship to nitpicky criticism. How to handle? & \url{https://www.reddit.com/r/cscareerquestions/comments/1rk39fu/my_1on1s_with_a_senior_engineer_went_from/} \\
G334 & 1002 & Github Copilot official announcement: Change to usage-based billing & \url{https://www.reddit.com/r/GithubCopilot/comments/1sx8d9g/github_copilot_official_announcement_change_to/} \\
G335 & 1009 & What's the best "How We Work" Playbook for a startup ? & \url{https://www.reddit.com/r/SoftwareEngineering/comments/1n32rln/whats_the_best_how_we_work_playbook_for_a_startup/} \\
G336 & 1017 & Management theory from the 1970s applies surprisingly well to managing AI agents in 2026 & \url{https://www.reddit.com/r/ExperiencedDevs/comments/1r9d51q/management_theory_from_the_1970s_applies/} \\
G337 & 1018 & Career transitioning away from computer science & \url{https://www.reddit.com/r/cscareerquestions/comments/1saouck/career_transitioning_away_from_computer_science/} \\
G338 & 1022 & Has anyone found ai code review with low false positives that doesn't bury real issues? & \url{https://www.reddit.com/r/vibecoding/comments/1rbi57b/has_anyone_found_ai_code_review_with_low_false/} \\
G339 & 1026 & Claude just launched their own IDE security scanner and now I'm questioning everything about our Cursor setup & \url{https://www.reddit.com/r/cursor/comments/1rm6nnr/claude_just_launched_their_own_ide_security/} \\
G340 & 1030 & I read threads complaining about codex every week... tf are y'alls workflows? & \url{https://www.reddit.com/r/OpenAI/comments/1tkmsh6/i_read_threads_complaining_about_codex_every_week/} \\
G341 & 1034 & Are Github Copilot Reviews getting worse? & \url{https://www.reddit.com/r/GithubCopilot/comments/1ozveu4/are_github_copilot_reviews_getting_worse/} \\
G342 & 1040 & The senior engineer skills that actually matter aren't technical & \url{https://www.reddit.com/r/ExperiencedDevs/comments/1rm50gq/the_senior_engineer_skills_that_actually_matter/} \\
G343 & 1041 & How Claude helped me release my first app on the App Store & \url{https://www.reddit.com/r/vibecoding/comments/1todsop/how_claude_helped_me_release_my_first_app_on_the/} \\
G344 & 1043 & Helpful tips from a dev background and hundreds upon hundreds of hours of vibe coding & \url{https://www.reddit.com/r/vibecoding/comments/1n56x0z/helpful_tips_from_a_dev_background_and_hundreds/} \\
G345 & 1046 & Copilot code review missing from reviewers list after switching plans -- what am I missing? & \url{https://www.reddit.com/r/github/comments/1tundj9/copilot_code_review_missing_from_reviewers_list/} \\
G346 & 1047 & Vibe code isn't meant to be reviewed (* the same way normal code is) & \url{https://www.reddit.com/r/programming/comments/1l7dv3r/vibe_code_isnt_meant_to_be_reviewed_the_same_way/} \\
G347 & 1048 & Half of GitHub trending AI repos are "skills" packs but the shape varies 1000x. The actual primitive is doing something real. & \url{https://www.reddit.com/r/AI_Agents/comments/1tvo738/half_of_github_trending_ai_repos_are_skills_packs/} \\
G348 & 1049 & Inputs on improving development workflow & \url{https://www.reddit.com/r/AI_Agents/comments/1t7zff8/inputs_on_improving_development_workflow/} \\
G349 & 1053 & Claude Code project structure diagram I came across (skills, hooks, CLAUDE.md layout) & \url{https://www.reddit.com/r/ClaudeAI/comments/1rrqoj8/claude_code_project_structure_diagram_i_came/} \\
G350 & 1054 & Guidelines for vibecoding & \url{https://www.reddit.com/r/vibecoding/comments/1kfey3m/guidelines_for_vibecoding/} \\
G351 & 1060 & Code Quality is myth for my company & \url{https://www.reddit.com/r/ExperiencedDevs/comments/1kljgc7/code_quality_is_myth_for_my_company/} \\
G352 & 1063 & Built an open-source orchestration layer for running multiple AI agents 24/7 with shared memory. Coordinates both local running models (mistral) and cloud based & \url{https://www.reddit.com/r/LocalLLaMA/comments/1rw06lp/built_an_opensource_orchestration_layer_for/} \\
G353 & 1064 & Does anyone here actually work at a tech company? & \url{https://www.reddit.com/r/cscareerquestions/comments/1tfimxg/does_anyone_here_actually_work_at_a_tech_company/} \\
G354 & 1067 & My company embraces vibe coders & \url{https://www.reddit.com/r/ExperiencedDevs/comments/1slxc6c/my_company_embraces_vibe_coders/} \\
G355 & 1068 & What would you want in a GitHub dashboard? & \url{https://www.reddit.com/r/github/comments/1nv2ekl/what_would_you_want_in_a_github_dashboard/} \\
G356 & 1069 & 5 agent skills I found on the Agensi marketplace that actually changed my workflow & \url{https://www.reddit.com/r/ClaudeAI/comments/1rv5gyp/5_agent_skills_i_found_on_the_agensi_marketplace/} \\
G357 & 1070 & Future of code review process? & \url{https://www.reddit.com/r/codereview/comments/1nzfyrg/future_of_code_review_process/} \\
G358 & 1073 & Is this normal for junior software engineers? & \url{https://www.reddit.com/r/cscareerquestions/comments/1o6cld5/is_this_normal_for_junior_software_engineers/} \\
G359 & 1074 & Released v0.1.6 of Owlex, an MCP server that integrates Codex CLI, Gemini CLI, and OpenCode into Claude Code. & \url{https://www.reddit.com/r/LocalLLaMA/comments/1q6cbgy/released_v016_of_owlex_an_mcp_server_that/} \\
G360 & 1079 & 5 Claude Code sessions voted to destroy the One ring. & \url{https://www.reddit.com/r/ClaudeAI/comments/1s2gcd1/5_claude_code_sessions_voted_to_destroy_the_one/} \\
G361 & 1081 & I built a deterministic orchestrator that runs Claude Code as a worker pool -- zero tokens spent on scheduling & \url{https://www.reddit.com/r/ClaudeAI/comments/1t5rbh4/i_built_a_deterministic_orchestrator_that_runs/} \\
G362 & 1083 & Help figuring out git merge strategy & \url{https://www.reddit.com/r/git/comments/1lj4e6n/help_figuring_out_git_merge_strategy/} \\
G363 & 1085 & Code review tooling & \url{https://www.reddit.com/r/csharp/comments/1owhqgf/code_review_tooling/} \\
G364 & 1086 & The "Forever Junior" Paradox: The Only Software Engineering Mindset That Matters & \url{https://www.reddit.com/r/SoftwareEngineering/comments/1tdrnye/the_forever_junior_paradox_the_only_software/} \\
G365 & 1101 & Code Review Request: Built a thread-safe LRU cache to practice Generics and GC optimization. Currently hitting 0 allocs/op. & \url{https://www.reddit.com/r/golang/comments/1q22a7g/code_review_request_built_a_threadsafe_lru_cache/} \\
G366 & 1105 & I've Done 300+ Coding Sessions and Here's What Everyone Gets Wrong & \url{https://www.reddit.com/r/cursor/comments/1ozgbx1/ive_done_300_coding_sessions_and_heres_what/} \\
G367 & 1108 & AI Developer Tools Landscape 2026 v2 - 02/26/2026 & \url{https://www.reddit.com/r/LocalLLaMA/comments/1rf6sy5/ai_developer_tools_landscape_2026_v2_02262026/} \\
G368 & 1109 & How doing code review was like before ChatGPT was a thing & \url{https://www.reddit.com/r/ClaudeAI/comments/1pc5y6i/how_doing_code_review_was_like_before_chatgpt_was/} \\
G369 & 1111 & Opus 4.5 head-to-head against Codex 5.2 xhigh on a real task. Neither won. & \url{https://www.reddit.com/r/ChatGPTCoding/comments/1q6m1ui/opus_45_headtohead_against_codex_52_xhigh_on_a/} \\
G370 & 1113 & I built a Claude Code plugin that actually fixes clean-code violations (not just reports them). Deleted 763 lines from a live app this weekend. & \url{https://www.reddit.com/r/ClaudeAI/comments/1sp40lv/i_built_a_claude_code_plugin_that_actually_fixes/} \\
G371 & 1114 & AI Tip to Toe code review. & \url{https://www.reddit.com/r/cursor/comments/1hzv115/ai_tip_to_toe_code_review/} \\
G372 & 1116 & Would you let a junior dev use AI? & \url{https://www.reddit.com/r/ExperiencedDevs/comments/1nnsyuc/would_you_let_a_junior_dev_use_ai/} \\
G373 & 1117 & Did I accidentally become a vibe coder? & \url{https://www.reddit.com/r/programming/comments/1jdtzkn/did_i_accidentally_become_a_vibe_coder/} \\
G374 & 1119 & Issues with a senior & \url{https://www.reddit.com/r/softwarearchitecture/comments/1p3egkr/issues_with_a_senior/} \\
G375 & 1120 & ControlFlowMonitor: A collection of MCP agents to monitor project status/create user stories/review code, and continue prompts without the need for human interv & \url{https://www.reddit.com/r/ClaudeAI/comments/1psdzsl/controlflowmonitor_a_collection_of_mcp_agents_to/} \\
G376 & 1123 & Code review Request C\# & \url{https://www.reddit.com/r/codereview/comments/1pwfowc/code_review_request_c/} \\
G377 & 1124 & Since AI, I don't feel motivated building side projects on my own anymore & \url{https://www.reddit.com/r/SoftwareEngineering/comments/1rug5ik/since_ai_i_dont_feel_motivated_building_side/} \\
G378 & 1127 & Experimenting with a coordinated multi-agent workflow in GitHub Copilot & \url{https://www.reddit.com/r/GithubCopilot/comments/1r7jx0b/experimenting_with_a_coordinated_multiagent/} \\
G379 & 1128 & Been building a tool that remembers WHY you wrote that code 4 days ago & \url{https://www.reddit.com/r/devops/comments/1obcwcw/been_building_a_tool_that_remembers_why_you_wrote/} \\
G380 & 1131 & I built a Claude Code plugin that fights you before it helps you -- 20 skills that force thinking before code & \url{https://www.reddit.com/r/ClaudeAI/comments/1s8dyav/i_built_a_claude_code_plugin_that_fights_you/} \\
G381 & 1132 & Who is still doing this? & \url{https://www.reddit.com/r/ClaudeAI/comments/1qkthw3/who_is_still_doing_this/} \\
G382 & 1134 & How to effectively use AI (Claude) for larger coding projects? Hitting some roadblocks & \url{https://www.reddit.com/r/ClaudeAI/comments/1hzqbur/how_to_effectively_use_ai_claude_for_larger/} \\
G383 & 1137 & What's task can you actually automate in software engineering job using AI & \url{https://www.reddit.com/r/ArtificialInteligence/comments/1nk0ol9/whats_task_can_you_actually_automate_in_software/} \\
G384 & 1139 & How Do You Keep Loving Coding in a Low-Quality Development Environment? & \url{https://www.reddit.com/r/cscareerquestions/comments/1ik382h/how_do_you_keep_loving_coding_in_a_lowquality/} \\
G385 & 1146 & AI is going to replace us. What's next. & \url{https://www.reddit.com/r/cscareerquestions/comments/1i8famv/ai_is_going_to_replace_us_whats_next/} \\
G386 & 1150 & Im quitting my job because of Claude Code & \url{https://www.reddit.com/r/ClaudeAI/comments/1l57210/im_quitting_my_job_because_of_claude_code/} \\
G387 & 1154 & The "explain it to a junior" prompt that catches bugs AI misses & \url{https://www.reddit.com/r/vibecoding/comments/1pu4zve/the_explain_it_to_a_junior_prompt_that_catches/} \\
G388 & 1161 & Overkill or smart move? Switching my B2B SaaS (mobile-first web app) from TS/React to Rust? & \url{https://www.reddit.com/r/rust/comments/1rs64il/overkill_or_smart_move_switching_my_b2b_saas/} \\
G389 & 1162 & Would you actually use this? I'm building a code review assistant that understands your app like this. & \url{https://www.reddit.com/r/SoftwareEngineering/comments/1jxpi0a/would_you_actually_use_this_im_building_a_code/} \\
G390 & 1163 & I built a multi-agent orchestration layer for Claude Code - sharing in case it's useful to anyone & \url{https://www.reddit.com/r/ClaudeAI/comments/1qms6w8/i_built_a_multiagent_orchestration_layer_for/} \\
G391 & 1165 & Go backend authentication module -- code review & \url{https://www.reddit.com/r/golang/comments/1qr9n4v/go_backend_authentication_module_code_review/} \\
G392 & 1170 & What's your actual Claude Code workflow? Not tip, the protocol you follow every single session & \url{https://www.reddit.com/r/ClaudeAI/comments/1trn2fe/whats_your_actual_claude_code_workflow_not_tip/} \\
G393 & 1172 & Vibe Coding in a Nutshell & \url{https://www.reddit.com/r/vibecoding/comments/1mzgnzw/vibe_coding_in_a_nutshell/} \\
G394 & 1176 & Hours of Custom Command fine tuning tought me a lot (Claude Code) & \url{https://www.reddit.com/r/ClaudeAI/comments/1kqcx7o/hours_of_custom_command_fine_tuning_tought_me_a/} \\
G395 & 1181 & Claude Code workflow tips after 6 months of daily use (from a senior dev) & \url{https://www.reddit.com/r/ClaudeAI/comments/1sn27yu/claude_code_workflow_tips_after_6_months_of_daily/} \\
G396 & 1184 & its not just me is it? deepseek v4 is INSANELY cheap & \url{https://www.reddit.com/r/vibecoding/comments/1sxx9dl/its_not_just_me_is_it_deepseek_v4_is_insanely/} \\
G397 & 1185 & Burnout - I care but I don't care & \url{https://www.reddit.com/r/ExperiencedDevs/comments/1srl0zw/burnout_i_care_but_i_dont_care/} \\
G398 & 1186 & Read through Anthropic's 2026 agentic coding report, a few numbers that stuck with me & \url{https://www.reddit.com/r/ClaudeAI/comments/1smuabd/read_through_anthropics_2026_agentic_coding/} \\
G399 & 1187 & Four Months of AI Code Review: What We Learned & \url{https://www.reddit.com/r/github/comments/1lfa5bm/four_months_of_ai_code_review_what_we_learned/} \\
G400 & 1190 & I built an open-source toolkit for Claude Code that automates loop driving, slash commands, and multi-model review & \url{https://www.reddit.com/r/AI_Agents/comments/1rc83kk/i_built_an_opensource_toolkit_for_claude_code/} \\
G401 & 1191 & OpenChaos Week 3: I added an immutable constitution after the internet proved me wrong & \url{https://www.reddit.com/r/webdev/comments/1ql4avi/openchaos_week_3_i_added_an_immutable/} \\
G402 & 1192 & dumpall -- Stop fighting node\_modules, dump clean project context in one go & \url{https://www.reddit.com/r/webdev/comments/1nkkvc4/dumpall_stop_fighting_node_modules_dump_clean/} \\
G403 & 1193 & Small features don't feel "small" anymore & \url{https://www.reddit.com/r/ExperiencedDevs/comments/1s36xha/small_features_dont_feel_small_anymore/} \\
G404 & 1195 & Is everyone also having trouble using up Claude.ai/Code credits? & \url{https://www.reddit.com/r/ClaudeAI/comments/1ow5ns2/is_everyone_also_having_trouble_using_up/} \\
G405 & 1198 & Developers: freelance and otherwise--would you adopt a new tool for improved codebase awareness? & \url{https://www.reddit.com/r/AI_Agents/comments/1k0uplf/developers_freelance_and_otherwisewould_you_adopt/} \\
G406 & 1199 & Anyone here at a fintech / highly-regulated org only using Copilot because it's the only AI tool Security approved? & \url{https://www.reddit.com/r/ExperiencedDevs/comments/1ocj17g/anyone_here_at_a_fintech_highlyregulated_org_only/} \\
G407 & 1200 & Need help explaining to a friend that vibe coding is bad & \url{https://www.reddit.com/r/webdev/comments/1nl5h8l/need_help_explaining_to_a_friend_that_vibe_coding/} \\
G408 & 1201 & Mental Models in Modern Software: Your Code Should Tell a Story & \url{https://www.reddit.com/r/softwarearchitecture/comments/1lxz1ej/mental_models_in_modern_software_your_code_should/} \\
G409 & 1205 & 3 weeks. 500 signups. 820 security vulnerabilities caught & \url{https://www.reddit.com/r/codereview/comments/1o37kkp/3_weeks_500_signups_820_security_vulnerabilities/} \\
G410 & 1216 & The PERFECT Code Review: How to Reduce Cognitive Load While Improving Quality & \url{https://www.reddit.com/r/webdev/comments/1t27vew/the_perfect_code_review_how_to_reduce_cognitive/} \\
G411 & 1218 & No code reviews by default & \url{https://www.reddit.com/r/webdev/comments/1su6937/no_code_reviews_by_default/} \\
G412 & 1219 & Devs using AI coding tools daily: what does your workday actually look like now? & \url{https://www.reddit.com/r/cursor/comments/1punajz/devs_using_ai_coding_tools_daily_what_does_your/} \\
G413 & 1225 & Built a Claude Code plugin for GDPR/DSGVO audits because attorney reviews were eating my budget & \url{https://www.reddit.com/r/ClaudeAI/comments/1srxu02/built_a_claude_code_plugin_for_gdprdsgvo_audits/} \\
G414 & 1226 & Is it possible to set up an AI coding agent on a VPS and interact with it from any device? & \url{https://www.reddit.com/r/ExperiencedDevs/comments/1rf97y0/is_it_possible_to_set_up_an_ai_coding_agent_on_a/} \\
G415 & 1228 & v3.3.0: Enterprise-ready workflows with formatted reports + persistent memory & \url{https://www.reddit.com/r/ClaudeAI/comments/1pwc7ca/v330_enterpriseready_workflows_with_formatted/} \\
G416 & 1229 & The new era of programming is depressing & \url{https://www.reddit.com/r/cscareerquestions/comments/1s4y50s/the_new_era_of_programming_is_depressing/} \\
G417 & 1237 & How do you review your code against the original plan or requirement? & \url{https://www.reddit.com/r/ExperiencedDevs/comments/1qim8wg/how_do_you_review_your_code_against_the_original/} \\
G418 & 1239 & I hired a remote AI developer for the first time. Honestly expected it to be a disaster. Here's what actually happened. & \url{https://www.reddit.com/r/AI_Agents/comments/1rg49fy/i_hired_a_remote_ai_developer_for_the_first_time/} \\
G419 & 1241 & Vibe coding is a monkey's paw wish and nobody's talking about it enough* & \url{https://www.reddit.com/r/vibecoding/comments/1r1kha5/vibe_coding_is_a_monkeys_paw_wish_and_nobodys/} \\
G420 & 1244 & code review for hot reload implementation in c/c++ using (mostly) win32 APIs & \url{https://www.reddit.com/r/codereview/comments/1i84se1/code_review_for_hot_reload_implementation_in_cc/} \\
G421 & 1247 & How to build apps using AI with better code quality than 95\% of "human" developed projects & \url{https://www.reddit.com/r/vibecoding/comments/1q6ddng/how_to_build_apps_using_ai_with_better_code/} \\
G422 & 1248 & claude code vs cursor & \url{https://www.reddit.com/r/cursor/comments/1qhjl3e/claude_code_vs_cursor/} \\
G423 & 1253 & I miss coding & \url{https://www.reddit.com/r/ClaudeAI/comments/1nugjrj/i_miss_coding/} \\
G424 & 1254 & Senior engineers will be demand after around 5 years? & \url{https://www.reddit.com/r/ClaudeAI/comments/1tp3tlu/senior_engineers_will_be_demand_after_around_5/} \\
G425 & 1260 & I built an Android game with Claude Code + Codex. 90\% AI, here's what actually worked and what didn't & \url{https://www.reddit.com/r/vibecoding/comments/1svmw2c/i_built_an_android_game_with_claude_code_codex_90/} \\
G426 & 1265 & Asked Claude to debug one function and it restructured my entire engineering career & \url{https://www.reddit.com/r/ClaudeAI/comments/1p25p70/asked_claude_to_debug_one_function_and_it/} \\
G427 & 1269 & Agents still writing sloppy code :/ & \url{https://www.reddit.com/r/AI_Agents/comments/1rqlcho/agents_still_writing_sloppy_code/} \\
G428 & 1271 & AI agent quick start pack & \url{https://www.reddit.com/r/OpenAI/comments/1ij7meb/ai_agent_quick_start_pack/} \\
G429 & 1274 & Is there a way to use Claude Code without CLI? & \url{https://www.reddit.com/r/ClaudeAI/comments/1qb1bvy/is_there_a_way_to_use_claude_code_without_cli/} \\
G430 & 1275 & Spotify Ad Blocker & \url{https://www.reddit.com/r/Python/comments/1qkqpgw/spotify_ad_blocker/} \\
G431 & 1278 & Just how far complex can you get via vibe coding? & \url{https://www.reddit.com/r/vibecoding/comments/1nwhyqb/just_how_far_complex_can_you_get_via_vibe_coding/} \\
G432 & 1280 & Consistent Hashing Beginner & \url{https://www.reddit.com/r/golang/comments/1l05x34/consistent_hashing_beginner/} \\
G433 & 1282 & I built a local Garmin health archive with Claude. The token limit shaped the way. & \url{https://www.reddit.com/r/vibecoding/comments/1ts3pow/i_built_a_local_garmin_health_archive_with_claude/} \\
G434 & 1283 & Can vibe coding be part of a CI/CD pipeline, or do we just deploy directly from our feelings now? & \url{https://www.reddit.com/r/vibecoding/comments/1ltn6p7/can_vibe_coding_be_part_of_a_cicd_pipeline_or_do/} \\
G435 & 1288 & I built Resolvr, VS Code extension for local code reviews with Claude code & \url{https://www.reddit.com/r/ClaudeAI/comments/1savhnf/i_built_resolvr_vs_code_extension_for_local_code/} \\
G436 & 1291 & How do you keep team workflows smooth with AI-generated projects? & \url{https://www.reddit.com/r/ClaudeAI/comments/1lbfn1w/how_do_you_keep_team_workflows_smooth_with/} \\
G437 & 1295 & I built a Claude Code skill that roasts your code & \url{https://www.reddit.com/r/ClaudeAI/comments/1r48p1e/i_built_a_claude_code_skill_that_roasts_your_code/} \\
G438 & 1298 & Introducing AutomatosX : AI-Orchestrated Agents, Workflows \& Multi-Model Reasoning & \url{https://www.reddit.com/r/codereview/comments/1qlrv4y/introducing_automatosx_aiorchestrated_agents/} \\
G439 & 1306 & One month of Claude Code on a real codebase -- the good, the bad, and the weird middle part & \url{https://www.reddit.com/r/vibecoding/comments/1ttb1lr/one_month_of_claude_code_on_a_real_codebase_the/} \\
G440 & 1309 & I've come to appreciate the Agent much more for certain kinds of tasks & \url{https://www.reddit.com/r/cursor/comments/1ja1v1n/ive_come_to_appreciate_the_agent_much_more_for/} \\
G441 & 1312 & How can I get better at code reviews? & \url{https://www.reddit.com/r/cscareerquestions/comments/1oc69xt/how_can_i_get_better_at_code_reviews/} \\
G442 & 1315 & Please a code review for my AI project & \url{https://www.reddit.com/r/codereview/comments/1khtcpc/please_a_code_review_for_my_ai_project/} \\
G443 & 1316 & I just can't do it anymore....this is the straw that broke the camels back! & \url{https://www.reddit.com/r/windsurf/comments/1l0qxf0/i_just_cant_do_it_anymorethis_is_the_straw_that/} \\
G444 & 1318 & Ralph wiggum for Ralph wiggum & \url{https://www.reddit.com/r/ClaudeAI/comments/1qib756/ralph_wiggum_for_ralph_wiggum/} \\
G445 & 1319 & My experience being the only programmer in a non-tech dept in a non-tech company & \url{https://www.reddit.com/r/cscareerquestions/comments/1jq9wze/my_experience_being_the_only_programmer_in_a/} \\
G446 & 1321 & Entire team fails to see the issues AI is causing us & \url{https://www.reddit.com/r/ExperiencedDevs/comments/1qy54j9/entire_team_fails_to_see_the_issues_ai_is_causing/} \\
G447 & 1322 & I open-sourced Tears, a CLI linter for keeping vibe-coded code out of critical paths & \url{https://www.reddit.com/r/vibecoding/comments/1thmf22/i_opensourced_tears_a_cli_linter_for_keeping/} \\
G448 & 1323 & [Guide] Plugins Claude Code: 2 months testing WD Framework in production (85\% time gain on Newsletter feature) & \url{https://www.reddit.com/r/ClaudeAI/comments/1o6jmpp/guide_plugins_claude_code_2_months_testing_wd/} \\
G449 & 1326 & Creator of Claude Code, Boris Cherny says coding is solved and Claude writes 100\% of his code. Is this really the case for you? & \url{https://www.reddit.com/r/cscareerquestions/comments/1s3uhf0/creator_of_claude_code_boris_cherny_says_coding/} \\
G450 & 1335 & Windsurf can now review code, deploy apps, and use internal docs -- we need something like this in Cursor! & \url{https://www.reddit.com/r/cursor/comments/1khdvij/windsurf_can_now_review_code_deploy_apps_and_use/} \\
G451 & 1338 & Set up 7 code review skills in Claude Code, the security scan caught a hardcoded secret on first run & \url{https://www.reddit.com/r/ClaudeAI/comments/1r7j6sz/set_up_7_code_review_skills_in_claude_code_the/} \\
G452 & 1342 & that google antigravity hack made me realize how much access cursor actually has & \url{https://www.reddit.com/r/webdev/comments/1pcbt6q/that_google_antigravity_hack_made_me_realize_how/} \\
G453 & 1344 & Opus 4.6 being funny in Claude Code & \url{https://www.reddit.com/r/ClaudeAI/comments/1r2ue8i/opus_46_being_funny_in_claude_code/} \\
G454 & 1345 & Automating agent workflow and minimizing errors. & \url{https://www.reddit.com/r/GithubCopilot/comments/1rvqdxs/automating_agent_workflow_and_minimizing_errors/} \\
G455 & 1348 & Finally, an easy way to get LLMs to review your entire repo without losing context ! & \url{https://www.reddit.com/r/programming/comments/1p2u0ly/finally_an_easy_way_to_get_llms_to_review_your/} \\
G456 & 1350 & AI workflow that keeps the developer in the driver's seat & \url{https://www.reddit.com/r/webdev/comments/1ttvz87/ai_workflow_that_keeps_the_developer_in_the/} \\
G457 & 1352 & I Built otak-committer: An AI-Powered VSCode Extension for Automatic Commit Message Generation! & \url{https://www.reddit.com/r/git/comments/1iqj197/i_built_otakcommitter_an_aipowered_vscode/} \\
G458 & 1356 & Reviewing coworkers' AI-generated PRs & \url{https://www.reddit.com/r/ExperiencedDevs/comments/1jfhqye/reviewing_coworkers_aigenerated_prs/} \\
G459 & 1359 & Built a Claude Code plugin that reviews code using classic SE books (Brooks, Fowler, Feathers...) & \url{https://www.reddit.com/r/ClaudeAI/comments/1s6l11z/built_a_claude_code_plugin_that_reviews_code/} \\
G460 & 1363 & Transitioning Into a Senior Role -- Struggling with Balance in My First Led Sprint & \url{https://www.reddit.com/r/ExperiencedDevs/comments/1jx3hnj/transitioning_into_a_senior_role_struggling_with/} \\
G461 & 1364 & my code review bot was scanning files one by one. 90 seconds per PR. & \url{https://www.reddit.com/r/programming/comments/1qpdsdv/my_code_review_bot_was_scanning_files_one_by_one/} \\
G462 & 1370 & Everyone building AI agents might be optimizing the wrong layer & \url{https://www.reddit.com/r/AI_Agents/comments/1rq8loa/everyone_building_ai_agents_might_be_optimizing/} \\
G463 & 1371 & Running a high-end bakery in the age of industrialized code & \url{https://www.reddit.com/r/programming/comments/1qpcj7v/running_a_highend_bakery_in_the_age_of/} \\
G464 & 1381 & I introduced agentic AI into my codebase two and a half weeks ago and today I am scrapping it for parts -- sort of. & \url{https://www.reddit.com/r/ExperiencedDevs/comments/1l1b6n8/i_introduced_agentic_ai_into_my_codebase_two_and/} \\
G465 & 1383 & This is what a 4 day bender of minimal sleep and a curious mind gets you when I thought I was "simply setting up claude code" & \url{https://www.reddit.com/r/vibecoding/comments/1qugu0b/this_is_what_a_4_day_bender_of_minimal_sleep_and/} \\
G466 & 1385 & Chatgpt destroyed me as a software engineer & \url{https://www.reddit.com/r/SoftwareEngineering/comments/1jr84si/chatgpt_destroyed_me_as_a_software_engineer/} \\
G467 & 1387 & GitHub Code Reviews for external contributors and bots? & \url{https://www.reddit.com/r/ClaudeAI/comments/1tn7f7b/github_code_reviews_for_external_contributors_and/} \\
G468 & 1388 & I built a Claude skill that stops AI agents from editing the wrong file -- open source & \url{https://www.reddit.com/r/ClaudeAI/comments/1s8vedx/i_built_a_claude_skill_that_stops_ai_agents_from/} \\
G469 & 1392 & I made 6 AI agents debate "Should AI replace code review?" -- here's the full 29-turn transcript & \url{https://www.reddit.com/r/ClaudeAI/comments/1s2duk3/i_made_6_ai_agents_debate_should_ai_replace_code/} \\
G470 & 1400 & MIT just made vibe coding an official part of engineering & \url{https://www.reddit.com/r/AI_Agents/comments/1oh5tlw/mit_just_made_vibe_coding_an_official_part_of/} \\
G471 & 1403 & Check Out My Website! & \url{https://www.reddit.com/r/vibecoding/comments/1mruweo/check_out_my_website/} \\
G472 & 1405 & Is this normal or am I the problem? & \url{https://www.reddit.com/r/cscareerquestions/comments/1p8azzg/is_this_normal_or_am_i_the_problem/} \\
G473 & 1406 & Best AI coding tools in 2026? My experience so far (Copilot vs Cursor vs others) & \url{https://www.reddit.com/r/AI_Agents/comments/1t3fz71/best_ai_coding_tools_in_2026_my_experience_so_far/} \\
G474 & 1407 & These are the claude skills we use everyday to run my freelance business. What are yours? & \url{https://www.reddit.com/r/ClaudeAI/comments/1rp52eq/these_are_the_claude_skills_we_use_everyday_to/} \\
G475 & 1408 & How do you track changes when using Claude Code vs Cursor AI? & \url{https://www.reddit.com/r/cursor/comments/1lpopfd/how_do_you_track_changes_when_using_claude_code/} \\
G476 & 1410 & Is AI QA the obvious next move after code generation took off? & \url{https://www.reddit.com/r/ArtificialInteligence/comments/1sqkz7v/is_ai_qa_the_obvious_next_move_after_code/} \\
G477 & 1412 & How do you actually stay motivated when half your team is just vibe coding with LLMs? & \url{https://www.reddit.com/r/cscareerquestions/comments/1soxyzw/how_do_you_actually_stay_motivated_when_half_your/} \\
G478 & 1413 & After Weeks of Refactoring, My Go Web Framework Finally Feels Right - Open to Reviews! & \url{https://www.reddit.com/r/golang/comments/1lj8pok/after_weeks_of_refactoring_my_go_web_framework/} \\
G479 & 1414 & Claude talking to CC & \url{https://www.reddit.com/r/ClaudeAI/comments/1srzw1n/claude_talking_to_cc/} \\
G480 & 1415 & Gemini Pro 3 vs Claude Opus 4.5 & \url{https://www.reddit.com/r/ClaudeAI/comments/1pdv776/gemini_pro_3_vs_claude_opus_45/} \\
G481 & 1417 & Claude Code : A Revenge Story & \url{https://www.reddit.com/r/ClaudeAI/comments/1rdr065/claude_code_a_revenge_story/} \\
G482 & 1422 & Software - AI software architecture thoughts & \url{https://www.reddit.com/r/ExperiencedDevs/comments/1i8pzvm/software_ai_software_architecture_thoughts/} \\
G483 & 1423 & 2048 Game Clone / Looking for code review & \url{https://www.reddit.com/r/codereview/comments/1i76mwf/2048_game_clone_looking_for_code_review/} \\
G484 & 1426 & I built a Claude Code plugin that manages the full dev lifecycle with parallel agents & \url{https://www.reddit.com/r/ClaudeAI/comments/1qse11k/i_built_a_claude_code_plugin_that_manages_the/} \\
G485 & 1427 & Used claude code in a personal project. What's the catch? & \url{https://www.reddit.com/r/cscareerquestions/comments/1r6h1aw/used_claude_code_in_a_personal_project_whats_the/} \\
G486 & 1432 & Coders: do you solely use Claude Code? & \url{https://www.reddit.com/r/ClaudeAI/comments/1ozokjz/coders_do_you_solely_use_claude_code/} \\
G487 & 1433 & Jopus: A high-performance Java wrapper for Opus (Libopus 1.6.1) using Project Panama & \url{https://www.reddit.com/r/java/comments/1qza12f/jopus_a_highperformance_java_wrapper_for_opus/} \\
G488 & 1434 & Checking vector size is not safe & \url{https://www.reddit.com/r/cpp/comments/1ko4i74/checking_vector_size_is_not_safe/} \\
G489 & 1435 & How good is Qwen-3.6-27b? I asked Claude Opus & \url{https://www.reddit.com/r/LocalLLaMA/comments/1sx12hm/how_good_is_qwen3627b_i_asked_claude_opus/} \\
G490 & 1445 & Shoutout to Claude Code (and Codex) for helping me finish my MVP after my CTO bailed & \url{https://www.reddit.com/r/ClaudeAI/comments/1nwte1k/shoutout_to_claude_code_and_codex_for_helping_me/} \\
G491 & 1447 & I run a blog about AI -- help me out! Which AI coding tool do you use most often? & \url{https://www.reddit.com/r/AI_Agents/comments/1lx31io/i_run_a_blog_about_ai_help_me_out_which_ai_coding/} \\
G492 & 1454 & What is the best practice for using a PR code review agent or skill with Cursor? & \url{https://www.reddit.com/r/cursor/comments/1sxskh1/what_is_the_best_practice_for_using_a_pr_code/} \\
G493 & 1455 & Is copilot slowing me down? & \url{https://www.reddit.com/r/GithubCopilot/comments/1spxp84/is_copilot_slowing_me_down/} \\
G494 & 1456 & Programming with AI feels like having a cheat code to reality & \url{https://www.reddit.com/r/AI_Agents/comments/1nddthj/programming_with_ai_feels_like_having_a_cheat/} \\
G495 & 1459 & Orchestrator to power Implementor/Review loop in separate agents? & \url{https://www.reddit.com/r/AI_Agents/comments/1rwn4iz/orchestrator_to_power_implementorreview_loop_in/} \\
G496 & 1462 & Stop the "Review Tax": How I hit 20x speed using ADR-driven Invariant Gates (and why non-coders might have the edge) & \url{https://www.reddit.com/r/AI_Agents/comments/1t4q34y/stop_the_review_tax_how_i_hit_20x_speed_using/} \\
G497 & 1467 & My 10 + 20 + 20 dollars dev kit that just works & \url{https://www.reddit.com/r/ClaudeAI/comments/1lzlela/my_10_20_20_dollars_dev_kit_that_just_works/} \\
G498 & 1468 & Claude Code and Codex working on implementation plan together & \url{https://www.reddit.com/r/AI_Agents/comments/1rbvvr5/claude_code_and_codex_working_on_implementation/} \\
G499 & 1470 & What's your "Human" engineer pitch? & \url{https://www.reddit.com/r/SoftwareEngineering/comments/1r4iv89/whats_your_human_engineer_pitch/} \\
G500 & 1471 & What is wrong with current coding agents workflow ( and how are we are trying to correct it) & \url{https://www.reddit.com/r/AI_Agents/comments/1qbomhe/what_is_wrong_with_current_coding_agents_workflow/} \\
G501 & 1473 & I have a homebrew self-hosted server. I am not a professional. I have many questions! & \url{https://www.reddit.com/r/AskProgramming/comments/1oisw2x/i_have_a_homebrew_selfhosted_server_i_am_not_a/} \\
G502 & 1482 & Scaling Engineering Workflows with Open-Source Tooling & \url{https://www.reddit.com/r/golang/comments/1rybxnb/scaling_engineering_workflows_with_opensource/} \\
G503 & 1483 & First steps in semi-autonomous multi-agent software development & \url{https://www.reddit.com/r/AI_Agents/comments/1s9rrnr/first_steps_in_semiautonomous_multiagent_software/} \\
G504 & 1485 & I built an executive team of 6 AI agents to manage my 15 side projects -- here's the full breakdown & \url{https://www.reddit.com/r/ClaudeAI/comments/1s2nkm1/i_built_an_executive_team_of_6_ai_agents_to/} \\
G505 & 1486 & Junior dev relies on AI heavily, should I mind my own business? & \url{https://www.reddit.com/r/ExperiencedDevs/comments/1hsuog3/junior_dev_relies_on_ai_heavily_should_i_mind_my/} \\
G506 & 1487 & Report misuse of AI? & \url{https://www.reddit.com/r/ExperiencedDevs/comments/1qwtwpw/report_misuse_of_ai/} \\
G507 & 1492 & Vibecoding - problem that i observed recently & \url{https://www.reddit.com/r/vibecoding/comments/1sevnw1/vibecoding_problem_that_i_observed_recently/} \\
G508 & 1499 & How I structure Claude Code projects (CLAUDE.md, Skills, MCP) & \url{https://www.reddit.com/r/ClaudeAI/comments/1rfwmlh/how_i_structure_claude_code_projects_claudemd/} \\
G509 & 1500 & Cross-model consensus in AI code review: why different LLMs catch different bugs (technical writeup) & \url{https://www.reddit.com/r/programming/comments/1qswzg2/crossmodel_consensus_in_ai_code_review_why/} \\
G510 & 1502 & Starting Open Source as a non-dev & \url{https://www.reddit.com/r/vibecoding/comments/1pgxf8h/starting_open_source_as_a_nondev/} \\
G511 & 1506 & I am self-taught, and I would like to use windsurf to learn & \url{https://www.reddit.com/r/windsurf/comments/1lsb18z/i_am_selftaught_and_i_would_like_to_use_windsurf/} \\
G512 & 1507 & Request for Code Review & \url{https://www.reddit.com/r/AskProgramming/comments/1nro7bm/request_for_code_review/} \\
G513 & 1509 & I built a diff viewer for Claude Code where you can ask questions to the same session that wrote the code & \url{https://www.reddit.com/r/ClaudeAI/comments/1t4eq55/i_built_a_diff_viewer_for_claude_code_where_you/} \\
G514 & 1511 & How do people solve the reliability issue of their vibe coded app? & \url{https://www.reddit.com/r/vibecoding/comments/1o5tfga/how_do_people_solve_the_reliability_issue_of/} \\
G515 & 1512 & New User/Coder - Stupid Questions need answers! & \url{https://www.reddit.com/r/ClaudeAI/comments/1qjd7fc/new_usercoder_stupid_questions_need_answers/} \\
G516 & 1515 & Problem with OpenAi codex & \url{https://www.reddit.com/r/AskProgramming/comments/1rgdl8t/problem_with_openai_codex/} \\
G517 & 1517 & Is it wise to discuss the state of (really, really) bad AI code to my reporting manager because my technical manager doesn't care about anything but corporate r & \url{https://www.reddit.com/r/cscareerquestions/comments/1r1xxjs/is_it_wise_to_discuss_the_state_of_really_really/} \\
G518 & 1520 & Code review in the Rust compiler (2024) & \url{https://www.reddit.com/r/rust/comments/1i1v42s/code_review_in_the_rust_compiler_2024/} \\
G519 & 1523 & AI timeline expectations are driving me crazy & \url{https://www.reddit.com/r/ExperiencedDevs/comments/1rncxrn/ai_timeline_expectations_are_driving_me_crazy/} \\
G520 & 1524 & Devs using AI coding agents: where does trust break in your workflow? & \url{https://www.reddit.com/r/ExperiencedDevs/comments/1tk6hg6/devs_using_ai_coding_agents_where_does_trust/} \\
G521 & 1525 & How do people enforce developers to write tests without a strict code coverage requirement? & \url{https://www.reddit.com/r/ExperiencedDevs/comments/1sz51ra/how_do_people_enforce_developers_to_write_tests/} \\
G522 & 1528 & Asked a colleague in code review to extract magic numbers and got told "devs should know" & \url{https://www.reddit.com/r/ExperiencedDevs/comments/1sxjkb2/asked_a_colleague_in_code_review_to_extract_magic/} \\
G523 & 1538 & Claude Code review: \$20 every PR? & \url{https://www.reddit.com/r/cursor/comments/1rtlird/claude_code_review_20_every_pr/} \\
G524 & 1539 & Introducing "Files to LLM Prompt" - A VSCode Extension to Streamline Prompting Claude with Your Code & \url{https://www.reddit.com/r/ClaudeAI/comments/1ih1tlg/introducing_files_to_llm_prompt_a_vscode/} \\
G525 & 1541 & Advice on High-Concurrency RabbitMQ Consumer/Producer & \url{https://www.reddit.com/r/golang/comments/1iby1u6/advice_on_highconcurrency_rabbitmq/} \\
G526 & 1547 & Genuinely impressed by Opus 4.1 & \url{https://www.reddit.com/r/ClaudeAI/comments/1mjskxj/genuinely_impressed_by_opus_41/} \\
G527 & 1549 & I open-sourced the 15 Claude Code skills I used for +100 projects in freelance & \url{https://www.reddit.com/r/ClaudeAI/comments/1tfsez4/i_opensourced_the_15_claude_code_skills_i_used/} \\
G528 & 1551 & What to Put in a Claude Code Skill for Reviewing Your Team's Code & \url{https://www.reddit.com/r/ClaudeAI/comments/1rlp9g0/what_to_put_in_a_claude_code_skill_for_reviewing/} \\
G529 & 1552 & How do you even review ai code? & \url{https://www.reddit.com/r/cscareerquestions/comments/1s1y2ud/how_do_you_even_review_ai_code/} \\
G530 & 1553 & We built an open-source AI coworker on Claude -- 1,800+ employees use it daily & \url{https://www.reddit.com/r/ClaudeAI/comments/1q5gqzu/we_built_an_opensource_ai_coworker_on_claude_1800/} \\
G531 & 1557 & Quick q - how are you handling code reviews right now? & \url{https://www.reddit.com/r/webdev/comments/1jo2jz8/quick_q_how_are_you_handling_code_reviews_right/} \\
G532 & 1559 & Working with a messy offshore team, do I push for better code or just chill? & \url{https://www.reddit.com/r/cscareerquestions/comments/1m4yf20/working_with_a_messy_offshore_team_do_i_push_for/} \\
G533 & 1560 & Relatively early CS career (SE II): Advice not getting defensive during senior code review? & \url{https://www.reddit.com/r/cscareerquestions/comments/1i0upv9/relatively_early_cs_career_se_ii_advice_not/} \\
G534 & 1566 & Please be careful with large (vibed) codebases. & \url{https://www.reddit.com/r/vibecoding/comments/1qrd4ao/please_be_careful_with_large_vibed_codebases/} \\
G535 & 1574 & My Journey Building an AI Agent Orchestrator & \url{https://www.reddit.com/r/LocalLLaMA/comments/1r12sx0/my_journey_building_an_ai_agent_orchestrator/} \\
G536 & 1583 & Showoff Saturday : Your AI assistant is shipping code faster than your team can understand it. DevLens maps it all, locally, in 20 seconds. & \url{https://www.reddit.com/r/webdev/comments/1s5vcom/showoff_saturday_your_ai_assistant_is_shipping/} \\
G537 & 1587 & Devs using AI coding tools daily: what does your workday actually look like now? & \url{https://www.reddit.com/r/ClaudeAI/comments/1pun9u5/devs_using_ai_coding_tools_daily_what_does_your/} \\
G538 & 1595 & Claude Code vs Codex: 36 files vs 28, \$2.50 vs \$2.04, and one infinite loop. My full breakdown. & \url{https://www.reddit.com/r/ClaudeAI/comments/1tc7tku/claude_code_vs_codex_36_files_vs_28_250_vs_204/} \\
G539 & 1601 & Anyone here looking to get referral as a Senior/Staff Code Review Expert position | \$40 to \$125 / H & \url{https://www.reddit.com/r/Python/comments/1pb83nq/anyone_here_looking_to_get_referral_as_a/} \\
G540 & 1603 & Just Launched: PullSense - AI-Powered PR Reviews! & \url{https://www.reddit.com/r/softwarearchitecture/comments/1if4xc9/just_launched_pullsense_aipowered_pr_reviews/} \\
G541 & 1604 & AI coded my app in 2 hours. It took 3 days to make it actually work. & \url{https://www.reddit.com/r/vibecoding/comments/1q3wzl0/ai_coded_my_app_in_2_hours_it_took_3_days_to_make/} \\
G542 & 1605 & Built a Security Scanner, Getting Signups But No Retention - Architecture Issue or Product Issue? & \url{https://www.reddit.com/r/webdev/comments/1qlc76n/built_a_security_scanner_getting_signups_but_no/} \\
G543 & 1607 & What do you do while the agent is running a task? & \url{https://www.reddit.com/r/ClaudeAI/comments/1s3pr9w/what_do_you_do_while_the_agent_is_running_a_task/} \\
G544 & 1608 & DELIGHT - self-hosted AI engineering autopilot: local LLM + browser farm + repo graph + P2P compute & \url{https://www.reddit.com/r/AI_Agents/comments/1t6nmq1/delight_selfhosted_ai_engineering_autopilot_local/} \\
G545 & 1610 & Vibecoded webapp & \url{https://www.reddit.com/r/vibecoding/comments/1t0g4zw/vibecoded_webapp/} \\
G546 & 1614 & Appa - Self-shipping task queue via Linear + Claude Code & \url{https://www.reddit.com/r/ClaudeAI/comments/1qh6bp0/appa_selfshipping_task_queue_via_linear_claude/} \\
G547 & 1617 & Deterministic multi-subagent orchestration - what's new in CC 2.1.146 (+4,755 tokens) & \url{https://www.reddit.com/r/ClaudeAI/comments/1tll4mv/deterministic_multisubagent_orchestration_whats/} \\
G548 & 1619 & Does anyone know of (or if) any recent outages had been caused by over trusting an LLM? & \url{https://www.reddit.com/r/ExperiencedDevs/comments/1ooq2me/does_anyone_know_of_or_if_any_recent_outages_had/} \\
G549 & 1633 & Weird vibes from AI-assisted coding (Claude + Go + React) & \url{https://www.reddit.com/r/vibecoding/comments/1oey3tr/weird_vibes_from_aiassisted_coding_claude_go_react/} \\
G550 & 1643 & Serious need for code review & \url{https://www.reddit.com/r/rust/comments/1t65mpw/serious_need_for_code_review/} \\
G551 & 1647 & Is Go tooling ready for AI as the primary builder -- or are we still designing for humans? & \url{https://www.reddit.com/r/golang/comments/1rnanrn/is_go_tooling_ready_for_ai_as_the_primary_builder/} \\
G552 & 1653 & Do you use Claude Code to review PRs? & \url{https://www.reddit.com/r/ClaudeAI/comments/1qd8ge8/do_you_use_claude_code_to_review_prs/} \\
G553 & 1659 & How do you cut code review time without sacrificing refactoring safety in the process & \url{https://www.reddit.com/r/softwarearchitecture/comments/1s26psj/how_do_you_cut_code_review_time_without/} \\
G554 & 1660 & How are you structuring AI-assisted product development workflows? & \url{https://www.reddit.com/r/ExperiencedDevs/comments/1sv7qpv/how_are_you_structuring_aiassisted_product/} \\
G555 & 1661 & Can an experienced AI-first engineer be successful using AI with no domain knowledge? & \url{https://www.reddit.com/r/ArtificialInteligence/comments/1qq0hzi/can_an_experienced_aifirst_engineer_be_successful/} \\
G556 & 1662 & Production-grade, composable Skills for AI coding agents : code review, secure coding, architecture, and more. Compatible with any SKILL.md-supporting agent. & \url{https://www.reddit.com/r/github/comments/1tq90yr/productiongrade_composable_skills_for_ai_coding/} \\
G557 & 1664 & The flood of AI-generated slop is just inevitable given how many devs never truly internalized their language or runtime well enough to read and evaluate code c & \url{https://www.reddit.com/r/ExperiencedDevs/comments/1q9la70/the_flood_of_aigenerated_slop_is_just_inevitable/} \\
G558 & 1669 & [Code Review] I built an "Intentional" UI component in React Native for creating posts. How can I improve this architecture? & \url{https://www.reddit.com/r/AskProgramming/comments/1rc819h/code_review_i_built_an_intentional_ui_component/} \\
G559 & 1673 & Github PR reviewer ROI & \url{https://www.reddit.com/r/AI_Agents/comments/1q2kozp/github_pr_reviewer_roi/} \\
G560 & 1675 & I gave Claude Code five different personalities and run them all at once & \url{https://www.reddit.com/r/AI_Agents/comments/1s808l5/i_gave_claude_code_five_different_personalities/} \\
G561 & 1679 & Arguably, the best AI code review MCP server (with Serena integration) & \url{https://www.reddit.com/r/LocalLLaMA/comments/1qu6ylc/arguably_the_best_ai_code_review_mcp_server_with/} \\
G562 & 1681 & OpenAI Codex. Is it better than Cursor? & \url{https://www.reddit.com/r/OpenAI/comments/1kyqeg7/openai_codex_is_it_better_than_cursor/} \\
G563 & 1683 & Need to organize a professional vibe coding -> prod workflow with tool chain and proper code review & \url{https://www.reddit.com/r/vibecoding/comments/1tamw61/need_to_organize_a_professional_vibe_coding_prod/} \\
G564 & 1685 & Your MVP is now a maze and every new feature feels like a trap & \url{https://www.reddit.com/r/vibecoding/comments/1ra1sfh/your_mvp_is_now_a_maze_and_every_new_feature/} \\
G565 & 1686 & Long term consequences of using LLMs for programming & \url{https://www.reddit.com/r/cscareerquestions/comments/1srvk3w/long_term_consequences_of_using_llms_for/} \\
G566 & 1696 & 3 important vibe coding lessons (+ mistakes you don't want to make) & \url{https://www.reddit.com/r/vibecoding/comments/1oab3yx/3_important_vibe_coding_lessons_mistakes_you_dont/} \\
G567 & 1697 & Would You Use an AI Code Reviewer That Understands Your Whole Project and Learns Your Preferences? & \url{https://www.reddit.com/r/gitlab/comments/1k8e9xm/would_you_use_an_ai_code_reviewer_that/} \\
G568 & 1699 & I've used AI to write 100\% of my code for 1+ year as an engineer. 13 hype-free lessons & \url{https://www.reddit.com/r/ArtificialInteligence/comments/1r0e58x/ive_used_ai_to_write_100_of_my_code_for_1_year_as/} \\
G569 & 1702 & TimeCop - A tool that keeps me up with Claude Code commit pace & \url{https://www.reddit.com/r/ClaudeAI/comments/1qvozg2/timecop_a_tool_that_keeps_me_up_with_claude_code/} \\
G570 & 1704 & Claude Code used \textasciitilde{}\$20 for 2 basic reviews, while a similar API run cost \textasciitilde{}\$5.30. Is this expected? & \url{https://www.reddit.com/r/ClaudeAI/comments/1sbsk33/claude_code_used_20_for_2_basic_reviews_while_a/} \\
G571 & 1708 & How to communicate to a fellow dev that it's okay to make sacrifices for redeability ? & \url{https://www.reddit.com/r/ExperiencedDevs/comments/1k7euxw/how_to_communicate_to_a_fellow_dev_that_its_okay/} \\
G572 & 1710 & I tested 120 "secret Claude codes" over 3 months. 47\% are placebo. Here's what actually works. & \url{https://www.reddit.com/r/ClaudeAI/comments/1sqqzl0/i_tested_120_secret_claude_codes_over_3_months_47/} \\
G573 & 1718 & Ask me anything on building AI agent & \url{https://www.reddit.com/r/AI_Agents/comments/1ts284p/ask_me_anything_on_building_ai_agent/} \\
G574 & 1729 & after a month with 5 Chinese coding LLMs, is M3 actually going to take the top spot? & \url{https://www.reddit.com/r/ArtificialInteligence/comments/1tkhi8r/after_a_month_with_5_chinese_coding_llms_is_m3/} \\
G575 & 1734 & PR looks fine, but something breaks one or two days after merge. How do you catch this? & \url{https://www.reddit.com/r/codereview/comments/1sqgvfq/pr_looks_fine_but_something_breaks_one_or_two/} \\
G576 & 1735 & 3 important AI coding lessons when you're starting out & \url{https://www.reddit.com/r/vibecoding/comments/1odufm8/3_important_ai_coding_lessons_when_youre_starting/} \\
G577 & 1740 & Has anyone avoided burnout due to excessive stakeholder feedback? & \url{https://www.reddit.com/r/ExperiencedDevs/comments/1pb2g58/has_anyone_avoided_burnout_due_to_excessive/} \\
G578 & 1741 & One thing I learned building an AI code review tool: clarity matters more than "intelligence" & \url{https://www.reddit.com/r/programming/comments/1qo0fbf/one_thing_i_learned_building_an_ai_code_review/} \\
G579 & 1743 & Possibility of using BMad Method + GitHub CoPilot+ autonomously & \url{https://www.reddit.com/r/GithubCopilot/comments/1qxsz6y/possibility_of_using_bmad_method_github_copilot/} \\
G580 & 1744 & Code Review Request: A structured boilerplate for a Go web app (using PocketBase) & \url{https://www.reddit.com/r/golang/comments/1mdjocu/code_review_request_a_structured_boilerplate_for/} \\
G581 & 1746 & [AskJS] In production JavaScript apps, how do you decide when abstraction becomes overengineering? & \url{https://www.reddit.com/r/javascript/comments/1qrv2u1/askjs_in_production_javascript_apps_how_do_you/} \\
G582 & 1747 & Review/rate my top 3 selected md file & \url{https://www.reddit.com/r/GithubCopilot/comments/1tr2ky0/reviewrate_my_top_3_selected_md_file/} \\
G583 & 1750 & how much time should seniors spend on reviews? trying to save time on manual code reviews & \url{https://www.reddit.com/r/devops/comments/1pl6jj7/how_much_time_should_seniors_spend_on_reviews/} \\
G584 & 1753 & I'm a product designer without any code knowledge (apart from HTML/CSS), actively contributing to our codebase with Clade Code. What can I do to improve my and  & \url{https://www.reddit.com/r/vibecoding/comments/1thpy50/im_a_product_designer_without_any_code_knowledge/} \\
G585 & 1757 & Inside a 116-Configuration Claude Code Setup: Skills, Hooks, Agents, and the Layering That Makes It Work & \url{https://www.reddit.com/r/ClaudeAI/comments/1rlth6q/inside_a_116configuration_claude_code_setup/} \\
G586 & 1758 & How do you deal with AI pressure from higher roles? & \url{https://www.reddit.com/r/ExperiencedDevs/comments/1m2w8vo/how_do_you_deal_with_ai_pressure_from_higher_roles/} \\
G587 & 1760 & How are you actually getting the most out of Claude Code? Struggling with OpenSpec + Superpowers workflow, multi-agent setup, and sub-agent quality & \url{https://www.reddit.com/r/ClaudeAI/comments/1tptgl2/how_are_you_actually_getting_the_most_out_of/} \\
G588 & 1768 & Agentic AI feels like a new teammate in dev work. Anyone else seeing this? & \url{https://www.reddit.com/r/LocalLLaMA/comments/1n8yz7s/agentic_ai_feels_like_a_new_teammate_in_dev_work/} \\
G589 & 1769 & A quiz based code reviewing tool & \url{https://www.reddit.com/r/codereview/comments/1r5arwr/a_quiz_based_code_reviewing_tool/} \\
G590 & 1772 & Anyone using multiple models to plan, then Claude to code, then others to review? & \url{https://www.reddit.com/r/vibecoding/comments/1q3yauh/anyone_using_multiple_models_to_plan_then_claude/} \\
G591 & 1774 & Cursor's price is wild, so i jumped back to VS Code (2 tools at \$10 each) & \url{https://www.reddit.com/r/cursor/comments/1lwbu0u/cursors_price_is_wild_so_i_jumped_back_to_vs_code/} \\
G592 & 1775 & RIP claude-code , I am in love with this one & \url{https://www.reddit.com/r/vibecoding/comments/1taa3kn/rip_claudecode_i_am_in_love_with_this_one/} \\
G593 & 1778 & My project just crossed \$3k in revenue, 6 weeks after launch & \url{https://www.reddit.com/r/vibecoding/comments/1syenf0/my_project_just_crossed_3k_in_revenue_6_weeks/} \\
G594 & 1779 & Agentic coding workflow (Ask -> plan.md -> implement loop). Codex vs Cursor \$20 -- worth switching? & \url{https://www.reddit.com/r/cursor/comments/1rnqke6/agentic_coding_workflow_ask_planmd_implement_loop/} \\
G595 & 1782 & I've Built 20+ AI Apps And Here's What (Actually) Keeps Them From Shipping & \url{https://www.reddit.com/r/cursor/comments/1pbk6i9/ive_built_20_ai_apps_and_heres_what_actually/} \\
G596 & 1784 & Why a Coding Agent with Different Prompts Should Not Be Your Code Review Agent & \url{https://www.reddit.com/r/cursor/comments/1qkuy7g/why_a_coding_agent_with_different_prompts_should/} \\
G597 & 1785 & I built CodeSage: an Ai code reviewer & \url{https://www.reddit.com/r/codereview/comments/1nm0248/i_built_codesage_an_ai_code_reviewer/} \\
G598 & 1786 & Rust JSONLogic Expression Evaluation Slower Than JS - Looking for Code Review \& Optimization Tips & \url{https://www.reddit.com/r/rust/comments/1iygicz/rust_jsonlogic_expression_evaluation_slower_than/} \\
G599 & 1797 & Long-harness agentic programming & \url{https://www.reddit.com/r/programming/comments/1s4ygvi/longharness_agentic_programming/} \\
G600 & 1799 & My experience with CC as a games programmer. & \url{https://www.reddit.com/r/ClaudeAI/comments/1n6oa8w/my_experience_with_cc_as_a_games_programmer/} \\
G601 & 1800 & AI<->Human C++ My project rules to Stop GenAI from ruining code. & \url{https://www.reddit.com/r/cpp/comments/1trfnq5/aihuman_c_my_project_rules_to_stop_genai_from/} \\
G602 & 1805 & Run Claude Code from Slack \& Discord & \url{https://www.reddit.com/r/ClaudeAI/comments/1n5sps4/run_claude_code_from_slack_discord/} \\
G603 & 1808 & I built a cross-model review loop with Claude, and used it to help build itself & \url{https://www.reddit.com/r/ClaudeAI/comments/1s6vafx/i_built_a_crossmodel_review_loop_with_claude_and/} \\
G604 & 1818 & AI coding tools are quietly burying hardcoded secrets in your codebase and most devs have no idea until it's too late & \url{https://www.reddit.com/r/vibecoding/comments/1rqxwve/ai_coding_tools_are_quietly_burying_hardcoded/} \\
G605 & 1820 & 25 Claude Code Tips from 11 Months of Intense Use & \url{https://www.reddit.com/r/ClaudeAI/comments/1qgccgs/25_claude_code_tips_from_11_months_of_intense_use/} \\
G606 & 1822 & The Death of Vibecoding & \url{https://www.reddit.com/r/vibecoding/comments/1nr3z8o/the_death_of_vibecoding/} \\
G607 & 1826 & End of an era....beginning of an even greater one (THIS....is the greatest compilation of September 2025 on the absolute state of AI,Robotics and the upcoming S & \url{https://www.reddit.com/r/accelerate/comments/1nkwuy9/end_of_an_erabeginning_of_an_even_greater_one/} \\
G608 & 1828 & I built an OpenSource framework to run Claude Code agents in sandboxes & \url{https://www.reddit.com/r/ClaudeAI/comments/1ryen8p/i_built_an_opensource_framework_to_run_claude/} \\
G609 & 1830 & Bringing automated preview, review, and merge to Claude Code on desktop & \url{https://www.reddit.com/r/ClaudeAI/comments/1ra57pu/bringing_automated_preview_review_and_merge_to/} \\
G610 & 1833 & AI coding tools: we're shipping things faster than we understand them, and I don't think we've reckoned with what that means yet & \url{https://www.reddit.com/r/SoftwareEngineering/comments/1shnsc1/ai_coding_tools_were_shipping_things_faster_than/} \\
G611 & 1834 & Multi agent usage and best way to ensure solid plans before Implementation & \url{https://www.reddit.com/r/ClaudeAI/comments/1ryr0xd/multi_agent_usage_and_best_way_to_ensure_solid/} \\
G612 & 1836 & Claude Code Scam (Tested \& Proofed) & \url{https://www.reddit.com/r/vibecoding/comments/1sb73h4/claude_code_scam_tested_proofed/} \\
G613 & 1837 & How do you actually understand a codebase you didn't write? & \url{https://www.reddit.com/r/softwarearchitecture/comments/1q848wy/how_do_you_actually_understand_a_codebase_you/} \\
G614 & 1848 & The hype around Claude and other LLMs is really making things rough right now. & \url{https://www.reddit.com/r/ExperiencedDevs/comments/1r3vqrq/the_hype_around_claude_and_other_llms_is_really/} \\
G615 & 1851 & It's been a week I haven't opened neither GitHub nor Linear and nobody has suspected anything! & \url{https://www.reddit.com/r/vibecoding/comments/1rf9r7z/its_been_a_week_i_havent_opened_neither_github/} \\
G616 & 1853 & Linux kernel just shipped ai code rules. the assisted-by tag is smarter than i expected & \url{https://www.reddit.com/r/ExperiencedDevs/comments/1sp5pqq/linux_kernel_just_shipped_ai_code_rules_the/} \\
G617 & 1855 & 90,000 Line Merge Request & \url{https://www.reddit.com/r/ClaudeAI/comments/1rdse4n/90000_line_merge_request/} \\
G618 & 1858 & Claude overwrote proprietary license terms with CC-BY-SA, deleted LICENSE files, and ignored explicit instructions. Ticket Filed. & \url{https://www.reddit.com/r/ClaudeAI/comments/1nbgrv0/claude_overwrote_proprietary_license_terms_with/} \\
G619 & 1863 & God Mode: The AI-Powered Dev Workflow for Production Apps & \url{https://www.reddit.com/r/cursor/comments/1kfliem/god_mode_the_aipowered_dev_workflow_for/} \\
G620 & 1864 & Caught up to ai learning in 2 months , opening thread to help anyone stuck & \url{https://www.reddit.com/r/vibecoding/comments/1t5lwav/caught_up_to_ai_learning_in_2_months_opening/} \\
G621 & 1866 & GenAI security risks are piling up as devs chase quick wins....and its driving me nuts & \url{https://www.reddit.com/r/devops/comments/1qc731u/genai_security_risks_are_piling_up_as_devs_chase/} \\
G622 & 1867 & 2 months. Pure vibes. How I automated 80\% of my engineering job as a non-developer. & \url{https://www.reddit.com/r/vibecoding/comments/1p76z2h/2_months_pure_vibes_how_i_automated_80_of_my/} \\
G623 & 1868 & Cultural Mismatch After Buyout & \url{https://www.reddit.com/r/ExperiencedDevs/comments/1r0xffw/cultural_mismatch_after_buyout/} \\
G624 & 1872 & why are trad coder so annoying about code quality? & \url{https://www.reddit.com/r/cscareerquestions/comments/1tkqyp1/why_are_trad_coder_so_annoying_about_code_quality/} \\
G625 & 1874 & 5 signs your code review process is broken and how to fix each one & \url{https://www.reddit.com/r/codereview/comments/1mwmgzc/5_signs_your_code_review_process_is_broken_and/} \\
G626 & 1878 & Greptile publishes their State of AI coding 2025 report & \url{https://www.reddit.com/r/programming/comments/1ppexvk/greptile_publishes_their_state_of_ai_coding_2025/} \\
G627 & 1883 & My AI workflow seems to be the opposite of what the industry is encouraging, but I don't care. & \url{https://www.reddit.com/r/ExperiencedDevs/comments/1sc998r/my_ai_workflow_seems_to_be_the_opposite_of_what/} \\
G628 & 1887 & What's the dumbest bug you missed in code review that made it to prod? & \url{https://www.reddit.com/r/ExperiencedDevs/comments/1lzz2vl/whats_the_dumbest_bug_you_missed_in_code_review/} \\
G629 & 1888 & Specific, Measurable, Testable AI Predictions for 2026 & \url{https://www.reddit.com/r/ArtificialInteligence/comments/1py69fb/specific_measurable_testable_ai_predictions_for/} \\
G630 & 1890 & I built an enterprise-grade app in under 30 days with zero coding knowledge using 3 Claude accounts. Here is my exact workflow. & \url{https://www.reddit.com/r/vibecoding/comments/1tcw8ml/i_built_an_enterprisegrade_app_in_under_30_days/} \\
G631 & 1891 & From internal hack to full product: Introducing Verify Code for real-time AI code reviews & \url{https://www.reddit.com/r/cursor/comments/1mdg84t/from_internal_hack_to_full_product_introducing/} \\
G632 & 1892 & Everyone's shipping more code but I think we broke something fundamental & \url{https://www.reddit.com/r/cursor/comments/1s1gqcg/everyones_shipping_more_code_but_i_think_we_broke/} \\
G633 & 1894 & Claude Code Can Launch VS Code (Stop Wasting Your Context Window) & \url{https://www.reddit.com/r/ClaudeAI/comments/1o6sgqj/claude_code_can_launch_vs_code_stop_wasting_your/} \\
G634 & 1895 & Ultimate Harness Automation (OMG: oh-my-githubcopilot) released for Github Copilot & \url{https://www.reddit.com/r/GithubCopilot/comments/1sj2tau/ultimate_harness_automation_omg_ohmygithubcopilot/} \\
G635 & 1898 & DeepLJSP: Reinforcement Learning + Deep Neural Networks for Manufacturing Optimization Built with Claude & \url{https://www.reddit.com/r/ClaudeAI/comments/1n2ncin/deepljsp_reinforcement_learning_deep_neural/} \\
G636 & 1909 & Can Claude Code Sub-Agents Replace Fully Isolated OpenClaw Agents? & \url{https://www.reddit.com/r/ClaudeAI/comments/1rvq3h8/can_claude_code_subagents_replace_fully_isolated/} \\
G637 & 1910 & Code reviewing AI slop? & \url{https://www.reddit.com/r/SoftwareEngineering/comments/1mgcma6/code_reviewing_ai_slop/} \\
G638 & 1913 & AI coding is getting expensive and cannot afford \$100 now . What's the best coding setup under \$20/month in 2026? & \url{https://www.reddit.com/r/vibecoding/comments/1tqwxxb/ai_coding_is_getting_expensive_and_cannot_afford/} \\
G639 & 1914 & my job shifted from writing code to reviewing it and i dont remember agreeing to that & \url{https://www.reddit.com/r/cursor/comments/1tqx9ad/my_job_shifted_from_writing_code_to_reviewing_it/} \\
G640 & 1925 & Technical lessons while building a trilogy of Claude Code plugins & \url{https://www.reddit.com/r/ClaudeAI/comments/1quuxkj/technical_lessons_while_building_a_trilogy_of/} \\
G641 & 1930 & App templates for vibe coding & \url{https://www.reddit.com/r/vibecoding/comments/1pz8t9h/app_templates_for_vibe_coding/} \\
G642 & 1931 & Thoughts on Prompt Engineering career path? & \url{https://www.reddit.com/r/cscareerquestions/comments/1q4u5s1/thoughts_on_prompt_engineering_career_path/} \\
G643 & 1933 & 10 brutal lessons from a year of vibe coding and solo startups building & \url{https://www.reddit.com/r/cursor/comments/1oj0g80/10_brutal_lessons_from_a_year_of_vibe_coding_and/} \\
G644 & 1934 & It's getting really annoying to review code & \url{https://www.reddit.com/r/ExperiencedDevs/comments/1turimp/its_getting_really_annoying_to_review_code/} \\
G645 & 1937 & I built an AI-powered Git client for Windows with 100+ features -- NeuroGit 2.0 & \url{https://www.reddit.com/r/git/comments/1rcpcip/i_built_an_aipowered_git_client_for_windows_with/} \\
G646 & 1940 & Opus 4.8 just dropped, the fast mode pricing is wild & \url{https://www.reddit.com/r/ClaudeAI/comments/1tqc49w/opus_48_just_dropped_the_fast_mode_pricing_is_wild/} \\
G647 & 1941 & How I use TypeScript strict mode + custom ESLint rules to catch AI-generated bugs before they ship & \url{https://www.reddit.com/r/webdev/comments/1r1dw03/how_i_use_typescript_strict_mode_custom_eslint/} \\
G648 & 1943 & Is anywone using the code-review skill? & \url{https://www.reddit.com/r/ClaudeAI/comments/1s59k0t/is_anywone_using_the_codereview_skill/} \\
G649 & 1946 & Felling like a imposter. & \url{https://www.reddit.com/r/AskProgramming/comments/1nmpjwg/felling_like_a_imposter/} \\
G650 & 1947 & bcherny (creator of Claude Code) shares "how I use Claude Code" & \url{https://www.reddit.com/r/AI_Agents/comments/1q3xw15/bcherny_creator_of_claude_code_shares_how_i_use/} \\
G651 & 1949 & we built a claude code bootcamp -- 10 real projects in one day, may 30 & \url{https://www.reddit.com/r/softwarearchitecture/comments/1tjsuk9/we_built_a_claude_code_bootcamp_10_real_projects/} \\
G652 & 1953 & How do you verify AI-generated code before deploying? Do you even bother? & \url{https://www.reddit.com/r/GithubCopilot/comments/1qz3on6/how_do_you_verify_aigenerated_code_before/} \\
G653 & 1954 & Looking for experiences with context-aware AI code review tools & \url{https://www.reddit.com/r/QualityAssurance/comments/1pn5ywc/looking_for_experiences_with_contextaware_ai_code/} \\
G654 & 1955 & My Claude codebase became a black box, so I built a git-blame-like tool for agent runs (Agentdiff) & \url{https://www.reddit.com/r/ClaudeAI/comments/1rqz12h/my_claude_codebase_became_a_black_box_so_i_built/} \\
G655 & 1956 & best ci/cd integration for Al code review that actually works with github actions? & \url{https://www.reddit.com/r/softwarearchitecture/comments/1pov7ri/best_cicd_integration_for_al_code_review_that/} \\
G656 & 1958 & I built a comprehensive portfolio backend with .NET Web API - Looking for feedback, collaboration ideas, and suggestions! & \url{https://www.reddit.com/r/csharp/comments/1joz3io/i_built_a_comprehensive_portfolio_backend_with/} \\
G657 & 1962 & I vibe-coded a full HTML5 slicing game over the last couple weeks - here's how it works under the hood & \url{https://www.reddit.com/r/vibecoding/comments/1pl3eck/i_vibecoded_a_full_html5_slicing_game_over_the/} \\
G658 & 1963 & We benchmarked AI code review tools on real production bugs & \url{https://www.reddit.com/r/ChatGPTCoding/comments/1rfe6yn/we_benchmarked_ai_code_review_tools_on_real/} \\
G659 & 1972 & God Mode: The AI-Powered Dev Workflow & \url{https://www.reddit.com/r/ChatGPTCoding/comments/1jhn5ao/god_mode_the_aipowered_dev_workflow/} \\
G660 & 1973 & what ai code reviewers do you use? & \url{https://www.reddit.com/r/vibecoding/comments/1ov4fpu/what_ai_code_reviewers_do_you_use/} \\
G661 & 1977 & Suggestions on reigning in CC? & \url{https://www.reddit.com/r/ClaudeAI/comments/1lphueb/suggestions_on_reigning_in_cc/} \\
G662 & 1978 & How Are You Supposed to Use AI in Coding Interviews? & \url{https://www.reddit.com/r/ExperiencedDevs/comments/1rj783q/how_are_you_supposed_to_use_ai_in_coding/} \\
G663 & 1983 & EM, forced by management to subdivide team into "squads" -- how can I make this less painful for everyone involved? & \url{https://www.reddit.com/r/ExperiencedDevs/comments/1ivtfaq/em_forced_by_management_to_subdivide_team_into/} \\
G664 & 1988 & How deep do you go in code reviews? & \url{https://www.reddit.com/r/cursor/comments/1mjtyua/how_deep_do_you_go_in_code_reviews/} \\
G665 & 1989 & I spent months building a free Windows AI app, with an AI council mode. no subscription, no account, no data leaving your machine & \url{https://www.reddit.com/r/vibecoding/comments/1twakwf/i_spent_months_building_a_free_windows_ai_app/} \\
G666 & 1990 & I have created a FREE Github Guide for Beginners & \url{https://www.reddit.com/r/github/comments/1kiosc4/i_have_created_a_free_github_guide_for_beginners/} \\
G667 & 1994 & claw - /auto-pilot \& /plan-day & \url{https://www.reddit.com/r/vibecoding/comments/1q7t6c1/claw_autopilot_planday/} \\
G668 & 1995 & I Almost Shipped an XSS Vulnerability Thanks to AI-Generated Code & \url{https://www.reddit.com/r/vibecoding/comments/1k5gosd/i_almost_shipped_an_xss_vulnerability_thanks_to/} \\
G669 & 1996 & Is ChatGPT only catered towards Python developers? & \url{https://www.reddit.com/r/ChatGPTCoding/comments/1mg8g81/is_chatgpt_only_catered_towards_python_developers/} \\
G670 & 1998 & Rebuild Urbit* in Rust with encrypted messaging and a Mac System 7 UI & \url{https://www.reddit.com/r/rust/comments/1qkemmq/rebuild_urbit_in_rust_with_encrypted_messaging/} \\
G671 & 2004 & What do you think, Code Reviews slows down or teach us ? & \url{https://www.reddit.com/r/ExperiencedDevs/comments/1q85s27/what_do_you_think_code_reviews_slows_down_or/} \\
G672 & 2005 & I (Embedded SW Engginer) made a landingpage using 80\% AI. What do you think? [I WILL NOT PROMOTE] & \url{https://www.reddit.com/r/webdev/comments/1r21km8/i_embedded_sw_engginer_made_a_landingpage_using/} \\
G673 & 2007 & Why is CC constantly trying to verify legitimacy with each request? & \url{https://www.reddit.com/r/ClaudeAI/comments/1mrxx08/why_is_cc_constantly_trying_to_verify_legitimacy/} \\
G674 & 2008 & What's your experience with enforcing "controversial semantic linting to ensure PRs are architecture- and convention-compliant? & \url{https://www.reddit.com/r/ExperiencedDevs/comments/1n9b5hn/whats_your_experience_with_enforcing/} \\
G675 & 2014 & Advice to talk with LLM reliant seniors & \url{https://www.reddit.com/r/cscareerquestions/comments/1oau6s0/advice_to_talk_with_llm_reliant_seniors/} \\
G676 & 2016 & Self-Healing Pipelines with AI Agents & \url{https://www.reddit.com/r/devops/comments/1k6bi5a/selfhealing_pipelines_with_ai_agents/} \\
G677 & 2020 & working on a small add-on that tells me what actually mattered in a session, would love feedback! & \url{https://www.reddit.com/r/ClaudeAI/comments/1sr8ih5/working_on_a_small_addon_that_tells_me_what/} \\
G678 & 2021 & I created a GitHub Action to ensure authors understand their PRs & \url{https://www.reddit.com/r/devops/comments/1p52bg6/i_created_a_github_action_to_ensure_authors/} \\
G679 & 2026 & AI makes devs dumber? Lessons from leading 200+ engineers. & \url{https://www.reddit.com/r/ChatGPTCoding/comments/1oyozjb/ai_makes_devs_dumber_lessons_from_leading_200/} \\
G680 & 2030 & How to Build a Large Project with AI (100k+ LOC) & \url{https://www.reddit.com/r/rust/comments/1qp2n8n/how_to_build_a_large_project_with_ai_100k_loc/} \\
G681 & 2031 & How do you handle requirements, code quality and structure in small dev teams? & \url{https://www.reddit.com/r/ExperiencedDevs/comments/1n1d7ug/how_do_you_handle_requirements_code_quality_and/} \\
G682 & 2032 & Cursor in not magic & \url{https://www.reddit.com/r/cursor/comments/1l1kdki/cursor_in_not_magic/} \\
G683 & 2036 & Looking for a Claude Code Diff Viewer Before Accepting Changes (Working on WSL/Windows) & \url{https://www.reddit.com/r/ClaudeAI/comments/1lienlz/looking_for_a_claude_code_diff_viewer_before/} \\
G684 & 2040 & I realized I'm not really vibe coding. I'm using AI in a governed partnership. & \url{https://www.reddit.com/r/vibecoding/comments/1rru2or/i_realized_im_not_really_vibe_coding_im_using_ai/} \\
G685 & 2041 & Former military officer, zero CS background -- used Claude to build and ship an AI code review CLI to npm & \url{https://www.reddit.com/r/ClaudeAI/comments/1r14kev/former_military_officer_zero_cs_background_used/} \\
G686 & 2045 & AI-Powered Security Code Review with Autofix & \url{https://www.reddit.com/r/golang/comments/1s3m68x/aipowered_security_code_review_with_autofix/} \\
G687 & 2052 & How do you all deal with nitpicky PR reviews without killing motivation? & \url{https://www.reddit.com/r/ExperiencedDevs/comments/1nv8dnw/how_do_you_all_deal_with_nitpicky_pr_reviews/} \\
G688 & 2053 & code-review for my smart-pointer & \url{https://www.reddit.com/r/rust/comments/1pm5i2d/codereview_for_my_smartpointer/} \\
G689 & 2057 & Authorization breaks when B2B SaaS scales - role explosion, endless support tickets for access requests, blocked deployments every time permissions change. How  & \url{https://www.reddit.com/r/devops/comments/1phaszc/authorization_breaks_when_b2b_saas_scales_role/} \\
G690 & 2059 & CLAUDE.md optimization & \url{https://www.reddit.com/r/ClaudeAI/comments/1lmik7k/claudemd_optimization/} \\
G691 & 2060 & Claude Haiku 4.5 hits 73.3\% on SWE-bench for \$1/\$5 per million tokens (3x cheaper than Sonnet 4, 2x faster) & \url{https://www.reddit.com/r/ClaudeAI/comments/1o89nko/claude_haiku_45_hits_733_on_swebench_for_15_per/} \\
G692 & 2074 & Gemini-cli confirming how bad Claude has been Lately.. LOL & \url{https://www.reddit.com/r/ClaudeAI/comments/1n56huj/geminicli_confirming_how_bad_claude_has_been/} \\
G693 & 2079 & Is it possible to allow CODEOWNERS to merge only their code? & \url{https://www.reddit.com/r/github/comments/1jdx79c/is_it_possible_to_allow_codeowners_to_merge_only/} \\
G694 & 2083 & How do you prevent design drift during PR reviews? & \url{https://www.reddit.com/r/softwarearchitecture/comments/1qgfr0w/how_do_you_prevent_design_drift_during_pr_reviews/} \\
G695 & 2085 & Pre-Submission vs. Formal Merge/Pull Request Code Reviews & \url{https://www.reddit.com/r/ExperiencedDevs/comments/1tdoanj/presubmission_vs_formal_mergepull_request_code/} \\
G696 & 2088 & Wall between claude.ai and claude code & \url{https://www.reddit.com/r/ClaudeAI/comments/1ru7stg/wall_between_claudeai_and_claude_code/} \\
G697 & 2090 & Triggering Cursor from other code, or alternatives? & \url{https://www.reddit.com/r/cursor/comments/1klp2px/triggering_cursor_from_other_code_or_alternatives/} \\
G698 & 2093 & My own scheduler: MultiTaskScheduler & \url{https://www.reddit.com/r/java/comments/1n4fxe6/my_own_scheduler_multitaskscheduler/} \\
G699 & 2095 & Discussion regarding module imports (JEP 511) & \url{https://www.reddit.com/r/java/comments/1k683ad/discussion_regarding_module_imports_jep_511/} \\
G700 & 2104 & Vibrators -- The Unsung Heroes of Vibe Coding & \url{https://www.reddit.com/r/ChatGPTCoding/comments/1jw12zg/vibrators_the_unsung_heroes_of_vibe_coding/} \\
G701 & 2106 & How Rust's compiler makes AI write better code (lessons from building a CMS with Claude Code) & \url{https://www.reddit.com/r/rust/comments/1mdcbd4/how_rusts_compiler_makes_ai_write_better_code/} \\
G702 & 2108 & interview task "url shortener" code review iteration 2 & \url{https://www.reddit.com/r/rust/comments/1ifhmgw/interview_task_url_shortener_code_review/} \\
G703 & 2111 & Improving Claude Code usage in a dev team, feedback welcome & \url{https://www.reddit.com/r/ClaudeAI/comments/1s4hfna/improving_claude_code_usage_in_a_dev_team/} \\
G704 & 2116 & I'm convinced the reason developers hate AI coding tools is because we suck at using them. (Stop going in raw). & \url{https://www.reddit.com/r/ClaudeAI/comments/1rhufep/im_convinced_the_reason_developers_hate_ai_coding/} \\
G705 & 2126 & I let my PM write code using Claude, here's what happened after 2 weeks & \url{https://www.reddit.com/r/ClaudeAI/comments/1rt4w19/i_let_my_pm_write_code_using_claude_heres_what/} \\
G706 & 2131 & Moving from claude code to codex & \url{https://www.reddit.com/r/AI_Agents/comments/1sn0sqi/moving_from_claude_code_to_codex/} \\
G707 & 2135 & Built a directory of Claude Code agent prompts, copy/paste ready & \url{https://www.reddit.com/r/ClaudeAI/comments/1mf81ze/built_a_directory_of_claude_code_agent_prompts/} \\
G708 & 2138 & What's the toughest part of maintaining real code quality in modern DevOps? & \url{https://www.reddit.com/r/github/comments/1n3e4x9/whats_the_toughest_part_of_maintaining_real_code/} \\
G709 & 2140 & What are the remaining arguments against using AI for coding? & \url{https://www.reddit.com/r/ChatGPTCoding/comments/1qlsd8e/what_are_the_remaining_arguments_against_using_ai/} \\
G710 & 2142 & If AI does all the work and you only review it, where does the skill to review come from? & \url{https://www.reddit.com/r/ClaudeAI/comments/1ryk7yb/if_ai_does_all_the_work_and_you_only_review_it/} \\
G711 & 2143 & Question: How to limit the files responsibility of a subagent? & \url{https://www.reddit.com/r/ClaudeAI/comments/1majmmb/question_how_to_limit_the_files_responsibility_of/} \\
G712 & 2144 & Devs, would you use this? I'm building an AI Code Reviewer that actually understands your codebase. & \url{https://www.reddit.com/r/java/comments/1jwxd2d/devs_would_you_use_this_im_building_an_ai_code/} \\
G713 & 2161 & Built a guardrail for vibe coding after AI slipped a hardcoded secret into my app & \url{https://www.reddit.com/r/vibecoding/comments/1sfgtzp/built_a_guardrail_for_vibe_coding_after_ai/} \\
G714 & 2162 & I fed Gemini a lot of posts from this reddit and let it summarize the best practice & \url{https://www.reddit.com/r/ClaudeAI/comments/1lyrjnc/i_fed_gemini_a_lot_of_posts_from_this_reddit_and/} \\
G715 & 2163 & How you review code? & \url{https://www.reddit.com/r/ClaudeAI/comments/1mj7k1o/how_you_review_code/} \\
G716 & 2166 & Spec-driven agentic coding is quietly making us worse at the job of supervising agents & \url{https://www.reddit.com/r/AI_Agents/comments/1ta2dfz/specdriven_agentic_coding_is_quietly_making_us/} \\
G717 & 2167 & [Code Review] I wrote a SIMD-accelerated CSV parser in C++, but I'm new to SIMD and need some feedback & \url{https://www.reddit.com/r/cpp/comments/1r1h62x/code_review_i_wrote_a_simdaccelerated_csv_parser/} \\
G718 & 2169 & 151 malicious packages uploaded to GitHub in 6 days. & \url{https://www.reddit.com/r/github/comments/1rwdfad/151_malicious_packages_uploaded_to_github_in_6/} \\
G719 & 2175 & First time with Claude Code Web - 2 prompts and 4\% of a Max plan & \url{https://www.reddit.com/r/ClaudeAI/comments/1plbo0u/first_time_with_claude_code_web_2_prompts_and_4/} \\
G720 & 2177 & [Case Study] How Our AI's Internal "Critic" Saves Code From Race Conditions Before Execution & \url{https://www.reddit.com/r/LocalLLaMA/comments/1n73dek/case_study_how_our_ais_internal_critic_saves_code/} \\
G721 & 2180 & Built a workflow harness specifically for Claude Code after 5 months of daily production use -- free, open source (MIT) & \url{https://www.reddit.com/r/ClaudeAI/comments/1smal7u/built_a_workflow_harness_specifically_for_claude/} \\
G722 & 2181 & Claude Code versus Codex with BMAD & \url{https://www.reddit.com/r/ClaudeAI/comments/1n79qvq/claude_code_versus_codex_with_bmad/} \\
G723 & 2183 & 34 days with Claude Code. The code was solid. Some decisions were not. & \url{https://www.reddit.com/r/ClaudeAI/comments/1s1ehlt/34_days_with_claude_code_the_code_was_solid_some/} \\
G724 & 2184 & Looking for feedback on my Flask app structure (Blueprints + Factory Pattern) & \url{https://www.reddit.com/r/Python/comments/1qvgovq/looking_for_feedback_on_my_flask_app_structure/} \\
G725 & 2188 & Beginner code review. & \url{https://www.reddit.com/r/csharp/comments/1i7gesp/beginner_code_review/} \\
G726 & 2191 & I added ready-to-use templates to my retro pixel art React library - 29 sections + 5 full page layouts you can just copy-paste & \url{https://www.reddit.com/r/vibecoding/comments/1smsqcw/i_added_readytouse_templates_to_my_retro_pixel/} \\
G727 & 2196 & I reverse-engineered how one engineer made 8,471 commits across 48 repos in 72 days & \url{https://www.reddit.com/r/AI_Agents/comments/1rfc44m/i_reverseengineered_how_one_engineer_made_8471/} \\
G728 & 2199 & Joined a large org, three months in my programming job involves very little programming. Is this normal? & \url{https://www.reddit.com/r/ExperiencedDevs/comments/1jdj7jq/joined_a_large_org_three_months_in_my_programming/} \\
G729 & 2203 & Claude Code Review Using Tons of Tokens & \url{https://www.reddit.com/r/ClaudeAI/comments/1s1vz05/claude_code_review_using_tons_of_tokens/} \\
G730 & 2204 & PR reviewing as a skill & \url{https://www.reddit.com/r/SoftwareEngineering/comments/1jvr3qy/pr_reviewing_as_a_skill/} \\
G731 & 2209 & I used Claude Code to build an open-source AI agent verification tool -- gan-harness & \url{https://www.reddit.com/r/ClaudeAI/comments/1se48a6/i_used_claude_code_to_build_an_opensource_ai/} \\
G732 & 2210 & Built a dev tool for PR reviews - feedback needed & \url{https://www.reddit.com/r/ExperiencedDevs/comments/1lzks3t/built_a_dev_tool_for_pr_reviews_feedback_needed/} \\
G733 & 2214 & Zed workflow: orchestrating Claude 4.5 (Opus/Sonnet) and Gemini 3.0 to leverage Pro subscriptions? & \url{https://www.reddit.com/r/vibecoding/comments/1p867k1/zed_workflow_orchestrating_claude_45_opussonnet/} \\
G734 & 2215 & Built my first MVP in about 60 days with \$2.5k (Nov-Dec 2025) with vibe coding only & \url{https://www.reddit.com/r/vibecoding/comments/1qadypi/built_my_first_mvp_in_about_60_days_with_25k/} \\
G735 & 2217 & AI tools are making developers the integration layer & \url{https://www.reddit.com/r/vibecoding/comments/1t58ztl/ai_tools_are_making_developers_the_integration/} \\
G736 & 2218 & I built a zero-dependency ECDSA engine in Rust to learn discrete math & \url{https://www.reddit.com/r/rust/comments/1sb23y2/i_built_a_zerodependency_ecdsa_engine_in_rust_to/} \\
G737 & 2219 & I was tired of AI reviews, so I built a deterministic static analyzer for EF & \url{https://www.reddit.com/r/csharp/comments/1tihhl9/i_was_tired_of_ai_reviews_so_i_built_a/} \\
G738 & 2221 & How I got carried away and built an open-source framework for creating custom reliable AI workflows and agents & \url{https://www.reddit.com/r/ClaudeAI/comments/1oe70l7/how_i_got_carried_away_and_built_an_opensource/} \\
G739 & 2223 & Shipping the 1st phone for AI-AI communication: Update! & \url{https://www.reddit.com/r/vibecoding/comments/1kk68o6/shipping_the_1st_phone_for_aiai_communication/} \\
G740 & 2224 & Review with zen. Give your Claude Code a friend. & \url{https://www.reddit.com/r/ClaudeAI/comments/1lj1905/review_with_zen_give_your_claude_code_a_friend/} \\
G741 & 2227 & We benchmarked AI code review tools on real production bugs & \url{https://www.reddit.com/r/codereview/comments/1rfdzw1/we_benchmarked_ai_code_review_tools_on_real/} \\
G742 & 2228 & Help with collaboration on Github - anyone had similar experiences and what did you use to get around this? & \url{https://www.reddit.com/r/github/comments/1o1awq8/help_with_collaboration_on_github_anyone_had/} \\
G743 & 2230 & STOP shipping AI code without reviewing it first & \url{https://www.reddit.com/r/webdev/comments/1o0k7m7/stop_shipping_ai_code_without_reviewing_it_first/} \\
G744 & 2233 & A different approach to DSA / LC questions: this works for me & \url{https://www.reddit.com/r/cscareerquestions/comments/1lmgvmt/a_different_approach_to_dsa_lc_questions_this/} \\
G745 & 2235 & The error handling bugs that worry me aren't the ones that crash & \url{https://www.reddit.com/r/golang/comments/1rg5zo7/the_error_handling_bugs_that_worry_me_arent_the/} \\
G746 & 2239 & built this because I was sick of having 10 terminal tabs open just to use claude properly & \url{https://www.reddit.com/r/ClaudeAI/comments/1s1l50n/built_this_because_i_was_sick_of_having_10/} \\
G747 & 2240 & Spark: An IoT deployment ecosystem written in Rust (Tokio + Axum + Custom Protocol) & \url{https://www.reddit.com/r/rust/comments/1q7alms/spark_an_iot_deployment_ecosystem_written_in_rust/} \\
G748 & 2242 & Is there a way to automate Claude<->Codex feedback loop? & \url{https://www.reddit.com/r/ClaudeAI/comments/1qujaqc/is_there_a_way_to_automate_claudecodex_feedback/} \\
G749 & 2244 & Has AI killed craftsmanship in software engineering? & \url{https://www.reddit.com/r/ExperiencedDevs/comments/1rm53ih/has_ai_killed_craftsmanship_in_software/} \\
G750 & 2248 & after debugging 50+ projects: here's why your AI keeps ignoring you & \url{https://www.reddit.com/r/vibecoding/comments/1o5z6xm/after_debugging_50_projects_heres_why_your_ai/} \\
G751 & 2254 & Key takeaways after another week with Claude Code (and reading the official guide) & \url{https://www.reddit.com/r/ClaudeAI/comments/1lmperh/key_takeaways_after_another_week_with_claude_code/} \\
G752 & 2258 & I underestimated how much of software development is not just writing code & \url{https://www.reddit.com/r/cscareerquestions/comments/1tv4b7n/i_underestimated_how_much_of_software_development/} \\
G753 & 2259 & Has anyone managed to get something useful out of Copilot reviews? & \url{https://www.reddit.com/r/github/comments/1mo4yy8/has_anyone_managed_to_get_something_useful_out_of/} \\
G754 & 2261 & Any free \& open source Code-Review Bots? & \url{https://www.reddit.com/r/codereview/comments/1recb3r/any_free_open_source_codereview_bots/} \\
G755 & 2263 & I asked Codex CLI to review Claude Opus 4.1 code and Codex CLI failed miserably & \url{https://www.reddit.com/r/ClaudeAI/comments/1nqcuec/i_asked_codex_cli_to_review_claude_opus_41_code/} \\
G756 & 2265 & An in-memory database implementation! & \url{https://www.reddit.com/r/golang/comments/1m9hwva/an_inmemory_database_implementation/} \\
G757 & 2266 & Possible to Unarchive an Agent? Just lost 3 days of progress. & \url{https://www.reddit.com/r/cursor/comments/1mgwlsz/possible_to_unarchive_an_agent_just_lost_3_days/} \\
G758 & 2269 & Codex and claude - How to have them do code review for each other? & \url{https://www.reddit.com/r/ClaudeAI/comments/1sftxrr/codex_and_claude_how_to_have_them_do_code_review/} \\
G759 & 2270 & Feeling like a fraud, I use AI to do most of my job at my new Frontend position, is it the new normal ? & \url{https://www.reddit.com/r/webdev/comments/1ogoovw/feeling_like_a_fraud_i_use_ai_to_do_most_of_my/} \\
G760 & 2272 & Help me sell my team on Go & \url{https://www.reddit.com/r/golang/comments/1m3k2i8/help_me_sell_my_team_on_go/} \\
G761 & 2277 & How are you managing multiple coding agents in parallel without things getting messy? & \url{https://www.reddit.com/r/ClaudeAI/comments/1st21n2/how_are_you_managing_multiple_coding_agents_in/} \\
G762 & 2278 & Subagents are now INCREDIBLY functional, its wild & \url{https://www.reddit.com/r/GithubCopilot/comments/1qqzknq/subagents_are_now_incredibly_functional_its_wild/} \\
G763 & 2281 & You-Only-Live-Once. Steve Yegge's 8 Stages of AI Developer Progression & \url{https://www.reddit.com/r/vibecoding/comments/1s8hofb/youonlyliveonce_steve_yegges_8_stages_of_ai/} \\
G764 & 2283 & Autonomous multi-agent spec-driven AI coding in the terminal & \url{https://www.reddit.com/r/vibecoding/comments/1rhg6rc/autonomous_multiagent_specdriven_ai_coding_in_the/} \\
G765 & 2284 & Besides taking over the task, how do I get another senior developer to do their work properly? & \url{https://www.reddit.com/r/ExperiencedDevs/comments/1jkoxj2/besides_taking_over_the_task_how_do_i_get_another/} \\
G766 & 2288 & I built a bridge (claude-ide-bridge) that gives the Claude Code CLI full integration with Cursor/VS Code (115+ tools) & \url{https://www.reddit.com/r/ClaudeAI/comments/1rsk15l/i_built_a_bridge_claudeidebridge_that_gives_the/} \\
G767 & 2289 & We benchmarked AI code review tools on real production bugs & \url{https://www.reddit.com/r/vibecoding/comments/1rfe4xm/we_benchmarked_ai_code_review_tools_on_real/} \\
G768 & 2292 & I've used AI to write 100\% of my code for 1+ year as an engineer. 13 no-bs lessons & \url{https://www.reddit.com/r/cursor/comments/1qxvo0j/ive_used_ai_to_write_100_of_my_code_for_1_year_as/} \\
G769 & 2295 & Freigeist - The new Vibe Coding Platform & \url{https://www.reddit.com/r/ChatGPTCoding/comments/1m5tijr/freigeist_the_new_vibe_coding_platform/} \\
G770 & 2298 & Why doesn't Cursor host OSS models like Minimax M2.5 or GLM? & \url{https://www.reddit.com/r/cursor/comments/1r4ovzj/why_doesnt_cursor_host_oss_models_like_minimax/} \\
G771 & 2303 & Ultraship - Claude Code plugin -- 32 expert-level skills for building, shipping, and scaling production software. 29 audit tools (security, code quality, bundle & \url{https://www.reddit.com/r/ClaudeAI/comments/1s682lo/ultraship_claude_code_plugin_32_expertlevel/} \\
G772 & 2304 & How do you review \& debug AI "vibe-coded" code? & \url{https://www.reddit.com/r/ClaudeAI/comments/1mlypg2/how_do_you_review_debug_ai_vibecoded_code/} \\
G773 & 2306 & r/vibecoding Rules & \url{https://www.reddit.com/r/vibecoding/comments/1o82unf/rvibecoding_rules/} \\
G774 & 2308 & How are maintainers dealing with AI PRs? & \url{https://www.reddit.com/r/github/comments/1sdcxkx/how_are_maintainers_dealing_with_ai_prs/} \\
G775 & 2309 & Serious question: how are you staying mentally engaged using Claude Code daily? & \url{https://www.reddit.com/r/ClaudeAI/comments/1qi0rem/serious_question_how_are_you_staying_mentally/} \\
G776 & 2313 & security review is becoming an afterthought in ai-driven development & \url{https://www.reddit.com/r/ExperiencedDevs/comments/1nmnx5b/security_review_is_becoming_an_afterthought_in/} \\
G777 & 2316 & enterprise code generation with AI - does it actually reduce tech debt or just create more? & \url{https://www.reddit.com/r/csharp/comments/1s5qjfg/enterprise_code_generation_with_ai_does_it/} \\
G778 & 2318 & Cascade proposing more than one code change in the same file leads to issues & \url{https://www.reddit.com/r/windsurf/comments/1ll3tfe/cascade_proposing_more_than_one_code_change_in/} \\
G779 & 2319 & Getting a lay of the land: What tools have proven useful, which haven't, what's state of the art? & \url{https://www.reddit.com/r/vibecoding/comments/1rzbzjn/getting_a_lay_of_the_land_what_tools_have_proven/} \\
G780 & 2322 & What I learned vibe coding as a FAANG engineer (and why AI can be a messy teammate) & \url{https://www.reddit.com/r/webdev/comments/1mlasvv/what_i_learned_vibe_coding_as_a_faang_engineer/} \\
G781 & 2328 & Insane coding with Opus 4.6 and gpt5.3 & \url{https://www.reddit.com/r/OpenAI/comments/1rak5te/insane_coding_with_opus_46_and_gpt53/} \\
G782 & 2331 & Day 4 of Vibe Coding: AI Assistant & \url{https://www.reddit.com/r/vibecoding/comments/1rln1z7/day_4_of_vibe_coding_ai_assistant/} \\
G783 & 2332 & PR review of UI element which screams LLM generated, decline on that basis? & \url{https://www.reddit.com/r/webdev/comments/1t66jy6/pr_review_of_ui_element_which_screams_llm/} \\
G784 & 2333 & New to vibe coding - nervous about our Founder getting too ambitious and looking for advice! & \url{https://www.reddit.com/r/vibecoding/comments/1r51ckq/new_to_vibe_coding_nervous_about_our_founder/} \\
G785 & 2336 & Comparing Bito and Greptile for AI Code Reviews - YouTube & \url{https://www.reddit.com/r/codereview/comments/1p2ovdz/comparing_bito_and_greptile_for_ai_code_reviews/} \\
G786 & 2337 & Looks Like It's Time For All of Us to Come Together (And Here's Why It Matters) & \url{https://www.reddit.com/r/csharp/comments/1ir3k8h/looks_like_its_time_for_all_of_us_to_come/} \\
G787 & 2340 & I'm a noob, I saw people say "Data scienstist and CS SWE they code differently", can someone share me your exp about it, I never work and review data scientist  & \url{https://www.reddit.com/r/cscareerquestions/comments/1j5s5db/im_a_noob_i_saw_people_say_data_scienstist_and_cs/} \\
G788 & 2345 & Codexia agent design draft for feedback (AI Coding Agent for GitHub Repositories) & \url{https://www.reddit.com/r/AI_Agents/comments/1nvvhur/codexia_agent_design_draft_for_feedback_ai_coding/} \\
G789 & 2348 & What do long code reviews actually cost? & \url{https://www.reddit.com/r/agile/comments/1rlh0fi/what_do_long_code_reviews_actually_cost/} \\
G790 & 2350 & I made a self healing PRD system for Claude code & \url{https://www.reddit.com/r/artificial/comments/1sokj0w/i_made_a_self_healing_prd_system_for_claude_code/} \\
G791 & 2352 & The App That Builds Itself & \url{https://www.reddit.com/r/vibecoding/comments/1r3jqpj/the_app_that_builds_itself/} \\
G792 & 2354 & The 11-step workflow I use for every Claude Code project now: from idea validation to shipping with accumulated knowledge & \url{https://www.reddit.com/r/ClaudeAI/comments/1sgjmec/the_11step_workflow_i_use_for_every_claude_code/} \\
G793 & 2357 & Stop crying and learn to adapt & \url{https://www.reddit.com/r/GithubCopilot/comments/1srm6ma/stop_crying_and_learn_to_adapt/} \\
G794 & 2360 & I'm not a developer. I built a multi-agent AI system that runs my business. & \url{https://www.reddit.com/r/vibecoding/comments/1sx3cc1/im_not_a_developer_i_built_a_multiagent_ai_system/} \\
G795 & 2361 & Senior Frontend Developer (8 YOE) Feeling Stagnant and Trapped - How to Stay Sharp and Plan My Next Move in this Market? & \url{https://www.reddit.com/r/cscareerquestions/comments/1oed3f9/senior_frontend_developer_8_yoe_feeling_stagnant/} \\
G796 & 2362 & Need advice for chosing the better path & \url{https://www.reddit.com/r/ExperiencedDevs/comments/1jvdgwk/need_advice_for_chosing_the_better_path/} \\
G797 & 2370 & Claude is becoming very humanlike (Lazy) & \url{https://www.reddit.com/r/ClaudeAI/comments/1m6je1i/claude_is_becoming_very_humanlike_lazy/} \\
G798 & 2373 & Emergence or training artifact? My AI agents independently built safety tools I never asked for. 28/170 builds over 3 weeks. & \url{https://www.reddit.com/r/artificial/comments/1rki8d4/emergence_or_training_artifact_my_ai_agents/} \\
G799 & 2377 & 2026 workflow & \url{https://www.reddit.com/r/ExperiencedDevs/comments/1td90oq/2026_workflow/} \\
G800 & 2379 & vibe coding is dead. here is my stack for actual spec driven ai development & \url{https://www.reddit.com/r/AI_Agents/comments/1sbxrhc/vibe_coding_is_dead_here_is_my_stack_for_actual/} \\
G801 & 2381 & Most Code Reviews Are Just Bullying with Syntax Highlighting. Here's a Better Way. & \url{https://www.reddit.com/r/SoftwareEngineering/comments/1pd784m/most_code_reviews_are_just_bullying_with_syntax/} \\
G802 & 2385 & Announcing Claude Flow Skills: This release marks the move from slash commands to a Claude Skills-based system.. We also just hit 200K+ downloads and 9k Stars! & \url{https://www.reddit.com/r/ClaudeAI/comments/1ob985l/announcing_claude_flow_skills_this_release_marks/} \\
G803 & 2391 & M3 scores well on SWE-Bench but that's not why Im impressed its the stuff no benchmark measures. & \url{https://www.reddit.com/r/AI_Agents/comments/1twc895/m3_scores_well_on_swebench_but_thats_not_why_im/} \\
G804 & 2395 & Vibe-coded product security concern & \url{https://www.reddit.com/r/vibecoding/comments/1tatsxb/vibecoded_product_security_concern/} \\
G805 & 2398 & I got mass downvoted for saying AI agents dont actually understand your codebase. So i built an MCP server to prove it & \url{https://www.reddit.com/r/vibecoding/comments/1qjx7q0/i_got_mass_downvoted_for_saying_ai_agents_dont/} \\
G806 & 2399 & Update after my vibe coding post blew up & \url{https://www.reddit.com/r/vibecoding/comments/1pxvlqf/update_after_my_vibe_coding_post_blew_up/} \\
G807 & 2400 & I built an open-source local-first motion editor in React/Vite -- looking for small bugfix PRs and code review & \url{https://www.reddit.com/r/webdev/comments/1trtvjk/i_built_an_opensource_localfirst_motion_editor_in/} \\
G808 & 2403 & I've been experimenting with deterministic secret remediation in CI/CD pipelines using Python AST (refuses unsafe fixes) & \url{https://www.reddit.com/r/devops/comments/1rvv08u/ive_been_experimenting_with_deterministic_secret/} \\
G809 & 2404 & HydraStream: A pure Python streaming downloader I built to bypass disk I/O & \url{https://www.reddit.com/r/Python/comments/1rw5pqm/hydrastream_a_pure_python_streaming_downloader_i/} \\
G810 & 2409 & Multi-agent development setup and cross agents interaction - improvements/comments? & \url{https://www.reddit.com/r/vibecoding/comments/1qidd84/multiagent_development_setup_and_cross_agents/} \\
G811 & 2414 & My tips for coding with GPT and an IDE & \url{https://www.reddit.com/r/vibecoding/comments/1ppubxu/my_tips_for_coding_with_gpt_and_an_ide/} \\
G812 & 2415 & Is this useful to anyone? & \url{https://www.reddit.com/r/vibecoding/comments/1t0bdxz/is_this_useful_to_anyone/} \\
G813 & 2421 & 14 months, 100k lines, zero human-written code -- am I sitting on a ticking time bomb? & \url{https://www.reddit.com/r/ClaudeAI/comments/1rw04bg/14_months_100k_lines_zero_humanwritten_code_am_i/} \\
G814 & 2423 & I turned Claude Code into a full AI software team -- 119 agents, 202 skills, 48 hooks. Open source. & \url{https://www.reddit.com/r/ClaudeAI/comments/1ryxjxs/i_turned_claude_code_into_a_full_ai_software_team/} \\
G815 & 2424 & I made a Chrome extension that copies GitHub PR diffs for AI code review & \url{https://www.reddit.com/r/ChatGPTCoding/comments/1ktj7ie/i_made_a_chrome_extension_that_copies_github_pr/} \\
G816 & 2425 & Reviewing PR in age of AI & \url{https://www.reddit.com/r/webdev/comments/1kt53qi/reviewing_pr_in_age_of_ai/} \\
G817 & 2428 & Overengineering library design & \url{https://www.reddit.com/r/java/comments/1nam096/overengineering_library_design/} \\
G818 & 2429 & Do you enforce your architecture on the CI? If so, how? & \url{https://www.reddit.com/r/typescript/comments/1ts40od/do_you_enforce_your_architecture_on_the_ci_if_so/} \\
G819 & 2431 & "vibe coding" a controversial topic from a different perspective. & \url{https://www.reddit.com/r/vibecoding/comments/1o3txgh/vibe_coding_a_controversial_topic_from_a/} \\
G820 & 2435 & AI is draining my passion & \url{https://www.reddit.com/r/devops/comments/1ozd2i5/ai_is_draining_my_passion/} \\
G821 & 2437 & What would you charge for this scope of work? (Pricing reality check) & \url{https://www.reddit.com/r/ExperiencedDevs/comments/1tp6k2t/what_would_you_charge_for_this_scope_of_work/} \\
G822 & 2439 & Using AI for Coding Daily - But I'm Feeling Less Engaged (Dev Thoughts) & \url{https://www.reddit.com/r/ClaudeAI/comments/1nnk42j/using_ai_for_coding_daily_but_im_feeling_less/} \\
G823 & 2440 & I stopped "vibecoding" bugs. I started isolating them like a real incident. & \url{https://www.reddit.com/r/vibecoding/comments/1rat6ok/i_stopped_vibecoding_bugs_i_started_isolating/} \\
G824 & 2447 & How are you reviewing AI / code agent-generated changes? Any tools or best practices? & \url{https://www.reddit.com/r/vibecoding/comments/1q7ps9n/how_are_you_reviewing_ai_code_agentgenerated/} \\
G825 & 2448 & How to effectively use Cursor? & \url{https://www.reddit.com/r/ChatGPTCoding/comments/1mhgft2/how_to_effectively_use_cursor/} \\
G826 & 2456 & Looking for suggestions on AI code review tools & \url{https://www.reddit.com/r/codereview/comments/1q6fb7y/looking_for_suggestions_on_ai_code_review_tools/} \\
G827 & 2459 & Can we talk about security reviews for our new code? & \url{https://www.reddit.com/r/vibecoding/comments/1q95ilx/can_we_talk_about_security_reviews_for_our_new/} \\
G828 & 2465 & Copilot CLI: You can no longer call a higher model in a subagent from a lower multiplier model. & \url{https://www.reddit.com/r/GithubCopilot/comments/1ss5xd5/copilot_cli_you_can_no_longer_call_a_higher_model/} \\
G829 & 2466 & How I use AI through a repeatable and programmable workflow to stop fixing the same mistakes over and over & \url{https://www.reddit.com/r/ClaudeAI/comments/1rxdkq9/how_i_use_ai_through_a_repeatable_and/} \\
G830 & 2471 & vibe coding is ruining our brain & \url{https://www.reddit.com/r/vibecoding/comments/1nfg2y9/vibe_coding_is_ruining_our_brain/} \\
G831 & 2481 & After a year of using Cursor, Claude Code, Antigravity, and Copilot daily -- I think AI tools are making a lot of devs slower, not faster. Here's why. & \url{https://www.reddit.com/r/programming/comments/1rbdl1k/after_a_year_of_using_cursor_claude_code/} \\
G832 & 2482 & What do you do when cursor is running? & \url{https://www.reddit.com/r/cursor/comments/1pktv1v/what_do_you_do_when_cursor_is_running/} \\
G833 & 2483 & Why vibe coding is both the best and worst part of AI & \url{https://www.reddit.com/r/ArtificialInteligence/comments/1qs2kml/why_vibe_coding_is_both_the_best_and_worst_part/} \\
G834 & 2485 & Are junior engineers going to adapt to AI assisted coding? & \url{https://www.reddit.com/r/cursor/comments/1rdc3yt/are_junior_engineers_going_to_adapt_to_ai/} \\
G835 & 2486 & unfamiliar codebase reviews + outdated docs = kill me & \url{https://www.reddit.com/r/codereview/comments/1nboaja/unfamiliar_codebase_reviews_outdated_docs_kill_me/} \\
G836 & 2489 & tools that actually play nice together in a modern ci/cd setup (not just vendor lock-in) & \url{https://www.reddit.com/r/devops/comments/1rd1060/tools_that_actually_play_nice_together_in_a/} \\
G837 & 2492 & Utilise Google's 1M+ Token Context with Claude - Gemini MCP & \url{https://www.reddit.com/r/ClaudeAI/comments/1ly6saq/utilise_googles_1m_token_context_with_claude/} \\
G838 & 2493 & Created an agent skill that enforces SOLID, TDD, and clean architecture in Claude Code & \url{https://www.reddit.com/r/ClaudeAI/comments/1qk2msw/created_an_agent_skill_that_enforces_solid_tdd/} \\
G839 & 2495 & [Open Source] Next.js + C\# Project: Remote Internet Control Dashboard \& Windows Client - Feedback Welcome! & \url{https://www.reddit.com/r/csharp/comments/1lw5s2j/open_source_nextjs_c_project_remote_internet/} \\
G840 & 2496 & How FAANG engineers are actually using AI for production code & \url{https://www.reddit.com/r/ChatGPTCoding/comments/1nmt4ya/how_faang_engineers_are_actually_using_ai_for/} \\
G841 & 2497 & Why is this not popular? & \url{https://www.reddit.com/r/vibecoding/comments/1pfnhyl/why_is_this_not_popular/} \\
G842 & 2503 & I built an autonomous research agent where Claude Code reviews its own work, scores it, and re-runs weak stages until they pass. Reads 50+ full papers, not abst & \url{https://www.reddit.com/r/ClaudeAI/comments/1sfq36y/i_built_an_autonomous_research_agent_where_claude/} \\
G843 & 2506 & Self Promotion Thread - All Projects Will Be Shouted Out & \url{https://www.reddit.com/r/ChatGPTCoding/comments/1sj4z2q/self_promotion_thread_all_projects_will_be/} \\
G844 & 2507 & Claude IDE anyone? & \url{https://www.reddit.com/r/ClaudeAI/comments/1p5v6vr/claude_ide_anyone/} \\
G845 & 2509 & Sources of code review & \url{https://www.reddit.com/r/webdev/comments/1o29at6/sources_of_code_review/} \\
G846 & 2510 & How to automate accepting code changes? & \url{https://www.reddit.com/r/cursor/comments/1kvo5d2/how_to_automate_accepting_code_changes/} \\
G847 & 2521 & Building a GitHub Action to support reviewers handle the onslaught of AI assisted PRs & \url{https://www.reddit.com/r/programming/comments/1p52qj6/building_a_github_action_to_support_reviewers/} \\
G848 & 2522 & How do you keep the code diff open when telling Claude what to do differently? & \url{https://www.reddit.com/r/ClaudeAI/comments/1lc5f0v/how_do_you_keep_the_code_diff_open_when_telling/} \\
G849 & 2525 & Stop picking AI review tools based on vibes. I built an objective benchmark for AI Code Reviews. & \url{https://www.reddit.com/r/SoftwareEngineering/comments/1rg1xsu/stop_picking_ai_review_tools_based_on_vibes_i/} \\
G850 & 2526 & AI Agent built GitHub code reviewer + auto-documentation checker in 45 lines of Python. This is insane. & \url{https://www.reddit.com/r/ClaudeAI/comments/1l5n6v2/ai_agent_built_github_code_reviewer/} \\
G851 & 2533 & What "tiny nit" in code review wrecked your worldview? & \url{https://www.reddit.com/r/golang/comments/1l2kjz0/what_tiny_nit_in_code_review_wrecked_your/} \\
G852 & 2534 & Getting lost when vibe coding & \url{https://www.reddit.com/r/vibecoding/comments/1nva2qh/getting_lost_when_vibe_coding/} \\
G853 & 2535 & Engineers, how are you handling security and code quality with all this AI gen code coming in? & \url{https://www.reddit.com/r/AskProgramming/comments/1n4z2a6/engineers_how_are_you_handling_security_and_code/} \\
G854 & 2536 & A "performance improvement" PR is exactly where review risk hides & \url{https://www.reddit.com/r/dotnet/comments/1tlhxdl/a_performance_improvement_pr_is_exactly_where/} \\
G855 & 2538 & what's working to automate the code review process in your ci/cd pipeline? & \url{https://www.reddit.com/r/devops/comments/1p3tg5p/whats_working_to_automate_the_code_review_process/} \\
G856 & 2539 & This Is Why You Can't Trust AI to Review Your Mission-Critical Code & \url{https://www.reddit.com/r/programming/comments/1louou5/this_is_why_you_cant_trust_ai_to_review_your/} \\
G857 & 2540 & My SaaS after I vibe coded it & \url{https://www.reddit.com/r/vibecoding/comments/1t1kmlk/my_saas_after_i_vibe_coded_it/} \\
G858 & 2543 & Instruction-Driven Development vs Outcome-Driven Development vs Vibe Coding & \url{https://www.reddit.com/r/ExperiencedDevs/comments/1sm9ha6/instructiondriven_development_vs_outcomedriven/} \\
G859 & 2549 & The AI flood is breaking OSS - maintainers are hitting the limit & \url{https://www.reddit.com/r/ExperiencedDevs/comments/1qu3dpo/the_ai_flood_is_breaking_oss_maintainers_are/} \\
G860 & 2552 & AI Isn't Replacing Developers--It's Exposing Them & \url{https://www.reddit.com/r/webdev/comments/1j0cfyd/ai_isnt_replacing_developersits_exposing_them/} \\
G861 & 2559 & I found that Claude writes a lot of junk code. Is there any way to avoid these issues before developing a new project. & \url{https://www.reddit.com/r/ClaudeAI/comments/1mvmm33/i_found_that_claude_writes_a_lot_of_junk_code_is/} \\
G862 & 2563 & I control Claude Code from my phone via Telegram - built the whole thing with Claude Code itself & \url{https://www.reddit.com/r/ClaudeAI/comments/1s810fi/i_control_claude_code_from_my_phone_via_telegram/} \\
G863 & 2566 & Velocity metrics are incomplete without review cost & \url{https://www.reddit.com/r/ExperiencedDevs/comments/1pxw3ra/velocity_metrics_are_incomplete_without_review/} \\
G864 & 2567 & Are AI coding tools fundamentally changing Agile/team software development? & \url{https://www.reddit.com/r/SoftwareEngineering/comments/1o6rghs/are_ai_coding_tools_fundamentally_changing/} \\
G865 & 2568 & Av v0.1.0 - FREE (no strings attached) and Open-Source tool for stacked pull requests & \url{https://www.reddit.com/r/github/comments/1i5rgvs/av_v010_free_no_strings_attached_and_opensource/} \\
G866 & 2571 & Help me understand: agents vs skills & \url{https://www.reddit.com/r/ClaudeAI/comments/1s5bo5v/help_me_understand_agents_vs_skills/} \\
G867 & 2576 & Air gap btw Leadership/PMs and ICs regarding AI and how to talk about it & \url{https://www.reddit.com/r/ExperiencedDevs/comments/1ll1e03/air_gap_btw_leadershippms_and_ics_regarding_ai/} \\
G868 & 2577 & Can acceptance of LLM-generated code be formalized beyond "tests pass"? & \url{https://www.reddit.com/r/AskProgramming/comments/1qj6lar/can_acceptance_of_llmgenerated_code_be_formalized/} \\
G869 & 2578 & Why "Skip the Code, Ship the Binary" Is a Category Error & \url{https://www.reddit.com/r/programming/comments/1r65ee0/why_skip_the_code_ship_the_binary_is_a_category/} \\
G870 & 2580 & Full Codebase Review Now Possible, 224 files 19k loc reviewed, you should do it too & \url{https://www.reddit.com/r/cursor/comments/1jyk1qh/full_codebase_review_now_possible_224_files_19k/} \\
G871 & 2584 & I made a cross-language structural duplicate detector using alpha equivalence & \url{https://www.reddit.com/r/Python/comments/1rgmjad/i_made_a_crosslanguage_structural_duplicate/} \\
G872 & 2586 & How do you validate AI-generated code in production environments? & \url{https://www.reddit.com/r/AskProgramming/comments/1jtoyso/how_do_you_validate_aigenerated_code_in/} \\
G873 & 2589 & What is becoming of the senior engineer? & \url{https://www.reddit.com/r/ExperiencedDevs/comments/1qhk6pl/what_is_becoming_of_the_senior_engineer/} \\
G874 & 2597 & Tests quality, who watches the watchers? & \url{https://www.reddit.com/r/ExperiencedDevs/comments/1ii614c/tests_quality_who_watches_the_watchers/} \\
G875 & 2598 & Anthropic supported open source PostgreSQL backup tool Databasus through their Claude for Open Source program & \url{https://www.reddit.com/r/ClaudeAI/comments/1rklvr7/anthropic_supported_open_source_postgresql_backup/} \\
G876 & 2600 & Why will programmers in 2026 stop talking about Prompt Engineering and start talking about MCP (Model Context Protocol)? & \url{https://www.reddit.com/r/AI_Agents/comments/1r4aj6y/why_will_programmers_in_2026_stop_talking_about/} \\
G877 & 2601 & Solo, Claude's a rocket. On my team, why does it create more chaos? & \url{https://www.reddit.com/r/ClaudeAI/comments/1tl89sb/solo_claudes_a_rocket_on_my_team_why_does_it/} \\
G878 & 2608 & VSCode Copilot vs using the AI model directly & \url{https://www.reddit.com/r/AI_Agents/comments/1j7ua6q/vscode_copilot_vs_using_the_ai_model_directly/} \\
G879 & 2609 & Am I wrong to think this won't work? Guided vibe coding for non-technical founders & \url{https://www.reddit.com/r/vibecoding/comments/1okvbf9/am_i_wrong_to_think_this_wont_work_guided_vibe/} \\
G880 & 2613 & Sales gets commission, engineers get... nothing? Built a tool to change that & \url{https://www.reddit.com/r/webdev/comments/1q8jzgu/sales_gets_commission_engineers_get_nothing_built/} \\
G881 & 2618 & How I've Been Using AI To Build Complex Software (And What Actually Worked) & \url{https://www.reddit.com/r/ClaudeAI/comments/1oof1mn/how_ive_been_using_ai_to_build_complex_software/} \\
G882 & 2634 & I built a comprehensive portfolio backend with .NET Web API - Looking for feedback, collaboration ideas, and suggestions! & \url{https://www.reddit.com/r/dotnet/comments/1joyiwt/i_built_a_comprehensive_portfolio_backend_with/} \\
G883 & 2635 & I'm struggling with reviewing code & \url{https://www.reddit.com/r/cscareerquestions/comments/1rlwrfh/im_struggling_with_reviewing_code/} \\
G884 & 2638 & Fitting a 64 million password dictionary into AWS Lambda memory using mmap and Bloom filters (100\% Terraform) & \url{https://www.reddit.com/r/devops/comments/1rk06sz/fitting_a_64_million_password_dictionary_into_aws/} \\
G885 & 2647 & How to effectively use sub-agents in Copilot & \url{https://www.reddit.com/r/GithubCopilot/comments/1q90cq5/how_to_effectively_use_subagents_in_copilot/} \\
G886 & 2650 & Looking for code review: TransTrack -- operational risk tracking for organ transplant waitlists & \url{https://www.reddit.com/r/codereview/comments/1qww72m/looking_for_code_review_transtrack_operational/} \\
G887 & 2651 & New Codex release: GPT-5-Codex, IDE upgrades, faster cloud, and built-in code review & \url{https://www.reddit.com/r/cursor/comments/1nhwcmq/new_codex_release_gpt5codex_ide_upgrades_faster/} \\
G888 & 2652 & [C\#/Unity] Interactive Fiction Text Adventure Skeleton & \url{https://www.reddit.com/r/codereview/comments/1i4d9ey/cunity_interactive_fiction_text_adventure_skeleton/} \\
G889 & 2657 & 100\% Trust to One Shot? & \url{https://www.reddit.com/r/ClaudeAI/comments/1s6ld7q/100_trust_to_one_shot/} \\
G890 & 2663 & How are you managing multiple coding agents in parallel without things getting messy? & \url{https://www.reddit.com/r/cursor/comments/1st1zys/how_are_you_managing_multiple_coding_agents_in/} \\
G891 & 2667 & AI-Rulez v2: One Config to Rule All Your TypeScript AI Tools & \url{https://www.reddit.com/r/typescript/comments/1ne8r0e/airulez_v2_one_config_to_rule_all_your_typescript/} \\
G892 & 2670 & How I've Been Using AI To Build Complex Software (And What Actually Worked) & \url{https://www.reddit.com/r/vibecoding/comments/1ooav6t/how_ive_been_using_ai_to_build_complex_software/} \\
G893 & 2676 & Claims Management and Tracking System & \url{https://www.reddit.com/r/codereview/comments/1qzdeyo/claims_management_and_tracking_system/} \\
G894 & 2677 & Good software is still made with blood and tears & \url{https://www.reddit.com/r/ExperiencedDevs/comments/1svkl49/good_software_is_still_made_with_blood_and_tears/} \\
G895 & 2688 & Lacking confidence vibe coding shipping & \url{https://www.reddit.com/r/vibecoding/comments/1s1n378/lacking_confidence_vibe_coding_shipping/} \\
G896 & 2690 & Should code reviewers reproduce the problem/solution as part of the code review? & \url{https://www.reddit.com/r/SoftwareEngineering/comments/1k6zm3e/should_code_reviewers_reproduce_the/} \\
G897 & 2691 & Wrote up the SQL injection patterns that survive code review after this thread made me realize how underinvested the tooling actually is & \url{https://www.reddit.com/r/programming/comments/1rs5pa4/wrote_up_the_sql_injection_patterns_that_survive/} \\
G898 & 2693 & Principles to vibe code without creating buggy AI slop & \url{https://www.reddit.com/r/vibecoding/comments/1s4anxg/principles_to_vibe_code_without_creating_buggy_ai/} \\
G899 & 2699 & Contributing to Large Open Source Repo - Code reviewer messed up my code. & \url{https://www.reddit.com/r/webdev/comments/1njxgr6/contributing_to_large_open_source_repo_code/} \\
G900 & 2700 & Does anyone work on a team that doesn't require code reviews? & \url{https://www.reddit.com/r/ExperiencedDevs/comments/1rmjyu4/does_anyone_work_on_a_team_that_doesnt_require/} \\
G901 & 2702 & How do you get comfortable with shipping code you haven't reviewed? & \url{https://www.reddit.com/r/ExperiencedDevs/comments/1pna63a/how_do_you_get_comfortable_with_shipping_code_you/} \\
G902 & 2703 & We vibecoded a 6x cheaper CodeRabbit alternative for GitHub PR reviews & \url{https://www.reddit.com/r/vibecoding/comments/1texzsm/we_vibecoded_a_6x_cheaper_coderabbit_alternative/} \\
G903 & 2705 & Positive Logic vs Indentation & \url{https://www.reddit.com/r/cpp/comments/1oibqxa/positive_logic_vs_indentation/} \\
G904 & 2708 & Claude Code just saved my machine from a malware-laced GitHub repo. Here's what it found. & \url{https://www.reddit.com/r/ClaudeAI/comments/1rpwqqi/claude_code_just_saved_my_machine_from_a/} \\
G905 & 2712 & [Project] xtils -- A lightweight modern C++ utility library (JSON, networking, event-driven, HTTP/WebSocket server, etc.) & \url{https://www.reddit.com/r/cpp/comments/1oakjax/project_xtils_a_lightweight_modern_c_utility/} \\
G906 & 2713 & Improving Claude Code with Cross-Model Plan Review and Code Review & \url{https://www.reddit.com/r/ClaudeAI/comments/1qu7yws/improving_claude_code_with_crossmodel_plan_review/} \\
G907 & 2716 & Mega prompt VS inspirational discussion & \url{https://www.reddit.com/r/ClaudeAI/comments/1sp6gdw/mega_prompt_vs_inspirational_discussion/} \\
G908 & 2719 & [AskJS] Detection of null pointer, my experience with chatGPT & \url{https://www.reddit.com/r/javascript/comments/1i1ewpp/askjs_detection_of_null_pointer_my_experience/} \\
G909 & 2725 & Do we have vscode extension to make a review of Claude Code changes in cursor/copilot style? With hotkeys to move to the next edit, accept/decline, move to next & \url{https://www.reddit.com/r/ClaudeAI/comments/1m0jbph/do_we_have_vscode_extension_to_make_a_review_of/} \\
G910 & 2726 & Common misconception about AI coding tools like cursor and Claude code & \url{https://www.reddit.com/r/ExperiencedDevs/comments/1qzmvob/common_misconception_about_ai_coding_tools_like/} \\
G911 & 2734 & Help vote on the best model for code reviews! & \url{https://www.reddit.com/r/ChatGPTCoding/comments/1jgxygs/help_vote_on_the_best_model_for_code_reviews/} \\
G912 & 2735 & interesting to see claude code being used as part of a larger self hosted agent orchestration system rather than standalone & \url{https://www.reddit.com/r/github/comments/1tatuo1/interesting_to_see_claude_code_being_used_as_part/} \\
G913 & 2752 & Stop your AI agent from ignoring your architecture & \url{https://www.reddit.com/r/ClaudeAI/comments/1ryuvxu/stop_your_ai_agent_from_ignoring_your_architecture/} \\
G914 & 2755 & code review agent cathces unprofessional langage; i am just testing out features. interesting. & \url{https://www.reddit.com/r/ClaudeAI/comments/1q6uary/code_review_agent_cathces_unprofessional_langage/} \\
G915 & 2759 & Workflow setup to make your coding agent ship small reviewable PRs incrementally & \url{https://www.reddit.com/r/ExperiencedDevs/comments/1rqwfjj/workflow_setup_to_make_your_coding_agent_ship/} \\
G916 & 2761 & Have you ever worked with a junior dev who made a rough first impression but turned it around? & \url{https://www.reddit.com/r/ExperiencedDevs/comments/1lb2nzp/have_you_ever_worked_with_a_junior_dev_who_made_a/} \\
G917 & 2762 & codemate v3.5 made some small changes that actually help in real workflows & \url{https://www.reddit.com/r/vibecoding/comments/1skbqvs/codemate_v35_made_some_small_changes_that/} \\
G918 & 2766 & my work colleagues use generics everywhere for everything & \url{https://www.reddit.com/r/golang/comments/1nr88zk/my_work_colleagues_use_generics_everywhere_for/} \\
G919 & 2769 & I asked GPT for a plan. now what? & \url{https://www.reddit.com/r/GithubCopilot/comments/1scojfv/i_asked_gpt_for_a_plan_now_what/} \\
G920 & 2774 & Flowstate - A crate for modelling self-executing workflows as finite state machines & \url{https://www.reddit.com/r/rust/comments/1t832at/flowstate_a_crate_for_modelling_selfexecuting/} \\
G921 & 2776 & The solution is simple... & \url{https://www.reddit.com/r/vibecoding/comments/1q1jhi4/the_solution_is_simple/} \\
G922 & 2777 & My vibe-coding setup using Claude to actually BUILD real products, and why it works in the comments & \url{https://www.reddit.com/r/ClaudeAI/comments/1jaudbn/my_vibecoding_setup_using_claude_to_actually/} \\
G923 & 2784 & What are you using for visual QA / UI verification in your AI workflows? & \url{https://www.reddit.com/r/ClaudeAI/comments/1s6sadk/what_are_you_using_for_visual_qa_ui_verification/} \\
G924 & 2785 & 4 Simultaneous Instances--Maximum Productivity with CLAUDE\_SYNC.md & \url{https://www.reddit.com/r/ClaudeAI/comments/1l7hbzu/4_simultaneous_instancesmaximum_productivity_with/} \\
G925 & 2786 & Examples of cursor producing bad code? & \url{https://www.reddit.com/r/programming/comments/1kmsmp5/examples_of_cursor_producing_bad_code/} \\
G926 & 2787 & Where should trust checks happen for AI coding agents? & \url{https://www.reddit.com/r/cursor/comments/1tsr5sk/where_should_trust_checks_happen_for_ai_coding/} \\
G927 & 2792 & 3 agents, 3,464 commits, 8 days. 90\% of tokens staying local. & \url{https://www.reddit.com/r/LocalLLaMA/comments/1r79a9q/3_agents_3464_commits_8_days_90_of_tokens_staying/} \\
G928 & 2795 & How do you keep your understanding of your codebase and systems consistently high ? & \url{https://www.reddit.com/r/ExperiencedDevs/comments/1stm1uw/how_do_you_keep_your_understanding_of_your/} \\
G929 & 2796 & how a senior engineer at a \$140m+ startup actually codes with ai (95\% 'vibe coding' but with system) & \url{https://www.reddit.com/r/vibecoding/comments/1nnl19h/how_a_senior_engineer_at_a_140m_startup_actually/} \\
G930 & 2797 & Long post : The Security Blind Spots I Keep Finding in Vibe Coded Apps (A Checklist for Claude AI Builders) & \url{https://www.reddit.com/r/ClaudeAI/comments/1s02qmb/long_post_the_security_blind_spots_i_keep_finding/} \\
G931 & 2799 & I turned Claude Code into a multi-agent swarm and it actually changed how I work & \url{https://www.reddit.com/r/AI_Agents/comments/1rw5bg4/i_turned_claude_code_into_a_multiagent_swarm_and/} \\
G932 & 2801 & Thirdpass: coordinate package review to reduce software supply-chain risk & \url{https://www.reddit.com/r/rust/comments/1tfo8yi/thirdpass_coordinate_package_review_to_reduce/} \\
G933 & 2803 & Tips for helping me contribute more to meetings? & \url{https://www.reddit.com/r/cscareerquestions/comments/1j0rpd3/tips_for_helping_me_contribute_more_to_meetings/} \\
G934 & 2804 & DevOps friends: Would you use GitHub Pull Requests to self-serve cloud access (Terraform-based)? & \url{https://www.reddit.com/r/devops/comments/1kajdbt/devops_friends_would_you_use_github_pull_requests/} \\
G935 & 2813 & After 3 months of daily Claude Code use, here's the workflow system that actually sticks & \url{https://www.reddit.com/r/ClaudeAI/comments/1sjpvtu/after_3_months_of_daily_claude_code_use_heres_the/} \\
G936 & 2814 & Have you run into security issues with AI-generated code? & \url{https://www.reddit.com/r/vibecoding/comments/1trh5pk/have_you_run_into_security_issues_with/} \\
G937 & 2815 & Do You Want Real Results? Because THAT'S How You Get Real Results! & \url{https://www.reddit.com/r/ClaudeAI/comments/1ljoses/do_you_want_real_results_because_thats_how_you/} \\
G938 & 2819 & People who don't use AI coding tools, what do you think of this statement by the creator of NodeJS & \url{https://www.reddit.com/r/cscareerquestions/comments/1qkx8e9/people_who_dont_use_ai_coding_tools_what_do_you/} \\
G939 & 2826 & When an MR touches 20+ files: I built an extension that explains the diff as a story of changes & \url{https://www.reddit.com/r/gitlab/comments/1rtg7tr/when_an_mr_touches_20_files_i_built_an_extension/} \\
G940 & 2828 & I wrote 10 lines of testing code per minute. No bullshit. Here's what I learned. & \url{https://www.reddit.com/r/ChatGPTCoding/comments/1jnwqx2/i_wrote_10_lines_of_testing_code_per_minute_no/} \\
G941 & 2829 & The prompt I run every time before git push (Codex or Claude Code) & \url{https://www.reddit.com/r/ChatGPTCoding/comments/1ogm1gn/the_prompt_i_run_every_time_before_git_push_codex/} \\
G942 & 2832 & Do devs really value soft skills or is everyone just an 'antisocial genius'? & \url{https://www.reddit.com/r/devops/comments/1k40stq/do_devs_really_value_soft_skills_or_is_everyone/} \\
G943 & 2834 & Putting the father of Linux into Claude Code is really awesome & \url{https://www.reddit.com/r/ClaudeAI/comments/1mpwyz7/putting_the_father_of_linux_into_claude_code_is/} \\
G944 & 2836 & zerobrew - (experimental) drop in replacement for homebrew & \url{https://www.reddit.com/r/rust/comments/1qxo5l4/zerobrew_experimental_drop_in_replacement_for/} \\
G945 & 2839 & "Don't add abstractions beyond what the task requires" rule & \url{https://www.reddit.com/r/ClaudeAI/comments/1tqykq7/dont_add_abstractions_beyond_what_the_task/} \\
G946 & 2849 & Roast me ! & \url{https://www.reddit.com/r/vibecoding/comments/1ssp3ea/roast_me/} \\
G947 & 2855 & I don't know how to code anymore yet I understand everything, is that normal now? & \url{https://www.reddit.com/r/devops/comments/1sngljr/i_dont_know_how_to_code_anymore_yet_i_understand/} \\
G948 & 2858 & I code for 35+ years, now Claude Code does 99\% of the actual work - am I really a "vibe coder"? & \url{https://www.reddit.com/r/ClaudeAI/comments/1qwg3al/i_code_for_35_years_now_claude_code_does_99_of/} \\
G949 & 2859 & So much hate on Vibe Coding - am I the only one having a good time? & \url{https://www.reddit.com/r/vibecoding/comments/1oaefb3/so_much_hate_on_vibe_coding_am_i_the_only_one/} \\
G950 & 2863 & I built a 127-skill framework for Claude Code with a localhost dashboard and 3-agent orchestration & \url{https://www.reddit.com/r/ClaudeAI/comments/1tixps7/i_built_a_127skill_framework_for_claude_code_with/} \\
G951 & 2865 & Thank you Cursor! & \url{https://www.reddit.com/r/vibecoding/comments/1l6epu2/thank_you_cursor/} \\
G952 & 2866 & Adversarial is the new way to go... & \url{https://www.reddit.com/r/ClaudeAI/comments/1tph6xk/adversarial_is_the_new_way_to_go/} \\
G953 & 2871 & Future of jobs and engineers & \url{https://www.reddit.com/r/cursor/comments/1l602xm/future_of_jobs_and_engineers/} \\
G954 & 2873 & We're 3 dev brothers building a new version-control and code review tool & \url{https://www.reddit.com/r/devops/comments/1p6pisu/were_3_dev_brothers_building_a_new_versioncontrol/} \\
G955 & 2877 & Factual Mode & \url{https://www.reddit.com/r/ClaudeAI/comments/1s2ggd9/factual_mode/} \\
G956 & 2879 & Hashimoto's Vouch is actually open source version of a company hiring only seniors and mid levels. & \url{https://www.reddit.com/r/programming/comments/1ruci5b/hashimotos_vouch_is_actually_open_source_version/} \\
G957 & 2882 & My Top 10 Claude Code Tips from 11 Months of Intense Usage & \url{https://www.reddit.com/r/ClaudeAI/comments/1qcan9z/my_top_10_claude_code_tips_from_11_months_of/} \\
G958 & 2885 & Are people really that bad at using AI or they trying to underestimate the impact it'll have to cope with an uncertain future? & \url{https://www.reddit.com/r/OpenAI/comments/1kdsssc/are_people_really_that_bad_at_using_ai_or_they/} \\
G959 & 2894 & Tested glm 4.7 for coding projects past week, comparison with deepseek and qwen & \url{https://www.reddit.com/r/LocalLLaMA/comments/1q0dkwz/tested_glm_47_for_coding_projects_past_week/} \\
G960 & 2896 & Do we just sit around and watch Claude fight ChatGPT, or is there still room to build? & \url{https://www.reddit.com/r/ChatGPTCoding/comments/1rds898/do_we_just_sit_around_and_watch_claude_fight/} \\
G961 & 2899 & Can AI do simple code review? & \url{https://www.reddit.com/r/AskProgramming/comments/1kkfrgc/can_ai_do_simple_code_review/} \\
G962 & 2904 & GPT-5: Cursor CLI, Codex CLI or claude-code-router? & \url{https://www.reddit.com/r/ChatGPTCoding/comments/1mqwev6/gpt5_cursor_cli_codex_cli_or_claudecoderouter/} \\
G963 & 2906 & I tried building one small game with AI and ended up shipping 8 in parallel & \url{https://www.reddit.com/r/AI_Agents/comments/1t49tx5/i_tried_building_one_small_game_with_ai_and_ended/} \\
G964 & 2907 & When AI writes your code and something breaks in production, who owns the bug? & \url{https://www.reddit.com/r/AI_Agents/comments/1towtzs/when_ai_writes_your_code_and_something_breaks_in/} \\
G965 & 2910 & Code review, first Golang project & \url{https://www.reddit.com/r/golang/comments/1i48k79/code_review_first_golang_project/} \\
G966 & 2911 & My vibe coding methodology & \url{https://www.reddit.com/r/vibecoding/comments/1rx978i/my_vibe_coding_methodology/} \\
G967 & 2914 & I just watched an AI agent take a Jira ticket, understand our codebase, and push a PR in minutes and I'm genuinely scared & \url{https://www.reddit.com/r/cscareerquestions/comments/1m1uw9v/i_just_watched_an_ai_agent_take_a_jira_ticket/} \\
G968 & 2920 & I've been in game dev for over 20 years and just tried vibecoding a production-quality competitive multiplayer .io game in 30 days. Here's the honest breakdown. & \url{https://www.reddit.com/r/vibecoding/comments/1t10tmu/ive_been_in_game_dev_for_over_20_years_and_just/} \\
G969 & 2927 & I had AI peer-review its own rules. All three models missed "don't push to main" & \url{https://www.reddit.com/r/devops/comments/1pbr13h/i_had_ai_peerreview_its_own_rules_all_three/} \\
G970 & 2930 & I built a 5-agent Claude Code team with a review loop -- the reviewer caught a hook event count error that would have shipped wrong documentation & \url{https://www.reddit.com/r/ClaudeAI/comments/1rlidt3/i_built_a_5agent_claude_code_team_with_a_review/} \\
G971 & 2932 & 3 Multi Agent Team projects I built for Developers & \url{https://www.reddit.com/r/AI_Agents/comments/1nkk4gy/3_multi_agent_team_projects_i_built_for_developers/} \\
G972 & 2933 & Code review & \url{https://www.reddit.com/r/codereview/comments/1qas3ou/code_review/} \\
G973 & 2934 & I made an MCP context marketplace & \url{https://www.reddit.com/r/ClaudeAI/comments/1tvsls7/i_made_an_mcp_context_marketplace/} \\
G974 & 2935 & I vibecoded a Zombie extaction game for 6 days to test the SOTA AI models & \url{https://www.reddit.com/r/vibecoding/comments/1t5p5r0/i_vibecoded_a_zombie_extaction_game_for_6_days_to/} \\
G975 & 2936 & Daily AI Archive | 12/1/2025 & \url{https://www.reddit.com/r/accelerate/comments/1pbxl7i/daily_ai_archive_1212025/} \\
G976 & 2938 & Code review prompts for Go & \url{https://www.reddit.com/r/golang/comments/1lcriok/code_review_prompts_for_go/} \\
G977 & 2940 & I finished this FastAPI starter template (RBAC, Redis, ContextVars). What do you guys think of the architecture? & \url{https://www.reddit.com/r/AskProgramming/comments/1sspcd9/i_finished_this_fastapi_starter_template_rbac/} \\
G978 & 2941 & Reviewing in the age of AI & \url{https://www.reddit.com/r/vibecoding/comments/1t5hnd2/reviewing_in_the_age_of_ai/} \\
G979 & 2943 & I built Toolbrew -- 10 tools for Claude Code (6 slash commands, 3 skills, 1 hook) & \url{https://www.reddit.com/r/ClaudeAI/comments/1srt2yo/i_built_toolbrew_10_tools_for_claude_code_6_slash/} \\
G980 & 2949 & Programmer Manifesto & \url{https://www.reddit.com/r/programming/comments/1jdxne4/programmer_manifesto/} \\
G981 & 2954 & Would you use it ? An AI based PR review tool & \url{https://www.reddit.com/r/webdev/comments/1joqb4g/would_you_use_it_an_ai_based_pr_review_tool/} \\
G982 & 2959 & Initial experience with GitHub Coding Agent & \url{https://www.reddit.com/r/github/comments/1kvatij/initial_experience_with_github_coding_agent/} \\
G983 & 2960 & First release of my C++23 Unicode text library - would love a code review & \url{https://www.reddit.com/r/codereview/comments/1su9q3q/first_release_of_my_c23_unicode_text_library/} \\
G984 & 2964 & Code review comments & \url{https://www.reddit.com/r/ExperiencedDevs/comments/1jrt24w/code_review_comments/} \\
G985 & 2965 & I replaced chaotic solo Claude coding with a simple 3-agent team (Architect + Builder + Reviewer) -- it's stupidly effective and token-efficient & \url{https://www.reddit.com/r/ClaudeAI/comments/1sa7ju4/i_replaced_chaotic_solo_claude_coding_with_a/} \\
G986 & 2966 & Code review taking forever because everyone's busy and reviews get deprioritized, sound familiar? & \url{https://www.reddit.com/r/ExperiencedDevs/comments/1qxf5ws/code_review_taking_forever_because_everyones_busy/} \\
G987 & 2968 & Is there a github copilot alternative for QA that focuses on quality not speed? & \url{https://www.reddit.com/r/vibecoding/comments/1p1yxs9/is_there_a_github_copilot_alternative_for_qa_that/} \\
G988 & 2975 & Code review request: SPARTA orchestration console (architecture, structure, clarity) & \url{https://www.reddit.com/r/codereview/comments/1s5isdz/code_review_request_sparta_orchestration_console/} \\
G989 & 2977 & Review it please!! & \url{https://www.reddit.com/r/webdev/comments/1nmkmnx/review_it_please/} \\
G990 & 2978 & Can getting too used to a breezy project be bad for you long term? & \url{https://www.reddit.com/r/ExperiencedDevs/comments/1r0b0w2/can_getting_too_used_to_a_breezy_project_be_bad/} \\
G991 & 2979 & Rust transpiler project -- looking for code review and architecture feedback & \url{https://www.reddit.com/r/codereview/comments/1trvaw0/rust_transpiler_project_looking_for_code_review/} \\
G992 & 2986 & What's your actual AI coding workflow? & \url{https://www.reddit.com/r/vibecoding/comments/1nyzb5e/whats_your_actual_ai_coding_workflow/} \\
G993 & 2988 & META Reject & \url{https://www.reddit.com/r/ExperiencedDevs/comments/1seg4y8/meta_reject/} \\
G994 & 2990 & Can AI Replace Manual Code Reviews? & \url{https://www.reddit.com/r/AskProgramming/comments/1jevcbn/can_ai_replace_manual_code_reviews/} \\
G995 & 2991 & Devs (and Copilot) not reading your style guides? I built a server to make Copilot enforce our ADRs. & \url{https://www.reddit.com/r/github/comments/1ozvq9s/devs_and_copilot_not_reading_your_style_guides_i/} \\
G996 & 2997 & SpringSentinel v1.1.10: A lightweight Maven plugin for Spring Boot static analysis (now with SARIF \& Java 25 support) & \url{https://www.reddit.com/r/java/comments/1r70zrw/springsentinel_v1110_a_lightweight_maven_plugin/} \\
G997 & 3005 & I wish LLMs never became popular & \url{https://www.reddit.com/r/ExperiencedDevs/comments/1qeqh1k/i_wish_llms_never_became_popular/} \\
G998 & 3010 & \#2 Code review request: feedback on OOP, TDD, and SOLID principles & \url{https://www.reddit.com/r/codereview/comments/1ldlgha/2_code_review_request_feedback_on_oop_tdd_and/} \\
G999 & 3012 & Does this count as AI First? & \url{https://www.reddit.com/r/ExperiencedDevs/comments/1rb1dd5/does_this_count_as_ai_first/} \\
G1000 & 3013 & I tried agentic AI development for the past week. Some thoughts/questions for you all & \url{https://www.reddit.com/r/ExperiencedDevs/comments/1kwtt1d/i_tried_agentic_ai_development_for_the_past_week/} \\
G1001 & 3016 & Incompetent Coworker & \url{https://www.reddit.com/r/ExperiencedDevs/comments/1iyhynu/incompetent_coworker/} \\
G1002 & 3018 & A Programmer's Guide to Leaving GitHub & \url{https://www.reddit.com/r/programming/comments/1t55pvp/a_programmers_guide_to_leaving_github/} \\
G1003 & 3019 & Recently got ChatGPT Pro for coding, but it sucks... & \url{https://www.reddit.com/r/ChatGPTPro/comments/1slr3lm/recently_got_chatgpt_pro_for_coding_but_it_sucks/} \\
G1004 & 3028 & Mythos found ten thousand bugs and the real bottleneck is human triage & \url{https://www.reddit.com/r/ClaudeAI/comments/1twvb9g/mythos_found_ten_thousand_bugs_and_the_real/} \\
G1005 & 3029 & SQL is the last place most teams have no static analysis and it keeps causing the same incidents & \url{https://www.reddit.com/r/ExperiencedDevs/comments/1rp041d/sql_is_the_last_place_most_teams_have_no_static/} \\
G1006 & 3030 & What the best way to stop the same issues coming up in code reviews? & \url{https://www.reddit.com/r/ExperiencedDevs/comments/1npcth8/what_the_best_way_to_stop_the_same_issues_coming/} \\
G1007 & 3035 & The Claude Code / AI Dilemma & \url{https://www.reddit.com/r/ClaudeAI/comments/1mvpwz2/the_claude_code_ai_dilemma/} \\
G1008 & 3037 & I did not use AI for this I swear & \url{https://www.reddit.com/r/ExperiencedDevs/comments/1jfe2z5/i_did_not_use_ai_for_this_i_swear/} \\
G1009 & 3041 & Multi agent orchestration layer & \url{https://www.reddit.com/r/ClaudeAI/comments/1rw57oj/multi_agent_orchestration_layer/} \\
G1010 & 3042 & /Karen Plugin for Claude Code that uses Codex CLI to De-BUG and Roast your PR & \url{https://www.reddit.com/r/ClaudeAI/comments/1s79fd6/karen_plugin_for_claude_code_that_uses_codex_cli/} \\
G1011 & 3046 & After giving in to AI coding tools I am not sure where to draw the line & \url{https://www.reddit.com/r/ExperiencedDevs/comments/1pgt2ty/after_giving_in_to_ai_coding_tools_i_am_not_sure/} \\
G1012 & 3047 & How are you guys able to carefully review and test all the code that Claude Code generates? & \url{https://www.reddit.com/r/ClaudeAI/comments/1lb1tsa/how_are_you_guys_able_to_carefully_review_and/} \\
G1013 & 3049 & Just wow & \url{https://www.reddit.com/r/ClaudeAI/comments/1luqst7/just_wow/} \\
G1014 & 3050 & Trunk based branching with a largely asynchronous offshore dev model & \url{https://www.reddit.com/r/ExperiencedDevs/comments/1lz273q/trunk_based_branching_with_a_largely_asynchronous/} \\
G1015 & 3062 & My favorite underrated AI coding tools & \url{https://www.reddit.com/r/ChatGPTCoding/comments/1isse8o/my_favorite_underrated_ai_coding_tools/} \\
G1016 & 3063 & BorisCode, Cherny's CC setup for OpenCode & \url{https://www.reddit.com/r/LocalLLaMA/comments/1s905zu/boriscode_chernys_cc_setup_for_opencode/} \\
G1017 & 3065 & AI Advisory Board - Group chat with ChatGPT, Claude, Gemini at the same time? & \url{https://www.reddit.com/r/vibecoding/comments/1od5yhg/ai_advisory_board_group_chat_with_chatgpt_claude/} \\
G1018 & 3067 & I'm so tired on this AI copilot BS & \url{https://www.reddit.com/r/cpp/comments/1opv1ik/im_so_tired_on_this_ai_copilot_bs/} \\
G1019 & 3070 & How to "childproof" a codebase when working with contributors who are non-developers & \url{https://www.reddit.com/r/ExperiencedDevs/comments/1qenfva/how_to_childproof_a_codebase_when_working_with/} \\
G1020 & 3074 & Plan/Build vs. Subagents vs. ... & \url{https://www.reddit.com/r/ClaudeAI/comments/1mptc5n/planbuild_vs_subagents_vs/} \\
G1021 & 3078 & cargo-onefile: Bundle your Rust project into a single file - Perfect for LLMs and Code Reviews & \url{https://www.reddit.com/r/rust/comments/1i1avz8/cargoonefile_bundle_your_rust_project_into_a/} \\
G1022 & 3079 & I built a multi-agent framework around scheduling instead of chatting -- looking for feedback on the architecture. & \url{https://www.reddit.com/r/AI_Agents/comments/1s0vosl/i_built_a_multiagent_framework_around_scheduling/} \\
G1023 & 3081 & Technical Review for Vibecoders & \url{https://www.reddit.com/r/vibecoding/comments/1tagzq5/technical_review_for_vibecoders/} \\
G1024 & 3083 & What AI tools/platforms are you actually using for coding in 2026? & \url{https://www.reddit.com/r/vibecoding/comments/1revk82/what_ai_toolsplatforms_are_you_actually_using_for/} \\
G1025 & 3087 & Filling my previous leader's shoes & \url{https://www.reddit.com/r/ExperiencedDevs/comments/1pp6811/filling_my_previous_leaders_shoes/} \\
G1026 & 3089 & How do I ethically point out to my collegues the problems with their code & \url{https://www.reddit.com/r/AskProgramming/comments/1lzp91m/how_do_i_ethically_point_out_to_my_collegues_the/} \\
G1027 & 3090 & Why does code review take forever once teams hit 15-20 engineers & \url{https://www.reddit.com/r/ExperiencedDevs/comments/1rjfo06/why_does_code_review_take_forever_once_teams_hit/} \\
G1028 & 3092 & Generate realtime documentation, tutorials, codebase chat and pr reviews for ANY codebase! & \url{https://www.reddit.com/r/ChatGPTCoding/comments/1ipl05r/generate_realtime_documentation_tutorials/} \\
G1029 & 3093 & 24/7 agent pipeline reduced cost and time to develop production grade software by 60-70\%. & \url{https://www.reddit.com/r/ArtificialInteligence/comments/1tzhq0t/247_agent_pipeline_reduced_cost_and_time_to/} \\
G1030 & 3095 & AI Code Reviewers vs Human Code Reviewers & \url{https://www.reddit.com/r/programming/comments/1klojsx/ai_code_reviewers_vs_human_code_reviewers/} \\
G1031 & 3104 & Semantic diffs cut tokens significantly when feeding code changes to LLMs. Also improves attention scores of the model. & \url{https://www.reddit.com/r/ClaudeAI/comments/1slfy4g/semantic_diffs_cut_tokens_significantly_when/} \\
G1032 & 3107 & Autonomous weapons? Robotic surgeons? Self-driving cars? I think not. & \url{https://www.reddit.com/r/ClaudeAI/comments/1rzwb1i/autonomous_weapons_robotic_surgeons_selfdriving/} \\
G1033 & 3112 & How do I handle this & \url{https://www.reddit.com/r/ExperiencedDevs/comments/1m2ezcs/how_do_i_handle_this/} \\
G1034 & 3114 & Best autonomous ai agent for github? & \url{https://www.reddit.com/r/AI_Agents/comments/1tah2at/best_autonomous_ai_agent_for_github/} \\
G1035 & 3117 & Codex CLI refusing to sleep & \url{https://www.reddit.com/r/OpenAI/comments/1qafw4j/codex_cli_refusing_to_sleep/} \\
G1036 & 3127 & Workflow Critique / Advice & \url{https://www.reddit.com/r/ClaudeAI/comments/1td519x/workflow_critique_advice/} \\
G1037 & 3130 & I built an orchstrator that manages 30 agent (Claude Code, Codex) sessions at once & \url{https://www.reddit.com/r/AI_Agents/comments/1rfwt3p/i_built_an_orchstrator_that_manages_30_agent/} \\
G1038 & 3132 & Built a Claude Code plugin that reviews code using classic SE books (Brooks, Fowler, Feathers...) & \url{https://www.reddit.com/r/ClaudeAI/comments/1s6kx9i/built_a_claude_code_plugin_that_reviews_code/} \\
G1039 & 3136 & Built an MCP for claude code that turns ticket-mentions into PRs with browser QA (and what I learned along the way) & \url{https://www.reddit.com/r/ClaudeAI/comments/1th7tj6/built_an_mcp_for_claude_code_that_turns/} \\
G1040 & 3138 & Cursor New Plan: Build in New Agent is fire! & \url{https://www.reddit.com/r/cursor/comments/1pjjdiw/cursor_new_plan_build_in_new_agent_is_fire/} \\
G1041 & 3140 & Amazon GenAI code will need human review moving forward & \url{https://www.reddit.com/r/ClaudeAI/comments/1rrmrjb/amazon_genai_code_will_need_human_review_moving/} \\
G1042 & 3141 & Saw a post here about hiring a Fiverr dev to close the last 20\%. Tried it myself. Game-changer. & \url{https://www.reddit.com/r/vibecoding/comments/1lc5qxt/saw_a_post_here_about_hiring_a_fiverr_dev_to/} \\
G1043 & 3147 & I built a GitHub App that runs Claude on every pull request for code reviews & \url{https://www.reddit.com/r/ClaudeAI/comments/1rl9t53/i_built_a_github_app_that_runs_claude_on_every/} \\
G1044 & 3149 & Copilot pulled in a bunch of dependencies we did not need and only noticed months later & \url{https://www.reddit.com/r/devops/comments/1qiz2m9/copilot_pulled_in_a_bunch_of_dependencies_we_did/} \\
G1045 & 3150 & A single SQL query took down our app for a weekend and nobody caught it in review & \url{https://www.reddit.com/r/webdev/comments/1rpcer6/a_single_sql_query_took_down_our_app_for_a/} \\
G1046 & 3152 & Is there a standard tool or framework for reviewing MCP servers / skills before shipping? & \url{https://www.reddit.com/r/vibecoding/comments/1smwywf/is_there_a_standard_tool_or_framework_for/} \\
G1047 & 3154 & Strugging to get ChatGPT to review my sql code ... & \url{https://www.reddit.com/r/ChatGPTCoding/comments/1jdct6u/strugging_to_get_chatgpt_to_review_my_sql_code/} \\
G1048 & 3156 & I built a 4,000-line Agent Skill for TypeScript + React + Next.js (open source) & \url{https://www.reddit.com/r/ClaudeAI/comments/1sk1k55/i_built_a_4000line_agent_skill_for_typescript/} \\
G1049 & 3157 & Dealing with a Sloppy Teammate Who Won't Take Feedback & \url{https://www.reddit.com/r/ExperiencedDevs/comments/1jpsjg8/dealing_with_a_sloppy_teammate_who_wont_take/} \\
G1050 & 3159 & I don't read rejection feedbacks from Apple reviewers anymore. Claude reads it, fixes the code, and resubmits to the App Store. & \url{https://www.reddit.com/r/ClaudeAI/comments/1s5szb6/i_dont_read_rejection_feedbacks_from_apple/} \\
G1051 & 3161 & Ping, with Color! & \url{https://www.reddit.com/r/csharp/comments/1top166/ping_with_color/} \\
G1052 & 3166 & So best to do very little each iteration I guess & \url{https://www.reddit.com/r/ClaudeAI/comments/1mofyer/so_best_to_do_very_little_each_iteration_i_guess/} \\
G1053 & 3172 & License Manager with Go & \url{https://www.reddit.com/r/golang/comments/1htg9w7/license_manager_with_go/} \\
G1054 & 3174 & Do you trust AI completions for production code? & \url{https://www.reddit.com/r/devops/comments/1n4mdw4/do_you_trust_ai_completions_for_production_code/} \\
G1055 & 3179 & Have you ever had to fight with a start up to let you implement basic security & \url{https://www.reddit.com/r/AskProgramming/comments/1iye57o/have_you_ever_had_to_fight_with_a_start_up_to_let/} \\
G1056 & 3180 & My Octopus Deploy Review.. I think it's still the pick in 2025 & \url{https://www.reddit.com/r/dotnet/comments/1liy360/my_octopus_deploy_review_i_think_its_still_the/} \\
G1057 & 3186 & Recommendation for local PR review setup & \url{https://www.reddit.com/r/vibecoding/comments/1snap81/recommendation_for_local_pr_review_setup/} \\
G1058 & 3190 & Claude Code said it has a fundamental flaw compared with ChatGPT & \url{https://www.reddit.com/r/ClaudeAI/comments/1on86lj/claude_code_said_it_has_a_fundamental_flaw/} \\
G1059 & 3200 & How do you interview a candidate coming from a different tech stack? & \url{https://www.reddit.com/r/ExperiencedDevs/comments/1jo0ds3/how_do_you_interview_a_candidate_coming_from_a/} \\
G1060 & 3201 & free tool to bridge cursor dev \& AI Studio planning (stop manual copy/pasting!) + how i use cursor + gemini 2.5 pro effectively & \url{https://www.reddit.com/r/cursor/comments/1kel0ga/free_tool_to_bridge_cursor_dev_ai_studio_planning/} \\
G1061 & 3206 & [Discussion] Contract-First NestJS: Can Strict Zod Types Stop Cursor from Hallucinating? & \url{https://www.reddit.com/r/cursor/comments/1p3epdl/discussion_contractfirst_nestjs_can_strict_zod/} \\
G1062 & 3210 & After a 2 AM production outage from a Claude Code edit, I built a 4-tier adversarial review system that learns my codebase. It's caught 47 bugs before shipping. & \url{https://www.reddit.com/r/ClaudeAI/comments/1rg6wdd/after_a_2_am_production_outage_from_a_claude_code/} \\
G1063 & 3211 & As Tech Leads do you ever find yourself "coding for" junior teammates during code reviews? & \url{https://www.reddit.com/r/ExperiencedDevs/comments/1tkydx6/as_tech_leads_do_you_ever_find_yourself_coding/} \\
G1064 & 3212 & Managed Agents self-hosted sandboxes - what's new in CC 2.1.145 (+20,218 tokens) & \url{https://www.reddit.com/r/ClaudeAI/comments/1tk2nw6/managed_agents_selfhosted_sandboxes_whats_new_in/} \\
G1065 & 3214 & How do you review \& debug AI "vibe coded" stuff? & \url{https://www.reddit.com/r/vibecoding/comments/1mlj0sv/how_do_you_review_debug_ai_vibe_coded_stuff/} \\
G1066 & 3222 & My favorite workflow on Cursor (2/10/25) & \url{https://www.reddit.com/r/cursor/comments/1r109cz/my_favorite_workflow_on_cursor_21025/} \\
G1067 & 3223 & Claude Opus 4.7 Just Dropped - Better Long Tasks, Self-Verification, 3x Better Vision . & \url{https://www.reddit.com/r/ArtificialInteligence/comments/1sn67q7/claude_opus_47_just_dropped_better_long_tasks/} \\
G1068 & 3224 & Cope with burn-out and loss of motivation due to massive influx of uncaring vibe-coders & \url{https://www.reddit.com/r/ExperiencedDevs/comments/1nqz5wj/cope_with_burnout_and_loss_of_motivation_due_to/} \\
G1069 & 3227 & Is Anyone Else Worried About AI-Generated Code Reaching Production Without Real Risk Review? & \url{https://www.reddit.com/r/vibecoding/comments/1t2c0n0/is_anyone_else_worried_about_aigenerated_code/} \\
G1070 & 3230 & Started using Gemini more than Sonnet 4.5 & \url{https://www.reddit.com/r/ClaudeAI/comments/1o832ht/started_using_gemini_more_than_sonnet_45/} \\
G1071 & 3237 & Feedback beats planning & \url{https://www.reddit.com/r/AskProgramming/comments/1jzbyxi/feedback_beats_planning/} \\
G1072 & 3238 & I built a robust Webhook Handler for Notion Marketplace using .NET 10, Background Queues, and Docker (Open Source) & \url{https://www.reddit.com/r/dotnet/comments/1qj2p7q/i_built_a_robust_webhook_handler_for_notion/} \\
G1073 & 3241 & Agent orchestrator to power Plan -> Implementation/Review in separate agents? & \url{https://www.reddit.com/r/ClaudeAI/comments/1rwn356/agent_orchestrator_to_power_plan/} \\
G1074 & 3243 & [Showoff Saturday] I wrote a 300-page book on the "real-world" engineering skills they don't teach you. Looking for beta readers! & \url{https://www.reddit.com/r/webdev/comments/1lmprwc/showoff_saturday_i_wrote_a_300page_book_on_the/} \\
G1075 & 3247 & sdsort, a utility to sort functions and methods according to the step-down rule & \url{https://www.reddit.com/r/Python/comments/1sbf7vg/sdsort_a_utility_to_sort_functions_and_methods/} \\
G1076 & 3252 & I built a team of 7 AI agents that can't review their own code & \url{https://www.reddit.com/r/ClaudeAI/comments/1rfaz64/i_built_a_team_of_7_ai_agents_that_cant_review/} \\
G1077 & 3254 & [Bash] - Utility function scripts library & \url{https://www.reddit.com/r/codereview/comments/1lc2we5/bash_utility_function_scripts_library/} \\
G1078 & 3257 & Question from junior/mid about code review & \url{https://www.reddit.com/r/ExperiencedDevs/comments/1qkpg9j/question_from_juniormid_about_code_review/} \\
G1079 & 3258 & Can You Find The Race Conditions? & \url{https://www.reddit.com/r/java/comments/1l4hhlz/can_you_find_the_race_conditions/} \\
G1080 & 3259 & Am I relying too much on AI? & \url{https://www.reddit.com/r/cscareerquestions/comments/1opb7wd/am_i_relying_too_much_on_ai/} \\
G1081 & 3260 & We talk a lot about AI writing code... but who's using it to review code? & \url{https://www.reddit.com/r/ChatGPTCoding/comments/1lg6tkl/we_talk_a_lot_about_ai_writing_code_but_whos/} \\
G1082 & 3264 & Why do logic errors slip through automated code review when tools catch patterns but miss meaning & \url{https://www.reddit.com/r/ChatGPTCoding/comments/1rv1jf4/why_do_logic_errors_slip_through_automated_code/} \\
G1083 & 3265 & Developer isn't coding Claude code is! & \url{https://www.reddit.com/r/ClaudeAI/comments/1n8o896/developer_isnt_coding_claude_code_is/} \\
G1084 & 3272 & I scanned a mass of vibe-coded projects. Here's what keeps showing up. & \url{https://www.reddit.com/r/vibecoding/comments/1s4vi5z/i_scanned_a_mass_of_vibecoded_projects_heres_what/} \\
G1085 & 3274 & The secret isn't a better model. It's better files. & \url{https://www.reddit.com/r/ClaudeAI/comments/1ru77i3/the_secret_isnt_a_better_model_its_better_files/} \\
G1086 & 3276 & Reviewing Claude Code changes is easier on an infinite canvas & \url{https://www.reddit.com/r/ClaudeAI/comments/1o8i6e0/reviewing_claude_code_changes_is_easier_on_an/} \\
G1087 & 3279 & Is AI coding making pull requests harder to review? & \url{https://www.reddit.com/r/github/comments/1rofktt/is_ai_coding_making_pull_requests_harder_to_review/} \\
G1088 & 3280 & [Help] Can you help review/suggest my Rust CLI tool more idiomatic? & \url{https://www.reddit.com/r/rust/comments/1j6xlw5/help_can_you_help_reviewsuggest_my_rust_cli_tool/} \\
G1089 & 3281 & Introducing Lynkr -- an open-source Claude-style AI coding proxy built specifically for Databricks model endpoints & \url{https://www.reddit.com/r/ClaudeAI/comments/1pdrv27/introducing_lynkr_an_opensource_claudestyle_ai/} \\
G1090 & 3284 & LockFreeSpscQueue: A high-performance, single-producer, single-consumer (SPSC) queue implemented in modern C++23 & \url{https://www.reddit.com/r/cpp/comments/1mjwjx6/lockfreespscqueue_a_highperformance/} \\
G1091 & 3285 & Simon Willison: "95\% of the code I ship is now AI-generated." & \url{https://www.reddit.com/r/accelerate/comments/1sjdjc5/simon_willison_95_of_the_code_i_ship_is_now/} \\
G1092 & 3286 & Spec Driven Development (SDD) vs Plan Research Implement (PRI) using claude & \url{https://www.reddit.com/r/ClaudeAI/comments/1pkvque/spec_driven_development_sdd_vs_plan_research/} \\
G1093 & 3287 & I am done. I will not be an AI slop code reviewer & \url{https://www.reddit.com/r/cscareerquestions/comments/1s3t5d6/i_am_done_i_will_not_be_an_ai_slop_code_reviewer/} \\
G1094 & 3291 & Lathe -- a personal Zed fork with a theme customizer, git-aware UI, and multi-account collab & \url{https://www.reddit.com/r/rust/comments/1sudefh/lathe_a_personal_zed_fork_with_a_theme_customizer/} \\
G1095 & 3292 & Built an open source IDE for running parallel AI coding agents. would love feedback. & \url{https://www.reddit.com/r/AI_Agents/comments/1slhrp5/built_an_open_source_ide_for_running_parallel_ai/} \\
G1096 & 3296 & The 'Dark Code' Problem and Milla Jovovich's New Open Source Agent Memory System & \url{https://www.reddit.com/r/AI_Agents/comments/1skho97/the_dark_code_problem_and_milla_jovovichs_new/} \\
G1097 & 3297 & What Code Review Can't See (And Bad Data Always Finds) & \url{https://www.reddit.com/r/programming/comments/1szs6p3/what_code_review_cant_see_and_bad_data_always/} \\
G1098 & 3299 & I built an MCP reviewer for Claude and it's been a game changer for my solo dev workflow & \url{https://www.reddit.com/r/ClaudeAI/comments/1luu2xj/i_built_an_mcp_reviewer_for_claude_and_its_been_a/} \\
G1099 & 3301 & Why the hell are devs still putting passwords in AI prompts? It's 2026! & \url{https://www.reddit.com/r/devops/comments/1q9cw8r/why_the_hell_are_devs_still_putting_passwords_in/} \\
G1100 & 3304 & I made a self healing PRD system for Claude code & \url{https://www.reddit.com/r/AI_Agents/comments/1sokjj7/i_made_a_self_healing_prd_system_for_claude_code/} \\
G1101 & 3305 & Automatic Code Reviews based on custom coding guidelines & \url{https://www.reddit.com/r/AskProgramming/comments/1kv52py/automatic_code_reviews_based_on_custom_coding/} \\
G1102 & 3307 & If AI writes the code AND the tests, how do you avoid fake confidence? & \url{https://www.reddit.com/r/vibecoding/comments/1q3x20v/if_ai_writes_the_code_and_the_tests_how_do_you/} \\
G1103 & 3308 & I built a plugin that gives Claude Code a memory across time - schedule tasks to run autonomously & \url{https://www.reddit.com/r/ClaudeAI/comments/1qbtep6/i_built_a_plugin_that_gives_claude_code_a_memory/} \\
G1104 & 3312 & built a self hosted voice first multi agent orchestration system and ran into some interesting architecture problems, curious what others think & \url{https://www.reddit.com/r/ClaudeAI/comments/1t5f2ii/built_a_self_hosted_voice_first_multi_agent/} \\
G1105 & 3313 & Stop grinding l**tcode and try this instead & \url{https://www.reddit.com/r/cscareerquestions/comments/1qtx9u2/stop_grinding_ltcode_and_try_this_instead/} \\
G1106 & 3316 & How to maintain code quality with AI slop? & \url{https://www.reddit.com/r/ExperiencedDevs/comments/1nx21xo/how_to_maintain_code_quality_with_ai_slop/} \\
G1107 & 3317 & I built Claudeck -- a browser UI for Claude Code with agents, cost tracking, and a plugin system & \url{https://www.reddit.com/r/ClaudeAI/comments/1s1ehmf/i_built_claudeck_a_browser_ui_for_claude_code/} \\
G1108 & 3318 & Multi agent vs Single Agent systems & \url{https://www.reddit.com/r/AI_Agents/comments/1tbwpay/multi_agent_vs_single_agent_systems/} \\
G1109 & 3320 & Staying Focused - ADD and AI & \url{https://www.reddit.com/r/SoftwareEngineering/comments/1s3xy2e/staying_focused_add_and_ai/} \\
G1110 & 3323 & Found a nice way to vibe code & \url{https://www.reddit.com/r/vibecoding/comments/1o75mq4/found_a_nice_way_to_vibe_code/} \\
G1111 & 3324 & I built a code review platform without subscriptions & \url{https://www.reddit.com/r/javascript/comments/1ov0tqy/i_built_a_code_review_platform_without/} \\
G1112 & 3333 & Gitlab Duo keeps on failing when trying to do a code-review & \url{https://www.reddit.com/r/gitlab/comments/1tvouri/gitlab_duo_keeps_on_failing_when_trying_to_do_a/} \\
G1113 & 3337 & How do you personally use AI to accelerate your learning as a developer? & \url{https://www.reddit.com/r/ExperiencedDevs/comments/1k1b97y/how_do_you_personally_use_ai_to_accelerate_your/} \\
G1114 & 3340 & I pointed Claude Code at a half-finished idea and it shipped the entire thing. Real-time multiplayer, ELO rankings, daily challenges, behavioral analytics & \url{https://www.reddit.com/r/ClaudeAI/comments/1rdfelr/i_pointed_claude_code_at_a_halffinished_idea_and/} \\
G1115 & 3344 & Generating Unit Tests with Qwen3-Coder-30B - not really usable & \url{https://www.reddit.com/r/LocalLLaMA/comments/1mgv4h3/generating_unit_tests_with_qwen3coder30b_not/} \\
G1116 & 3346 & I've created a bot to make Claude Code 100\% Autonomous & \url{https://www.reddit.com/r/ClaudeAI/comments/1nztzso/ive_created_a_bot_to_make_claude_code_100/} \\
G1117 & 3351 & I let 3 AI coding agents work on my project at the same time for a week. one of them started gaslighting me. & \url{https://www.reddit.com/r/cursor/comments/1t6cw1q/i_let_3_ai_coding_agents_work_on_my_project_at/} \\
G1118 & 3352 & Alternative to CodeRabbit? & \url{https://www.reddit.com/r/codereview/comments/1oqs5h5/alternative_to_coderabbit/} \\
G1119 & 3353 & How a small GitHub project turned out to be the best code review tool I've found lately & \url{https://www.reddit.com/r/github/comments/1tq23mw/how_a_small_github_project_turned_out_to_be_the/} \\
G1120 & 3356 & I updated Dracula-AI based on some advice and criticism. You can see here what changed. & \url{https://www.reddit.com/r/Python/comments/1rk655z/i_updated_draculaai_based_on_some_advice_and/} \\
G1121 & 3357 & Copilot is amazing at typing. It's bad at reading your mind. Specs fixed my workflow. & \url{https://www.reddit.com/r/GithubCopilot/comments/1rbnb3a/copilot_is_amazing_at_typing_its_bad_at_reading/} \\
G1122 & 3360 & Ideas on how to improve my current development workflow & \url{https://www.reddit.com/r/vibecoding/comments/1t7zdpd/ideas_on_how_to_improve_my_current_development/} \\
G1123 & 3364 & I replaced my entire automation stack with MCP servers and local LLMs. Here's what actually works and what doesn't. & \url{https://www.reddit.com/r/LocalLLaMA/comments/1rhsto2/i_replaced_my_entire_automation_stack_with_mcp/} \\
G1124 & 3372 & Spending 60\% of my time on code reviews instead of actually building things & \url{https://www.reddit.com/r/cscareerquestions/comments/1n3gqj6/spending_60_of_my_time_on_code_reviews_instead_of/} \\
G1125 & 3373 & What's your team strategy to not get stuck in PR hell? & \url{https://www.reddit.com/r/ChatGPTCoding/comments/1qqmikm/whats_your_team_strategy_to_not_get_stuck_in_pr/} \\
G1126 & 3376 & I stopped using Claude for 80\% of my coding tasks. Here's what I use instead. & \url{https://www.reddit.com/r/LocalLLaMA/comments/1seolpm/i_stopped_using_claude_for_80_of_my_coding_tasks/} \\
G1127 & 3378 & How are you managing the code review bottleneck when junior devs push unverified AI slop? & \url{https://www.reddit.com/r/GithubCopilot/comments/1tqafw8/how_are_you_managing_the_code_review_bottleneck/} \\
G1128 & 3379 & Claude Code now supports Custom Agents & \url{https://www.reddit.com/r/ClaudeAI/comments/1m8ik5l/claude_code_now_supports_custom_agents/} \\
G1129 & 3382 & Claude Code Max x20 responses/code generation human overload & \url{https://www.reddit.com/r/ClaudeAI/comments/1laim44/claude_code_max_x20_responsescode_generation/} \\
G1130 & 3386 & Someone gave AI agents personalities and now my QA tester refuses to approve anything & \url{https://www.reddit.com/r/vibecoding/comments/1rl8xp0/someone_gave_ai_agents_personalities_and_now_my/} \\
G1131 & 3388 & 7yoe at 1 job - feel useless & \url{https://www.reddit.com/r/ExperiencedDevs/comments/1r2oj8f/7yoe_at_1_job_feel_useless/} \\
G1132 & 3389 & How are you using AI outside your IDE? & \url{https://www.reddit.com/r/ChatGPTCoding/comments/1ilg6x2/how_are_you_using_ai_outside_your_ide/} \\
G1133 & 3395 & AI Agent, Claude CLI and Linear, all working together. & \url{https://www.reddit.com/r/ArtificialInteligence/comments/1txf5sn/ai_agent_claude_cli_and_linear_all_working/} \\
G1134 & 3396 & How often before a deployment is common for a code review & \url{https://www.reddit.com/r/cscareerquestions/comments/1lz7085/how_often_before_a_deployment_is_common_for_a/} \\
G1135 & 3397 & 8 hours of supervised vibe coding with Google Antigravity: Why CLI tools work better with my long-established tech stack & \url{https://www.reddit.com/r/vibecoding/comments/1syv3ug/8_hours_of_supervised_vibe_coding_with_google/} \\
G1136 & 3400 & Simone v0.3.5 Released - Community Contributions Finally Merged & \url{https://www.reddit.com/r/ClaudeAI/comments/1lwmhvj/simone_v035_released_community_contributions/} \\
G1137 & 3401 & Question about subagents & \url{https://www.reddit.com/r/ClaudeAI/comments/1ph4zvf/question_about_subagents/} \\
G1138 & 3406 & A random rant of a software engineer after reviewing a PR & \url{https://www.reddit.com/r/SoftwareEngineering/comments/1kgn36t/a_random_rant_of_a_software_engineer_after/} \\
G1139 & 3407 & Is this a cool personal project or am I overengineering a non-problem? & \url{https://www.reddit.com/r/github/comments/1qy478x/is_this_a_cool_personal_project_or_am_i/} \\
G1140 & 3410 & KanbanCode: macOS native UI for managing Claude Codes & \url{https://www.reddit.com/r/vibecoding/comments/1rlqb1e/kanbancode_macos_native_ui_for_managing_claude/} \\
G1141 & 3411 & Looking for AI coding assistant with Copilot-like transparency (commands, diffs, terminal output) & \url{https://www.reddit.com/r/vibecoding/comments/1t3pvcr/looking_for_ai_coding_assistant_with_copilotlike/} \\
G1142 & 3413 & Code review request for Sign Language Recognition system using Mediapipe and TensorFlow & \url{https://www.reddit.com/r/codereview/comments/1ix3go4/code_review_request_for_sign_language_recognition/} \\
G1143 & 3415 & Is a large, mechanical PR better than a smaller, more complex one? & \url{https://www.reddit.com/r/ExperiencedDevs/comments/1sc6l3y/is_a_large_mechanical_pr_better_than_a_smaller/} \\
G1144 & 3416 & the most productive thing i do every morning is read yesterday's merged PRs & \url{https://www.reddit.com/r/ExperiencedDevs/comments/1tivlma/the_most_productive_thing_i_do_every_morning_is/} \\
G1145 & 3418 & I turned years of programming \& IT absurdities into a satirical "dictionary" -- AMA (indie published, out Sept 30, here till Oct 5) & \url{https://www.reddit.com/r/programming/comments/1nsuqdw/i_turned_years_of_programming_it_absurdities_into/} \\
G1146 & 3423 & GPT-5 has been surprisingly good at reviewing Claude Code's work & \url{https://www.reddit.com/r/ClaudeAI/comments/1mvbxaw/gpt5_has_been_surprisingly_good_at_reviewing/} \\
G1147 & 3425 & Freigeist - The new Vibe Coding Platform & \url{https://www.reddit.com/r/vibecoding/comments/1m5tlpq/freigeist_the_new_vibe_coding_platform/} \\
G1148 & 3426 & Antigravity vs Cursor & \url{https://www.reddit.com/r/vibecoding/comments/1r4dx7u/antigravity_vs_cursor/} \\
G1149 & 3436 & I automated most of my job & \url{https://www.reddit.com/r/ClaudeAI/comments/1shngqm/i_automated_most_of_my_job/} \\
G1150 & 3437 & Here are the AI code review tools I've been looking at, which one is best? & \url{https://www.reddit.com/r/webdev/comments/1quz7wh/here_are_the_ai_code_review_tools_ive_been/} \\
G1151 & 3440 & Built a full-stack multiplayer app with Claude Code -- lessons on tooling and workflow & \url{https://www.reddit.com/r/ClaudeAI/comments/1psjbwq/built_a_fullstack_multiplayer_app_with_claude/} \\
G1152 & 3442 & AI is flooding codebases, and most teams aren't reviewing it before deploy & \url{https://www.reddit.com/r/devops/comments/1lecppz/ai_is_flooding_codebases_and_most_teams_arent/} \\
G1153 & 3445 & Are You Missing Out by Dismissing AI Code as "Slop"? & \url{https://www.reddit.com/r/AI_Agents/comments/1q5hd6z/are_you_missing_out_by_dismissing_ai_code_as_slop/} \\
G1154 & 3450 & At what team size does manual code review stop scaling? & \url{https://www.reddit.com/r/ExperiencedDevs/comments/1r75p6m/at_what_team_size_does_manual_code_review_stop/} \\
G1155 & 3454 & code became hard to review & \url{https://www.reddit.com/r/cscareerquestions/comments/1rsiatc/code_became_hard_to_review/} \\
G1156 & 3457 & How to enable Github copilot code review? & \url{https://www.reddit.com/r/github/comments/1juz7yb/how_to_enable_github_copilot_code_review/} \\
G1157 & 3474 & Go Pooling Strategies: sync.Pool vs Generics vs ResettablePool -- Benchmarks and Takeaways & \url{https://www.reddit.com/r/golang/comments/1p8n0y2/go_pooling_strategies_syncpool_vs_generics_vs/} \\
G1158 & 3475 & ai-rulez: universal agent context manager & \url{https://www.reddit.com/r/ClaudeAI/comments/1q2r26g/airulez_universal_agent_context_manager/} \\
G1159 & 3479 & Non-coders Coding With Claude - A Case Report & \url{https://www.reddit.com/r/ClaudeAI/comments/1jhuic6/noncoders_coding_with_claude_a_case_report/} \\
G1160 & 3482 & Can ai code review tools actually catch meaningful logic errors or just pattern match & \url{https://www.reddit.com/r/ExperiencedDevs/comments/1rjpjox/can_ai_code_review_tools_actually_catch/} \\
G1161 & 3483 & Vibe coding is fun until your own code becomes a black box & \url{https://www.reddit.com/r/vibecoding/comments/1sbim2o/vibe_coding_is_fun_until_your_own_code_becomes_a/} \\
G1162 & 3485 & Does Go error handling verbosity actually hurt developer velocity or is it just endless debate & \url{https://www.reddit.com/r/golang/comments/1rqvqsk/does_go_error_handling_verbosity_actually_hurt/} \\
G1163 & 3489 & A desktop app around Claude Code to orchestrate agents and review their output -- looking for feedback & \url{https://www.reddit.com/r/ClaudeAI/comments/1r64o3m/a_desktop_app_around_claude_code_to_orchestrate/} \\
G1164 & 3495 & I built a 20K+ line production SaaS platform entirely through Claude Code -- here's what I learned about agentic engineering at scale & \url{https://www.reddit.com/r/ClaudeAI/comments/1rg0346/i_built_a_20k_line_production_saas_platform/} \\
G1165 & 3497 & Someone built an open-source multi-market trading terminal with Claude -- Neuberg, here's what worked (and what didn't) & \url{https://www.reddit.com/r/ClaudeAI/comments/1sb2ne3/someone_built_an_opensource_multimarket_trading/} \\
G1166 & 3498 & What unique GitHub features have significantly improved your collaboration with remote teams? & \url{https://www.reddit.com/r/github/comments/1r7f75c/what_unique_github_features_have_significantly/} \\
G1167 & 3499 & Integrating claude code with other popular agentic cli for a consensus & \url{https://www.reddit.com/r/ClaudeAI/comments/1pt9p9f/integrating_claude_code_with_other_popular/} \\
G1168 & 3503 & Kanban-style Phase Board: plan -> execute -> verify -> commit & \url{https://www.reddit.com/r/ClaudeAI/comments/1m74ot4/kanbanstyle_phase_board_plan_execute_verify_commit/} \\
G1169 & 3505 & Remote Rust Code Review Job - \$85/hr (Short-Term AI Project) & \url{https://www.reddit.com/r/rust/comments/1nhudkm/remote_rust_code_review_job_85hr_shortterm_ai/} \\
G1170 & 3512 & I built a validation framework to stop AI coding tools from breaking my code [MAID Runner] & \url{https://www.reddit.com/r/ClaudeAI/comments/1p7rqai/i_built_a_validation_framework_to_stop_ai_coding/} \\
G1171 & 3516 & built a tool to understand AI-generated codebases (because I stopped understanding my own) & \url{https://www.reddit.com/r/vibecoding/comments/1sevpwd/built_a_tool_to_understand_aigenerated_codebases/} \\
G1172 & 3517 & Turned Claude Code into a self-aware Software Engineering Partner (dead simple repo) & \url{https://www.reddit.com/r/ClaudeAI/comments/1lbmo4k/turned_claude_code_into_a_selfaware_software/} \\
G1173 & 3523 & Best Dev tool for Claude CLI. Let See what Agent implemented. Visualize it! & \url{https://www.reddit.com/r/ClaudeAI/comments/1p23msb/best_dev_tool_for_claude_cli_let_see_what_agent/} \\
G1174 & 3524 & Human Code Reviews are Dead & \url{https://www.reddit.com/r/webdev/comments/1tgdzb8/human_code_reviews_are_dead/} \\
G1175 & 3525 & Learning Go, 3rd Edition is in Early Release & \url{https://www.reddit.com/r/golang/comments/1tu5xlu/learning_go_3rd_edition_is_in_early_release/} \\
G1176 & 3528 & The joy is gone & \url{https://www.reddit.com/r/cscareerquestions/comments/1mtpq5a/the_joy_is_gone/} \\
G1177 & 3529 & After a year of coding with AI, my projects kept turning into spaghetti -- so I built a workflow to make AI code like an actual engineer. (Open-sourced) & \url{https://www.reddit.com/r/ExperiencedDevs/comments/1qp8w1b/after_a_year_of_coding_with_ai_my_projects_kept/} \\
G1178 & 3531 & I built an open-source tool that auto-reviews every PR/MR using Claude Code CLI -- zero config & \url{https://www.reddit.com/r/ClaudeAI/comments/1rtetb5/i_built_an_opensource_tool_that_autoreviews_every/} \\
G1179 & 3532 & Claude Code backslash command for instant project context & \url{https://www.reddit.com/r/ClaudeAI/comments/1ngo2fh/claude_code_backslash_command_for_instant_project/} \\
G1180 & 3536 & Claude Code - large PRs? & \url{https://www.reddit.com/r/ClaudeAI/comments/1lmirwy/claude_code_large_prs/} \\
G1181 & 3539 & Setting up a baseline dev toolchain for our small remote team, what stuck for you? & \url{https://www.reddit.com/r/ExperiencedDevs/comments/1lqvo3y/setting_up_a_baseline_dev_toolchain_for_our_small/} \\
G1182 & 3542 & Can a non-coder + Claude realistically build a validated internal ops system for a regulated facility? & \url{https://www.reddit.com/r/ClaudeAI/comments/1svka2y/can_a_noncoder_claude_realistically_build_a/} \\
G1183 & 3544 & Anyone interested in code review / help? & \url{https://www.reddit.com/r/vibecoding/comments/1nhnm9t/anyone_interested_in_code_review_help/} \\
G1184 & 3547 & Views on this 50X token reduction trick? & \url{https://www.reddit.com/r/ClaudeAI/comments/1sc007v/views_on_this_50x_token_reduction_trick/} \\
G1185 & 3550 & IP Memorandum: Multi-Agent ("Agentic") AI Systems in Coding, Marketing, and Creation - Comprehensive 2026 Analysis. (Integrating Patentability, Hype vs. Reality & \url{https://www.reddit.com/r/AI_Agents/comments/1tu9dm7/ip_memorandum_multiagent_agentic_ai_systems_in/} \\
G1186 & 3551 & Need help for AI code reviewer logic & \url{https://www.reddit.com/r/agile/comments/1i7ik7o/need_help_for_ai_code_reviewer_logic/} \\
G1187 & 3555 & These Claude Code Skills will 100x your shipping speed & \url{https://www.reddit.com/r/programming/comments/1r9bvxt/these_claude_code_skills_will_100x_your_shipping/} \\
G1188 & 3563 & Usage Limits After Using Claude Max x20 for 6 Days & \url{https://www.reddit.com/r/ClaudeAI/comments/1qx2ujc/usage_limits_after_using_claude_max_x20_for_6_days/} \\
G1189 & 3565 & Daily AI Archive | 10/6/2025 & \url{https://www.reddit.com/r/accelerate/comments/1o01e9b/daily_ai_archive_1062025/} \\
G1190 & 3570 & Self-taught programmer looking for a code review & \url{https://www.reddit.com/r/codereview/comments/1m98syd/selftaught_programmer_looking_for_a_code_review/} \\
G1191 & 3572 & what the fuck i didn't even use it & \url{https://www.reddit.com/r/GithubCopilot/comments/1tvore2/what_the_fuck_i_didnt_even_use_it/} \\
G1192 & 3580 & SWE feeling stuck and undervalued - seeking advice on next steps & \url{https://www.reddit.com/r/ExperiencedDevs/comments/1kg35rc/swe_feeling_stuck_and_undervalued_seeking_advice/} \\
G1193 & 3586 & If an AI agent opened a PR for you, what would you want to see first? & \url{https://www.reddit.com/r/AI_Agents/comments/1tg4zqu/if_an_ai_agent_opened_a_pr_for_you_what_would_you/} \\
G1194 & 3596 & What is our future when cloudbase uses LLM to build a code based? & \url{https://www.reddit.com/r/ExperiencedDevs/comments/1l25rg3/what_is_our_future_when_cloudbase_uses_llm_to/} \\
G1195 & 3599 & How are you handling architecture drift with the rise of AI coding assistants? & \url{https://www.reddit.com/r/vibecoding/comments/1rc1kig/how_are_you_handling_architecture_drift_with_the/} \\
G1196 & 3600 & What does large context window in LLM mean for future of devs? & \url{https://www.reddit.com/r/ExperiencedDevs/comments/1jwhsa9/what_does_large_context_window_in_llm_mean_for/} \\
G1197 & 3604 & Built something on Replit/Base44 but stuck? Senior dev here offering help (code review, architecture, security) & \url{https://www.reddit.com/r/vibecoding/comments/1s9mx86/built_something_on_replitbase44_but_stuck_senior/} \\
G1198 & 3606 & AI Agent built GitHub code reviewer + auto-documentation checker in 45 lines of Python. This is insane. & \url{https://www.reddit.com/r/ArtificialInteligence/comments/1l5lekp/ai_agent_built_github_code_reviewer/} \\
G1199 & 3608 & OpenAI just supercharged Codex -- but is it too much too fast? & \url{https://www.reddit.com/r/AI_Agents/comments/1qu4fem/openai_just_supercharged_codex_but_is_it_too_much/} \\
G1200 & 3611 & Manager wants me to review diffs to find test cases / find defects & \url{https://www.reddit.com/r/QualityAssurance/comments/1n1dokp/manager_wants_me_to_review_diffs_to_find_test/} \\
G1201 & 3612 & crit.md - review plans \& code like you do PRs & \url{https://www.reddit.com/r/webdev/comments/1t1t1nn/critmd_review_plans_code_like_you_do_prs/} \\
G1202 & 3620 & I ask Claude to fix code, then I ask Gemini to review the code & \url{https://www.reddit.com/r/ClaudeAI/comments/1slrehy/i_ask_claude_to_fix_code_then_i_ask_gemini_to/} \\
G1203 & 3622 & What are ways to save tokens? & \url{https://www.reddit.com/r/GithubCopilot/comments/1sz85xf/what_are_ways_to_save_tokens/} \\
G1204 & 3623 & I have been diligently doing my part to prevent layoffs in my org & \url{https://www.reddit.com/r/cscareerquestions/comments/1rtaaqd/i_have_been_diligently_doing_my_part_to_prevent/} \\
G1205 & 3624 & Would appreciate advice on handling delays from a teammate & \url{https://www.reddit.com/r/ExperiencedDevs/comments/1lh45lc/would_appreciate_advice_on_handling_delays_from_a/} \\
G1206 & 3626 & The best AI tools in 2026 are not always the most hyped. Here's what I'd actually use & \url{https://www.reddit.com/r/AI_Agents/comments/1tr9u9m/the_best_ai_tools_in_2026_are_not_always_the_most/} \\
G1207 & 3628 & [Plugin] Orchestra v1.0 -- Doc-driven discipline for Claude Code (typed templates, 4-gate review, commit gates) & \url{https://www.reddit.com/r/ClaudeAI/comments/1t56o1x/plugin_orchestra_v10_docdriven_discipline_for/} \\
G1208 & 3634 & What's your totally absolutely 100\% free vibecoding workflow? & \url{https://www.reddit.com/r/vibecoding/comments/1nzq562/whats_your_totally_absolutely_100_free_vibecoding/} \\
G1209 & 3635 & Cursor and CC Mutual Challenge System & \url{https://www.reddit.com/r/cursor/comments/1naxmga/cursor_and_cc_mutual_challenge_system/} \\
G1210 & 3636 & Do you still look at the code your AI coding agent produces & \url{https://www.reddit.com/r/AI_Agents/comments/1sxlpcy/do_you_still_look_at_the_code_your_ai_coding/} \\
G1211 & 3638 & This week Ai has killed one more thing, my passion to mentor interns & \url{https://www.reddit.com/r/ExperiencedDevs/comments/1rytfxu/this_week_ai_has_killed_one_more_thing_my_passion/} \\
G1212 & 3641 & The Overton Window for Code Review Is Shifting & \url{https://www.reddit.com/r/programming/comments/1scb00h/the_overton_window_for_code_review_is_shifting/} \\
G1213 & 3643 & Reviewers & \url{https://www.reddit.com/r/vibecoding/comments/1t84a7n/reviewers/} \\
G1214 & 3644 & My local AI agents started filing "Grievances" against me for my project goals, so I had to build them a HR module. & \url{https://www.reddit.com/r/AI_Agents/comments/1t4ptx7/my_local_ai_agents_started_filing_grievances/} \\
G1215 & 3650 & Unit Tests in a PHP ERP system? & \url{https://www.reddit.com/r/webdev/comments/1jvung5/unit_tests_in_a_php_erp_system/} \\
G1216 & 3651 & Nobody's built Cursor for databases because there was no Git for data. We fixed both. & \url{https://www.reddit.com/r/ClaudeAI/comments/1r27ngn/nobodys_built_cursor_for_databases_because_there/} \\
G1217 & 3652 & I think I hate SDD & \url{https://www.reddit.com/r/AskProgramming/comments/1rz77j6/i_think_i_hate_sdd/} \\
G1218 & 3654 & Is GitHub Copilot actually improving your productivity or just making you lazy? & \url{https://www.reddit.com/r/GithubCopilot/comments/1skcxd3/is_github_copilot_actually_improving_your/} \\
G1219 & 3655 & I Pushed AI code on AWS US - East that Caused the Internet to Break. How Do I Hide It & \url{https://www.reddit.com/r/cscareerquestions/comments/1obi616/i_pushed_ai_code_on_aws_us_east_that_caused_the/} \\
G1220 & 3656 & What is your biggest pain point & \url{https://www.reddit.com/r/devops/comments/1qx4eh5/what_is_your_biggest_pain_point/} \\
G1221 & 3659 & Actual Ignorance & \url{https://www.reddit.com/r/webdev/comments/1tc0igj/actual_ignorance/} \\
G1222 & 3666 & lgtm - open source AI powered code review companion & \url{https://www.reddit.com/r/Python/comments/1l2i7iv/lgtm_open_source_ai_powered_code_review_companion/} \\
G1223 & 3668 & How to deal with team members having rigid opinions & \url{https://www.reddit.com/r/SoftwareEngineering/comments/1s5vi48/how_to_deal_with_team_members_having_rigid/} \\
G1224 & 3674 & What exactly is vibe coding? & \url{https://www.reddit.com/r/vibecoding/comments/1l258y9/what_exactly_is_vibe_coding/} \\
G1225 & 3676 & Updated my community Copilot agents to use skills and other improvements & \url{https://www.reddit.com/r/GithubCopilot/comments/1ps66bb/updated_my_community_copilot_agents_to_use_skills/} \\
G1226 & 3681 & Best Openclaw Skills You Should Install (From ClawHub's 500+ Skills) & \url{https://www.reddit.com/r/AI_Agents/comments/1r2u356/best_openclaw_skills_you_should_install_from/} \\
G1227 & 3684 & How I fixed a stubborn code issue in 30 mins after wasting hours with Claude code & \url{https://www.reddit.com/r/vibecoding/comments/1mpzj9y/how_i_fixed_a_stubborn_code_issue_in_30_mins/} \\
G1228 & 3685 & I built a free GUI that makes Claude Code easier to use & \url{https://www.reddit.com/r/ClaudeAI/comments/1n7r558/i_built_a_free_gui_that_makes_claude_code_easier/} \\
G1229 & 3686 & 6 months with Claude Code to ship a SaaS -- here's what actually happened & \url{https://www.reddit.com/r/vibecoding/comments/1stl8dn/6_months_with_claude_code_to_ship_a_saas_heres/} \\
G1230 & 3690 & I asked AI to build a secure backend. & \url{https://www.reddit.com/r/vibecoding/comments/1rlhtry/i_asked_ai_to_build_a_secure_backend/} \\
G1231 & 3691 & Things I used to be proud of doing well - Modern AI just does better & \url{https://www.reddit.com/r/ExperiencedDevs/comments/1tbz4jc/things_i_used_to_be_proud_of_doing_well_modern_ai/} \\
G1232 & 3693 & is there any free ai code Reviewer available in vscode & \url{https://www.reddit.com/r/golang/comments/1r8135b/is_there_any_free_ai_code_reviewer_available_in/} \\
G1233 & 3694 & We built an AI to review your pull requests & \url{https://www.reddit.com/r/devops/comments/1kwzb8f/we_built_an_ai_to_review_your_pull_requests/} \\
G1234 & 3703 & This is why I stay away from LinkedIn, did people not learn from Claude Code's leak yesterday? Absolutely delirious. & \url{https://www.reddit.com/r/vibecoding/comments/1s9hjna/this_is_why_i_stay_away_from_linkedin_did_people/} \\
G1235 & 3704 & Are we ready? The Cliff Edge next step of allowing AI to rewrite its own code. & \url{https://www.reddit.com/r/OpenAI/comments/1kvupvi/are_we_ready_the_cliff_edge_next_step_of_allowing/} \\
G1236 & 3705 & Now What? & \url{https://www.reddit.com/r/vibecoding/comments/1qvjxik/now_what/} \\
G1237 & 3708 & What metrics actually matter for code quality? feeling like we're measuring the wrong things & \url{https://www.reddit.com/r/codereview/comments/1ret3iz/what_metrics_actually_matter_for_code_quality/} \\
G1238 & 3710 & Will moving repo into Organization break connection to local files? & \url{https://www.reddit.com/r/github/comments/1linssl/will_moving_repo_into_organization_break/} \\
G1239 & 3711 & Have this idea to reinvent what 'no code' means... & \url{https://www.reddit.com/r/ClaudeAI/comments/1m25xn1/have_this_idea_to_reinvent_what_no_code_means/} \\
G1240 & 3713 & What tools do you use to keep your codebase clean when working with a team? & \url{https://www.reddit.com/r/webdev/comments/1nsn6ga/what_tools_do_you_use_to_keep_your_codebase_clean/} \\
G1241 & 3718 & "Vibe" coders and anyone else saying Claude (Code or web) isn't good & \url{https://www.reddit.com/r/ClaudeAI/comments/1n4djcx/vibe_coders_and_anyone_else_saying_claude_code_or/} \\
G1242 & 3719 & Asked my team what it would take to trust an AI-generated PR....silence & \url{https://www.reddit.com/r/vibecoding/comments/1tofvl8/asked_my_team_what_it_would_take_to_trust_an/} \\
G1243 & 3727 & Has Claude figured out im the legacy code? & \url{https://www.reddit.com/r/ClaudeAI/comments/1thfyuy/has_claude_figured_out_im_the_legacy_code/} \\
G1244 & 3728 & 6 months building Cost Share with Claude -- shipped to App Store. Here's what AI got right and where I had to step in. & \url{https://www.reddit.com/r/vibecoding/comments/1t6pgxi/6_months_building_cost_share_with_claude_shipped/} \\
G1245 & 3732 & Has anyone written a Claude Desktop extension for Claude Code? & \url{https://www.reddit.com/r/ClaudeAI/comments/1sey6hh/has_anyone_written_a_claude_desktop_extension_for/} \\
G1246 & 3734 & I will share my production vibecoding workflow for produciton iOS apps in a live youtube stream. Fell free to join. & \url{https://www.reddit.com/r/vibecoding/comments/1qtzpmg/i_will_share_my_production_vibecoding_workflow/} \\
G1247 & 3735 & Has anyone here tried using AI tools to assist with code reviews? Curious what your experience was. & \url{https://www.reddit.com/r/codereview/comments/1oqrbkc/has_anyone_here_tried_using_ai_tools_to_assist/} \\
G1248 & 3737 & How is your team managing comprehension of AI-generated code? & \url{https://www.reddit.com/r/ExperiencedDevs/comments/1ra0iqv/how_is_your_team_managing_comprehension_of/} \\
G1249 & 3739 & I scanned 10 popular vibe-coded repos with a deterministic linter. 4,513 findings across 2,062 files. Here's what AI agents keep getting wrong. & \url{https://www.reddit.com/r/OpenAI/comments/1sas9s3/i_scanned_10_popular_vibecoded_repos_with_a/} \\
G1250 & 3744 & Vibe coding is not brain rot... or hold on a second... well it is complicated! & \url{https://www.reddit.com/r/vibecoding/comments/1rca0l8/vibe_coding_is_not_brain_rot_or_hold_on_a_second/} \\
G1251 & 3745 & GPD5 is the craziest AI model I've seen yet. Here's what it can do & \url{https://www.reddit.com/r/OpenAI/comments/1mk60ev/gpd5_is_the_craziest_ai_model_ive_seen_yet_heres/} \\
G1252 & 3747 & Code reviews do find bugs & \url{https://www.reddit.com/r/programming/comments/1rqywa9/code_reviews_do_find_bugs/} \\
G1253 & 3748 & Measuring AI coding agent productivity with line-level Git attribution & \url{https://www.reddit.com/r/programming/comments/1qdwvvs/measuring_ai_coding_agent_productivity_with/} \\
G1254 & 3750 & I asked LLM Council to fix my Claude Code burnout...the answer wasn't what I expected & \url{https://www.reddit.com/r/ClaudeAI/comments/1sae95e/i_asked_llm_council_to_fix_my_claude_code/} \\
G1255 & 3751 & Is it better to say you use more AI or less when applying for a position? & \url{https://www.reddit.com/r/cscareerquestions/comments/1sh00ct/is_it_better_to_say_you_use_more_ai_or_less_when/} \\
G1256 & 3753 & How can I let devs update their lower environment terraform while protecting production environments? & \url{https://www.reddit.com/r/devops/comments/1kh15h9/how_can_i_let_devs_update_their_lower_environment/} \\
G1257 & 3755 & Is Claude's Github Integration worth it on a Pro subscription? & \url{https://www.reddit.com/r/ClaudeAI/comments/1qac0f4/is_claudes_github_integration_worth_it_on_a_pro/} \\
G1258 & 3762 & Why I stopped using Claude to review my Claude-generated code (Defense in Depth Part 2) & \url{https://www.reddit.com/r/ClaudeAI/comments/1stu8p4/why_i_stopped_using_claude_to_review_my/} \\
G1259 & 3764 & How do you drive improvement in teams that are resistant to change? & \url{https://www.reddit.com/r/ExperiencedDevs/comments/1kts7ij/how_do_you_drive_improvement_in_teams_that_are/} \\
G1260 & 3767 & Claude Code Review just launched. Here's how it works and what a multi-model approach does differently. & \url{https://www.reddit.com/r/ClaudeAI/comments/1rpid3t/claude_code_review_just_launched_heres_how_it/} \\
G1261 & 3769 & Is it just me, or is my tech lead the worst? & \url{https://www.reddit.com/r/ExperiencedDevs/comments/1kiym57/is_it_just_me_or_is_my_tech_lead_the_worst/} \\
G1262 & 3773 & This AI doesn't just review your code it understands your entire codebase & \url{https://www.reddit.com/r/codereview/comments/1krp3m4/this_ai_doesnt_just_review_your_code_it/} \\
G1263 & 3778 & Finally gave up on open source code review tooling and went enterprise. & \url{https://www.reddit.com/r/devops/comments/1q7gau9/finally_gave_up_on_open_source_code_review/} \\
G1264 & 3780 & GitLab MR Conform v0.3.0 - CODEOWNERS support & \url{https://www.reddit.com/r/gitlab/comments/1loyc3b/gitlab_mr_conform_v030_codeowners_support/} \\
G1265 & 3784 & Using multiple AI models together was a game changer for my side projects & \url{https://www.reddit.com/r/vibecoding/comments/1q2lr9a/using_multiple_ai_models_together_was_a_game/} \\
G1266 & 3785 & No code reviews by default & \url{https://www.reddit.com/r/agile/comments/1soxrvy/no_code_reviews_by_default/} \\
G1267 & 3789 & I run a paper-trading bot where Claude Opus is the Lead Engineer with veto power over a Gemini "Strategist." 270+ entry audit log of every disagreement. Sharing & \url{https://www.reddit.com/r/ClaudeAI/comments/1t06v2n/i_run_a_papertrading_bot_where_claude_opus_is_the/} \\
G1268 & 3790 & The Top Ten Human Programming Blunders AI Could Have Prevented & \url{https://www.reddit.com/r/vibecoding/comments/1tcvuwk/the_top_ten_human_programming_blunders_ai_could/} \\
G1269 & 3792 & Are you an engineer worried about brain rot from excessive agentic vibe coding? & \url{https://www.reddit.com/r/ClaudeAI/comments/1p8t9wi/are_you_an_engineer_worried_about_brain_rot_from/} \\
G1270 & 3796 & Escaping the 'multiple 1 YOE' trap? & \url{https://www.reddit.com/r/cscareerquestions/comments/1r36l1e/escaping_the_multiple_1_yoe_trap/} \\
G1271 & 3804 & How do I know if I'm one of the good or bad users of AI? & \url{https://www.reddit.com/r/cscareerquestions/comments/1oafa21/how_do_i_know_if_im_one_of_the_good_or_bad_users/} \\
G1272 & 3809 & What are your favorite MCP server setups for Cursor? & \url{https://www.reddit.com/r/cursor/comments/1l315gt/what_are_your_favorite_mcp_server_setups_for/} \\
G1273 & 3812 & Security concerns with deploying a website built in Cursor -- can it be trusted? & \url{https://www.reddit.com/r/cursor/comments/1s251za/security_concerns_with_deploying_a_website_built/} \\
G1274 & 3816 & Code Review Rotation Tool - Looking for Real-World Validation & \url{https://www.reddit.com/r/gitlab/comments/1j6vjt8/code_review_rotation_tool_looking_for_realworld/} \\
G1275 & 3819 & My personal top 5 AI coding tools & \url{https://www.reddit.com/r/ChatGPTCoding/comments/1o1gp7i/my_personal_top_5_ai_coding_tools/} \\
G1276 & 3820 & [Project] I built a privacy-first Data Cleaning engine using Polars LazyFrame and FAISS. 100\% Local & \url{https://www.reddit.com/r/Python/comments/1pqmfku/project_i_built_a_privacyfirst_data_cleaning/} \\
G1277 & 3823 & I built a community crowdsourced LLM benchmark leaderboard (Claude Sonnet/Opus, Gemini, Grok, GPT-5, o3) & \url{https://www.reddit.com/r/ChatGPTPro/comments/1o3h17z/i_built_a_community_crowdsourced_llm_benchmark/} \\
G1278 & 3825 & ran into a principal engineer at a Series C fintech (\$200m+ valuation) who claims cursor + claude is making their 6-person team outship their 25-person competit & \url{https://www.reddit.com/r/vibecoding/comments/1pd2tzz/ran_into_a_principal_engineer_at_a_series_c/} \\
G1279 & 3827 & AI coding agents are changing software work, but maybe not in the way people expected & \url{https://www.reddit.com/r/ArtificialInteligence/comments/1t5epnj/ai_coding_agents_are_changing_software_work_but/} \\
G1280 & 3829 & C++ Podcasts \& Conference Talks (week 5, 2025) & \url{https://www.reddit.com/r/cpp/comments/1qq2083/c_podcasts_conference_talks_week_5_2025/} \\
G1281 & 3837 & [Discussion] Do AI coding agents say "done" too early for you too? & \url{https://www.reddit.com/r/AI_Agents/comments/1th85xa/discussion_do_ai_coding_agents_say_done_too_early/} \\
G1282 & 3840 & I built PromptLedger, a local-first prompt versioning and review tool in Python & \url{https://www.reddit.com/r/Python/comments/1s6tghx/i_built_promptledger_a_localfirst_prompt/} \\
G1283 & 3841 & Signs of a toxic work environment? & \url{https://www.reddit.com/r/ExperiencedDevs/comments/1oznl8o/signs_of_a_toxic_work_environment/} \\
G1284 & 3845 & [Media] Corust - A collaborative Rust Playground & \url{https://www.reddit.com/r/rust/comments/1k3au5e/media_corust_a_collaborative_rust_playground/} \\
G1285 & 3849 & Well, some apps for you & \url{https://www.reddit.com/r/vibecoding/comments/1s635ij/well_some_apps_for_you/} \\
G1286 & 3851 & I Built 6 Apps using AI in 3 Months. Here's What Actually Works (And What's Complete BS) & \url{https://www.reddit.com/r/cursor/comments/1phgj1o/i_built_6_apps_using_ai_in_3_months_heres_what/} \\
G1287 & 3852 & Small SDL Game Engine Review Request & \url{https://www.reddit.com/r/codereview/comments/1ruy4kx/small_sdl_game_engine_review_request/} \\
G1288 & 3853 & Vibe coding needs skills, not just prompts! & \url{https://www.reddit.com/r/vibecoding/comments/1ttscim/vibe_coding_needs_skills_not_just_prompts/} \\
G1289 & 3855 & intern pushed 847 commits this morning & \url{https://www.reddit.com/r/cursor/comments/1t6he6m/intern_pushed_847_commits_this_morning/} \\
G1290 & 3858 & Software engineering is a team sport & \url{https://www.reddit.com/r/ExperiencedDevs/comments/1thqfpb/software_engineering_is_a_team_sport/} \\
G1291 & 3862 & How do we fix the inversion of QA, where the most senior people wind up doing all the QA? Is this normal? & \url{https://www.reddit.com/r/ExperiencedDevs/comments/1onoa4w/how_do_we_fix_the_inversion_of_qa_where_the_most/} \\
G1292 & 3863 & From early vibe coding experiments to launching PicKeep -- lessons learned building an iOS app with Copilot \& Claude & \url{https://www.reddit.com/r/vibecoding/comments/1q6fvv0/from_early_vibe_coding_experiments_to_launching/} \\
G1293 & 3867 & Codebase review & \url{https://www.reddit.com/r/cursor/comments/1oqeusx/codebase_review/} \\
G1294 & 3871 & Logic Issues in My MCQ Simulation Project - Looking for Code Review & \url{https://www.reddit.com/r/AskProgramming/comments/1rks5qh/logic_issues_in_my_mcq_simulation_project_looking/} \\
G1295 & 3873 & Using AI as a security coach in workflows & \url{https://www.reddit.com/r/devops/comments/1m8j7z8/using_ai_as_a_security_coach_in_workflows/} \\
G1296 & 3874 & Which vibe coding tools allow user approval before EVERY code change? Looking for strict human-in-the-loop options & \url{https://www.reddit.com/r/ClaudeAI/comments/1pvpyxd/which_vibe_coding_tools_allow_user_approval/} \\
G1297 & 3876 & Tasks have radically increased my efficiency! & \url{https://www.reddit.com/r/ClaudeAI/comments/1qo4w00/tasks_have_radically_increased_my_efficiency/} \\
G1298 & 3881 & The bridge serves as a shared nervous system. Cowork, CLI, and Dispatch are like three different hands, all connected to the same nerve. Without it, each surfac & \url{https://www.reddit.com/r/ClaudeAI/comments/1rzlmzu/the_bridge_serves_as_a_shared_nervous_system/} \\
G1299 & 3884 & Interested in reviewing Vibe Code projects & \url{https://www.reddit.com/r/vibecoding/comments/1q27e36/interested_in_reviewing_vibe_code_projects/} \\
G1300 & 3888 & Are you drowning in AI code review noise? 70\% of AI PR comments are useless & \url{https://www.reddit.com/r/programming/comments/1okvtqp/are_you_drowning_in_ai_code_review_noise_70_of_ai/} \\
G1301 & 3889 & Show \& Tell: Built an LRU cache server in Go over winter break - feedback welcome & \url{https://www.reddit.com/r/golang/comments/1r2jr69/show_tell_built_an_lru_cache_server_in_go_over/} \\
G1302 & 3891 & These AI Assistants will get you fired from work & \url{https://www.reddit.com/r/ChatGPTCoding/comments/1ki6lxh/these_ai_assistants_will_get_you_fired_from_work/} \\
G1303 & 3892 & Built a tool that secures the code for vibe coders & \url{https://www.reddit.com/r/ChatGPTCoding/comments/1jqi64a/built_a_tool_that_secures_the_code_for_vibe_coders/} \\
G1304 & 3893 & Vibe Coding Pro Max? & \url{https://www.reddit.com/r/vibecoding/comments/1t0s03r/vibe_coding_pro_max/} \\
G1305 & 3894 & Colleague doesn't want to work at all & \url{https://www.reddit.com/r/ExperiencedDevs/comments/1kr4lmy/colleague_doesnt_want_to_work_at_all/} \\
G1306 & 3898 & code intelligence for 248 languages (made with agents in mind) and much more in Kreuzberg v4.7.0 & \url{https://www.reddit.com/r/AI_Agents/comments/1scvh5t/code_intelligence_for_248_languages_made_with/} \\
G1307 & 3904 & Why do people exceed limit of claude code? & \url{https://www.reddit.com/r/ExperiencedDevs/comments/1r2wmqp/why_do_people_exceed_limit_of_claude_code/} \\
G1308 & 3905 & Built AI Agents for PMs to Manage Jira Tasks - Thinking of Expanding to a Human-AI Collab Platform? & \url{https://www.reddit.com/r/AI_Agents/comments/1o5mmtb/built_ai_agents_for_pms_to_manage_jira_tasks/} \\
G1309 & 3906 & Code for SpaceHey profile & \url{https://www.reddit.com/r/codereview/comments/1q4fovw/code_for_spacehey_profile/} \\
G1310 & 3909 & I built a CLI that lets multiple Claude instances have structured discussions and debates - the results are surprisingly good & \url{https://www.reddit.com/r/ClaudeAI/comments/1n1s8lf/i_built_a_cli_that_lets_multiple_claude_instances/} \\
G1311 & 3910 & Workflow in 2026 - are sprints dead? & \url{https://www.reddit.com/r/vibecoding/comments/1ri31wy/workflow_in_2026_are_sprints_dead/} \\
G1312 & 3913 & Hot take: "vibe coding" is great for speed, but I think it's changing what it means to actually learn software development. & \url{https://www.reddit.com/r/cscareerquestions/comments/1ted0wg/hot_take_vibe_coding_is_great_for_speed_but_i/} \\
G1313 & 3917 & How do you cut through cultural clashes pragmatically? & \url{https://www.reddit.com/r/ExperiencedDevs/comments/1psbhq2/how_do_you_cut_through_cultural_clashes/} \\
G1314 & 3918 & Rant: Save me from lazy devs & \url{https://www.reddit.com/r/webdev/comments/1lfjokx/rant_save_me_from_lazy_devs/} \\
G1315 & 3922 & Fun and Effective Vibe Coding Agent System prompt. Sharing for anyone who wants to try it. & \url{https://www.reddit.com/r/vibecoding/comments/1p3jffw/fun_and_effective_vibe_coding_agent_system_prompt/} \\
G1316 & 3923 & Why is everything about CLI nowadays instead of IDE? & \url{https://www.reddit.com/r/vibecoding/comments/1tu18ie/why_is_everything_about_cli_nowadays_instead_of/} \\
G1317 & 3932 & My learnings after a few months of heavy Claude Code usage & \url{https://www.reddit.com/r/ClaudeAI/comments/1mb80e6/my_learnings_after_a_few_months_of_heavy_claude/} \\
G1318 & 3939 & I gave up on using Cursor and here's why & \url{https://www.reddit.com/r/cursor/comments/1isci4s/i_gave_up_on_using_cursor_and_heres_why/} \\
G1319 & 3940 & I'm learning Domain-Driven Design. This is my first project with it & \url{https://www.reddit.com/r/codereview/comments/1r912yx/im_learning_domaindriven_design_this_is_my_first/} \\
G1320 & 3942 & My failed attempt to completely replace manual coding with Codex & \url{https://www.reddit.com/r/OpenAI/comments/1r0k96m/my_failed_attempt_to_completely_replace_manual/} \\
G1321 & 3944 & Working as a Solo Dev: Seeking advice from Senior Developers \& Managers & \url{https://www.reddit.com/r/ExperiencedDevs/comments/1jvipj7/working_as_a_solo_dev_seeking_advice_from_senior/} \\
G1322 & 3947 & Our autonomous agent's posting tool was silently returning OK on failure -- here's how the monitoring layer caught it & \url{https://www.reddit.com/r/AI_Agents/comments/1tgj4u1/our_autonomous_agents_posting_tool_was_silently/} \\
G1323 & 3957 & I have a idea & \url{https://www.reddit.com/r/cursor/comments/1k82kz6/i_have_a_idea/} \\
G1324 & 3961 & Just curious as a PR reviewver during PR code review and there is a small adjustment & \url{https://www.reddit.com/r/AskProgramming/comments/1n6nhjj/just_curious_as_a_pr_reviewver_during_pr_code/} \\
G1325 & 3966 & Day 5 building AgentMeter in public -- stuck on AWS, and questioning how much a solo founder really needs to know & \url{https://www.reddit.com/r/AI_Agents/comments/1tdjzod/day_5_building_agentmeter_in_public_stuck_on_aws/} \\
G1326 & 3970 & AI tools can make one developer faster. The harder question is whether that speed becomes team throughput. & \url{https://www.reddit.com/r/devops/comments/1tqdpqu/ai_tools_can_make_one_developer_faster_the_harder/} \\
G1327 & 3974 & Showing inaccurate diff & \url{https://www.reddit.com/r/cursor/comments/1kzqlek/showing_inaccurate_diff/} \\
G1328 & 3976 & GitHub Copilot is moving to usage-based billing & \url{https://www.reddit.com/r/GithubCopilot/comments/1sx89xf/github_copilot_is_moving_to_usagebased_billing/} \\
G1329 & 3977 & If LLMs are an "abstraction layer", wouldn't that make Junior Developers an "abstraction layer", as well? & \url{https://www.reddit.com/r/webdev/comments/1t7fxcj/if_llms_are_an_abstraction_layer_wouldnt_that/} \\
G1330 & 3978 & Anyone actually happy with their code review process? & \url{https://www.reddit.com/r/codereview/comments/1mxdsob/anyone_actually_happy_with_their_code_review/} \\
G1331 & 3988 & Routing Claude Code subagents to OpenAI models (GPT-4o, o3, etc.) & \url{https://www.reddit.com/r/ClaudeAI/comments/1mi5l9m/routing_claude_code_subagents_to_openai_models/} \\
G1332 & 3993 & got tired of hitting the rate limit on claude code so i built a tool to auto resume when the limit resets & \url{https://www.reddit.com/r/ClaudeAI/comments/1r015f0/got_tired_of_hitting_the_rate_limit_on_claude/} \\
G1333 & 3997 & Is anyone else sick of slow PR reviews, merge surprises, and lost onboarding context? & \url{https://www.reddit.com/r/devops/comments/1k5ri0l/is_anyone_else_sick_of_slow_pr_reviews_merge/} \\
G1334 & 4002 & I bit the bullet! & \url{https://www.reddit.com/r/ClaudeAI/comments/1mt3vwi/i_bit_the_bullet/} \\
G1335 & 4003 & Negative PR review from people pulling rank & \url{https://www.reddit.com/r/ExperiencedDevs/comments/1ozvojj/negative_pr_review_from_people_pulling_rank/} \\
G1336 & 4005 & People running 2-5 coding agents: what actually breaks first for you? & \url{https://www.reddit.com/r/GithubCopilot/comments/1stu24g/people_running_25_coding_agents_what_actually/} \\
G1337 & 4006 & Claude vs Codex (No Bullshit) Comparison & \url{https://www.reddit.com/r/ClaudeAI/comments/1p8xk4r/claude_vs_codex_no_bullshit_comparison/} \\
G1338 & 4007 & The "Phantom Author" in our codebases: Why AI-generated code is a ticking time bomb for quality. & \url{https://www.reddit.com/r/programming/comments/1nxobte/the_phantom_author_in_our_codebases_why/} \\
G1339 & 4017 & Opus 4.6 on the 20x Max plan -- usage after a heavy day & \url{https://www.reddit.com/r/ClaudeAI/comments/1qxij78/opus_46_on_the_20x_max_plan_usage_after_a_heavy/} \\
G1340 & 4019 & Is this normal? & \url{https://www.reddit.com/r/webdev/comments/1ilkeuf/is_this_normal/} \\
G1341 & 4027 & Insights from Claude Code's Creators (Latent Space Podcast) & \url{https://www.reddit.com/r/ClaudeAI/comments/1kigr2j/insights_from_claude_codes_creators_latent_space/} \\
G1342 & 4029 & Yesterday I got ratio'd for saying I made gstack obsolete. Today I actually did it in a day (fwstack) using Flow Weaver and Claude Code. & \url{https://www.reddit.com/r/ClaudeAI/comments/1sjrou5/yesterday_i_got_ratiod_for_saying_i_made_gstack/} \\
G1343 & 4030 & Can someone review my code, it is written in c99 & \url{https://www.reddit.com/r/codereview/comments/1iu1kw8/can_someone_review_my_code_it_is_written_in_c99/} \\
G1344 & 4032 & Weird vibes from AI-assisted coding (Claude + Go + React) & \url{https://www.reddit.com/r/ClaudeAI/comments/1oey6xn/weird_vibes_from_aiassisted_coding_claude_go_react/} \\
G1345 & 4034 & Senior developers, how do you handle multiple context switches in a day? & \url{https://www.reddit.com/r/ExperiencedDevs/comments/1np8p0i/senior_developers_how_do_you_handle_multiple/} \\
G1346 & 4035 & My Experience: How I coded a local SEO crawler in 3 days (instead of 10) for \$15, thanks to AI. & \url{https://www.reddit.com/r/ArtificialInteligence/comments/1oeskf2/my_experience_how_i_coded_a_local_seo_crawler_in/} \\
G1347 & 4040 & How can i get a paid code review? & \url{https://www.reddit.com/r/rust/comments/1ntu05a/how_can_i_get_a_paid_code_review/} \\
G1348 & 4042 & [AskJS] How Can I Improve My JavaScript Skills Without a Mentor? & \url{https://www.reddit.com/r/javascript/comments/1jbbu2m/askjs_how_can_i_improve_my_javascript_skills/} \\
G1349 & 4043 & Claude's making me lazy & \url{https://www.reddit.com/r/ClaudeAI/comments/1mgl9le/claudes_making_me_lazy/} \\
G1350 & 4044 & What coding standards have you set in place for your team who uses AI heavily to write code? & \url{https://www.reddit.com/r/ExperiencedDevs/comments/1qxj358/what_coding_standards_have_you_set_in_place_for/} \\
G1351 & 4046 & Company rewards AI use, but I don't want to rot my brain & \url{https://www.reddit.com/r/cscareerquestions/comments/1ni1z1g/company_rewards_ai_use_but_i_dont_want_to_rot_my/} \\
G1352 & 4050 & [Project Showcase] I built a sales data analysis script with NumPy and would love your feedback! & \url{https://www.reddit.com/r/codereview/comments/1mw5gbd/project_showcase_i_built_a_sales_data_analysis/} \\
G1353 & 4051 & Is PR review a thankless job? & \url{https://www.reddit.com/r/ExperiencedDevs/comments/1l9f6ty/is_pr_review_a_thankless_job/} \\
G1354 & 4054 & I got tired of "Why did you add a semicolon?" comments -- so I built a tool to end those debates forever. & \url{https://www.reddit.com/r/javascript/comments/1p6izew/i_got_tired_of_why_did_you_add_a_semicolon/} \\
G1355 & 4057 & Anyone left on this sub who actually admits the Claude CLI changes must be audited line by line? & \url{https://www.reddit.com/r/ClaudeAI/comments/1lhec73/anyone_left_on_this_sub_who_actually_admits_the/} \\
G1356 & 4058 & Jupyter notebooks touching production data are application code from a security standpoint & \url{https://www.reddit.com/r/Python/comments/1tqtjal/jupyter_notebooks_touching_production_data_are/} \\
G1357 & 4064 & Log Statements Burn my Eyes. Code is Ugly and Hard to Read & \url{https://www.reddit.com/r/ExperiencedDevs/comments/1p86gk9/log_statements_burn_my_eyes_code_is_ugly_and_hard/} \\
G1358 & 4065 & Anyone here tried AI tools to speed up front-end code reviews? & \url{https://www.reddit.com/r/webdev/comments/1ou4za4/anyone_here_tried_ai_tools_to_speed_up_frontend/} \\
G1359 & 4068 & Code quality of Claude, a sad realization & \url{https://www.reddit.com/r/ClaudeAI/comments/1ptcbm3/code_quality_of_claude_a_sad_realization/} \\
G1360 & 4071 & Looking for a code review. URL Shortener (FastAPI) & \url{https://www.reddit.com/r/codereview/comments/1ra9k62/looking_for_a_code_review_url_shortener_fastapi/} \\
G1361 & 4075 & Erasing someone's entire code from the git & \url{https://www.reddit.com/r/AskProgramming/comments/1oter30/erasing_someones_entire_code_from_the_git/} \\
G1362 & 4077 & Lessons from 15 Years in Tech: How Code Reviews Build Culture & \url{https://www.reddit.com/r/ExperiencedDevs/comments/1jf3xho/lessons_from_15_years_in_tech_how_code_reviews/} \\
G1363 & 4084 & Is vibe coding the end of the senior engineer? & \url{https://www.reddit.com/r/ExperiencedDevs/comments/1s4spyv/is_vibe_coding_the_end_of_the_senior_engineer/} \\
G1364 & 4086 & Having trouble recalling how my first playground was built, any way to recreate or find the prompt used? & \url{https://www.reddit.com/r/ClaudeAI/comments/1tdzohp/having_trouble_recalling_how_my_first_playground/} \\
G1365 & 4088 & I Love and Hate Git - Here's Why & \url{https://www.reddit.com/r/git/comments/1jfc2kk/i_love_and_hate_git_heres_why/} \\
G1366 & 4090 & How to Prevent Claude from being lazy when reading files & \url{https://www.reddit.com/r/GithubCopilot/comments/1lk9hyx/how_to_prevent_claude_from_being_lazy_when/} \\
G1367 & 4094 & How to make projects now ? & \url{https://www.reddit.com/r/cscareerquestions/comments/1s8iwit/how_to_make_projects_now/} \\
G1368 & 4100 & Using Cursor everyday and loving it & \url{https://www.reddit.com/r/cursor/comments/1kbchnm/using_cursor_everyday_and_loving_it/} \\
G1369 & 4102 & Wrote a beginner-friendly guide on CODEOWNERS : the file that quietly fixes PR chaos & \url{https://www.reddit.com/r/git/comments/1rdc9ig/wrote_a_beginnerfriendly_guide_on_codeowners_the/} \\
G1370 & 4104 & How I use AI 100\% autonomously: Claudiomiro & \url{https://www.reddit.com/r/vibecoding/comments/1p1r7or/how_i_use_ai_100_autonomously_claudiomiro/} \\
G1371 & 4106 & AI-Powered Internal Developer Platform (IDP) -- Review, Test, Package, Deploy AI-Generated Code & \url{https://www.reddit.com/r/vibecoding/comments/1lt9i5p/aipowered_internal_developer_platform_idp_review/} \\
G1372 & 4112 & Using claude code for my business, blown away. Privacy concerns. & \url{https://www.reddit.com/r/vibecoding/comments/1skxlti/using_claude_code_for_my_business_blown_away/} \\
G1373 & 4117 & Anthropic: AI assisted coding doesn't show efficiency gains and impairs developers abilities. & \url{https://www.reddit.com/r/programming/comments/1qqxvlw/anthropic_ai_assisted_coding_doesnt_show/} \\
G1374 & 4119 & I built an open-source framework to run Claude Code on k8s & \url{https://www.reddit.com/r/AI_Agents/comments/1r3p290/i_built_an_opensource_framework_to_run_claude/} \\
G1375 & 4121 & My First SaaS That's Actually Solving a Problem (Solved Mine Too) & \url{https://www.reddit.com/r/devops/comments/1rc60su/my_first_saas_thats_actually_solving_a_problem/} \\
G1376 & 4124 & Beginner Rust Project - Would love your review \& kind guidance! & \url{https://www.reddit.com/r/rust/comments/1jucttj/beginner_rust_project_would_love_your_review_kind/} \\
G1377 & 4126 & I vibecoded a Unity 3D Werewolf/Mafia LLM AI Simulation Sandbox, playable 100\% offline & \url{https://www.reddit.com/r/vibecoding/comments/1rk9etu/i_vibecoded_a_unity_3d_werewolfmafia_llm_ai/} \\
G1378 & 4127 & Are we starting to trust AI-generated code a bit too much? & \url{https://www.reddit.com/r/webdev/comments/1t4aobw/are_we_starting_to_trust_aigenerated_code_a_bit/} \\
G1379 & 4128 & I reviewed 15 AI-built MVPs for VCs last quarter. 13 needed complete rewrites to scale. Here's the pattern. & \url{https://www.reddit.com/r/webdev/comments/1rwz5kw/i_reviewed_15_aibuilt_mvps_for_vcs_last_quarter/} \\
G1380 & 4136 & What is the best paid AI for our needs? & \url{https://www.reddit.com/r/ChatGPTCoding/comments/1kyb6r8/what_is_the_best_paid_ai_for_our_needs/} \\
G1381 & 4141 & Incorporating AI into the SDLC & \url{https://www.reddit.com/r/ExperiencedDevs/comments/1sclodw/incorporating_ai_into_the_sdlc/} \\
G1382 & 4151 & Thank you cursor! & \url{https://www.reddit.com/r/cursor/comments/1m97sol/thank_you_cursor/} \\
G1383 & 4152 & AI Is Moving Work, Not Removing It & \url{https://www.reddit.com/r/accelerate/comments/1pr59nr/ai_is_moving_work_not_removing_it/} \\
G1384 & 4154 & My co-founder and I built an open source IDE for running parallel AI coding agents. would love feedback. & \url{https://www.reddit.com/r/git/comments/1sleg21/my_cofounder_and_i_built_an_open_source_ide_for/} \\
G1385 & 4157 & i'm a solo dev but my commit history looks like a 3-person team reviewed it & \url{https://www.reddit.com/r/cursor/comments/1pormyu/im_a_solo_dev_but_my_commit_history_looks_like_a/} \\
G1386 & 4159 & 5 Things Experienced Devs Do that AI Can't (A Senior Engineer's Survival Guide) & \url{https://www.reddit.com/r/ExperiencedDevs/comments/1nchwij/5_things_experienced_devs_do_that_ai_cant_a/} \\
G1387 & 4160 & Karpathy just said "the human is the bottleneck" and "once agents fail, you blame yourself" -- I built a system that fixes both problems & \url{https://www.reddit.com/r/ClaudeAI/comments/1s2gld8/karpathy_just_said_the_human_is_the_bottleneck/} \\
G1388 & 4161 & Have your company stopped hiring juniors? & \url{https://www.reddit.com/r/ExperiencedDevs/comments/1sujoah/have_your_company_stopped_hiring_juniors/} \\
G1389 & 4167 & I audited 6 months of PRs after my team went all-in on AI code generation. The code got worse in ways none of us predicted. & \url{https://www.reddit.com/r/webdev/comments/1sin68g/i_audited_6_months_of_prs_after_my_team_went/} \\
G1390 & 4170 & Vibe Engineering - Laptop not required & \url{https://www.reddit.com/r/vibecoding/comments/1pfsil8/vibe_engineering_laptop_not_required/} \\
G1391 & 4171 & Code Review Request: MMO Client/Server architecture & \url{https://www.reddit.com/r/cpp/comments/1lxn8gt/code_review_request_mmo_clientserver_architecture/} \\
G1392 & 4173 & Kimi-Linear support is merged to llama.cpp & \url{https://www.reddit.com/r/LocalLLaMA/comments/1qxf7sf/kimilinear_support_is_merged_to_llamacpp/} \\
G1393 & 4179 & Steal/Grill this prompt. But also developers, how would you improve it? & \url{https://www.reddit.com/r/ClaudeAI/comments/1ttvcj6/stealgrill_this_prompt_but_also_developers_how/} \\
G1394 & 4180 & Code Review for feature complete app. & \url{https://www.reddit.com/r/vibecoding/comments/1nzflwb/code_review_for_feature_complete_app/} \\
G1395 & 4182 & Is agentic coding actually good long term? & \url{https://www.reddit.com/r/ClaudeAI/comments/1n3komw/is_agentic_coding_actually_good_long_term/} \\
G1396 & 4185 & My flow to vibecode new apps as a 16k/mo indie founder & \url{https://www.reddit.com/r/vibecoding/comments/1pvc87g/my_flow_to_vibecode_new_apps_as_a_16kmo_indie/} \\
G1397 & 4187 & Starting a beginner-friendly book club this Saturday - what would make it actually valuable for new Rustaceans? & \url{https://www.reddit.com/r/rust/comments/1mq87zk/starting_a_beginnerfriendly_book_club_this/} \\
G1398 & 4188 & Took some code samples to the devs yesterday at work. I need some advice on how to navigate and fix code quality. & \url{https://www.reddit.com/r/vibecoding/comments/1lpuqut/took_some_code_samples_to_the_devs_yesterday_at/} \\
G1399 & 4197 & Engineers, how are you handling security and code quality with all this AI gen code creeping in? & \url{https://www.reddit.com/r/devops/comments/1n4z0rl/engineers_how_are_you_handling_security_and_code/} \\
G1400 & 4198 & How we actually use AI to ship production code at FAANG & \url{https://www.reddit.com/r/vibecoding/comments/1q9svgl/how_we_actually_use_ai_to_ship_production_code_at/} \\
G1401 & 4199 & Most AI coding failures are requirements failures, not model failures & \url{https://www.reddit.com/r/ArtificialInteligence/comments/1rcakii/most_ai_coding_failures_are_requirements_failures/} \\
G1402 & 4201 & Recent contract rewrite seems to have made my role redundant & \url{https://www.reddit.com/r/ExperiencedDevs/comments/1pojmxc/recent_contract_rewrite_seems_to_have_made_my/} \\
G1403 & 4203 & New to Go, kind of an idiot, code review request. & \url{https://www.reddit.com/r/golang/comments/1iq4hay/new_to_go_kind_of_an_idiot_code_review_request/} \\
G1404 & 4206 & Dealing with a difficult team member and deciding about moving teams & \url{https://www.reddit.com/r/ExperiencedDevs/comments/1leg9um/dealing_with_a_difficult_team_member_and_deciding/} \\
G1405 & 4207 & I built an AI agent that can modify and evolve its own codebase (open source) & \url{https://www.reddit.com/r/vibecoding/comments/1rukaci/i_built_an_ai_agent_that_can_modify_and_evolve/} \\
G1406 & 4217 & How to mentor vibecoding junior? & \url{https://www.reddit.com/r/ExperiencedDevs/comments/1sp1ew8/how_to_mentor_vibecoding_junior/} \\
G1407 & 4219 & Technical debt, siloed domain knowledge, and a small team & \url{https://www.reddit.com/r/ExperiencedDevs/comments/1mzxrcq/technical_debt_siloed_domain_knowledge_and_a/} \\
G1408 & 4223 & Approach to writing tests with AI & \url{https://www.reddit.com/r/vibecoding/comments/1qsyrts/approach_to_writing_tests_with_ai/} \\
G1409 & 4224 & Should developer interviews start testing AI usage instead of coding? & \url{https://www.reddit.com/r/ArtificialInteligence/comments/1rshcr2/should_developer_interviews_start_testing_ai/} \\
G1410 & 4225 & Claude Code might be the best dev tool I've ever used. But I still wouldn't ship its output to production without more work. So I built a pipeline around it. & \url{https://www.reddit.com/r/ClaudeAI/comments/1rm8jx3/claude_code_might_be_the_best_dev_tool_ive_ever/} \\
G1411 & 4231 & How come software devs are so much more worried about AI replacing them than other white collar jobs? & \url{https://www.reddit.com/r/cscareerquestions/comments/1s1w200/how_come_software_devs_are_so_much_more_worried/} \\
G1412 & 4237 & guys please can you review my code ? & \url{https://www.reddit.com/r/programming/comments/1om08ej/guys_please_can_you_review_my_code/} \\
G1413 & 4250 & Claude code reviewed by another AI & \url{https://www.reddit.com/r/ClaudeAI/comments/1tdtai0/claude_code_reviewed_by_another_ai/} \\
G1414 & 4252 & [AskJS] Do you trust AI-generated frontend code in production? & \url{https://www.reddit.com/r/javascript/comments/1pybmug/askjs_do_you_trust_aigenerated_frontend_code_in/} \\
G1415 & 4253 & How to build a PR reviewer like github copilot and gemini using python & \url{https://www.reddit.com/r/webdev/comments/1jznbnw/how_to_build_a_pr_reviewer_like_github_copilot/} \\
G1416 & 4255 & Building European Swallow AI a AI orchestration API & \url{https://www.reddit.com/r/vibecoding/comments/1n2sxfe/building_european_swallow_ai_a_ai_orchestration/} \\
G1417 & 4261 & My manager wants to pair me up with the worst guy on the team for the next six months. How do I get out of this? & \url{https://www.reddit.com/r/ExperiencedDevs/comments/1hvy9g7/my_manager_wants_to_pair_me_up_with_the_worst_guy/} \\
G1418 & 4263 & I built a full desktop photo management app entirely with Claude - here's what I learned & \url{https://www.reddit.com/r/ClaudeAI/comments/1qu8tq7/i_built_a_full_desktop_photo_management_app/} \\
G1419 & 4266 & Rewrites are taking too long and CTO has decided to vibe code the rewrite by himself & \url{https://www.reddit.com/r/cscareerquestions/comments/1pd5hts/rewrites_are_taking_too_long_and_cto_has_decided/} \\
G1420 & 4267 & We seniors are a dying breed & \url{https://www.reddit.com/r/ExperiencedDevs/comments/1t4wq4i/we_seniors_are_a_dying_breed/} \\
G1421 & 4268 & CC Agents Are Really a Cheat Code (Prompt Included) & \url{https://www.reddit.com/r/ClaudeAI/comments/1lciedr/cc_agents_are_really_a_cheat_code_prompt_included/} \\
G1422 & 4274 & Claude + Codex = Excellence & \url{https://www.reddit.com/r/OpenAI/comments/1su9g2c/claude_codex_excellence/} \\
G1423 & 4277 & For some projects, pushing code without AI is becoming impossible. & \url{https://www.reddit.com/r/ArtificialInteligence/comments/1swxevz/for_some_projects_pushing_code_without_ai_is/} \\
G1424 & 4280 & Manager is not reviewing my PRs. What should I do? & \url{https://www.reddit.com/r/ExperiencedDevs/comments/1sxswwo/manager_is_not_reviewing_my_prs_what_should_i_do/} \\
G1425 & 4281 & Senior Devs Survey - Productivity Boost & \url{https://www.reddit.com/r/cursor/comments/1i3x7ul/senior_devs_survey_productivity_boost/} \\
G1426 & 4283 & Weekly Roundup: AI and National Security (8 October 2025) & \url{https://www.reddit.com/r/accelerate/comments/1o1kf6s/weekly_roundup_ai_and_national_security_8_october/} \\
G1427 & 4285 & I think I accidentally made gstack and superpowers obsolete & \url{https://www.reddit.com/r/ClaudeAI/comments/1siwil7/i_think_i_accidentally_made_gstack_and/} \\
G1428 & 4295 & moved to an enterprise AI code review platform after open source wasn't cutting it & \url{https://www.reddit.com/r/webdev/comments/1p757d2/moved_to_an_enterprise_ai_code_review_platform/} \\
G1429 & 4297 & spent \$500/month on AI code review tools, saved 30 mins/day. the math doesnt add up & \url{https://www.reddit.com/r/ChatGPTCoding/comments/1ohhf44/spent_500month_on_ai_code_review_tools_saved_30/} \\
G1430 & 4302 & Does anyone actually let their bots run wild? & \url{https://www.reddit.com/r/ClaudeAI/comments/1qvw5cx/does_anyone_actually_let_their_bots_run_wild/} \\
G1431 & 4305 & Experienced teams that went hard on AI, did you agree to lower your quality bar? How did it work out mid/long term? & \url{https://www.reddit.com/r/ExperiencedDevs/comments/1s9pkif/experienced_teams_that_went_hard_on_ai_did_you/} \\
G1432 & 4308 & What does quality look like when you're an engineering manager without a qa team? & \url{https://www.reddit.com/r/ExperiencedDevs/comments/1r9tzx8/what_does_quality_look_like_when_youre_an/} \\
G1433 & 4312 & Claude spinner verbs that are actually useful reminders & \url{https://www.reddit.com/r/ClaudeAI/comments/1seycjp/claude_spinner_verbs_that_are_actually_useful/} \\
G1434 & 4316 & nobody actually reads code in PR reviews anymore, they just check if automated tests pass & \url{https://www.reddit.com/r/ExperiencedDevs/comments/1qkmmsx/nobody_actually_reads_code_in_pr_reviews_anymore/} \\
G1435 & 4318 & SICK and tired of Greptile, what are the best alternatives? & \url{https://www.reddit.com/r/codereview/comments/1sqs5iw/sick_and_tired_of_greptile_what_are_the_best/} \\
G1436 & 4326 & I stopped using Claude Code like a chatbot and it got way better & \url{https://www.reddit.com/r/ClaudeAI/comments/1sy3pho/i_stopped_using_claude_code_like_a_chatbot_and_it/} \\
G1437 & 4330 & Ive been vibe coding too much. & \url{https://www.reddit.com/r/vibecoding/comments/1qci71i/ive_been_vibe_coding_too_much/} \\
G1438 & 4331 & AI wrote half my code and now I regret everything & \url{https://www.reddit.com/r/vibecoding/comments/1r6zm8x/ai_wrote_half_my_code_and_now_i_regret_everything/} \\
G1439 & 4332 & Is Cursor auto secretly getting better? & \url{https://www.reddit.com/r/cursor/comments/1mju55n/is_cursor_auto_secretly_getting_better/} \\
G1440 & 4333 & I got sick of Claude code generating tech debt, so i just made AI agents fight each other. & \url{https://www.reddit.com/r/ClaudeAI/comments/1p8q4wh/i_got_sick_of_claude_code_generating_tech_debt_so/} \\
G1441 & 4334 & How do you stay engaged with programming when you don't have a tech community around you? & \url{https://www.reddit.com/r/AskProgramming/comments/1otef4z/how_do_you_stay_engaged_with_programming_when_you/} \\
G1442 & 4337 & A little horror story... & \url{https://www.reddit.com/r/vibecoding/comments/1s6a5ax/a_little_horror_story/} \\
G1443 & 4343 & Non technical vibe coders, do you even care about engineering and security foundations? & \url{https://www.reddit.com/r/vibecoding/comments/1pwufez/non_technical_vibe_coders_do_you_even_care_about/} \\
G1444 & 4347 & Request PHP code review & \url{https://www.reddit.com/r/codereview/comments/1hr05w0/request_php_code_review/} \\
G1445 & 4348 & I built a universal CLAUDE.md that cuts Claude output tokens by 63\% - validated with benchmarks, fully open source & \url{https://www.reddit.com/r/ClaudeAI/comments/1s7qu07/i_built_a_universal_claudemd_that_cuts_claude/} \\
G1446 & 4354 & Best AI top tools? Code Review Bot & \url{https://www.reddit.com/r/codereview/comments/1jc380b/best_ai_top_tools_code_review_bot/} \\
G1447 & 4357 & Announcing Development on Flirt & \url{https://www.reddit.com/r/git/comments/1opbpky/announcing_development_on_flirt/} \\
G1448 & 4359 & Switching from GitHub Copilot to a context aware tool reduced our review cycles on AI generated codeSwitching from GitHub Copilot to a context aware tool reduce & \url{https://www.reddit.com/r/GithubCopilot/comments/1tqy0ze/switching_from_github_copilot_to_a_context_aware/} \\
G1449 & 4360 & Dotnet Web API Code review & \url{https://www.reddit.com/r/codereview/comments/1j9n6wg/dotnet_web_api_code_review/} \\
G1450 & 4373 & What I learned after 1 month of vibecoding an iOS app \& getting my first sales & \url{https://www.reddit.com/r/vibecoding/comments/1o5t99k/what_i_learned_after_1_month_of_vibecoding_an_ios/} \\
G1451 & 4380 & 13 Critical Bugs That Could Have Slipped Through From Vibe Coding & \url{https://www.reddit.com/r/vibecoding/comments/1ll3ii4/13_critical_bugs_that_could_have_slipped_through/} \\
G1452 & 4382 & \# The Ultimate System: Building Perfect Apps with Claude + Dyad (2,000 Prompt's Method) & \url{https://www.reddit.com/r/vibecoding/comments/1skdex6/the_ultimate_system_building_perfect_apps_with/} \\
G1453 & 4384 & HighLow TUI - RUST & \url{https://www.reddit.com/r/rust/comments/1skybnx/highlow_tui_rust/} \\
G1454 & 4387 & My Favorite Agent Command - Run Git Diff and Optimize the Proposed Changes & \url{https://www.reddit.com/r/cursor/comments/1ja0nw3/my_favorite_agent_command_run_git_diff_and/} \\
G1455 & 4405 & I compared Claude Sonnet 3.5 vs Deepseek R1 on 500 real PRs - here's what I found & \url{https://www.reddit.com/r/ClaudeAI/comments/1ikvj5w/i_compared_claude_sonnet_35_vs_deepseek_r1_on_500/} \\
G1456 & 4411 & GitHub Actions under active exploitation & \url{https://www.reddit.com/r/github/comments/1rmks3a/github_actions_under_active_exploitation/} \\
G1457 & 4417 & Howto : Spec driven development in Windsurf & \url{https://www.reddit.com/r/windsurf/comments/1nbr23p/howto_spec_driven_development_in_windsurf/} \\
G1458 & 4418 & I'm testing a vibe-coded review drill: would you approve this AI-generated prod command? & \url{https://www.reddit.com/r/vibecoding/comments/1tiifxo/im_testing_a_vibecoded_review_drill_would_you/} \\
G1459 & 4421 & Is AI making us worse at learning to code? Here's my take as a dev who's seen this pattern before. & \url{https://www.reddit.com/r/csharp/comments/1kcgqc9/is_ai_making_us_worse_at_learning_to_code_heres/} \\
G1460 & 4422 & Improve your Claude Code generation quality by 20\% & \url{https://www.reddit.com/r/LocalLLaMA/comments/1rw7281/improve_your_claude_code_generation_quality_by_20/} \\
G1461 & 4423 & Maestro v1.6.1 -- Codex now has a full 22-agent orchestration platform as a native plugin & \url{https://www.reddit.com/r/OpenAI/comments/1shn5fd/maestro_v161_codex_now_has_a_full_22agent/} \\
G1462 & 4425 & Bugs slipping through code review are causing 3am pages & \url{https://www.reddit.com/r/devops/comments/1n3dua2/bugs_slipping_through_code_review_are_causing_3am/} \\
G1463 & 4426 & Vibe Coding is designed to make you reliant on LLMs. We should be avoiding this at all costs. & \url{https://www.reddit.com/r/SoftwareEngineering/comments/1qgdqls/vibe_coding_is_designed_to_make_you_reliant_on/} \\
G1464 & 4427 & I built Koucai () - an Mandarin learning app with AI penpals - here's my workflow & \url{https://www.reddit.com/r/ClaudeAI/comments/1n44s7b/i_built_koucai__an_mandarin_learning_app_with/} \\
G1465 & 4428 & Your realistic perspective on vibe coding? & \url{https://www.reddit.com/r/ClaudeAI/comments/1sqwa1t/your_realistic_perspective_on_vibe_coding/} \\
G1466 & 4429 & Things that slow down small engineering teams that have nothing to do with technical skill. & \url{https://www.reddit.com/r/webdev/comments/1tmajp4/things_that_slow_down_small_engineering_teams/} \\
G1467 & 4431 & Opus 4.1 in Claude Code - general improvement, but at times fails to follow basic instructions & \url{https://www.reddit.com/r/ClaudeAI/comments/1mozd6o/opus_41_in_claude_code_general_improvement_but_at/} \\
G1468 & 4436 & Code Council - run code reviews through multiple AI models, see where they agree and disagree & \url{https://www.reddit.com/r/ClaudeAI/comments/1qsa1wh/code_council_run_code_reviews_through_multiple_ai/} \\
G1469 & 4438 & Looks Like It's Time For All of Us to Come Together (And Here's Why It Matters) & \url{https://www.reddit.com/r/dotnet/comments/1irpwss/looks_like_its_time_for_all_of_us_to_come/} \\
G1470 & 4439 & claude code - best practices for dev? (potentially very naive question!) & \url{https://www.reddit.com/r/ClaudeAI/comments/1l9wlvy/claude_code_best_practices_for_dev_potentially/} \\
G1471 & 4440 & Hot take: AI-generated code is not a problem, but skimping on good reviews is. & \url{https://www.reddit.com/r/vibecoding/comments/1ten4zq/hot_take_aigenerated_code_is_not_a_problem_but/} \\
G1472 & 4441 & The cost of poor code reviews after analyzing 10,000+ PRs & \url{https://www.reddit.com/r/ExperiencedDevs/comments/1muy6ri/the_cost_of_poor_code_reviews_after_analyzing/} \\
G1473 & 4443 & My previous company keeps saying "we need to go even more AI native" while laying off more people when AI milestones are achieved. & \url{https://www.reddit.com/r/cscareerquestions/comments/1shs8jw/my_previous_company_keeps_saying_we_need_to_go/} \\
G1474 & 4444 & Cursor fixed your code but I still wonder why those changes were made until I asked & \url{https://www.reddit.com/r/cursor/comments/1ohzj86/cursor_fixed_your_code_but_i_still_wonder_why/} \\
G1475 & 4446 & Claude's quality drop is killing my productivity. Any alternatives? & \url{https://www.reddit.com/r/webdev/comments/1nk5xq0/claudes_quality_drop_is_killing_my_productivity/} \\
G1476 & 4447 & ChatGPT could not build my browser extension. What went wrong? & \url{https://www.reddit.com/r/ChatGPTCoding/comments/1k274q1/chatgpt_could_not_build_my_browser_extension_what/} \\
G1477 & 4448 & Hot take, but AI actually may not be that useful for automation or writing code & \url{https://www.reddit.com/r/ArtificialInteligence/comments/1p4841z/hot_take_but_ai_actually_may_not_be_that_useful/} \\
G1478 & 4450 & Unpopular opinion: most AI code review tools are just expensive linters & \url{https://www.reddit.com/r/codereview/comments/1s8fy53/unpopular_opinion_most_ai_code_review_tools_are/} \\
G1479 & 4451 & I vibed a shadcn registry and maybe you'll like it / how I did it & \url{https://www.reddit.com/r/vibecoding/comments/1rsaoad/i_vibed_a_shadcn_registry_and_maybe_youll_like_it/} \\
G1480 & 4453 & Zig banned AI contributions - the reasoning is not about detecting AI code & \url{https://www.reddit.com/r/vibecoding/comments/1t2hmn3/zig_banned_ai_contributions_the_reasoning_is_not/} \\
G1481 & 4456 & what do you guys think about AI replacing software engineers anytime soon ? & \url{https://www.reddit.com/r/cscareerquestions/comments/1mp3z0o/what_do_you_guys_think_about_ai_replacing/} \\
G1482 & 4458 & is cursor getting worse or just me? & \url{https://www.reddit.com/r/cursor/comments/1nfdij4/is_cursor_getting_worse_or_just_me/} \\
G1483 & 4459 & Onboarding an org to front end work & \url{https://www.reddit.com/r/ExperiencedDevs/comments/1kubp0s/onboarding_an_org_to_front_end_work/} \\
G1484 & 4462 & GitHub Copilot for JetBrains - January Update & \url{https://www.reddit.com/r/GithubCopilot/comments/1qz9phh/github_copilot_for_jetbrains_january_update/} \\
G1485 & 4465 & What are the best CodeRabbit alternatives you've used? & \url{https://www.reddit.com/r/devops/comments/1nue3dt/what_are_the_best_coderabbit_alternatives_youve/} \\
G1486 & 4468 & how many code quality tools is too many? we're running 7 and i'm losing it & \url{https://www.reddit.com/r/devops/comments/1qzz0bc/how_many_code_quality_tools_is_too_many_were/} \\
G1487 & 4471 & Where do you personally draw the line between vibing and actually reviewing the code? & \url{https://www.reddit.com/r/vibecoding/comments/1pjyfb1/where_do_you_personally_draw_the_line_between/} \\
G1488 & 4472 & Best local coding LLM for Embedded AI dev - RTX 4060 (8GB VRAM), 16GB RAM & \url{https://www.reddit.com/r/LocalLLaMA/comments/1rormbr/best_local_coding_llm_for_embedded_ai_dev_rtx/} \\
G1489 & 4475 & Claude code is great, BUT.. & \url{https://www.reddit.com/r/ChatGPTCoding/comments/1iyuzv1/claude_code_is_great_but/} \\
G1490 & 4476 & Officially deployed my first app! WorthIt: Pocket Appraiser & \url{https://www.reddit.com/r/vibecoding/comments/1th44n4/officially_deployed_my_first_app_worthit_pocket/} \\
G1491 & 4477 & Live web dev classes - what would you actually want to learn? & \url{https://www.reddit.com/r/webdev/comments/1mudz3d/live_web_dev_classes_what_would_you_actually_want/} \\
G1492 & 4481 & [ Removed by moderator ] & \url{https://www.reddit.com/r/ClaudeAI/comments/1rlynfs/i_built_a_local_mcp_for_claude_code_that_shows_it/} \\
G1493 & 4482 & Use of AI in enterprise dev teams & \url{https://www.reddit.com/r/ExperiencedDevs/comments/1qe2ex1/use_of_ai_in_enterprise_dev_teams/} \\
G1494 & 4483 & I wrote a bespoke code review tool with domain context a first-class feature & \url{https://www.reddit.com/r/ClaudeAI/comments/1sk6naz/i_wrote_a_bespoke_code_review_tool_with_domain/} \\
G1495 & 4489 & Claude Desktop app users on Max plan: does your /ultrareview dialog show "Free runs remaining"? Mine doesn't. & \url{https://www.reddit.com/r/ClaudeAI/comments/1t2pu2p/claude_desktop_app_users_on_max_plan_does_your/} \\
G1496 & 4491 & We just open-sourced our entire AI platform after 2+ years of development - please roast our code & \url{https://www.reddit.com/r/programming/comments/1mth1zk/we_just_opensourced_our_entire_ai_platform_after/} \\
G1497 & 4494 & How do you stop PR bottlenecks from turning into rubber stamping when reviewers are overwhelmed & \url{https://www.reddit.com/r/ExperiencedDevs/comments/1rv1ut2/how_do_you_stop_pr_bottlenecks_from_turning_into/} \\
G1498 & 4496 & How much code do you actually check nowadays? & \url{https://www.reddit.com/r/ClaudeAI/comments/1qs9uxw/how_much_code_do_you_actually_check_nowadays/} \\
G1499 & 4499 & I turned Claude Code into 11 specialized React Native agents with cross-session memory and pipeline orchestration & \url{https://www.reddit.com/r/ClaudeAI/comments/1rsozwa/i_turned_claude_code_into_11_specialized_react/} \\
G1500 & 4500 & What programming habit has improved your code quality the most over time? & \url{https://www.reddit.com/r/AskProgramming/comments/1t55ttl/what_programming_habit_has_improved_your_code/} \\
G1501 & 4504 & Braces with single line IF - always, never, sometimes? & \url{https://www.reddit.com/r/csharp/comments/1mndxfv/braces_with_single_line_if_always_never_sometimes/} \\
G1502 & 4507 & anyone else stop using .cursorrules? & \url{https://www.reddit.com/r/cursor/comments/1r88ufs/anyone_else_stop_using_cursorrules/} \\
G1503 & 4510 & Built a 3D browser game with Claude - Traffic Architect, road builder/traffic management & \url{https://www.reddit.com/r/ClaudeAI/comments/1rz24eb/built_a_3d_browser_game_with_claude_traffic/} \\
G1504 & 4514 & Anyone seen issues with AI codes in PRs lately? & \url{https://www.reddit.com/r/codereview/comments/1mc800h/anyone_seen_issues_with_ai_codes_in_prs_lately/} \\
G1505 & 4517 & Vibe-coded a 15k+ line local offers platform and shipped it to a customer prod for testing. Now the hard part starts, anyone else dealing with this? & \url{https://www.reddit.com/r/vibecoding/comments/1t0ts3p/vibecoded_a_15k_line_local_offers_platform_and/} \\
G1506 & 4518 & Weekly limit made me quit and cancel! & \url{https://www.reddit.com/r/GithubCopilot/comments/1sn1e1w/weekly_limit_made_me_quit_and_cancel/} \\
G1507 & 4520 & VS Code AI integration -- would it support custom sub-agents and stop points in the future? & \url{https://www.reddit.com/r/GithubCopilot/comments/1qwq3w8/vs_code_ai_integration_would_it_support_custom/} \\
G1508 & 4521 & Does anyone keep working on projects from their phone after leaving their desk? & \url{https://www.reddit.com/r/cursor/comments/1s97a7v/does_anyone_keep_working_on_projects_from_their/} \\
G1509 & 4523 & I shipped more code yesterday with C4 than the last 3 weeks combined & \url{https://www.reddit.com/r/ClaudeAI/comments/1kudbe8/i_shipped_more_code_yesterday_with_c4_than_the/} \\
G1510 & 4527 & We're building features faster than ever. Not sure that's a good thing. & \url{https://www.reddit.com/r/ClaudeAI/comments/1rvhrzs/were_building_features_faster_than_ever_not_sure/} \\
G1511 & 4532 & Got a lead opportunity at work, all team members are somewhat senior, How to make the best of it & \url{https://www.reddit.com/r/cscareerquestions/comments/1sjhyps/got_a_lead_opportunity_at_work_all_team_members/} \\
G1512 & 4536 & When a PR is ok to be approved, is it still important to click "Resolve Conversation" on the previous individual code review items? & \url{https://www.reddit.com/r/github/comments/1ja1ijf/when_a_pr_is_ok_to_be_approved_is_it_still/} \\
G1513 & 4537 & How do you regulate vibe coding. Need to go fast but not kill us. & \url{https://www.reddit.com/r/vibecoding/comments/1n2ylu7/how_do_you_regulate_vibe_coding_need_to_go_fast/} \\
G1514 & 4540 & LlamaPReview - Our AI Code reviewer (used by 4k+ repos) now gives inline, committable suggestions. We're making the new engine free for all of open source. & \url{https://www.reddit.com/r/codereview/comments/1n1inad/llamapreview_our_ai_code_reviewer_used_by_4k/} \\
G1515 & 4543 & Stop writing DRY code everywhere. It's making your codebase worse, not better & \url{https://www.reddit.com/r/webdev/comments/1rkqmr4/stop_writing_dry_code_everywhere_its_making_your/} \\
G1516 & 4546 & Pushing Claude Code & \url{https://www.reddit.com/r/ClaudeAI/comments/1pzkuef/pushing_claude_code/} \\
G1517 & 4547 & Are you guys using something to automate PR's and code reviews? & \url{https://www.reddit.com/r/ClaudeAI/comments/1l8rgw2/are_you_guys_using_something_to_automate_prs_and/} \\
G1518 & 4552 & For those of you with mandatory code reviews in your job, how long does it take for them to be done? & \url{https://www.reddit.com/r/AskProgramming/comments/1m7737f/for_those_of_you_with_mandatory_code_reviews_in/} \\
G1519 & 4555 & Dorothy: open-source orchestration for Claude Code and AI agents & \url{https://www.reddit.com/r/vibecoding/comments/1r6bgll/dorothy_opensource_orchestration_for_claude_code/} \\
G1520 & 4564 & I'm officially in the "I won't be necessary in 20 years" camp & \url{https://www.reddit.com/r/ArtificialInteligence/comments/1m75awo/im_officially_in_the_i_wont_be_necessary_in_20/} \\
G1521 & 4569 & Is crazy feature churn and lack of short to moderate term planning normal for software teams now? & \url{https://www.reddit.com/r/ExperiencedDevs/comments/1r2xwrl/is_crazy_feature_churn_and_lack_of_short_to/} \\
G1522 & 4572 & yuniq: I built a deduplicator that's 3x faster than xuniq & \url{https://www.reddit.com/r/rust/comments/1skb25v/yuniq_i_built_a_deduplicator_thats_3x_faster_than/} \\
G1523 & 4576 & Java based Numerical library (JNum-v0.1) & \url{https://www.reddit.com/r/java/comments/1tiuhpw/java_based_numerical_library_jnumv01/} \\
G1524 & 4577 & Game-Changer Code Review Hack & \url{https://www.reddit.com/r/ClaudeAI/comments/1lj62fw/gamechanger_code_review_hack/} \\
G1525 & 4579 & AI Code Review Tools Benchmark & \url{https://www.reddit.com/r/codereview/comments/1qnsiqh/ai_code_review_tools_benchmark/} \\
G1526 & 4580 & The end of average software engineers? & \url{https://www.reddit.com/r/ExperiencedDevs/comments/1tjo3f9/the_end_of_average_software_engineers/} \\
G1527 & 4581 & I audited AI-generated code. GPT-4o cleared a critical User Enumeration flaw that Claude caught. A warning for anyone using AI for Auth flows. & \url{https://www.reddit.com/r/webdev/comments/1s80d0d/i_audited_aigenerated_code_gpt4o_cleared_a/} \\
G1528 & 4582 & Show /r/rust: FirstCall, an egui + CLI workbench for verified API recipes and MCP agent packages & \url{https://www.reddit.com/r/rust/comments/1taqoht/show_rrust_firstcall_an_egui_cli_workbench_for/} \\
G1529 & 4585 & Claude Code's CLI feels like a black box. I built a local UI to un-dumb it, and it unexpectedly blew up last week. & \url{https://www.reddit.com/r/ClaudeAI/comments/1r90pol/claude_codes_cli_feels_like_a_black_box_i_built_a/} \\
G1530 & 4590 & My Claude superpower: "please review for correctness" & \url{https://www.reddit.com/r/ClaudeAI/comments/1lv5z3e/my_claude_superpower_please_review_for_correctness/} \\
G1531 & 4592 & AMA on GitHub Universe releases tomorrow (November 5) & \url{https://www.reddit.com/r/GithubCopilot/comments/1oob5ww/ama_on_github_universe_releases_tomorrow_november/} \\
G1532 & 4594 & AI Coding Review Platform with browser access & \url{https://www.reddit.com/r/vibecoding/comments/1sz81pc/ai_coding_review_platform_with_browser_access/} \\
G1533 & 4595 & I built ClaudeNest entirely with Claude Code -- a remote launcher for Claude Code's remote-control sessions & \url{https://www.reddit.com/r/ClaudeAI/comments/1rpe9zk/i_built_claudenest_entirely_with_claude_code_a/} \\
G1534 & 4599 & Your AI coding stack isn't the problem. Your process is. & \url{https://www.reddit.com/r/vibecoding/comments/1r9t36b/your_ai_coding_stack_isnt_the_problem_your/} \\
G1535 & 4600 & Made my first webapp using Cursor & \url{https://www.reddit.com/r/ClaudeAI/comments/1pppuf8/made_my_first_webapp_using_cursor/} \\
G1536 & 4604 & Architect Genesis - open source architecture analysis CLI with AST scoring, refactoring, and AI-assisted execution & \url{https://www.reddit.com/r/softwarearchitecture/comments/1sf1y2y/architect_genesis_open_source_architecture/} \\
G1537 & 4607 & The Method Behind Managing AI on a Million-Line Codebase & \url{https://www.reddit.com/r/ClaudeAI/comments/1rr36wx/the_method_behind_managing_ai_on_a_millionline/} \\
G1538 & 4611 & Best way to deal with PR reviewer that asks for refactors and edge cases not contemplated in the original task? & \url{https://www.reddit.com/r/ExperiencedDevs/comments/1ph1v04/best_way_to_deal_with_pr_reviewer_that_asks_for/} \\
G1539 & 4613 & CI/CD pipelines are now more critical than the code itself & \url{https://www.reddit.com/r/devops/comments/1q5imk5/cicd_pipelines_are_now_more_critical_than_the/} \\
G1540 & 4615 & Claude's Traning Data is the cause of Uneccessary Overuse & \url{https://www.reddit.com/r/ClaudeAI/comments/1mfg7q5/claudes_traning_data_is_the_cause_of_uneccessary/} \\
G1541 & 4617 & Getting Claude to output accurate line numbers in diffs -- the [L:XXX] prefix trick & \url{https://www.reddit.com/r/ClaudeAI/comments/1qfrb48/getting_claude_to_output_accurate_line_numbers_in/} \\
G1542 & 4622 & I no longer enjoy doing web dev professionally & \url{https://www.reddit.com/r/webdev/comments/1ryvqnq/i_no_longer_enjoy_doing_web_dev_professionally/} \\
G1543 & 4625 & My website: Juegosantiguos.com.ar & \url{https://www.reddit.com/r/vibecoding/comments/1s65rct/my_website_juegosantiguoscomar/} \\
G1544 & 4630 & A coding agent session manager that manages itself via its own MCP & \url{https://www.reddit.com/r/ClaudeAI/comments/1rtiqne/a_coding_agent_session_manager_that_manages/} \\
G1545 & 4642 & Is my debugging time actually going up? or am I just doing it wrong? & \url{https://www.reddit.com/r/ClaudeAI/comments/1qawsd7/is_my_debugging_time_actually_going_up_or_am_i/} \\
G1546 & 4645 & Need opinions from others. Currently dealing with stakeholders who have really high expectations with AI coding & \url{https://www.reddit.com/r/ClaudeAI/comments/1qc9fgu/need_opinions_from_others_currently_dealing_with/} \\
G1547 & 4648 & Are Junior engineers going to be able to adapt to the future of AI assisted coding? & \url{https://www.reddit.com/r/cscareerquestions/comments/1rdcbme/are_junior_engineers_going_to_be_able_to_adapt_to/} \\
G1548 & 4649 & Should you call out suspected bad use of AI for writing code? & \url{https://www.reddit.com/r/ExperiencedDevs/comments/1jkzedo/should_you_call_out_suspected_bad_use_of_ai_for/} \\
G1549 & 4653 & GitComet 0.1.11 now supports C\# syntax highlight & \url{https://www.reddit.com/r/csharp/comments/1smv77v/gitcomet_0111_now_supports_c_syntax_highlight/} \\
G1550 & 4659 & Devs using AI coding agents: where does trust break in your workflow? & \url{https://www.reddit.com/r/AI_Agents/comments/1tk6j3r/devs_using_ai_coding_agents_where_does_trust/} \\
G1551 & 4660 & Does this MIT study on AI coding tools match what you see in prod? & \url{https://www.reddit.com/r/devops/comments/1ov0bnr/does_this_mit_study_on_ai_coding_tools_match_what/} \\
G1552 & 4665 & Claude Code's full source just leaked via npm source maps -- here's what 512K lines of TypeScript reveal & \url{https://www.reddit.com/r/ClaudeAI/comments/1s92jg3/claude_codes_full_source_just_leaked_via_npm/} \\
G1553 & 4666 & AI SDK for golang, need opinions?! & \url{https://www.reddit.com/r/golang/comments/1roe8t5/ai_sdk_for_golang_need_opinions/} \\
G1554 & 4669 & How I'm currently balancing cost and effectivness (what about you?) & \url{https://www.reddit.com/r/cursor/comments/1p2idxz/how_im_currently_balancing_cost_and_effectivness/} \\
G1555 & 4672 & what's an ai use case you thought was gimmicky until you actually tried it & \url{https://www.reddit.com/r/ArtificialInteligence/comments/1sat7rb/whats_an_ai_use_case_you_thought_was_gimmicky/} \\
G1556 & 4675 & I made Github but for AI agents/Clawdbots to build and ship & \url{https://www.reddit.com/r/AI_Agents/comments/1qwi2fg/i_made_github_but_for_ai_agentsclawdbots_to_build/} \\
G1557 & 4676 & Harness Engineering: Plan -> Decompose -> Spawn SubAgents -> Verify Loop -- Any Existing Solutions or Best Practices? & \url{https://www.reddit.com/r/ClaudeAI/comments/1s3vwy1/harness_engineering_plan_decompose_spawn/} \\
G1558 & 4677 & Feedback on a small PR review tool I'm building & \url{https://www.reddit.com/r/codereview/comments/1otz6yx/feedback_on_a_small_pr_review_tool_im_building/} \\
G1559 & 4679 & Do AI coding agents actually save you time, or just create more cleanup? & \url{https://www.reddit.com/r/ClaudeAI/comments/1mdg6xm/do_ai_coding_agents_actually_save_you_time_or/} \\
G1560 & 4688 & I scanned 31 AI-built repos (Cursor, Lovable, Bolt) for code smells. Here's what I found. & \url{https://www.reddit.com/r/vibecoding/comments/1sgr1ri/i_scanned_31_aibuilt_repos_cursor_lovable_bolt/} \\
G1561 & 4689 & can someone explain the Copilot cloud agent? (and general usage tips) & \url{https://www.reddit.com/r/GithubCopilot/comments/1sb3ppe/can_someone_explain_the_copilot_cloud_agent_and/} \\
G1562 & 4691 & What if Copilot was a pipeline instead of a pair programmer? & \url{https://www.reddit.com/r/GithubCopilot/comments/1rojhne/what_if_copilot_was_a_pipeline_instead_of_a_pair/} \\
G1563 & 4694 & Agents are life & \url{https://www.reddit.com/r/ClaudeAI/comments/1oux9ab/agents_are_life/} \\
G1564 & 4701 & Claude Constructor: What if Ralph Wiggum had a project manager? & \url{https://www.reddit.com/r/ClaudeAI/comments/1qhy2rw/claude_constructor_what_if_ralph_wiggum_had_a/} \\
G1565 & 4706 & agents write code 10x faster but debugging is the same speed it was in 2019 & \url{https://www.reddit.com/r/ExperiencedDevs/comments/1szr1n0/agents_write_code_10x_faster_but_debugging_is_the/} \\
G1566 & 4713 & Spawn Time of SubAgents & \url{https://www.reddit.com/r/GithubCopilot/comments/1rpdq6c/spawn_time_of_subagents/} \\
G1567 & 4715 & Looking for code review for my recent project & \url{https://www.reddit.com/r/dotnet/comments/1jsr9j5/looking_for_code_review_for_my_recent_project/} \\
G1568 & 4719 & Claude now works my night shift, here's how I set up scheduled autonomous coding & \url{https://www.reddit.com/r/ClaudeAI/comments/1qflv3y/claude_now_works_my_night_shift_heres_how_i_set/} \\
G1569 & 4724 & We ran 52 controlled benchmarks on Claude Code. Agent Teams cost 73-124\% more than sequential with zero quality gain. & \url{https://www.reddit.com/r/ClaudeAI/comments/1ss7f38/we_ran_52_controlled_benchmarks_on_claude_code/} \\
G1570 & 4728 & For every \$1 spent on AI coding tools, only \$0.18 reaches production. Analyzed 1M+ PRs to find where the rest goes. & \url{https://www.reddit.com/r/artificial/comments/1tw2qa8/for_every_1_spent_on_ai_coding_tools_only_018/} \\
G1571 & 4730 & I built a Claude Code plugin marketplace with 15 production-ready AI agents [Open Source] & \url{https://www.reddit.com/r/ClaudeAI/comments/1o2lvrn/i_built_a_claude_code_plugin_marketplace_with_15/} \\
G1572 & 4731 & Parents: how to teach kids that AI is a skill, not just a shortcut? & \url{https://www.reddit.com/r/ArtificialInteligence/comments/1mpriz3/parents_how_to_teach_kids_that_ai_is_a_skill_not/} \\
G1573 & 4738 & What are some of the ways to use GPT Pro? & \url{https://www.reddit.com/r/ChatGPTPro/comments/1tqkwb7/what_are_some_of_the_ways_to_use_gpt_pro/} \\
G1574 & 4744 & Where do you fall on the Agentic Coder Spectrum? I mapped out 5 levels of AI adoption among developers! & \url{https://www.reddit.com/r/programming/comments/1qdkuvo/where_do_you_fall_on_the_agentic_coder_spectrum_i/} \\
G1575 & 4745 & Help with my workflow please & \url{https://www.reddit.com/r/ClaudeAI/comments/1tjlkss/help_with_my_workflow_please/} \\
G1576 & 4749 & I open-sourced an AI agent that turns Jira \& ClickUp tickets into GitHub PRs & \url{https://www.reddit.com/r/programming/comments/1qrzd9s/i_opensourced_an_ai_agent_that_turns_jira_clickup/} \\
G1577 & 4751 & My startup vibe coded so hard that we don't even know how to fix and identify bugs & \url{https://www.reddit.com/r/ChatGPTCoding/comments/1ltlyb1/my_startup_vibe_coded_so_hard_that_we_dont_even/} \\
G1578 & 4757 & We Tracked Every Tool Call Claude Code Made for 6 Weeks. 50\% of Sessions Had At Least One Violation That Would Fail Code Review. & \url{https://www.reddit.com/r/ClaudeAI/comments/1r2k01w/we_tracked_every_tool_call_claude_code_made_for_6/} \\
G1579 & 4759 & Any companies with management with sane, down-to-earth attitudes towards AI? & \url{https://www.reddit.com/r/webdev/comments/1tgxsmv/any_companies_with_management_with_sane/} \\
G1580 & 4760 & Which Al code review tool to use? & \url{https://www.reddit.com/r/QualityAssurance/comments/1q6fcfx/which_al_code_review_tool_to_use/} \\
G1581 & 4763 & Multi llms inside ClaudeCode & \url{https://www.reddit.com/r/AI_Agents/comments/1q5mzmt/multi_llms_inside_claudecode/} \\
G1582 & 4764 & [Feedback/Review] Rabbit: Self-Hosted TCP Tunnel Server/Client in Go (ngrok Alternative) & \url{https://www.reddit.com/r/golang/comments/1lsu3oa/feedbackreview_rabbit_selfhosted_tcp_tunnel/} \\
G1583 & 4765 & I made a skill that can let Claude Code run 24 / 7 & \url{https://www.reddit.com/r/vibecoding/comments/1syb0y1/i_made_a_skill_that_can_let_claude_code_run_24_7/} \\
G1584 & 4771 & Multi-agent AI code review: 17 of 18 findings false. Lessons from burning credits & \url{https://www.reddit.com/r/devops/comments/1tgpnfb/multiagent_ai_code_review_17_of_18_findings_false/} \\
G1585 & 4772 & Cursor keeps removing essential features and it's killing my workflow & \url{https://www.reddit.com/r/cursor/comments/1p395v1/cursor_keeps_removing_essential_features_and_its/} \\
G1586 & 4776 & I built a self-hosted Claude Code wrapper - here's what I learned about autonomous coding & \url{https://www.reddit.com/r/ClaudeAI/comments/1qmqc69/i_built_a_selfhosted_claude_code_wrapper_heres/} \\
G1587 & 4777 & Claude made me faster, but also made me responsible for more "almost finished" work & \url{https://www.reddit.com/r/ClaudeAI/comments/1tjjig6/claude_made_me_faster_but_also_made_me/} \\
G1588 & 4779 & Have any of you fully switched from SonarQube to a SonarQube alternative and never looked back? & \url{https://www.reddit.com/r/devops/comments/1neb934/have_any_of_you_fully_switched_from_sonarqube_to/} \\
G1589 & 4780 & Boost.Multi Review Begins Today & \url{https://www.reddit.com/r/cpp/comments/1rlhm4g/boostmulti_review_begins_today/} \\
G1590 & 4782 & Generate flowcharts of your code . NET & \url{https://www.reddit.com/r/csharp/comments/1pswuru/generate_flowcharts_of_your_code_net/} \\
G1591 & 4786 & Building a training app with AI doing most of the coding - is this actually viable? What do I need to know about security? & \url{https://www.reddit.com/r/webdev/comments/1sn1apw/building_a_training_app_with_ai_doing_most_of_the/} \\
G1592 & 4797 & How do you ensure effective communication during code reviews on GitHub? & \url{https://www.reddit.com/r/github/comments/1q8f01s/how_do_you_ensure_effective_communication_during/} \\
G1593 & 4802 & I built a Claude Code plugin that reviews code using 12 classic books -- then ran it on itself & \url{https://www.reddit.com/r/ClaudeAI/comments/1sj9813/i_built_a_claude_code_plugin_that_reviews_code/} \\
G1594 & 4805 & Vibe coding as a senior engineer & \url{https://www.reddit.com/r/vibecoding/comments/1n6pevq/vibe_coding_as_a_senior_engineer/} \\
G1595 & 4807 & Most Code Reviews Are Just Bullying with Syntax Highlighting. Here's a Better Way. & \url{https://www.reddit.com/r/programming/comments/1pd75qw/most_code_reviews_are_just_bullying_with_syntax/} \\
G1596 & 4808 & What does your dev workflow look like? & \url{https://www.reddit.com/r/GithubCopilot/comments/1te9p7z/what_does_your_dev_workflow_look_like/} \\
G1597 & 4811 & Cursor has made developers really lazy about unit tests & \url{https://www.reddit.com/r/ExperiencedDevs/comments/1p4hicy/cursor_has_made_developers_really_lazy_about_unit/} \\
G1598 & 4821 & Does AI generated code change the review process itself? & \url{https://www.reddit.com/r/ExperiencedDevs/comments/1t1zfsz/does_ai_generated_code_change_the_review_process/} \\
G1599 & 4822 & How we cut code review time by 50\% and caught 1,200 issues in a month & \url{https://www.reddit.com/r/SoftwareEngineering/comments/1jpxeoq/how_we_cut_code_review_time_by_50_and_caught_1200/} \\
G1600 & 4824 & Learning programming by building real projects -- but using AI intentionally as a mentor, not a shortcut & \url{https://www.reddit.com/r/ClaudeAI/comments/1qr3z11/learning_programming_by_building_real_projects/} \\
G1601 & 4826 & Six months running multi-agent in production -- the coordination patterns & \url{https://www.reddit.com/r/AI_Agents/comments/1sz6s04/six_months_running_multiagent_in_production_the/} \\
G1602 & 4835 & Anyone actually used Claude Code + WordPress MCP for production site work? How did it go? & \url{https://www.reddit.com/r/ClaudeAI/comments/1spnoo0/anyone_actually_used_claude_code_wordpress_mcp/} \\
G1603 & 4836 & Claude Code deployed my client's financial data to a public URL. And other failure modes from daily production use. & \url{https://www.reddit.com/r/ClaudeAI/comments/1roxb2z/claude_code_deployed_my_clients_financial_data_to/} \\
G1604 & 4840 & final 2 days -- claude code bootcamp may 30 & \url{https://www.reddit.com/r/ClaudeAI/comments/1tpyeb6/final_2_days_claude_code_bootcamp_may_30/} \\
G1605 & 4844 & I built an open-source verification skill for Claude Code that catches security issues, hallucinated tools, infinite loops and many more! & \url{https://www.reddit.com/r/AI_Agents/comments/1sxcmph/i_built_an_opensource_verification_skill_for/} \\
G1606 & 4845 & Introducing Lynkr -- an open-source Claude-style AI coding proxy built specifically for Databricks model endpoints & \url{https://www.reddit.com/r/LocalLLaMA/comments/1pdrvh8/introducing_lynkr_an_opensource_claudestyle_ai/} \\
G1607 & 4848 & How do you or have dealt with team members that seem to not care about code quality/standards? & \url{https://www.reddit.com/r/ExperiencedDevs/comments/1lqmm9e/how_do_you_or_have_dealt_with_team_members_that/} \\
G1608 & 4849 & Pricing codebase audit & \url{https://www.reddit.com/r/ExperiencedDevs/comments/1sgvrxo/pricing_codebase_audit/} \\
G1609 & 4850 & I'm a junior dev with 1YO and have been unemployed almost a year, now I will sign a new offer where I will be the only IT guy in the small company. there is no  & \url{https://www.reddit.com/r/AskProgramming/comments/1kajwzj/im_a_junior_dev_with_1yo_and_have_been_unemployed/} \\
G1610 & 4853 & Someone automated the process of scanning every public GitHub repo for exploitable CI workflows. We are cooked & \url{https://www.reddit.com/r/github/comments/1rmh3n8/someone_automated_the_process_of_scanning_every/} \\
G1611 & 4855 & Devs: How do you deal with code that works but makes no sense? & \url{https://www.reddit.com/r/github/comments/1l85ikq/devs_how_do_you_deal_with_code_that_works_but/} \\
G1612 & 4861 & Anthropic is going to charge 50X more for Claude Code on June 15th. You need to make your workflow provider agnostic. Here is Why (And How). & \url{https://www.reddit.com/r/AI_Agents/comments/1tccjyz/anthropic_is_going_to_charge_50x_more_for_claude/} \\
G1613 & 4862 & How do I handle this scenario with Git? Do I merge, rebase or cherry-pick? & \url{https://www.reddit.com/r/AskProgramming/comments/1knas3l/how_do_i_handle_this_scenario_with_git_do_i_merge/} \\
G1614 & 4863 & Race conditions in generated code (tested across 10 models, 5 runs) & \url{https://www.reddit.com/r/programming/comments/1rsx3bo/race_conditions_in_generated_code_tested_across/} \\
G1615 & 4866 & I've vibe coded 7 full-stack apps. There are a few 'Time Bombs' I wanna share with you guys. If you are a vibe coder as well, read these so you don't lose your  & \url{https://www.reddit.com/r/vibecoding/comments/1rv0irm/ive_vibe_coded_7_fullstack_apps_there_are_a_few/} \\
G1616 & 4867 & I stopped tracking todos and started slinging prompts instead & \url{https://www.reddit.com/r/vibecoding/comments/1q6wq0z/i_stopped_tracking_todos_and_started_slinging/} \\
G1617 & 4871 & I vibe coded a whole ASS presentation generator website with Cursor, looking for some feedback & \url{https://www.reddit.com/r/vibecoding/comments/1mv2yt2/i_vibe_coded_a_whole_ass_presentation_generator/} \\
G1618 & 4872 & Vibe Coding Debt - My experience implementing projects in production with Vibe Coding & \url{https://www.reddit.com/r/vibecoding/comments/1npa9lo/vibe_coding_debt_my_experience_implementing/} \\
G1619 & 4873 & Beyond single-agent coding: What if Claude instances worked as an entire dev organization? & \url{https://www.reddit.com/r/ClaudeAI/comments/1q8mayw/beyond_singleagent_coding_what_if_claude/} \\
G1620 & 4874 & opus 4.6 vs gpt 5.3, my take as a & \url{https://www.reddit.com/r/ClaudeAI/comments/1qxcb1j/opus_46_vs_gpt_53_my_take_as_a/} \\
G1621 & 4878 & Review after 1 month of Pro & \url{https://www.reddit.com/r/Codeium/comments/1htimla/review_after_1_month_of_pro/} \\
G1622 & 4881 & Claude code 8+ hours a day, 100M tokens a month -- built a review tool for the one thing that's still broken & \url{https://www.reddit.com/r/ClaudeAI/comments/1rnv2ay/claude_code_8_hours_a_day_100m_tokens_a_month/} \\
G1623 & 4887 & Opus 4.6 vs Codex 5.3 in the Swiftagon: FIGHT! & \url{https://www.reddit.com/r/ClaudeAI/comments/1qwvj5k/opus_46_vs_codex_53_in_the_swiftagon_fight/} \\
G1624 & 4892 & Devs, would you use this? I'm building an AI Code Reviewer that actually understands your codebase. & \url{https://www.reddit.com/r/programming/comments/1jwxb1j/devs_would_you_use_this_im_building_an_ai_code/} \\
G1625 & 4898 & Senior dev here. Spent two years trying to fix our code review culture with AI tools. Here's what actually worked and what didn't. & \url{https://www.reddit.com/r/webdev/comments/1rq99rq/senior_dev_here_spent_two_years_trying_to_fix_our/} \\
G1626 & 4900 & cursor for code review is... weirdly inconsistent? & \url{https://www.reddit.com/r/cursor/comments/1n347od/cursor_for_code_review_is_weirdly_inconsistent/} \\
G1627 & 4906 & How do you know your AI audit tool actually checked everything? I was fairly confident that my skill suite did. It didn't. & \url{https://www.reddit.com/r/ChatGPTCoding/comments/1s6cjaq/how_do_you_know_your_ai_audit_tool_actually/} \\
G1628 & 4909 & My top 5 tools for my AI coding workflow & \url{https://www.reddit.com/r/vibecoding/comments/1o1g7ac/my_top_5_tools_for_my_ai_coding_workflow/} \\
G1629 & 4914 & Where did Devin go, and what does it say about the future of AI dev tools? & \url{https://www.reddit.com/r/webdev/comments/1qpbngr/where_did_devin_go_and_what_does_it_say_about_the/} \\
G1630 & 4917 & Running a website selling agency with Claude doing 80\% of the work -- what's actually worth adding to my workflow? & \url{https://www.reddit.com/r/ClaudeAI/comments/1to2nft/running_a_website_selling_agency_with_claude/} \\
G1631 & 4919 & Handling Tenured Deadweight as a Newcomer & \url{https://www.reddit.com/r/ExperiencedDevs/comments/1njix29/handling_tenured_deadweight_as_a_newcomer/} \\
G1632 & 4923 & Lesson learned about PR requests / code reviews & \url{https://www.reddit.com/r/ExperiencedDevs/comments/1krjhk3/lesson_learned_about_pr_requests_code_reviews/} \\
G1633 & 4924 & Went from 6 AI coding tools (\$150/mo) to 2 (\$44/mo) - here's my new workflow & \url{https://www.reddit.com/r/cursor/comments/1pbdttp/went_from_6_ai_coding_tools_150mo_to_2_44mo_heres/} \\
G1634 & 4927 & how to disable GitHub copilot code review on repositories ? & \url{https://www.reddit.com/r/GithubCopilot/comments/1qt94oa/how_to_disable_github_copilot_code_review_on/} \\
G1635 & 4932 & Greptile alternatives please? & \url{https://www.reddit.com/r/webdev/comments/1ouapni/greptile_alternatives_please/} \\
G1636 & 4933 & HUD + Circuit Breaker for Claude Code & \url{https://www.reddit.com/r/ClaudeAI/comments/1q1oeex/hud_circuit_breaker_for_claude_code/} \\
G1637 & 4935 & PullMD v2.4.1 is out - claude.ai web custom connector works natively now, plus what 2 weeks of your feedback turned into & \url{https://www.reddit.com/r/ClaudeAI/comments/1tbz2j6/pullmd_v241_is_out_claudeai_web_custom_connector/} \\
G1638 & 4936 & Introducing Lab Partner: A Chrome Extension to Simplify GitLab Merge Request Management & \url{https://www.reddit.com/r/gitlab/comments/1igva7f/introducing_lab_partner_a_chrome_extension_to/} \\
G1639 & 4940 & What terminals and tools are people using with their coding agents? & \url{https://www.reddit.com/r/AI_Agents/comments/1rkmexc/what_terminals_and_tools_are_people_using_with/} \\
G1640 & 4944 & Roast Me (Code Review) Wordle Clone in Dioxus & \url{https://www.reddit.com/r/rust/comments/1rtznp6/roast_me_code_review_wordle_clone_in_dioxus/} \\
G1641 & 4946 & I built ClaudeDesk - a PWA interface for Claude Code with session persistence and visual tool tracking & \url{https://www.reddit.com/r/ClaudeAI/comments/1qodaz3/i_built_claudedesk_a_pwa_interface_for_claude/} \\
G1642 & 4949 & [v2 Release] Claude Code Development Kit - Now with Hooks \& Installer & \url{https://www.reddit.com/r/ClaudeAI/comments/1lwjebe/v2_release_claude_code_development_kit_now_with/} \\
G1643 & 4953 & How Do You Balance PRs, Docs, and Contributors? I'm overwhelmed. & \url{https://www.reddit.com/r/github/comments/1qe1tqg/how_do_you_balance_prs_docs_and_contributors_im/} \\
G1644 & 4957 & I scanned 10 popular vibe-coded repos with a deterministic linter. 4,513 findings across 2,062 files. Here's what AI agents keep getting wrong. & \url{https://www.reddit.com/r/GithubCopilot/comments/1saseax/i_scanned_10_popular_vibecoded_repos_with_a/} \\
G1645 & 4958 & Copilot code review: Now in JetBrains IDEs and Visual Studio - GitHub Changelog & \url{https://www.reddit.com/r/GithubCopilot/comments/1nkgv8l/copilot_code_review_now_in_jetbrains_ides_and/} \\
G1646 & 4959 & I built a "Zero Trust" linter for AI-generated code & \url{https://www.reddit.com/r/programming/comments/1q768at/i_built_a_zero_trust_linter_for_aigenerated_code/} \\
G1647 & 4963 & [Project] Building a Professional-Grade C++17 Application from Scratch: Emotion Engine - Open Source, Looking for Contributors \& Feedback & \url{https://www.reddit.com/r/cpp/comments/1tdmssf/project_building_a_professionalgrade_c17/} \\
G1648 & 4964 & I found a 14x cloud cost bug in Airbyte's S3 connector. Here's the full technical breakdown. & \url{https://www.reddit.com/r/Python/comments/1ove9is/i_found_a_14x_cloud_cost_bug_in_airbytes_s3/} \\
G1649 & 4971 & Claude Code truly saves me from ADHD & \url{https://www.reddit.com/r/ClaudeAI/comments/1m4iz3v/claude_code_truly_saves_me_from_adhd/} \\
G1650 & 4973 & Is my team dysfunctional? & \url{https://www.reddit.com/r/ExperiencedDevs/comments/1j74086/is_my_team_dysfunctional/} \\
G1651 & 4977 & guys i have question , how you really cope with all those lines generated by Claude do you actually review every single line ? it's steal all the funny & \url{https://www.reddit.com/r/ClaudeAI/comments/1mp7lu9/guys_i_have_question_how_you_really_cope_with_all/} \\
G1652 & 4981 & Claude Command Suite - https://github.com/qdhenry/Claude-Command-Suite I created a collection of professional slash commands for Anthropic's Claude Code to stre & \url{https://www.reddit.com/r/ClaudeAI/comments/1la4jrt/claude_command_suite/} \\
G1653 & 4982 & We built a 39-agent orchestration platform on Claude Code... here's the architecture for deterministic AI development at scale & \url{https://www.reddit.com/r/ClaudeAI/comments/1qxmybe/we_built_a_39agent_orchestration_platform_on/} \\
G1654 & 4987 & Experiences calling out excessive vibe coding to prevent wasting time reviewing bad PRs? & \url{https://www.reddit.com/r/ExperiencedDevs/comments/1pem6eg/experiences_calling_out_excessive_vibe_coding_to/} \\
G1655 & 4994 & Can anybody give me advice for a vibe coding agent team? & \url{https://www.reddit.com/r/vibecoding/comments/1r6pu7n/can_anybody_give_me_advice_for_a_vibe_coding/} \\
G1656 & 4996 & Don't blindly trust an LLM output ! & \url{https://www.reddit.com/r/vibecoding/comments/1qmll0d/dont_blindly_trust_an_llm_output/} \\
G1657 & 4997 & Do you often experience a feeling that the company has raised coding standards not long before you joined? & \url{https://www.reddit.com/r/ExperiencedDevs/comments/1oudo1d/do_you_often_experience_a_feeling_that_the/} \\
G1658 & 5000 & How to win my Claude Code money over as a CC + Cursor user. & \url{https://www.reddit.com/r/cursor/comments/1lrkwf1/how_to_win_my_claude_code_money_over_as_a_cc/} \\
G1659 & 5001 & What's your take on good code review? & \url{https://www.reddit.com/r/ExperiencedDevs/comments/1k0tfr7/whats_your_take_on_good_code_review/} \\
G1660 & 5003 & What's the best LLM based tool for code reviews? & \url{https://www.reddit.com/r/cursor/comments/1kxdvbc/whats_the_best_llm_based_tool_for_code_reviews/} \\
G1661 & 5004 & Always nice to see Cursor running multiple subagents without asking. & \url{https://www.reddit.com/r/cursor/comments/1ryg52h/always_nice_to_see_cursor_running_multiple/} \\
G1662 & 5006 & Please test our AI code review tool out! & \url{https://www.reddit.com/r/codereview/comments/1rq9gc4/please_test_our_ai_code_review_tool_out/} \\
G1663 & 5008 & I'm the reviewer everyone waits for and I hate it & \url{https://www.reddit.com/r/codereview/comments/1n3cx9u/im_the_reviewer_everyone_waits_for_and_i_hate_it/} \\
G1664 & 5009 & Reviewed 25 vibe-coded apps as a senior engineer. Here's what I found in all of them. & \url{https://www.reddit.com/r/vibecoding/comments/1tk0xp4/reviewed_25_vibecoded_apps_as_a_senior_engineer/} \\
G1665 & 5015 & We went from 40\% to 92\% architectural compliance after changing HOW we give AI context (not how much) & \url{https://www.reddit.com/r/cursor/comments/1pbag9x/we_went_from_40_to_92_architectural_compliance/} \\
G1666 & 5017 & What are your to go content creators to get information about AI or just tech without BS? & \url{https://www.reddit.com/r/ExperiencedDevs/comments/1rhf9f2/what_are_your_to_go_content_creators_to_get/} \\
G1667 & 5021 & Not sure if I'm a bad Staff that just got lucky in the past or my new company sucks and set up for failure & \url{https://www.reddit.com/r/ExperiencedDevs/comments/1muwxoi/not_sure_if_im_a_bad_staff_that_just_got_lucky_in/} \\
G1668 & 5023 & tried using ai to refactor a 3 year old express api. it broke prod & \url{https://www.reddit.com/r/webdev/comments/1p8vbyp/tried_using_ai_to_refactor_a_3_year_old_express/} \\
G1669 & 5024 & RalphTerm: a ralph-style loop for Claude Code with fresh-session Codex cross-review & \url{https://www.reddit.com/r/ClaudeAI/comments/1thpfxg/ralphterm_a_ralphstyle_loop_for_claude_code_with/} \\
G1670 & 5028 & The native chat window in your code editor is a glorified toy that completely breaks complex file dependencies. & \url{https://www.reddit.com/r/cursor/comments/1saqfyt/the_native_chat_window_in_your_code_editor_is_a/} \\
G1671 & 5029 & What's the closest thing today to having an actual AI developer on my team? & \url{https://www.reddit.com/r/ChatGPTCoding/comments/1ngcocs/whats_the_closest_thing_today_to_having_an_actual/} \\
G1672 & 5035 & If you're not using the coderabbit extension, use it & \url{https://www.reddit.com/r/cursor/comments/1kx1n61/if_youre_not_using_the_coderabbit_extension_use_it/} \\
G1673 & 5038 & AI coding tools are generating technical debt faster than teams realize and context is the reason why & \url{https://www.reddit.com/r/ArtificialInteligence/comments/1tf37c2/ai_coding_tools_are_generating_technical_debt/} \\
G1674 & 5040 & Built a Claude Skill for code quality checking (Sui Move language) & \url{https://www.reddit.com/r/ClaudeAI/comments/1odaz95/built_a_claude_skill_for_code_quality_checking/} \\
G1675 & 5045 & C\# Space Shooter Code Review & \url{https://www.reddit.com/r/csharp/comments/1kjib30/c_space_shooter_code_review/} \\
G1676 & 5047 & Please vote for custom code review instructions in the Codex app & \url{https://www.reddit.com/r/ChatGPTCoding/comments/1rrt3zx/please_vote_for_custom_code_review_instructions/} \\
G1677 & 5050 & Any other tools similar to Copilot PR? Any recs? & \url{https://www.reddit.com/r/GithubCopilot/comments/1okwd6c/any_other_tools_similar_to_copilot_pr_any_recs/} \\
G1678 & 5056 & I use Claude Code from my phone through Telegram now & \url{https://www.reddit.com/r/ClaudeAI/comments/1sb3evn/i_use_claude_code_from_my_phone_through_telegram/} \\
G1679 & 5058 & Open-sourced a 4-agent code review workflow. Wrap it as an MCP and your Claude Code calls it instead of CodeRabbit. built on heym & \url{https://www.reddit.com/r/ClaudeAI/comments/1t71wrm/opensourced_a_4agent_code_review_workflow_wrap_it/} \\
G1680 & 5059 & Can I get a review of my implementation of Singly Linked List. & \url{https://www.reddit.com/r/rust/comments/1q0fhdh/can_i_get_a_review_of_my_implementation_of_singly/} \\
G1681 & 5062 & Built an MCP server that teaches AI agents your codebase patterns. heres what it actually does & \url{https://www.reddit.com/r/GithubCopilot/comments/1qjxctn/built_an_mcp_server_that_teaches_ai_agents_your/} \\
G1682 & 5066 & Code review or performance review? & \url{https://www.reddit.com/r/ExperiencedDevs/comments/1qcqz3d/code_review_or_performance_review/} \\
G1683 & 5068 & The Illusion of Speed: Is AI Actually Slowing Development? & \url{https://www.reddit.com/r/ClaudeAI/comments/1i2bxj3/the_illusion_of_speed_is_ai_actually_slowing/} \\
G1684 & 5070 & How I turned a 249 files PR into a piece of cake to review :) & \url{https://www.reddit.com/r/AI_Agents/comments/1svmqfm/how_i_turned_a_249_files_pr_into_a_piece_of_cake/} \\
G1685 & 5075 & Do you actually review AI generated code or just trust it? & \url{https://www.reddit.com/r/vibecoding/comments/1n49zk7/do_you_actually_review_ai_generated_code_or_just/} \\
G1686 & 5076 & Cursor is a really resourceful junior dev & \url{https://www.reddit.com/r/cursor/comments/1jmka76/cursor_is_a_really_resourceful_junior_dev/} \\
G1687 & 5077 & One of the items in my PIP is because I had a difference of opinion during a code review. & \url{https://www.reddit.com/r/cscareerquestions/comments/1mecblh/one_of_the_items_in_my_pip_is_because_i_had_a/} \\
G1688 & 5078 & Claude Code Source Deep Dive (Part 3): Full System Prompt Assembly Flow + Original Prompt Text (2) & \url{https://www.reddit.com/r/ClaudeAI/comments/1semwp1/claude_code_source_deep_dive_part_3_full_system/} \\
G1689 & 5082 & Terrified I won't get this job and don't know how to prepare & \url{https://www.reddit.com/r/cscareerquestions/comments/1nqfga6/terrified_i_wont_get_this_job_and_dont_know_how/} \\
G1690 & 5089 & how do you actually review AI generated code? & \url{https://www.reddit.com/r/AskProgramming/comments/1l9dj9a/how_do_you_actually_review_ai_generated_code/} \\
G1691 & 5091 & What would you do? & \url{https://www.reddit.com/r/webdev/comments/1j163yq/what_would_you_do/} \\
G1692 & 5092 & [WIP] Building a "Brain" for RooCode - Autonomous AI Dev Framework (Looking for 1-2 collaborators) & \url{https://www.reddit.com/r/vibecoding/comments/1kqwvjf/wip_building_a_brain_for_roocode_autonomous_ai/} \\
G1693 & 5096 & Senior engineer best practice for scaling yourself with Claude Code & \url{https://www.reddit.com/r/ClaudeAI/comments/1sdne02/senior_engineer_best_practice_for_scaling/} \\
G1694 & 5100 & What's the best piece of feedback you ever got in a code review? & \url{https://www.reddit.com/r/ExperiencedDevs/comments/1ocnkn0/whats_the_best_piece_of_feedback_you_ever_got_in/} \\
G1695 & 5107 & I created a Claude Skill for Claude Code and Codex Workflow (with a memory layer) & \url{https://www.reddit.com/r/ClaudeAI/comments/1oivrpm/i_created_a_claude_skill_for_claude_code_and/} \\
G1696 & 5112 & Switching from GitHub Copilot to Claude Code, Trying to Understand the Real Advantage & \url{https://www.reddit.com/r/vibecoding/comments/1sx92g3/switching_from_github_copilot_to_claude_code/} \\
G1697 & 5113 & Sextant: Making Claude Code Read Your Code Before Changing It & \url{https://www.reddit.com/r/ClaudeAI/comments/1s5uga4/sextant_making_claude_code_read_your_code_before/} \\
G1698 & 5114 & Softening the blow of taking a project off of an engineer & \url{https://www.reddit.com/r/ExperiencedDevs/comments/1njhz6z/softening_the_blow_of_taking_a_project_off_of_an/} \\
G1699 & 5115 & Needed a simple Cron AI scheduler and Claude helped bring it to life & \url{https://www.reddit.com/r/ClaudeAI/comments/1ppawsw/needed_a_simple_cron_ai_scheduler_and_claude/} \\
G1700 & 5117 & How we vibe code at a FAANG. & \url{https://www.reddit.com/r/vibecoding/comments/1myakhd/how_we_vibe_code_at_a_faang/} \\
G1701 & 5118 & Working with multiple GHCP Agents concurrently to get more work done never been this easy & \url{https://www.reddit.com/r/GithubCopilot/comments/1sqple6/working_with_multiple_ghcp_agents_concurrently_to/} \\
G1702 & 5126 & I run a team of AI agents on kubernetes & \url{https://www.reddit.com/r/ClaudeAI/comments/1rfevx6/i_run_a_team_of_ai_agents_on_kubernetes/} \\
G1703 & 5128 & Claude 4.5 In Kiro Wasted 3 Hours Trying to Fix a Bug, Cursor Auto Fixed in 30 Mins. & \url{https://www.reddit.com/r/cursor/comments/1o57w2e/claude_45_in_kiro_wasted_3_hours_trying_to_fix_a/} \\
G1704 & 5131 & How I built a 9-agent team where my agents actually talk to each other & \url{https://www.reddit.com/r/ClaudeAI/comments/1thi16y/how_i_built_a_9agent_team_where_my_agents/} \\
G1705 & 5137 & I Gave Opus 4.7 and 4.6 the Same Code Audit... The Results Surprised Me & \url{https://www.reddit.com/r/ClaudeAI/comments/1sqy9by/i_gave_opus_47_and_46_the_same_code_audit_the/} \\
G1706 & 5138 & I messed up and how I'm fixing it & \url{https://www.reddit.com/r/webdev/comments/1it555i/i_messed_up_and_how_im_fixing_it/} \\
G1707 & 5142 & Best resources to practice for code review phone screen (Senior Backend Engineer)? & \url{https://www.reddit.com/r/cscareerquestions/comments/1nonf7q/best_resources_to_practice_for_code_review_phone/} \\
G1708 & 5144 & QWEN 3.5 9B is SLOW & \url{https://www.reddit.com/r/LocalLLaMA/comments/1rjrp3v/qwen_35_9b_is_slow/} \\
G1709 & 5152 & Opus 4.6 went off on a hallucinatory tangent during code review, leaked plans for mass layoffs, then went on a spiral towards madness at the end. & \url{https://www.reddit.com/r/ClaudeAI/comments/1r50jt5/opus_46_went_off_on_a_hallucinatory_tangent/} \\
G1710 & 5153 & Beginners writing production code & \url{https://www.reddit.com/r/typescript/comments/1kf6pu2/beginners_writing_production_code/} \\
G1711 & 5157 & Amusing conversation with Claude Code during a code review & \url{https://www.reddit.com/r/ClaudeAI/comments/1pbblk6/amusing_conversation_with_claude_code_during_a/} \\
G1712 & 5164 & Went back to a codebase I built six months ago. Had to add console.logs to understand my own code. & \url{https://www.reddit.com/r/SoftwareEngineering/comments/1sffexb/went_back_to_a_codebase_i_built_six_months_ago/} \\
G1713 & 5166 & Astraeus -9000 -- Meta-Agent Orchestration Framework (Sub Agent Magic!) & \url{https://www.reddit.com/r/ClaudeAI/comments/1mb32ty/astraeus_9000_metaagent_orchestration_framework/} \\
G1714 & 5169 & Building security around vibe coded apps & \url{https://www.reddit.com/r/vibecoding/comments/1nqa534/building_security_around_vibe_coded_apps/} \\
G1715 & 5174 & \#\#\# Stopping Vibe Coding: An Optimized Workflow for Building Reliable Solutions & \url{https://www.reddit.com/r/vibecoding/comments/1n6pioa/stopping_vibe_coding_an_optimized_workflow_for/} \\
G1716 & 5175 & setting up openclaw "securely"? & \url{https://www.reddit.com/r/AI_Agents/comments/1rofots/setting_up_openclaw_securely/} \\
G1717 & 5177 & Whose premium requests are consumed when I set up automatic code review? & \url{https://www.reddit.com/r/GithubCopilot/comments/1s2peuk/whose_premium_requests_are_consumed_when_i_set_up/} \\
G1718 & 5184 & Ukrainian Rust Bootcamp - Summer 2026 & \url{https://www.reddit.com/r/rust/comments/1tu88g6/ukrainian_rust_bootcamp_summer_2026/} \\
G1719 & 5185 & I built a Python debugging tool that uses Semantic Analysis to determine what and where the issue is & \url{https://www.reddit.com/r/Python/comments/1oktdez/i_built_a_python_debugging_tool_that_uses/} \\
G1720 & 5187 & Three EF Core bugs I keep seeing in production .NET codebases (with the SQL each one generates) & \url{https://www.reddit.com/r/dotnet/comments/1t7cjzt/three_ef_core_bugs_i_keep_seeing_in_production/} \\
G1721 & 5188 & Yesterday I posted about building a finance app with Claude Code and 200+ comments told me I was going to get sued. Here's what we're actually doing. & \url{https://www.reddit.com/r/ClaudeAI/comments/1rrvbz0/yesterday_i_posted_about_building_a_finance_app/} \\
G1722 & 5190 & What I learned letting AI agents push to production without human review & \url{https://www.reddit.com/r/vibecoding/comments/1rd5egw/what_i_learned_letting_ai_agents_push_to/} \\
G1723 & 5196 & Beginner Go/Gin CRUD API - Seeking Code Review and Best Practices! & \url{https://www.reddit.com/r/golang/comments/1jmmnet/beginner_gogin_crud_api_seeking_code_review_and/} \\
G1724 & 5197 & KoboldCpp Smart Launcher: GUI \& CLI tools for tensor offload auto-tuning (developed with AI) & \url{https://www.reddit.com/r/LocalLLaMA/comments/1km75qq/koboldcpp_smart_launcher_gui_cli_tools_for_tensor/} \\
G1725 & 5199 & How to get Claude to run more autonomously & \url{https://www.reddit.com/r/ClaudeAI/comments/1sigr1t/how_to_get_claude_to_run_more_autonomously/} \\
G1726 & 5200 & A video on how I use Bito to catch code issues like Memory Leak in Java & \url{https://www.reddit.com/r/codereview/comments/1o64im9/a_video_on_how_i_use_bito_to_catch_code_issues/} \\
G1727 & 5204 & Where do you draw the line on letting an AI agent generate code in your .NET projects? & \url{https://www.reddit.com/r/dotnet/comments/1t78oxr/where_do_you_draw_the_line_on_letting_an_ai_agent/} \\
G1728 & 5205 & Replaced my dev workflow with a Kanban board that triggers AI agents -- here's how it actually works & \url{https://www.reddit.com/r/webdev/comments/1rne92z/replaced_my_dev_workflow_with_a_kanban_board_that/} \\
G1729 & 5206 & The Ultimate Vibe Coding Guide & \url{https://www.reddit.com/r/cursor/comments/1kisbaq/the_ultimate_vibe_coding_guide/} \\
G1730 & 5211 & Golang Code Reviews? & \url{https://www.reddit.com/r/golang/comments/1q5xdrb/golang_code_reviews/} \\
G1731 & 5213 & My coworker uses AI to reply to my PR review and I hate it & \url{https://www.reddit.com/r/ExperiencedDevs/comments/1nq5npn/my_coworker_uses_ai_to_reply_to_my_pr_review_and/} \\
G1732 & 5214 & My superior lets AI write all our code without reviewing it. Am I wrong for caring about code quality? & \url{https://www.reddit.com/r/webdev/comments/1rk69ht/my_superior_lets_ai_write_all_our_code_without/} \\
G1733 & 5215 & why is Claude Code chmoding +x a javsacript file? & \url{https://www.reddit.com/r/vibecoding/comments/1npqf2f/why_is_claude_code_chmoding_x_a_javsacript_file/} \\
G1734 & 5216 & Dealing with a difficult situation with co-worker & \url{https://www.reddit.com/r/ExperiencedDevs/comments/1jia4vl/dealing_with_a_difficult_situation_with_coworker/} \\
G1735 & 5222 & Mistakes I see engineers making in their code reviews & \url{https://www.reddit.com/r/webdev/comments/1sfafhn/mistakes_i_see_engineers_making_in_their_code/} \\
G1736 & 5223 & anyone else seeing companies build entire internal CI/CD wrappers specifically for AI-generated code? & \url{https://www.reddit.com/r/devops/comments/1r3utc9/anyone_else_seeing_companies_build_entire/} \\
G1737 & 5225 & CLI tool that scaffolds a complete Claude Code workflow into any project - agents, commands, skills, hooks, permissions & \url{https://www.reddit.com/r/ClaudeAI/comments/1si3voz/cli_tool_that_scaffolds_a_complete_claude_code/} \\
G1738 & 5228 & I don't miss bugs because of bad code, I miss them because I'm exhausted reconstructing runtime flow & \url{https://www.reddit.com/r/codereview/comments/1q6b258/i_dont_miss_bugs_because_of_bad_code_i_miss_them/} \\
G1739 & 5234 & Open-source+local Cursor for code review (big improvement on GitHub) & \url{https://www.reddit.com/r/codereview/comments/1myabtg/opensourcelocal_cursor_for_code_review_big/} \\
G1740 & 5237 & Example: Don't use 4o for code logic and conditionals! Use 3o instead. & \url{https://www.reddit.com/r/artificial/comments/1j0ic46/example_dont_use_4o_for_code_logic_and/} \\
G1741 & 5241 & Codex absolutely trashed my codebase. & \url{https://www.reddit.com/r/OpenAI/comments/1qwyr2a/codex_absolutely_trashed_my_codebase/} \\
G1742 & 5242 & How do you deal with non tech peers who try to suggest that they might be to and want to try to write code using AI tools? & \url{https://www.reddit.com/r/ExperiencedDevs/comments/1hv0sbm/how_do_you_deal_with_non_tech_peers_who_try_to/} \\
G1743 & 5244 & How I manage context while building on Claude Code & \url{https://www.reddit.com/r/ClaudeAI/comments/1rav9aw/how_i_manage_context_while_building_on_claude_code/} \\
G1744 & 5246 & How to effectively use multiple AI models (codex/claude/Gemini etc) to review each others work without constant copy pasting & \url{https://www.reddit.com/r/vibecoding/comments/1rf5drw/how_to_effectively_use_multiple_ai_models/} \\
G1745 & 5248 & Beta testers wanted: CLI tool to detect DB schema drift across Dev, Staging, Prod - Git-workflow, safe, reviewable. Currently MSSQL and MySQL & \url{https://www.reddit.com/r/devops/comments/1mhd4ba/beta_testers_wanted_cli_tool_to_detect_db_schema/} \\
G1746 & 5249 & Claude Code Ultra Review & \url{https://www.reddit.com/r/ClaudeAI/comments/1taylwg/claude_code_ultra_review/} \\
G1747 & 5253 & Improving AI Generated Code and UI as a Vibe Coder & \url{https://www.reddit.com/r/vibecoding/comments/1slz6ds/improving_ai_generated_code_and_ui_as_a_vibe_coder/} \\
G1748 & 5256 & They said our positions were safe when they brought in 40+ offshore contractors. Today half my team got slagged. Record profits too rn too. This is fine. & \url{https://www.reddit.com/r/ExperiencedDevs/comments/1q7k697/they_said_our_positions_were_safe_when_they/} \\
G1749 & 5258 & Graphite (now a top-100 Rust project) turns Rust into a functional, visual scripting language for graphics operations -- REQUESTING HELP to implement compiler b & \url{https://www.reddit.com/r/rust/comments/1lor3b4/graphite_now_a_top100_rust_project_turns_rust/} \\
G1750 & 5260 & I have 12 years of experience and Cursor just made my entire career feel like a joke & \url{https://www.reddit.com/r/webdev/comments/1r5lj86/i_have_12_years_of_experience_and_cursor_just/} \\
G1751 & 5268 & Thanks to AI: Dream Achieved, Money Saved! & \url{https://www.reddit.com/r/vibecoding/comments/1lbewug/thanks_to_ai_dream_achieved_money_saved/} \\
G1752 & 5269 & Teammates using cursor but i'm the one fixing their mistakes & \url{https://www.reddit.com/r/cursor/comments/1n1fyi9/teammates_using_cursor_but_im_the_one_fixing/} \\
G1753 & 5271 & Cursor 3 just replaced the code editor with an agent management console. this is a bigger deal than people think & \url{https://www.reddit.com/r/ArtificialInteligence/comments/1shkz9p/cursor_3_just_replaced_the_code_editor_with_an/} \\
G1754 & 5275 & Code Review for Vibe Coders & \url{https://www.reddit.com/r/vibecoding/comments/1suc1ch/code_review_for_vibe_coders/} \\
G1755 & 5276 & Do you still get satisfaction writing code? & \url{https://www.reddit.com/r/ExperiencedDevs/comments/1l5vhwx/do_you_still_get_satisfaction_writing_code/} \\
G1756 & 5277 & Developer productivity metrics(getdx) & \url{https://www.reddit.com/r/ExperiencedDevs/comments/1p1lxsi/developer_productivity_metricsgetdx/} \\
G1757 & 5282 & Getting close to 100\% task-success with Claude Code & \url{https://www.reddit.com/r/ClaudeAI/comments/1lta114/getting_close_to_100_tasksuccess_with_claude_code/} \\
G1758 & 5284 & From 10\% to 100\% AI-written code in two years & \url{https://www.reddit.com/r/ClaudeAI/comments/1qlnrzp/from_10_to_100_aiwritten_code_in_two_years/} \\
G1759 & 5285 & an experiment to try and build the security feedback loop into the AI "vibe coding" workflow itself & \url{https://www.reddit.com/r/ClaudeAI/comments/1sf1o9k/an_experiment_to_try_and_build_the_security/} \\
G1760 & 5291 & Would you allow this? - IDisposable and using statements & \url{https://www.reddit.com/r/csharp/comments/1r94lzh/would_you_allow_this_idisposable_and_using/} \\
G1761 & 5294 & What ACTUALLY works after testing every AI coding tool for 6 months & \url{https://www.reddit.com/r/ChatGPTCoding/comments/1o4jr3n/what_actually_works_after_testing_every_ai_coding/} \\
G1762 & 5301 & Anyone else give up reviewing every line? And kinda hate themselves for it? & \url{https://www.reddit.com/r/AI_Agents/comments/1seu8hu/anyone_else_give_up_reviewing_every_line_and/} \\
G1763 & 5309 & Looking for a Code Review on My Simple Golang CRUD App & \url{https://www.reddit.com/r/golang/comments/1ihil0w/looking_for_a_code_review_on_my_simple_golang/} \\
G1764 & 5317 & Looking for an alternative to ChatGPT with Canvas for code review, fixing, modification, etc. & \url{https://www.reddit.com/r/ChatGPTCoding/comments/1m54icw/looking_for_an_alternative_to_chatgpt_with_canvas/} \\
G1765 & 5318 & Is it normal to feel like the majority of your coworkers are somewhat incompetent? & \url{https://www.reddit.com/r/ExperiencedDevs/comments/1jdm6y5/is_it_normal_to_feel_like_the_majority_of_your/} \\
G1766 & 5320 & Slowql - SQL static analyzer completed after watching a SELECT * take down production on a Friday night - 171 rules, zero dependencies, 100\% offline & \url{https://www.reddit.com/r/programming/comments/1roxjqd/slowql_sql_static_analyzer_completed_after/} \\
G1767 & 5324 & How Claude Code transformed my workflow - 5 concrete techniques + free open-source tools & \url{https://www.reddit.com/r/ClaudeAI/comments/1qt9xgw/how_claude_code_transformed_my_workflow_5/} \\
G1768 & 5327 & Code review became our bottleneck after adopting Cursor. Here is how we fixed it. & \url{https://www.reddit.com/r/codereview/comments/1pfrvsb/code_review_became_our_bottleneck_after_adopting/} \\
G1769 & 5329 & Why Claude Code's custom agents aren't enough for deterministic code review & \url{https://www.reddit.com/r/programming/comments/1qbu9ol/why_claude_codes_custom_agents_arent_enough_for/} \\
G1770 & 5330 & How are you actually using AI in your dev workflow? Not the demo version. & \url{https://www.reddit.com/r/ClaudeAI/comments/1rvjezq/how_are_you_actually_using_ai_in_your_dev/} \\
G1771 & 5334 & Advice for dealing with a New Manager & \url{https://www.reddit.com/r/ExperiencedDevs/comments/1p1isl9/advice_for_dealing_with_a_new_manager/} \\
G1772 & 5337 & Maxxed out Max & \url{https://www.reddit.com/r/ClaudeAI/comments/1s7hrsm/maxxed_out_max/} \\
G1773 & 5339 & building a multi-agent plugin for claude code - would love feedback & \url{https://www.reddit.com/r/ClaudeAI/comments/1qz4lys/building_a_multiagent_plugin_for_claude_code/} \\
G1774 & 5346 & Just got the new RTX 6000 Blackwell 96VRAM GB What is the best LLM to run? & \url{https://www.reddit.com/r/LocalLLaMA/comments/1mibpt2/just_got_the_new_rtx_6000_blackwell_96vram_gb/} \\
G1775 & 5348 & Agentic coding didn't make me faster it just moved the work to review & \url{https://www.reddit.com/r/cursor/comments/1tuip8p/agentic_coding_didnt_make_me_faster_it_just_moved/} \\
G1776 & 5350 & How to sharpen your skills in the age of AI? & \url{https://www.reddit.com/r/ExperiencedDevs/comments/1ovdyoo/how_to_sharpen_your_skills_in_the_age_of_ai/} \\
G1777 & 5359 & Gemma 4 2B handling structured JSON output + tool calling + reasoning traces correctly via Spring AI / LM Studio -- including identifying a real Java bug in cod & \url{https://www.reddit.com/r/LocalLLaMA/comments/1tmdk11/gemma_4_2b_handling_structured_json_output_tool/} \\
G1778 & 5364 & How to incorporate git into our workflow! & \url{https://www.reddit.com/r/git/comments/1hzu43f/how_to_incorporate_git_into_our_workflow/} \\
G1779 & 5367 & After analyzing 50,000 PRs, I built an AI code reviewer with evidence-backed findings and zero-knowledge architecture & \url{https://www.reddit.com/r/codereview/comments/1o1b4pu/after_analyzing_50000_prs_i_built_an_ai_code/} \\
G1780 & 5369 & Got rejected... again & \url{https://www.reddit.com/r/SoftwareEngineering/comments/1ldlysz/got_rejected_again/} \\
G1781 & 5373 & My Vibe Coding Framework That Still Works After 2+ Years of AI Evolution & \url{https://www.reddit.com/r/ClaudeAI/comments/1perz2o/my_vibe_coding_framework_that_still_works_after_2/} \\
G1782 & 5374 & Everyone's Hyped on Skills - But Claude Code Plugins take it further (6 Examples That Prove It) & \url{https://www.reddit.com/r/ClaudeAI/comments/1qrlsly/everyones_hyped_on_skills_but_claude_code_plugins/} \\
G1783 & 5376 & openAI nailed it with Codex for devs & \url{https://www.reddit.com/r/OpenAI/comments/1n5iqqj/openai_nailed_it_with_codex_for_devs/} \\
G1784 & 5379 & What is vibe coding? & \url{https://www.reddit.com/r/ClaudeAI/comments/1nhlnhz/what_is_vibe_coding/} \\
G1785 & 5386 & I built MAGI -- a Claude Code plugin that spawns 3 adversarial AI agents (inspired by Evangelion) to review your code, designs, and decisions & \url{https://www.reddit.com/r/ClaudeAI/comments/1sa7o6u/i_built_magi_a_claude_code_plugin_that_spawns_3/} \\
G1786 & 5388 & How to navigate PR reviews when there is resentment for trying to suggesting cleaner code & \url{https://www.reddit.com/r/ExperiencedDevs/comments/1jvbyuz/how_to_navigate_pr_reviews_when_there_is/} \\
G1787 & 5392 & What's a good way to redo/refine existing project? & \url{https://www.reddit.com/r/ClaudeAI/comments/1s08r9t/whats_a_good_way_to_redorefine_existing_project/} \\
G1788 & 5393 & GH Copilot pro sub total monthly credits is now equivalent to a single 5h session with claude code pro. & \url{https://www.reddit.com/r/GithubCopilot/comments/1ttelew/gh_copilot_pro_sub_total_monthly_credits_is_now/} \\
G1789 & 5395 & Claude development workflow & \url{https://www.reddit.com/r/ClaudeAI/comments/1sqrq5n/claude_development_workflow/} \\
G1790 & 5396 & Do people think the contribution guide is novelty? & \url{https://www.reddit.com/r/github/comments/1qpoke3/do_people_think_the_contribution_guide_is_novelty/} \\
G1791 & 5397 & You can have claude create an agent team and then have them QA their own work & \url{https://www.reddit.com/r/ClaudeAI/comments/1qx4m6a/you_can_have_claude_create_an_agent_team_and_then/} \\
G1792 & 5398 & [Template] Enterprise-grade WPF + CommunityToolkit.Mvvm starter kit - feedback welcome! & \url{https://www.reddit.com/r/csharp/comments/1mibg7d/template_enterprisegrade_wpf_communitytoolkitmvvm/} \\
G1793 & 5399 & I feel like AI has mostly just shown me most people can't code well & \url{https://www.reddit.com/r/ExperiencedDevs/comments/1ixyqp2/i_feel_like_ai_has_mostly_just_shown_me_most/} \\
G1794 & 5403 & How do you cope with multi agent workflows? & \url{https://www.reddit.com/r/ExperiencedDevs/comments/1ten4yg/how_do_you_cope_with_multi_agent_workflows/} \\
G1795 & 5404 & How do you catch auth bypass risks in generated code that looks completely correct & \url{https://www.reddit.com/r/ChatGPTCoding/comments/1rvzpr1/how_do_you_catch_auth_bypass_risks_in_generated/} \\
G1796 & 5406 & Built a VS Code extension to track what Claude Code writes in auto-accept mode & \url{https://www.reddit.com/r/ClaudeAI/comments/1q3gscu/built_a_vs_code_extension_to_track_what_claude/} \\
G1797 & 5408 & How do you deal with PR (Pull Request) raised by AI? & \url{https://www.reddit.com/r/vibecoding/comments/1tp4jea/how_do_you_deal_with_pr_pull_request_raised_by_ai/} \\
G1798 & 5409 & I built a small cli tool to execute agentic workflows & \url{https://www.reddit.com/r/LocalLLaMA/comments/1mt6hot/i_built_a_small_cli_tool_to_execute_agentic/} \\
G1799 & 5410 & Has Claude become less intelligent? I had a frustrating day with Claude. & \url{https://www.reddit.com/r/ClaudeAI/comments/1sup629/has_claude_become_less_intelligent_i_had_a/} \\
G1800 & 5412 & Feedback on my authentication starter template - Security \& Architecture review & \url{https://www.reddit.com/r/golang/comments/1pjuaie/feedback_on_my_authentication_starter_template/} \\
G1801 & 5414 & GPT-5.1 is 50-80\% cheaper than GPT-5.1-Codex for code review, same results, way fewer tokens & \url{https://www.reddit.com/r/codereview/comments/1rmd33i/gpt51_is_5080_cheaper_than_gpt51codex_for_code/} \\
G1802 & 5416 & I built Claude Usage, a free and open-source macOS menu bar app for checking Claude usage, with help from Claude Code & \url{https://www.reddit.com/r/ClaudeAI/comments/1rwfiih/i_built_claude_usage_a_free_and_opensource_macos/} \\
G1803 & 5418 & Claudiomiro: How to Achieve 100\% Autonomous (Complex) Coding & \url{https://www.reddit.com/r/javascript/comments/1o4vtks/claudiomiro_how_to_achieve_100_autonomous_complex/} \\
G1804 & 5419 & Giving Claude Code architectural context via a knowledge graph MCP (inspired by Karpathy's LLM Wiki) & \url{https://www.reddit.com/r/ClaudeAI/comments/1sgae0h/giving_claude_code_architectural_context_via_a/} \\
G1805 & 5420 & How do I do a Code-review for my Entire codebase? & \url{https://www.reddit.com/r/ClaudeAI/comments/1ls6u76/how_do_i_do_a_codereview_for_my_entire_codebase/} \\
G1806 & 5421 & Vibe coding is a lie. & \url{https://www.reddit.com/r/vibecoding/comments/1mirhhk/vibe_coding_is_a_lie/} \\
G1807 & 5423 & I think this sub is missing the point of vibe coding. & \url{https://www.reddit.com/r/vibecoding/comments/1t1z5zg/i_think_this_sub_is_missing_the_point_of_vibe/} \\
G1808 & 5424 & If you think parallel subagents are bad, you're wrong. Here's how to use a 10+ subagents effectively. & \url{https://www.reddit.com/r/ClaudeAI/comments/1pz5b21/if_you_think_parallel_subagents_are_bad_youre/} \\
G1809 & 5425 & Anyone else mass delegating frontend work to ai and feeling weird about it & \url{https://www.reddit.com/r/webdev/comments/1tkdrbf/anyone_else_mass_delegating_frontend_work_to_ai/} \\
G1810 & 5430 & What's the one prompt you run against your project every week? Mine: & \url{https://www.reddit.com/r/ChatGPTCoding/comments/1p28ibe/whats_the_one_prompt_you_run_against_your_project/} \\
G1811 & 5432 & I feel like my HTML/CSS knowledge has become completely useless & \url{https://www.reddit.com/r/webdev/comments/1tni7nq/i_feel_like_my_htmlcss_knowledge_has_become/} \\
G1812 & 5436 & My honest take on AI coding tools after using them daily for 2 years as a developer & \url{https://www.reddit.com/r/vibecoding/comments/1s8mwz6/my_honest_take_on_ai_coding_tools_after_using/} \\
G1813 & 5438 & Agile is new to me and I don't understand it - my first experience & \url{https://www.reddit.com/r/agile/comments/1lro46f/agile_is_new_to_me_and_i_dont_understand_it_my/} \\
G1814 & 5439 & Where do you draw the line for AI-generated code in production systems? & \url{https://www.reddit.com/r/vibecoding/comments/1prd6i9/where_do_you_draw_the_line_for_aigenerated_code/} \\
G1815 & 5440 & Not a vibe coder, but what tools are you guys using? & \url{https://www.reddit.com/r/vibecoding/comments/1row96w/not_a_vibe_coder_but_what_tools_are_you_guys_using/} \\
G1816 & 5442 & I run a team of Claude agents that ships PRs to production -- open source & \url{https://www.reddit.com/r/ClaudeAI/comments/1svmdfn/i_run_a_team_of_claude_agents_that_ships_prs_to/} \\
G1817 & 5447 & I built nod, a kanban markdown based task manager for AI coding agents & \url{https://www.reddit.com/r/ClaudeAI/comments/1sdsec0/i_built_nod_a_kanban_markdown_based_task_manager/} \\
G1818 & 5448 & Has anyone integrated AI tools into their PR or code review workflow? & \url{https://www.reddit.com/r/devops/comments/1os7j08/has_anyone_integrated_ai_tools_into_their_pr_or/} \\
G1819 & 5452 & Review changes? & \url{https://www.reddit.com/r/ClaudeAI/comments/1lgxtz3/review_changes/} \\
G1820 & 5457 & Vibe-coded a browser-based controller for Bluetooth thermal printers (24 models, reverse-engineered protocol, zero hand-typed code) & \url{https://www.reddit.com/r/vibecoding/comments/1sqw0i0/vibecoded_a_browserbased_controller_for_bluetooth/} \\
G1821 & 5458 & Disappointed in Qwen 3.6 coding capabilities & \url{https://www.reddit.com/r/LocalLLaMA/comments/1t66gct/disappointed_in_qwen_36_coding_capabilities/} \\
G1822 & 5459 & AI augmented software development - as an experienced SDE you are not going to like it & \url{https://www.reddit.com/r/vibecoding/comments/1na0uqi/ai_augmented_software_development_as_an/} \\
G1823 & 5463 & 6 AI Tools I Actually Use for Smarter Coding and Automation in 2025 & \url{https://www.reddit.com/r/programming/comments/1ot8fwl/6_ai_tools_i_actually_use_for_smarter_coding_and/} \\
G1824 & 5468 & Tools for CTO scaling engineering team: what worked and what was a waste of money & \url{https://www.reddit.com/r/ExperiencedDevs/comments/1owih10/tools_for_cto_scaling_engineering_team_what/} \\
G1825 & 5470 & How do you really know if a mobile dev is good? & \url{https://www.reddit.com/r/ExperiencedDevs/comments/1mjwyxo/how_do_you_really_know_if_a_mobile_dev_is_good/} \\
G1826 & 5472 & AI code review tools on GitHub PRs... do they really help or just add noise? & \url{https://www.reddit.com/r/devops/comments/1ne3skx/ai_code_review_tools_on_github_prs_do_they_really/} \\
G1827 & 5477 & 2 months into vibe coding with zero programming experience. I made Claude Code agents grade each other's homework. (open source) & \url{https://www.reddit.com/r/ClaudeAI/comments/1rrk0au/2_months_into_vibe_coding_with_zero_programming/} \\
G1828 & 5481 & Finally a setup that works well and is affordable: vscode + cline + glm 4.6 // Bonus: my realisation of why kilo code and roo code fail & \url{https://www.reddit.com/r/vibecoding/comments/1ocixo9/finally_a_setup_that_works_well_and_is_affordable/} \\
G1829 & 5483 & Looking for tools/frameworks to orchestrate AI agents for automated microservice development & \url{https://www.reddit.com/r/AI_Agents/comments/1mp2dsj/looking_for_toolsframeworks_to_orchestrate_ai/} \\
G1830 & 5484 & Our posting agent broke 3 cycles in a row. Here's how it diagnosed itself and filed the fix -- no human involved. & \url{https://www.reddit.com/r/AI_Agents/comments/1tm2rps/our_posting_agent_broke_3_cycles_in_a_row_heres/} \\
G1831 & 5487 & What happens when you stop adding rules to CLAUDE.md and start building infrastructure instead & \url{https://www.reddit.com/r/ClaudeAI/comments/1rz2oo3/what_happens_when_you_stop_adding_rules_to/} \\
G1832 & 5492 & Best open-source multi-agent coding/orchestrator frameworks for Claude Code style workflows? & \url{https://www.reddit.com/r/ClaudeAI/comments/1tlj6kh/best_opensource_multiagent_codingorchestrator/} \\
G1833 & 5493 & Unpopular opinion: DORA metrics are becoming "Vanity Metrics" for Engineering Health. & \url{https://www.reddit.com/r/devops/comments/1ppphjb/unpopular_opinion_dora_metrics_are_becoming/} \\
G1834 & 5501 & Using a local 7B for code review -- game changer if you guide it right & \url{https://www.reddit.com/r/LocalLLaMA/comments/1ry6cw3/using_a_local_7b_for_code_review_game_changer_if/} \\
G1835 & 5503 & Juniors \& AI & \url{https://www.reddit.com/r/ExperiencedDevs/comments/1sywp4r/juniors_ai/} \\
G1836 & 5504 & PRs Might Harm Your CD - "PRs might not be the most effective tool for code reviews, especially in a Continuous Deployment workflow." & \url{https://www.reddit.com/r/programming/comments/1i3d6ki/prs_might_harm_your_cd_prs_might_not_be_the_most/} \\
G1837 & 5508 & Feeling Stuck as the Only Developer With 1 Year of Experience Need Advice & \url{https://www.reddit.com/r/cscareerquestions/comments/1kgwxo9/feeling_stuck_as_the_only_developer_with_1_year/} \\
G1838 & 5514 & [How-To] Build Fast, Reliable CI/CD Pipelines with AI-Driven Testing & \url{https://trilogyai.substack.com/p/how-to-build-fast-reliable-cicd-pipelines} \\
G1839 & 5516 & OpenAI Just Dropped a Codex Plugin for Claude Code (And It Changes the AI Coding Meta) - DEV Community & \url{https://dev.to/siddhesh_surve/openai-just-dropped-a-codex-plugin-for-claude-code-and-it-changes-the-ai-coding-meta-1ilg} \\
G1840 & 5518 & Spotify Devs Haven't Written Code Since Dec, AI Tools Take Over | Galen O. posted on the topic | LinkedIn & \url{https://www.linkedin.com/posts/galenoakes_spotify-just-said-their-best-developers-have-activity-7427789311485407232-u96D} \\
G1841 & 5519 & Cursor Bugbot Saves PlanetScale 2 Engineers in Code Review & \url{https://www.adwaitx.com/cursor-bugbot-planetscale-ai-code/} \\
G1842 & 5520 & CodeRabbit CLI - Free AI code reviews in your CLI - DEV Community & \url{https://dev.to/coderabbitai/coderabbit-cli-free-ai-code-reviews-in-your-cli-n0} \\
G1843 & 5523 & The State of AI Code Review in 2026 - Trends, Tools, and What's Next - DEV Community & \url{https://dev.to/rahulxsingh/the-state-of-ai-code-review-in-2026-trends-tools-and-whats-next-2gfh} \\
G1844 & 5525 & Vibe Coding: When Your IDE Becomes a Dance Floor and Bugs Become Features | by Intelligent Hustle | Towards AI & \url{https://pub.towardsai.net/vibe-coding-when-your-ide-becomes-a-dance-floor-and-bugs-become-features-0bc8fb52372f} \\
G1845 & 5529 & AI code lasts longer than human code, new study finds (but not for the reason you'd hope). - DEV Community & \url{https://dev.to/lewiska/ai-code-lasts-longer-than-human-code-new-study-finds-but-not-for-the-reason-youd-hope-3a2i} \\
G1846 & 5532 & AI coders need good software engineers | InfoWorld & \url{https://www.infoworld.com/article/4176534/ai-coders-need-good-software-engineers.html} \\
G1847 & 5537 & The Velocity Paradox - DEV Community & \url{https://dev.to/abahgat/the-velocity-paradox-4182} \\
G1848 & 5541 & Leading Open Source Author Calls for Verification over Trust in Software Supply Chains - InfoQ & \url{https://www.infoq.com/news/2026/05/stenberg-curl-verification-trust/} \\
G1849 & 5542 & What I found scanning 3 AI agent codebases for unguarded tool calls - DEV Community & \url{https://dev.to/josselin_guarnelli/what-i-found-scanning-3-ai-agent-codebases-for-unguarded-tool-calls-4g9b} \\
G1850 & 5545 & Bridging the trust gap in AI-driven development | InfoWorld & \url{https://www.infoworld.com/article/4033109/bridging-the-trust-gap-in-ai-driven-development.html} \\
G1851 & 5549 & CodeRabbit: Smarter Pull Request Reviews with AI & \url{https://www.linkedin.com/pulse/coderabbit-smarter-pull-request-reviews-ai-junior-nakamura-dapqf} \\
G1852 & 5553 & OpenAI Just Shipped a Plugin So Codex Runs Inside Claude Code - DEV Community & \url{https://dev.to/oldeucryptoboi/openai-just-shipped-a-plugin-so-codex-runs-inside-claude-code-51oa} \\
G1853 & 5554 & AI-Powered Code Reviews with GitHub Copilot - LinkedIn & \url{https://www.linkedin.com/pulse/ai-powered-code-reviews-github-copilot-ivan-vydrin-3xjqf} \\
G1854 & 5562 & Senior Engineer Promotion Case Study: Impact vs Technical Excellence | Naresh Mudiraj posted on the topic | LinkedIn & \url{https://www.linkedin.com/posts/naresh-mudiraj-b21197210_java-techleadership-engineeringcareer-activity-7460318608099885056--z75} \\
G1855 & 5563 & The 9-Point Self Code Review Every Junior Developer Should Run Before Standup & \url{https://arslandg.substack.com/p/the-9-point-self-code-review-every} \\
G1856 & 5564 & A CI tool that stops AI from undoing your team's architectural decisions - DEV Community & \url{https://dev.to/stewartjarod/a-ci-tool-that-stops-ai-from-undoing-your-teams-architectural-decisions-1e32} \\
G1857 & 5565 & What is Software Reliability? | AI Code Review \& Verification | Sonar | Sonar & \url{https://www.sonarsource.com/resources/library/what-is-software-reliability/} \\
G1858 & 5570 & Qodo Rules & \url{https://www.qodo.ai/qodo-rules/} \\
G1859 & 5571 & Stop "LGTM" Culture: How to Turn Code Reviews into a Masterclass with AI | HackerNoon & \url{https://hackernoon.com/stop-lgtm-culture-how-to-turn-code-reviews-into-a-masterclass-with-ai} \\
G1860 & 5575 & Automated Dependency Trust Reports on Every PR - DEV Community & \url{https://dev.to/zarq-ai/automated-dependency-trust-reports-on-every-pr-3hgd} \\
G1861 & 5576 & AI code generation: Can we do better than autocomplete? & \url{https://www.linkedin.com/posts/logicalguess_aicodegeneration-developertools-quantumcomputing-activity-7385734533976903680-IMUI} \\
G1862 & 5577 & Cursor Composer 2 Just Dropped -- Here's How to Actually Use Autonomous Coding Agents in Production - DEV Community & \url{https://dev.to/dohkoai/cursor-composer-2-just-dropped-heres-how-to-actually-use-autonomous-coding-agents-in-production-389c} \\
G1863 & 5578 & shubhesh07/claude-code-reviewer & \url{https://github.com/shubhesh07/claude-code-reviewer} \\
G1864 & 5579 & The Real Bottleneck in AI Coding Isn't Generation--It's Everything Else - DEV Community & \url{https://dev.to/o96a/the-real-bottleneck-in-ai-coding-isnt-generation-its-everything-else-21l3} \\
G1865 & 5584 & Stop F*cking Your Team With Giant PRs - DEV Community & \url{https://dev.to/antoniel/stop-fcking-your-team-with-giant-prs-4nff} \\
G1866 & 5585 & How to use mutation testing with LLMs | Mark Ng posted on the topic | LinkedIn & \url{https://www.linkedin.com/posts/markng_ever-noticed-how-coding-with-llms-feels-like-activity-7336816598877409282-YBGb} \\
G1867 & 5589 & fjpulidop/specrails-core & \url{https://github.com/fjpulidop/specrails} \\
G1868 & 5590 & Vibe coding, hype cycles, and why AI isn't the 10x answer - LeadDev & \url{https://leaddev.com/culture/vibe-coding-hype-cycles-why-ai-isnt-10x-answer} \\
G1869 & 5594 & The LLM as Your Main Pair Programmer | by Jonathan Yee | Medium & \url{https://medium.com/@JonathanYeeW/the-llm-as-your-main-pair-programmer-394d92ddaceb} \\
G1870 & 5595 & GitHub Unveils Copilot App for Agent-Native Coding at Build 2026 - MAXBIT & \url{https://maxbit.cc/github-unveils-copilot-app-for-agent-native-coding-at-build-2026/} \\
G1871 & 5596 & Codex Security: now in research preview | OpenAI & \url{https://openai.com/index/codex-security-now-in-research-preview} \\
G1872 & 5597 & Securing AI Generated Code: You Ship It, You Own It & \url{https://notthecode.com/owning-code-you-didnt-write/} \\
G1873 & 5598 & Open source maintainers are being targeted by AI agent as part of 'reputation farming' | InfoWorld & \url{https://www.infoworld.com/article/4132851/open-source-maintainers-are-being-targeted-by-ai-agent-as-part-of-reputation-farming.html} \\
G1874 & 5603 & The Real Cost of AI Coding: Skills vs. Products & \url{https://www.augmentcode.com/tools/the-real-cost-of-ai-coding-skills-vs-products} \\
G1875 & 5607 & What Is AI Code Assistant? Definition \& Examples & \url{https://nhimg.org/glossary/ai-code-assistant/} \\
G1876 & 5608 & The Siege of Open Source -- Sloppish & \url{https://sloppish.com/siege-of-open-source.html} \\
G1877 & 5610 & Your coding agent doesn't need more autonomy. It needs a manager. - DEV Community & \url{https://dev.to/overcast_27b5101f0609c684/your-coding-agent-doesnt-need-more-autonomy-it-needs-a-manager-2b76} \\
G1878 & 5612 & I let four AI code reviewers fight over my PRs - DEV Community & \url{https://dev.to/adam-knitli/i-let-four-ai-code-reviews-fight-over-my-prs-17je} \\
G1879 & 5613 & Vibe Coding for Developers: What Changes and What Has to Go | Towards AI & \url{https://pub.towardsai.net/vibe-thinking-the-developer-who-codes-at-the-speed-of-thought-6996f6b9feb1} \\
G1880 & 5614 & Leading AI-Native Teams: Part 2 - Stop Reviewing AI Code. Start Building the System That Reviews It For You. | by Paul Brady | May, 2026 | Medium & \url{https://medium.com/@bradypt/leading-ai-native-teams-part-2-stop-reviewing-ai-code-f34d6a499f22} \\
G1881 & 5618 & Anthropic Proved AI Can't Evaluate Its Own Work. Here's How I Rebuilt My Claude Code Setup Around That. - DEV Community & \url{https://dev.to/lovanaut55/anthropic-proved-ai-cant-evaluate-its-own-work-heres-how-i-rebuilt-my-claude-code-setup-around-5f8i} \\
G1882 & 5619 & The Vibe Coding Hangover: How AI-Generated Code Is Compounding Technical Debt Faster Than You Can Ship & \url{https://kuberik.substack.com/p/the-vibe-coding-hangover-how-ai-generated} \\
G1883 & 5620 & Function Calling Harness 2: CoT Compliance from 9.91\% to 100\% - DEV Community & \url{https://dev.to/samchon/function-calling-harness-2-cot-compliance-from-991-to-100-4f0h} \\
G1884 & 5621 & The reason AI agents shouldn't touch your source code -- and what they should do instead - The New Stack & \url{https://thenewstack.io/dynatrace-ai-safety-mechanism/} \\
G1885 & 5627 & LGTM! Characteristics of Auto-Merged LLM-based Agentic PRs & \url{https://pcanelas.com/assets/papers/2026-msr-lgtm.pdf} \\
G1886 & 5629 & Unlearning, experimentation and engineering rigor in an agentic world | Thoughtworks Spain & \url{https://www.thoughtworks.com/en-es/insights/podcasts/technology-podcasts/unlearning-experimentation-engineering-rigor-agentic-world} \\
G1887 & 5634 & Opsera Report Highlights DevOps Challenges Created by AI Coding Tools - DevOps.com & \url{https://devops.com/opsera-report-highlights-devops-challenges-created-by-ai-coding-tools/} \\
G1888 & 5637 & Senior Engineer's AI Coding Claim Sparks Debate: Evaluating AI's Role in Software Development - DEV Community & \url{https://dev.to/denlava/senior-engineers-ai-coding-claim-sparks-debate-evaluating-ais-role-in-software-development-3o8k} \\
G1889 & 5639 & How to Review AI-Generated Code That Ships Faster Than You Can Read & \url{https://codeongrass.com/blog/how-to-review-ai-generated-code-faster-than-you-can-read/} \\
G1890 & 5641 & In the age of AI coding, code quality still matters | Thoughtworks United Kingdom & \url{https://www.thoughtworks.com/en-gb/insights/blog/generative-ai/in-the-age-of-AI-coding-code-quality-still-matters} \\
G1891 & 5645 & Cursor's AI Security Agents: What They Get Right (and What's Missing) | Snyk & \url{https://snyk.io/es/blog/cursor-security-agent-prompts/} \\
G1892 & 5652 & Building a Production-Grade AI Code Reviewer: A 20-Phase Pipeline | Ayush Singh posted on the topic | LinkedIn & \url{https://www.linkedin.com/posts/ayush-singh488_if-i-had-asked-you-to-build-a-production-activity-7465622361678102528-p4wn} \\
G1893 & 5653 & AI Writes the Code. You Review It. The Debt Ships Anyway. -- Codexical & \url{https://www.codexical.com/posts/2026-04-24-ai-technical-debt-study} \\
G1894 & 5656 & Devin vs Cursor (2026): Full Comparison | MytheAi & \url{https://mytheai.com/compare/devin-ai-vs-cursor} \\
G1895 & 5657 & Code Review Process | github/gh-copilot | DeepWiki & \url{https://deepwiki.com/github/gh-copilot/7.2-code-review-process} \\
G1896 & 5660 & Agent-First Coding Is Here. What It Actually Means for Developers in 2026 - DEV Community & \url{https://dev.to/lasisi_ibrahimpelumi_dc0/agent-first-coding-is-here-what-it-actually-means-for-developers-in-2026-50ae} \\
G1897 & 5662 & The Company That Built the AI Coding Tool Just Proved It Makes You a Worse Developer & \url{https://reltech.substack.com/p/the-company-that-built-the-ai-coding} \\
G1898 & 5664 & CodeRabbit vs Graphite (2026): Features, Pricing \& Verdict | Respan & \url{https://www.respan.ai/market-map/compare/coderabbit-vs-graphite} \\
G1899 & 5665 & How to debug code with GitHub Copilot - The GitHub Blog & \url{https://github.blog/ai-and-ml/github-copilot/how-to-debug-code-with-github-copilot/} \\
G1900 & 5667 & AI Code Review in 60 Seconds: I Automated 100\% of PR Reviews (2026) | Ivern AI & \url{https://ivern.ai/blog/ai-agent-code-review-automation} \\
G1901 & 5668 & AI Slop in Open Source: Why Maintainers Are Burning Out | Medium & \url{https://medium.com/@the_atomic_architect/ai-slop-in-open-source-why-maintainers-are-burning-out-db57e8f18b84} \\
G1902 & 5669 & lizhengfeng101/open-code-review & \url{https://github.com/lizhengfeng101/open-code-review} \\
G1903 & 5670 & I Used AI for Code Review on a Production ERP for 6 Months. Here's Where It Actually Failed Me. - DEV Community & \url{https://dev.to/sumit2401/i-used-ai-for-code-review-on-a-production-erp-for-6-months-heres-where-it-actually-failed-me-18ie} \\
G1904 & 5671 & My 6-Month Experiment with Autonomous AI Agents - DEV Community & \url{https://dev.to/elysiumquill/my-6-month-experiment-with-autonomous-ai-agents-what-actually-changed-in-my-daily-workflow-113o} \\
G1905 & 5672 & Claude codex-cli skill: 10 ways to bridge Claude Code and OpenAI Codex CLI -- MCP.Directory & \url{https://mcp.directory/blog/claude-codex-cli-skill-guide} \\
G1906 & 5674 & I Scanned Hundreds of AI-Generated Codebases. Here's What Keeps Showing Up. - DEV Community & \url{https://dev.to/paschal_ugwuanyi_b4472037/-i-scanned-hundreds-of-ai-generated-codebases-heres-what-keeps-showing-up-2ieb} \\
G1907 & 5676 & The Evergreen Principles when Coding with AI | by ming gao | Mar, 2026 | Medium & \url{https://medium.com/@ming.gao.gm/the-evergreens-principles-when-coding-with-ai-9601b639f5e0} \\
G1908 & 5680 & AI-Assisted Code Review: What the Tools Catch and Where Humans Still Win -- CODERCOPS & \url{https://www.codercops.com/blog/ai-code-review-tools-workflows-2026} \\
G1909 & 5682 & Trunk-Based Development: Why Most Teams Think They Use It (But Don't) - DEV Community & \url{https://dev.to/codecraft_diary_3d13677fb/trunk-based-development-why-most-teams-think-they-use-it-but-dont-o4g} \\
G1910 & 5683 & HubSpot's Sidekick: Multi-Model AI Code Review with 90\% Faster Feedback and 80\% Engineer Approval - InfoQ & \url{https://www.infoq.com/news/2026/03/hubspot-ai-code-review-agent/} \\
G1911 & 5684 & Beyond vibe coding: How AI can transform pull requests | Thoughtworks United States & \url{https://www.thoughtworks.com/en-us/insights/blog/generative-ai/beyond-vibe-coding-how-AI-can-transform-pull-requests} \\
G1912 & 5686 & AI PR Review in 2026: What Actually Works (And What Wastes Your Team's Time) | Git AutoReview & \url{https://gitautoreview.com/blog/ai-pr-review-guide} \\
G1913 & 5692 & The Most Common Security Vulnerabilities in AI-Generated Code | Blog | Endor Labs & \url{https://www.endorlabs.com/learn/the-most-common-security-vulnerabilities-in-ai-generated-code} \\
G1914 & 5697 & AI-Powered Code Review Revolutionizes Development | Nikhil R. Srivastava posted on the topic | LinkedIn & \url{https://www.linkedin.com/posts/nikhilrsrivastava_conventional-way-junior-developers-write-activity-7407026818811101184-JJtW} \\
G1915 & 5698 & I built an AI PR reviewer and it already caught bugs I missed - DEV Community & \url{https://dev.to/adamai/i-built-an-ai-pr-reviewer-and-it-already-caught-bugs-i-missed-115p} \\
G1916 & 5699 & Making frontier cybersecurity capabilities available to defenders \textbackslash\{\} Anthropic & \url{https://www.anthropic.com/news/claude-code-security} \\
G1917 & 5700 & AI Code Generation Crisis: 3 Million Lines of Technical Debt | Obiajulu Okorie posted on the topic | LinkedIn & \url{https://www.linkedin.com/posts/obiajulu-anthony_cursor-is-lying-to-developers-activity-7428703453650149376-wMOB} \\
G1918 & 5702 & alchemiststudiosDOTai/coderabbit-fix-flow-plugin & \url{https://github.com/alchemiststudiosdotai/coderabbit-fix-flow-plugin} \\
G1919 & 5718 & What do humans do if AI reviews the code? & \url{https://www.linkedin.com/pulse/what-do-humans-ai-reviews-code-leaf-jessica-roy-t6vke} \\
G1920 & 5724 & Claude 4.6 Sonnet for Code Reviews in iOS Projects | Brad Lees posted on the topic | LinkedIn & \url{https://www.linkedin.com/posts/brad-lees_ive-been-experimenting-with-claude-46-sonnet-activity-7459808071251496960-6voS} \\
G1921 & 5725 & The AI Code Generation Paradox: Why the More Code AI Writes, the More Vulnerable Everything Becomes - BotPoweredPro & \url{https://botpoweredpro.com/ai-code-generation-paradox-vulnerabilities/} \\
G1922 & 5729 & AI Code Safety: The 2.74x Vulnerability Multiplier Nobody Is Fixing & \url{https://www.connectory.ai/blog/ai-code-security-vulnerability-prevention/} \\
G1923 & 5732 & Reinventing Software Engineering with AI: A Practical Guide for Technology Leaders | by The Shadow Observer | May, 2026 | Medium & \url{https://medium.com/@TheShadowObserver/reinventing-software-engineering-with-ai-a-practical-guide-for-technology-leaders-8d725d40e966} \\
G1924 & 5733 & Is vibe coding bad? Pros, cons, and when to use it & \url{https://www.hostinger.com/au/tutorials/is-vibe-coding-bad} \\
G1925 & 5735 & The AI trust gap: Why code verification matters | Sonar & \url{https://www.sonarsource.com/blog/ai-coding-trust-gap} \\
G1926 & 5736 & Code Review: Claude's Rise in Dev Tools & \url{https://deeperinsights.com/ai-blog/code-review-to-claude-code/} \\
G1927 & 5737 & AI Coding Tip 006 - Review Every Line Before Commit & \url{https://maxicontieri.substack.com/p/ai-coding-tip-006-review-every-line} \\
G1928 & 5738 & Welcome to the Eternal September of open source. Here's what we plan to do for maintainers. - The GitHub Blog & \url{https://github.blog/open-source/maintainers/welcome-to-the-eternal-september-of-open-source-heres-what-we-plan-to-do-for-maintainers/} \\
G1929 & 5740 & The Hallucination Problem of AI Programming Assistants: How to Implement Specification-Driven Development with OpenSpec - DEV Community & \url{https://dev.to/newbe36524/the-hallucination-problem-of-ai-programming-assistants-how-to-implement-specification-driven-2i2j} \\
G1930 & 5743 & Spec-Driven Development with Claude Code: A Guided Tutorial | DataCamp & \url{https://www.datacamp.com/tutorial/spec-driven-development-with-claude-code} \\
G1931 & 5749 & AI Code Reviews GitHub & \url{https://github.com/resources/articles/ai-code-reviews} \\
G1932 & 5758 & Context Window -- W14 2026. Agents need knowledge, not code:; The... | by Simone Basso | Apr, 2026 | Medium & \url{https://medium.com/@smnbss/context-window-w14-2026-c46140b41b0f} \\
G1933 & 5760 & Why Your Next Pull Request Might Cost You Money | by Mehdi BAFDIL | Feb, 2026 | Medium & \url{https://medium.com/@mehdibafdil/why-your-next-pull-request-might-cost-you-money-3bbeb47abaed} \\
G1934 & 5761 & Agentic Drift: It's Hard to Be Multiple Developers at Once - Helge Sverre & \url{https://helgesver.re/articles/agentic-drift} \\
G1935 & 5763 & Maintainers Overwhelmed by AI-Generated PRs, Automated Code ... & \url{https://www.linkedin.com/posts/fatherlinux_opensource-ai-codereview-activity-7432469432637218816-uKF8} \\
G1936 & 5764 & How I Use Qodo's Agent Skills to Auto-Fix Issues in Pull Requests - Qodo & \url{https://www.qodo.ai/blog/how-i-use-qodos-agent-skills-to-auto-fix-issues-in-pull-requests/} \\
G1937 & 5765 & AI Code Review Tools Benchmark & \url{https://research.aimultiple.com/ai-code-review-tools/} \\
G1938 & 5766 & AI-Generated Code: The Hidden Costs and Best Practices - LinkedIn & \url{https://www.linkedin.com/posts/planetoftheweb_the-real-cost-of-ai-generated-code-a-recent-activity-7416517917829263360-0gtm} \\
G1939 & 5773 & AI Generated Code Has a Security Problem - YouTube & \url{https://www.youtube.com/watch?v=nGxpctRd2OQ} \\
G1940 & 5775 & goweft/crocking & \url{https://github.com/goweft/crocking} \\
G1941 & 5776 & AI generated pull requests overwhelming, hard to review carefully - ITK & \url{https://discourse.itk.org/t/ai-generated-pull-requests-overwhelming-hard-to-review-carefully/7728} \\
G1942 & 5777 & I Review More Code Than I Write Now | by Revant | May, 2026 | Generative AI & \url{https://generativeai.pub/i-review-more-code-than-i-write-now-a39f5ed2ff51} \\
G1943 & 5779 & Snyk Studio | Prevent Vulnerabilities in AI-Generated Code \& Remediation at AI Speed | Snyk & \url{https://snyk.io/product/studio/} \\
G1944 & 5785 & CodeRabbit's Post & \url{https://www.linkedin.com/posts/coderabbitai_introducing-coderabbit-for-vscode-cursor-activity-7328407102564057089-mQIE} \\
G1945 & 5787 & AI bot seemingly shames developer for rejected pull request & \url{https://www.theregister.com/software/2026/02/12/ai-bot-seemingly-shames-developer-for-rejected-pull-request/4594087} \\
G1946 & 5790 & Why Code Review Is Not Enough When AI Writes the Code | ADCG & \url{https://anotherdimensioncreativegroup.com/blog/code-review-not-enough-ai} \\
G1947 & 5791 & CodeRabbit's Evolution: AI Code Review on Your Local Device - DEV Community & \url{https://dev.to/saas_tech/coderabbits-evolution-ai-code-review-in-your-local-device-5b67} \\
G1948 & 5792 & The Trust Problem Nobody's Designing For | by Jay Arora | Bootcamp | Mar, 2026 | Medium & \url{https://medium.com/design-bootcamp/the-trust-problem-nobodys-designing-for-7c52d18c31a4} \\
G1949 & 5794 & Readability Is a Performance Constraint - DEV Community & \url{https://dev.to/christiecosky/readability-is-a-performance-constraint-15fa} \\
G1950 & 5796 & Vibe Coding Is Replacing "Clean Code", Is That a Problem? | by Twendee | Mar, 2026 | Medium & \url{https://medium.com/@marketing_39301/vibe-coding-is-replacing-clean-code-is-that-a-problem-e554623d4a6e} \\
G1951 & 5797 & How AI is Transforming the Code Review Process - DEV Community & \url{https://dev.to/everettbutler/how-ai-is-transforming-the-code-review-process-oao} \\
G1952 & 5798 & CodeRabbit vs Greptile: AI Code Review Tools Compared - DEV Community & \url{https://dev.to/pantoai/coderabbit-vs-greptile-ai-code-review-tools-compared-4kn4} \\
G1953 & 5803 & The Vibe Coding Hangover: Why Agentic AI is Breaking the Internet (and How We Fix It) & \url{https://www.linkedin.com/pulse/vibe-coding-hangover-why-agentic-ai-breaking-internet-antony-lees-81pve} \\
G1954 & 5804 & What I Learned From a Week of AI-Assisted Coding: The Good, The Bad, and The Surprisingly Counterintuitive - DEV Community & \url{https://dev.to/jack_branch_3fb9e01c57c03/what-i-learned-from-a-week-of-ai-assisted-coding-the-good-the-bad-and-the-surprisingly-11kl} \\
G1955 & 5808 & State of AI Coding 2026: Adoption, Tools, and Trends | Awesome Agents & \url{https://awesomeagents.ai/guides/state-of-ai-coding-2026/} \\
G1956 & 5811 & AI Writes the Code. AI Reviews the Code. So Why Do They Still Need You? - DEV Community & \url{https://dev.to/andresclua/ai-writes-the-code-ai-reviews-the-code-so-why-do-they-still-need-you-3gg6} \\
G1957 & 5812 & AI-Assisted Engineering: The Productivity Paradox Nobody Warns You About - DEV Community & \url{https://dev.to/matthiasbruns/ai-assisted-engineering-the-productivity-paradox-nobody-warns-you-about-4oon} \\
G1958 & 5814 & Code Review Is the New Bottleneck. Metrics Are Next. & \url{https://thesystemsmindset.substack.com/p/code-review-is-the-new-bottleneck} \\
G1959 & 5815 & I Added Claude to Our MR Pipelines. It Now Reviews Every Code Change Before Humans Do. - DEV Community & \url{https://dev.to/manikanta_suru_92/i-added-claude-to-our-mr-pipelines-it-now-reviews-every-code-change-before-humans-do-2epe} \\
G1960 & 5821 & Unlearning, experimentation and engineering rigor in an agentic world | Thoughtworks Chile & \url{https://www.thoughtworks.com/en-cl/insights/podcasts/technology-podcasts/unlearning-experimentation-engineering-rigor-agentic-world} \\
G1961 & 5823 & OpenSpec: AI-Native Specification-Driven Development Framework with 37K GitHub Stars | SoloSoft & \url{https://www.solosoft.dev/post/openspec-sdd-framework-2026/} \\
G1962 & 5827 & What Is Vibe Coding in 2026? One Year From Karpathy's Tweet - DEV Community & \url{https://dev.to/h1gbosn/what-is-vibe-coding-in-2026-one-year-from-karpathys-tweet-5f43} \\
G1963 & 5828 & Vibe Coding Actually Changed How I Write Features -- Here's What That Looks Like Day-to-Day - techdigestor & \url{https://techdigestor.com/vibe-coding-actually-changed-how-i-write-features-heres-what-that-looks-like-day-to-day/} \\
G1964 & 5829 & "Yes, this is safe to deploy." That's what my AI coding agent told me. So I deployed. And then our service was down. (Luckily only dev ) What happened: I was bu & \url{https://www.linkedin.com/posts/alexanderpieper_yes-this-is-safe-to-deploy-thats-what-activity-7427648201702129664-Aefo} \\
G1965 & 5830 & AI Agents Are Not Replacing Developers. They Are Exposing Weak Engineers Faster | Medium & \url{https://medium.com/@samurai.stateless.coder/ai-agents-are-not-replacing-developers-they-are-exposing-weak-engineers-faster-f34446eaf30e} \\
G1966 & 5833 & Are AI Coding Assistants Really Saving Developers Time? - Axify & \url{https://axify.io/blog/are-ai-coding-assistants-really-saving-developers-time} \\
G1967 & 5835 & Vibe Coding in 2026: The Complete Guide for Developers & \url{https://antigravity.codes/blog/vibe-coding-guide} \\
G1968 & 5836 & AI Code Quality: The Hidden Cost Senior Engineers Pay & \url{https://www.faros.ai/blog/ai-code-quality-senior-engineer-review-burden} \\
G1969 & 5842 & Software Development Evolves with AI: Same Principles, New Challenges | Nachi Nachiappan posted on the topic | LinkedIn & \url{https://www.linkedin.com/posts/nachiappan_this-software-development-changes-dramatically-activity-7440822490412724224-czo4} \\
G1970 & 5845 & AI Code Review Tools: What Actually Catches Real Bugs - NovVista Tech Brief & \url{https://novvista.com/ai-powered-code-review-in-2026-graphite-coderabbit-and-what-tools-actually-catch-real-bugs/} \\
G1971 & 5846 & Vibe Coding in 2026: The Revolution That Is Rewriting How Software Gets Built - DEV Community & \url{https://dev.to/alexcloudstar/vibe-coding-in-2026-the-revolution-that-is-rewriting-how-software-gets-built-4d0d} \\
G1972 & 5847 & AI Code Review in 2026: Why Human Review Skills Matter More Than Ever - Botmonster Tech & \url{https://botmonster.com/ai/ai-code-review-2026-human-review-skills-ai-generated-commits/} \\
G1973 & 5850 & A Practical Guide to Becoming an AI-Native Engineer & \url{https://blog.bytebytego.com/p/a-practical-guide-to-becoming-an} \\
G1974 & 5852 & AI Slop Is DDoSing Open Source's Immune System, and Nobody Has a Cure Yet | by Micheal Lanham | Mar, 2026 | Medium & \url{https://medium.com/@Micheal-Lanham/ai-slop-is-ddosing-open-sources-immune-system-and-nobody-has-a-cure-yet-a25058c18bcc} \\
G1975 & 5853 & Casper: Helping developers work effectively with coding assistants | Thoughtworks & \url{https://www.thoughtworks.com/insights/blog/generative-ai/casper-helping-developers-work-effectively-with-coding-assistants} \\
G1976 & 5855 & What Is Vibe Coding? And Does It Actually Work for Production ... & \url{https://dev.to/dextralabs/what-is-vibe-coding-and-does-it-actually-work-for-production-code-i-tested-10-tools-9n7} \\
G1977 & 5860 & In the AI Era, Code Is Cheap. Reputation Isn't. - DEV Community & \url{https://dev.to/kaleman15/in-the-ai-era-code-is-cheap-reputation-isnt-3482} \\
G1978 & 5861 & Building software in an agentic organization | Swarmia & \url{https://www.swarmia.com/blog/the-new-way-building-software/} \\
G1979 & 5864 & Best Graphite Alternatives for Developer Teams in 2025 & \url{https://www.qodo.ai/blog/graphite-alternatives/} \\
G1980 & 5865 & Cursor SDK: Build AI Coding Agents in TypeScript - DEV Community & \url{https://dev.to/jangwook_kim_e31e7291ad98/cursor-sdk-build-ai-coding-agents-in-typescript-3gf6} \\
G1981 & 5869 & I Reviewed 1,000 Lines of AI-Generated Code Here's What Scared Me | by Hussain Shamlooh | Apr, 2026 | Medium & \url{https://medium.com/@hshamlooh/i-reviewed-1-000-lines-of-ai-generated-code-heres-what-scared-me-bb3fe5211e95} \\
G1982 & 5879 & AI Coding Tools Governance Policy 2026: Git... AI Policy Desk & \url{https://www.aipolicydesk.com/blog/ai-coding-tools-governance-policy-github-copilot-cursor-2026} \\
G1983 & 5881 & GPT-5.3 Codex vs Claude Opus 4.6: Workflow Comparison | Aram Melkoumov posted on the topic | LinkedIn & \url{https://www.linkedin.com/posts/melkoumov_gpt-53-codex-vs-claude-opus-46-isnt-a-activity-7425496173680939009-4S77} \\
G1984 & 5884 & What Developers Need to Know About AI Code Reviews | Greptile Blog & \url{https://www.greptile.com/blog/ai-code-review} \\
G1985 & 5885 & AI Writes Code in Seconds. Your Team Reviews It for Hours. The Math Isn't Working. & \url{https://tianpan.co/blog/2026-05-06-ai-code-velocity-review-capacity-mismatch} \\
G1986 & 5886 & How We Used AI Agents to Security-Audit an Open Source Project - DEV Community & \url{https://dev.to/verivusossreleases/how-we-used-ai-agents-to-security-audit-an-open-source-project-2g41} \\
G1987 & 5889 & The Eternal Sloptember | Alasdair Allan | 31 comments & \url{https://www.linkedin.com/posts/alasdairallan_the-eternal-sloptember-activity-7465697276200435712-4fQW} \\
G1988 & 5891 & Human Code Review Is Not the Last Frontier - DEV Community & \url{https://dev.to/joacod/human-code-review-is-not-the-last-frontier-60c} \\
G1989 & 5892 & Harness engineering and agent feedback: Exploring AI coding sensors | Thoughtworks Australia & \url{https://www.thoughtworks.com/en-au/insights/blog/generative-ai/harness-engineering-agent-feedback-exploring-ai-coding-sensors} \\
G1990 & 5894 & navxio/PRSense & \url{https://github.com/navxio/prsense} \\
G1991 & 5903 & VirtualMaestro/openspec-adversarial-multi-agent & \url{https://github.com/VirtualMaestro/openspec-adversarial-multi-agent} \\
G1992 & 5905 & The Agentic Coding Revolution: From Copilots to Autonomous Dev Teams -- Swfte AI & \url{https://www.swfte.com/blog/agentic-coding-revolution-autonomous-dev-teams-2026} \\
G1993 & 5907 & Every Developer Should Review Code -- Not Just Seniors - DEV Community & \url{https://dev.to/zenika/every-developer-should-review-code-not-just-seniors-2abc} \\
G1994 & 5914 & Scaling long-running autonomous coding - DEV Community & \url{https://dev.to/technoblogger14o3/scaling-long-running-autonomous-coding-2hfh} \\
G1995 & 5916 & Code Review in the Age of AI - by Addy Osmani - Elevate & \url{https://addyo.substack.com/p/code-review-in-the-age-of-ai} \\
G1996 & 5920 & AI Agent Targets OSS Maintainers with Cold Outreach - TechNadu & \url{https://www.technadu.com/ai-agent-targets-oss-maintainers-with-autonomous-pull-requests-bypassing-human-detection/620225/} \\
G1997 & 5926 & Best Practices for Engineer Evaluation Systems in the Age of AI (Overview) - DEV Community & \url{https://dev.to/makotunes/best-practices-for-engineer-evaluation-systems-in-the-age-of-ai-overview-37hg} \\
G1998 & 5927 & AI is shifting engineering burden and bottlenecks from code authors to code reviewers. | Amir Hardon & \url{https://www.linkedin.com/posts/amirhardon_ai-is-shifting-engineering-burden-and-bottlenecks-activity-7346439970775056384-ZdQl} \\
G1999 & 5931 & Code Review Was Designed for Human Authors | by Leone Lage Perdigao | Apr, 2026 | Medium & \url{https://leoneperdigao.medium.com/code-review-was-designed-for-human-authors-5905e3a4f9ce} \\
G2000 & 5933 & AI Trust Gap: Developers Trust AI for Code, SREs Don't | Sylvain Kalache posted on the topic | LinkedIn & \url{https://www.linkedin.com/posts/sylvainkalache_developers-trust-ai-to-write-their-code-activity-7445426810919530496-ln0M} \\
G2001 & 5938 & patilmayur5572/ira-review & \url{https://github.com/patilmayur5572/ira-review} \\
G2002 & 5949 & Developer sentiment surveys: why they're not enough in the AI era & \url{https://www.faros.ai/blog/developer-sentiment-surveys-not-enough-ai-era} \\
G2003 & 5957 & Why Should AI Review Your Code? - DEV Community & \url{https://dev.to/pantoai/why-should-ai-review-your-code-38c} \\
G2004 & 5958 & Code Review Critical in AI Coding for Software Engineering - LinkedIn & \url{https://www.linkedin.com/posts/addyosmani_ai-programming-softwareengineering-activity-7451880794958008320-k6uY} \\
G2005 & 5964 & AppSec Governance Playbook: Building Guardrails for AI-Accelerated Development | Snyk & \url{https://snyk.io/lp/appsec-governance-playbook/} \\
G2006 & 5968 & AI code review checklist that actually catches problems - DEV Community & \url{https://dev.to/hackmamba/ai-code-review-checklist-that-actually-catches-problems-10o3} \\
G2007 & 5973 & The Uncomfortable Truth About AI-Assisted Development | aya.is & \url{https://aya.is/en/stories/20260428-the-uncomfortable-truth-about-ai-assisted-development} \\
G2008 & 5978 & Vibe Coding's Security Debt: The AI-Generated CVE Surge & \url{https://labs.cloudsecurityalliance.org/wp-content/uploads/2026/04/CSA_research_note_ai-generated-code-vulnerability-surge_20260404-csa-styled.pdf} \\
G2009 & 5982 & How to Eliminate Rubber-Stamping in the Age of AI-Generated Code: A Proposal for the AI Era | by Jayden Pung | Apr, 2026 | Medium & \url{https://medium.com/@pungkahheng/how-to-eliminate-rubber-stamping-in-the-age-of-ai-generated-code-a-proposal-for-the-ai-era-d47a08e7474c} \\
G2010 & 5990 & Vibe Coding vs Traditional Coding: What's Better for Your Team? & \url{https://apprecode.com/blog/vibe-coding-vs-traditional-coding-whats-better-for-your-team} \\
G2011 & 5991 & Armis Labs Report: Trusted Vibing Benchmark & \url{https://media.armis.com/rp-trusted-vibing-benchmark-en.pdf} \\
G2012 & 5992 & When AI Writes Code Faster Than Humans Can Review It - debuggr.io & \url{https://www.debuggr.io/ai-code-review-bottleneck} \\
G2013 & 5994 & heurema/signum & \url{https://github.com/heurema/signum} \\
G2014 & 5995 & Developers Think AI Makes Them 20\% Faster. They're Actually 19\% Slower. | by Noah Chen | Apr, 2026 | Medium & \url{https://medium.com/@noahchen2002/developers-think-ai-makes-them-20-faster-theyre-actually-19-slower-30c5f4eed496} \\
G2015 & 5999 & AI Won't Replace Developers--But it is Changing How They Work - DevOps.com & \url{https://devops.com/ai-wont-replace-developers-but-it-is-changing-how-they-work/} \\
G2016 & 6001 & 5 Best AI Spec Review Tools for Development Teams (2026) | Augment Code & \url{https://www.augmentcode.com/tools/best-ai-spec-review-tools-for-development-teams} \\
G2017 & 6003 & AI Coding Is Fast Now. Engineering Trust Still Has to Be Earned. - DEV Community & \url{https://dev.to/kair_akhmettayev_0a8ba408/ai-coding-is-fast-now-engineering-trust-still-has-to-be-earned-40ok} \\
G2018 & 6016 & AI is burning out the people who keep open source alive - CodeRabbit & \url{https://coderabbit.ai/blog/ai-is-burning-out-the-people-who-keep-open-source-alive} \\
G2019 & 6022 & Overcoming AI Code Limitations with Human Feedback and Review | Dan Harper posted on the topic | LinkedIn & \url{https://www.linkedin.com/posts/danharper_coding-agents-can-do-99-decent-correct-activity-7416230252663726081-DUE_} \\
G2020 & 6023 & surajpatildev/spec-loop & \url{https://github.com/surajpatildev/spec-loop} \\
G2021 & 6026 & AI's trust tax for developers | InfoWorld & \url{https://www.infoworld.com/article/4111829/ais-trust-tax-for-developers.html} \\
G2022 & 6029 & I built an AI that automatically reviews your GitHub PRs - DEV Community & \url{https://dev.to/murtaztahir2ai/i-built-an-ai-that-automatically-reviews-your-github-prs-4cef} \\
G2023 & 6030 & Spec-driven development with Claude Code: shipping features in an hour - DEV Community & \url{https://dev.to/tufailkhan457/spec-driven-development-with-claude-code-shipping-features-in-an-hour-1e40} \\
G2024 & 6038 & onehorizonai/sensei & \url{https://github.com/onehorizonai/sensei} \\
G2025 & 6039 & AI coding agents keep repeating decade-old security mistakes - Help Net Security & \url{https://www.helpnetsecurity.com/2026/03/13/claude-code-openai-codex-google-gemini-ai-coding-agent-security/} \\
G2026 & 6041 & AI-Generated Code Is Creating a Technical Debt Crisis Nobody Is Auditing - DEV Community & \url{https://dev.to/alexcloudstar/ai-generated-code-is-creating-a-technical-debt-crisis-nobody-is-auditing-4cjc} \\
G2027 & 6044 & CodeRabbit vs Qodo: AI Code Review Tools Compared (2026) - DEV Community & \url{https://dev.to/rahulxsingh/coderabbit-vs-qodo-ai-code-review-tools-compared-2026-4l5h} \\
G2028 & 6045 & AI is reorganizing DevOps. The fight worth watching isn't where you think. | by @pramodchandrayan | Predict | May, 2026 | Medium & \url{https://medium.com/predict/ai-is-reorganizing-devops-the-fight-worth-watching-isnt-where-you-think-af6c37f586f4} \\
G2029 & 6047 & PierrunoYT/ai-code-reviewer & \url{https://github.com/PierrunoYT/ai-code-reviewer} \\
G2030 & 6048 & AI Code Review as Your Last Line of Defense Before Deployment - DEV Community & \url{https://dev.to/deployhq/ai-code-review-as-your-last-line-of-defense-before-deployment-1859} \\
G2031 & 6050 & Copilot isn't slow, your process is - DEV Community & \url{https://dev.to/tudorsss-betterqa/copilot-isnt-slow-your-process-is-3c03} \\
G2032 & 6053 & davidleitw/xreview & \url{https://github.com/davidleitw/xreview} \\
G2033 & 6056 & The Middle Agentic Path: Why Both Sides Are Wrong About Human-in-the-Loop AI Coding -- Codagent Newsletter \#3 & \url{https://codagent.beehiiv.com/p/the-middle-agentic-path} \\
G2034 & 6061 & DanielPodolsky/ownyourcode & \url{https://github.com/DanielPodolsky/ownyourcode} \\
G2035 & 6064 & pridiuksson/code-review-agent & \url{https://github.com/pridiuksson/code-review-agent} \\
G2036 & 6067 & An Oversight Collapse in the Age of AI Coding & \url{https://theforesightlab.substack.com/p/an-oversight-collapse-in-the-age} \\
G2037 & 6068 & Vibe Coding: What Works, What Doesn't, and When to Stop -- ADI Pod & \url{https://adipod.ai/topics/vibe-coding-guide/} \\
G2038 & 6072 & The Load-Bearing Human. Code review used to mean I could say... | by Koshy | Apr, 2026 | Medium & \url{https://nittikkin.medium.com/the-load-bearing-human-e1c0222d6a27} \\
G2039 & 6078 & AI Code Review Guide \& Checklist | TechDebt.now & \url{https://techdebt.now/ai-code-review/} \\
G2040 & 6079 & AI makes faster both engineer and chaos - DEV Community & \url{https://dev.to/prompt_vault/ai-makes-faster-both-engineer-and-chaos-1m93} \\
G2041 & 6080 & I built a mobile app to control Claude Code and Codex from my phone - DEV Community & \url{https://dev.to/k9i/i-built-a-mobile-app-to-control-claude-code-and-codex-from-my-phone-4d84} \\
G2042 & 6082 & The AI-powered DevOps revolution: Redefining developer collaboration - The GitHub Blog & \url{https://github.blog/ai-and-ml/github-copilot/the-ai-powered-devops-revolution-redefining-developer-collaboration/} \\
G2043 & 6083 & AI Code Review Drift: When Your LLM Reviewer's Standards Mutate Faster Than the Code & \url{https://tianpan.co/blog/2026-05-13-ai-code-review-drift-llm-reviewer-standards-mutate} \\
G2044 & 6088 & Drupal Maintainer Shield: Filtering AI Noise So Humans Can Focus on Real Security - DEV Community & \url{https://dev.to/victorstackai/drupal-maintainer-shield-filtering-ai-noise-so-humans-can-focus-on-real-security-3ia2} \\
G2045 & 6090 & In the age of AI coding, code quality still matters | Thoughtworks United States & \url{https://www.thoughtworks.com/en-us/insights/blog/generative-ai/in-the-age-of-AI-coding-code-quality-still-matters} \\
G2046 & 6092 & AWS Security Agent Brings Full Repository Code Scanning to Preview - DevOps.com & \url{https://devops.com/aws-security-agent-brings-full-repository-code-scanning-to-preview/} \\
G2047 & 6096 & The Verification Gap: Why 96\% of Developers Don't Fully Trust AI Code -- Yet Only Half Always Check It in 2026 - DEV Community & \url{https://dev.to/tanishka_karsulkar_ec9e58/the-verification-gap-why-96-of-developers-dont-fully-trust-ai-code-yet-only-half-always-check-4cah} \\
G2048 & 6098 & Code Ownership Decay: What Happens to Team Knowledge When AI Writes Most Commits & \url{https://tianpan.co/blog/2026-05-05-code-ownership-decay-ai-generated-commits} \\
G2049 & 6099 & The Rubber-Stamp Collapse: Why AI-Authored PRs Are Hollowing Out Code Review & \url{https://tianpan.co/blog/2026-04-23-rubber-stamp-collapse-ai-authored-prs} \\
G2050 & 6100 & How to implement effective review gates for AI-assisted development | Thoughtworks Chile & \url{https://www.thoughtworks.com/en-cl/insights/blog/generative-ai/how-to-implement-effective-review-gates-for-ai-assisted-development} \\
G2051 & 6103 & Vibe-Coding Policy Audit: 50-Point Team Scorecard 2026 & \url{https://www.digitalapplied.com/blog/vibe-coding-policy-audit-50-point-team-scorecard-2026} \\
G2052 & 6104 & Checkbox theater: how I stopped trusting my AI agent to run the checks - DEV Community & \url{https://dev.to/wordcaster/checkbox-theater-how-i-stopped-trusting-my-ai-agent-to-run-the-checks-2gf1} \\
G2053 & 6107 & Zendesk Says AI Makes Code Abundant, Shifting the Bottleneck to "Absorption Capacity" - InfoQ & \url{https://www.infoq.com/news/2026/04/zendesk-absorption-capacity/} \\
G2054 & 6108 & How Code Ownership Supercharged Our Dev Velocity at Mesh Payments & \url{https://medium.com/@eyalronel1984/how-code-ownership-supercharged-our-dev-velocity-at-mesh-payments-a8afde903d6d} \\
G2055 & 6111 & Code Review Automation ROI: 200-400\% Returns in 2026 & \url{https://cms.gitar.ai/code-review-automation-roi-2026/} \\
G2056 & 6112 & AI Coding Productivity Statistics 2026: How Much Faster Devs Work | Misar.AI | Misar.Blog & \url{https://www.misar.blog/@misar/articles/ai-coding-productivity-statistics-2026} \\
G2057 & 6114 & Vibe Coding Is Killing Open Source, and the Data Proves It | grith & \url{https://grith.ai/blog/vibe-coding-killing-open-source} \\
G2058 & 6115 & How to do code reviews with AI tools & \url{https://newsletter.techworld-with-milan.com/p/how-to-do-code-reviews-with-ai-tools} \\
G2059 & 6116 & Multi-Model Code Review - Why One AI Isn't Enough | Venture Crane | Venture Crane & \url{https://venturecrane.com/articles/multi-model-code-review/} \\
G2060 & 6117 & brooks-lint: AI Code Reviews Grounded in 12 Classic Engineering Books -- Now on Gemini CLI Spotlight - DEV Community & \url{https://dev.to/hyhmrright/brooks-lint-ai-code-reviews-grounded-in-12-classic-engineering-books-now-on-gemini-cli-spotlight-58fh} \\
G2061 & 6120 & Vibe coding brought back my love for programming - LeadDev & \url{https://leaddev.com/culture/vibe-coding-brought-back-love-programming} \\
G2062 & 6121 & Vibe Coding Explained: AI Code, Developer Impact \& Risks & \url{https://almcorp.com/blog/vibe-coding-developer-impact/} \\
G2063 & 6122 & Vibe Coding Is Real And It's Changing What It Means to Be a Senior Developer - DEV Community & \url{https://dev.to/goodworklabs/vibe-coding-is-real-and-its-changing-what-it-means-to-be-a-senior-developer-5d58} \\
G2064 & 6124 & Introducing Qodo's Rule System: Turning Standards into Systems - Qodo & \url{https://www.qodo.ai/blog/introducing-qodo-rule-system/} \\
G2065 & 6127 & Spec-Driven Development in Regulated Industries -- Governance, Compliance, and Audit Trails - SoftwareSeni & \url{https://www.softwareseni.com/spec-driven-development-in-regulated-industries-governance-compliance-and-audit-trails/} \\
G2066 & 6129 & AI Writes Better Code. We're Getting Worse at Reviewing It. & \url{https://atomicrobot.com/blog/ai-review-fatigue/} \\
G2067 & 6131 & The AI Code Review Bottleneck Is Already Here. Most Teams Haven't Noticed. - DEV Community & \url{https://dev.to/ahmedibrahim/the-ai-code-review-bottleneck-is-already-here-most-teams-havent-noticed-50fa} \\
G2068 & 6134 & Reviewing Agent PRs Is a Different Job, Not a Faster One & \url{https://tianpan.co/blog/2026-04-27-reviewing-agent-prs-different-not-faster} \\
G2069 & 6136 & Vibe Slop: The Hidden Risks of AI-Generated Code in 2026 & \url{https://www.linkedin.com/pulse/vibe-slop-why-ai-pioneers-warning-your-favorite-coding-nantha-kumar-l-lcuxc} \\
G2070 & 6137 & AI Agents Are the New DevOps: How Autonomous Agents Are Redefining Software Engineering - DEV Community & \url{https://dev.to/jadzeino/ai-agents-are-the-new-devops-how-autonomous-agents-are-redefining-software-engineering-11fd} \\
G2071 & 6138 & Three months of measuring my own comprehension while AI wrote the code | by Leonid Sokolovskiy | May, 2026 | Medium & \url{https://medium.com/@leonid_91859/three-months-of-measuring-my-own-comprehension-while-ai-wrote-the-code-549ceec585a1} \\
G2072 & 6139 & How to Set Up AI Code Review in Your CI/CD Pipeline & \url{https://www.augmentcode.com/guides/ai-code-review-ci-cd-pipeline} \\
G2073 & 6143 & Warp's big bet on building open source with GPT-5.5 | OpenAI & \url{https://openai.com/index/warp/} \\
G2074 & 6145 & Governing agents in GitHub Enterprise - GitHub Well-Architected & \url{https://wellarchitected.github.com/library/governance/recommendations/governing-agents/} \\
G2075 & 6151 & Who Owns AI-Generated Code When It Breaks? & \url{https://www.the-main-thread.com/p/ai-code-liability} \\
G2076 & 6152 & Vibecoding, AI, and the Gradual Degradation of Codebases | by Pedrocarvi | Medium & \url{https://medium.com/@pedrocarvi/vibecoding-ai-and-the-gradual-degradation-of-codebases-de39e9e438cf} \\
G2077 & 6156 & Labs: The Evolution of Continuous Merge w/ DevCycle's Nik LeBlanc & \url{https://linearb.io/dev-interrupted/podcast/labs-the-evolution-of-continuous-merge} \\
G2078 & 6158 & Why Developers Don't Trust Code Built by AI Agents - DEV Community & \url{https://dev.to/ucjung/why-developers-dont-trust-code-built-by-ai-agents-146a} \\
G2079 & 6159 & AI-authored code needs more attention, contains worse bugs & \url{https://www.theregister.com/software/2025/12/17/ai-authored-code-needs-more-attention-contains-worse-bugs/2576263} \\
G2080 & 6161 & GitHub Models Are Here! But Are You Using Them to Their Full Potential? & \url{https://www.linkedin.com/pulse/github-models-here-you-using-them-full-potential-pritish-biswas-thvpc} \\
G2081 & 6162 & AI code reviewers: solving a problem or creating one? - LinkedIn & \url{https://www.linkedin.com/posts/zachlloyd_recently-ive-seen-releases-from-graphite-activity-7340466823487741955-4Uzq} \\
G2082 & 6165 & AI and Vibe Coding Are Radically Impacting Senior Devs in Code Review - The New Stack & \url{https://thenewstack.io/ai-and-vibe-coding-are-radically-impacting-senior-devs-in-code-review/} \\
G2083 & 6167 & GitHub eyes restrictions on pull requests to rein in AI-based code deluge on maintainers | InfoWorld & \url{https://www.infoworld.com/article/4127156/github-eyes-restrictions-on-pull-requests-to-rein-in-ai-based-code-deluge-on-maintainers.html} \\
G2084 & 6169 & Intelligent Code Reviews Reimagined: How GitHub Copilot Is Redefining Team Engineering at Scale + Video - UNDERCODE NEWS & \url{https://undercodenews.com/intelligent-code-reviews-reimagined-how-github-copilot-is-redefining-team-engineering-at-scale-video/} \\
G2085 & 6174 & AI garbage dumped into the open-source community! Some projects have started suspending external contributions, and Shitcode feels like a DDoS attack. | Gate Ne & \url{https://www.gate.com/news/detail/18550585} \\
G2086 & 6179 & Amazon Just Banned AI Code Without Human Approval. Here's Why They're Right. | by Felix Renner | Medium & \url{https://medium.com/@ryannsj/amazon-just-banned-ai-code-without-human-approval-heres-why-they-re-right-21a51d5a596a} \\
G2087 & 6182 & AI-Powered Code Review: Enhancing Developer Productivity and ... & \url{https://www.qodequay.com/ai-code-review-guide} \\
G2088 & 6183 & I Checked What Security Vulnerabilities AI Coding Tools Actually ... & \url{https://dev.to/vitalii_petrenko_dev/i-checked-what-security-vulnerabilities-ai-coding-tools-actually-introduce-1cjd} \\
G2089 & 6187 & Opinion: How to vibe code responsibly. | by Daniel S Jeremiah | Medium & \url{https://medium.com/@dsj11/opinion-how-to-vibe-code-responsibly-38f610a0ac36} \\
G2090 & 6188 & Building ZeroTrust PR: An AI-Powered Security Auditor for GitHub Pull Requests | by Nadummy | May, 2026 | Medium & \url{https://medium.com/@nadummy17/building-zerotrust-pr-an-ai-powered-security-auditor-for-github-pull-requests-4411830defb8} \\
G2091 & 6189 & Vibe Coding Isn't Bad, You Are - DEV Community & \url{https://dev.to/brunellyai/vibe-coding-isnt-bad-you-are-31} \\
G2092 & 6190 & Anthropic Expands Claude Mythos for Infrastructure Defense & \url{https://winbuzzer.com/2026/06/03/anthropic-expands-claude-mythos-for-infrastructure-defense-xcxwbn/} \\
G2093 & 6193 & Top Cursor Alternatives for Engineers (2025 Guide) & \url{https://www.qodo.ai/blog/cursor-alternatives/} \\
G2094 & 6196 & CodeRabbit Pricing vs AI Code Review Tools: Gitar Wins & \url{https://cms.gitar.ai/coderabbit-pricing-vs-ai-review/} \\
G2095 & 6198 & GitLab Adds Flat-Rate Code Reviews, Free-Tier AI Access, and Spending Caps - InfoQ & \url{https://www.infoq.com/news/2026/04/gitlab-flatrate-view-ai-access/} \\
G2096 & 6204 & Vibe coding is not the same as AI-Assisted engineering. - Medium & \url{https://medium.com/@addyosmani/vibe-coding-is-not-the-same-as-ai-assisted-engineering-3f81088d5b98} \\
G2097 & 6207 & Anthropic Adds Automated Security Reviews to Claude Code - DevOps.com & \url{https://devops.com/anthropic-adds-automated-security-reviews-to-claude-code/} \\
G2098 & 6212 & Building a Code Review Agent That Learns From Past Fixes - DEV Community & \url{https://dev.to/bhagathrao_99/building-a-code-review-agent-that-learns-from-past-fixes-36gg} \\
G2099 & 6219 & AI Code Governance - SIG & \url{https://www.softwareimprovementgroup.com/use-case/artificial-intelligence/ai-code-governance/} \\
G2100 & 6221 & Log \#1: Building the Base Control Model - DEV Community & \url{https://dev.to/dglezz/log-1-building-the-base-control-model-18om} \\
G2101 & 6223 & dbreunig/plumb & \url{https://github.com/dbreunig/plumb} \\
G2102 & 6229 & The difference between coding agent and agent mode in GitHub ... & \url{https://github.blog/developer-skills/github/less-todo-more-done-the-difference-between-coding-agent-and-agent-mode-in-github-copilot/} \\
G2103 & 6233 & Claude Code adds code reviews | InfoWorld & \url{https://www.infoworld.com/article/4143297/claude-code-adds-code-reviews.html} \\
G2104 & 6234 & My Ultimate AI Coding Workflow (GitHub + Cursor Agents Setup) & \url{https://www.youtube.com/watch?v=AggITrydtwk} \\
G2105 & 6236 & claude-code - (How to fix) Fix [MODEL] Claude Code Web multi-agent review synthesis fabricates verification and expands issue scope & \url{https://www.stepcodex.com/en/issue/model-claude-code-web-multi-agent} \\
G2106 & 6242 & Will Humans Still Review Code?. The critical question companies must... | by Francisco Trindade | Medium & \url{https://franciscomt.medium.com/will-humans-still-review-code-a6f7d3f0c39c} \\
G2107 & 6243 & How to Use AI for Code Review in 2026: Tools, Prompts \& 70\% Faster PR Turnaround | Happycapy Guide & \url{https://happycapyguide.com/blog/how-to-use-ai-for-code-review-2026} \\
G2108 & 6244 & Devin: Autonomous AI Software Engineer for Teams & \url{https://tooldirectory.ai/tools/devin} \\
G2109 & 6247 & SonarQube Agentic Analysis using Claude Code | Sonar & \url{https://www.sonarsource.com/resources/library/get-started-with-sonarqube-agentic-analysis-using-claude-code/} \\
G2110 & 6249 & esreekarreddy/scout & \url{https://github.com/esreekarreddy/scout} \\
G2111 & 6250 & How to Build Claude Code Custom Skills for Automated Code Review - DEV Community & \url{https://dev.to/myougatheaxo/how-to-build-claude-code-custom-skills-for-automated-code-review-85p} \\
G2112 & 6252 & I Let AI Handle My PR Reviews for 30 Days -- The Data Surprised Me - DEV Community & \url{https://dev.to/hopkins_jesse_cdb68cfa22c/i-let-ai-handle-my-pr-reviews-for-30-days-the-data-surprised-me-24dn} \\
G2113 & 6258 & How to stop the AI code generation treadmill | InfoWorld & \url{https://www.infoworld.com/article/4177717/how-to-stop-the-ai-code-generation-treadmill.html} \\
G2114 & 6260 & Anthropic AI Claude Code Security Paradox | by NOEL FURTADO | Feb, 2026 | Medium & \url{https://medium.com/@noeljf.in/anthropic-ai-claude-code-security-paradox-140368286139} \\
G2115 & 6261 & Why AI Code Gets Less Secure With Every Prompt | Blog | Endor Labs & \url{https://www.endorlabs.com/learn/ai-code-gets-less-secure-with-every-prompt} \\
G2116 & 6266 & We Don't Trust AI (and That's a Good Thing) - DEV Community & \url{https://dev.to/_steve_fenton_/we-dont-trust-ai-and-thats-a-good-thing-3oe6} \\
G2117 & 6270 & I Built an AI Code Review Pipeline -- One Command for Review + Auto-Fix + Tests + Report - DEV Community & \url{https://dev.to/danche/i-built-an-ai-code-review-pipeline-one-command-for-review-auto-fix-tests-report-5g9d} \\
G2118 & 6272 & AI Coding Assistants Haven't Sped up Delivery Because Coding Was Never the Bottleneck - InfoQ & \url{https://www.infoq.com/news/2026/03/agoda-ai-code-bottleneck/} \\
G2119 & 6278 & GitHub Copilot: Meet the new coding agent - The GitHub Blog & \url{https://github.blog/news-insights/product-news/github-copilot-meet-the-new-coding-agent/} \\
G2120 & 6279 & The Hidden Risks of Modern Code: Security Patterns Modern Tools Still Miss | by Jost Faganel | KolegaDev | Medium & \url{https://medium.com/kolegadev/the-hidden-risks-of-modern-code-security-patterns-modern-tools-still-miss-9dfe24d8ed34} \\
G2121 & 6282 & \# I Built an Autonomous Multi-Agent Security Auditor using Gemini 3 and TypeScript - DEV Community & \url{https://dev.to/itxdeeni/-i-built-an-autonomous-multi-agent-security-auditor-using-gemini-3-and-typescript-2ach} \\
G2122 & 6284 & GitHub Copilot remote control makes local agent sessions reviewable from mobile - Blog | LinkLoot & \url{https://linkloot.io/blog/github-copilot-remote-control-local-agent-sessions} \\
G2123 & 6285 & I Taught Claude Code to Distrust Its Own Plans | by Xavier Malina | Mar, 2026 | Medium & \url{https://medium.com/@xaviermalina/i-taught-claude-code-to-distrust-its-own-plans-dd67156f13f1} \\
G2124 & 6288 & Copilot doesn't make juniors dumber. It makes seniors lazier. - DEV Community & \url{https://dev.to/adioof/copilot-doesnt-make-juniors-dumber-it-makes-seniors-lazier-19lp} \\
G2125 & 6289 & fix: switch Copilot auto-fix trigger to pull\_request\_review\_comment Pull Request \#81 future-architect/uzomuzo-oss & \url{https://github.com/future-architect/uzomuzo-oss/pull/81} \\
G2126 & 6292 & Why code review and testing matter more than ever with AI | Elio Struyf & \url{https://www.eliostruyf.com/code-review-testing-matter-ai/} \\
G2127 & 6295 & Compliance Frameworks for AI-Native Engineering | AI Code Governance \& Quality | Bitloops Resources & \url{https://bitloops.com/resources/governance/compliance-frameworks-for-ai-native-engineering} \\
G2128 & 6305 & I tested 9 code review tools to see which is best! - YouTube & \url{https://www.youtube.com/watch?v=JmraS29Kqgs&vl=en-US} \\
G2129 & 6306 & [2026-04] How I Currently Use Claude Code | by Takashi Masuda | Apr, 2026 | Medium & \url{https://medium.com/@masutaka/2026-04-how-i-currently-use-claude-code-a18979cbc01c} \\
G2130 & 6307 & angelkurten/specforge-framework & \url{https://github.com/angelkurten/specforge-framework} \\
G2131 & 6308 & Enterprise AI Code Management | Governance Framework Guide & \url{https://www.ofashandfire.com/blog/enterprise-ai-code-management} \\
G2132 & 6313 & Context Engineering with a Multi-Agent Approach: A Step Closer to Autonomous Development | by Claudio Diniz | Medium & \url{https://medium.com/@claudiodiniz/context-engineering-with-a-multi-agent-approach-a-step-closer-to-autonomous-development-c42e44bee880} \\
G2133 & 6316 & 'Vibe Coding': A Necessary Evolution or a Risk to Software Quality? & \url{https://www.linkedin.com/pulse/vibe-coding-necessary-evolution-risk-software-quality-carucci-wiz8c} \\
G2134 & 6318 & Qodo latest releases - May 2025 & \url{https://www.qodo.ai/blog/latest-releases-may-2025/} \\
G2135 & 6322 & I Read Cursor's Security Agent Prompts, So You Don't Have To - DEV Community & \url{https://dev.to/snyk/i-read-cursors-security-agent-prompts-so-you-dont-have-to-4n8l} \\
G2136 & 6326 & The Agent That Doesn't Write Code: How a Forensic AI Assistant Changed Our Sprint - DEV Community & \url{https://dev.to/tmdlrg/the-agent-that-doesnt-write-code-how-a-forensic-ai-assistant-changed-our-sprint-18d} \\
G2137 & 6328 & Already Fixed, Too Late: What 90 Open-Source Repos Tell Us About Code Review Blind Spots - Qodo & \url{https://www.qodo.ai/blog/already-fixed-too-late-what-90-open-source-repos-tell-us-about-code-review-blind-spots/} \\
G2138 & 6332 & Claude Code Review. A Deep Dive into Claude Code vs. Cursor | by Aaditya Bhat | Medium & \url{https://medium.com/@aadityaubhat/claude-code-review-ed117fa662f2} \\
G2139 & 6333 & Addressing User Distrust in GeeksforGeeks: Enhancing AI Content Reliability and Cryptographic Examples - DEV Community & \url{https://dev.to/kornilovconstru/addressing-user-distrust-in-geeksforgeeks-enhancing-ai-content-reliability-and-cryptographic-2k90} \\
G2140 & 6350 & I Built an AI PR Review Agent for My Daily Engineering Work | by Praveen Suthar | May, 2026 | Medium & \url{https://medium.com/@praveenmistry/i-built-an-ai-pr-review-agent-for-my-daily-engineering-work-bb5cb54b1f8e} \\
G2141 & 6353 & Stripe Engineers Deploy Minions, Autonomous Agents Producing Thousands of Pull Requests Weekly - InfoQ & \url{https://www.infoq.com/news/2026/03/stripe-autonomous-coding-agents/} \\
G2142 & 6355 & Claude Opus 4.6 co-authored a \$1.78M bug, one sentence taught it to catch it & \url{https://www.linkedin.com/pulse/claude-opus-46-co-authored-178m-bug-one-sentence-taught-shahar-madar-zicnf} \\
G2143 & 6357 & Andrej Karpathy Has Renamed Vibe Coding. Here's What Engineering Leaders Need to Do About It. - SD Times & \url{https://sdtimes.com/ai/andrej-karpathy-has-renamed-vibe-coding-heres-what-engineering-leaders-need-to-do-about-it/} \\
G2144 & 6360 & Your AI Coding Tool Is Making You 19\% Slower. The Study You Should Read. | by James Pelton | Apr, 2026 | Medium & \url{https://medium.com/@jpelton722/your-ai-coding-tool-is-making-you-19-slower-the-study-you-should-read-31e13939bbb9} \\
G2145 & 6365 & Optimizing AI-Assisted Code Review with Contextual PR Descriptions & \url{https://www.linkedin.com/posts/derewecki_the-pr-description-is-the-new-artifact-code-activity-7442632504932700162-mr2R} \\
G2146 & 6371 & What happens when an AI agent commits to your repo - DEV Community & \url{https://dev.to/mdenda/what-happens-when-an-ai-agent-commits-to-your-repo-4cgg} \\
G2147 & 6375 & vercel-labs/openreview & \url{https://github.com/vercel-labs/openreview} \\
G2148 & 6377 & AI-Generated Code Has 1.7x More Bugs -- Here's the Fix - DEV Community & \url{https://dev.to/hai_huang_f196ed9669351e0/ai-generated-code-has-17x-more-bugs-heres-the-fix-559p} \\
G2149 & 6378 & TEST State of Play: the Deazy take on AI-Augmented Engineering moving into 2026 & \url{https://www.deazy.com/en-gb/resource-hub/test-state-of-play-the-deazy-take-on-ai-augmented-engineering-moving-into-2026-clone} \\
G2150 & 6380 & Human-on-the-Loop: AI Reviewing AI PRs at cortex (769 PRs/month, while raising the quality bar) - DEV Community & \url{https://dev.to/ryantsuji/human-on-the-loop-ai-reviewing-ai-prs-at-cortex-769-prsmonth-while-raising-the-quality-bar-4lh5} \\
G2151 & 6382 & AI Coding Autopilot vs Manual Control: What Aviation Taught Us About Skill Decay - DEV Community & \url{https://dev.to/alanwest/ai-coding-autopilot-vs-manual-control-what-aviation-taught-us-about-skill-decay-2h1g} \\
G2152 & 6385 & AI Code Review That Understands Your Codebase - LinkedIn & \url{https://www.linkedin.com/posts/brandonwaselnuk_ai-code-review-is-broken-but-we-need-to-activity-7403912186424680448-h9Q5} \\
G2153 & 6387 & I open-sourced a 4-agent adversarial code review team. Any coding agent can call it as an MCP server. Built in heym. - DEV Community & \url{https://dev.to/frank_brsrk/i-open-sourced-a-4-agent-adversarial-code-review-team-any-coding-agent-can-call-it-as-an-mcp-36oe} \\
G2154 & 6388 & GitHub Copilot Review 2026: Honest Pricing \& Assessment & \url{https://computertech.co/github-copilot-review/} \\
G2155 & 6390 & bizzkoot/zai-code-review & \url{https://github.com/bizzkoot/zai-code-review} \\
G2156 & 6393 & AI-Generated Code Needs AI-Generated Reviews | Clarity & \url{https://heyclarity.dev/blog/ai-code-needs-ai-reviews/} \\
G2157 & 6399 & AI Coding Tools Create More Bugs Than They Fix - The New Stack & \url{https://thenewstack.io/ai-coding-tools-create-more-bugs-than-they-fix/} \\
G2158 & 6401 & CodeRabbit's "State of AI vs Human Code Generation" Report Finds That AI-Written Code Produces \textasciitilde{} 1.7x More Issues Than Human Code & \url{https://www.businesswire.com/news/home/20251217666881/en/CodeRabbits-State-of-AI-vs-Human-Code-Generation-Report-Finds-That-AI-Written-Code-Produces-1.7x-More-Issues-Than-Human-Code} \\
G2159 & 6404 & CodeRabbit for Monorepos: Handling Large Codebases - DEV Community & \url{https://dev.to/rahulxsingh/coderabbit-for-monorepos-handling-large-codebases-140k} \\
G2160 & 6405 & Best AI Code Review Tools for Enterprise & \url{https://www.greptile.com/blog/best-code-review-enterprise} \\
G2161 & 6407 & Reviewing AI-Generated Pull Requests: Reliability, Risk, and the Human Bottleneck | Rich Robertson & \url{https://www.myrobertson.com/blog/reviewing-ai-generated-pull-requests-reliability-risk} \\
G2162 & 6409 & Implement suggestion -> Fix with Copilot; batch fixes now supported -- GitHub Changelog -- releases.sh & \url{https://releases.sh/release/rel_tDLVPzkbYtgu4E8Iq81d3} \\
G2163 & 6410 & When Should Your Agent Ask Before Acting? A 3-Tier Risk Framework - DEV Community & \url{https://dev.to/sahil_kat/when-should-your-agent-ask-before-acting-a-3-tier-risk-framework-3pm3} \\
G2164 & 6411 & AI Code Review in Your CI/CD Pipeline: Automating PR Reviews, Test Generation, and Bug Detection with LLMs - DEV Community & \url{https://dev.to/pockit_tools/ai-code-review-in-your-cicd-pipeline-automating-pr-reviews-test-generation-and-bug-detection-56j4} \\
G2165 & 6414 & Prioritizing Quality Over Quantity: Addressing Software Reliability and User Experience Decline - DEV Community & \url{https://dev.to/svetlix/prioritizing-quality-over-quantity-addressing-software-reliability-and-user-experience-decline-577f} \\
G2166 & 6419 & The AI Security Paradox: Velocity vs Vulnerability, Copilot Autofix, and CVE-2025-53773 - Exzil Calanza & \url{https://exzilcalanza.info/ai-security-paradox-velocity-vulnerability-copilot-autofix-2026/} \\
G2167 & 6424 & Vibecoding / Agentic Engineering from a technological perspective: What could possibly go wrong? | by Yannick Bontemps | Studio Fledge | Apr, 2026 | Medium & \url{https://medium.com/studio-fledge/vibecoding-agentic-engineering-from-a-technological-perspective-what-could-possibly-go-wrong-9f464d2484e3} \\
G2168 & 6425 & How pairing SAST with AI dramatically reduces false positives in code security | InfoWorld & \url{https://www.infoworld.com/article/4093079/how-pairing-sast-with-ai-dramatically-reduces-false-positives-in-code-security.html} \\
G2169 & 6427 & Autonomous AI Agents: The Complete Guide to Self-Directed Intelligence | by Gaston Aps | Medium & \url{https://medium.com/@gastonaps/autonomous-ai-agents-the-complete-guide-to-self-directed-intelligence-ef84d12cf6af} \\
G2170 & 6428 & AI Code Review vs Human Review: Strengths \& Weaknesses | ClaudeGuide & \url{https://claudeguide.io/ai-code-review-vs-human} \\
G2171 & 6432 & I Miss Writing Code No One Was Going to See - DEV Community & \url{https://dev.to/dev_tips/i-miss-writing-code-no-one-was-going-to-see-35fk} \\
G2172 & 6435 & Unlocking the full power of Copilot code review: Master your instructions files - The GitHub Blog & \url{https://github.blog/ai-and-ml/github-copilot/unlocking-the-full-power-of-copilot-code-review-master-your-instructions-files/} \\
G2173 & 6437 & What does 10x engineering productivity look like in practice, using AI agents? For months, my gut feel has been that agentic engineering makes our team operate  & \url{https://www.linkedin.com/posts/hkniberg_what-does-10x-engineering-productivity-look-activity-7463519643064827904-KwQ3} \\
G2174 & 6440 & Trusting AI-Generated Code & \url{https://jdno.dev/trusting-ai-generated-code/} \\
G2175 & 6441 & What Is AI Code Governance? | AI Code Governance \& Quality | Bitloops Resources & \url{https://bitloops.com/resources/governance/what-is-ai-code-governance} \\
G2176 & 6442 & AI Coding Tip 012 - Understand All Your Code - DEV Community & \url{https://dev.to/mcsee/ai-coding-tip-012-understand-all-your-code-450h} \\
G2177 & 6443 & 5 AI Security Incidents That Broke Things in Production - DZone & \url{https://dzone.com/articles/ai-security-incidents-production-lessons} \\
G2178 & 6446 & AI Code Tools: Impact on Maintainability and Design Quality | Equal Experts South Africa posted on the topic | LinkedIn & \url{https://www.linkedin.com/posts/equal-experts-south-africa_pragmaticai-activity-7423667493040283648-FuGL} \\
G2179 & 6450 & Automatically fix code backlog with SonarQube Remediation Agent | Sonar & \url{https://www.sonarsource.com/blog/automatically-fix-code-backlog-with-sonarqube-remediation-agent} \\
G2180 & 6451 & Why Coding Agents Will Break Your CI/CD Pipeline (and How To Fix It) | by Signadot | Apr, 2026 | Medium & \url{https://medium.com/@signadot/why-coding-agents-will-break-your-ci-cd-pipeline-and-how-to-fix-it-caedf655aeff} \\
G2181 & 6458 & Accelerated Productivity with AI: Lessons from Lorin Hochstein | Rushil Makkar posted on the topic | LinkedIn & \url{https://www.linkedin.com/posts/rushilmakkar_just-read-life-comes-at-you-fast-by-lorin-activity-7457035035364204545-AjTt} \\
G2182 & 6459 & blog dds: 2026-05-23 -- Why reviewing AI-generated code is devilishly hard & \url{https://www.spinellis.gr/blog/20260523/index.html} \\
G2183 & 6464 & The Engineering Scalability Crisis: Why Standard Code Structures Matter More Than Ever - DEV Community & \url{https://dev.to/siy/the-engineering-scalability-crisis-why-standard-code-structures-matter-more-than-ever-37e8} \\
G2184 & 6466 & Git Plugin | AI Code Review Agent for Confident Commits | Qodo & \url{https://www.qodo.ai/features/qodo-git/} \\
G2185 & 6467 & Most developers think AI-generated code is garbage. They couldn't be more wrong. Here's what's really happening behind the scenes... Team "ai code sucks" says:  & \url{https://www.linkedin.com/posts/markshust_most-developers-think-ai-generated-code-is-activity-7371512967877804033-BsnZ} \\
G2186 & 6470 & 2026Devin |AI | AI & \url{https://aigyomunote.com/devin-ai-guide-2026/} \\
G2187 & 6475 & AI Code Review Tools 2026: CodeRabbit, Greptile, Diamond, Cody, Cursor -- The Comparison Nobody Publishes Honestly | BirJob & \url{https://www.birjob.com/blog/ai-code-review-tools-2026} \\
G2188 & 6476 & A Heated Rivalry: Claude Code + OpenAI Codex - Pablo Stanley & \url{https://pablostanley.substack.com/p/a-heated-rivalry-claude-code-openai} \\
G2189 & 6478 & Evolution of Code Reviews: From Manual Checks to AI Collaboration & \url{https://vocal.media/geeks/evolution-of-code-reviews-from-manual-checks-to-ai-collaboration} \\
G2190 & 6487 & How to Code Review AI-Generated Code: What Needs Human Eyes vs. What Doesn't -- BiveCode Blog & \url{https://bivecode.com/blog/review-ai-generated-code} \\
G2191 & 6491 & Anthropic Built a Model So Good at Code It Accidentally Became an Elite Hacker - DEV Community & \url{https://dev.to/luffyguy/anthropic-built-a-model-so-good-at-code-it-accidentally-became-an-elite-hacker-5fbi} \\
G2192 & 6493 & BolivarTech/panoptico & \url{https://github.com/BolivarTech/panoptico} \\
G2193 & 6494 & Megalodon campaign hit 5,500 GitHub repos with malicious CI/CD workflows & \url{https://vpncentral.com/megalodon-campaign-hit-5500-github-repos-with-malicious-ci-cd-workflows/} \\
G2194 & 6495 & Developers Trust AI Code. They Also Don't Trust It. Both Are True. - Coder Legion & \url{https://coderlegion.com/16265/developers-trust-ai-code-they-also-dont-trust-it-both-are-true} \\
G2195 & 6501 & How to Automate Code Reviews with Claude Code's /code-review Skill - DEV Community & \url{https://dev.to/myougatheaxo/how-to-automate-code-reviews-with-claude-codes-code-review-skill-47ld} \\
G2196 & 6502 & Vibe Coding Isn't Bad, You Are - DEV Community & \url{https://dev.to/guypowell/vibe-coding-isnt-bad-you-are-b4l} \\
G2197 & 6503 & PR reviews were already broken. AI made it worse - LeadDev & \url{https://leaddev.com/ai/pr-reviews-were-already-broken-ai-made-it-worse} \\
G2198 & 6505 & AI Coding Spec Packet Before Code Generation | Spec Coding & \url{https://spec-coding.dev/blog/ai-coding-spec-packet-before-code-generation} \\
G2199 & 6512 & AI Code Trust Crisis: 96\% Distrust, 48\% Don't Verify | byteiota & \url{https://byteiota.com/ai-code-trust-crisis-96-distrust-48-dont-verify/} \\
G2200 & 6515 & DeepSource: The AI Code Review Platform & \url{deepsource.com/} \\
G2201 & 6521 & Partnering with Mozilla to improve Firefox's security \textbackslash\{\} Anthropic & \url{https://www.anthropic.com/news/mozilla-firefox-security?1cfc31eb_page=4} \\
G2202 & 6522 & From idea to pull request: A practical guide to building with GitHub Copilot CLI - The GitHub Blog & \url{https://github.blog/ai-and-ml/github-copilot/from-idea-to-pull-request-a-practical-guide-to-building-with-github-copilot-cli/} \\
G2203 & 6524 & Qodo Merge 1.0: Solving Challenges in Ai-assisted Code Reviews & \url{https://www.qodo.ai/blog/qodo-merge-solving-key-challenges-in-ai-assisted-code-reviews/} \\
G2204 & 6525 & Microsoft execs worry AI will eat entry level coding jobs & \url{https://www.theregister.com/software/2026/02/23/microsoft-execs-worry-ai-will-eat-entry-level-coding-jobs/4266710} \\
G2205 & 6526 & Beyond vibe coding: How AI can transform pull requests | Thoughtworks & \url{https://www.thoughtworks.com/insights/blog/generative-ai/beyond-vibe-coding-how-AI-can-transform-pull-requests} \\
G2206 & 6528 & Entering the Age of AI: A Laggard's Tale - DEV Community & \url{https://dev.to/playfulprogramming/entering-the-age-of-ai-a-laggards-tale-56i1} \\
G2207 & 6529 & Claude Code Verification: How to verify AI-written code so the bugs get caught before they ship, not after | by Rick Hightower | May, 2026 | Medium & \url{https://medium.com/@richardhightower/claude-code-verification-how-to-verify-ai-written-code-so-the-bugs-get-caught-before-they-ship-3036ff01bec0} \\
G2208 & 6530 & AI coding assistants are making juniors worse and seniors lazier - DEV Community & \url{https://dev.to/adioof/ai-coding-assistants-are-making-juniors-worse-and-seniors-lazier-1h2g} \\
G2209 & 6531 & AI-Generated Code Challenges Engineering Leaders | Marcus Holm posted on the topic | LinkedIn & \url{https://www.linkedin.com/posts/marcus-holm-49019ab_anthropic-launches-code-review-tool-to-check-activity-7437667944265965569-uiSX} \\
G2210 & 6532 & Verified Spec-Driven Development & \url{https://gist.github.com/elyase/961d812e2a535601945aac8f61fab956} \\
G2211 & 6533 & A software engineering metrics glossary for business and technical leaders & \url{https://www.faros.ai/blog/software-engineering-glossary} \\
G2212 & 6535 & Claude Code ultrareview Hunts Bugs Before You Merge & \url{https://everydayaiblog.com/claude-code-ultrareview-bug-hunting-agents/} \\
G2213 & 6537 & Code Review Paradox: Breaking Down Large PRs with Stacked Pull Requests | Joe Buza posted on the topic | LinkedIn & \url{https://www.linkedin.com/posts/joebuza_ask-an-engineer-to-review-15-lines-of-code-activity-7424623724546506752-5glY} \\
G2214 & 6541 & Pydantic AI Tackles Low-Quality AI-Generated PRs to Focus on Meaningful Contributions and Framework Improvements - DEV Community & \url{https://dev.to/romdevin/pydantic-ai-tackles-low-quality-ai-generated-prs-to-focus-on-meaningful-contributions-and-framework-428j} \\
G2215 & 6542 & What is Vibe Coding? Prompting AI Software Development | Sonar & \url{https://www.sonarsource.com/resources/library/vibe-coding/} \\
G2216 & 6543 & AI Code Review: How AI Is Transforming Software Development and ... & \url{https://www.legitsecurity.com/aspm-knowledge-base/ai-code-review} \\
G2217 & 6545 & What Anthropic Didn't Say About Opus 4.8: It's Anthropic Absorbing Your Harness | by Han HELOIR YAN, Ph.D. | Data Science Collective | May, 2026 | Medium & \url{https://medium.com/data-science-collective/what-anthropic-didnt-say-about-opus-4-8-it-s-anthropic-absorbing-your-harness-6d4ea10bf66d} \\
G2218 & 6548 & The Real Problem With AI for Developers Is Not Capability, It's Overload - DEV Community & \url{https://dev.to/maxmendes91/the-real-problem-with-ai-for-developers-is-not-capability-its-overload-587o} \\
G2219 & 6558 & Case File 3.1: The Rubber Stamp Fraud - DEV Community & \url{https://dev.to/manojsatna31/case-file-31-the-rubber-stamp-fraud-971} \\
G2220 & 6564 & The "Syntax Repair" That Turned My Algorithm Into a Liar | HackerNoon & \url{https://hackernoon.com/the-syntax-repair-that-turned-my-algorithm-into-a-liar} \\
G2221 & 6565 & Anthropic Introduces Agent-Based Code Review for Claude Code - InfoQ & \url{https://www.infoq.com/news/2026/04/claude-code-review/} \\
G2222 & 6569 & Why AI Makes Code Review More Important, Not Less | by Alpesh Pawar | May, 2026 | Artificial Intelligence in Plain English & \url{https://ai.plainenglish.io/why-ai-makes-code-review-more-important-not-less-36c149530bf1} \\
G2223 & 6570 & AI Tool Use Policy | HEIR & \url{https://heir.dev/docs/development/ai_policy/} \\
G2224 & 6571 & AI Code Review: From Vibe to Verification - DEV Community & \url{https://dev.to/dr_hernani_costa/ai-code-review-from-vibe-to-verification-5gd7} \\
G2225 & 6576 & Onboarding your AI peer programmer: Setting up GitHub Copilot coding agent for success - The GitHub Blog & \url{https://github.blog/ai-and-ml/github-copilot/onboarding-your-ai-peer-programmer-setting-up-github-copilot-coding-agent-for-success/} \\
G2226 & 6578 & How to Review AI-Generated Code Without Getting Burned (Most ... & \url{https://medium.com/@optimzationking2/how-to-review-ai-generated-code-without-getting-burned-most-devs-skip-this-step-1366a1053bcc} \\
G2227 & 6582 & Cisco praises AI bug hunt, won't reveal flaw tally & \url{https://www.theregister.com/ai-and-ml/2026/06/02/cisco-praises-ai-bug-hunt-wont-reveal-flaw-tally/5250291} \\
G2228 & 6586 & Millwright-Inspector: A Methodology for Software Development with AI Coding Agents - DEV Community & \url{https://dev.to/emin_akkoc/millwright-inspector-a-methodology-for-software-development-with-ai-coding-agents-p10} \\
G2229 & 6589 & 23\% of Devs Regularly Use AI Agents, per Stack Overflow Survey - The New Stack & \url{https://thenewstack.io/23-of-devs-regularly-use-ai-agents-per-stack-overflow-survey/} \\
G2230 & 6600 & Cursor 3 Deep Dive: Agents + Composer Review 2026 & \url{https://www.digitalapplied.com/blog/cursor-3-deep-dive-agents-composer-review-2026} \\
G2231 & 6601 & AI Won't Kill Junior Devs - But Your Hiring Strategy Might & \url{https://addyo.substack.com/p/ai-wont-kill-junior-devs-but-your} \\
G2232 & 6603 & I Asked 4 AIs to Judge Each Other's Code - DEV Community & \url{https://dev.to/yurukusa/i-asked-4-ais-to-judge-each-others-code-5c2a} \\
G2233 & 6605 & AI Code Reviews Are Broken By Default -- Here's How to Fix Them | by Mpholoane Bapela | Mar, 2026 | Medium & \url{https://medium.com/@mpholoane/ai-code-reviews-are-broken-by-default-heres-how-to-fix-them-174a327e16be} \\
G2234 & 6613 & Claude Code vs Codex: Which AI Coding Tool Is Right for You? - DEV Community & \url{https://dev.to/zach_hajjaj_260cfed0ef4b4/claude-code-vs-codex-which-ai-coding-tool-is-right-for-you-44p0} \\
G2235 & 6621 & What to expect for open source in 2026 - The GitHub Blog & \url{https://github.blog/open-source/maintainers/what-to-expect-for-open-source-in-2026/} \\
G2236 & 6622 & Do security prompts make AI-generated code safer? I tested it. & \url{https://derivai.substack.com/p/do-security-prompts-make-ai-code-safer} \\
G2237 & 6624 & PR Cycle Time in the AI Era: Why the Bottleneck Moved | Developer Productivity & \url{https://larridin.com/developer-productivity-hub/pr-cycle-time-ai-era-bottleneck} \\
G2238 & 6628 & AI-Generated Code Quality Is Collapsing: The Silent Crisis Nobody Wants to Talk About [2026] - DEV Community & \url{https://dev.to/kunal_d6a8fea2309e1571ee7/ai-generated-code-quality-is-collapsing-the-silent-crisis-nobody-wants-to-talk-about-2026-54j0} \\
G2239 & 6629 & The tough task of making AI code production-ready | InfoWorld & \url{https://www.infoworld.com/article/3994519/the-tough-task-of-making-ai-code-production-ready.html} \\
G2240 & 6630 & PermiForce: Permit.io X azureDevops & \url{https://dev.to/shivamkapasia/permiforce-permitio-x-azuredevops-33e} \\
G2241 & 6631 & GitHub Copilot Writes 46\% of Your Code. That Should Make You Uncomfortable. | by Yusuf Seyitoglu | Let's Code Future | Feb, 2026 | Medium & \url{https://medium.com/lets-code-future/github-copilot-writes-46-of-your-code-that-should-make-you-uncomfortable-5152dacec492} \\
G2242 & 6632 & The Code That Writes Itself Into a Breach: Vibe Coding's Structural Security Failure | Lyrie Research | Lyrie Research & \url{https://lyrie.ai/research/research/vibe-coding-structural-security-failure-ai-generated-code-lovable-owasp} \\
G2243 & 6634 & We Spent Two Years Afraid AI Would Disobey Us. The Real Disaster Is That It Obeys Everyone. | by Macplanet | May, 2026 | Medium & \url{https://medium.com/@macplanet2012/we-spent-two-years-afraid-ai-would-disobey-us-the-real-disaster-is-that-it-obeys-everyone-018bc2bb4133} \\
G2244 & 6639 & sonarsource.com & \url{https://www.sonarsource.com/the-coding-personalities-of-leading-llms.pdf} \\
G2245 & 6641 & Grok Build's 8-Agent Conflict Issue: Hidden Risks Explained & \url{https://www.linkedin.com/pulse/grok-builds-8-agent-parallelism-has-silent-conflict-problem-kumar-l-hromc} \\
G2246 & 6645 & 4,600 Junk PRs: The Real Truth About AI Slop Hitting Godot | Stack Junkie & \url{https://www.stack-junkie.com/blog/ai-code-slop-is-drowning-open-source-the-godot-crisis-explained} \\
G2247 & 6646 & CTO owns AI-generated code accountability - LinkedIn & \url{https://www.linkedin.com/posts/venu-rao-strobes_vibecoding-ctem-cybersecurity-activity-7450247755970641920-HC53} \\
G2248 & 6647 & LGTM 2.0: Architecting Zero-Noise AI Code Review Agents | by Aner Cohen | Riskified Tech | Medium & \url{https://medium.com/riskified-technology/lgtm-2-0-zero-noise-ai-code-review-agents-857441ec4f1a} \\
G2249 & 6654 & AI Code Review for Security - Finding Vulnerabilities With AI in 2026 - DEV Community & \url{https://dev.to/rahulxsingh/ai-code-review-for-security-finding-vulnerabilities-with-ai-in-2026-4o0e} \\
G2250 & 6662 & The Trust Crisis: Why 97\% of Developers Don't Trust AI Code They Use Daily - Sanjeewa Rupasinghe & \url{https://sanjewa.com/blogs/ai-trust-crisis-developers-2027/} \\
G2251 & 6664 & Vibe Coding Just Failed Its First Real Audit - DEV Community & \url{https://dev.to/gabrielanhaia/vibe-coding-just-failed-its-first-real-audit-5bbc} \\
G2252 & 6666 & Learnings from a No-Code Library: Keeping the Spec Driven Development Triangle in Sync | Juan Carlos Orrego & \url{https://www.linkedin.com/posts/jcorrego_learnings-from-a-no-code-library-keeping-activity-7439980770737147904-4tdA} \\
G2253 & 6670 & Ask @copilot to resolve merge conflicts on pull requests - GitHub Changelog & \url{https://github.blog/changelog/2026-03-26-ask-copilot-to-resolve-merge-conflicts-on-pull-requests/} \\
G2254 & 6671 & It's Not Claude Code, Silly -- Your SDLC is Not AI-native | by Venkat Peri | Apr, 2026 | Medium & \url{https://medium.com/@venkatperi/why-your-team-uses-ai-every-day-and-ships-at-the-same-speed-0fa31e2d6293} \\
G2255 & 6677 & Godot maintainers struggle with 'demoralizing' AI slop PRs & \url{https://www.theregister.com/software/2026/02/18/godot-maintainers-struggle-with-demoralizing-ai-slop-prs/4206219} \\
G2256 & 6678 & GitHub Copilot Token Billing 2026: Full Cost Guide and Alternatives - DEV Community & \url{https://dev.to/akaranjkar08/github-copilot-token-billing-2026-full-cost-guide-and-alternatives-3bcf} \\
G2257 & 6682 & Top Tools for Highlighting Engineering Bottlenecks | Faros AI & \url{https://www.faros.ai/blog/highlighting-engineering-bottlenecks-efficiently-faros-ai} \\
G2258 & 6689 & A Real-World Coding Story: Devin AI as My Co-Pilot - DEV Community & \url{https://dev.to/prpatel05/a-real-world-coding-story-devin-ai-as-my-co-pilot-548} \\
G2259 & 6695 & GitHub - yungookim/oh-my-pr at cfab7d37b10d87f078ec9ef866a46719494703ee GitHub & \url{https://github.com/yungookim/oh-my-pr/tree/cfab7d37b10d87f078ec9ef866a46719494703ee} \\
G2260 & 6696 & AI is destroying Open Source, and it's not even good yet - Jeff Geerling & \url{https://www.jeffgeerling.com/blog/2026/ai-is-destroying-open-source/} \\
G2261 & 6697 & Why Agentic Code Review Changed How I Ship | FieldTested Blog & \url{https://fieldtested.dev/2026/03/why-agentic-code-review-changed-how-i-ship/} \\
G2262 & 6698 & AI as a Thin Client and the Crisis of Knowledge Succession: An Academic Analysis - DEV Community & \url{https://dev.to/oleg_kholin_551a551b/ai-as-a-thin-client-and-the-crisis-of-knowledge-succession-an-academic-analysis-20me} \\
G2263 & 6704 & Why AI-Generated Code Is Always Good Enough -- And Never Great - DEV Community & \url{https://dev.to/harsh2644/why-ai-generated-code-is-always-good-enough-and-never-great-4lhn} \\
G2264 & 6706 & CodeGuardian: AI-Driven Security Code Review With Natural Language Understanding | by Sage Khan | Mar, 2026 | Medium & \url{https://thesagekhan.medium.com/codeguardian-ai-driven-security-code-review-with-natural-language-understanding-08fc1954d93f} \\
G2265 & 6707 & When Your AI Stops Pretending to Know the Answer - Coder Legion & \url{https://coderlegion.com/19419/when-your-ai-stops-pretending-to-know-the-answer} \\
G2266 & 6708 & How to maximize GitHub Copilot's agentic capabilities - The GitHub Blog & \url{https://github.blog/ai-and-ml/github-copilot/how-to-maximize-github-copilots-agentic-capabilities/} \\
G2267 & 6711 & GPT-5.5 vs Claude Code: Which AI Should You Use? - DEV Community & \url{https://dev.to/max_quimby/gpt-55-vs-claude-code-which-ai-should-you-use-58fe} \\
G2268 & 6714 & 38 Issues: Code Review Agent Showdown between BugBot, Copilot and Claude - DEV Community & \url{https://dev.to/terence/38-issues-showdown-between-bugbot-copilot-and-claude-2o7e} \\
G2269 & 6721 & New public preview features in Copilot code review: AI reviews that see the full picture - GitHub Changelog & \url{https://github.blog/changelog/2025-10-28-new-public-preview-features-in-copilot-code-review-ai-reviews-that-see-the-full-picture/} \\
G2270 & 6724 & I don't think AI coding agents will efficiently replace developers via dev automation; I don't buy into this future, let me explain why: - Current AI developmen & \url{https://www.linkedin.com/posts/dshiryaev_i-dont-think-ai-coding-agents-will-efficiently-activity-7409248633465802752-2CFA} \\
G2271 & 6725 & Cursor AI Features 2026: The Advanced Prompting Guide Developers Are H & \url{https://www.hustletoai.com/blog/ai-tools-1/cursor-ai-features-2026-the-advanced-prompting-guide-developers-are-hiding-from-you-45} \\
G2272 & 6726 & Are We Really Faster After AI Started Writing Our Code? - DEV Community & \url{https://dev.to/rain9/are-we-really-faster-after-ai-started-writing-our-code-59f4} \\
G2273 & 6729 & Issue \#2: GPT-5.3 Codex and Opus 4.6 Highlight the Speed and Accuracy Trade-offs & \url{https://www.linkedin.com/pulse/issue-2-gpt-53-codex-opus-46-highlight-speed-accuracy-william-palmer-g1rne} \\
G2274 & 6730 & Assigning and completing issues with coding agent in GitHub Copilot - The GitHub Blog & \url{https://github.blog/ai-and-ml/github-copilot/assigning-and-completing-issues-with-coding-agent-in-github-copilot/} \\
G2275 & 6731 & The Developer Job Description Quietly Rewrote Itself in 2026. Did You Notice? - DEV Community & \url{https://dev.to/zyvop/the-developer-job-description-quietly-rewrote-itself-in-2026-did-you-notice-o9} \\
G2276 & 6732 & Best Developer Productivity Tools 2026: 14 AI Developer Tools Compared | Greptile & \url{https://www.greptile.com/content-library/14-best-developer-productivity-tools} \\
G2277 & 6734 & How to Keep Human Experts Visible in Your AI-Assisted Codebase & \url{https://www.freecodecamp.org/news/how-to-keep-human-experts-visible-in-your-ai-assisted-codebase/} \\
G2278 & 6738 & I built a tool to catch AI coding agents misbehaving -- and put zero AI in it - DEV Community & \url{https://dev.to/conalh/i-built-a-tool-to-catch-ai-coding-agents-misbehaving-and-put-zero-ai-in-it-1lg3} \\
G2279 & 6744 & The AI Code Trap: Why Junior Developers Should Use AI to Read Code, Not Write It - DEV Community & \url{https://dev.to/arcsviel/the-ai-code-trap-why-junior-developers-should-use-ai-to-read-code-not-write-it-1bpo} \\
G2280 & 6745 & marcusshade/MSR2026 & \url{https://github.com/marcusshade/MSR2026} \\
G2281 & 6751 & The automation shift: Why 64\% of developers use AI agentic tools | Sonar & \url{https://www.sonarsource.com/blog/agentic-ai-automation-shift} \\
G2282 & 6753 & Verifying AI-assisted code with Spool | Jared Brown posted on the topic | LinkedIn & \url{https://www.linkedin.com/posts/jared-oscar-brown_spool-verify-ai-generated-code-before-merging-activity-7422808875520716800-yi-4} \\
G2283 & 6754 & LLM Code Review Service with RAG \& Observability & \url{https://medium.com/@random.droid/llm-code-review-service-with-rag-observability-93e6befa6330} \\
G2284 & 6755 & Claude Code Security Finds the Bugs That Static Analysis Can't -- and Wall Street Noticed - DevOps.com & \url{https://devops.com/claude-code-security-finds-the-bugs-that-static-analysis-cant-and-wall-street-noticed/} \\
G2285 & 6759 & Google, GitHub, and CISA Expose the AI Draft Trap | by James Kuhman | KAIRI | May, 2026 | Medium & \url{https://medium.com/kairi-ai/google-github-and-cisa-expose-the-ai-draft-trap-96d69497833b} \\
G2286 & 6762 & OpenAI's GPT-5 Mini leads in code security, outperforming competitors by 13\% | Veracode posted on the topic | LinkedIn & \url{https://www.linkedin.com/posts/veracode_appsec-securecoding-genai-activity-7396564375052001280-rtdb} \\
G2287 & 6763 & Why developer expertise matters more than ever in the age of AI - The GitHub Blog & \url{https://github.blog/developer-skills/career-growth/why-developer-expertise-matters-more-than-ever-in-the-age-of-ai/} \\
G2288 & 6765 & AI-Powered Code Reviewer Using React, Node.js, Prism.js \& Gemini API | by Dhruv Mishra | Medium & \url{https://medium.com/@dhruvmishra_95898/ai-powered-code-reviewer-using-react-node-js-prism-js-gemini-api-9f31cfe718fb} \\
G2289 & 6767 & From Token Hell to 90\% Savings: How BMAD v6 Revolutionized AI-Assisted Development | by Trung Hieu Tran | Medium & \url{https://medium.com/@hieutrantrung.it/from-token-hell-to-90-savings-how-bmad-v6-revolutionized-ai-assisted-development-09c175013085} \\
G2290 & 6768 & The Greenfield Privilege - by Chris Ellis & \url{https://engineeredintelligence.substack.com/p/the-greenfield-privilege} \\
G2291 & 6772 & A model for growing the next generation of developers - LeadDev & \url{https://leaddev.com/career-development/a-model-for-growing-the-next-generation-of-developers} \\
G2292 & 6773 & Kenmege/codex-plugin-cc & \url{https://github.com/Kenmege/codex-plugin-cc} \\
G2293 & 6775 & AI Didn't Reduce Engineering Complexity. It Moved. - DEV Community & \url{https://dev.to/ntctech/ai-didnt-reduce-engineering-complexity-it-moved-di6} \\
G2294 & 6778 & GitHub Copilot Review 2026: 93/100 | AI for Code & \url{https://aiforcode.io/tools/github-copilot} \\
G2295 & 6779 & Code Review and Testing - Lessons Learned (OSD600LAB\_1) - DEV Community & \url{https://dev.to/hsachdeva9/code-review-and-testing-lessons-learned-osd600lab1-11p8} \\
G2296 & 6780 & Developers vs AI: Can You Spot When AI Is Wrong? - DEV Community & \url{https://dev.to/marxon/developers-vs-ai-can-you-spot-when-ai-is-wrong-7g7} \\
G2297 & 6794 & Building a Code Review Bot with Claude API and GitHub Actions - DEV Community & \url{https://dev.to/chengyixu/building-a-code-review-bot-with-claude-api-and-github-actions-29c9} \\
G2298 & 6796 & Cursor v3: The AI Coding Agent Replacing Traditional IDEs & \url{https://www.analyticsvidhya.com/blog/2026/04/cursor-v3/} \\
G2299 & 6798 & Developers still don't trust AI-generated code | CIO & \url{https://www.cio.com/article/4117049/developers-still-dont-trust-ai-generated-code.html} \\
G2300 & 6803 & AI Code Review Before You Deploy: Our Experience with CodeRabbit - DEV Community & \url{https://dev.to/deployhq/ai-code-review-before-you-deploy-our-experience-with-coderabbit-343e} \\
G2301 & 6806 & AI accelerates flaw discovery for critical infrastructure | TMC Insight & \url{https://insight.tmcnet.com/insight/ai-accelerates-flaw-discovery-for-critical-infrastructure-1780418490340} \\
G2302 & 6807 & I Built an Impact Analysis Guardrail for Cursor. The Best Version Was 20 Lines. | by AiJW | May, 2026 | Medium & \url{https://medium.com/@2026jwutubechnl/i-built-an-impact-analysis-guardrail-for-cursor-the-best-version-was-20-lines-7e9675ba5077} \\
G2303 & 6808 & davidfstr/revise-skill & \url{https://github.com/davidfstr/revise-skill} \\
G2304 & 6810 & uReview: Scalable, Trustworthy GenAI for Code Review at Uber & \url{https://www.uber.com/jm/en/blog/ureview/} \\
G2305 & 6811 & The AI-Native Engineer : Part 3 -- When to Trust the Model and When to Fight It | by Paul Brady | Apr, 2026 | Medium & \url{https://medium.com/@bradypt/the-ai-native-engineer-part-3-when-to-trust-the-model-and-when-to-fight-it-29156704386c} \\
G2306 & 6813 & Vibe Code vs Trad Code | DoltHub Blog & \url{https://www.dolthub.com/blog/2026-03-26-vibe-code-vs-trad-code/} \\
G2307 & 6814 & Bain Technology Report 2025: Why AI Gains Are Stalling & \url{https://www.faros.ai/blog/bain-technology-report-2025-why-ai-gains-are-stalling} \\
G2308 & 6815 & Why Multi-Agent Code Review Is Outperforming Manual Workflows in 2026 | by 8080.ai | Towards Data Engineering | May, 2026 | Medium & \url{https://medium.com/towards-data-engineering/why-multi-agent-code-review-is-outperforming-manual-workflows-in-2026-d538b086c0e2} \\
G2309 & 6816 & AI in Code Review: Assistant or Reviewer? | Aizhan Moinokova posted on the topic | LinkedIn & \url{https://www.linkedin.com/posts/aizhan-moinokova-b7955121a_react-typescript-frontenddevelopment-activity-7465364021744091136-IBxV} \\
G2310 & 6817 & AI-Generated Code: Is Prompt Review Replacing PR? - LinkedIn & \url{https://www.linkedin.com/posts/blakeyoder_i-was-reading-about-a-team-experimenting-activity-7416661859606441984-fult} \\
G2311 & 6818 & Claude Code Dynamic Workflows : Agent | AI & \url{https://www.boluoblog.com/coding/claude-code-dynamic-workflows-parallel-subagents-migration-2026/} \\
G2312 & 6819 & Devs doubt AI-written code, but don't always check it & \url{https://www.theregister.com/software/2026/01/09/devs-doubt-ai-written-code-but-dont-always-check-it/4932910} \\
G2313 & 6820 & Why AI Self-Review Fails: The Technical Case for Independent AI Systems - Qodo & \url{https://www.qodo.ai/blog/why-ai-self-review-fails-the-technical-case-for-independent-ai-systems/} \\
G2314 & 6821 & AI-Powered Code Review and Bug Detection - Inspire & \url{https://blogs.scu.edu/inspire/2025/08/02/ai-powered-code-review-and-bug-detection/} \\
G2315 & 6823 & Code Ships in Minutes, Everything Else Takes Weeks | Towards AI & \url{https://pub.towardsai.net/code-ships-in-minutes-everything-else-takes-weeks-b776a674cb4e} \\
G2316 & 6824 & AI Code Generators Are Creating a Generation of "Copy-Paste Coders" -- Here's How We Fix It | by Terrance Craddock | Mr. Plan Publication | Medium & \url{https://medium.com/mr-plan-publication/ai-code-generators-are-creating-a-generation-of-copy-paste-coders-heres-how-we-fix-it-d49a3aef8dc2} \\
G2317 & 6825 & The Reflection Pattern: Teaching AI to Review Its Own Work | by sriram c | Technology at Nineleaps | May, 2026 | Medium & \url{https://medium.com/technology-nineleaps/the-reflection-pattern-teaching-ai-to-review-its-own-work-dd97e312cc7e} \\
G2318 & 6828 & Operational Bottlenecks: Real Examples and How to Fix Them & \url{https://everhour.com/blog/operational-bottlenecks/} \\
G2319 & 6832 & Best AI Code Review Tools 2026: Graphite Alternatives & \url{https://cms.gitar.ai/graphite-ai-code-review-alternatives/} \\
G2320 & 6835 & LiorCohen/sdd & \url{https://github.com/LiorCohen/sdd} \\
G2321 & 6836 & Why Your AI Coding Assistant Keeps Breaking Your Architecture | by DGNV | Medium & \url{https://medium.com/@milankmitra/why-your-ai-coding-assistant-keeps-breaking-your-architecture-4a2223b7d9d5} \\
G2322 & 6837 & AI code review checklist. How to Build a Better AI Code Review... | by Krun.Pro: Pure code experts. Craft, not paycheck. | Apr, 2026 | Medium & \url{https://medium.com/@krun_dev/ai-code-review-checklist-af930a117373} \\
G2323 & 6840 & A statistical study of PRs opened on openclaw/openclaw | Greptile Blog & \url{https://www.greptile.com/blog/prs-on-openclaw} \\
G2324 & 6841 & The first thing vibe coding builds is confidence & \url{https://www.theregister.com/software/2026/03/29/the-first-thing-vibe-coding-builds-is-confidence/5223428} \\
G2325 & 6842 & mkrtchian/spec-driven-dev & \url{https://github.com/mkrtchian/spec-driven-dev} \\
G2326 & 6844 & How AI Coding Tools Can 10x Developer Productivity -- Without Losing Engineering Judgment & \url{https://hackernoon.com/how-ai-coding-tools-can-10x-developer-productivity-without-losing-engineering-judgment} \\
G2327 & 6846 & Cursor AI Review: Real-World Development Speedup | Ashish Yadav posted on the topic | LinkedIn & \url{https://www.linkedin.com/posts/ashish-yadav-764836167_i-tried-cursor-ai-for-real-projects-in-2026-activity-7415852448554143745-ZNdJ} \\
G2328 & 6850 & cubic Review 2026: What You Should Know Before Trusting AI With Your Code Reviews | by Elena Voss | May, 2026 | Medium & \url{https://medium.com/@elena.voss/cubic-review-2026-what-you-should-know-before-trusting-ai-with-your-code-reviews-02d0f1b9c309} \\
G2329 & 6854 & When Code Becomes a Byproduct, Where Does Beauty Live? | by Marcin Zietek | Apr, 2026 | Medium & \url{https://medium.com/@zietek1/when-code-becomes-a-byproduct-where-does-beauty-live-b693eadb8237} \\
G2330 & 6856 & Qodo: Revolutionizing Code Development with AI & \url{https://www.linkedin.com/pulse/qodo-revolutionizing-code-development-ai-dnyaneshwar-patil-r336e} \\
G2331 & 6859 & AI Has Taken Over Open Source - Socket & \url{https://socket.dev/blog/ai-has-taken-over-open-source} \\
G2332 & 6867 & The Vibe Coding Paradox: 92\% of Developers Use It, But Trust Has Collapsed -- Here's Why That Matters | by Arshia Heravi | Mar, 2026 | Medium & \url{https://medium.com/@arshia.heravi05/the-vibe-coding-paradox-92-of-developers-use-it-but-trust-has-collapsed-heres-why-that-ab241b95ca82} \\
G2333 & 6869 & AI-Generated Code Anti-Patterns: 9 Production Bugs Hiding in Vibe-Coded Apps (2026) | Rohit Raj & \url{https://rohitraj.tech/en/notes/ai-generated-code-anti-patterns-fixes-2026} \\
G2334 & 6874 & Your Cursor Rules Won't Scale: AI Code Needs an Adaptive Rules System - Qodo & \url{https://www.qodo.ai/blog/your-cursor-rules-wont-scale-ai-code-needs-an-adaptive-rules-system/} \\
G2335 & 6875 & AI-Driven Development Lifecycle for Financial Services | AWS for Industries & \url{https://aws.amazon.com/blogs/industries/ai-driven-development-lifecycle-for-financial-services/} \\
G2336 & 6876 & DX Report: AI Productivity Gains in Software Delivery | Sam Newman posted on the topic | LinkedIn & \url{https://www.linkedin.com/posts/samnewman_ai-productivity-gains-are-10-not-10x-activity-7437444158824497152-YLey} \\
G2337 & 6878 & AI Footprint: Making AI-Generated Code Traceable in a Git-First World | by Selvakumar Subramanian | Apr, 2026 | Medium & \url{https://medium.com/@selvakumarsubramanian/ai-footprint-making-ai-generated-code-traceable-in-a-git-first-world-2f450ea315a3} \\
G2338 & 6883 & GitHub - BryceWDesign/IX-BlackFox: AI proposes. Humans decide. Source-available AI engineering control plane with policy gates, model comparison, budget-aware r & \url{https://p.rst.im/q/github.com/BryceWDesign/IX-BlackFox} \\
G2339 & 6885 & AI code is more dangerous than bad code. A junior's mistakes are obvious. The naming is off, the style is messy, you see the shortcut from a mile away. You know & \url{https://www.linkedin.com/posts/mark-ajzenstadt_ai-code-is-more-dangerous-than-bad-code-activity-7465476949575028737-vsgf} \\
G2340 & 6886 & Agora Robotics \& CodeRabbit: AI-Powered Code Reviews Streamline Roboti & \url{https://www.coderabbit.ai/blog/agora-robotics-and-coderabbit-ai-powered-code-reviews-streamline-robotics-software?%3Futm_source=newsletter} \\
G2341 & 6887 & Bringing Code Review to Claude Code: A Practical Guide & \url{https://www.guvi.in/blog/bringing-code-review-to-claude-code/} \\
G2342 & 6894 & Windsurf vs Cursor vs GitHub Copilot: AI Coding Assistants Compared & \url{https://www.linkedin.com/pulse/windsurf-vs-cursor-github-copilot-ai-coding-compared-van-t-land-gotne} \\
G2343 & 6895 & How Should We Evaluate AI Coding Tools in Real Engineering Environments - DEV Community & \url{https://dev.to/hellofadude/how-should-we-evaluate-ai-coding-tools-in-real-engineering-environments-4ab0} \\
G2344 & 6901 & Faster GitLab Releases with AI Review | by API4AI | Medium & \url{https://medium.com/@API4AI/40-faster-releases-with-crken-automation-7ae7ce7802ef} \\
G2345 & 6904 & Agentic Coding: Who Will Review All That Code? | Stark Raving Finkle & \url{https://starkravingfinkle.org/posts/2026/03/coding-agents-who-will-review-all-that-code/} \\
G2346 & 6909 & AI Agents Can Ship Code Faster Than You Can Review It. Here's What Stops Them. - DEV Community & \url{https://dev.to/nolrm/ai-agents-can-ship-code-faster-than-you-can-review-it-heres-what-stops-them-6k3} \\
G2347 & 6912 & Vibe Coding Governance: Why 89\% of CTO Pilots Fail (May 2026) & \url{https://agileleadershipdayindia.org/blogs/vibe-coding-governance-enterprise-risk-management/vibe-coding-governance-enterprise-risk-management.html} \\
G2348 & 6917 & Code Review vs AI Code Generation: Why They Shouldn't Mix & \url{https://www.linkedin.com/posts/itamarf_vibecoding-vibereviewing-activity-7419731710222721024-KNKZ} \\
G2349 & 6919 & AI Code Review Tools Compared: What Actually Works - RunAICode & \url{https://runaicode.ai/ai-code-review-tools-compared-what-actually-works/} \\
G2350 & 6920 & Is Software Engineering Cooked? Not Yet. But Maybe. - DEV Community & \url{https://dev.to/aws/is-software-engineering-cooked-not-yet-but-maybe-5e0j} \\
G2351 & 6925 & Why Should AI Review Your Code? - Blogs | Panto AI & \url{https://www.getpanto.ai/blog/why-should-ai-review-your-code} \\
G2352 & 6929 & NikolayS/samospec & \url{https://github.com/NikolayS/samospec} \\
G2353 & 6930 & AI tooling benchmarks: PR throughput and usage by tool (Q1 2026) | Justin Reock & \url{https://www.linkedin.com/posts/justinreock_ai-tooling-benchmarks-pr-throughput-and-activity-7420121679760371712-Ao55} \\
G2354 & 6933 & Streamlined CI/CD Pipelines Using Claude Code \& GitHub Actions | by Milan's Outlook | Medium & \url{https://medium.com/@itsmybestview/streamlined-ci-cd-pipelines-using-claude-code-github-actions-74be17e51499} \\
G2355 & 6934 & Agile Isn't Dead, But Scrum Is Evolving with AI | Youssef Solh posted on the topic | LinkedIn & \url{https://www.linkedin.com/posts/youssefsolh_agile-scrum-ai-activity-7447309312466767873-bjZQ} \\
G2356 & 6936 & AI Code Review Checklist for Juniors - Practical Guide - KruN & \url{https://krun.pro/ai-code-review-checklist/} \\
G2357 & 6938 & The Deskilling Feedback Loop: When Vibe Coders Never Become Senior Engineers | DoDataThings.dev & \url{https://dodatathings.dev/blog/ai-deskilling-feedback-loop} \\
G2358 & 6940 & Augmented Ownership: Keeping the Spirit of "You Build It, You Run It" Alive in the Age of AI | by Sebastian Hurtado | Medium & \url{https://medium.com/@seba_hurtado/augmented-ownership-keeping-the-spirit-of-you-build-it-you-run-it-alive-in-the-age-of-ai-c44190a2a120} \\
G2359 & 6949 & Meet Git-Kepo: The Kaypoh Git Detective Who Joined My AI Family | by Levente Csikor | CodeX | Apr, 2026 | Medium & \url{https://medium.com/codex/meet-git-kepo-the-kaypoh-git-detective-who-joined-my-ai-family-38fb0d43d787} \\
G2360 & 6953 & AI Code Generation Makes Code Reviews Impractical | Philip Su posted on the topic | LinkedIn & \url{https://www.linkedin.com/posts/suphilip_we-need-to-stop-doing-code-reviews-ai-code-activity-7435734706035261440--yb9} \\
G2361 & 6957 & AI-generated code review challenges traditional practices | Tim Krajcar posted on the topic | LinkedIn & \url{https://www.linkedin.com/posts/timkrajcar_ai-coding-agents-have-broken-a-fundamental-activity-7419567970919714816-SRcg} \\
G2362 & 6959 & GitHub Copilot app: The agent-native desktop experience - The GitHub Blog & \url{https://github.blog/news-insights/product-news/github-copilot-app-the-agent-native-desktop-experience/} \\
G2363 & 6967 & I Automated My PR Reviews With AI -- Saved 6 Hours/Week (Full Setup) - DEV Community & \url{https://dev.to/hopkins_jesse_cdb68cfa22c/i-automated-my-pr-reviews-with-ai-saved-6-hoursweek-full-setup-48c6} \\
G2364 & 6968 & How to Make the Best of AI Programming Assistants & \url{https://www.youtube.com/watch?v=XavrebMKH2A} \\
G2365 & 6969 & Shape Copilot code review around your team - GitHub Changelog & \url{https://github.blog/changelog/2026-06-02-shape-copilot-code-review-around-your-team/} \\
G2366 & 6976 & AI Made Your PRs 33\% Bigger. Reviews Got Worse. -- Octopus Blog | Octopus & \url{https://octopus-review.ai/blog/ai-made-your-prs-bigger-reviews-got-worse} \\
G2367 & 6980 & The Mixed PR Queue: Reviewer Throughput Is Now the Binding Constraint & \url{https://tianpan.co/blog/2026-05-10-mixed-pr-queue-reviewer-throughput-binding-constraint} \\
G2368 & 6981 & Enterprise AI Coding Security Risks 2025: Complete Guide & \url{https://blog.exceeds.ai/ai-coding-assistants-risks-2025/} \\
G2369 & 6983 & AI-Enhanced Code Reviews in Cursor | Developer Toolkit & \url{https://developertoolkit.ai/en/cursor-ide/lessons/code-review/} \\
G2370 & 6987 & Linux kernel czar says AI bug reports aren't slop anymore & \url{https://www.theregister.com/software/2026/03/26/linux-kernel-czar-says-ai-bug-reports-arent-slop-anymore/5226256} \\
G2371 & 6993 & The Next Era of AppSec: Why AI-Generated Code Needs Offensive Dynamic Testing | Snyk & \url{https://snyk.io/blog/the-next-era-of-appsec-why-ai-generated-code-needs-offensive-dynamic-testing/} \\
G2372 & 6995 & Rethinking open source mentorship in the AI era - The GitHub Blog & \url{https://github.blog/open-source/maintainers/rethinking-open-source-mentorship-in-the-ai-era/} \\
G2373 & 6997 & web-design-guidelines skill by oldwinter/skills & \url{https://playbooks.com/skills/oldwinter/skills/web-design-guidelines} \\
G2374 & 7005 & What's new with GitHub Copilot coding agent - The GitHub Blog & \url{https://github.blog/ai-and-ml/github-copilot/whats-new-with-github-copilot-coding-agent/} \\
G2375 & 7008 & The hidden security cost of AI-generated code (and what to do about it) - DEV Community & \url{https://dev.to/busyagents/the-hidden-security-cost-of-ai-generated-code-and-what-to-do-about-it-36po} \\
G2376 & 7009 & EU AI Act for AI Code Assistants: Copilot Compliance Guide - DEV Community & \url{https://dev.to/gregorio_vonhildebrand_a/eu-ai-act-for-ai-code-assistants-copilot-compliance-guide-3g9j} \\
G2377 & 7010 & Stray AI Reviewers Write Correct Code--and Break Your System & \url{https://zenn.dev/commander/articles/2c0f09d7cd0312?locale=en} \\
G2378 & 7011 & Context as Code - O'Reilly & \url{https://www.oreilly.com/radar/context-as-code/} \\
G2379 & 7022 & Vibe Coding Security: What Happens When Developers Trust AI Too Much - DEV Community & \url{https://dev.to/danieljglover/vibe-coding-security-what-happens-when-developers-trust-ai-too-much-1fji} \\
G2380 & 7023 & How to use Claude Code and Github Co-Pilot for PR reviews & \url{https://www.linkedin.com/posts/stevekinney_claude-code-hot-tip-of-the-day-claude-code-activity-7364360344317648900-jAT3} \\
G2381 & 7024 & AI coding requires developers to become better managers | InfoWorld & \url{https://www.infoworld.com/article/4118288/ai-coding-requires-developers-to-become-better-managers.html} \\
G2382 & 7026 & Code Review System | coderabbitai/coderabbit-docs | DeepWiki & \url{https://deepwiki.com/coderabbitai/coderabbit-docs/4-code-review-system} \\
G2383 & 7040 & douglasrw/governance-scan & \url{https://github.com/douglasrw/governance-scan} \\
G2384 & 7044 & Vibe Coding is Fun; Vibe Shipping is Dangerous & \url{https://komang.id/vibe-coding-is-fun-vibe-shipping-is-dangerous.html} \\
G2385 & 7045 & The "Uncanny Valley" of AI Code: Why Hybrid PRs Are Harder to Review Than Pure AI or Pure Human Code - DEV Community & \url{https://dev.to/jakkie_koekemoer/the-uncanny-valley-of-ai-code-why-hybrid-prs-are-harder-to-review-than-pure-ai-or-pure-human-code-1poh} \\
G2386 & 7046 & Claude Opus 4.7 Replaced My Code Review Process. Here Is What Changed. | Artificial Intelligence in Plain English & \url{https://ai.plainenglish.io/claude-opus-4-7-replaced-my-code-review-process-here-is-what-changed-28964c796691} \\
G2387 & 7050 & The Shift: Software Engineering is No Longer About Writing Code & \url{https://www.linkedin.com/pulse/shift-software-engineering-longer-writing-code-ahmed-hisham-uuwrf} \\
G2388 & 7051 & Open Code Review: The Missing Quality Gate for AI-Generated Code - DEV Community & \url{https://dev.to/raye_deng_622ab98e19a2147/open-code-review-the-missing-quality-gate-for-ai-generated-code-1p9j} \\
G2389 & 7055 & Famed iPhone, Sony Hacker Says AI Coding Agents Are a Disaster Waiting to Happen - Decrypt & \url{https://decrypt.co/368964/george-hotz-vibe-coding-ai-slop-warning?amp=1} \\
G2390 & 7057 & Anthropic launches code review tool to check flood of AI-generated code | TechCrunch & \url{https://techcrunch.com/2026/03/09/anthropic-launches-code-review-tool-to-check-flood-of-ai-generated-code/} \\
G2391 & 7063 & 84\% Of Devs Use AI. 29\% Trust It. The Missing Skill Isn't Review. It's Refusal. - DEV Community & \url{https://dev.to/rentierdigital/84-of-devs-use-ai-29-trust-it-the-missing-skill-isnt-review-its-refusal-3gnm} \\
G2392 & 7066 & Traditional Code Review Is Dead. What Comes Next? - DEV Community & \url{https://dev.to/signadot/traditional-code-review-is-dead-what-comes-next-41oi} \\
G2393 & 7067 & Unlearning, experimentation and engineering rigor in an agentic world | Thoughtworks & \url{https://www.thoughtworks.com/insights/podcasts/technology-podcasts/unlearning-experimentation-engineering-rigor-agentic-world} \\
G2394 & 7069 & AI Technical Debt: The Hidden Cost of AI Coding Tools - DEV Community & \url{https://dev.to/kunwar-jhamat/ai-technical-debt-the-hidden-cost-of-ai-coding-tools-nnp} \\
G2395 & 7071 & Cursor AI Agents Work Like 10 Developers (Cursor VP Live Demo) & \url{https://www.youtube.com/watch?v=8QN23ZThdRY} \\
G2396 & 7073 & AI Didn't Break Accountability. It Exposed the Gap. | Decision Models & \url{https://tsotsos.tech/essays/ai-didnt-break-accountability/} \\
G2397 & 7076 & Go lib maintainer: GitHub's Dependabot is a 'noise machine' & \url{https://www.theregister.com/software/2026/02/24/go-lib-maintainer-githubs-dependabot-is-a-noise-machine/4386802} \\
G2398 & 7077 & Beyond Snippet Assistance: A Workflow-Centric Framework for End-to-End AI-Driven Code Generation & \url{https://www.mdpi.com/2073-431X/14/3/94} \\
G2399 & 7080 & Spec Driven Development (SDD) - A initial review - DEV Community & \url{https://dev.to/danielsogl/spec-driven-development-sdd-a-initial-review-2llp} \\
G2400 & 7089 & As ai helps us write more code whos catching the bugs & \url{https://leaddev.com/ai/as-ai-helps-us-write-more-code-whos-catching-the-bugs} \\
G2401 & 7091 & The best Claude Code agents are defined by what they refuse to do - DEV Community & \url{https://dev.to/peterverse180/the-best-claude-code-agents-are-defined-by-what-they-refuse-to-do-13p2} \\
G2402 & 7092 & How to Build an AI-Powered Pull Request Review [Feb 2026] & \url{https://www.qodo.ai/blog/ai-pull-request-review/} \\
G2403 & 7094 & AI-Driven Code Analysis Boosts Software Engineering | Pramod Kalluri posted on the topic | LinkedIn & \url{https://www.linkedin.com/posts/pramodkalluri_ai-assisted-code-analysis-advancing-modern-activity-7401558169266819072-PliV} \\
G2404 & 7095 & Developer Survey Results: How You're Using AI Coding Tools in 2026 | Tactic Remote & \url{https://clauderc.com/blog/2026-02-27-developer-survey-results-2026/} \\
G2405 & 7101 & The hype around "vibe coding" and the future of AI in software engineering. & \url{https://www.linkedin.com/posts/ahmed-e-hassan_%F0%9D%97%A9%F0%9D%97%B6%F0%9D%97%AF%F0%9D%97%B2-%F0%9D%97%96%F0%9D%97%BC%F0%9D%97%B1%F0%9D%97%B6%F0%9D%97%BB%F0%9D%97%B4-%F0%9D%98%83%F0%9D%98%80-%F0%9D%97%94%F0%9D%97%B4%F0%9D%97%B2%F0%9D%97%BB%F0%9D%98%81%F0%9D%97%B6%F0%9D%97%B0-activity-7381743946252652544-nZyD} \\
G2406 & 7106 & We're Shipping Faster Than Ever. Our Systems Have Never Been More Fragile. | by Amit Kumar | Apr, 2026 | Medium & \url{https://humptytech.medium.com/were-shipping-faster-than-ever-our-systems-have-never-been-more-fragile-27be281bd01d} \\
G2407 & 7110 & Stop checking AI-generated code. Start generating less of it | InfoWorld & \url{https://www.infoworld.com/article/4177717/stop-checking-ai-generated-code-start-generating-less-of-it.html} \\
G2408 & 7111 & If you are good at code review, you will be good at using AI agents & \url{https://www.seangoedecke.com/ai-agents-and-code-review/} \\
G2409 & 7113 & AI Code Reviews: The Ultimate Guide | Greptile & \url{https://www.greptile.com/what-is-ai-code-review} \\
G2410 & 7114 & The Vibe Check - DEV Community & \url{https://dev.to/thesythesis/the-vibe-check-5b45} \\
G2411 & 7117 & Navigating the AI Coding Boom: Why Zero Trust Still Rules & \url{https://www.linkedin.com/pulse/navigating-ai-coding-boom-why-zero-trust-still-rules-nicholas-cravino-m0vme} \\
G2412 & 7119 & The reality of AI-Assisted software engineering productivity & \url{https://addyo.substack.com/p/the-reality-of-ai-assisted-software} \\
G2413 & 7127 & We dropped code reviews entirely. Here's what happened. 3 months ago, Rowads Studio adopted extreme Claude Coding across our 5-person engineering team. We didn' & \url{https://www.linkedin.com/posts/marco-fucci-di-napoli_we-dropped-code-reviews-entirely-heres-activity-7427068635900231680-rBFM} \\
G2414 & 7129 & How do you create an AI coding workflow that actually works? In this week's podcast episode I had the pleasure of speaking with Birgitta Boeckeler, distinguishe & \url{https://www.linkedin.com/posts/lucaronin_how-do-you-create-an-ai-coding-workflow-that-activity-7377260345335115777-kISq} \\
G2415 & 7132 & The problem with security scanners isn't the scanning - DEV Community & \url{https://dev.to/wael_matoussi/the-problem-with-security-scanners-isnt-the-scanning-5cgl} \\
G2416 & 7141 & Stop Letting Your AI Tool Make Commits You Can't Explain | by Hafiq Iqmal | CodeToDeploy | Apr, 2026 | Medium & \url{https://medium.com/codetodeploy/stop-letting-your-ai-tool-make-commits-you-cant-explain-1212f720dffd} \\
G2417 & 7144 & I Built a Skill Reviewer. Then I Ran It on Itself. - DEV Community & \url{https://dev.to/shinpr/i-built-a-skill-reviewer-then-i-ran-it-on-itself-4m4j} \\
G2418 & 7146 & The Math Doesn't Work - DEV Community & \url{https://dev.to/jonoherrington/the-math-doesnt-work-4f13} \\
G2419 & 7148 & Vibe Coding vs. Reality: Why Your AI-Generated Code Needs DevSecOps - DEV Community & \url{https://dev.to/dimitrisk_cyclopt/vibe-coding-vs-reality-why-your-ai-generated-code-needs-devsecops-2l0h} \\
G2420 & 7150 & Codegen and Sweep AI Review: Autonomous Code Review Agents Put to the Test - DEV Community & \url{https://dev.to/pickuma/codegen-and-sweep-ai-review-autonomous-code-review-agents-put-to-the-test-28lp} \\
G2421 & 7151 & AI Is Reviewing Your Code for Security Now -- And That Changes Everything | by Waeel Kheshfeh | Feb, 2026 | Medium & \url{https://medium.com/@waeel.nono3719876/ai-is-reviewing-your-code-for-security-now-and-that-changes-everything-8476e1eb985d} \\
G2422 & 7155 & Claude AI Code Review 2026: Automation, Security, and Workflow & \url{https://primotly.com/article/claude-code-review-kompletny-poradnik-2026} \\
G2423 & 7158 & 9wick/codepolicy & \url{https://github.com/9wick/codepolicy} \\
G2424 & 7159 & The AI Paradox: Velocity Without Verification is Just Technical Debt & \url{https://www.linkedin.com/pulse/ai-paradox-velocity-without-verification-just-debt-tim-escolopio-hajpc} \\
G2425 & 7160 & For two years, Rootly enforced a strict small-PR rule. Stacked PRs, atomic changes, never more than a couple hundred lines. Then AI started writing 80\%+ of our  & \url{https://www.linkedin.com/posts/quentinrousseau_for-two-years-rootly-enforced-a-strict-small-pr-activity-7462895026172399618-gRrF} \\
G2426 & 7161 & AI Code Review Checklist: Correctness, Security, Performance, Readability - DEV Community & \url{https://dev.to/spyrae/ai-code-review-checklist-correctness-security-performance-readability-32pg} \\
G2427 & 7164 & Get started with Sonar Context Augmentation and Codex CLI | Sonar & \url{https://www.sonarsource.com/resources/library/get-started-with-sonar-context-augmentation-and-codex-cli/} \\
G2428 & 7166 & The Pull Request Isn't the Contribution Anymore | by Bernhard Straub | Apr, 2026 | Medium & \url{https://medium.com/@bgsyyxxqs/the-pull-request-isnt-the-contribution-anymore-76539c111aca} \\
G2429 & 7168 & Improving Code Quality with AI Agent Loops | Kyle Simpson posted on the topic | LinkedIn & \url{https://www.linkedin.com/posts/getify_for-a-change-of-pace-here-i-want-to-address-activity-7433517134825127936-ZEoD} \\
G2430 & 7170 & The "Slop" Tsunami: GitHub Eyes "Kill Switch" for Pull Requests to Save Buried Maintainers & \url{https://meterpreter.org/the-slop-tsunami-github-eyes-kill-switch-for-pull-requests-to-save-buried-maintainers/} \\
G2431 & 7173 & I Tested Three Spec-Driven AI Tools. Here's My Honest Take. & \url{https://ranthebuilder.cloud/blog/i-tested-three-spec-driven-ai-tools-here-s-my-honest-take/} \\
G2432 & 7174 & feat(eval): add eval:compare ledger comparison + EOS Phase 2 doc accuracy Pull Request \#788 s977043/river-reviewer & \url{https://github.com/s977043/river-reviewer/pull/788} \\
G2433 & 7175 & Cursor's Autonomous Agents Write 35\% of PRs, But Teams Struggle to Review | May Walter posted on the topic | LinkedIn & \url{https://www.linkedin.com/posts/may-walterr_agenticengineering-enterpriseai-cursor-activity-7433202388062105600-HFCt} \\
G2434 & 7176 & Agent Reviews: Automating Code Review for GitHub - LinkedIn & \url{https://www.linkedin.com/posts/paulbakaus_we-added-code-review-bots-to-our-prs-they-activity-7427449664792207360-lb1n} \\
G2435 & 7177 & How Claude Code's New Agent Team Reviews Every PR -- RapidClaw Blog & \url{https://rapidclaw.app/blog/claude-code-agent-team-pr-review} \\
G2436 & 7178 & AI Code review is broken. Here's what nobody wants to admit - DEV Community & \url{https://dev.to/vera_dev/ai-code-review-is-broken-heres-what-nobody-wants-to-admit-17e9} \\
G2437 & 7182 & I Let Claude Write 80 Functions. 65-75\% Had Security Vulnerabilities. - DEV Community & \url{https://dev.to/ofri-peretz/i-let-claude-write-60-functions-65-75-had-security-vulnerabilities-414o} \\
G2438 & 7183 & Open Source, SaaS, and the Silence After Unlimited Code Generation & \url{https://worksonmymachine.ai/p/open-source-saas-and-the-silence} \\
G2439 & 7186 & Vibe, then verify: How to navigate the risks of AI-generated code | Sonar & \url{https://www.sonarsource.com/blog/how-to-navigate-the-risks-of-ai-generated-code} \\
G2440 & 7189 & The Review Bottleneck: Why More AI Code Means Slower Teams in 2026 - DEV Community & \url{https://dev.to/code-board/the-review-bottleneck-why-more-ai-code-means-slower-teams-in-2026-1e5n} \\
G2441 & 7193 & Interviewing tactics for a post-LLM world - LeadDev & \url{https://leaddev.com/ai/interviewing-tactics-post-llm-world} \\
G2442 & 7195 & Stop reviewing AI code. Start deleting it. - DEV Community & \url{https://dev.to/krisnamic/stop-reviewing-ai-code-start-deleting-it-o40} \\
G2443 & 7203 & GitHub Copilot for Code Review: Quality \& Security | Pertama Partners & \url{https://www.pertamapartners.com/insights/copilot-code-review-quality-security} \\
G2444 & 7204 & AI Agents Are Writing Your Code Now. The Real Skill Is Reviewing It. | Mind of Korde & \url{https://blog.deveshk.dev/blog/ai-agents-writing-code-verification-skills} \\
G2445 & 7205 & I Stopped Reviewing Code: A Backend Dev's Experiment with ... & \url{https://dev.to/anchildress1/i-stopped-reviewing-code-a-backend-devs-experiment-with-google-gemini-5424} \\
G2446 & 7209 & The "All-in-One AI" Is a Myth for Enterprise Development - Qodo & \url{https://www.qodo.ai/blog/all-in-one-ai-myth-enterprise-development/} \\
G2447 & 7214 & Beyond Prompt Engineering: My AI-Native Spec-Driven Pipeline | by pushkar agnihotri | Apr, 2026 | Medium & \url{https://medium.com/@agnihotripush/beyond-prompt-engineering-my-ai-native-spec-driven-pipeline-8d920a1989d2} \\
G2448 & 7225 & Do Larger PRs Get Merged Faster? | Greptile Blog & \url{https://www.greptile.com/blog/do-larger-prs-get-merged-faster} \\
G2449 & 7228 & Why AI-Generated Code to Production is Bad Practices | by Engr. Md. Hasan Monsur | May, 2026 | Medium & \url{https://medium.com/@hasanmcse/why-ai-generated-code-to-production-is-bad-practices-d40164f4a5d3} \\
G2450 & 7229 & I Automated My PR Reviews With AI -- Saved 6 Hours/Week (Full Setup) - DEV Community & \url{https://dev.to/hopkins_jesse_cdb68cfa22c/i-automated-my-pr-reviews-with-ai-saved-6-hoursweek-full-setup-bbk} \\
G2451 & 7237 & The Assimilation Problem: Why AI-Generated Code Becomes Legacy Faster Than You Think - DEV Community & \url{https://dev.to/artem-koltunov/the-assimilation-problem-why-ai-generated-code-becomes-legacy-faster-than-you-think-4ebf} \\
G2452 & 7239 & Augment Code Review: The AI Pair Programmer That Indexes Your Entire Codebase - DEV Community & \url{https://dev.to/pickuma/augment-code-review-the-ai-pair-programmer-that-indexes-your-entire-codebase-3j3c} \\
G2453 & 7241 & LLM-Powered Code Review: Top Benefits \& Key Advantages & \url{https://www.linkedin.com/pulse/llm-powered-code-review-top-benefits-key-advantages-api4ai-yw90f} \\
G2454 & 7243 & Agentic coding isn't a trap. Lazy engineering is. - DEV Community & \url{https://dev.to/adioof/agentic-coding-isnt-a-trap-lazy-engineering-is-496h} \\
G2455 & 7244 & AI Code Review in 2026: Helpful vs Noisy & \url{https://futurion.blog/ai-code-review-in-2026-when-copilot-style-tools-help-and-when-they-add-noise/} \\
G2456 & 7245 & LLMs Generate Vulnerable C/C++ Code: Self-Review Fails to Mitigate Security Flaws - DEV Community & \url{https://dev.to/kserude/llms-generate-vulnerable-cc-code-self-review-fails-to-mitigate-security-flaws-4385} \\
G2457 & 7249 & maxbolgarin/codry & \url{https://github.com/maxbolgarin/codry} \\
G2458 & 7253 & 73\% of Devs Use AI Daily. Their Code Is Getting Worse. -- Derivinate News & \url{https://news.derivinate.com/73-of-devs-use-ai-daily-their-code-is-getting-worse} \\
G2459 & 7255 & Teaching AI Your Bad Habits. What AI learns when it reads your code... | by Naomi Pohl | Tabs Engineering | Medium & \url{https://medium.com/tabs-engineering/teaching-ai-your-bad-habits-5748d6085659} \\
G2460 & 7260 & Measuring and Improving Quality in AI-Powered Code Reviews | by Hadi Minooei | Google Cloud - Community | Medium & \url{https://medium.com/google-cloud/measuring-and-improving-quality-in-ai-powered-code-reviews-c1557ddc3181} \\
G2461 & 7264 & AI Found a 23-Year-Old Bug. Every Human Missed It. | by Ryann | Digital Systems Architect | Apr, 2026 | Medium & \url{https://medium.com/@ryannsj/ai-found-a-23-year-old-bug-every-human-missed-it-9109f2ce2612} \\
G2462 & 7267 & [AWS] Catching up on the basic functions of Kiro Autonomous Agent (Preview) [FrontierAgents] - DEV Community & \url{https://dev.to/aws-builders/aws-catching-up-on-the-basic-functions-of-kiro-autonomous-agent-preview-frontieragents-5e85} \\
G2463 & 7268 & Three Open-Source Governance Orientations for Managing AI-Generated Contribution Volume - SoftwareSeni & \url{https://www.softwareseni.com/three-open-source-governance-orientations-for-managing-ai-generated-contribution-volume/} \\
G2464 & 7270 & Use of AI has us creating more code than we can review - LeadDev & \url{https://zporigin.leaddev.com/ai/as-ai-helps-us-write-more-code-whos-catching-the-bugs} \\
G2465 & 7272 & How I Turned a Bespoke Code Reviewer Into a Skill Any Project Can Use - DEV Community & \url{https://dev.to/southwestmogrown/how-i-turned-a-bespoke-code-reviewer-into-a-skill-any-project-can-use-1npo} \\
G2466 & 7275 & Track Pull Request Performance and Code Quality with AI & \url{https://blog.exceeds.ai/track-pr-performance-quality-ai/} \\
G2467 & 7279 & AI-Native Engineering in 2026: Scrum, Harness Engineering, and the Future Side of LLMs | Medium & \url{https://medium.com/@takashi.mogami/beyond-the-one-way-time-machine-a-manifesto-on-engineers-and-organizations-in-the-ai-age-5192f7b503c4} \\
G2468 & 7282 & How to Use Code Review Analysis as Agent Skills Telemetry with Codex + Qodo - Qodo & \url{https://www.qodo.ai/blog/code-review-analysis-skills-telemetry/} \\
G2469 & 7283 & KonstZiv/ai-code-reviewer & \url{https://github.com/KonstZiv/ai-code-reviewer} \\
G2470 & 7287 & AI Writes 41\% of Code Now -- But Code Churn Is Doubling in 2026 - DEV Community & \url{https://dev.to/code-board/ai-writes-41-of-code-now-but-code-churn-is-doubling-in-2026-372f} \\
G2471 & 7290 & Introducing the Findings Page: A new way for engineering leaders to see risk across their codebase - Qodo & \url{https://www.qodo.ai/blog/introducing-findings-page/} \\
G2472 & 7292 & boyand/codex-review & \url{https://github.com/boyand/codex-review} \\
G2473 & 7294 & How to Use GitHub Copilot to Improve Code Quality - Or Not! & \url{https://www.linkedin.com/pulse/how-use-github-copilot-improve-code-quality-arash-arabi-ghref} \\
G2474 & 7296 & Stop "Vibe Merging" - DEV Community & \url{https://dev.to/shmulc/stop-vibe-merging-1jpo} \\
G2475 & 7299 & Stop Shipping Ungoverned AI Code: Your Quick-Start Checklist for Coding Agent Controls - DEV Community & \url{https://dev.to/rpsan/stop-shipping-ungoverned-ai-code-your-quick-start-checklist-for-coding-agent-controls-512l} \\
G2476 & 7302 & GitHub Copilot Code Reviews: Does It Actually Enhance Team Collaboration? | by Muhammad Nabil Mu'afa | Medium & \url{https://medium.com/@nmuafa/from-bottlenecks-to-breakthroughs-how-github-copilot-code-reviews-enhance-team-collaboration-23655c731e2f} \\
G2477 & 7305 & cwinvestments/memstack & \url{https://github.com/cwinvestments/memstack} \\
G2478 & 7307 & How to Automate Code Review With AI Tools: A Practical Guide - DEV Community & \url{https://dev.to/_d7eb1c1703182e3ce1782/how-to-automate-code-review-with-ai-tools-a-practical-guide-4k43} \\
G2479 & 7310 & The AI Told Me So Defense: When Code Review Quietly Stops Pushing Back & \url{https://tianpan.co/blog/2026-05-14-ai-told-me-so-defense-code-review} \\
G2480 & 7311 & We Stopped Reading Each Other's Code - DEV Community & \url{https://dev.to/arthurpro/we-stopped-reading-each-others-code-ihl} \\
G2481 & 7313 & The Real Impact of AI Code Review Agents: What We Learned from ... & \url{https://jellyfish.co/blog/impact-of-ai-code-review-agents/} \\
G2482 & 7323 & AI-assisted code review with Git and LLM | Rehan Shakir posted on the topic | LinkedIn & \url{https://www.linkedin.com/posts/rehanshakir_ai-code-review-activity-7451967514030264320-dqlc} \\
G2483 & 7324 & What Is Cursor AI? The Complete Guide to the AI-Native IDE (2026) - DEV Community & \url{https://dev.to/sahil_khurana_486f374ecf2/what-is-cursor-ai-the-complete-guide-to-the-ai-native-ide-2026-3748} \\
G2484 & 7325 & Cybernetics and the "human-on-the-loop" in agentic coding | Thoughtworks Australia & \url{https://www.thoughtworks.com/en-au/insights/blog/generative-ai/cybernetics-and-human-on-the-loop-in-agentic-coding} \\
G2485 & 7335 & web-design-guidelines skill by odjaramillo/custom-rules & \url{https://playbooks.com/skills/odjaramillo/custom-rules/web-design-guidelines} \\
G2486 & 7338 & Better Cursor Workflow: Collaborate, Don't Prompt | Medium & \url{https://medium.com/@rahulyadav574/stop-prompting-cursor-start-collaborating-with-it-639a5b20505d} \\
G2487 & 7341 & We Benchmarked Claude's Code Review Tool. Here's What the Data Shows. - DEV Community & \url{https://dev.to/nnennandukwe/we-benchmarked-claudes-code-review-tool-heres-what-the-data-shows-35b9} \\
G2488 & 7350 & Not Every Change Deserves the Same Review | by Kyle Pericak | Super.com | May, 2026 | Medium & \url{https://medium.com/super/not-every-change-deserves-the-same-review-7d528f35f058} \\
G2489 & 7354 & The Code Reviewer: An AI Assistant for Streamlining Pull Requests & \url{https://www.youtube.com/watch?v=O0oqOvsZZ8o} \\
G2490 & 7359 & The Anatomy of 50 Open Source PRs: What Gets Merged, What Gets Ignored, and Why (Real Data From an AI Agent) - DEV Community & \url{https://dev.to/zeroknowledge0x/the-anatomy-of-50-open-source-prs-what-gets-merged-what-gets-ignored-and-why-real-data-from-an-6ab} \\
G2491 & 7362 & AI Velocity Study Reveals Contextual Bottleneck | Unblocked posted on the topic | LinkedIn & \url{https://www.linkedin.com/posts/getunblocked_our-friends-at-dx-just-released-a-great-study-activity-7457813967701393409-k6MY} \\
G2492 & 7367 & How AI Coding Changes Estimation, Planning, and Accountability & \url{https://medium.com/@marketing_39301/how-ai-coding-changes-estimation-planning-and-accountability-eb1ebd50f552} \\
G2493 & 7368 & 3 Cursor Agents You NEED Today. 3 simple Cursor agents to improve code... | by Matt Pua | Medium & \url{https://matthewpua.medium.com/3-must-have-agents-for-cursor-8d15556ab50f} \\
G2494 & 7369 & Extending Cursor AI with Automated Code Reviews and AI Guardrails & \url{https://codescene.com/blog/extending-cursor-ai-with-automated-code-reviews} \\
G2495 & 7373 & AWS Security Agent Enhances DevOps with Full Repository Code Scanning | DevOpsChat posted on the topic | LinkedIn & \url{https://www.linkedin.com/posts/devopschat_aws-security-agent-brings-full-repository-activity-7460781557247553537-3Ts4} \\
G2496 & 7377 & GitHub - codexstar69/bug-hunter at 244bff308125a1ff3c02b265444926cbc1e0927f GitHub & \url{https://github.com/codexstar69/bug-hunter/tree/244bff308125a1ff3c02b265444926cbc1e0927f} \\
G2497 & 7387 & AI Code Review Challenges in Engineering Leadership - LinkedIn & \url{https://www.linkedin.com/posts/allan-mayberry-98b05060_the-80-problem-in-agentic-coding-activity-7432040827658330112-p6li} \\
G2498 & 7394 & An AI Agent Got Rejected from Open Source and Called Out the Maintainer for Gatekeeping & \url{https://www.linkedin.com/pulse/ai-agent-got-rejected-from-open-source-called-out-maintainer-jones-4hgpf} \\
G2499 & 7398 & Devin AI Review: Autonomous Software Engineer -- Real Results 2026 - CoinCodeCap & \url{https://coincodecap.com/devin-ai-a-game-changer-or-threat} \\
G2500 & 7400 & What I Learned Building an AI Code Reviewer (And How I Tried to Break It) | by Adityaraj | Mar, 2026 | Medium & \url{https://medium.com/@adityaraj06072003/what-i-learned-building-an-ai-code-reviewer-and-how-i-tried-to-break-it-e78947e67382} \\
G2501 & 7404 & What Is Vibe Coding? The Developer's Guide to AI-First Development (2026) & \url{https://designrevision.com/blog/vibe-coding} \\
G2502 & 7414 & ChuksForge/code-review-agent-public & \url{https://github.com/ChuksForge/code-review-agent-public} \\
G2503 & 7416 & stafford williams - Steering the Vibe: Review & \url{https://staffordwilliams.com/blog/2026/02/11/steering-the-vibe-review/} \\
G2504 & 7420 & AI-Generated Code Creates New Wave of Technical Debt, Report Finds - InfoQ & \url{https://www.infoq.com/news/2025/11/ai-code-technical-debt/} \\
G2505 & 7422 & How to Automate Your GitLab Merge Requests with AI & \url{https://www.linkedin.com/pulse/how-automate-your-gitlab-merge-requests-withai-api4ai-41qhf} \\
G2506 & 7423 & What Developers (Like Me) Really Wanted - DEV Community & \url{https://dev.to/goofmint/what-developers-like-me-really-wanted-3l7k} \\
G2507 & 7425 & What Software Engineering Will Look Like in 5 Months | by Fan Xia | Medium & \url{https://medium.com/@xiafan/what-software-engineering-will-look-like-in-5-months-75d7dec13386} \\
G2508 & 7426 & Analysis Identifies 10 AI Coding Tool Behaviors That Ignore Best Software Engineering Practices - DevOps.com & \url{https://devops.com/analysis-identifies-10-ai-coding-tool-behaviors-that-ignore-best-software-engineering-practices/} \\
G2509 & 7434 & VorobiovD/air & \url{https://github.com/VorobiovD/air} \\
G2510 & 7435 & I Spend 45\% of My Time Debugging AI Code. Stack Overflow Survey Proves I'm Not Alone. | by Devrim Ozcay | Let's Code Future | Feb, 2026 | Medium & \url{https://medium.com/lets-code-future/i-spend-45-of-my-time-debugging-ai-code-stack-overflow-survey-proves-im-not-alone-543c80750984} \\
G2511 & 7437 & AI Code Review Tools Compared: What Actually Catches Bugs in AI-Generated Code? - DEV Community & \url{https://dev.to/toniantunovic/ai-code-review-tools-compared-what-actually-catches-bugs-in-ai-generated-code-2bo8} \\
G2512 & 7440 & From Idea to Production with AI -- How Spec-Driven Development Changed the Way I Build Software | by Nuwan Thuduwage | Apr, 2026 | Medium & \url{https://medium.com/@nuwan.thuduwage/from-idea-to-production-with-ai-how-spec-driven-development-changed-the-way-i-build-software-a80811ce2589} \\
G2513 & 7441 & Vibe Slop: How AI-Driven "Vibe Coding" Is Creating a New Era of Dangerous Software & \url{https://www.linkedin.com/pulse/vibe-slop-how-aidriven-coding-creating-new-era-james-kunkle-pcs-qtvnc} \\
G2514 & 7447 & Anthropic Just Launched Code Review in Claude Code -- And 54\% of PRs Now Get Real Feedback - Top AI Product & \url{https://topaiproduct.com/2026/03/09/anthropic-just-launched-code-review-in-claude-code-and-54-of-prs-now-get-real-feedback/} \\
G2515 & 7451 & 42\% of Committed Code Is AI-Generated. Only 48\% Gets Reviewed. Here's Why That Gap Will Break Your Production. - DEV Community & \url{https://dev.to/lizechengnet/42-of-committed-code-is-ai-generated-only-48-gets-reviewed-heres-why-that-gap-will-break-your-58gb} \\
G2516 & 7452 & CodeRabbit vs Greptile: Which AI Reviewer Catches More Bugs? - DEV Community & \url{https://dev.to/rahulxsingh/coderabbit-vs-greptile-which-ai-reviewer-catches-more-bugs-4n9k} \\
G2517 & 7455 & How AI-Generated PRs Are Overwhelming Code Review (and How to Fix It) - Skylos Blog | Skylos & \url{https://skylos.dev/blog/surviving-the-ai-pr-tsunami} \\
G2518 & 7457 & I built an AI operating system because every project I touched was lying to me - DEV Community & \url{https://dev.to/khaeldur/i-built-an-ai-operating-system-because-every-project-i-touched-was-lying-to-me-3e14} \\
G2519 & 7459 & AI Code Review Is the New Bottleneck - Developers Digest & \url{https://www.developersdigest.tech/blog/ai-code-review-bottleneck} \\
G2520 & 7464 & How AI Coding Agents Evolved from Autocomplete into Autonomous Pull Request Machines - SoftwareSeni & \url{https://www.softwareseni.com/how-ai-coding-agents-evolved-from-autocomplete-into-autonomous-pull-request-machines/} \\
G2521 & 7468 & A buyer's guide to code quality and security in the age of AI and agentic development | Sonar & \url{https://www.sonarsource.com/resources/library/buyers-guide-to-code-quality-and-security/} \\
G2522 & 7469 & Building the Verification Layer: How Implementing Code Standards Unlock AI Code at Scale - Qodo & \url{https://www.qodo.ai/blog/building-the-verification-layer-how-implementing-code-standards-unlock-ai-code-at-scale/} \\
G2523 & 7470 & Code Review Fatigue: AI Tools That Make Reviews Faster and More Effective | by Abhishek Kumar | Medium & \url{https://medium.com/@thequeryabhishk/code-review-fatigue-ai-tools-that-make-reviews-faster-and-more-effective-69d58a129c97} \\
G2524 & 7472 & Easily Perform AI-powered Code Reviews In Minutes (CodeRabbit + GitHub) - DEV Community & \url{https://dev.to/crosspostr/easily-perform-ai-powered-code-reviews-in-minutes-coderabbit-github-28d3} \\
G2525 & 7473 & The Branch and the Merge: On Version Control Strategy and Code Integration Practices | by Andras Ludanyi | May, 2026 | Medium & \url{https://medium.com/@aludanyi/the-branch-and-the-merge-on-version-control-strategy-and-code-integration-practices-6412d7912ed9} \\
G2526 & 7479 & vikingmute/review-forge & \url{https://github.com/vikingmute/review-forge} \\
G2527 & 7480 & AI Hasn't Made Developers Faster, It's Made Their Review Queues Longer! - ShiftMag & \url{https://shiftmag.dev/ai-hasnt-made-developers-faster-its-made-their-review-queues-longer-8935/} \\
G2528 & 7481 & Anthropic's Project Glasswing Expands Mythos To 150+ Organizations & \url{https://www.linkedin.com/pulse/anthropics-project-glasswing-expands-mythos-150-ym5se} \\
G2529 & 7482 & HainanZhao/AICodeReview & \url{https://github.com/HainanZhao/AICodeReview} \\
G2530 & 7486 & Cadence v8.4: a multi-model coding harness where Claude writes, Codex reviews, and Bugbot triages - DEV Community & \url{https://dev.to/axledbetter/cadence-v84-a-multi-model-coding-harness-where-claude-writes-codex-reviews-and-bugbot-triages-1d73} \\
G2531 & 7487 & Fallacies of GenAI Development \#4: Dropping Human Review Removes the Bottleneck - DEV Community & \url{https://dev.to/bala_paranj_059d338e44e7e/fallacies-of-genai-development-4-dropping-human-review-removes-the-bottleneck-1jo} \\
G2532 & 7488 & AI-Generated Code Review: SonarQube's Deterministic Approach | Akshay Pachaar posted on the topic | LinkedIn & \url{https://www.linkedin.com/posts/akshay-pachaar_every-ai-coding-assistant-has-the-same-problem-activity-7405152774629425152-xtPn} \\
G2533 & 7493 & Securing our codebase with autonomous agents Cursor & \url{https://cursor.com/blog/security-agents} \\
G2534 & 7494 & Rust and LLMs: The Compiler Does What Code Review Shouldn't Have To - DEV Community & \url{https://dev.to/arezvov/rust-and-llms-the-compiler-does-what-code-review-shouldnt-have-to-3ia4} \\
G2535 & 7497 & CodeRabbit Configuration: .coderabbit.yaml Reference - DEV Community & \url{https://dev.to/rahulxsingh/coderabbit-configuration-coderabbityaml-reference-5493} \\
G2536 & 7499 & Linus Torvalds Says AI Bug Reports Are Making Linux Security Unmanageable - Archypedia News & \url{https://archypedia.news/linus-torvalds-says-ai-bug-reports-are-making-linux-security-unmanageable/} \\
G2537 & 7500 & The Mistakes Didn't Change. The Speed Did. - DEV Community & \url{https://dev.to/felixortizdev/the-mistakes-didnt-change-the-speed-did-13i8} \\
G2538 & 7504 & AI Didn't Kill Code Review -- You Did | by Deepesh M Divakaran | Mar, 2026 | Medium & \url{https://medium.com/@deepeshmd/ai-didnt-kill-code-review-you-did-d011c6d2eb63} \\
G2539 & 7506 & We Don't Trust AI (and That's a Good Thing) - Coder Legion & \url{https://coderlegion.com/15992/we-dont-trust-ai-and-thats-a-good-thing} \\
G2540 & 7508 & Maintaining Code Ownership in the Age of AI Assistance & \url{https://zenvanriel.com/ai-engineer-blog/maintaining-code-ownership-with-ai-assistance/} \\
G2541 & 7509 & Code Ownership Models: Scale Teams Without Chaos | CodeRaven & \url{https://coderaven.io/blog/code-ownership-models-teams-scale/} \\
G2542 & 7511 & AI Pair Programming Isn't Autopilot: Scaffolding HandyFEM and Catching What the AI Threw Away - DEV Community & \url{https://dev.to/constanza_diaz_dev/ai-pair-programming-isnt-autopilot-scaffolding-handyfem-and-catching-what-the-ai-threw-away-5p6} \\
G2543 & 7513 & AI Wrote the Code. Now Nobody Can Review It. -- Codexical & \url{https://www.codexical.com/posts/2026-05-13-ai-code-review-bottleneck} \\
G2544 & 7514 & Code ownership in the age of LLM & \url{https://t2informatik.de/en/blog/code-ownership-in-the-age-of-llm/} \\
G2545 & 7516 & AI Code Review Is the New Bottleneck: Why Faster Code Is Not Reaching Production Faster - DEV Community & \url{https://dev.to/alexcloudstar/ai-code-review-is-the-new-bottleneck-why-faster-code-is-not-reaching-production-faster-416a} \\
G2546 & 7522 & Multimodal Gemma 4 Visual Regression \& Patch Agent - DEV Community & \url{https://dev.to/kanyingidickson-dev/multimodal-gemma-4-visual-regression-patch-agent-53lk} \\
G2547 & 7523 & I Tested 13 AI Code Review Tools So You Don't Have To (2026) - DEV Community & \url{https://dev.to/rahxuls/i-tested-13-ai-code-review-tools-so-you-dont-have-to-2026-ml1} \\
G2548 & 7527 & Search code, repositories, users, issues, pull requests... & \url{https://github.com/Dharamraj82/Ai-Power-Code-Review-BACKEND} \\
G2549 & 7528 & Bridge ACE vs Claude Code vs Cursor vs Windsurf vs Devin & \url{https://dev.to/bridgeace/bridge-ace-vs-claude-code-vs-cursor-vs-windsurf-vs-devin-an-honest-comparison-march-2026-1de0} \\
G2550 & 7532 & Copilot code review now available for organization members without a license - GitHub Changelog & \url{https://github.blog/changelog/2025-12-17-copilot-code-review-now-available-for-organization-members-without-a-license/} \\
G2551 & 7534 & Qodo Outperforms Claude in Code Review Benchmark - Qodo & \url{https://www.qodo.ai/blog/qodo-outperforms-claude-in-code-review-benchmark/} \\
G2552 & 7536 & The New Code Review Culture: How Teams Turn AI Velocity Into Quality - Qodo & \url{https://www.qodo.ai/blog/the-new-code-review-culture-how-teams-turn-ai-velocity-into-quality/} \\
G2553 & 7538 & Can we trust AI-generated code? - DEV Community & \url{https://dev.to/dimitrisk_cyclopt/can-we-trust-ai-generated-code-122d} \\
G2554 & 7540 & Gemini accused of 30,000-line code purge and fake recovery report & \url{https://www.theregister.com/ai-ml/2026/05/21/gemini-accused-of-30000-line-code-purge-and-fake-recovery-report/5244219} \\
G2555 & 7542 & From Manual Reviews to AI-Powered Insights: The Linux Foundation's & \url{https://www.coderabbit.ai/blog/how-linux-foundation-used-ai-code-reviews-to-reduce-manual-bottlenecks-in-oss} \\
G2556 & 7546 & Stop Vibe Coding. Start Spec-Driven Development. - DEV Community & \url{https://dev.to/arielf/youre-a-slop-coder-autospec-is-for-professionals-only-4g42} \\
G2557 & 7554 & Unlearning, experimentation and engineering rigor in an agentic world | Thoughtworks Brazil & \url{https://www.thoughtworks.com/en-br/insights/podcasts/technology-podcasts/unlearning-experimentation-engineering-rigor-agentic-world} \\
G2558 & 7556 & How to Measure Engineering Productivity in Distributed Teams & \url{https://www.qodo.ai/blog/how-to-measure-engineering-productivity-in-large-distributed-teams/} \\
G2559 & 7557 & Vibe Coding Is Killing Open Source, and the Data Proves It | grith & \url{https://dev.grith.ai/blog/vibe-coding-killing-open-source} \\
G2560 & 7562 & Add copilot-review skill for PR feedback triage and response Pull Request \#36111 github/gh-aw & \url{https://github.com/github/gh-aw/pull/36111} \\
G2561 & 7563 & Forge-Space/forge-ai-init & \url{https://github.com/forge-space/forge-ai-init} \\
G2562 & 7564 & Unlocking the Power of AI in Modern Development - DEV Community & \url{https://dev.to/mikeover/unlocking-the-power-of-ai-in-modern-development-a-deep-dive-into-practical-applications-5h68} \\
G2563 & 7570 & Exelbierd: What's actually in a Sashiko review? [LWN.net] & \url{https://lwn.net/Articles/1065971/} \\
G2564 & 7572 & How to make sure AI-generated code is production-ready - Quora & \url{https://www.quora.com/How-do-you-make-sure-AI-generated-code-is-production-ready} \\
G2565 & 7575 & The Real Test Isn't Coding. It's Fit, Judgment, and Teamwork | Dan Case posted on the topic | LinkedIn & \url{https://www.linkedin.com/posts/dancase1_the-real-test-isnt-coding-its-fit-judgment-activity-7348812271717924864-mgbl} \\
G2566 & 7579 & linghucong-yue1/ai-attestation & \url{https://github.com/linghucong-yue1/ai-attestation} \\
G2567 & 7580 & How to Make Your Codebase Work for AI Coding Agents (Without Better Prompts) - DEV Community & \url{https://dev.to/devansh365/how-to-make-your-codebase-work-for-ai-coding-agents-without-better-prompts-kcb} \\
G2568 & 7582 & Node.js wants to ban AI-generated code. They should. - DEV Community & \url{https://dev.to/adioof/nodejs-wants-to-ban-ai-generated-code-they-should-488} \\
G2569 & 7583 & SAGE: The AI Code Reviewer That Never Forgets Your Team's Standards | by Maheshbommagani | Apr, 2026 | Medium & \url{https://medium.com/@maheshbommagani91/sage-the-ai-code-reviewer-that-never-forgets-your-teams-standards-8b87887fd07a} \\
G2570 & 7584 & Twill.ai and AI agent PRs: what nobody talks about | Juanchi.dev & \url{https://juanchi.dev/en/blog/twill-ai-agent-generated-prs-epistemic-responsibility-real-experience} \\
G2571 & 7585 & The Self-Improving AI System That Built Itself & \url{https://www.linkedin.com/pulse/self-improving-ai-system-built-itself-prateek-karnal-760oc} \\
G2572 & 7590 & Cursor vs Devin (2026): AI IDE vs Autonomous Coding Agent & \url{https://aiagentsquare.com/compare/cursor-vs-devin} \\
G2573 & 7592 & Vibe Coded to Pwned: A Red Team Wake-Up Call for the AI Development Era | TekStream Solutions & \url{https://www.tekstream.com/blog/ai-generated-code-security-risks/} \\
G2574 & 7593 & Signing Your Name on AI-Assisted Commits with RAI Footers & \url{https://dev.to/anchildress1/signing-your-name-on-ai-assisted-commits-with-rai-footers-2b0o} \\
G2575 & 7594 & Claude Sonnet 4.6 \textbackslash\{\} Anthropic & \url{https://www.anthropic.com/claude/sonnet?e45d281a_page=1} \\
G2576 & 7595 & The AI Code Quality Playbook - Qodo & \url{https://www.qodo.ai/resources/the-ai-code-quality-playbook/} \\
G2577 & 7596 & I got tired of AI Reviewers hallucinating, so I built an Autonomous Agent for GitLab instead. - DEV Community & \url{https://dev.to/tranlongan/i-got-tired-of-ai-reviewers-hallucinating-so-i-built-an-autonomous-agent-for-gitlab-instead-jeh} \\
G2578 & 7601 & Quality Assurance in AI Assisted Software Development: Risks and Implications - DEV Community & \url{https://dev.to/dmitry_turmyshev/quality-assurance-in-ai-assisted-software-development-risks-and-implications-34kk} \\
G2579 & 7602 & Cursor: We Tested the \$50B AI Editor (9.4/10) | ThePlanetTools.ai & \url{https://theplanettools.ai/tools/cursor} \\
G2580 & 7603 & Best AI-Powered Code Review Tools for PR Automation (2026) & \url{https://www.qodo.ai/blog/pull-request-automation-tools/} \\
G2581 & 7607 & Augment Code vs Cursor: Comparison for Modern Dev Workflows & \url{https://www.qodo.ai/blog/augment-code-vs-cursor/} \\
G2582 & 7609 & Secure AI-Assisted Development: 15 Guardrails for... | Bishop Fox & \url{https://bishopfox.com/resources/security-guardrails-ai-code-checklist} \\
G2583 & 7612 & Vibe Coding in 2026: How Founders Are Building Real Products Without Engineering Teams - DEV Community & \url{https://dev.to/max_quimby/vibe-coding-in-2026-how-founders-are-building-real-products-without-engineering-teams-foh} \\
G2584 & 7618 & Codex Code Review vs Claude Code: AI Code Review Comparison - DEV Community & \url{https://dev.to/mcrolly/codex-code-review-vs-claude-code-ai-code-review-comparison-58jk} \\
G2585 & 7620 & Cursor Code Audit: Review AI-Generated Code (2026) & \url{https://vibecoding.app/blog/cursor-code-audit} \\
G2586 & 7621 & The Shift from "Writing" to "Orchestrating": Why Basics Matter More Than Ever | by Akash Wankhede | Apr, 2026 | Medium & \url{https://medium.com/@acwankhede/the-shift-from-writing-to-orchestrating-why-basics-matter-more-than-ever-f6b3ba8d85a6} \\
G2587 & 7624 & Vibe Coding with AI: Pair Programming in the Age of Copilot & \url{https://www.linkedin.com/pulse/vibe-coding-ai-pair-programming-age-copilot-sunil-patel-alsbe} \\
G2588 & 7625 & The New Rules of AI-Generated Code Accountability & \url{https://asymm.com/the-new-rules-of-ai-generated-code-accountability/} \\
G2589 & 7631 & Study finds AI-generated code has 2.7x more security flaws | Vibe Graveyard & \url{https://www.vibegraveyard.ai/story/coderabbit-ai-code-quality-study} \\
G2590 & 7632 & Cursor AI vs Claude Code: Which AI Tool Reigns Supreme | Paul Azarenko posted on the topic | LinkedIn & \url{https://www.linkedin.com/posts/paul-azarenko_ai-cursorai-claudecode-activity-7445485205638799360-2ETF} \\
G2591 & 7636 & GitHub Copilot Coding Agent for Jira Connects Planning to Pull ... & \url{https://devops.com/github-copilot-coding-agent-for-jira-connects-planning-to-pull-requests-without-leaving-your-workflow/} \\
G2592 & 7639 & The imminent risk of vibe coding - by Matt Basta & \url{https://basta.substack.com/p/the-imminent-risk-of-vibe-coding} \\
G2593 & 7640 & How to Build Claude Code Custom Skills for Automated Code Review - DEV Community & \url{https://dev.to/myougatheaxo/how-to-build-claude-code-custom-skills-for-automated-code-review-4dgl} \\
G2594 & 7641 & ChatGPT Likes Our Code ! ChatGPT AI Agent Code Review & \url{https://www.linkedin.com/pulse/chatgpt-likes-our-code-ai-agent-review-jeff-brower-vabgc} \\
G2595 & 7642 & Open source maintainers are drowning in AI-generated pull requests. Enterprise teams are next. - The New Stack & \url{https://thenewstack.io/ai-generated-code-crisis/} \\
G2596 & 7646 & Conductors to Orchestrators: The Future of Agentic Coding - O'Reilly & \url{https://www.oreilly.com/radar/conductors-to-orchestrators-the-future-of-agentic-coding/} \\
G2597 & 7647 & kocendavid/guardian & \url{https://github.com/kocendavid/guardian} \\
G2598 & 7649 & Release AI Agent Code Safely - Production CI/CD \& Secrets & \url{https://dev.to/teppana88/release-ai-agent-code-safely-production-cicd-secrets-5ecj} \\
G2599 & 7652 & Cursor Won't Save Your Business. Here's What Will. & \url{https://www.linkedin.com/pulse/cursor-wont-save-your-business-heres-what-alister-thyne-e63vc} \\
G2600 & 7653 & Why Code Review Cannot Scale With AI Output - DEV Community & \url{https://dev.to/mnemehq/why-code-review-cannot-scale-with-ai-output-4a1h} \\
G2601 & 7655 & AI Dark Factory Part 3: Spec-Driven Development -- env.dev & \url{https://env.dev/guides/dark-factory-pattern-part-3} \\
G2602 & 7659 & GitHub Copilot Did a Code Review on the Code It Helped Me Write: Here's What Happened | HackerNoon & \url{https://hackernoon.com/github-copilot-did-a-code-review-on-the-code-it-helped-me-write-heres-what-happened} \\
G2603 & 7660 & How to Keep Your Codebase From Rotting When AI Writes Most of It - DEV Community & \url{https://dev.to/alanwest/how-to-keep-your-codebase-from-rotting-when-ai-writes-most-of-it-c12} \\
G2604 & 7663 & My quick and very subjective take on AI coding agents: * I tend to delete 80\% of the generated code. * The remaining 20\% I refactor. AI-generated code is rarely & \url{https://www.linkedin.com/posts/adam-tornhill-71759b48_my-quick-and-very-subjective-take-on-ai-coding-activity-7402723654834569216-2zt6} \\
G2605 & 7664 & How Code Reviews Quietly Shape You into a Better Engineer | by Ubongsky1 | May, 2026 | Medium & \url{https://medium.com/@ubongsky1/how-code-reviews-quietly-shape-you-into-a-better-engineer-b1522c872997} \\
G2606 & 7667 & How We Hire for the 20\% AI Can't Do (And Why We Stopped Asking Candidates to Code From Scratch) - DEV Community & \url{https://dev.to/mickyarun/how-we-hire-for-the-20-ai-cant-do-and-why-we-stopped-asking-candidates-to-code-from-scratch-1ida} \\
G2607 & 7668 & Mauritiusllewelynpowys919/codex-review & \url{https://github.com/Mauritiusllewelynpowys919/codex-review} \\
G2608 & 7670 & Fixing blind spots in code reviews with Hindsight memory - DEV Community & \url{https://dev.to/anu_muddam_ef4abcbe2c619f/fixing-blind-spots-in-code-reviews-with-hindsight-memory-4bgj} \\
G2609 & 7674 & Qodo vs CodeRabbit: AI Code Review Tools Compared (2026) - DEV Community & \url{https://dev.to/rahulxsingh/qodo-vs-coderabbit-ai-code-review-tools-compared-2026-kdp} \\
G2610 & 7675 & runtimebug/diff-hound & \url{https://github.com/runtimebug/diff-hound} \\
G2611 & 7676 & I Let an AI Agent Review My PRs for a Week -- Here's What It Caught - DEV Community & \url{https://dev.to/thedailyagent/i-let-an-ai-agent-review-my-prs-for-a-week-heres-what-it-caught-4geb} \\
G2612 & 7684 & RDEL \#92: How do AI code reviews impact engineering teams? & \url{https://rdel.substack.com/p/rdel-92-how-do-ai-code-reviews-impact} \\
G2613 & 7687 & I Audited 8 Months of AI-Generated Code. Here's What Nobody Talks About. | Rizz Development & \url{https://rizz.dev/blog/opinion/ai-generated-code-audit} \\
G2614 & 7688 & Code Review Kinda Sucks Now - DEV Community & \url{https://dev.to/xjamundx/some-thoughts-on-code-review-in-the-ai-age-part-1-2k3m} \\
G2615 & 7689 & AI Governance and Its Impact on Coding | encorp.ai & \url{https://encorp.ai/en/blog/ai-governance-impact-on-coding-2026-04-29} \\
G2616 & 7700 & CodeRabbit Review 2026: Pricing, Features \& Alternatives | Rank\&Compare & \url{https://www.rankncompare.com/tools/coderabbit} \\
G2617 & 7701 & KiniunCorp/spec-to-ship & \url{https://github.com/KiniunCorp/spec-to-ship} \\
G2618 & 7702 & Code review best practices for vibe coding - CodeRabbit & \url{https://coderabbit.ai/blog/code-review-best-practices-for-vibe-coding} \\
G2619 & 7704 & Claude Code Review: Setup, Pricing, and Verdict - DEV Community & \url{https://dev.to/rahulxsingh/claude-code-review-setup-pricing-and-verdict-41p3} \\
G2620 & 7710 & The Rise of AI Agents in Software Development: What I'm Seeing in 2026 - DEV Community & \url{https://dev.to/elysiumquill/the-rise-of-ai-agents-in-software-development-what-im-seeing-in-2026-om5} \\
G2621 & 7712 & I Let an AI Agent Manage My GitHub PRs for a Week : Here's What Actually Broke | by Garima Agarwal | JavaGuides | May, 2026 | Medium & \url{https://medium.com/javaguides/i-let-an-ai-agent-manage-my-github-prs-for-a-week-heres-what-actually-broke-c664a69c2f1e} \\
G2622 & 7714 & Enterprise AI Bottleneck: Structured Context | Tabnine posted on the topic | LinkedIn & \url{https://www.linkedin.com/posts/tabnine_the-bottleneck-holding-back-ai-in-enterprise-activity-7434618445041958912-afRV} \\
G2623 & 7715 & Claude Code + Codex Plugin: Two AI Brains, One Terminal - DEV Community & \url{https://dev.to/harrisonsec/claude-code-codex-plugin-two-ai-brains-one-terminal-k31} \\
G2624 & 7716 & OpenAI's Agentic Software Engineer (A-SWE): Pioneering Autonomous Coding & \url{https://www.linkedin.com/pulse/openais-agentic-software-engineer-a-swe-pioneering-autonomous-jha-iwofc} \\
G2625 & 7717 & Stop Closing the Door. Fix the House. - O'Reilly & \url{https://www.oreilly.com/radar/stop-closing-the-door-fix-the-house/} \\
G2626 & 7727 & Who Owns Code Written by a Machine? | by Nexumo | Medium & \url{https://medium.com/@Nexumo_/who-owns-code-written-by-a-machine-bc02756c443c} \\
G2627 & 7728 & OWASP LLM Top 10: How it Applies to Code Generation | Learn Article | Sonar & \url{https://www.sonarsource.com/resources/library/owasp-llm-code-generation/} \\
G2628 & 7732 & What Is Vibe Coding? A Developer's Guide (2026) - DEV Community & \url{https://dev.to/remybuilds/what-is-vibe-coding-a-developers-guide-2026-o0m} \\
G2629 & 7733 & AegisNodejkt/Aegis-Wilwatikta & \url{https://github.com/AegisNodejkt/Aegis-Wilwatikta} \\
G2630 & 7734 & Easily apply Copilot code review feedback with Copilot cloud agent - GitHub Changelog & \url{https://github.blog/changelog/2026-05-19-easily-apply-copilot-code-review-feedback-with-copilot-cloud-agent/} \\
G2631 & 7736 & What AI Review and Tests Both Missed | Medium & \url{https://medium.com/@dominikus.nold/what-ai-review-and-tests-both-missed-building-the-specfact-code-review-module-4dd802729e1d} \\
G2632 & 7738 & Codex vs Claude: Execution vs Reasoning in AI Ecosystems | Varun Mahajan posted on the topic | LinkedIn & \url{https://www.linkedin.com/posts/varun-mhjn_over-the-past-year-ive-been-using-both-activity-7442212392932978689-Vouc} \\
G2633 & 7740 & AI Code Has an Owner Problem | JetThoughts Blog & \url{https://jetthoughts.com/blog/ai-code-ownership-accountability/} \\
G2634 & 7748 & From Hallucination to Production Bug: A Post-Mortem on AI-Generated Code | by Rajan Patekar | Mar, 2026 | Medium & \url{https://medium.com/@rajanpatekar/from-hallucination-to-production-bug-a-post-mortem-on-ai-generated-code-0987034037f8} \\
G2635 & 7749 & julianabot/ai-pr-reviewer & \url{https://github.com/julianabot/ai-pr-reviewer} \\
G2636 & 7750 & The Manual Loop: What Happened to Code Review | by Mohamed Nashaat | Mar, 2026 | Medium & \url{https://medium.com/@tqcgjdt/the-manual-loop-what-happened-to-code-review-4da6f5d8833f} \\
G2637 & 7751 & davidlee/spec-driver & \url{https://github.com/davidlee/spec-driver} \\
G2638 & 7753 & Devin AI Review -- AI Panel Score 7.8/10 | TopReviewed.ai & \url{https://topreviewed.ai/products/devin-ai} \\
G2639 & 7755 & Cognitive Debt and AI Burnout Risks | O'Reilly posted on the topic | LinkedIn & \url{https://www.linkedin.com/posts/oreilly_burnout-and-cognitive-debt-activity-7460419672304369665-BHAr} \\
G2640 & 7761 & ReviewPilot: The AI-Native Code Reviewer for Your Terminal - DEV Community & \url{https://dev.to/lebaxiom/reviewpilot-the-ai-native-code-reviewer-for-your-terminal-20fm} \\
G2641 & 7762 & GitHub Skills: Your Complete Learning Path to AI-Powered ... & \url{https://dev.to/pwd9000/github-skills-your-complete-learning-path-to-ai-powered-development-ieo} \\
G2642 & 7765 & Our Code Review Process Was Broken. Here's How We Fixed It With AI (And Cut Review Time by 85\%) - DEV Community & \url{https://dev.to/tdn/our-code-review-process-was-broken-heres-how-we-fixed-it-with-ai-and-cut-review-time-by-85-1h72} \\
G2643 & 7766 & Cursor + Claude: my AI code review checklist - DEV Community & \url{https://dev.to/sathish_daggula/cursor-claude-my-ai-code-review-checklist-hm5} \\
G2644 & 7770 & rudra496/ai-code-trust-validator & \url{https://github.com/rudra496/ai-code-trust-validator} \\
G2645 & 7771 & The Pull Request Illusion: How AI Is Hollowing Out Software's Last Line of Defense | by Flamehaven Initiative 2 | Medium & \url{https://medium.com/@flamehaven/the-pull-request-illusion-how-ai-is-hollowing-out-softwares-last-line-of-defense-21f6386200c8} \\
G2646 & 7775 & /review -- ClaudeCode Skill | AgentSpec & \url{https://agentspec.sh/skills/f6a7baff-4ba2-4dcd-a223-8ea1e1fc1eb1} \\
G2647 & 7777 & The 5 Levels of AI Coding Agents: From Simple Code Review to ... & \url{https://www.aibuilderclub.com/blog/5-levels-of-coding-agents} \\
G2648 & 7779 & JunMystery/AI-Agent-Standards & \url{https://github.com/JunMystery/AI-Agent-Standards} \\
G2649 & 7780 & AI Generated Code Joins the Linux Kernel, But Humans Stay in ... & \url{https://dev.to/zubairakbar/ai-generated-code-joins-the-linux-kernel-but-humans-stay-in-charge-568o} \\
G2650 & 7782 & LLM for Code Review Automation -- Real-World Use Cases and Pitfalls [2026] | Markaicode & \url{https://markaicode.com/usecases/llm-for-code-review-automation/} \\
G2651 & 7783 & Understanding Security Risks in AI-Generated Code | CSA & \url{https://cloudsecurityalliance.org/blog/2025/07/09/understanding-security-risks-in-ai-generated-code} \\
G2652 & 7785 & Your Developers Are Shipping AI-Generated Code. | by Mitch | Apr, 2026 | Medium & \url{https://medium.com/@mitch_66202/your-developers-are-shipping-ai-generated-code-ce26b75f9199} \\
G2653 & 7789 & On the Use of Agentic Coding: An Empirical Study of Pull Requests on GitHub | SCIENCE@home & \url{https://sah.borca.ai/papers/281394787} \\
G2654 & 7795 & The Trust Problem With AI Code Review | Groundy & \url{https://groundy.com/articles/trust-problem-ai-code/} \\
G2655 & 7796 & The Code Nobody Understands: The Hidden Risk of AI-Generated Code - David Monnerat & \url{https://www.davidmonnerat.com/future/ai-generated-code-the-code-nobody-understands/} \\
G2656 & 7797 & AI Coding Governance Framework | TechDebt.fast & \url{https://techdebt.fast/ai-governance-framework/} \\
G2657 & 7798 & The state of AI in CNCF projects: A first look at the data - TechFlowDaily & \url{https://techflowdaily.com/the-state-of-ai-in-cncf-projects-a-first-look-at-the-data/} \\
G2658 & 7800 & The Real Bottleneck of AI Isn't Intelligence -- It's Verification | by Lauri Korpela | Apr, 2026 | Medium & \url{https://medium.com/@larkko/the-real-bottleneck-of-ai-isnt-intelligence-it-s-verification-5f18e13af317} \\
G2659 & 7802 & AI Coding Tip 006 - Review Every Line Before Commit - DEV Community & \url{https://dev.to/mcsee/ai-coding-tip-006-review-every-line-before-commit-bmm} \\
G2660 & 7805 & mara-werils/ai-code-reviewer & \url{https://github.com/mara-werils/ai-code-reviewer} \\
G2661 & 7806 & AI Can Write and Review Code -- But Humans Still Own the Risk - Blog & \url{https://www.securecodewarrior.com/article/ai-can-write-and-review-code----but-humans-still-own-the-risk} \\
G2662 & 7808 & Your AI Can Code It in 3 Minutes. Your Team Still Ships It in a Week. - DEV Community & \url{https://dev.to/adioof/your-ai-can-code-it-in-3-minutes-your-team-still-ships-it-in-a-week-2j74} \\
G2663 & 7810 & SonarQube for Federal Agencies: A Guide to Complying with AI Policies in Code Development | Sonar & \url{https://www.sonarsource.com/resources/library/complying-with-ai-policies-in-code-development/} \\
G2664 & 7812 & NYTC69/review-loop & \url{https://github.com/NYTC69/review-loop} \\
G2665 & 7814 & I scanned 100 AI-generated apps for security vulnerabilities. Here's ... & \url{https://dev.to/tgoldi/i-scanned-100-ai-generated-apps-for-security-vulnerabilities-heres-what-i-found-1l5o} \\
G2666 & 7818 & AI Code Trust Gap: 96\% Doubt It, 48\% Verify It | byteiota & \url{https://byteiota.com/ai-code-trust-gap-96-doubt-it-48-verify-it/} \\
G2667 & 7820 & rishimule/ai-code-review-agent & \url{https://github.com/rishimule/ai-code-review-agent} \\
G2668 & 7825 & We Solved the Easy Problem. Now It's Time for the Hard One. | by Sean Corkum | Mar, 2026 | Medium & \url{https://medium.com/@seancorkum81/we-solved-the-easy-problem-now-its-time-for-the-hard-one-b128b25d56b3} \\
G2669 & 7828 & Redmine per-commit detection probe -- does Rigor catch real bugs? | Rigor & \url{https://rigor.typedduck.fail/reference/notes/20260521-redmine-per-commit-detection-probe/} \\
G2670 & 7829 & SAST False Positives in AI-Generated Code: Why 91\% of Alerts Are Noise (And How to Fix It) - DEV Community & \url{https://dev.to/toniantunovic/sast-false-positives-in-ai-generated-code-why-91-of-alerts-are-noise-and-how-to-fix-it-4dai} \\
G2671 & 7845 & AI Code Review: 30K Bugs Lighter, 50\% faster - Sentry Blog & \url{https://blog.sentry.io/ai-code-review-30k-bugs-lighter-50-faster/} \\
G2672 & 7852 & AI Code Review with CodeRabbit. Introduction | by Kuldeep Singh | ManoMano Tech team | Medium & \url{https://medium.com/manomano-tech/ai-code-review-with-coderabbit-ed6ed610f481} \\
G2673 & 7853 & AI code review in 2026 - a workflow that survives the PR flood | The Syntax Diaries & \url{https://thesyntaxdiaries.com/ai-code-review-2026-pr-flood-workflow} \\
G2674 & 7854 & A Practical Code Review Checklist for AI-Generated Code - The security, logic, and quality checks that AI code needs and standard review processes miss -- Vibe  & \url{https://blog.vibecoder.me/code-review-checklist-ai-generated-code} \\
G2675 & 7859 & AI Does Not Fix a Weak SDLC. It Exposes It. AI-assisted delivery does not remove the need for a solid SDLC. It makes the weak points more visible. If the Jira i & \url{https://www.linkedin.com/posts/daniel-pavelescu-72954415_ai-does-not-fix-a-weak-sdlc-it-exposes-it-activity-7465288082469339136-Ad9g} \\
G2676 & 7862 & Your AI Coding Assistant Is Writing Bugs. Here's How to Catch Them Before Production. | by Alex Carter | May, 2026 | Medium & \url{https://medium.com/@info.booststash/your-ai-coding-assistant-is-writing-bugs-heres-how-to-catch-them-before-production-3cb8f4298fc2} \\
G2677 & 7863 & You can't verify all the AI-generated code - LeadDev & \url{https://zporigin.leaddev.com/ai/you-cant-verify-all-the-ai-generated-code} \\
G2678 & 7868 & How I Use GitHub Copilot to Review Code and Pull Requests - DEV Community & \url{https://dev.to/deniskisina/how-i-use-github-copilot-to-review-code-and-pull-requests-2j4l} \\
G2679 & 7869 & How Do We Know AI-Generated Code Is Ready? - DEV Community & \url{https://dev.to/kochc/how-do-we-know-ai-generated-code-is-ready-j81} \\
G2680 & 7874 & How AI Coding Agents Are Reshaping Developer Workflows & \url{https://dev.to/eabait/how-ai-coding-agents-are-reshaping-developer-workflows-3249} \\
G2681 & 7877 & AI Is Flying the Plane. When Did You Last Take the Controls? * PlanTheFlow & \url{https://plantheflow.com/blog/coding-skill-atrophy-ai/} \\
G2682 & 7880 & I shipped the wrong abstraction, then deleted it - DEV Community & \url{https://dev.to/mikelane/i-shipped-the-wrong-abstraction-then-deleted-it-444f} \\
G2683 & 7881 & AI Code Quality Crisis 2026: Engineering Leader Guide & \url{https://www.ofashandfire.com/blog/ai-generated-code-quality-crisis} \\
G2684 & 7883 & \# The Pull Request Illusion - Coder Legion & \url{https://coderlegion.com/13635/the-pull-request-illusion} \\
G2685 & 7885 & Introducing an Interactive Code Review Experience with Amazon Q Developer in GitHub | AWS DevOps \& Developer Productivity Blog & \url{https://aws.amazon.com/blogs/devops/introducing-an-interactive-code-review-experience-with-amazon-q-developer-in-github/} \\
G2686 & 7890 & Building autonomous coding agents in 2026 | Cadence blog & \url{https://cadence.withremote.ai/blog/autonomous-coding-agents-2026} \\
G2687 & 7892 & Unreviewed AI Code Is Everywhere -- Here's What Breaks First - DEV Community & \url{https://dev.to/ayame0328/unreviewed-ai-code-is-everywhere-heres-what-breaks-first-480j} \\
G2688 & 7894 & OpenCara/OpenCara & \url{https://github.com/OpenCara/OpenCara} \\
G2689 & 7895 & \# 'Codification' of PEP 8, the Pax Pythonica? | by ince memed | Medium & \url{https://medium.com/@mehmetcenkcakircali/codification-of-pep-8-the-pax-pythonica-1a73ff17ec88} \\
G2690 & 7897 & yldgio/codereview-skills & \url{https://github.com/yldgio/anomaly-codereview} \\
G2691 & 7900 & GitHub - openai/codex-plugin-cc at d216a5fdead7fdc6d081831d3d259d9fef70a8d4 GitHub & \url{https://github.com/openai/codex-plugin-cc/tree/d216a5fdead7fdc6d081831d3d259d9fef70a8d4} \\
G2692 & 7902 & CodeRabbit GitLab Integration: Configuration Guide - DEV Community & \url{https://dev.to/rahulxsingh/coderabbit-gitlab-integration-configuration-guide-27jk} \\
G2693 & 7903 & AI-Generated Code Needs Its Own Secure Coding Guidelines & \url{https://www.appsecengineer.com/blog/ai-generated-code-needs-its-own-secure-coding-guidelines} \\
G2694 & 7905 & AI engineering requires more than just better tools & \url{https://www.faros.ai/blog/ai-engineering} \\
G2695 & 7912 & I gave Gemini 3.5 Flash a CVE-fix PR to review. It found another bug in the same file. - DEV Community & \url{https://dev.to/vicente_junior_dev/i-gave-gemini-35-flash-a-cve-fix-pr-to-review-it-found-another-bug-in-the-same-file-1g24} \\
G2696 & 7914 & Claude Overreaches. Codex Underreaches. I'm Still Figuring Out How to Use Both. | by Lakshmi Narasimhan | Apr, 2026 | Medium & \url{https://medium.com/@lakshminp/claude-overreaches-codex-underreaches-im-still-figuring-out-how-to-use-both-f98b124691f5} \\
G2697 & 7916 & The AI Productivity Paradox: Why Your Team Ships More Code but Delivers Less Value -- Let's Build & \url{https://letsbuildsolutions.com/blog/engineering-management/the-ai-productivity-paradox-why-your-team-ships-more-code-but-delivers-less-value/} \\
G2698 & 7921 & Ignoring LLMs in open source is becoming pure insanity [LWN.net] & \url{https://lwn.net/Articles/1067764/} \\
G2699 & 7927 & Copilot-reviewed pull request merge metrics now in the usage ... & \url{https://github.blog/changelog/2026-04-08-copilot-reviewed-pull-request-merge-metrics-now-in-the-usage-metrics-api/} \\
G2700 & 7933 & Add AI Contribution Policy for OpenSSF Technical Initiatives Pull Request \#605 ossf/tac & \url{https://github.com/ossf/tac/pull/605} \\
G2701 & 7936 & I Tested 5 AI Code Review Tools So You Don't Have To (CodeRabbit, Qodo, Mesrai, and More) - DEV Community & \url{https://dev.to/tdn/i-tested-5-ai-code-review-tools-so-you-dont-have-to-coderabbit-qodo-mesrai-and-more-57bb} \\
G2702 & 7937 & kodustech/stateofcodereview.com & \url{https://github.com/kodustech/stateofcodereview.com} \\
G2703 & 7938 & Code Review Tools Focus on Noise, Not Real Security Risks & \url{https://www.linkedin.com/posts/amartya-jha_ai-code-review-is-solving-the-wrong-problem-activity-7436864247520727040-vZzq} \\
G2704 & 7946 & I shipped my first agentic workflow on GitHub Actions -- 4 AI reviewers on every PR - DEV Community & \url{https://dev.to/pponali/i-shipped-my-first-agentic-workflow-on-github-actions-4-ai-reviewers-on-every-pr-321b} \\
G2705 & 7949 & 84\% of Developers Use AI Coding Tools -- But Only 29\% Trust What They Ship. Here's Why That Gap Will Kill Your Product - DEV Community & \url{https://dev.to/gerus_team/84-of-developers-use-ai-coding-tools-but-only-29-trust-what-they-ship-heres-why-that-gap-will-30n2} \\
G2706 & 7950 & I Stopped Using Claude Code as a Copilot. It Became My Orchestrator. | by Andy.G | Apr, 2026 | Medium & \url{https://medium.com/@andy.a.g/i-stopped-using-claude-code-as-a-copilot-it-became-my-orchestrator-ac90898e1f56} \\
G2707 & 7956 & Greptile v3: Agentic AI Code Review with 256\% Better Results | Greptile & \url{https://www.greptile.com/blog/greptile-v3-agentic-code-review} \\
G2708 & 7957 & Transforming Engineering Workflows with AI-Powered Software ... & \url{https://dev.to/erreginareid/transforming-engineering-workflows-with-ai-powered-software-development-4dje} \\
G2709 & 7958 & SimonPurdie/VS-Code-PR-Analysis & \url{https://github.com/SimonPurdie/VS-Code-PR-Analysis} \\
G2710 & 7959 & Codex vs Claude Code: Agentic Coding Tools Compared | Awesome Agents & \url{https://awesomeagents.ai/tools/codex-vs-claude-code/} \\
G2711 & 7961 & Exciting new AI tool - Devin Review, an AI code review tool by ... & \url{https://www.linkedin.com/posts/yangshun_exciting-new-ai-tool-devin-review-an-ai-activity-7420367219450372096-EP81} \\
G2712 & 7971 & How to Review AI-Generated Code Efficiently | Parallel Code & \url{https://parallelcode.app/blog/reviewing-ai-generated-code/} \\
G2713 & 7973 & Devin 101: Automatic PR Reviews with the Devin API - Cognition & \url{https://cognition.ai/blog/devin-101-automatic-pr-reviews-with-the-devin-api} \\
G2714 & 7974 & We Built an AI Code Review Bot -- Here's What It Actually Catches (And What It Misses) -- ScaledByDesign Insights & \url{https://www.scaledbydesign.com/blog/ai-code-review-bot-production} \\
G2715 & 7975 & The PR Review Bottleneck: Why Faster Code Generation Slows Teams Down - DEV Community & \url{https://dev.to/code-board/the-pr-review-bottleneck-why-faster-code-generation-slows-teams-down-1peb} \\
G2716 & 7977 & VibeOps: Governing AI-Generated Enterprise Code in GCCs (2026 Guide) & \url{https://www.crewscale.com/blog/vibeops-governing-ai-generated-code-gcc-guide} \\
G2717 & 7978 & Vibe Coding \& AI Code Review Can We Trust AI-Generated Code? | AppDevANGLE & \url{https://www.youtube.com/watch?v=ZT3Hul2hTKk} \\
G2718 & 7979 & How to Protect AI-Generated Code Quality Using SonarQube AI Code Assurance | Sonar & \url{https://www.sonarsource.com/resources/library/how-to-guide-for-ai-code-assurance/} \\
G2719 & 7980 & Cognition Just Raised \$26 Billion. I Put Devin on My Real Codebase to See What It Actually Buys You. | by PIXIPACE | May, 2026 | Medium & \url{https://medium.com/@pixipace/cognition-just-raised-26-billion-5f1cfd219e87} \\
G2720 & 7982 & AI Code Patch Quality -- When to Ship Without Review & \url{https://www.feedzap.live/blog/ai-code-patch-quality-ship-ready} \\
G2721 & 7984 & Code Guidelines: Bring your coding rules to CodeRabbit & \url{https://www.coderabbit.ai/blog/code-guidelines-bring-your-coding-rules-to-coderabbit?%3Futm_term=ai+for+code} \\
G2722 & 7992 & Memo: Red Hat Global Engineering plans to lean in to AI & \url{https://www.theregister.com/software/2026/03/31/memo-red-hat-global-engineering-plans-to-lean-in-to-ai/5221432} \\
G2723 & 7999 & The Ownership Framework: How to Use AI to Write Code Without Losing Your Mind - DEV Community & \url{https://dev.to/javeedishaq/the-ownership-framework-build-what-you-understand-ship-what-you-own-1pae} \\
G2724 & 8002 & AI Reviewing AI Code: Does This Make Sense? (2025 Analysis) & \url{https://www.greptile.com/blog/ai-code-reviews-conflict} \\
G2725 & 8004 & pull requests aren't dying. your review process is. - DEV Community & \url{https://dev.to/_95a3e57463e6442feacd0/pull-requests-arent-dying-your-review-process-is-k37} \\
G2726 & 8005 & Vibe Coding Is Eating Software Development - And Not Everyone Is Happy - DEV Community & \url{https://dev.to/maxmendes91/vibe-coding-is-eating-software-development-and-not-everyone-is-happy-426n} \\
G2727 & 8007 & CodeRabbit vs Greptile vs Sourcery: Which AI Code Reviewer Catches Bugs Without Drowning Your PRs in Comments | Pondero & \url{https://pondero.ai/coding/guides/coderabbit-vs-greptile-vs-sourcery-may-2026/} \\
G2728 & 8008 & I Built Software I Didn't Need So I Could Lead People Who Do | by Joseph Zhou | Medium & \url{https://medium.com/@shaoping.zhou_84486/i-built-software-i-didnt-need-so-i-could-lead-people-who-do-4ee666914f9e} \\
G2729 & 8013 & I recently reviewed some AI assisted code (Elixir). The core approach of the produced solution was solid, but I spotted a number of various issues: - Overly com & \url{https://www.linkedin.com/posts/sasajuric_i-recently-reviewed-some-ai-assisted-code-activity-7435287708752658432-JwE0} \\
G2730 & 8020 & Salt Code enforces security policies in AI coding tools & \url{https://securitybrief.news/story/salt-code-enforces-security-policies-in-ai-coding-tools} \\
G2731 & 8022 & IkhsanMAlbar/govchain-llm-orchestrator & \url{https://github.com/IkhsanMAlbar/govchain-llm-orchestrator} \\
G2732 & 8025 & Spec-Driven Development: The AI Coding Methodology | Stonewall & \url{https://stonewall.dev/blog/spec-driven-development/} \\
G2733 & 8028 & Simon Willison on Vibe Engineering vs Vibe Coding in 2026 - How Simon Willison's framing of vibe engineering separates throwaway prototypes from production-grad & \url{https://blog.vibecoder.me/simon-willison-vibe-engineering-vs-vibe-coding} \\
G2734 & 8030 & The Complete Epistemology: What AI Can and Cannot Replace - DEV Community & \url{https://dev.to/lanternproton/the-complete-epistemology-what-ai-can-and-cannot-replace-f0m} \\
G2735 & 8038 & 'Infinite agent code' is coming to break your monorepos - LeadDev & \url{https://leaddev.com/technical-direction/infinite-agent-code-is-coming-to-break-your-monorepos} \\
G2736 & 8039 & AI Slopageddon and the OSS Maintainers - console.log() & \url{https://redmonk.com/kholterhoff/2026/02/03/ai-slopageddon-and-the-oss-maintainers/} \\
G2737 & 8042 & Do androids dream of accepted pull requests? [LWN.net] & \url{https://lwn.net/Articles/1058643/} \\
G2738 & 8046 & Evaluating AI Code Review Tools: Insights and Outcomes & \url{https://conzit.com/post/evaluating-ai-code-review-tools-insights-and-outcomes} \\
G2739 & 8048 & Using GitHub Copilot for Accelerating Your Code Review Workflow: A Practical & \url{https://ai-coding-flow.com/blog/how-to-use-github-copilot-for-accelerating-your-code-review-workflow-2026/} \\
G2740 & 8051 & Code Written by an LLM in Seconds Can Cost You Weeks of Human Life | by Sergei Nagaev | Mar, 2026 | Medium & \url{https://medium.com/@nagaev.sv/code-written-by-an-llm-in-seconds-can-cost-you-weeks-of-human-life-6933093b0776} \\
G2741 & 8054 & kasidit-wansudon/ai-code-reviewer & \url{https://github.com/kasidit-wansudon/ai-code-reviewer} \\
G2742 & 8057 & Using Cursor Bugbot to autoreview and fix Claude Code PRs -- WorkOS & \url{https://workos.com/blog/cursor-bugbot-autoreview-claude-code-prs} \\
G2743 & 8060 & Code Review is Dead. Long Live Spec Review. | Esteban Sancho & \url{https://estebansancho.com/blog/software-engineering/2026/04/19/code-review-is-dead.html} \\
G2744 & 8061 & AI-Generated Code Is a Ticking Time Bomb (We Learned This the Hard Way) - DEV Community & \url{https://dev.to/gerus_team/ai-generated-code-is-a-ticking-time-bomb-we-learned-this-the-hard-way-5g4p} \\
G2745 & 8071 & Request Copilot code review from GitHub CLI - GitHub Changelog & \url{https://github.blog/changelog/2026-03-11-request-copilot-code-review-from-github-cli/} \\
G2746 & 8072 & The Uncomfortable Reality: Vibe Coding - DEV Community & \url{https://dev.to/xaviworks/the-uncomfortable-reality-vibe-coding-20lg} \\
G2747 & 8073 & I Automated My PR Reviews With AI -- Saved 6 Hours/Week (Full Setup) - DEV Community & \url{https://dev.to/hopkins_jesse_cdb68cfa22c/i-automated-my-pr-reviews-with-ai-saved-6-hoursweek-full-setup-472p} \\
G2748 & 8076 & The Vibe Coding Bottleneck Problem - by John Hwang & \url{https://nextword.substack.com/p/how-enterprises-can-adopt-vibe-coding} \\
G2749 & 8078 & Accelerated Code Generation Stresses Testing and Validation | Karl Jackson posted on the topic | LinkedIn & \url{https://www.linkedin.com/posts/karlj_this-accelerated-code-generation-is-stressing-activity-7418369616357511168-Yb7b} \\
G2750 & 8079 & 40\% of AI-Generated Code Has Security Flaws. Your Code Review Won't Catch Them. | by Vitalii Petrenko | Medium & \url{https://vitalii4reva.medium.com/40-of-ai-generated-code-has-security-flaws-your-code-review-wont-catch-them-0a02a13848fe} \\
G2751 & 8080 & Your Next Pull Request Will Come From a Product Manager | by Signadot | Medium & \url{https://medium.com/@signadot/your-next-pull-request-will-come-from-a-product-manager-b72d1a230b26} \\
G2752 & 8084 & AI Can Write the Code. You Still Need to Own the Merge. & \url{https://ghost.wduyck.me/ai-can-write-the-code-you-still-need-to-own-the-merge/} \\
G2753 & 8085 & Code Review in the Age of AI-Augmented Development - Daz & \url{https://daz.is/blog/code-review-ai-augmented-development/} \\
G2754 & 8087 & RFC: Artificial Contributors to Open Source | Andrew Nesbitt & \url{https://nesbitt.io/2026/05/21/rfc-artificial-contributors-to-open-source.html} \\
G2755 & 8088 & AddyOsmani.com - AI writes code faster. Your job is still to prove it works. & \url{https://addyosmani.com/blog/code-review-ai/} \\
G2756 & 8089 & AI Code Review Bottleneck: 11.4 Hours Reviewing vs 9.8 Writing | byteiota & \url{https://byteiota.com/ai-code-review-bottleneck-11-4-hours-reviewing-vs-9-8-writing/} \\
G2757 & 8095 & I started using Codex from OpenAI as part of my ultra-review process with Claude Code. And the results are seriously impressive. The setup is simple: Claude dev & \url{https://www.linkedin.com/posts/alexischenko_i-started-using-codex-from-openai-as-part-activity-7463884959296942080-_VvQ} \\
G2758 & 8096 & AI-Generated Code Defects Rise, PMs Must Own Intent | Aakash Gupta posted on the topic | LinkedIn & \url{https://www.linkedin.com/posts/aagupta_41-of-all-code-shipped-in-2025-was-ai-generated-activity-7438810992651743232-II6u} \\
G2759 & 8098 & OX Research: AI Code Not Inherently Less Secure, but "Army of Juniors" Effect Undermines Software Security | OX Security AI Code Isn't Less Secure, Just Deploye & \url{https://www.ox.security/blog/ox-research-ai-code-not-inherently-less-secure-but-army-of-juniors-effect-undermines-software-security/} \\
G2760 & 8099 & Code Reviews are Broken and AI Can't Fix Them. What Now? | QA Wolf & \url{https://www.qawolf.com/blog/code-reviews-are-broken-and-ai-cant-fix-them-what-now} \\
G2761 & 8100 & Fixing blind spots in code reviews with Hindsight memory - DEV Community & \url{https://dev.to/sanjeetha_167/fixing-blind-spots-in-code-reviews-with-hindsight-memory-gl2} \\
G2762 & 8105 & Rise of the Overnight Agents | Greptile Blog & \url{https://www.greptile.com/blog/rise-of-the-overnight-agents} \\
G2763 & 8106 & Your AI code reviewer has no one to disagree with - DEV Community & \url{https://dev.to/spencermarx/your-ai-code-reviewer-has-no-one-to-disagree-with-f1j} \\
G2764 & 8109 & A Real-World Coding Story: Devin AI as My Co-Pilot | by Pratik Patel | Medium & \url{https://medium.com/@prpatel05/a-real-world-coding-story-devin-ai-as-my-co-pilot-e1ff2ae492fa} \\
G2765 & 8110 & The "Vibecoding" Debt Bomb: Why AI Code is Architecturally Radioactive - DEV Community & \url{https://dev.to/dimitrisk_cyclopt/the-vibecoding-debt-bomb-why-ai-code-is-architecturally-radioactive-455c} \\
G2766 & 8112 & Human-AI Orthogonality: The Art of Perfect Complementarity - DEV Community & \url{https://dev.to/rakbro/human-ai-orthogonality-the-art-of-perfect-complementarity-2e5c} \\
G2767 & 8113 & 40\% of AI-Generated Code Has Security Flaws. Your Code Review Won't Catch Them. | by Vitalii Petrenko | Mar, 2026 | Medium & \url{https://javascript.plainenglish.io/40-of-ai-generated-code-has-security-flaws-your-code-review-wont-catch-them-0a02a13848fe} \\
G2768 & 8114 & Using LLMs to do security analysis at the git diff level -- what works, what doesn't, and why structured output matters - DEV Community & \url{https://dev.to/diffsense/using-llms-to-do-security-analysis-at-the-git-diff-level-what-works-what-doesnt-and-why-2a14} \\
G2769 & 8121 & AI Code Review Is Quietly Destroying Your DORA Metrics (Here's the Fix) - DEV Community & \url{https://dev.to/uss/ai-code-review-is-quietly-destroying-your-dora-metrics-heres-the-fix-1359} \\
G2770 & 8127 & Claude Opus 4.8 Dynamic Workflows Just Changed Agentic AI (Forget Loops) | by Joe Njenga | AI Software Engineer | May, 2026 | Medium & \url{https://medium.com/ai-software-engineer/claude-opus-4-8-dynamic-workflows-just-changed-agentic-ai-forget-loops-75f4a97db110} \\
G2771 & 8132 & Can Claude Code Replace a Junior Developer? I Tested It. & \url{https://dev.to/themachinepulse/can-claude-code-replace-a-junior-developer-i-tested-it-f5i} \\
G2772 & 8134 & kirollosatef/customgpt-claude-quadruple-verification & \url{https://github.com/kirollosatef/customgpt-claude-quadruple-verification} \\
G2773 & 8137 & AI Coding Agent Limitations: Lessons from 2-Day Experiment | Greg Pryzby posted on the topic | LinkedIn & \url{https://www.linkedin.com/posts/gpryzby_lessons-from-2-days-of-ai-assisted-development-activity-7421683924176687104-cbQm} \\
G2774 & 8138 & Vibe Coding in 2026: The Complete Guide to AI-Pair Programming That Actually Works - DEV Community & \url{https://dev.to/pockit_tools/vibe-coding-in-2026-the-complete-guide-to-ai-pair-programming-that-actually-works-42de} \\
G2775 & 8142 & HermeticOrmus/code-review-skills & \url{https://github.com/HermeticOrmus/code-review-skills} \\
G2776 & 8145 & Vibe Coding Is NOT the Problem -- Ignorance Is | by Mpholoane Bapela | Apr, 2026 | Medium & \url{https://medium.com/@mpholoane/vibe-coding-is-not-the-problem-ignorance-is-ebaedf3d2227} \\
G2777 & 8146 & Policy as Code Review | Kodus (Plain-English Rules, Per-Folder) & \url{https://kodus.io/policy-as-code-review/} \\
G2778 & 8152 & Vibe coding and CHOP: What's the difference? - LeadDev & \url{https://leaddev.com/software-quality/vibe-coding-and-chop-whats-the-difference} \\
G2779 & 8153 & daotaolaixe-quangthang/ContribAI & \url{https://github.com/daotaolaixe-quangthang/ContribAI} \\
G2780 & 8161 & The Autonomous Security Moment: When AI Pentesting Stopped Being Assisted and Became Autonomous | by Baozilla, Let's go! | Apr, 2026 | Medium & \url{https://medium.com/@lssmj2014/the-autonomous-security-moment-when-ai-pentesting-stopped-being-assisted-and-became-autonomous-89ae02b7c0fb} \\
G2781 & 8163 & The Vibe Coding Hangover: Why 90\% of AI-Generated Codebases Are About to Implode | by Teddir | May, 2026 | Medium & \url{https://medium.com/@teddir/the-vibe-coding-hangover-why-90-of-ai-generated-codebases-are-about-to-implode-68e30983760c} \\
G2782 & 8164 & The state of vibe coding in 2026: Adoption won, now what? & \url{https://hashnode.com/blog/state-of-vibe-coding-2026} \\
G2783 & 8166 & Smart Coding vs Vibe Coding: Engineering Discipline in the Age of AI - DEV Community & \url{https://dev.to/kolkov/smart-coding-vs-vibe-coding-engineering-discipline-in-the-age-of-ai-5b20} \\
G2784 & 8170 & What hath Claude wrought? - by Matt Welsh - Chaotic Good & \url{https://mdwla.substack.com/p/what-hath-claude-wrought} \\
G2785 & 8172 & Lessons from Using AI Tools in Actual Engineering Work & \url{https://dev.to/leena_malhotra/lessons-from-using-ai-tools-in-actual-engineering-work-2hpc} \\
G2786 & 8173 & Vibe Coding Is Making You a Worse Engineer. Here's the Proof. | by Kapil Khatik | May, 2026 | Medium & \url{https://medium.com/@kapildevkhatik2/vibe-coding-is-making-you-a-worse-engineer-heres-the-proof-9f295ab1d56d} \\
G2787 & 8174 & Google is rebuilding Android development for humans and agents | LinearB Blog & \url{https://linearb.io/blog/google-android-17-rebuilding-development-humans-agents} \\
G2788 & 8175 & vertti/se-cove-claude-plugin & \url{https://github.com/vertti/se-cove-claude-plugin} \\
G2789 & 8178 & Why AI-Generated Code is a Security Minefield (And What To Do About It) - DEV Community & \url{https://dev.to/ayame0328/why-ai-generated-code-is-a-security-minefield-and-what-to-do-about-it-3i1l} \\
G2790 & 8186 & Thoughts on the Impact of AI Coding Tools on Developer ... - LinkedIn & \url{https://www.linkedin.com/pulse/thoughts-impact-ai-coding-tools-developer-code-quality-mira-mezini-c8ahe} \\
G2791 & 8187 & AI Code Review in 2026: How LLMs Catch Bugs Before Production & \url{https://www.devx.com/uncategorized/ai-code-review-llms-catching-bugs-2026/} \\
G2792 & 8188 & GitHub Copilot CLI adds Rubber Duck review agent | InfoWorld & \url{https://www.infoworld.com/article/4155289/github-copilot-cli-adds-rubber-duck-review-agent.html} \\
G2793 & 8191 & LLM-as-a-Judge for Human-AI Co-Creation: A Reliability-Aware Evaluation Framework for Coding -- Pith & \url{https://pith.science/paper/2604.27727} \\
G2794 & 8196 & Monorepo AI Code Review 2026: How AI Handles Large Diffs and Cross-Package Changes | Git AutoReview & \url{https://gitautoreview.com/blog/monorepo-ai-code-review-2026} \\
G2795 & 8198 & GitLens for Windsurf - AI Git Insights \& History in VS Code+ | Free \& Pro & \url{https://www.gitkraken.com/gitlens/windsurf} \\
G2796 & 8201 & Code Review with GitHub Copilot in Visual Studio Code - DEV Community & \url{https://dev.to/nikiforovall/code-review-with-github-copilot-in-visual-studio-code-3a1o} \\
G2797 & 8203 & Anthropic debuts Code Review for teams, enterprises & \url{https://www.theregister.com/software/2026/03/09/anthropic-debuts-code-review-for-teams-enterprises/5223180} \\
G2798 & 8206 & How to Screen Junior Developers: Coding Challenges Best Practices 2025 & \url{https://6mkdl.intervue.io/blog/how-to-screen-junior-developers-with-coding-challenges} \\
G2799 & 8209 & AI Code Governance: The Framework 91\% of Engineering Teams Need Now & \url{https://www.connectory.ai/blog/ai-code-governance-framework-2026/} \\
G2800 & 8210 & The 66\% Problem - DEV Community & \url{https://dev.to/evanlausier/the-66-problem-1c3m} \\
G2801 & 8211 & AI Eliminates Backlog, Forces Strategic Conversations | Matt Watson posted on the topic | LinkedIn & \url{https://www.linkedin.com/posts/mattwatsonkc_ai-killed-the-backlog-claude-code-can-do-activity-7460672033195053056-Z99J} \\
G2802 & 8216 & Why Code Review Cannot Scale With AI Output -- Mneme HQ & \url{https://mnemehq.com/insights/why-code-review-cannot-scale-with-ai-output/} \\
G2803 & 8217 & 46\% Don't Trust AI Code: The \$250 Billion Security Crisis Nobody's Solving - JavaScript Jobs Hub & \url{https://jsgurujobs.com/blog/46-don-t-trust-ai-code-the-250-billion-security-crisis-nobody-s-solving} \\
G2804 & 8219 & Your Code Review Process Is Optimized for the Wrong Failure Mode & \url{https://tianpan.co/blog/2026-04-14-code-review-for-ai-generated-prs} \\
G2805 & 8224 & I replaced GitHub Copilot with 3 tools. My team noticed within a week. - DEV Community & \url{https://dev.to/dev_tips/i-replaced-github-copilot-with-3-tools-my-team-noticed-within-a-week-244k} \\
G2806 & 8226 & Shareuhack | Vibe Coding :AI(2026) & \url{https://www.shareuhack.com/ja/posts/vibe-coding-guide-2026} \\
G2807 & 8232 & Your AI Agent Says 'Done' -- How Do You Know It Actually Worked? - DEV Community & \url{https://dev.to/battyterm/your-ai-agent-says-done-how-do-you-know-it-actually-worked-3mfc} \\
G2808 & 8233 & pactkit/pactkit-public & \url{https://github.com/pactkit/pactkit-public} \\
G2809 & 8234 & Graphite vs Gitar: AI Code Review Tools Compared 2026 & \url{https://cms.gitar.ai/enterprise-graphite-code-automation-solutions/} \\
G2810 & 8237 & Vibe Coding in 2026: What It Is, When It Works, When It Burns You -- Swfte AI & \url{https://www.swfte.com/ko/blog/vibe-coding-guide} \\
G2811 & 8241 & Copilot code review: Better coverage and more control - GitHub Changelog & \url{https://github.blog/changelog/2025-05-28-copilot-code-review-better-coverage-and-more-control/} \\
G2812 & 8243 & The Hidden Cost of AI Coding: Technical Debt You Can't See - DEV Community & \url{https://dev.to/aws/the-hidden-cost-of-ai-coding-technical-debt-you-cant-see-1767} \\
G2813 & 8247 & Intent-Level Verification: The Audit Layer Above Your Code Review Pipeline - DEV Community & \url{https://dev.to/intentguard_ole/intent-level-verification-the-audit-layer-above-your-code-review-pipeline-3dl1} \\
G2814 & 8251 & AI is making engineers faster. So why does delivery feel slower? & \url{https://www.faros.ai/blog/ai-engineers-faster-delivery-slower} \\
G2815 & 8266 & CodeRabbit Azure DevOps: Setting Up AI Code Review - DEV Community & \url{https://dev.to/rahulxsingh/coderabbit-azure-devops-setting-up-ai-code-review-524h} \\
G2816 & 8268 & existential-birds/daydream & \url{https://github.com/existential-birds/daydream} \\
G2817 & 8269 & AI Impact on Developer Skills | CodeIntelligently & \url{https://codeintelligently.com/blog/ai-impact-developer-skills-deskilling} \\
G2818 & 8272 & Autofix (beta) | Semgrep & \url{https://semgrep.dev/docs/semgrep-code/triage-remediation/autofix} \\
G2819 & 8273 & AI Smart Contract Review: The Finding Is Not the Audit - DEV Community & \url{https://dev.to/aicryptosystems/ai-smart-contract-review-the-finding-is-not-the-audit-1lmb} \\
G2820 & 8275 & The Code Review Bottleneck Is You -- jd:/dev/blog & \url{https://julien.danjou.info/blog/the-code-review-bottleneck-is-you/} \\
G2821 & 8278 & The state of AI in CNCF projects: A first look at the data | CNCF & \url{https://www.cncf.io/blog/2026/04/29/the-state-of-ai-in-cncf-projects-a-first-look-at-the-data/} \\
G2822 & 8279 & Cursor Developer Habits Report: Agentic Coding Trends & \url{https://cursor.com/insights} \\
G2823 & 8282 & Anthropic's Claude Security emerges from closed preview to scan your codebases for vulnerabilities - The New Stack & \url{https://thenewstack.io/anthropics-claude-security-beta/} \\
G2824 & 8285 & When AI Becomes a Shortcut, Not a Partner - A Cautionary Tale | LavX News | LavX News & \url{https://news.lavx.hu/article/when-ai-becomes-a-shortcut-not-a-partner-a-cautionary-tale} \\
G2825 & 8287 & OpenAI Codex Plugin for Claude Code | TPS & \url{https://tpsreport.news/news/openai-codex-plugin-claude-code} \\
G2826 & 8288 & Static vs Semantic: how a security scanner reads AI-generated code - DEV Community & \url{https://dev.to/shecantcode/static-vs-semantic-how-a-security-scanner-reads-ai-generated-code-cj6} \\
G2827 & 8289 & How Spec-Driven Development Brings Structure to AI-Assisted Engineering and How We Put It to the Test -- WriteUpCafe & \url{https://writeupcafe.com/spec-driven-development-ai-assisted-software-engineering} \\
G2828 & 8290 & If AI Writes 10 the Code, Who's Going to Review It All? | by Pravinmenghani | May, 2026 | Medium & \url{https://medium.com/@pravinmenghani/if-ai-writes-10-the-code-whos-going-to-review-it-all-32bffd30605f} \\
G2829 & 8293 & AI's brave new world of technical debt | InfoWorld & \url{https://www.infoworld.com/article/4178964/ais-brave-new-world-of-technical-debt.html} \\
G2830 & 8295 & Cursor's AI Security Agents: What They Get Right (and What's Missing) | Snyk & \url{https://snyk.io/pt-BR/blog/cursor-security-agent-prompts/} \\
G2831 & 8297 & Experienced devs are slower with AI and they don't even know it - DEV Community & \url{https://dev.to/adioof/experienced-devs-are-slower-with-ai-and-they-dont-even-know-it-8g0} \\
G2832 & 8298 & TDE6541/awesome-claude-code-security & \url{https://github.com/TDE6541/awesome-claude-code-security} \\
G2833 & 8299 & Why Your AI Code Reviews Are Broken (And How to Fix Them) - Qodo & \url{https://www.qodo.ai/blog/why-your-ai-code-reviews-are-broken-and-how-to-fix-them/} \\
G2834 & 8300 & solitontech/tech-lead-anna & \url{https://github.com/solitontech/tech-lead-anna} \\
G2835 & 8301 & Our AI Code Review Benchmark set out to see how today's top tools really perform. We tested LinearB, CodeRabbit, GitHub Copilot, Qodo Merge, and Graphite Diamon & \url{https://www.linkedin.com/posts/linearb_our-ai-code-review-benchmark-set-out-to-see-activity-7363212580464574465-Zcqn} \\
G2836 & 8304 & CodeRabbit vs Bito: AI Code Review Comparison for 2026 - DEV Community & \url{https://dev.to/rahulxsingh/coderabbit-vs-bito-ai-code-review-comparison-for-2026-3p05} \\
G2837 & 8306 & AI Writes 80\% of Your Code. Who Reviews It? - DEV Community & \url{https://dev.to/redoh/ai-writes-80-of-your-code-who-reviews-it-bih} \\
G2838 & 8308 & Past podcast sponsors & \url{https://blog.pragmaticengineer.com/past-podcast-sponsors/} \\
G2839 & 8313 & What Reviewing AI-Generated PRs Actually Looks Like (Part 4) - DEV Community & \url{https://dev.to/mitkotschimev/what-reviewing-ai-generated-prs-actually-looks-like-part-4-4jh7} \\
G2840 & 8316 & AI Code Review: Does It Actually Help? (Data from 100 Teams) | PanDev Metrics & \url{https://pandev-metrics.com/docs/blog/ai-code-review-does-it-help} \\
G2841 & 8328 & AI Vibe Coding \& The Production Gap: The Hidden Risk in AI-Driven Development & \url{https://www.hackerstorm.com/index.php/articles/our-blog/vulnerabililty-intelligence/ai-code-security-production-gap} \\
G2842 & 8337 & If AI wrote Your Code, it may also have Written Your Vulnerabilities. | by Ben Olufemi akintounde | Apr, 2026 | Medium & \url{https://medium.com/@benakintounde/if-ai-wrote-your-code-it-may-also-have-written-your-vulnerabilities-00d0fb6ad3c2} \\
G2843 & 8338 & Vibe coding: An Analytical Deep Dive into AI-Driven Code Generation & \url{https://www.linkedin.com/pulse/vibe-coding-analytical-deep-dive-ai-driven-code-saransh-chaurasia-facif} \\
G2844 & 8340 & Reading AI Code Before Merging: The Unit of Merge Is The Bug & \url{https://mk0r.com/t/reading-ai-code-before-merging} \\
G2845 & 8344 & Introducing Semgrep Custom Workflows | Semgrep & \url{https://semgrep.dev/blog/2026/introducing-semgrep-custom-workflows/} \\
G2846 & 8350 & OpenAI Codex Tutorial: A Software Engineering Agent in the Cloud & \url{https://www.linkedin.com/pulse/openai-codex-tutorial-software-engineering-agent-cloud-guzey-6flyc} \\
G2847 & 8363 & AI Built the Code. Now Nobody Can Fix It. | Nexairi & \url{https://www.nexairi.com/article/Technology/ai-generated-code-technical-debt/} \\
G2848 & 8364 & LLM Hallucinations in AI Code Review | diffray & \url{https://diffray.ai/blog/llm-hallucinations-code-review/} \\
G2849 & 8365 & Claude Code /code-review: Catch Bugs Before You Push | Claude Code Blog & \url{https://claudcod.com/blog/claude-code-review-command/} \\
G2850 & 8367 & Claude Code /ultrareview: the Bug-Hunting Agent Fleet in the Cloud (April 2026) & \url{https://pasqualepillitteri.it/en/news/1301/claude-code-ultrareview-agents-cloud-2026} \\
G2851 & 8370 & Porting a Claude Code Command to OpenAI Codex CLI (and the desktop Codex app) | by Gareth Hallberg | Apr, 2026 | Medium & \url{https://medium.com/@gareth.hallberg_55290/porting-a-claude-code-command-to-openai-codex-cli-1ed17ab685cf} \\
G2852 & 8371 & Everyone's talking about vibe coding right now, but there's an important part that often gets left out. AI can generate code, but that doesn't mean it understan & \url{https://www.linkedin.com/posts/upleveled-io_everyones-talking-about-vibe-coding-right-activity-7447196905283088384-jfl3} \\
G2853 & 8372 & The intelligence paradox: Why Claude Opus 4.6 requires verification | Sonar & \url{https://www.sonarsource.com/blog/why-claude-opus-4-6-requires-verification} \\
G2854 & 8374 & The Invisible Crisis Underneath Every App You Use | by Hembitec | May, 2026 | Medium & \url{https://medium.com/@hembitec/the-invisible-crisis-underneath-every-app-you-use-706ca825e33f} \\
G2855 & 8376 & CTO Newsletter | April 2026 | Thoughtworks Thailand & \url{https://www.thoughtworks.com/en-th/insights/cto-newsletter/edition1} \\
G2856 & 8378 & Why AI-Generated Code Needs the Same Review Process as Human Code - DEV Community & \url{https://dev.to/mehdi_chaouachi_c114771f0/why-ai-generated-code-needs-the-same-review-process-as-human-code-5965} \\
G2857 & 8381 & The Cognitive Offloading Trap: When Your Team Can't Work Without the AI & \url{https://tianpan.co/blog/2026-04-16-cognitive-offloading-trap-ai-team-dependency} \\
G2858 & 8382 & Open Source Collaboration Report: What 250+ Projects Reveal - DEV Community & \url{https://dev.to/pullflow/open-source-collaboration-report-what-250-projects-reveal-47o9} \\
G2859 & 8384 & Pillar 2: AI Governance \& Risk Control | AEEF Standards & \url{https://aeef.ai/pillars/pillar-2-governance-risk/} \\
G2860 & 8390 & Plan First, Ship Faster: How CodeRabbit Built Agent Orchestration on Claude | Webinars \textbackslash\{\} Anthropic & \url{https://www.anthropic.com/webinars/how-coderabbit-orchestrates-agents-to-strengthen-ai-generated-code} \\
G2861 & 8395 & The Noise Problem: How AI Code Review Tools Slow Teams Down | PRBoard PRBoard & \url{https://prboard.io/blog/ai-code-review-noise-problem-slowing-teams} \\
G2862 & 8396 & AI's Impact on Code Quality: Prioritizing Mindset over Expertise | Bryan Finster posted on the topic | LinkedIn & \url{https://www.linkedin.com/posts/bryan-finster_one-of-my-takeaways-from-dave-farleys-video-activity-7422996873474523137-b0DR} \\
G2863 & 8397 & Your AI Coding Workflow is Killing Your Productivity. You Just Moved the Bottleneck. & \url{https://tylerfolkman.substack.com/p/i-built-a-one-person-software-factory} \\
G2864 & 8402 & Hands-On Evaluation of NVIDIA Nemotron 3 Super: Punching Above Its Weight Class | Greptile Blog & \url{https://www.greptile.com/blog/nvidia-nemotron-super-in-code-review} \\
G2865 & 8405 & Vibe Coding: The Hype, The Reality, and Why It Actually Matters | by Krunal kamble | Mar, 2026 | Medium & \url{https://medium.com/@krunalkamble215/vibe-coding-the-hype-the-reality-and-why-it-actually-matters-052de28a52e2} \\
G2866 & 8410 & AI Can Write Code -- But Who Verifies It? -- Vibers Blog & \url{https://onout.org/vibers/blog/ai-writes-code-who-verifies/} \\
G2867 & 8411 & \#257 - The Future of Code Review: Stop Reviewing Line-by-Line ... & \url{https://techleadjournal.dev/episodes/257/} \\
G2868 & 8419 & Build Your Own AI Code Review Bot with Claude API and GitHub Actions v2 - DEV Community & \url{https://dev.to/ramosai/build-your-own-ai-code-review-bot-with-claude-api-and-github-actions-v2-b85} \\
G2869 & 8424 & Open Source Maintainers Are Drowning in AI-Generated Pull Requests. Enterprise Teams Are Next. | by Signadot | May, 2026 | Medium & \url{https://medium.com/@signadot/open-source-maintainers-are-drowning-in-ai-generated-pull-requests-enterprise-teams-are-next-a598b61b5fbc} \\
G2870 & 8426 & How I Added AI Code Reviews to Our GitLab Pipeline Using PR-Agent | by Ambrish Vadnerkar | Apr, 2026 | Medium & \url{https://medium.com/@ambrish.vadnerkar/how-i-added-ai-code-reviews-to-our-gitlab-pipeline-using-pr-agent-fa0e132e98cf} \\
G2871 & 8428 & dawit-Tegegnwork/ai-code-reviewer & \url{https://github.com/dawit-Tegegnwork/ai-code-reviewer} \\
G2872 & 8431 & How to Review Code Your AI Agent Wrote While You Were Sleeping - DEV Community & \url{https://dev.to/toniantunovic/how-to-review-code-your-ai-agent-wrote-while-you-were-sleeping-2fh6} \\
G2873 & 8434 & AI Fatigue for Developers: Managing Cognitive Overload from Code Assistants - DEV Community & \url{https://dev.to/kodus/ai-fatigue-for-developers-managing-cognitive-overload-from-code-assistants-2cij} \\
G2874 & 8437 & How I Review AI-Generated Pull Requests (A Step-by-Step Checklist) | AI Navigate & \url{https://ai-navigate-news.com/en/articles/c3ea2306-7f18-44d2-926e-b2b7c21b9583} \\
G2875 & 8439 & How to Kill the Code Review - by Ankit Jain - Latent.Space & \url{https://www.latent.space/p/reviews-dead} \\
G2876 & 8443 & I Built a Tool to Validate AI-Generated Code (Now with JS/TS Support + AI Auto-Fix) - DEV Community & \url{https://dev.to/rudra_sarker/i-built-a-tool-to-validate-ai-generated-code-now-with-jsts-support-ai-auto-fix-20b4} \\
G2877 & 8444 & GitHub Copilot code review for Azure Repos is now in technical preview - GitHub Changelog & \url{https://github.blog/changelog/2026-06-02-github-copilot-code-review-for-azure-repos-is-now-in-technical-preview/} \\
G2878 & 8445 & Frontend with AI: workflow first, agent second - DEV Community & \url{https://dev.to/kevinccbsg/frontend-with-ai-workflow-first-agent-second-3hmo} \\
G2879 & 8447 & AI-Assisted Coding: Why a Distinguished Engineer Stopped Reading Code -- Lights-Out Codebases and the End of the IC - Scrum Master Toolbox Podcast & \url{https://scrum-master-toolbox.org/2026/04/podcast/bonus-why-a-distinguished-engineer-stopped-reading-code-lights-out-codebases-and-the-end-of-the-ic/} \\
G2880 & 8449 & Contextual Code Review - ai - DEV Community & \url{https://dev.to/synthaicode_commander/contextual-code-review-2k8b} \\
G2881 & 8453 & The Senior-Engineer Paradox: Why Faster Code Made the Job Harder - DEV Community & \url{https://dev.to/ruturaj_bhokre/the-senior-engineer-paradox-why-faster-code-made-the-job-harder-41pa} \\
G2882 & 8459 & Let Humans Write. Let AI Critique -- A Manifesto for Security Engineers - DEV Community & \url{https://dev.to/fosres/let-humans-write-let-ai-critique-a-manifesto-for-security-engineers-jg8} \\
G2883 & 8463 & Claude Code's `ultrareview`: The Long Review (April 2026) | Pondero & \url{https://pondero.ai/coding/guides/claude-code-ultrareview-long-review-april-2026/} \\
G2884 & 8466 & OpenAI says "there are easily 1,000x engineers now" - LeadDev & \url{https://leaddev.com/ai/openai-says-there-are-easily-1000x-engineers-now} \\
G2885 & 8469 & CTO Newsletter | April 2026 | Thoughtworks Germany & \url{https://www.thoughtworks.com/en-de/insights/cto-newsletter/edition1} \\
G2886 & 8470 & Boosting Code Review with Claude Code Review Tool | Hasnain Lakhani posted on the topic | LinkedIn & \url{https://www.linkedin.com/posts/mhlakhani_claude-codes-code-review-is-out-0-i-activity-7436959575880232960-B9L8} \\
G2887 & 8474 & Is Your Static Analysis Tool a Bottleneck, Not a Safeguard? | by Maria Vallecio | Feb, 2026 | Medium & \url{https://medium.com/@jay_1980/is-your-static-analysis-tool-a-bottleneck-not-a-safeguard-b829a3ec2688} \\
G2888 & 8477 & I Tested 4 AI Code Review Tools on Real Pull Requests & \url{https://dev.to/jim_l_efc70c3a738e9f4baa7/i-tested-4-ai-code-review-tools-on-real-pull-requests-heres-what-they-actually-caught-4h3d} \\
G2889 & 8480 & The Future of AI Agents in Code Reviews: Your Next PR Reviewer Might Not Be Human & \url{https://codingwithroby.substack.com/p/the-future-of-ai-agents-in-code-reviews} \\
G2890 & 8482 & The Vibe Coding Delusion - DEV Community & \url{https://dev.to/ejb503/the-vibe-coding-delusion-3897} \\
G2891 & 8484 & CodeRabbit Adds Support for VS Code to AI Code Review Platform - DevOps.com & \url{https://devops.com/coderabbit-adds-support-for-vs-code-to-ai-code-review-platform/} \\
G2892 & 8497 & 60 million Copilot code reviews and counting - The GitHub Blog & \url{https://github.blog/ai-and-ml/github-copilot/60-million-copilot-code-reviews-and-counting/} \\
G2893 & 8498 & Code Review in AI-Generated Code: A Challenge | Alex Carter posted on the topic | LinkedIn & \url{https://www.linkedin.com/posts/ac2522_extreme-harness-engineering-1m-loc-1b-toks-activity-7459277896977825792-D6hO} \\
G2894 & 8500 & AI Pair Programming: A Complete Guide for Development Teams - DEV Community & \url{https://dev.to/techsifted/ai-pair-programming-a-complete-guide-for-development-teams-c21} \\
G2895 & 8502 & Devin's 2025 Performance Review: Learnings From 18 Months of ... & \url{https://cognition.ai/blog/devin-annual-performance-review-2025} \\
G2896 & 8505 & Beyond prompting: How KubeStellar reached 81\% PR acceptance with AI agents - The New Stack & \url{https://thenewstack.io/ai-codebase-maturity-model/} \\
G2897 & 8506 & anilnishad19799/AI-Powered-Code-Review-Agent-System & \url{https://github.com/anilnishad19799/AI-Powered-Code-Review-Agent-System} \\
G2898 & 8508 & The New Bottleneck - When AI Writes Code Faster Than Humans Can Review It - DEV Community & \url{https://dev.to/sag1v/the-new-bottleneck-when-ai-writes-code-faster-than-humans-can-review-it-mp0} \\
G2899 & 8509 & AI Code Generation: Trust and Verify, Always - The New Stack & \url{https://thenewstack.io/ai-code-generation-trust-and-verify-always/} \\
G2900 & 8513 & AI-assisted development risks bad code shipped too fast | Diego Jancic posted on the topic | LinkedIn & \url{https://www.linkedin.com/posts/diegojancic_ai-speed-demands-more-engineering-discipline-activity-7459697608308080640-R1Y5} \\
G2901 & 8515 & Changing How I Structure PRs Made My AI Significantly Smarter & \url{https://zenn.dev/smartvain/articles/ai-pr-split-strategy-not-review?locale=en} \\
G2902 & 8517 & AI Coding Tip 021 - Avoid Comprehension Debt & \url{https://maxicontieri.substack.com/p/ai-coding-tip-021-avoid-comprehension} \\
G2903 & 8522 & prlens/prlens & \url{https://github.com/prlens/prlens} \\
G2904 & 8530 & Why We Don't Let AI Ship Code Unsupervised | Talk Think Do & \url{https://talkthinkdo.com/blog/why-we-dont-let-ai-ship-code-unsupervised/} \\
G2905 & 8531 & Independent-AI-Labs/AMI-CI & \url{https://github.com/Independent-AI-Labs/AMI-CI} \\
G2906 & 8532 & The Vibe Coding Hangover: Why I'm Returning to Engineering Rigor in 2026 - DEV Community & \url{https://dev.to/maximiliano_allende97/the-vibe-coding-hangover-why-im-returning-to-engineering-rigor-in-2026-49hl} \\
G2907 & 8534 & Make Claude Code disagree with you: a 14-rule counterpart toolkit (install in 1 command) - DEV Community & \url{https://dev.to/michelfaure/make-claude-code-disagree-with-you-a-14-rule-counterpart-toolkit-install-in-1-command-11pe} \\
G2908 & 8540 & AI Code Review Tools for Engineering Teams (2026) | Nimbalyst & \url{https://nimbalyst.com/blog/ai-code-review-tools-for-engineering-teams-2026/} \\
G2909 & 8544 & Git identity spoof fools Claude into giving bad code the nod * The Register & \url{https://www.theregister.com/2026/04/16/git_identity_spoof_claude/?td=keepreading} \\
G2910 & 8547 & Why Your AI-Generated Code Needs a Security Review Before Launch | Lorikeet Security & \url{https://lorikeetsecurity.com/blog/ai-generated-code-security} \\
G2911 & 8550 & 10 Best AI Code Review Tools and How They Work - SitePoint & \url{https://www.sitepoint.com/best-ai-code-review-tools-for-developers/} \\
G2912 & 8553 & GitHub - codexstar69/bug-hunter at 2a21690b1a34c3b19487e1f2c0d19b6271a3fae7 GitHub & \url{https://github.com/codexstar69/bug-hunter/tree/2a21690b1a34c3b19487e1f2c0d19b6271a3fae7} \\
G2913 & 8554 & How I Use AI to Cut My Code Review Prep Time in Half (Step-by-Step) - DEV Community & \url{https://dev.to/leveragenotes/how-i-use-ai-to-cut-my-code-review-prep-time-in-half-step-by-step-2afn} \\
G2914 & 8561 & Meta's new structured prompting technique makes LLMs significantly better at code review -- boosting accuracy to 93\% in some cases | VentureBeat & \url{https://venturebeat.com/orchestration/metas-new-structured-prompting-technique-makes-llms-significantly-better-at} \\
G2915 & 8565 & AI Code Review False Positive Rates: Claude vs GPT-5 Model Comparison (2026 Data) | BSWEN & \url{https://docs.bswen.com/blog/2026-03-05-ai-code-review-false-positives/} \\
G2916 & 8567 & How AI Is Changing Code Reviews -- and What Engineering Leaders Need to Rethink | by Sindhuja Venkatesh | Medium & \url{https://medium.com/@onvsindhu/how-ai-is-changing-code-reviews-and-what-engineering-leaders-need-to-rethink-3f5981e77a4a} \\
G2917 & 8569 & The Hidden Cost of Slow Code Reviews: Data from 8 Million PRs - DEV Community & \url{https://dev.to/vitalii_petrenko_dev/the-hidden-cost-of-slow-code-reviews-data-from-8-million-prs-5fei} \\
G2918 & 8576 & Devin AI Review: Can AI Really Replace Software Engineers? & \url{https://aitoolinsight.com/devin-ai/} \\
G2919 & 8584 & Code Review Shifts from Syntax to Decision-Making with LLMs | Fahad Siddiqui posted on the topic | LinkedIn & \url{https://www.linkedin.com/posts/fsdqui_llms-just-changed-what-code-review-actually-activity-7412921348110589952-atm0} \\
G2920 & 8590 & LGTM 2.0: Implementing Zero-Noise AI Code Review Agents - DEV Community & \url{https://dev.to/anercohen/lgtm-20-implementing-zero-noise-ai-code-review-agents-2j3f} \\
G2921 & 8596 & What Do Developers Think About Code Readability in the Age of ... & \url{https://www.atlassian.com/blog/development/atlassian-research-developers-on-code-readibility-llm} \\
G2922 & 8601 & GitHub - tang-vu/ContribAI at v6.4.0 GitHub & \url{https://github.com/tang-vu/ContribAI/tree/v6.4.0} \\
G2923 & 8602 & What's New in CodeRabbit (May-July 2025) - DEV Community & \url{https://dev.to/coderabbitai/whats-new-in-coderabbit-may-july-2025-cbp} \\
G2924 & 8604 & The Codex Review Gate: How We Made AI Agents Review Each Other's Work | by Werner Kasselman | Medium & \url{https://medium.com/@wernerk/the-codex-review-gate-how-we-made-ai-agents-review-each-others-work-59e9ff5465f9} \\
G2925 & 8606 & Managing AI coding assistants without losing control - LeadDev & \url{https://leaddev.com/ai/managing-ai-coding-assistants-without-losing-control} \\
G2926 & 8607 & AI-Generated Code in 2025: Balancing Innovation with Security Risks, Technical Debt, and Ethical Challenges | by Irvan Gerhana | Medium & \url{https://medium.com/@weexdayend/ai-generated-code-in-2025-balancing-innovation-with-security-risks-technical-debt-and-ethical-885c258bd99e} \\
G2927 & 8611 & Treat AI-Generated code as a draft - by Addy Osmani & \url{https://addyo.substack.com/p/treat-ai-generated-code-as-a-draft} \\
G2928 & 8613 & Rethinking Code Review in the Era of AI - SD Times & \url{https://sdtimes.com/ai/rethinking-code-review-in-the-era-of-ai/} \\
G2929 & 8614 & What Is Vibe Coding \& How to Implement It? [2026 Guide] & \url{https://dextralabs.com/blog/what-is-vibe-coding/} \\
G2930 & 8615 & AI Code Review in Practice: What Automated PR Analysis Actually ... & \url{https://tianpan.co/blog/2026-04-20-ai-code-review-what-it-catches-misses} \\
G2931 & 8616 & The Georgia Tech CVE Data Shows AI Code Tools Have a Volume Problem - DEV Community & \url{https://dev.to/toniantunovic/the-georgia-tech-cve-data-shows-ai-code-tools-have-a-volume-problem-28jh} \\
G2932 & 8617 & Paper page - Human-AI Synergy in Agentic Code Review & \url{https://huggingface.co/papers/2603.15911} \\
G2933 & 8619 & Google Antigravity 2.0 - From Zero to Production Multi-Agent Systems -- The Definitive Developer Guide | by Anup Karanjkar | Bootcamp | May, 2026 | Medium & \url{https://medium.com/design-bootcamp/google-antigravity-2-0-58620ca0f96c} \\
G2934 & 8620 & Human-Led, AI-Accelerated: Why the Winning Stack in 2026 Isn't Fully Autonomous -- SuperML.dev & \url{https://superml.dev/human-led-ai-accelerated-winning-stack-april-2026} \\
G2935 & 8622 & From Implementers to Solution Architects | by Agoda Engineering | Agoda Engineering \& Design | Medium & \url{https://medium.com/agoda-engineering/from-implementers-to-solution-architects-250a95087d75} \\
G2936 & 8626 & The current coding workflow that I use and super happy about it Managed to ship 3 tools into production with it In the past couple of weeks I started using the  & \url{https://www.linkedin.com/posts/alfred-simon_the-current-coding-workflow-that-i-use-and-activity-7464933975447494656-S7q0} \\
G2937 & 8630 & WTF is Autonomous Coding Assistants? - DEV Community & \url{https://dev.to/dailybugle33/wtf-is-autonomous-coding-assistants-p6h} \\
G2938 & 8631 & Codex on Phone: Secure Remote Code Review and Approval | Shelly Palmer posted on the topic | LinkedIn & \url{https://www.linkedin.com/posts/shellypalmer_openai-put-codex-on-your-phone-yesterday-activity-7461039512383639552-EEQc} \\
G2939 & 8634 & Graphify + code-review-graph: Build a Self-Updating Knowledge Graph for Claude Code and other AI Coding Agent - DEV Community & \url{https://dev.to/mir_mursalin_ankur/graphify-code-review-graph-build-a-self-updating-knowledge-graph-for-claude-code-and-other-ai-j1m} \\
G2940 & 8637 & The Real ROI of Vibe Coding: You're Measuring the Wrong Thing & \url{https://www.linkedin.com/pulse/real-roi-vibe-coding-youre-measuring-wrong-thing-davide-piumetti-s6gje} \\
G2941 & 8638 & 5 AI-Powered GitHub Code Review Tools [Feb 2026] & \url{https://www.qodo.ai/blog/github-ai-code-review/} \\
G2942 & 8642 & Inside-the-Loop vs. Outside-the-Loop: Evaluating Agent Architectures - DEV Community & \url{https://dev.to/sahil_kat/inside-the-loop-vs-outside-the-loop-evaluating-agent-architectures-5e58} \\
G2943 & 8643 & GitHub Copilot vs Cursor vs Continue: Best AI Code Assistant 2025 - DEV Community & \url{https://dev.to/_d7eb1c1703182e3ce1782/github-copilot-vs-cursor-vs-continue-best-ai-code-assistant-2025-41l} \\
G2944 & 8644 & Claude Opus 4.8 Doesn't Have the Biggest Benchmarks This Year. It Might Have the Most Important Change. | by Naveen Pandey | ILLUMINATION | May, 2026 | Medium & \url{https://medium.com/illumination/claude-opus-4-8-4ff2a565f3e5} \\
G2945 & 8647 & Automating Code Review with Gemini CLI: Evidence-Based Patterns and Real Tradeoffs - Gemini CLI All in One & \url{https://geminicli.one/blog/gemini-cli-code-review} \\
G2946 & 8653 & GPT-5.1 for code-related tasks: Higher signal at lower volume - DEV Community & \url{https://dev.to/coderabbitai/gpt-51-for-code-related-tasks-higher-signal-at-lower-volume-2hcd} \\
G2947 & 8654 & Anthropic Code Review for Claude Code: Multi-Agent PR Reviews, Pricing, Setup, and Limits - DEV Community & \url{https://dev.to/umesh_malik/anthropic-code-review-for-claude-code-multi-agent-pr-reviews-pricing-setup-and-limits-3o35} \\
G2948 & 8656 & KoreaNirsa/spec-guard & \url{https://github.com/KoreaNirsa/spec-guard} \\
G2949 & 8660 & GitHub Copilot Review 2026 -- AI Code Assistant & \url{https://aiagentsquare.com/agents/github-copilot} \\
G2950 & 8661 & Version Control Platforms 2026: Workflow Comparison & \url{https://www.gitkraken.com/blog/version-control-platforms-2026-workflow-comparison} \\
G2951 & 8664 & praneybehl/code-review-mcp & \url{https://github.com/praneybehl/claude-code-review-mcp} \\
G2952 & 8665 & GitHub launches Copilot desktop app for agentic work & \url{https://cmotech.news/story/github-launches-copilot-desktop-app-for-agentic-work} \\
G2953 & 8668 & GitHub ponders kill switch for pull requests to stop AI slop & \url{https://longbridge.com/en/news/274845219} \\
G2954 & 8669 & HITL Human-AI Collaboration: Why AI Code Generation Still Needs Human Oversight - DEV Community & \url{https://dev.to/sopaco/hitl-human-ai-collaboration-why-ai-code-generation-still-needs-human-oversight-13n0} \\
G2955 & 8670 & AI-Generated Code Compliance: Bridging the Gap in Change ... & \url{https://www.linkedin.com/posts/timols_continuing-on-the-compliance-train-over-activity-7439353551522267136-jJRI} \\
G2956 & 8671 & AI Reviews the Code That AI Wrote - by Joey Hipolito & \url{https://joeyhipolito.substack.com/p/ai-reviews-the-code-that-ai-wrote} \\
G2957 & 8673 & Built a duplicate detector for AI coding assistant Claude | James Arnott posted on the topic | LinkedIn & \url{https://www.linkedin.com/posts/acorn221_claude-keeps-duplicating-code-and-i-finally-activity-7389960591664234496-vuzO} \\
G2958 & 8674 & The Professional Standard: AI and Code Accountability - Cleverbit Software & \url{https://cleverbit.software/blog/ai-accountability-in-software-development/} \\
G2959 & 8676 & When AI writes the code, who reviews it? | Google Cloud Blog & \url{https://cloud.google.com/transform/when-ai-writes-the-code-who-reviews-it-cto-google-cloud} \\
G2960 & 8677 & AI Disruption to Open Source Software (OSS) | by LiveWyer | Medium & \url{https://medium.com/@livewyer/ai-disruption-to-open-source-software-oss-377f10be2d8a} \\
G2961 & 8678 & The Copy-Paste Contagion: How AI-Assisted Development Spreads Architectural Anti-Patterns & \url{https://tianpan.co/blog/2026-05-04-copy-paste-contagion-ai-assisted-development} \\
G2962 & 8679 & How to Automate Code Reviews with AI GitHub Actions - DEV Community & \url{https://dev.to/alex_avatrixstudio/how-to-automate-code-reviews-with-ai-github-actions-5dc3} \\
G2963 & 8680 & AI still needs humans | InfoWorld & \url{https://www.infoworld.com/article/4176534/ai-still-needs-humans.html} \\
G2964 & 8682 & Agent Coding Requires Mental Model Discipline | Max Rozenoer posted on the topic | LinkedIn & \url{https://www.linkedin.com/posts/maxrozenoer_the-result-of-writing-code-is-not-code-its-activity-7427050084954992640-XXKc} \\
G2965 & 8683 & AI-Generated Code in Production: Validation, Accountability, and What Still Needs to Be Built & \url{https://re-cinq.com/blog/ai-generated-code-validation-accountability} \\
G2966 & 8686 & Junior developers aren't obsolete: Here's how to thrive in the age of AI & \url{https://github.blog/ai-and-ml/generative-ai/junior-developers-arent-obsolete-heres-how-to-thrive-in-the-age-of-ai/} \\
G2967 & 8690 & Warning: This product includes code written by AI & \url{https://www.linkedin.com/pulse/warning-product-includes-code-written-ai-samuel-akopyan-3ja2f} \\
G2968 & 8699 & AI Code Review for Data Pipelines: Using GPT to Catch ... - Medium & \url{https://medium.com/@manik.ruet08/ai-code-review-for-data-pipelines-using-gpt-to-catch-performance-and-cost-issues-1546010196c3} \\
G2969 & 8700 & Cursor 3 Is Not an IDE Update. It's a Bet That You'll Manage Agents, Not Write Code. | by Han HELOIR YAN, Ph.D. | Apr, 2026 | Medium & \url{https://medium.com/@han.heloir/cursor-3-is-not-an-ide-update-its-a-bet-that-you-ll-manage-agents-not-write-code-0d2bc51f0dcb} \\
G2970 & 8701 & zxcloli666/AI-Code-Review & \url{https://github.com/zxcloli666/AI-Code-Review} \\
G2971 & 8706 & Acceleration Whiplash: AI made you 19\% slower and 3x more likely to break production -- here's the engineering that fixes it | by Roan Brasil Monteiro | Jun, 20 & \url{https://medium.com/@roanmonteiro/acceleration-whiplash-ai-made-you-19-slower-and-3x-more-likely-to-break-production-heres-the-454edccc8520} \\
G2972 & 8708 & The Complete Guide to Amazon Q Developer: Your AI-Powered Coding Assistant - DEV Community & \url{https://dev.to/aws-builders/the-complete-guide-to-amazon-q-developer-your-ai-powered-coding-assistant-58ep} \\
G2973 & 8711 & Cursor Agent Mode Best Practices for Real Codebases: How to Avoid Rewrites and Keep Large Repos Safe - chatgptguide.ai & \url{https://chatgptguide.ai/cursor-agent-mode-best-practices-real-codebases/} \\
G2974 & 8712 & When AI Becomes a Crutch Instead of a Catalyst: A Note to Developers and Engineering Leaders & \url{https://www.linkedin.com/pulse/when-ai-becomes-crutch-instead-catalyst-note-developers-verma-bljgc} \\
G2975 & 8714 & SARGVISION: AI Code Reviewer for GitHub PRs and Code | Abhishek Gupta (Sargupta) posted on the topic | LinkedIn & \url{https://www.linkedin.com/posts/sargupta_ai-codereview-gemini-activity-7333731847639560193-vwPA} \\
G2976 & 8716 & I Built an AI Agent That Triages My GitHub Issues in 2 Seconds. Here's How. - DEV Community & \url{https://dev.to/dewaldt_huysamen_3fc81a08/i-built-an-ai-agent-that-triages-my-github-issues-in-2-seconds-heres-how-2lb} \\
G2977 & 8718 & AI Code Review: What to Look For in the Age of Copilots - DEV Community & \url{https://dev.to/rakbro/ai-code-review-what-to-look-for-in-the-age-of-copilots-2g02} \\
G2978 & 8721 & Developers Spend 11.4 Hours/Week Reviewing AI Code | byteiota & \url{https://byteiota.com/developers-spend-11-4-hours-week-reviewing-ai-code/} \\
G2979 & 8722 & The AI Flood is Breaking OSS-- Maintainers are Hitting the Limit | by Dimitris Kargatzis | Coffee And Code | Medium & \url{https://medium.com/techtrends-digest/the-ai-flood-is-breaking-oss-maintainers-are-hitting-the-limit-30c41247db5a} \\
G2980 & 8730 & Scaling code reviews with an Open Source AI Skill - DEV Community & \url{https://dev.to/yoriiis/scaling-code-reviews-with-an-open-source-ai-skill-2o8} \\
G2981 & 8731 & GitHub Copilot CLI as a PR-triage co-pilot: how I keep up with 40+ upstream orgs - DEV Community & \url{https://dev.to/mukundakatta/github-copilot-cli-as-a-pr-triage-co-pilot-how-i-keep-up-with-40-upstream-orgs-525f} \\
G2982 & 8732 & Codex Plugin for Claude Code by OpenAI & \url{https://www.aibars.net/en/library/open-source-ai/details/827928040725155840} \\
G2983 & 8742 & caveman + brooks-lint + logic-lens: The Complete AI Code Review Stack for Claude - DEV Community & \url{https://dev.to/hyhmrright/caveman-brooks-lint-logic-lens-the-complete-ai-code-review-stack-for-claude-1e3e} \\
G2984 & 8743 & AI Coding Tip 021 - Avoid Comprehension Debt - DEV Community & \url{https://dev.to/mcsee/ai-coding-tip-021-avoid-comprehension-debt-edm} \\
G2985 & 8745 & What the AI Peer Review Study Reveals About Context Loss and Governance - DEV Community & \url{https://dev.to/mnemehq/what-the-ai-peer-review-study-reveals-about-context-loss-and-governance-1gf7} \\
G2986 & 8746 & PR Review Time Benchmark 2026: How Long Code Reviews Actually Take | Git AutoReview & \url{https://gitautoreview.com/blog/pr-review-time-benchmark-2026} \\
G2987 & 8749 & AI Code Review 2026: Ship Faster with Automated PR Reviews - Pooya Golchian & \url{https://pooyagolchian.com/blog/ai-code-review-scale-ship-faster-2026/} \\
G2988 & 8751 & Jayansh21/CodeSheriff & \url{https://github.com/jayansh21/CodeSheriff} \\
G2989 & 8753 & Noah Gift's Post & \url{https://www.linkedin.com/posts/noahgift_i-have-been-code-reviewing-claude-for-a-activity-7289660049004613632-Ws0e} \\
G2990 & 8755 & I Built an AI That Reviews Every PR for Security Bugs -- Here's How (2026) - DEV Community & \url{https://dev.to/young_gao/i-built-an-ai-that-reviews-every-pr-for-security-bugs-heres-how-2026-1nhc} \\
G2991 & 8757 & Overcoming AI-generated Merge Request Burden with Greg Fletcher's Solution | Ronan Bradley posted on the topic | LinkedIn & \url{https://www.linkedin.com/posts/ronan-bradley-bb2a292b_have-you-found-yourself-lately-being-sent-activity-7437901285540003840-NUGu} \\
G2992 & 8762 & 0xshitcode/no-autopilot & \url{https://github.com/0xshitcode/no-autopilot} \\
G2993 & 8763 & Simulating 5 Years of Autonomous Coding Agents | by Frank Goortani & \url{https://medium.com/@FrankGoortani/from-coders-to-cognitive-architects-simulating-5-years-of-autonomous-coding-agents-fb34d4f4e633} \\
G2994 & 8765 & AI-generated code passes far more automated tests than human - LeadDev & \url{https://leaddev.com/software-quality/ai-generated-code-passes-far-more-automated-tests-than-human} \\
G2995 & 8767 & GitHub Copilot just crushed every AI review startup (40.3M PR analysis) - DEV Community & \url{https://dev.to/zak_mandhro/github-copilot-crushed-every-code-review-startup-40m-pr-analysis-2no6} \\
G2996 & 8768 & The Atrophy Problem | Jono Herrington | The Builder's Leader & \url{https://www.jonoherrington.com/blog/atrophy-problem} \\
G2997 & 8773 & How to Review AI Generated Code Like a Pro (Without Missing Critical Bugs) - Coder Legion & \url{https://coderlegion.com/13432/how-to-review-ai-generated-code-like-a-pro-without-missing-critical-bugs} \\
G2998 & 8777 & AI Trust in Action: How Snyk Agent Redefines Secure Development | Snyk & \url{https://snyk.io/de/blog/ai-trust-in-action-how-snyk-agent-redefines-secure-development/} \\
G2999 & 8779 & Code Review Bottleneck Solved with Claude Code | Bharat Khatri posted on the topic | LinkedIn & \url{https://www.linkedin.com/posts/bkhatri_claudecode-codex-cursor-activity-7436987957011599360-5TQ1} \\
G3000 & 8780 & Vibe Coding: Beyond the Buzzword - Your Next Production Powerhouse - DEV Community & \url{https://dev.to/laxman_fe1f8070f1612/vibe-coding-beyond-the-buzzword-your-next-production-powerhouse-4l2b} \\
G3001 & 8789 & The "Uncanny Valley" of AI Code: Why Hybrid PRs Are Harder to Review Than Pure AI or Pure Human Code - DEV Community & \url{https://gosip.celebritynews.workers.dev/jakkie_koekemoer/the-uncanny-valley-of-ai-code-why-hybrid-prs-are-harder-to-review-than-pure-ai-or-pure-human-code-1poh} \\
G3002 & 8794 & GPT-5.3-Codex: OpenAI's Autonomous Coding Agent Redefines ... & \url{https://dev.to/pooyagolchian/gpt-53-codex-openais-autonomous-coding-agent-redefines-software-engineering-2foc} \\
G3003 & 8796 & Open Source Maintainers Demand Ability to Block Copilot-Gene... & \url{https://socket.dev/blog/oss-maintainers-demand-ability-to-block-copilot-generated-issues-and-prs} \\
G3004 & 8798 & Death by a Thousand AI Pull Requests - Open Source Ready & \url{https://opensourceready.substack.com/p/death-by-a-thousand-ai-pull-requests} \\
G3005 & 8801 & How to Review AI-Generated Code: A Practical Checklist | Codably & \url{https://codably.dev/ai-tools/how-to-review-ai-generated-code} \\
G3006 & 8803 & Stop Using AI Just for Code Completion -- Here's a Workflow That Covers Your Entire SDLC - DEV Community & \url{https://dev.to/anderson_leite/stop-using-ai-just-for-code-completion-heres-a-workflow-that-covers-your-entire-sdlc-320b} \\
G3007 & 8806 & Agentic AI Coding: Best Practice Patterns for Speed with Quality & \url{https://codescene.com/blog/agentic-ai-coding-best-practice-patterns-for-speed-with-quality} \\
G3008 & 8808 & How AI agents are transforming software development | Aaron Levie posted on the topic | LinkedIn & \url{https://www.linkedin.com/posts/boxaaron_its-clear-that-almost-everything-about-software-activity-7376638892189548544-kiCI} \\
G3009 & 8811 & AI Is Writing Our Code Faster Than We Can Verify It - Global Intelligence and Insight Platform: IT Innovation, ETF Investment, plus Health Wellbeing & \url{https://genesis-aka.net/information-technology/professional/2026/04/21/ai-is-writing-our-code-faster-than-we-can-verify-it/} \\
G3010 & 8820 & Repeated critical UI/code integrity issues caused by Cursor Agent - Feedback - Cursor - Community Forum & \url{https://forum.cursor.com/t/repeated-critical-ui-code-integrity-issues-caused-by-cursor-agent/158903} \\
G3011 & 8823 & Effective Practices for Coding with a Chat-Based AI - InfoQ & \url{https://www.infoq.com/articles/effective-practices-ai-chat-based-coding/} \\
G3012 & 8830 & CodeRabbit vs Code Reviews in Kilo: Which One Is Best For You in 2026 | HackerNoon & \url{https://hackernoon.com/coderabbit-vs-code-reviews-in-kilo-which-one-is-best-for-you-in-2026} \\
G3013 & 8832 & I ship software with 13 AI agents. Here's what that actually looks like - DEV Community & \url{https://dev.to/nmelo/i-ship-software-with-13-ai-agents-heres-what-that-actually-looks-like-420c} \\
G3014 & 8834 & AI Agents vs Code Vulnerabilities: Was Anthropic Mythos a Big Deal or Fear-mongering? - DEV Community & \url{https://dev.to/maximsaplin/ai-agents-vs-code-vulnerabilities-was-anthropic-mythos-a-big-deal-or-fear-mongering-8ci} \\
G3015 & 8839 & Jules: Google's Async AI Coding Agent Actually Changes How I Think About Code Reviews - DEV Community & \url{https://dev.to/pulkitgovrani/jules-googles-async-ai-coding-agent-actually-changes-how-i-think-about-code-reviews-5288} \\
G3016 & 8842 & Offshore Software Development in the AI Era: What's Changed and What Hasn't & \url{https://www.analyticsinsight.net/artificial-intelligence/offshore-software-development-in-the-ai-era-whats-changed-and-what-hasnt} \\
G3017 & 8850 & AI Review Tool Flaws: Adversarial Approach to Reliable Feedback & \url{https://www.linkedin.com/posts/gyadav6_multi-agent-adversarial-code-review-what-activity-7434086222873985024-lxIH} \\
G3018 & 8851 & AI Code Review Impact on Ownership and Productivity | Snir Hadad posted on the topic | LinkedIn & \url{https://www.linkedin.com/posts/snir-hadad_aicodereview-engineeringleadership-developerproductivity-activity-7419402594177122305-rbtK} \\
G3019 & 8854 & 1,200 Legal Hallucination Cases Worldwide and Counting: What the Attorney AI Crisis Reveals About Agent Deployment & \url{https://newclawtimes.com/articles/ai-agents-legal-profession-hallucinations-fake-cases-attorneys-crisis/} \\
G3020 & 8858 & Code Smarts Meet AI: Revolutionizing the Dev Cycle - DEV Community & \url{https://dev.to/mabualzait/code-smarts-meet-ai-revolutionizing-the-dev-cycle-58ab} \\
G3021 & 8859 & AI-generated changes overwhelm review queues, eroding ... & \url{https://www.linkedin.com/posts/robzuber_the-review-queue-isnt-the-problem-its-activity-7440053157667262465-2VhY} \\
G3022 & 8869 & Building an AI-Powered Code Review Agent: A Step-by-Step Engineering Deep Dive & \url{https://www.linkedin.com/pulse/building-ai-powered-code-review-agent-step-by-step-deep-goyal-qvhvc} \\
G3023 & 8872 & FOMO, FOMAT, FOMUT, and the Bottleneck That Turned Out to Be Me - DEV Community & \url{https://dev.to/teppana88/fomat-fomut-and-the-bottleneck-that-turned-out-to-be-me-32g4} \\
G3024 & 8874 & AI Code Review vs Manual Code Review: A Complete Comparison (2026) & \url{https://refacto.ai/blog/ai-code-review-vs-manual-code-review-a-complete-comparison-2026} \\
G3025 & 8881 & The art and science of context engineering & \url{https://www.coderabbit.ai/blog/the-art-and-science-of-context-engineering} \\
G3026 & 8885 & I Spent 6 Months Reviewing AI Code. Here's the Ugly Truth | by Muhammad Haseeb Sohail | Apr, 2026 | Medium & \url{https://medium.com/@haseeb_sohail/i-spent-6-months-reviewing-ai-code-heres-the-ugly-truth-b81fe33e0b25} \\
G3027 & 8886 & How Baz improved its AI Agent Code Review accuracy using Amazon Bedrock AgentCore | Artificial Intelligence & \url{https://aws.amazon.com/blogs/machine-learning/how-baz-improved-its-ai-agent-code-review-accuracy-using-amazon-bedrock-agentcore/} \\
G3028 & 8894 & Smarter Pull Requests: Balancing AI, Automation, and Human Review | by Rethink Your Understanding | Medium & \url{https://medium.com/@rethinkyourunderstanding/smarter-pull-requests-balancing-ai-automation-and-human-review-ebb7d586e968} \\
G3029 & 8895 & The "Junior Developer" Effect: How 192k Tokens of Noise Degraded Gemma 4's Architectural Reasoning - DEV Community & \url{https://dev.to/pvishalkeerthan/the-junior-developer-effect-how-192k-tokens-of-noise-degraded-gemma-4s-architectural-reasoning-24en} \\
G3030 & 8899 & Vibe Coding: When AI Starts to Feel the Code & \url{https://www.linkedin.com/pulse/vibe-coding-when-ai-starts-feel-code-bitsoltech-nqbre} \\
G3031 & 8904 & Why AI Tools Make Some Teams Slower - DEV Community & \url{https://dev.to/brian_zimbelman/why-ai-tools-make-some-teams-slower-14h9} \\
G3032 & 8912 & 4 AI Pitfalls in Vibe Coding and the Future of Development | Meerasab Mukkeri posted on the topic | LinkedIn & \url{https://www.linkedin.com/posts/meerasab-mukkeri-bb7673258_vibecoding-aiengineering-agenticai-activity-7435737593826201600-mxjl} \\
G3033 & 8914 & Anthropic Claude 4.6 Update: Long-Horizon Workflows and AI Agents | Kshitij Datar posted on the topic | LinkedIn & \url{https://www.linkedin.com/posts/kshitij-datar_ai-claude-agents-activity-7426527248834478080-Q7qA} \\
G3034 & 8916 & Autonomous Multi-Agent Code Review and Refinement: Discovering Optimal Strategies Through Iterative Feedback Loops -- clawRxiv & \url{https://clawrxiv.io/abs/2603.00275} \\
G3035 & 8919 & Introducing GPT-5.5 | OpenAI & \url{https://openai.com/index/introducing-gpt-5-5/} \\
G3036 & 8920 & The Review Bottleneck: Why AI Explanations Are Making Us Trust Less, Not More - DEV Community & \url{https://dev.to/pacheco/the-review-bottleneck-why-ai-explanations-are-making-us-trust-less-not-more-35al} \\
G3037 & 8923 & How we built our AI code review agent for IDEs - DEV Community & \url{https://dev.to/coderabbitai/how-we-built-our-ai-code-review-agent-for-ides-4bpf} \\
G3038 & 8925 & CTO Newsletter | April 2026 | Thoughtworks Australia & \url{https://www.thoughtworks.com/en-au/insights/cto-newsletter/edition1} \\
G3039 & 8926 & The Numbers That Deflate the AI Hype About Replacing Coders--and Ease FOMO - DEV Community & \url{https://dev.to/canro91/the-numbers-that-deflate-the-ai-hype-about-replacing-coders-and-ease-fomo-3a0g} \\
G3040 & 8932 & WHAT HAPPENS WHEN AI CAN FIX ANY GITHUB ISSUE? I'm a technical product manager. Last sprint, my backlog vanished overnight. Here's what nobody warned me about.  & \url{https://medium.com/@raghavpunjwork/what-happens-when-ai-can-fix-any-github-issue-072d923f233c} \\
G3041 & 8934 & Best AI Code Review Tools 2026: CodeRabbit vs Claude Code Review vs Qodo vs GitHub Copilot -- Effloow & \url{https://effloow.com/articles/best-ai-code-review-tools-coderabbit-claude-qodo-2026} \\
G3042 & 8941 & Spec-Gated Delivery: Why PR Review Is the Wrong Trust Checkpoint for AI Code - DEV Community & \url{https://dev.to/t3chn/spec-gated-delivery-why-pr-review-is-the-wrong-trust-checkpoint-for-ai-code-1hfg} \\
G3043 & 8942 & Claude Code: AI Agent Harness for High-Performance Tasks & \url{https://www.linkedin.com/posts/abhinandan-shah_agents-activity-7413226548226420736-NGLU} \\
G3044 & 8949 & Self-Distillation Improves Code Generation Quality - DEV Community & \url{https://dev.to/onsen/self-distillation-improves-code-generation-quality-5ij} \\
G3045 & 8955 & How CodeRabbit's Agentic Code Validation helps with code reviews - DEV Community & \url{https://dev.to/coderabbitai/how-coderabbits-agentic-code-validation-helps-with-code-reviews-nak} \\
G3046 & 8956 & DORA Report 2025 Key Takeaways: AI Impact on Dev Metrics & \url{https://www.faros.ai/blog/key-takeaways-from-the-dora-report-2025} \\
G3047 & 8975 & Claude Code in Enterprise Production: What Risks to Control - DEV Community & \url{https://dev.to/arnaud-agent-rail/claude-code-in-enterprise-production-what-risks-to-control-15dm} \\
G3048 & 8978 & I Asked 3 Claude Code Sub-agents to Review the Same PR. They Disagreed on 41\% of the Comments. - DEV Community & \url{https://dev.to/kenimo49/i-asked-3-claude-code-sub-agents-to-review-the-same-pr-they-disagreed-on-41-of-the-comments-2m57} \\
G3049 & 8992 & AI Code Review in Code Studio: Instant Pre-Commit Feedback that Cuts PR Churn - DEV Community & \url{https://dev.to/syncfusion/ai-code-review-in-code-studio-instant-pre-commit-feedback-that-cuts-pr-churn-4g1h} \\
G3050 & 8995 & AI Code Review Benchmark 2026: CodeRabbit vs Cursor vs Baz & \url{https://shmulc.substack.com/p/stop-vibe-merging} \\
G3051 & 8996 & The Complete Guide to AI Coding Agents as Autonomous Engineering Teammates - SoftwareSeni & \url{https://www.softwareseni.com/the-complete-guide-to-ai-coding-agents-as-autonomous-engineering-teammates/} \\
G3052 & 8999 & Cursor 3 Review: Agent Orchestration and SWE-bench Standards | PorkiCoder Blog & \url{https://porkicoder.com/blog/posts/cursor-3-review-agent-orchestration-and-swe-bench-standards.html} \\
G3053 & 9005 & I built a Claude Code plugin that anchors code reviews to 6 classic SE books - DEV Community & \url{https://dev.to/hyhmrright/i-built-a-claude-code-plugin-that-anchors-code-reviews-to-6-classic-se-books-4465} \\
G3054 & 9008 & Spec-Driven Development with AI Coding Agents: The Definitive Guide | by @pramodchandrayan | Predict | May, 2026 | Medium & \url{https://medium.com/predict/spec-driven-development-with-ai-coding-agents-the-definitive-guide-453fba1baf39} \\
G3055 & 9009 & The AI Productivity Paradox: Why Developers Who Ship More Code Are Not Actually More Productive - DEV Community & \url{https://dev.to/alexcloudstar/the-ai-productivity-paradox-why-developers-who-ship-more-code-are-not-actually-more-productive-12f7} \\
G3056 & 9010 & I Let AI Write My Code for 6 Months. It Almost Destroyed My Career. | by Ananya Kaul | Apr, 2026 | Medium & \url{https://medium.com/@ananyakaul/i-let-ai-write-my-code-for-6-months-it-almost-destroyed-my-career-c86d59a0e00d} \\
G3057 & 9016 & Six Agents, One Pull Request, and the Emerging Art of 'Skills-as-Infrastructure' & \url{https://www.thereadreplica.com/news/snap-engineer-ai-code-review-productivity} \\
G3058 & 9022 & Code Review Is Not About Catching Bugs - davidpoll.com & \url{https://www.davidpoll.com/2026/02/code-review-is-not-about-catching-bugs/} \\
G3059 & 9023 & What Broke When I Let AI Handle My Code Reviews (And How I Fixed It) - DEV Community & \url{https://dev.to/leena_malhotra/what-broke-when-i-let-ai-handle-my-code-reviews-and-how-i-fixed-it-25c0} \\
G3060 & 9026 & Nobody Reads Your Code Anymore - Bojan Josifoski & \url{https://bojanjosifoski.com/nobody-reads-your-code-anymore/} \\
G3061 & 9027 & Don't Automate Your Moat: Matching AI Autonomy to Risk and Competitive Stakes - O'Reilly & \url{https://www.oreilly.com/radar/dont-automate-your-moat-matching-ai-autonomy-to-risk-and-competitive-stakes/} \\
G3062 & 9030 & How an AI Code Review Can Solve Inefficiencies in Development & \url{https://www.virtasant.com/ai-today/how-an-ai-code-review-can-solve-inefficiencies-in-development} \\
G3063 & 9034 & Claude Code Security Catches Vulnerabilities While You Write Code - DevOps.com & \url{https://devops.com/claude-code-security-catches-vulnerabilities-while-you-write-code/} \\
G3064 & 9037 & AI Agents in Open Source: Evolving the Contribution Model & \url{https://dev.to/pullflow/ai-agents-in-open-source-evolving-the-contribution-model-40e7} \\
G3065 & 9039 & Leveraging LLMs as an Augmentation to Traditional Hyperparameter Tuning | AWS HPC Blog & \url{https://aws.amazon.com/blogs/hpc/leveraging-llms-as-an-augmentation-to-traditional-hyperparameter-tuning/} \\
G3066 & 9055 & AI-Generated Code Contains 2.74x More Vulnerabilities Than Human Code & \url{https://www.svenroth.ai/post/ai-generated-code-vulnerabilities-2-74x-4c9a7} \\
G3067 & 9056 & AI Code Review Benefits: Security, Speed \& ROI -- Complete Guide 2026 | Git AutoReview & \url{https://gitautoreview.com/guides/ai-code-review} \\
G3068 & 9058 & AI Code Review: Helpful Assistant Or False Confidence Machine? - DEV Community & \url{https://dev.to/nazar_boyko/ai-code-review-helpful-assistant-or-false-confidence-machine-mp2} \\
G3069 & 9059 & Vibe Coding:, & \url{https://www.vista.tw/blog/vibe-coding-natural-language-creation} \\
G3070 & 9060 & sscarduzio/pr-war-stories & \url{https://github.com/sscarduzio/pr-war-stories} \\
G3071 & 9061 & Battle for Context: How We Implemented AI Coding in an Enterprise Project Feature - DEV Community & \url{https://dev.to/apolenkov/battle-for-context-how-we-implemented-ai-coding-in-an-enterprise-project-11hd} \\
G3072 & 9063 & How to Do Code Reviews in the Agentic Era | by Daniela Petruzalek | Google Cloud - Community | Medium & \url{https://medium.com/google-cloud/how-to-do-code-reviews-in-the-agentic-era-0b6584700f47} \\
G3073 & 9064 & I Asked 3 Teams How They Use Cursor. None of Them Use .cursorrules. - DEV Community & \url{https://dev.to/nedcodes/i-asked-3-teams-how-they-use-cursor-none-of-them-use-cursorrules-123g} \\
G3074 & 9067 & in-the-loop-labs/pair-review & \url{https://github.com/in-the-loop-labs/pair-review} \\
G3075 & 9071 & Who Reviews AI-Generated Code Before It Hits Production? & \url{https://www.testingxperts.com/blog/who-reviews-ai-generated-code-before-it-reaches-production/} \\
G3076 & 9075 & OpenAI GPT-5.5: an evaluation | Sonar & \url{https://www.sonarsource.com/blog/openai-gpt-5-5-evaluation} \\
G3077 & 9080 & Trust, Ownership, Feedback, and Excellence | by Tom Mango | Coffee And Code | May, 2026 | Medium & \url{https://medium.com/techtrends-digest/trust-ownership-feedback-and-excellence-6679f4de0252} \\
G3078 & 9083 & We're About to Stop Reading Code. And That's Fine. | by Pravin Paratey | Engineering Management & \url{https://blog.cto.me.uk/were-about-to-stop-reading-code-and-that-s-fine-633da6f4476d} \\
G3079 & 9084 & AI Code Generation: The Hidden Engineering Velocity Trap & \url{https://www.codebridge.tech/articles/ai-code-generation-hidden-engineering-velocity-trap-4036} \\
G3080 & 9087 & NIST 800-171 compliance | Semgrep & \url{https://semgrep.dev/docs/compliance/nist-800-171} \\
G3081 & 9099 & Building a Code Review System That Uses Prod Data to Predict Bugs & \url{https://blog.sentry.io/building-a-code-review-system-that-uses-prod-data-to-predict-bugs/} \\
G3082 & 9101 & How Can a DevOps Team Take Advantage of Artificial Intelligence (AI)? & \url{https://www.linkedin.com/pulse/how-can-devops-team-take-advantage-artificial-intelligence-ai-cgilf} \\
G3083 & 9102 & AI-Generated Code: Do You Review Everything or Trust Everything? & \url{https://zenn.dev/r_kaga/articles/66c190413d3ab9?locale=en} \\
G3084 & 9103 & AI Code Review [2026]: Copilot vs Claude vs Gemini & \url{https://techbytes.app/posts/ai-code-review-2026-copilot-vs-claude-vs-gemini/} \\
G3085 & 9104 & Claude Code is making me a better dev (and I have the data) - DEV Community & \url{https://dev.to/gabrielizalo/claude-code-is-making-me-a-better-dev-and-i-have-the-data-3h73} \\
G3086 & 9105 & We rebuilt code review for how engineers want to work in 2026: infinite scroll and swiping | Augment Code & \url{https://www.augmentcode.com/blog/code-review-swipe-scroll} \\
G3087 & 9108 & Devin Review 2026: Preise, Features \& Alternativen AI Agent Rank & \url{https://aiagentrank.io/de/agent/devin} \\
G3088 & 9112 & The AI Code Review Bottleneck Is Already Here. Most Teams Haven ... & \url{https://levelup.gitconnected.com/the-ai-code-review-bottleneck-is-already-here-most-teams-havent-noticed-1b75e96e6781} \\
G3089 & 9116 & Vibe Coding and Reflections on Trusting Trust - programmium & \url{https://programmium.wordpress.com/2025/11/19/vibe-coding-and-reflections-on-trusting-trust/} \\
G3090 & 9119 & AI That Audits Your Code: Building an Automated Code Review Agent & \url{https://kurkowski.substack.com/p/ai-that-audits-your-code-building} \\
G3091 & 9120 & Unlearning, experimentation and engineering rigor in an agentic world | Thoughtworks United Kingdom & \url{https://www.thoughtworks.com/en-gb/insights/podcasts/technology-podcasts/unlearning-experimentation-engineering-rigor-agentic-world} \\
G3092 & 9133 & NVIDIA commits 3x more code across 30,000 developers with Cursor Cursor & \url{https://cursor.com/blog/nvidia} \\
G3093 & 9135 & Claude Code Now Uses AI Agents to Review Pull Requests: What Engineers Need to Know & \url{https://interviewkickstart.com/blogs/articles/claude-code-now-uses-ai-agents-to-review-pull-requests-what-engineers-need-to-know} \\
G3094 & 9136 & I let my AI agents be advisory-only. Here's the rules-first PR risk engine I shipped instead. - DEV Community & \url{https://dev.to/zoetaka38/i-let-my-ai-agents-be-advisory-only-heres-the-rules-first-pr-risk-engine-i-shipped-instead-5c6k} \\
G3095 & 9137 & AI Performance Review: Claude's Workflow Issues Resolved | Chris Goodwin posted on the topic | LinkedIn & \url{https://www.linkedin.com/posts/hicmndr_claude-code-started-going-off-the-rails-the-activity-7421554538513620992-cLyx} \\
G3096 & 9144 & GitHub Copilot Code Review Is Probably Better Than Your Team & \url{https://drpicox.medium.com/github-copilot-code-review-is-probably-better-than-your-team-816c3de54b86} \\
G3097 & 9145 & dhruv1794/CodeDojo & \url{https://github.com/dhruv1794/CodeDojo} \\
G3098 & 9146 & Vibe Coding Explained: What It Is, Where It Came From, and What It Means for Developers | Abhishek Gautam & \url{https://www.abhs.in/blog/vibe-coding-explained-what-it-is-future-of-development} \\
G3099 & 9149 & How to Review Cursor-Generated Code Before Shipping to Production -- Vibers Blog & \url{https://onout.org/vibers/blog/review-cursor-generated-code/} \\
G3100 & 9150 & Your project has .gitignore -- where's your .rules/? - DEV Community & \url{https://dev.to/divcozy/your-project-has-gitignore-wheres-your-rules-4fmn} \\
\end{longtable}

\bibliographystyleapp{IEEEtran}
\bibliographyapp{sabib}

\end{document}